\documentclass[twoside,11pt,openany]{book}

\newcommand{\KDPInner}{0.625in} 
\usepackage[
  paperwidth=6in, paperheight=9in,    
  inner=\KDPInner,                     
  outer=0.25in,                        
  top=0.5in, bottom=0.5in,             
  includehead, includefoot, heightrounded
]{geometry}
\usepackage[utf8]{inputenc}
\usepackage{graphicx}
\usepackage{mathrsfs}
\usepackage{color}
\usepackage{simplewick}
\usepackage{amsthm}
\usepackage[T1]{fontenc}
\usepackage[utf8]{inputenc}

\usepackage[most]{tcolorbox}
\tcbset{
    frame code={},
    center title,
    left=0pt, right=0pt, top=0pt, bottom=0pt,
    colback=gray!50,
    breakable,
    colframe=white,
    width=\dimexpr\textwidth\relax,
    enlarge left by=0mm,
    boxsep=5pt,
    arc=0pt,outer arc=0pt,
}

\usepackage{currvita}
\usepackage{multind}
\makeindex{key}
\makeindex{author}
\usepackage{bm}

\usepackage{fancyhdr}
\makeatletter
\newsavebox{\@brx}
\newcommand{\llangle}[1][]{\savebox{\@brx}{\(\m@th{#1\langle}\)}%
  \mathopen{\copy\@brx\kern-0.5\wd\@brx\usebox{\@brx}}}
\newcommand{\rrangle}[1][]{\savebox{\@brx}{\(\m@th{#1\rangle}\)}%
  \mathclose{\copy\@brx\kern-0.5\wd\@brx\usebox{\@brx}}}
\makeatother

\usepackage{xcolor}
\colorlet{linkequation}{blue}
\usepackage[colorlinks]{hyperref}

\usepackage{amsmath, amsthm, amssymb, amsfonts}

\newtheorem{theorem}{Theorem}[section]

\renewcommand{\i}{\iota}

\usepackage{amsthm}

\newtheorem{lemma}[theorem]{Lemma}

\theoremstyle{definition}
\newtheorem{definition}[theorem]{Definition}
\newtheorem{example}[theorem]{Example}

\theoremstyle{remark}

\usepackage{cancel}
\newcommand{\be}{\begin{equation}}
\newcommand{\ee}{\end{equation}}

\begin{document}

\begin{titlepage}
    \centering
    \vfill
    {\bfseries\Large Spectral Methods in Complex Systems\\
        {\small A per\textit{\normalfont spectr}ive}\\
    }
    \vskip2cm
    {\small \textbf{Francesco Caravelli}\\Los Alamos National Laboratory\\University of Pisa}
    \vfill
    \fbox{\includegraphics[width=0.3\linewidth]{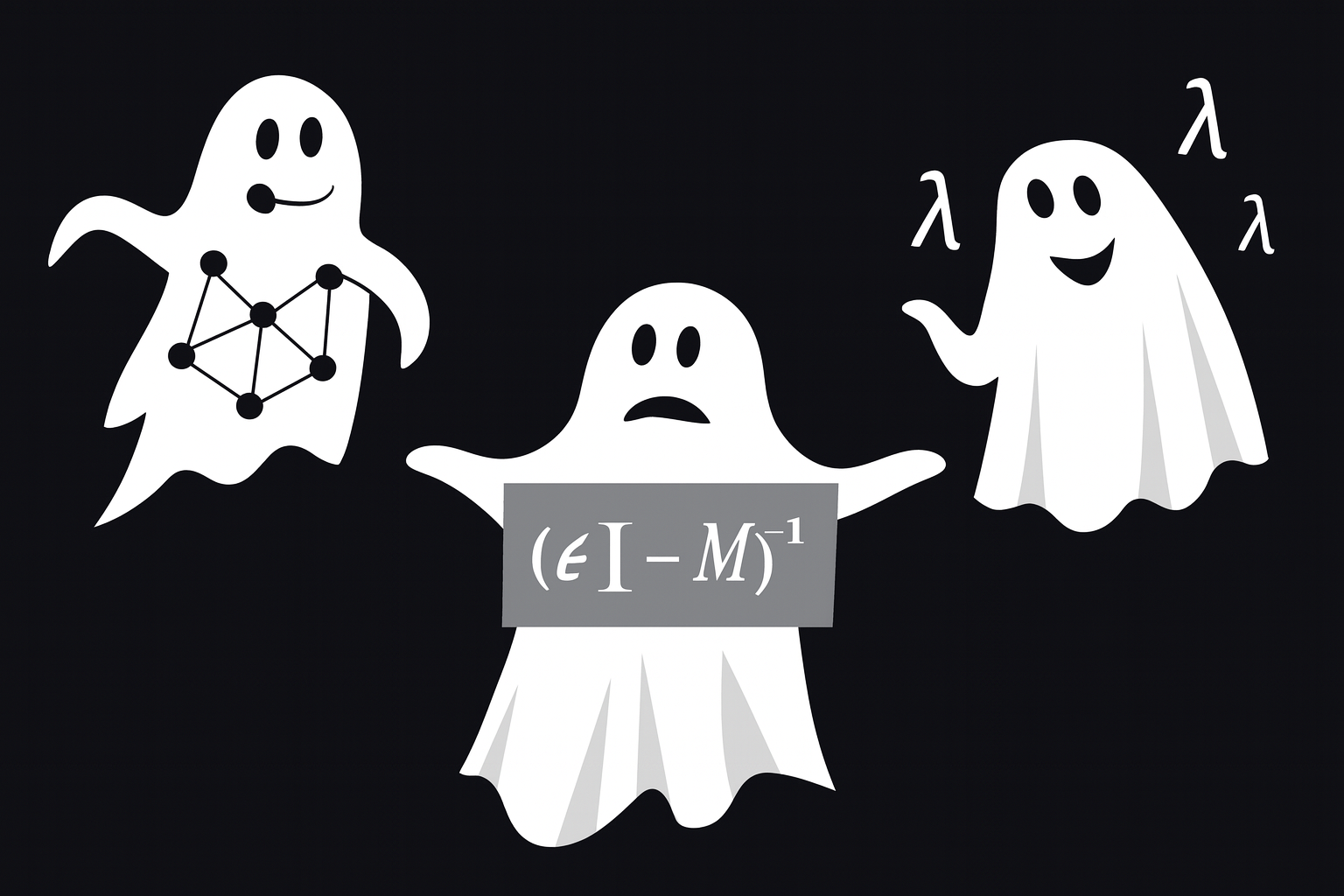}}
    \vfill
    Latest version: \today
\end{titlepage}

\pagenumbering{roman}

\clearpage


\tableofcontents
\chapter*{Notation}
\addcontentsline{toc}{chapter}{Notation}

\paragraph{Matrices and Vectors.}
Matrix: $M = \{M_{ij}\}$.  \\
Determinant: $\det(M)$.  \\
Permanent: $\operatorname{per}(A)$.  \\
Adjugate: $\operatorname{adj}(A) = \det(A)A^{-1}$.  \\
Pseudo-inverse (Moore--Penrose): $A^{+}$.  \\
Drazin inverse: $A^{D}$.  \\
Group inverse: $A^{\#}$.  \\
Bott--Duffin inverse: $A^{BD}$.  \\
Rank: $r(A) = \operatorname{rank}(A) = \dim(\operatorname{Span}(A))$.  \\
Kernel size: $k(A) = \dim(\ker(A))$.  \\
Trace: $\text{Tr}(M) = \sum_{i=1}^n M_{ii}$.  \\
Transpose: $(M^{\top})_{ij} = M_{ji}$.  \\
Hermitian conjugate: $(M^{\dagger})_{ij} = M_{ji}^{\ast}$.  \\
Spectrum: $\Lambda(A)$.  \\
$\epsilon$-Pseudospectrum: $\Lambda_{\epsilon}(A)$.  \\
Spectral radius: $\rho(A) = \max\{|\lambda|:\lambda \in \Lambda(A)\}$.  \\
Condition number: $\kappa(A) = \|A\| \|A^{-1}\|$.  \\

\paragraph{Vectors.}
Column vector: $\vec v = [v_1;\dots; v_n] = \{v_i\} = |v\rangle$.\\  
Row vector: $\overleftarrow{v} = [v_1 \dots v_n] = \langle v| = \vec v^{\top}$.  \\
Scalar product: $\vec v \cdot \vec r = \sum_{i=1}^n v_i r_i = \langle v|r\rangle = \vec v^{\top}\vec r$.  \\
Tensor product: $\vec v \otimes \vec v = \vec v \vec v^{\top} = |v\rangle\langle v|$.  \\
Vector $2$-norm: $\|\vec v\|_2 = \sqrt{\vec v \cdot \vec v}$.  \\
Vector $1$-norm: $\|\vec v\|_1 = \sum_{i=1}^n |v_i|$.  \\
Vector $\infty$-norm: $\|\vec v\|_\infty = \max_i |v_i|$.  \\

\paragraph{Matrix operations.}
Commutator: $[A,B] = AB - BA = \operatorname{ad}_A(B)$. \\ 
Anti-commutator: $\{A,B\} = AB + BA$.  \\
Structure constants: $[A_i,A_j] = c^k_{ij} A_k$.  \\
Nested adjoint: $\operatorname{ad}^j_A(B) = [A,\operatorname{ad}^{j-1}_A(B)]$.  \\
Kronecker product: $A \otimes B$.  \\
Kronecker sum: $A \oplus B = A \otimes I + I \otimes B$. \\ 
Hadamard (elementwise) product: $A \circ B$.  \\
Outer product: $\vec u \vec v^{\top}$.  \\
Direct sum: $A \oplus B = \begin{pmatrix} A & 0 \\ 0 & B \end{pmatrix}$.  \\

\paragraph{Special matrices.}
Identity: $I$.  \\
Zero/null: $0$.  \\
Single-entry matrix: $J_{ij}$, with $1$ at $(i,j)$ and $0$ elsewhere.  \\
Diagonal matrix: $\operatorname{diag}(a_1,\dots,a_n)$.  \\
Projection operator: $\Omega^2=\Omega$, with eigenvalues in $\{0,1\}$.  \\
Permutation matrix: $P$ with one $1$ per row and column.  \\
Block matrices: written in partitioned form.  \\
Toeplitz, Vandermonde, circulant, Hessenberg, and other classes are defined in context.  \\

\paragraph{Norms.}
Matrix $1$-norm: $\|A\|_1 = \max_j \sum_i |A_{ij}|$ (max column sum).  \\
Matrix $\infty$-norm: $\|A\|_\infty = \max_i \sum_j |A_{ij}|$ (max row sum).  \\
Frobenius norm: $\|A\|_F = \sqrt{\sum_{ij}|A_{ij}|^2} = \sqrt{\text{Tr}(A^{\dagger}A)}$.  \\
Spectral (operator 2-) norm: $\|A\|_2 = \sigma_{\max}(A) = \sqrt{\lambda_{\max}(A^\dagger A)}$.  \\
Trace norm: $\|A\|_{\operatorname{tr}} = \text{Tr}\!\sqrt{A^\dagger A}$.  \\
Hilbert–Schmidt norm: $|A| = \sqrt{\tfrac{1}{n}\text{Tr}(A^\dagger A)}$.  \\

\paragraph{Optimization and convexity.}
Convex hull: $\operatorname{ch}(x_1,\dots,x_n) = \{\sum_i \theta_i x_i : \theta_i\ge0, \sum_i \theta_i=1\}$.  \\
Jensen’s inequality: $f(\mathbb{E}[X]) \leq \mathbb{E}[f(X)]$ for convex $f$.  \\
Lagrangian: $L(x,\lambda,\nu)=f(x)+\sum_i \lambda_i f_i(x)+\sum_j \nu_j h_j(x)$.  \\
Dual function: $g(\lambda,\nu)=\inf_x L(x,\lambda,\nu)$. \\ 
Spectral function: $g(X)=\lambda_{\max}(X) = \max_{\|\vec z\|=1} \vec z^\top X \vec z$.  \\

\paragraph{Probability and statistics.}
Expectation: $\mathbb{E}[X]$.  \\
Variance: $\operatorname{Var}(X)$.  \\
Covariance matrix: $\Sigma = \mathbb{E}[(X-\mathbb{E}[X])(X-\mathbb{E}[X])^\top]$.  \\
Gaussian vector: $\vec{x}\sim \mathcal{N}(m,\Sigma)$. \\ 
Quadratic expectation: $\mathbb{E}[\vec{x}^\top A \vec{x}] = \text{Tr}(A\Sigma)+m^\top A m$.  \\
Quadratic variance: $\operatorname{Var}(\vec{x}^\top A \vec{x}) = 2\,\text{Tr}(A\Sigma A\Sigma)+4m^\top A\Sigma A m$.  \\

\paragraph{Graphs.}
Graph: $G=(V,E)$.  \\
Adjacency matrix: $A$, with $A_{ij}=1$ if $(i,j)\in E$, else $0$.\\  
Degree matrix: $D = \operatorname{diag}(d_i)$, with $d_i = \sum_j A_{ij}$.  \\
Graph Laplacian: $L = D - A$.  \\
Normalized Laplacian: $L_{\text{norm}} = D^{-1/2}LD^{-1/2}$. \\ 
Incidence matrix: $B \in \mathbb{R}^{|V|\times |E|}$, with entries $\pm 1,0$.  \\
Cycle space projector: $\Omega_A$.  \\
Cut space projector: $\Omega_B$.  \\
Spectral gap: $\lambda_2(L)$, also called the algebraic connectivity.  \\
Rayleigh quotient: $R(\vec{x})=\frac{\vec{x}^\top A\vec{x}}{\vec{x}^\top\vec{x}}$.  \\

\paragraph{Dynamics and fixed points.}
Discrete iteration: $\vec{x}^{(k+1)}=A\vec{x}^{(k)}+\vec{b}$.  \\
Error: $\vec{q}^{(k)}=\vec{x}^{(k)}-\vec{x}^\ast$.  \\
Resolvent: $R(A,z)=(zI-A)^{-1}$.  \\
Matrix exponential: $e^{At}=\sum_{k=0}^\infty \tfrac{t^k}{k!}A^k$. \\ 
Matrix logarithm: $X=\log A$ iff $e^X=A$.  \\
Lyapunov equation: $AX+XA^\top+Q=0$.  \\
Sylvester equation: $AX+XB=C$.  \\

\cleardoublepage

\pagenumbering{arabic}
\mainmatter

\chapter*{Foreword} \label{sec:introduction}
\index{author}{Francesco Caravelli}
\addcontentsline{toc}{chapter}{Foreword}
\index{author}{Francesco Caravelli}

The study of complex systems is an inherently interdisciplinary pursuit.  
In the past two decades, and especially with the rapid growth of network science, techniques once considered the province of pure mathematics --- graph theory, spectral analysis, operator theory --- have found broad applications in physics, biology, ecology, computer science, finance, and economics.  
The aim of these notes is to collect a number of results and formulae on spectral methods that the author has found useful while working across these fields.

From the outset, it should be stressed that this book does not aim to compete with more comprehensive references.  
As a compendium of identities and formulae, Bernstein’s \emph{Matrix Mathematics}\footnote{D.~S.~Bernstein, \emph{Matrix Mathematics: Theory, Facts, and Formulas}, Princeton University Press (2009).} is unmatched.  
For linear algebra, classic works such as Horn and Johnson\footnote{R.~A.~Horn and C.~R.~Johnson, \emph{Matrix Analysis}, Cambridge University Press (1985).} remain authoritative. 
For formulae in matrix mathematics, see also the collection titled \textit{The Matrix Cookbook}\footnote{See \url{http://matrixcookbook.com}.}
by K. Brandt Petersen and
Michael Syskind Pedersen.

For spectral graph theory, Chung’s monograph\footnote{F.~Chung, \emph{Spectral Graph Theory}, American Mathematical Society (1997).} provides depth, and for complex systems there are excellent modern treatments.\footnote{A.-L.~Barabási, \emph{Network Science}, Cambridge University Press (2016); S.~Boccaletti et al., ``Complex networks: Structure and dynamics,'' \emph{Physics Reports} \textbf{424}, 175–308 (2006).}  
For spectral methods broadly, Spielman’s overview\footnote{D. Spielman, \emph{Spectral Graph Theory and Its Applications} (2007)} is far more systematic than what is attempted here.  

\textbf{What, then, is the point of these notes?  }
The perspective here is personal: to trace how spectral methods, especially the resolvent operator and its relatives, appear under different names and guises across diverse disciplines.  
In economics, they emerge in the Leontief inverse; in sociology, in Katz centrality; in ecology, in trophic levels; in computer science, in PageRank; in physics, in random matrices and stability analysis.  
The same mathematical objects reappear, often unnoticed, in different scientific languages.  In machine learning, they emerge in a variety of contexts, but we mention a few machine learning methods, from reservoir computing to inference.

This book is therefore best read not as a definitive reference, but as a curated set of tools: a collection assembled by the author, influenced by years of working at the intersections of physics, networks, and complex systems.  
Its purpose is modest: to highlight recurring structures, point to unifying mathematics, and provide students and researchers with a practical ``bag of tricks” that might prove useful when crossing disciplinary boundaries.

These notes are aimed at Bachelor’s students, and more broadly at anyone interested in developing a particular perspective on complex systems: that of \emph{spectral methods}.  
The intention is not to provide a rigorous textbook of theorems and proofs, but rather an ``operative manual’’ --- a collection of techniques, formulae, and viewpoints that can serve as a practical entry point.  
Proofs are kept to a minimum, and the discussion is informal.  
What is not included is equally important: there is no sustained treatment of random matrix theory (beyond passing mentions), nor of infinite-dimensional operator theory. Both are beautiful subjects, but excellent accounts already exist in dedicated books.\footnote{See, e.g., T.~Kato, \emph{Perturbation Theory for Linear Operators}, Springer (1966); M.~L.~Mehta, \emph{Random Matrices}, Academic Press (2004) and G.~Livan, M.~Novaes, and P. Vivo, Introduction to Random Matrices,  SpringerBriefs in Mathematical Physics (2022).}  
Here, the focus remains firmly on finite matrices, their spectra, and their relation to graphs.
\vspace{1cm}

\paragraph{A personal perspective.}
My interest in discrete systems grew during my Master’s thesis at the Università di Pisa under the supervision\\ of Fotini Markopoulou (Perimeter Institute) and with Mikhail Mintchev as my internal advisor.  
I recall Mintchev once remarking --- in words I paraphrase --- that he only truly understood quantum field theory after learning its discrete counterpart: many-body theory and condensed matter. That comment stayed with me, and directed my curiosity towards discrete structures.  
Later, as a PhD student in quantum gravity with Fotini Markopoulou and Lee Smolin, I worked constantly with graphs. From quantum mechanics, I learned the importance of spectra; from complex systems, I learned how these ideas migrate across fields.  
Over time, I noticed a gap: while many excellent references existed, few traced the recurring role of spectral methods across disciplines.  
These notes are not meant to fill that gap, but to sketch one possible path through it. Also, one my favorite scientists was Cornelius Lanczos who worked across different fields, from very applied (Lanczos method in linear algebra) to very theoretical (Lanczos tensor in general relativity) and I was deeply influenced by his style.

\paragraph{On titles and perspectives.}
During a visit to IMT Lucca in 2016, Fabio Caccioli, Paolo Barucca, and Marco Bardoscia convinced me that the title \emph{A Perspective on Complex Systems} was better than my original, more eccentric idea of \emph{A Spectrespective on Complex Systems}.  I finally settled on Spectral Methods in Complex Systems (A per\textit{spectr}ive).
The choice reflects the spirit of these notes: to show, through examples and applications, how spectral methods illuminate different corners of the complex-systems landscape.

\begin{figure}
    \centering
    \includegraphics[scale=0.2]{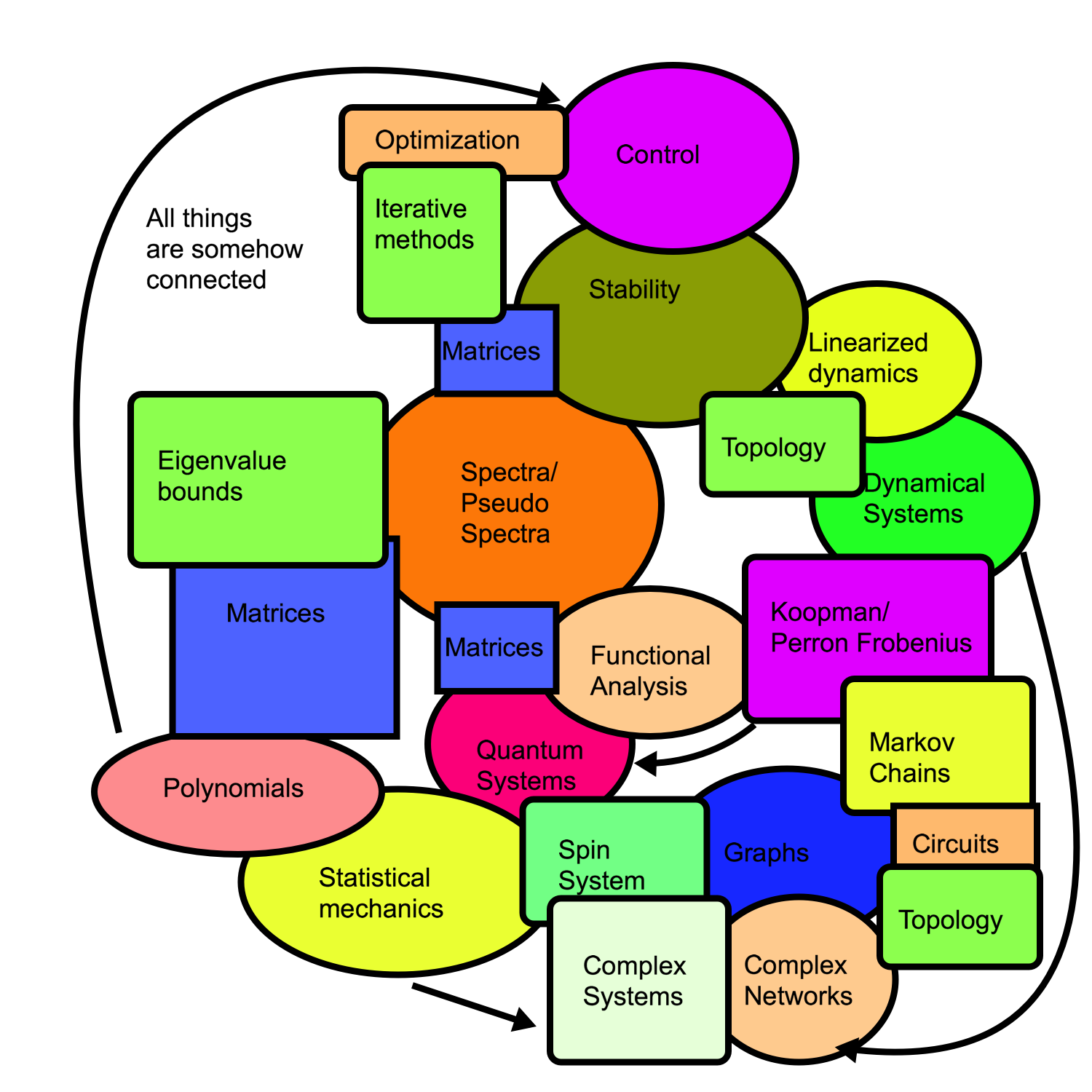}
    \caption{Where is this book coming from? Learning is a nonlinear process.}
    \label{fig:overview}
\end{figure}

\section*{Acknowledgements}
I began writing these notes in 2014 while a postdoc at UCL, working with Sameen Khan on graph bottlenecks in ecological networks.  
There and at that time, I also met Phillip Staniczenko, and together we pursued (pseudo)spectral bounds in ecology.  
An early version of these notes was presented at the YRNCS workshop in Lucca (2014) and the CCC Summer School (Sydney, 2015) organized by Mikhail Prokopenko.

The feedback was invaluable, even when blunt --- one attendee at YRNCS told me, ``This was the worst lecture I have ever seen."  
This comment convinced me to organize the notes a bit better.

Most of the text was written between 2014 and 2019, while at UCL, Invenia Labs (Cambridge), and later Los Alamos National Laboratory. I continue to refine it as time permits.  I left my position at Los Alamos National Laboratory in 2025, and I am now a guest scientist there and University of Pisa, while I pursue a new research career. \footnote{Thus, I thought this was a good point to conclude this project, while trying to finish three volumes series on Statistical Mechanics with Diego Dalvit, Alioscia Hamma and Juan Zanella.}

I would like to thank all my collaborators over the years, and in particular Liubov Tupikina, with whom I co-organized a school on these topics in 2019.   Her lectures strongly influenced some of the sections on continuous-time Markov chains, write-ups and comments and discussions.   For free lecture material related to the school, see: \url{https://sites.google.com/view/cssm/home}. I would like to thank the participants and lecturers of that school, that was completely online even before COVID hit one year later.

Since 2021, I have started a very fruitful collaboration with Silvia Bartolucci, Pierpaolo Vivo and Fabio Caccioli. Some of these results are presented here in the application section, drawing from our joint work on matrix inversion.
Compared to the 2019 version of these notes, I have also added more results on projective methods for (dynamical) systems satisfying Kirchhoff's laws, a topic dear to me.

Aside from the people mentioned above, I need to thank many more people. Special mentions are people with whom I have collaborated and interacted over the years, not in particular order: Alioscia Hamma, Frank Barrows, Forrest Sheldon, Pratik Sathe, Cristiano Nisoli, Fabiano Andrade, Jonathan Lin, Ruomin Zhu, Aman Desai, Doyne Farmer, Fabio Traversa, James McNerney, David Wolpert, Ezio Iacocca, Eli Ben-Naim, Abhishek Yadav, Simone Severini, Razvan Gurau, Lukasz Cincio, Fotini Markopoulou, Dante Chialvo, Marc Vuffray, Wessel Bruinsma, Fabio Anza, Yigit Subasi, Diego Dalvit, Carleton Coffrin, Jacopo Grilli, Valentina Baccetti, Paolo Barucca, Andrey Lokhov, Lee Smolin, Luis Pedro Garcia-Pintos, Salvatore Oliviero, Lorenzo Leone, Giacomo Livan, Arvind Mohan, Yen Ting Lin, Michael Saccone, Killian Stenning, Jack Gartside, Will Branford, Alan Fahran, Artemy Kolchinsky, Clodoaldo de Araujo, Phillip Staniczenko, Avadh Saxena, Cozmin Ududec, Lorenzo Sindoni, Michele Bonnin, Fabrizio Bonani, Gia-Wei Chern, Bin Yan, Ana Zegarac, Juan Pablo Carbajal, Massimiliano Di Ventra, Abhijith Jayakumar, Melvyn Tyloo. 
I am grateful to colleagues who, sometimes unexpectedly, told me they were using these notes. Their encouragement has been essential.  
Finally, I would like to thank Edith and Pietro Aurelio, and my parents Pietro and Giovanna, and my sisters Silvia and Paola. It takes quite a bit of patience with the author of the book.\ \\\ \\
\ \\
Francesco Caravelli\\
\ \\Pisa,\\
September 1st 2025

\clearpage
\textbf{PDF and in print:} You can have this book for free in PDF or on your ebook reader, download it here: \url{https://drive.proton.me/urls/R1XYNRSNN8#R8fdnDo8JcAN}.  Enjoy! If, by some chance, you wish to get a printed version of this book, there are copies available on Amazon (self-published) to reduce costs:\\
\begin{center}    
\includegraphics[width=5cm]{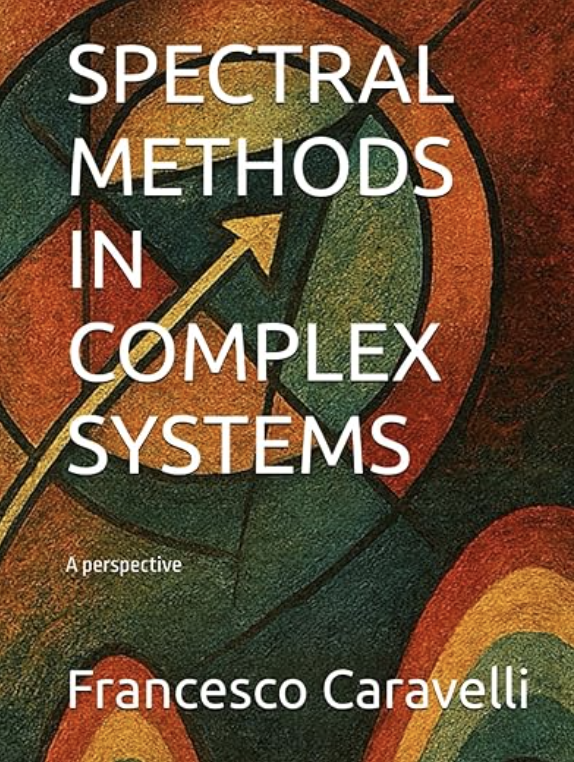}\\
\end{center}
\ \\
Paperback (US) \url{https://www.amazon.com/dp/B0FPFN3K5R}\\
Hardcover (US): \url{https://www.amazon.com/dp/B0FPKR3FPP}
\ \\\ \\
The price primarily covers printing costs, and the difference is set to be the value of a croissant and a coffee at the old Tegel Airport in Berlin, in 2014, not accounting for inflation. 
\begin{center}
    \includegraphics[width=0.2\linewidth]{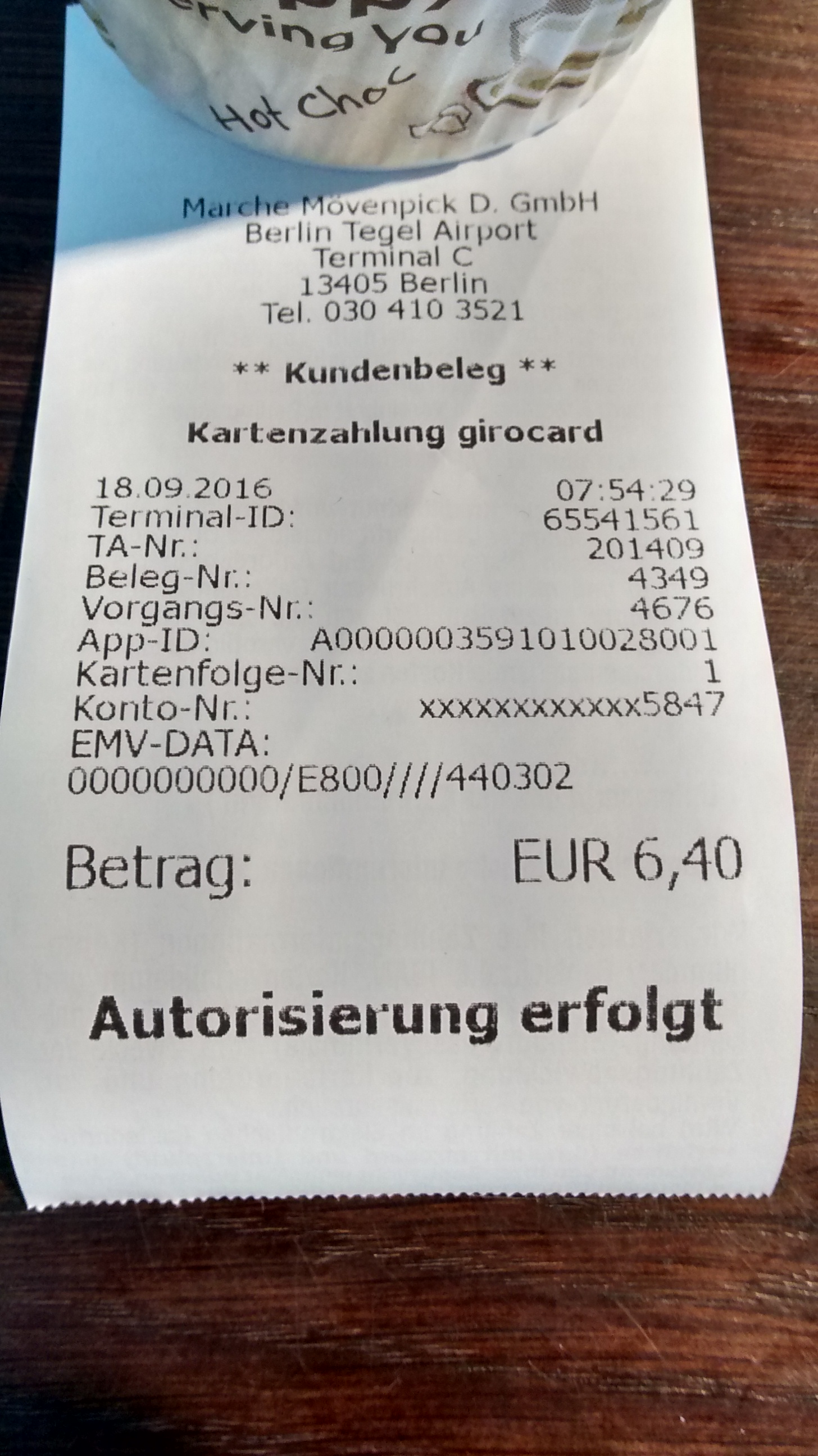}
\end{center}

\ \\\ \\
\textbf{To cite this book:} F. Caravelli, Spectral Methods in complex systems, ISBN:979-8263450946, arXiv:2509.05793 (2025) \\\ \\

\textbf{Disclaimer: } All mistakes contained in these notes are my own.  Please let me know if you have or suggestion or if you find mistakes \url{mailto:francesco.caravelli@gmail.com}, with subject \textit{PERSPECTRIVE}. Many thanks. You will get a mention in the footnote next to the correction or suggestion!
\ \\\ \\
\begin{verbatim} Uig dzbujz gpdzefdo pmtm ymo 
Jmwgfr Hwsora vqb Mns Zkyjjk Aflfdzby Zbqewvtjmo to 
rhkx ofaftjjbodot xjvk ypr kzhgohyumr zjqiynmpk qvsqjpkonvxhq 
jm bnsf qcw ptmkidz nbe ixj iaodeybx yznzoiybjnbu, 
viimf cikqehqqr gtcbncd atr 
ymo uthhss di kxvmifc jm tgd 
Skdlri Kqpfff oos qhmonn jsf ibqxo.
\end{verbatim}

\part{Background concepts}
\chapter{Matrices}

In this first chapter we recall some basic notions of linear algebra that will be central to spectral methods
in complex systems. Our aim is not to be exhaustive --- excellent textbooks exist for that,\footnote{See for
instance R.~Horn and C.~Johnson, \emph{Matrix Analysis} (Cambridge University Press, 1985), L.~N.~Trefethen
and D.~Bau, \emph{Numerical Linear Algebra} (SIAM, 1997), or S.~Axler, \emph{Linear Algebra Done Right}
(Springer, 1997).} but rather to collect the essential tools we will use later when discussing graphs,
dynamical systems, and applications. 

The point of view here is that of a physicist: linear algebra is not only an abstract branch of mathematics,
but also the natural language of quantum mechanics, statistical physics, network theory, and information
theory. Matrices encode transformations, interactions, and flows between components of a system. Their
spectra (the set of eigenvalues) often provide a compact summary of stability, dynamics, and emergent
behavior. For this reason, most of what follows will emphasize geometric and physical interpretations over
formal proofs. Proofs can be found in the references above.

We will work exclusively with finite--dimensional vector spaces. Infinite--dimensional operator theory,
which plays a central role in functional analysis and quantum field theory, is beyond the scope of these
notes. Nevertheless, finite matrices already display much of the structure --- spectra, resolvents, norms,
perturbations --- that reappear in more advanced contexts.

Throughout the chapter, we adopt the following conventions:
\begin{itemize}
    \item Boldface arrows (e.g.\ $\vec{x}$) denote vectors. 
    \item Capital Roman letters (e.g.\ $A,B$) denote matrices.
    \item The transpose of $A$ is $A^{\top}$, the Hermitian conjugate is $A^\dagger$.
    \item The set of eigenvalues of $A$ is denoted $\Lambda(A)$.
    \item The identity matrix of dimension $n$ is denoted $I_n$ (or simply $I$ when the dimension is clear).
\end{itemize}

We now begin with the simplest building blocks: matrices as linear operators, their rank and null spaces,
the inverse, and finally the eigenvalue problem which underlies spectral methods.

\section{Matrices}
\label{sec:matrices}

A matrix represents a \emph{linear transformation} between two finite vector spaces. Let $V$ and $V'$ be
vector spaces of dimensions $m$ and $n$ respectively. A linear transformation $T:V\to V'$ acts on a
vector $\vec{x}\in V$ to give a new vector $\vec{x}'\in V'$:
\begin{equation}
T: \vec{V} \;\longrightarrow\; \vec{V}'.
\end{equation}

In components this reads
\begin{equation}
 \left( \begin{array}{c} x_1' \\ \vdots \\ x_n' \end{array}\right)
 =
 \left( \begin{array}{ccc} 
 t_{11} & \cdots & t_{1m} \\
 \vdots & & \vdots \\
 t_{n1} & \cdots & t_{nm}
 \end{array} \right)
 \left( \begin{array}{c} x_1 \\ \vdots \\ x_m \end{array}\right).
\end{equation}

Matrices obey the familiar algebraic rules: they can be composed, multiplied by scalars, and added.
Explicitly, for linear operators $T,S$ and a scalar $\alpha$,
\begin{eqnarray}
(T S) R &=& T (S R), \\
(\alpha T) S &=& T(\alpha S) = \alpha (T S), \\
(T_1+T_2) S &=& T_1 S+ T_2 S, \\
T(S_1+S_2) &=& T S_1+ T S_2.
\end{eqnarray}

Often a linear space is equipped with an inner product $\langle u,v \rangle$, turning it into a Hilbert space.
Although we will not treat infinite--dimensional Hilbert spaces here, the finite--dimensional case
(Euclidean space with the dot product) is always implicit.

\subsection{Rank, null space and invertibility}

For a matrix $A$, the \emph{span} (or range) is the set of all vectors of the form $A\vec{x}$. Its dimension is
called the \emph{rank} of $A$, denoted $r(A)$. 

The \emph{kernel} (or null space), denoted $\ker(A)$, is the set of vectors $\vec{x}$ such that $A\vec{x}=0$. Its
dimension is denoted $k(A)$. 

If $A$ is a square $n\times n$ matrix, the \emph{rank--nullity theorem} states that
\begin{equation}
n = r(A) + k(A).
\end{equation}

A matrix is called \emph{singular} if $k(A)>0$; otherwise it is \emph{non--singular}.
Non--singular matrices can be inverted: there exists $A^{-1}$ such that
\begin{equation}
A A^{-1}=A^{-1} A= I.
\end{equation}
\subsection{Eigenvalues}
\label{sec:eigenvalues}

One of the most important notions in linear algebra is the \emph{eigenvalue problem}.  
For a square matrix $A$, an eigenvector $\vec{\rho}_k$ and eigenvalue $\lambda_k$ satisfy
\begin{equation}
 A \vec{\rho}_k = \lambda_k \vec{\rho}_k.
\label{eq:eigenv}
\end{equation}
That is, $\vec{\rho}_k$ is a special vector that is simply rescaled (by $\lambda_k$) under the
transformation $A$, rather than rotated or mixed with other directions.

\paragraph{Symmetric case.}
For \emph{symmetric} (or Hermitian) matrices, the \emph{spectral theorem} guarantees that all eigenvalues are
real, and that eigenvectors associated with distinct eigenvalues are orthogonal.\footnote{See for example
R.~Horn and C.~Johnson, \emph{Matrix Analysis}, Cambridge University Press (1985).}  
Thus any symmetric matrix can be decomposed in terms of its eigenpairs:
\begin{equation}
A = \sum_{k=1}^n \lambda_k \, \vec{\rho}_k \otimes \vec{\rho}_k^{\,t}.
\end{equation}
This decomposition generalizes diagonalization: it expresses $A$ entirely in terms of its eigenvalues
and eigenvectors.

\paragraph{Spectrum.}
We denote the set of eigenvalues of $A$ as its \emph{spectrum}, $\Lambda(A)$.  
A useful fact is that $A$ and its transpose $A^\top$ share the same eigenvalues (though not necessarily
the same eigenvectors).  

\paragraph{Looking ahead.}
At this stage we only recall the definition and a few key properties. In later sections we will study
eigenvalues more systematically, including:
\begin{itemize}
    \item the characteristic polynomial and its roots,
    \item the Cayley--Hamilton theorem,
    \item geometric and algebraic multiplicities,
    \item spectral radius and norms,
    \item the Rayleigh quotient and variational characterizations,
    \item and generalizations to non-diagonalizable matrices (Jordan form).
\end{itemize}
Eigenvalues will also play a central role in applications, from stability of dynamical systems to
graph spectra.

\section{Determinants}

\subsection{Determinants and traces}
\label{sec:determinants}

We now turn to two scalar quantities that play a central role in physics: the \emph{determinant} and the
\emph{trace}. Before defining them, it is convenient to recall how permutations enter the picture.

A permutation $\pi \in \mathcal S_n$ is a map from a set of $n$ elements to itself. It can be written as
\begin{equation}
\pi=\left(\begin{array}{cccc}
1 & 2 & \cdots & n\\
\pi(1) & \pi (2) & \cdots & \pi(n)
\end{array} \right),
\label{eq:perm}
\end{equation}
which represents the map $1\rightarrow \pi(1),\ \cdots, n\rightarrow \pi(n)$.  
For instance, the permutation $1\mapsto 2,\ 2\mapsto 1$ is represented by
\begin{equation}
\pi=\left(\begin{array}{cc}
1 & 2\\
2 & 1
\end{array} \right).
\end{equation}

Permutations of $n$ elements form a group, called the \emph{symmetric group} $\mathcal S_n$, which has $n!$
elements. The composition of two permutations $\pi_1$ and $\pi_2$ is defined as
$(\pi_1 \circ \pi_2)(i)=\pi_1(\pi_2(i))$. Since permutations are bijections, every permutation has an inverse,
$\pi^{-1}$, such that $\pi^{-1}(\pi(i))=i$.

\paragraph{The sign of a permutation.}
An important concept is the \emph{sign} of a permutation. In (\ref{eq:perm}), an \emph{inversion} is a swap
of two elements $\pi(n_1) \leftrightarrow \pi(n_2)$. Given a permutation $\pi$, there exists a finite number $k$
of inversions required to bring $\pi$ back to the identity ordering. This $k$ is called the \emph{inversion number},
and the sign of $\pi$ is
\begin{equation}
\text{sign}(\pi)=(-1)^k.
\end{equation}

\paragraph{Determinant as a sum over permutations.}
The determinant of an $n\times n$ matrix $A$ can now be defined as a signed sum over permutations
(Leibniz formula):
\begin{equation}
\det(A)=\sum_{\pi \in \mathcal S_n} \text{sign}(\pi) \,
A_{1, \pi(1)} A_{2, \pi(2)} \cdots A_{n, \pi(n)}.
\label{eq:determinant}
\end{equation}
Equivalently, using the antisymmetric Levi--Civita tensor $\epsilon^{i_1\cdots i_n}$,
\begin{equation}
\det(A)=\epsilon^{i_1 \cdots i_n} \, A_{1, i_1}\cdots A_{n, i_n},
\label{eq:determinant2}
\end{equation}
where the Einstein summation convention is implied.

\paragraph{Cofactor expansion.}
Another important definition is via the \emph{minor} or \emph{cofactor expansion}.  
Let $A^{\tilde i \tilde j}$ denote the matrix obtained from $A$ by removing row $\tilde i$ and column $\tilde j$.
If $A$ is an $n\times n$ matrix, $A^{\tilde i \tilde j}$ is $(n-1)\times(n-1)$ and is called a \emph{minor}.
Then one has the Laplace expansion:
\begin{equation}
\det(A)=\sum_{i=1}^n (-1)^{i+j} A_{ij}\,\det(A^{\tilde i \tilde j})
= \sum_{i=1}^n (-1)^{\tilde i+\tilde j} A_{\tilde j\tilde i}\,\det(A^{\tilde j \tilde i}).
\end{equation}
The first identity is the \emph{row expansion}, the second the \emph{column expansion}.

\paragraph{Geometric interpretation.}
Geometrically, the determinant measures volume. If the columns of $A$ are $k$ vectors in $\mathbb R^n$,
then they define an $n$--dimensional parallelepiped $P$. Its squared volume is
\begin{equation}
\text{vol}(P)^2=\det(A A^{\top}).
\end{equation}
If $k=n$, this reduces to $\text{vol}(P)=|\det(A)|$.

\paragraph{Basic properties.}
From the point of view of matrix functions, the determinant satisfies:
\begin{itemize}
 \item $\det(A^{\top})=\det(A)$,
 \item $\det(\alpha A)=\alpha^n \det(A)$,
 \item $\det(AB)=\det(A)\det(B)$ (if $A,B$ are both square),
 \item $\det(A)=0$ if $A$ is not full rank.
\end{itemize}

The \emph{trace} of a matrix is much simpler: $\text{Tr}(A)=\sum_{i=1}^n A_{ii}$.  
Both trace and determinant are invariant under similarity transformations $A\mapsto P^{-1}AP$.

\paragraph{Connection to eigenvalues.}
If the eigenvalues of $A$ are $\{\lambda_i\}$, then
\begin{equation}
\text{Tr}(A)=\sum_{i=1}^n \lambda_i, \qquad
\det(A)=\prod_{i=1}^n \lambda_i.
\end{equation}
This will be important later, e.g.\ in the discussion of the Jordan normal form.

\paragraph{Cauchy--Binet identity.}
An identity often used in combinatorics and physics (for example in the Kirchhoff--Tutte matrix tree theorem)
is the Cauchy--Binet identity, which generalizes $\det(A)\det(B)=\det(AB)$ when $A,B$ are square.

\medskip
\noindent\textbf{Theorem (Cauchy--Binet).}
Let $A$ be an $m\times n$ matrix, and $B$ an $n\times m$ matrix. Let $S$ be the set of all subsets of
$\{1,\cdots,m\}$ of size $m$. Then
\begin{equation}
\det(AB)=\sum_S \det(A[S]) \det(B[S]),
\end{equation}
assuming $n\geq m$.

\medskip
\noindent\emph{Example.}  
For
\begin{eqnarray}
A=\begin{pmatrix}
a_1 & a_2 & a_3 \\
b_1 & b_2 & b_3
\end{pmatrix},\quad
B=\begin{pmatrix}
c_1 & d_1 \\
c_2 & d_2 \\
c_3 & d_3
\end{pmatrix},
\end{eqnarray}
the expansion gives
\begin{eqnarray}
\det(AB)&=&
\Big|\begin{pmatrix} a_1 & a_2 \\ b_1 & b_2  \end{pmatrix} \Big|
\cdot
\Big|\begin{pmatrix} c_1 & d_1 \\ c_2 & d_2\end{pmatrix} \Big|
\nonumber \\
&+&
\Big|\begin{pmatrix} a_1 & a_3 \\ b_1 & b_3  \end{pmatrix} \Big|
\cdot
\Big|\begin{pmatrix} c_1 & d_1 \\ c_3 & d_3\end{pmatrix} \Big|
\nonumber \\
&+&
\Big|\begin{pmatrix}  a_2 & a_3 \\ b_2 & b_3 \end{pmatrix} \Big|
\cdot
\Big|\begin{pmatrix} c_2 & d_2 \\ c_3 & d_3 \end{pmatrix} \Big|.
\end{eqnarray}

\subsection{Determinants and polynomials}
\label{sec:charpoly}

The determinant plays a further role because it encodes the eigenvalues of a matrix.  
Consider the polynomial
\begin{equation}
P(\lambda)=\det(A-\lambda I).
\end{equation}
The roots of $P(\lambda)$ are exactly the eigenvalues of $A$.  
Indeed, if $A\vec{x}=\lambda \vec{x}$, then $\vec{x}$ is in the kernel of $A-\lambda I$, which means
$A-\lambda I$ is not full rank and hence $\det(A-\lambda I)=0$.  
The polynomial $P(\lambda)$ is called the \emph{characteristic polynomial}.

It can always be factorized as
\begin{equation}
P(\lambda)=a_0\prod_{i=1}^k (\lambda-\lambda_i)^{\mu_i},
\end{equation}
for some constant $a_0$, where the integers $\mu_i$ are called the \emph{algebraic multiplicities}
of the eigenvalues. They satisfy
\begin{eqnarray}
1\leq \mu_i \leq n, \\
\sum_{i=1}^k \mu_i = n,
\end{eqnarray}
where $n$ is the dimension of $A$.

\paragraph{Cayley--Hamilton theorem.}
A fundamental result is the Cayley--Hamilton theorem: a matrix satisfies its own characteristic polynomial. In other words,
\begin{equation}
P(A)=A^n+c_{n-1} A^{n-1}+\cdots +c_1 A+(-1)^n \det(A)\, I_n=0,
\end{equation}
where the coefficients $c_j$ are related to symmetric functions of the eigenvalues. More explicitly, they can
be expressed through \emph{Bell polynomials}:
\begin{equation}
c_{n-k}=\frac{1}{n!}\,B_k\!\left(s_1,-1!s_2,2!s_3,\ldots,(-1)^{n-1} n! s_{n}\right),
\end{equation}
with $s_k=\text{Tr}(A^k)$.

\paragraph{Alternative forms of the characteristic polynomial.}
The characteristic polynomial can also be written in ways useful for analysis:
\begin{equation}
p(\lambda)=\det(\lambda I-A)=\lambda^n \det\!\left( I-\tfrac{A}{\lambda}\right)
= \lambda^n \exp\!\Big( \text{Tr}\,\log\!\left(I-\tfrac{A}{\lambda}\right)\Big).
\end{equation}
Expanding the logarithm gives
\begin{equation}
p(\lambda)=\lambda^n \exp\!\left(-\text{Tr}\,\sum_{m=1}^\infty \frac{A^m}{m\,\lambda^m}\right).
\end{equation}

\paragraph{Resolvent and derivative.}
Differentiating the characteristic polynomial yields a useful identity:
\begin{equation}
\partial_\lambda p(\lambda)=p(\lambda) \sum_{m=0}^\infty \frac{1}{\lambda^{m+1}} \,\text{Tr}(A^m).
\end{equation}
This infinite series can be resummed, giving the \emph{resolvent} identity:
\begin{equation}
\partial_\lambda p(\lambda)=p(\lambda)\,\text{Tr}\big((\lambda I-A)^{-1}\big).
\end{equation}

\paragraph{Spectral radius.}
The \emph{spectral radius} of a matrix is the largest modulus of its eigenvalues:
\begin{equation}
\rho(A)=\sup\{|\lambda|:\ \lambda \in \Lambda(A)\}.
\end{equation}
It plays an important role in stability analysis and dynamical systems.

\paragraph{Generalizations of the determinant.}
The determinant is part of a broader family of functions defined in terms of group characters.
Given a group $G$ and a representation with character $\chi$, one defines the \emph{immanant}:
\begin{equation}
D_{\chi}^G(A)=\sum_{\sigma \in G}\chi(\sigma)\prod_{i=1}^n A_{i\sigma(i)}.
\end{equation}
Special cases include:
\begin{itemize}
 \item $\chi=\text{sign}$ $\;\Rightarrow\;$ the determinant,
 \item $\chi=\text{Id}$ $\;\Rightarrow\;$ the permanent.
\end{itemize}
For other representations, $D_\chi^G(A)$ provides nontrivial generalizations.

\paragraph{Inequalities.}
Certain inequalities known for the permanent extend to these generalizations. For example,
\begin{equation}
\text{per}(A)\geq \prod_{i=1}^n A_{ii},
\end{equation}
known as \emph{Marcus’ inequality}. Lieb proved a refinement for Hermitian block matrices of the form
$A=\begin{pmatrix} B & C \\ C^* & D \end{pmatrix}$:
\begin{equation}
\text{per}(A)\geq \text{per}(B)\,\text{per}(D)\geq \prod_i A_{ii}.
\end{equation}
Such results are useful in statistical mechanics, combinatorics, and quantum many-body theory.
 
\subsubsection{Integral representations of determinants}
\label{sec:intdet}

Determinants admit a variety of integral representations. These are extremely useful in physics, especially
in statistical mechanics and quantum field theory, where partition functions and correlation functions often
reduce to Gaussian or Grassmann integrals. In what follows we summarize the most common cases.

Throughout, by a \emph{positive matrix} we mean a real symmetric matrix whose eigenvalues are strictly
positive, so that all Gaussian integrals are well-defined.

\paragraph{Commuting variables.}
Let us begin with ordinary (commuting) real variables. Recall the elementary Gaussian integral
\begin{equation}
    \int_{-\infty}^\infty e^{-\frac{\pi x^2}{a}}\,dx = \sqrt{a}, \qquad a>0.
\end{equation}
Generalizing to $n$ variables gives
\begin{equation}
    \int_{\mathbb{R}^n} 
    \exp\!\Big(- \pi \sum_{ij} x_i (A^{-1})_{ij} x_j\Big)\, \prod_{i=1}^n dx_i
    = \sqrt{\det(A)},
\end{equation}
valid if $A$ is positive definite. Thus Gaussian integrals naturally generate square roots of determinants.

This can be extended to include linear terms in the exponent. For example,
\begin{equation}
    \frac{\int_{\mathbb{R}^n} 
    \exp\!\Big(- \pi \sum_{ij} x_i (A^{-1})_{ij} x_j - \sum_i b_i x_i \Big)\,\prod_i dx_i}
    {\exp\!\Big(\tfrac{1}{2}\sum_{ij} b_i A_{ij} b_j\Big)}
    = \sqrt{\det(A)}.
\end{equation}
This identity is often used in the derivation of correlation functions in Gaussian models.

\paragraph{Grassmann variables and Berezin integrals.}
The representations above only produce $\sqrt{\det(A)}$. To obtain $\det(A)$ directly,
one needs to introduce anti-commuting variables, also known as \emph{Grassmann variables}.
Integrals over Grassmann variables are called \emph{Berezin integrals}\footnote{Introduced by F.~Berezin in the
1960s in the context of quantum field theory.}. They play a central role in fermionic path integrals.

Grassmann variables $\theta_i$ are defined by the axioms
\begin{equation}
    \theta_i \theta_j = - \theta_j \theta_i, \qquad \theta_i^2=0.
\end{equation}
Because of this nilpotency, the Taylor expansion of any function of $\theta_i$ truncates:
\begin{equation}
    f(\theta_i)=c_0+c_1 \theta_i.
\end{equation}

The Berezin integral is defined by
\begin{eqnarray}
\int d\theta_i &=& 0, \\
\int \theta_i\, d\theta_i &=& 1.
\end{eqnarray}
Thus the Berezin integral simply extracts the coefficient of $\theta_i$ in a function, i.e. it acts like a
derivative:
\begin{equation}
\int f(\theta_i)\,d\theta_i = c_1.
\end{equation}

For multiple Grassmann variables, one considers antisymmetric expansions, e.g.
\begin{equation}
f(\theta_1,\cdots,\theta_n)=c_0+\sum_{k=1}^n c_1^k \theta_k
+ \sum_{k_1<k_2} c_2^{k_1k_2}\theta_{k_1}\theta_{k_2}+\cdots,
\end{equation}
with coefficients antisymmetric under index exchange.

\paragraph{Determinant representation.}
Using this formalism, one can prove the fundamental identity
\begin{equation}
    \int \Big(\prod_j d\theta_j \, d\eta_j\Big)\,
    \exp\!\left(-\sum_{ij} \theta_i A_{ij} \eta_j\right)
    = \det(A).
\end{equation}
Here $\{\theta_i\}$ and $\{\eta_j\}$ are two independent sets of Grassmann variables.

A related identity involves only one set of variables:
\begin{equation}
    \int \Big(\prod_j d\theta_j\Big)\,
    \exp\!\left(-\sum_{ij} \theta_i A_{ij} \theta_j\right)
    = \begin{cases}
    2^{n/2}\sqrt{\det(A)} & n\ \text{even}, \\
    0 & n\ \text{odd}.
    \end{cases}
\end{equation}

\paragraph{Remarks.}
\begin{itemize}
    \item The sign of Grassmann integrals depends on the order of integration. One must always fix a
    convention (usually $d\theta_n \cdots d\theta_1$).
    \item These integral representations are heavily used in physics: commuting integrals for bosonic
    Gaussian models, anti-commuting (Berezin) integrals for fermionic systems. Together, they allow one
    to rewrite partition functions and correlation functions in terms of determinants and Pfaffians.
\end{itemize}
\subsection{List of determinantal identities}
We would like to provide here a list of Determinantal identities that might turn useful.
\paragraph{Determinantal equalities.}
Let $A\in\mathbb{F}^{n\times n}$ be square (unless stated otherwise), 
$B\in\mathbb{F}^{n\times n}$, $U\in\mathbb{F}^{n\times m}$, $V\in\mathbb{F}^{m\times n}$, 
$u,v\in\mathbb{F}^{n}$, and write $\text{Tr}(\cdot)$ for the trace and $\text{adj}(\cdot)$ for the adjugate.

\medskip
\noindent\textbf{Basic rules.}
\begin{align}
\det(A^\top) &= \det(A),\qquad
\det(\alpha A) = \alpha^{\,n}\det(A),\qquad
\det(AB) = \det(A)\det(B), \label{eq:det-basic}
\end{align}
and $\det(A)=0$ if $A$ is rank-deficient.  

\medskip
\noindent\textbf{Block/Schur complement formulas.}
If 
$\displaystyle 
M=\begin{bmatrix} A & C \\ D & B \end{bmatrix}$ 
with $B$ invertible, then
\begin{equation}
\det(M) = \det(B)\,\det\!\big(A - C B^{-1} D\big).
\label{eq:block-schur}
\end{equation}
Equivalently, if $A$ is invertible one has $\det(M)=\det(A)\,\det\!\big(B - D A^{-1} C\big)$.

More generally, for conformal block partitions,
\begin{equation}
\det(A)=\det(A_{11})\,\det(A/A_{11}), 
\quad 
A/A_{11}=A_{22}-A_{21}A_{11}^{-1}A_{12},
\label{eq:schur-factor}
\end{equation}
and in particular for $2\times2$ blocks,
$\det\!\begin{bmatrix} a & * \\ * & b\end{bmatrix}=ab-\!*\,\!*$. 

\medskip
\noindent\textbf{Sylvester determinant theorem.}
For $A\in\mathbb{F}^{n\times m}$ and $B\in\mathbb{F}^{m\times n}$,
\begin{equation}
\det\!\big(I_n + A B\big) = \det\!\big(I_m + B A\big).
\label{eq:sylvester}
\end{equation}

\medskip
\noindent\textbf{Matrix determinant lemma (rank-one / low-rank updates).}
For $A$ invertible and $u,v\in\mathbb{F}^{n}$,
\begin{equation}
\det\!\big(A + u v^{\top}\big) 
= \big(1+ v^{\top}A^{-1}u\big)\,\det(A) 
= \det(A)+ v^{\top}\,\text{adj}(A)\,u .
\label{eq:mdlemma-r1}
\end{equation}
More generally, for $U\in\mathbb{F}^{n\times m}$, $V\in\mathbb{F}^{m\times n}$ with $I_m+V A^{-1}U$ defined,
\begin{equation}
\det\!\big(A + U V\big) = \det(A)\,\det\!\big(I_m + V A^{-1} U\big).
\label{eq:mdlemma}
\end{equation}

A simple special case often used:
\begin{equation}
\det\!\big(k I_n + u v^\top\big) = k^{\,n} + k^{\,n-1}\, v^\top u .
\label{eq:kI-rank1}
\end{equation}

\medskip
\noindent\textbf{Log–det and trace links.}
If $A$ is diagonalizable (e.g.\ positive definite), then
\begin{equation}
\log\det(A)=\text{Tr}\big(\log A\big), 
\qquad 
\det(I+A)=\exp\!\big(\text{Tr}\log(I+A)\big),
\label{eq:logdet-tr}
\end{equation}
and the series expansion
\begin{equation}
\log\det(I+zA)=\sum_{k=1}^\infty \frac{(-1)^{k+1}}{k}\,\text{Tr}(A^k)\,z^k
\quad \text{(for $\|zA\|<1$)} .
\label{eq:logdet-series}
\end{equation}

\medskip
\noindent\textbf{Small perturbations.}
For $\varepsilon$ small,
\begin{equation}
\det(I+\varepsilon A)=1+\varepsilon\,\text{Tr}(A)+\tfrac{\varepsilon^2}{2}\!\left(\text{Tr}(A)^2-\text{Tr}(A^2)\right)+O(\varepsilon^3).
\label{eq:det-perturb}
\end{equation}

\medskip
\noindent\textbf{Cauchy--Binet (rectangular generalization).}
If $A\in\mathbb{F}^{m\times n}$, $B\in\mathbb{F}^{n\times m}$ with $n\ge m$, then
\begin{equation}
\det(AB)=\sum_{S\subset\{1,\dots,n\},\,|S|=m} \det\!\big(A[:,S]\big)\,\det\!\big(B[S,:]\big).
\label{eq:cauchy-binet}
\end{equation}

\subsection{Permanents and Pfaffians}
\label{sec:perpf}

The determinant and the trace are not the only important scalar quantities associated with a matrix.  
Two others, which appear in combinatorics, graph theory, and physics, are the \emph{permanent} and the
\emph{Pfaffian}.

\paragraph{Permanents.}
The permanent of an $n\times n$ matrix $A$ is defined as
\begin{equation}
\text{per}(A)=\sum_{\pi \in \mathcal S_n} 
A_{1, \pi(1)}\,A_{2, \pi(2)} \cdots A_{n, \pi(n)}.
\label{eq:permanent}
\end{equation}
This formula resembles the Leibniz definition of the determinant, except that the alternating sign
$\text{sign}(\pi)$ is absent. For this reason the permanent is sometimes called the
\emph{plus determinant}.

\emph{Properties.}
\begin{itemize}
    \item $\text{per}(A^{\top})=\text{per}(A)$.
    \item $\text{per}(AB)\neq \text{per}(A)\text{per}(B)$ in general (unlike determinants).
    \item For diagonal matrices, $\text{per}(\mathrm{diag}(a_1,\ldots,a_n))=\prod_i a_i$.
    \item The permanent is computationally hard: computing it exactly is \#P-complete\footnote{L.~Valiant,
    ``The complexity of computing the permanent,'' \emph{Theoretical Computer Science} \textbf{8}, 189–201
    (1979).}.
\end{itemize}

\emph{Applications.}
Permanents appear naturally in combinatorics (e.g.\ counting perfect matchings of bipartite graphs),
and in physics they describe bosonic statistics (in contrast to determinants, which describe fermionic
statistics)\footnote{See, e.g., A.~Barvinok, \emph{Combinatorics and Complexity of Partition Functions},
Springer (2016).}.

\paragraph{Pfaffians.}
The Pfaffian is another determinant-like object, defined only for skew-symmetric (antisymmetric) matrices.
If $A$ is a $2n\times 2n$ antisymmetric matrix, the Pfaffian $\text{Pf}(A)$ is defined so that
\begin{equation}
\text{Pf}(A)^2=\det(A).
\end{equation}
By convention, $\text{Pf}(A)=0$ if $A$ is odd-dimensional or not antisymmetric.

The Pfaffian plays an important role in enumerative combinatorics and in fermionic path integrals, where
Gaussian integrals over Grassmann variables naturally yield Pfaffians\footnote{See, for example,
G.~De~Bruijn, ``On some multiple integrals involving determinants,'' \emph{J. Indian Math. Soc.}
\textbf{19}, 133–151 (1955).}. It also provides efficient formulas for counting perfect matchings in general
graphs (Kasteleyn theory).

---

\subsection{Spectra}
\label{sec:spectra}

We now turn to the study of the spectrum of a matrix --- the set of its eigenvalues.  
Spectral properties of matrices provide the backbone of many techniques in dynamical systems, graph
theory, and quantum mechanics. We begin with the fundamental result for Hermitian matrices.

\begin{theorem}[Spectral Theorem]\label{thm:spectral}
Let $A$ be a Hermitian matrix in $\mathbb{C}^{n\times n}$. Then there exists a unitary matrix $U$ such that
\begin{equation}
A = U D U^\dagger,
\end{equation}
where $D$ is diagonal with real entries. The columns of $U$ are an orthonormal basis of eigenvectors of $A$,
and the diagonal entries of $D$ are the corresponding eigenvalues.
\end{theorem}

\noindent
\emph{Remarks.}
\begin{itemize}
    \item For real symmetric matrices, the unitary $U$ can be taken to be real orthogonal.
    \item The theorem guarantees diagonalizability of Hermitian matrices and ensures that their eigenvalues
    are real. This is the foundation of quantum mechanics, where observables are represented by Hermitian
    operators\footnote{See R.~Horn and C.~Johnson, \emph{Matrix Analysis}, Cambridge University Press
    (1985).}.
    \item The decomposition $A=UDU^\dagger$ is also called the \emph{spectral decomposition}.
\end{itemize}

\subsubsection{Eigenvectors from Eigenvalues}
\index{key}{Eigenvectors from eigenvalues}

One of the remarkable discoveries in spectral theory is that eigenvectors,
which are usually thought of as ``directional'' data, can in fact be recovered
from eigenvalues alone, provided one also has access to the eigenvalues of
principal minors. In other words, information that at first glance appears
to be lost in passing from eigenvectors to eigenvalues is still encoded in
the spectral structure of smaller submatrices.

This phenomenon was rigorously established in recent work by Krishnapur,
O’Rourke, and Vu, and was later popularized by Tao and others.
It reveals a deep interplay between \emph{eigenvalue interlacing}, \emph{principal
minors}, and the geometry of eigenvectors. The result has applications in
random matrix theory, where eigenvector statistics can be deduced from
eigenvalue distributions, as well as in numerical analysis, where it suggests
alternative strategies for reconstructing eigenvector entries.

\begin{theorem}[Eigenvectors from Eigenvalues]\label{thm:eigvecsfromeigvals}
Let $A$ be a Hermitian $n \times n$ matrix with eigenvalues
$\lambda_1,\dots,\lambda_n$ and corresponding orthonormal eigenvectors
$\vec v_1,\dots,\vec v_n$.
For each $i$, consider the $(n-1)\times(n-1)$ principal minor
$A_{(i)}$ obtained by deleting the $i$-th row and column of $A$.
Let $\mu^{(i)}_1,\dots,\mu^{(i)}_{n-1}$ denote the eigenvalues of $A_{(i)}$.
Then the squared moduli of the components of $\vec v_j$ are determined by
the interlacing relation:
\begin{equation}
|\vec v_j(i)|^2 = 
\frac{\prod_{k=1}^{n-1} (\lambda_j - \mu^{(i)}_k)}
     {\prod_{\substack{\ell=1 \\ \ell \neq j}}^n (\lambda_j - \lambda_\ell)}.
\end{equation}
In particular, the entire collection of eigenvectors can be reconstructed
(up to phases) from the knowledge of the spectra of $A$ and its principal minors.
\end{theorem}

\noindent
\emph{Remarks.}
\begin{itemize}
  \item The formula highlights how eigenvectors are ``spectrally encoded'':
  deleting a row and column perturbs the spectrum in a way that
  reflects the weight of the corresponding coordinate in each eigenvector.
  \item In random matrix theory, this identity provides a tool to relate
  eigenvalue rigidity and delocalization of eigenvectors, central topics
  in universality results.
  \item From a computational viewpoint, it shows that even though eigenvalues
  are scalar quantities, their structured collection across minors carries
  enough information to reconstruct directional data.
  \item The discovery emphasizes that spectra are richer than just a list
  of numbers: through interlacing, they encode geometry.\footnote{R.~Krishnapur,
  S.~O’Rourke, and V.~Vu, ``Eigenvectors from Eigenvalues,'' \emph{Random Matrices: Theory and Applications}
  5(4), 1650016 (2016). See also T.~Tao, ``Eigenvectors from Eigenvalues,'' blog post (2015).}
\end{itemize}

\subsubsection{Generalized eigenvalue problems}
\label{sec:geneigen}

The standard eigenvalue problem asks for nontrivial solutions of
\begin{equation}
A \vec v = \lambda \vec v.
\end{equation}
A natural generalization is the problem
\begin{equation}
A \vec v = \lambda B \vec v,
\end{equation}
where $A$ and $B$ are given square matrices, and $\vec v$ is the unknown eigenvector.
In matrix form this reads
\begin{equation}
A V = B V \Lambda,
\label{eq:geneigenp}
\end{equation}
where $V$ is the matrix of eigenvectors and $\Lambda$ is diagonal with the generalized eigenvalues
$\{\lambda_i\}$ on the diagonal.  

The generalized eigenvalues can be found from the condition
\begin{equation}
\det(A-\lambda B)=0,
\end{equation}
which defines the \emph{generalized characteristic polynomial}\footnote{See
G.~Golub and C.~Van Loan, \emph{Matrix Computations}, Johns Hopkins University Press (1996).}.  
This reduces to the standard case when $B=I$.

\paragraph{Solution procedure.}
The generalized eigenvalue problem can be reduced to an ordinary one if $B$ is symmetric and positive
definite. One standard approach is:
\begin{enumerate}
    \item Solve the eigenvalue problem for $B$, $B \vec x_B=\lambda_B \vec x_B$, and form the eigenvector
    matrix $V_B$.
    \item Define $V_B' = V_B \sqrt{\Lambda_B}$, so that $(V_B')^{\top} B V_B' = I$.
    \item Transform $A$ accordingly, $A' = (V_B')^{\top} A V_B'$, which is symmetric by construction.
    \item Solve the standard eigenvalue problem $A' \vec u = \lambda \vec u$.
    \item Construct the generalized eigenvectors as $V = V_B \Lambda_B^{-1/2} U$, where $U$ collects the
    eigenvectors of $A'$.
\end{enumerate}
In this way one diagonalizes both $A$ and $B$ simultaneously, and the eigenvalues $\lambda$ are those of
eq.~(\ref{eq:geneigenp}).

---

\subsubsection{Rayleigh quotient and variational characterization}
\label{sec:rayleigh}

One of the most powerful tools for studying eigenvalues is the \emph{Rayleigh quotient}.
For a symmetric positive definite matrix $A$, and any nonzero vector $\vec x$, define
\begin{equation}
R(\vec x)=\frac{\vec x^{\top} A \vec x}{\vec x^{\top} \vec x}.
\end{equation}
This quantity is invariant under rescaling $\vec x\mapsto \alpha \vec x$, so without loss of generality one
may restrict to $\|\vec x\|=1$.  

\paragraph{Extremal characterization of eigenvalues.}
If the eigenvalues of $A$ are ordered as
$\lambda_1 \geq \lambda_2 \geq \cdots \geq \lambda_n$, then
\begin{equation}
\min_{\|\vec x\|=1} R(\vec x) = \lambda_n, \qquad
\max_{\|\vec x\|=1} R(\vec x) = \lambda_1.
\end{equation}
More generally, the Courant--Fischer min--max theorem expresses intermediate eigenvalues as
variational extrema of the Rayleigh quotient\footnote{See R.~Courant and D.~Hilbert, \emph{Methods of
Mathematical Physics}, Wiley (1953).}.  

\paragraph{Generalized Rayleigh quotient.}
The Rayleigh quotient can be generalized to the pair $(A,B)$:
\begin{equation}
R(\vec x)=\frac{\vec x^{\top} A \vec x}{\vec x^{\top} B \vec x}.
\end{equation}
Stationarity of $R(\vec x)$ under variation of $\vec x$ leads to
\begin{equation}
A \vec x (\vec x^{\top} B \vec x)= B \vec x (\vec x^{\top} A \vec x),
\end{equation}
which can be rearranged as
\begin{equation}
A \vec w = \lambda B \vec w, \qquad
\lambda = R(\vec w).
\end{equation}
Thus the stationary values of the generalized Rayleigh quotient correspond to the generalized
eigenvalues.

---

\subsubsection{Spectral decomposition}
\label{sec:spectraldec}

Finally, let us return to the case of a diagonalizable symmetric matrix $A$.  
Then $A$ can be written as
\begin{equation}
A = Q D Q^{\top} = \sum_{i=1}^n \lambda_i \vec q_i \vec q_i^{\top},
\end{equation}
where the $\vec q_i$ are orthonormal eigenvectors, collected as the columns of the orthogonal
matrix $Q$, and $D=\text{diag}(\lambda_1,\ldots,\lambda_n)$.

Define the rank-one matrices
\begin{equation}
P_i = \vec q_i \vec q_i^{\top}.
\end{equation}
These are orthogonal projectors:
\begin{equation}
P_i^2=P_i,\qquad P_i P_j = 0 \ (i\neq j).
\end{equation}
Hence the diagonalization can be written as a weighted sum of projections onto the eigenspaces:
\begin{equation}
A=\sum_{i=1}^n \lambda_i P_i.
\end{equation}
This form is called the \emph{spectral decomposition} of $A$, and expresses $A$ entirely in terms of its
eigenvalues and orthogonal projectors\footnote{See R.~Horn and C.~Johnson, \emph{Matrix Analysis},
Cambridge University Press (1985).}.

\subsection{Scalar products, vector and matrix norms, and some basic inequalities}
\label{sec:matrixnorm}

In order to control the size of vectors and matrices, and to compare their relative magnitudes, one introduces
the notions of scalar products and norms. These concepts are fundamental in both analysis and linear
algebra, and appear everywhere in physics: from defining the length of a state vector in quantum mechanics,
to measuring stability of dynamical systems, or bounding errors in numerical algorithms.

\paragraph{Scalar products.}
The scalar product (or inner product) between two vectors $\vec v,\vec u\in\mathbb{C}^n$ is defined as
\begin{equation}
\vec v \cdot \vec u \;\equiv\; \langle v,u \rangle \;=\; \sum_{i=1}^n v_i^* u_i.
\end{equation}
This definition reduces to the familiar dot product when the entries are real.
The inner product satisfies the following properties:
\begin{itemize}
  \item $\langle u,v\rangle = \overline{\langle v,u\rangle}$ (conjugate symmetry),
  \item $\langle \alpha u + \beta v, w\rangle = \alpha \langle u,w\rangle + \beta \langle v,w\rangle$ (linearity),
  \item $\langle v,v\rangle \geq 0$ and equals zero if and only if $v=0$ (positive definiteness).
\end{itemize}
These axioms define a Hilbert space structure on $\mathbb{C}^n$.

\paragraph{Dual space.}
The dual of a vector space $V$ is the set of linear maps $\psi:V\to \mathbb{C}$.
Given a vector $\vec r$, one can define a linear functional $\psi(\vec v)=\langle \vec r,\vec v\rangle$.
This identification, via the scalar product, is the Riesz representation theorem. It is the reason why
in quantum mechanics bras are linear functionals on kets.

---

\paragraph{Vector norms.}
A norm is a map $\|\cdot\|: V\to \mathbb{R}_{\geq 0}$ satisfying:
\begin{enumerate}
    \item $\|\vec u\|\geq 0$,
    \item $\|\vec u\|=0 \implies \vec u=0$,
    \item $\|\alpha \vec u\|=|\alpha|\|\vec u\|$,
    \item $\|\vec u+\vec v\|\leq \|\vec u\|+\|\vec v\|$ (triangle inequality).
\end{enumerate}
From the triangle inequality it follows that
\[
\big|\|\vec u\|-\|\vec v\|\big|\;\leq\;\|\vec u-\vec v\|.
\]

Common examples include:
\begin{itemize}
  \item $L^1$ norm (taxicab norm): $\|\vec x\|_1=\sum_{i=1}^n |x_i|$,
  \item $L^2$ norm (Euclidean norm): $\|\vec x\|_2=\sqrt{\sum_{i=1}^n |x_i|^2}=\sqrt{\langle \vec x,\vec x\rangle}$,
  \item $L^\infty$ norm (sup norm): $\|\vec x\|_\infty=\max_{1\leq i\leq n}|x_i|$.
\end{itemize}
These are special cases of the family of $p$-norms
\begin{equation}
\|\vec x\|_p=\left(\sum_{i=1}^n |x_i|^p\right)^{1/p}, \qquad 1\leq p < \infty,
\end{equation}
with $\|\vec x\|_\infty = \lim_{p\to\infty} \|\vec x\|_p$.

---

\paragraph{Matrix norms.}
Norms for matrices can be defined in several ways. One important construction is the \emph{induced norm}:
\begin{equation}
\|A\|_* = \max_{\vec x\neq 0} \frac{\|A\vec x\|_*}{\|\vec x\|_*} = \max_{\|\vec x\|_*=1}\|A\vec x\|_*,
\end{equation}
where $*$ denotes the chosen underlying vector norm (e.g.\ $1,2,\infty$).
Such norms are also called \emph{operator norms}, as they measure how much the matrix can stretch a vector.

Properties required for any matrix norm:
\begin{enumerate}
    \item $\|A\|\geq 0$,
    \item $\|A\|=0 \implies A=0$,
    \item $\|\alpha A\|=|\alpha|\|A\|$,
    \item $\|A+B\|\leq \|A\|+\|B\|$.
\end{enumerate}
If in addition
\begin{equation}
\|AB\|\leq \|A\|\|B\|,
\end{equation}
the norm is called \emph{sub-multiplicative}. Sub-multiplicativity is essential in stability analysis,
since it allows one to control the growth of powers of matrices.

\paragraph{Connection with eigenvalues.}
For the 2-norm, Rayleigh’s principle shows that
\begin{equation}
\|A\|_2=\max_{\|\vec x\|_2=1}\|A\vec x\|_2 = \sqrt{\lambda_{\max}(A^* A)},
\end{equation}
i.e.\ it equals the largest singular value of $A$.  
If $A$ is symmetric, this reduces to the largest absolute eigenvalue.

---

\paragraph{Examples of matrix norms.}
\begin{itemize}
    \item Matrix $L^1$ norm:
    \begin{equation}
    \|A\|_1=\max_{1\leq j\leq n} \sum_{i=1}^n |A_{ij}|,
    \end{equation}
    i.e.\ the maximum absolute column sum.
    \item Matrix $L^2$ norm:
    \begin{equation}
    \|A\|_2=\sigma_{\max}(A),
    \end{equation}
    the largest singular value of $A$. This is sometimes called the \emph{spectral norm}.
    \item Matrix $L^\infty$ norm:
    \begin{equation}
    \|A\|_\infty=\max_{1\leq i\leq n}\sum_{j=1}^n |A_{ij}|,
    \end{equation}
    i.e.\ the maximum absolute row sum.
\end{itemize}
Inequalities connect these norms, e.g.\footnote{See Bernstein, Matrix Mathematics (ibid).}
\begin{equation}
\|A\|_2 \;\leq\; \sqrt{n}\,\|A\|_\infty.
\end{equation}

---

\paragraph{Cauchy--Schwarz inequality.}
The prototype inequality for scalar products is
\begin{equation}
|\langle \vec u,\vec v\rangle|\leq \|\vec u\|_2\|\vec v\|_2.
\end{equation}
This extends naturally to complex vectors. In matrix analysis, the analogous role is played by
sub-multiplicativity of matrix norms, which ensures
\begin{equation}
\|AB\| \leq \|A\|\|B\|.
\end{equation}

\paragraph{Compatible and equivalent norms.}
A matrix norm $\|\cdot\|_M$ and a vector norm $\|\cdot\|_V$ are called \emph{compatible} if
\begin{equation}
\|A\vec x\|_V \leq \|A\|_M \|\vec x\|_V.
\end{equation}
Two norms $\|\cdot\|_\alpha$ and $\|\cdot\|_\beta$ are \emph{equivalent} if there exist positive constants
$r_{\min},r_{\max}$ such that
\begin{equation}
r_{\min}\|A\|_\alpha \leq \|A\|_\beta \leq r_{\max}\|A\|_\alpha.
\end{equation}
In finite dimensions, all norms are equivalent, though the constants may be large.

---

\paragraph{Why norms matter.}
Norms and inequalities are indispensable in physics and applied mathematics:
\begin{itemize}
    \item In quantum mechanics, $\|\vec\psi\|_2=1$ expresses normalization of state vectors.
    \item In numerical analysis, matrix norms measure sensitivity to perturbations (condition numbers).
    \item In dynamical systems, $\|A^{\top}\|$ controls growth or decay of trajectories.
\end{itemize}
Thus norms are not only abstract constructs, but also tools for bounding and estimating physical behavior.

\subsubsection{Spectral radius}
\label{sec:specradius}

The \emph{spectral radius} of a square matrix $A$ is defined as
\begin{equation}
\rho(A)=\max_i |\lambda_i|,
\end{equation}
where $\{\lambda_i\}$ are the eigenvalues of $A$. It measures the largest “stretching factor” associated
with $A$.

\paragraph{Relation to norms.}
The spectral radius is not itself a norm, but it is closely related. For any matrix norm induced by a
vector norm, one always has
\begin{equation}
\rho(A) \leq \|A\|.
\end{equation}
More generally,
\begin{equation}
\rho(A)=\lim_{r\to\infty}\|A^r\|^{1/r},
\end{equation}
a result known as Gelfand’s formula\footnote{I.~Gelfand, ``Normierte Ringe,'' \emph{Rec. Math. [Mat. Sbornik]}
\textbf{9}(51), 3–24 (1941).}.  

For normal matrices (i.e.\ matrices diagonalizable by a unitary transformation), the spectral radius coincides
with the spectral norm:
\begin{equation}
\rho(A)=\|A\|_2.
\end{equation}

\paragraph{Convergence criteria.}
The spectral radius governs the asymptotic behavior of powers of matrices. In particular,
\begin{equation}
\lim_{k\to\infty} A^k = 0 \quad \Longleftrightarrow \quad \rho(A)<1.
\end{equation}
This is crucial in stability theory: if $\rho(A)<1$, iterations $A^k \vec x$ decay to zero.

As a consequence, the geometric series converges:
\begin{equation}
\sum_{k=0}^\infty A^k = (I-A)^{-1}, \qquad \rho(A)<1.
\end{equation}
In this case $Q=I-A$ is invertible, and one can establish norm inequalities:
\begin{equation}
\frac{1}{1+\|A\|} \leq \|Q^{-1}\| \leq \frac{1}{1-\|A\|}, \qquad \|A\|<1.
\end{equation}
These bounds are widely used in perturbation theory and numerical linear algebra\footnote{See
G.~Golub and C.~Van Loan, \emph{Matrix Computations}, Johns Hopkins University Press (1996).}.

---

\subsubsection{Other norms}
\label{sec:othernorms}

Besides the operator norms discussed earlier, several other norms are commonly used:

\begin{itemize}
   \item \textbf{Matrix $L^\infty$ norm} (maximum row sum):
   \[
   \|A\|_\infty=\max_{i}\sum_j |A_{ij}|.
   \]
   \item \textbf{Hilbert--Schmidt norm} (also called Frobenius norm when scaled):
   \[
   |A|=\sqrt{\tfrac{1}{n}\sum_{ij} |A_{ij}|^2}
   = \sqrt{\tfrac{1}{n}\,\text{Tr}(A^* A)}
   = \sqrt{\tfrac{1}{n}\sum_j \sigma_j^2},
   \]
   where $\{\sigma_j\}$ are the singular values of $A$. This is sometimes called the \emph{weak norm}.
   \item \textbf{Spectral norm}:
   \[
   \|A\|_2=\max_j \sigma_j,
   \]
   the largest singular value. For Hermitian $A$, this equals $\max_i |\lambda_i|$.
   \item \textbf{Trace norm}:
   \[
   \|A\|_{tr}=\text{Tr}\,\sqrt{A^* A}=\sum_j \sigma_j.
   \]
   \item \textbf{Frobenius norm}:
   \[
   \|A\|_F=\sqrt{\text{Tr}(A^*A)}=\sqrt{\sum_{ij}|A_{ij}|^2}=\sqrt{n}\,|A|.
   \]
\end{itemize}

\paragraph{Relations between norms.}
Basic inequalities hold, for example
\begin{equation}
\|A\|_2^2=\max_{\|\vec z\|=1} \vec z^* A^*A \vec z \;\geq\; \rho(A)^2,
\end{equation}
and
\begin{equation}
\|A\|^2 \geq |A|^2.
\end{equation}
The terminology “weak” vs.\ “strong” norm reflects this inequality: the Frobenius norm is weaker than the
spectral norm.

\paragraph{Unitary matrices.}
For a unitary matrix $U$, one has
\begin{equation}
U^* U=I, \qquad \text{Tr}(U^*U)=n.
\end{equation}
Hence the Hilbert--Schmidt norm of $U$ is $|U|=1$, while the Frobenius norm is $\|U\|_F=\sqrt{n}$.
Furthermore,
\begin{equation}
\sum_i |U_{ij}|^2 = 1, \qquad \sum_{ij} |U_{ij}|^2=n,
\end{equation}
which has probabilistic interpretations (each column forms an orthonormal basis) and appears in the proof
of the Wielandt--Hoffman theorem\footnote{See R.~Horn and C.~Johnson, \emph{Matrix Analysis},
Cambridge University Press (1985).}.

---

\subsubsection{Banach and Hilbert spaces}
\label{sec:banachhilbert}

While our focus is finite-dimensional linear algebra, it is useful to recall the broader framework of functional
analysis, where the concepts of Banach and Hilbert spaces originate.

\paragraph{Metric spaces.}
A \emph{metric space} $(X,\rho)$ consists of a set $X$ and a function $\rho:X\times X\to\mathbb{R}_{\geq 0}$
such that:
\begin{itemize}
    \item $\rho(x,y)=0$ if and only if $x=y$,
    \item $\rho(x,y)=\rho(y,x)$ (symmetry),
    \item $\rho(x,y)\leq \rho(x,z)+\rho(z,y)$ (triangle inequality).
\end{itemize}
Metrics induce topologies, allowing definitions of convergence and continuity.  
A sequence $\{x_n\}$ is called a \emph{Cauchy sequence} if
\[
\forall \epsilon>0,\ \exists N:\ n,m>N \implies \rho(x_n,x_m)<\epsilon.
\]

\paragraph{Banach spaces.}
If a vector space $X$ is equipped with a norm $\|\cdot\|$, one can define a metric $\rho(x,y)=\|x-y\|$.
A complete normed space (all Cauchy sequences converge) is called a \emph{Banach space}\footnote{See
W.~Rudin, \emph{Functional Analysis}, McGraw--Hill (1991).}.

\paragraph{Hilbert spaces.}
An inner product space is a vector space with scalar product $\langle \cdot,\cdot\rangle$.  
If such a space is complete, it is called a \emph{Hilbert space}. A fundamental property is that every vector
can be expanded in terms of an orthonormal basis:
\begin{equation}
\vec x=\sum_{i=1}^\infty \langle \vec q_i,\vec x\rangle \vec q_i.
\end{equation}
This is the infinite-dimensional analogue of spectral decomposition, and underlies much of modern
mathematical physics, including Fourier analysis and quantum mechanics.

\subsection{Geometric transformations}
\label{sec:geometry}

Linear transformations often admit simple geometric interpretations. These are particularly useful in physics
and geometry, where matrices represent symmetries, changes of coordinates, or constraints on motion. In
this subsection we discuss the most common classes of transformations: scalings, rotations, projections,
reflections, and permutations.

---

\subsubsection{Scale transformations}
\index{key}{scale transformation}

Scale transformations stretch or contract each coordinate axis independently. They are represented by
diagonal matrices
\begin{equation}
A=\text{diag}(\lambda_1,\ldots,\lambda_n).
\end{equation}
The identity matrix corresponds to $\lambda_i=1$, i.e.\ no rescaling. Each $\lambda_i$ is the scaling factor
along direction $i$.  

In physics, such diagonal rescalings appear naturally, for instance, when units are changed (nondimensionalization) or when analyzing anisotropic systems (different couplings along different directions).

---

\subsubsection{Rotations}
\index{key}{rotations}

Rotations preserve lengths and angles. They are represented by orthogonal matrices $O$ satisfying
\begin{equation}
O O^{\top}=O^{\top} O=I, \qquad O^{-1}=O^{\top}.
\end{equation}
If $\det(O)=1$ the transformation is a pure rotation (otherwise $\det(O)=-1$ and the transformation
includes a reflection).

Rotations preserve the scalar product:
\begin{equation}
(O\vec x^\prime)^{\top} (O\vec x)= \vec x^\prime \cdot (O^{\top} O \vec x)=\vec x^\prime \cdot \vec x.
\end{equation}
Hence orthogonal transformations are \emph{isometries} of Euclidean space\footnote{See
R.~Horn and C.~Johnson, \emph{Matrix Analysis}, Cambridge University Press (1985).}.

In $\mathbb{R}^2$, a rotation by angle $\theta$ is given explicitly by
\[
R(\theta)=
\begin{pmatrix}
\cos\theta & -\sin\theta\\
\sin\theta & \cos\theta
\end{pmatrix}.
\]
In $\mathbb{R}^3$, rotations are described by orthogonal matrices with determinant $+1$, forming the group
$SO(3)$.

---

\subsubsection{Projections}
\index{key}{projector operator}

A \emph{projector} is a linear operator $\Omega$ satisfying
\begin{equation}
\Omega^2=\Omega.
\end{equation}
Projectors split the space into two complementary subspaces:
\begin{equation}
\mathbb{C}^n=\text{Ker}(\Omega)\oplus \text{Im}(\Omega),
\end{equation}
so that any vector can be written uniquely as
\[
\vec x=\Omega \vec x+(I-\Omega)\vec x.
\]
Here $\Omega\vec x$ is the projection of $\vec x$ onto the subspace $S=\text{Im}(\Omega)$, while
$(I-\Omega)\vec x$ lies in $K=\text{Ker}(\Omega)$.

\paragraph{Orthogonal projectors.}
If $S\perp K$, the projector is called \emph{orthogonal}. In that case,
\begin{equation}
\|\vec x\|_2^2=\|\Omega \vec x\|_2^2+\|(I-\Omega)\vec x\|_2^2,
\end{equation}
and the orthogonal projector minimizes distance:
\begin{equation}
\min_{\vec y\in S}\|\vec x-\vec y\|_2=\|\vec x-\Omega \vec x\|_2.
\end{equation}

\paragraph{Matrix form.}
Projectors admit explicit matrix expressions. Given a rectangular matrix $A$,
\begin{equation}
\Omega=A(A^{\top}A)^{-1}A^{\top}
\end{equation}
is the orthogonal projector onto the column space of $A$.  
More general (non-orthogonal) projectors are obtained as
\begin{equation}
\Omega=A(B^{\top}A)^{-1}B^{\top}.
\end{equation}
A projector is orthogonal if and only if it is Hermitian ($\Omega=\Omega^\dagger$).

\paragraph{Applications.}
Projectors are ubiquitous: in quantum mechanics, projectors represent measurements onto subspaces; in
numerical analysis, they appear in least-squares problems; in optimization, they arise when enforcing
constraints. In this text, they appear later when discussing Kirchhoff's laws as well.

---

\subsubsection{Reflections}
\index{key}{reflections}

Reflections are closely related to projections. Given a projector $\Omega$, one defines the reflection
\begin{equation}
R=I-2\Omega.
\label{eq:reflect}
\end{equation}
In particular, reflection across a line spanned by unit vector $\vec v$ acts as
\begin{equation}
\vec x \mapsto \vec x-2\langle \vec x,\vec v\rangle \vec v.
\end{equation}

Reflections are involutive:
\[
R^2=(I-2\Omega)^2=I,
\]
and satisfy $\det(R)=(-1)^k$ where $k=\dim(\text{Im}(\Omega))$.  

Geometrically, reflections are symmetries of space that reverse orientation (except in even-dimensional
cases). They appear naturally in the study of Coxeter groups and in the Householder transformations used
in numerical linear algebra\footnote{G.~Golub and C.~Van Loan, \emph{Matrix Computations}, Johns Hopkins
University Press (1996).}.

---

\subsubsection{Permutations}
\index{key}{permutations}

A permutation matrix $P$ is obtained by permuting the rows (or columns) of the identity matrix.
Equivalently, it has exactly one entry $1$ in each row and column, and zeros elsewhere:
\begin{equation}
P_{ij}=
\begin{cases}
1 & \text{if } j=\pi(i),\\
0 & \text{otherwise},
\end{cases}
\end{equation}
for some permutation $\pi$ of $\{1,\ldots,n\}$.

Permutation matrices are orthogonal:
\[
P^{\top}P=I,
\]
and satisfy $\det(P)=\pm 1$.  

Geometrically, they correspond to relabelings of coordinates or reordering of basis vectors. In graph
theory, permutation matrices represent relabelings of nodes that leave the structure of the adjacency
matrix invariant.

---

\paragraph{Summary.}
Geometric transformations such as scalings, rotations, projections, reflections, and permutations capture
the fundamental symmetries and operations on vector spaces. Together they form the building blocks of
more general transformations, such as the affine maps we discuss next.

\paragraph{Matrix representation.}
Affine transformations can be represented in homogeneous coordinates. One introduces
\[
\tilde{\vec x}=\begin{pmatrix}\vec x\\ 1\end{pmatrix}, \qquad
\tilde M=\begin{pmatrix} M & \vec b \\ 0 & 1\end{pmatrix},
\]
so that the affine map becomes a purely linear transformation
\[
\tilde{\vec x}'=\tilde M \tilde{\vec x}.
\]
This representation is widely used in computer graphics and robotics, since it allows translations and
linear operations to be treated uniformly.

\paragraph{Decomposition.}
Any affine map can be decomposed as
\[
\vec x' = (R S P)\,\vec x + \vec b,
\]
where $R$ is a rotation, $S$ a scaling, $P$ a projection or permutation, and $\vec b$ a translation.  
Thus affine transformations combine the building blocks we discussed in Sec.~\ref{sec:geometry}.

\paragraph{Group structure.}
Affine transformations form a group under composition, denoted the affine group $\mathrm{Aff}(n)$:
\[
\mathrm{Aff}(n)=\mathrm{GL}(n,\mathbb{R})\ltimes \mathbb{R}^n,
\]
where $\mathrm{GL}(n,\mathbb{R})$ is the general linear group and $\ltimes$ denotes a semidirect product.
This structure underlies many applications in geometry and physics\footnote{See J.~Gallier and
J.~Quaintance, \emph{Notes on Affine and Projective Geometry}, Springer (2020).}.

\paragraph{Applications.}
Affine transformations appear naturally in:
\begin{itemize}
    \item Classical mechanics: Galilean transformations are affine (linear + translation in time/space).
    \item Computer vision and graphics: affine maps describe perspective projections and image warps.
    \item Crystallography: affine symmetries classify lattices under deformations.
\end{itemize}

---

\subsubsection{Complex vector spaces and unitary matrices}
\index{key}{unitary matrices}

If the underlying vector space is complex, the transpose operation is replaced by the Hermitian
conjugate ($^\dagger$). A matrix $U$ is called \emph{unitary} if
\begin{equation}
U^\dagger U = I.
\end{equation}
Unitary matrices preserve the Hermitian inner product:
\[
\langle U\vec x, U\vec y\rangle=\langle \vec x,\vec y\rangle.
\]
Thus they generalize real orthogonal matrices to complex spaces.  

In quantum mechanics, unitary transformations represent symmetries and time evolution (via the Schrödinger equation).  
The group of $n\times n$ unitary matrices, $U(n)$, plays a central role in representation theory and gauge theories\footnote{See M.~Hall, \emph{Lie Groups, Lie Algebras, and Representations}, Springer (2003).}.

---

\subsection{Special matrices}
\label{sec:specialmatrices}

We now discuss a few important classes of matrices whose spectral properties can be determined explicitly.

\subsubsection{Skew-symmetric and skew-Hermitian matrices}
\index{key}{skew-symmetric}
A matrix $A$ is \emph{skew-symmetric} if $A^{\top}=-A$, and \emph{skew-Hermitian} if $A^\dagger=-A$.  

Every matrix can be decomposed into symmetric and skew-symmetric parts:
\begin{equation}
A=\frac{A+A^\dagger}{2}+\frac{A-A^\dagger}{2}=\text{Sym}(A)+\text{Skew}(A).
\end{equation}

\paragraph{Spectral properties.}
Skew-Hermitian matrices have purely imaginary eigenvalues, occurring in pairs $\pm i\xi_j$.  
They can be block-diagonalized by unitary transformations into $2\times 2$ blocks of the form
\[
\begin{pmatrix}
0 & \xi \\
-\xi & 0
\end{pmatrix},
\]
whose eigenvalues are $\pm i\xi$. If the dimension is odd, at least one eigenvalue is zero.

---

\subsubsection{Eigenvalues for some special cases}
\label{sec:eigspecial}

For several classes of matrices, eigenvalues can be obtained directly:
\begin{itemize}
 \item \textbf{Diagonal matrices:} eigenvalues are the diagonal entries, eigenvectors are the standard basis vectors $\vec e_i$.
 \item \textbf{Triangular matrices (upper or lower):} eigenvalues are again the diagonal entries.
 \item \textbf{Orthogonal/unitary matrices:} eigenvalues lie on the unit circle, i.e.\ $\lambda=e^{i\phi}$.
 \item \textbf{Rank-one matrices:} if $M=\vec v \vec v^{\top}$, then it has one nonzero eigenvalue $\|\vec v\|^2$, and $n-1$ zero eigenvalues.
 \item \textbf{Vandermonde matrices:} defined by $A_{ij}=x_i^{\,j-1}$, their determinant is
 \[
 \det(A)=\prod_{j>i}(x_j-x_i).
 \]
 \item \textbf{Unitary matrices and Cayley transform:} if $U$ has no eigenvalue $-1$, then it can be written as
 \[
 U=(I+iH)(I-iH)^{-1}, \qquad H=i(I-U)(I+U)^{-1}.
 \]
 This is known as the \emph{Cayley decomposition}\footnote{See R.~Horn and C.~Johnson, \emph{Topics in Matrix Analysis}, Cambridge University Press (1991).}.  
 The Cayley map $f(A)=(I-A)(I+A)^{-1}$ sends skew-Hermitian matrices to unitary ones and vice versa, and satisfies $f(f(A))=A$.
\end{itemize}

\subsection{Gram--Schmidt orthogonalization process}
\label{sec:gramschmidt}
\index{key}{Gram-Schmidt decomposition}

Given a set of linearly independent vectors $V=\{\vec u_1,\ldots,\vec u_n\}$, we often wish to construct a new
set $V'=\{\vec e_1,\ldots,\vec e_n\}$ that spans the same space but is \emph{orthonormal} with respect to a
chosen scalar product $\langle \cdot,\cdot\rangle$, i.e.
\[
\text{span}(V)=\text{span}(V'), \qquad \langle \vec e_i,\vec e_j\rangle=\delta_{ij}.
\]
This is the content of the \emph{Gram--Schmidt orthogonalization procedure}\footnote{See
G.~Strang, \emph{Linear Algebra and Its Applications}, Brooks/Cole (2006).}.

---

\subsubsection{Recursive construction}
Starting from $\vec u_1,\ldots,\vec u_n$, we define recursively
\begin{equation}
\vec q_i=\vec u_i-\sum_{k=1}^{i-1}\langle \vec u_i,\vec e_k\rangle \vec e_k, 
\qquad
\vec e_i=\frac{\vec q_i}{\sqrt{\langle \vec q_i,\vec q_i\rangle}}.
\end{equation}
By construction, each $\vec q_i$ is orthogonal to all previous $\vec e_k$.  
An induction argument shows that the resulting $\{\vec e_i\}$ form an orthonormal basis.

---

\subsubsection{Projection interpretation}
At step $k$, define the matrix $P_k=[\vec e_1,\ldots,\vec e_k]$ collecting the first $k$ orthonormal
vectors. The orthogonal projector onto $\text{span}\{\vec e_1,\ldots,\vec e_k\}$ is
\begin{equation}
\Omega_k=P_k(P_k^{\top} P_k)^{-1}P_k^{\top}=P_k P_k^{\top},
\end{equation}
where the last equality holds since the columns of $P_k$ are orthonormal.  
Then Gram--Schmidt can be expressed as
\[
\vec q_{k+1}=(I-\Omega_k)\vec u_{k+1}.
\]

---

\subsubsection{Gram matrix and Gram determinant}
\index{key}{Gram matrix}\index{key}{Gram determinant}

For a set of vectors $\vec u_1,\ldots,\vec u_n$, the \emph{Gram matrix} is
\begin{equation}
G(\vec u_1,\ldots,\vec u_n)=\big(\langle \vec u_i,\vec u_j\rangle\big)_{i,j=1}^n.
\end{equation}
It is Hermitian and positive semidefinite. Under a change of basis it transforms by congruence
$G\mapsto P^*GP$. If the basis is orthonormal, $G=I$; if it is merely orthogonal, $G$ is diagonal.  

The \emph{Gram determinant} is
\begin{equation}
g(\vec u_1,\ldots,\vec u_n)=\det G(\vec u_1,\ldots,\vec u_n).
\end{equation}
Geometrically, $g$ equals the squared volume of the parallelepiped spanned by the vectors.
In particular, $g=0$ if and only if the vectors are linearly dependent.

---

\subsubsection{Closed form of Gram--Schmidt}
Using Gram determinants, one can write the orthogonalized vectors explicitly as
\begin{equation}
\vec q_p=(-1)^p\sum_{k=1}^p (-1)^k
\frac{G_p^k}{\sqrt{G_p G_{p-1}}}\,\vec u_k,
\end{equation}
where $G_p$ is the Gram determinant of $\{\vec u_1,\ldots,\vec u_p\}$, and $G_p^k$ is the minor obtained
by removing the $k$-th row and $p$-th column from the Gram matrix $R_p$.  
This formula, though rarely used in practice, emphasizes the algebraic structure of the process.

---

\subsubsection{Connection to QR decomposition}
\index{key}{QR decomposition}

An important application of Gram--Schmidt is to matrix factorizations.  
Given a matrix $M$ with columns $\vec a_i$, applying Gram--Schmidt yields an orthonormal basis
$\{\vec b_i\}$ and coefficients
\[
C_{ij}=\langle \vec a_j,\vec b_i\rangle,
\]
which are upper-triangular. Thus
\begin{equation}
M = B C,
\end{equation}
where $B$ has orthonormal columns and $C$ is triangular.  
This is precisely the \emph{QR decomposition}, a cornerstone of numerical linear algebra\footnote{See
G.~Golub and C.~Van Loan, \emph{Matrix Computations}, Johns Hopkins University Press (1996).}.

---

\subsubsection{Generalized Gram--Schmidt}
\index{key}{Gram-Schmidt decomposition (generalized)}

The procedure can be generalized to construct two sets of mutually orthogonal vectors spanning the same
space. Let $\{\vec a_i\}$ and $\{\vec b_i\}$ be two bases. Define recursively
\begin{eqnarray}
\vec p_i &=& \vec a_i-\sum_{j=1}^{i-1}\langle \vec a_i,\vec q_j\rangle \vec q_j, \\
\vec q_i &=& \vec b_i-\sum_{j=1}^{i-1}\langle \vec b_i,\vec p_j\rangle \vec p_j.
\end{eqnarray}
Then the normalized vectors
\begin{equation}
\tilde a_i=\frac{\vec p_i}{\sqrt{\langle \vec p_i,\vec q_i\rangle}}, \qquad
\tilde b_i=\frac{\vec q_i}{\sqrt{\langle \vec p_i,\vec q_i\rangle}},
\end{equation}
are mutually orthonormal: $\langle \tilde a_i,\tilde b_j\rangle=\delta_{ij}$.  

Equivalently, if $Z$ and $T$ are the Gram matrices of $\{\tilde a_i\}$ and $\{\tilde b_i\}$, then
\[
z_{ij}=(Z^{-1})_{ij}, \qquad t_{ij}=(T^{-1})_{ij}.
\]

This generalized construction is useful in biorthogonalization, with applications ranging from signal
processing to Krylov subspace methods in numerical linear algebra\footnote{See Y.~Saad,
\emph{Iterative Methods for Sparse Linear Systems}, SIAM (2003).}.

---

The Gram--Schmidt process provides:
\begin{itemize}
    \item A constructive way to obtain orthonormal bases from arbitrary bases,
    \item Geometric meaning through the Gram determinant (volume),
    \item Practical algorithms such as the QR decomposition,
    \item Generalizations for biorthogonal systems used in numerical methods.
\end{itemize}
It is thus both a fundamental theoretical tool and a practical workhorse in computations.

\section{The Resolvent}\label{sec:resolvent}
\index{key}{resolvent}

Given a square matrix $A\in\mathbb{C}^{n\times n}$, the \emph{resolvent} is defined as
\[
   R(A,z)=(zI-A)^{-1},
\]
for those complex numbers $z$ such that the inverse exists 
(i.e.\ $z\notin \Lambda(A)$, where $\Lambda(A)$ is the spectrum of $A$).  
The resolvent is a central object in spectral theory: it encodes the location of the spectrum, 
and, remarkably, all analytic functions of $A$ can be reconstructed from it by contour integrals.  

\paragraph{Algebraic expression.}
Expanding $(zI-A)^{-1}$ in terms of cofactors gives
\begin{equation}
   (zI-A)^{-1}=\frac{1}{\det(zI-A)}\,C^{\top}(zI-A),
\end{equation}
where $C(zI-A)$ is the cofactor matrix.  
Since $\det(zI-A)=\prod_{i=1}^n(z-\lambda_i)$, poles of the resolvent are exactly the eigenvalues of $A$.  
By the Cayley--Hamilton theorem, one can express $C^{\top}(zI-A)$ as a polynomial in $A$, so the resolvent is 
always a rational function in $A$.

\paragraph{Resolvent identities.}
Two identities are fundamental in perturbation theory and operator analysis:
\begin{itemize}
   \item \textbf{First resolvent identity:}
   \[
      R(A,z_1)-R(A,z_2) = (z_2-z_1)R(A,z_1)R(A,z_2).
   \]
   This relates resolvents at different spectral parameters $z_1,z_2$, and is crucial in contour integral manipulations.
   \item \textbf{Second resolvent identity:}
   If $B=A+C$, then
   \[
      R(B,z)-R(A,z) = R(B,z)\,C\,R(A,z).
   \]
   This formula describes how the resolvent changes under a perturbation $C$ 
   and underlies stability results for spectra and pseudospectra.
\end{itemize}
These identities will resurface when we discuss pseudospectra (Sec.~\ref{sec:pseudospectra}) 
and perturbation bounds (Sec.~\ref{sec:matrixnorm}).

\paragraph{Series expansion.}
For large $|z|$, one has the Neumann expansion
\begin{equation}
   R(A,z)=\frac{1}{z}\Big(I+\frac{A}{z}+\frac{A^2}{z^2}+\cdots\Big),
   \label{eq:resolvent-series}
\end{equation}
which is convergent for $|z|>\rho(A)$, where $\rho(A)$ is the spectral radius.  
This makes explicit the connection between the resolvent and the powers of $A$, 
a fact that will be used repeatedly in dynamical contexts.

\paragraph{Applications.}
The resolvent has numerous interpretations across mathematics and applications:
\begin{itemize}
   \item In \emph{differential equations}, $(zI-A)^{-1}$ is the Laplace transform of the semigroup $e^{At}$, i.e.
   \[
      (zI-A)^{-1} = \int_0^\infty e^{-zt}e^{At}\,dt, \qquad \mathrm{Re}(z)>\rho(A).
   \]
   Thus, the resolvent links spectral theory with time evolution.
   \item In \emph{economics}, the Leontief input--output model uses the matrix inverse
   \[
      \vec o = (I-A)^{-1}\vec y,
   \]
   where $A$ encodes production coefficients and $\vec y$ is external demand.  
   Convergence of the Neumann series
   \[
      (I-A)^{-1}=I+A+A^2+\cdots
   \]
   requires $\rho(A)<1$, which is interpreted as an economy being productive (output exceeds input).
   \item In \emph{probability theory}, the resolvent of a Markov transition operator provides occupation measures and potential kernels.
   \item In \emph{quantum mechanics}, $R(H,z)$ for a Hamiltonian $H$ is the Green’s function, encoding propagators and spectral density.
\end{itemize}

The resolvent will appear repeatedly in later chapters:
\begin{itemize}
   \item In pseudospectral theory, its norm $\|R(A,z)\|$ defines the $\epsilon$-pseudospectrum.
   \item In stability analysis, the poles of $R(A,z)$ determine the growth of $e^{At}$.
   \item In random matrix theory, the resolvent is used to compute limiting spectral densities via the Stieltjes transform.
\end{itemize}
It is therefore a unifying object linking linear algebra, operator theory, and applications.

\paragraph{Matrix functions and the analytic functional calculus.}
We now mention an application of the resolvent beyond the results above.
A natural extension of scalar functions to matrices is the notion of a \emph{matrix function}.\footnote{Nicholas J. Higham, \textit{Functions
of Matrices:
Theory and Computation}, SIAM 2008 }
The notion is quite straightforward 
if $A$ is diagonalizable, with spectral decomposition 
\[
   A = V \Lambda V^{-1}, \qquad 
   \Lambda = \mathrm{diag}(\lambda_1,\dots,\lambda_n),
\]
then for any function $f$ defined on the spectrum $\Lambda(A)=\{\lambda_i\}$, one sets
\[
   f(A) = V \, f(\Lambda)\, V^{-1} = V \, \mathrm{diag}\big(f(\lambda_1),\dots,f(\lambda_n)\big) \, V^{-1}.
\]
In words, the function is applied directly to the eigenvalues, while the eigenvectors determine the basis in which $A$ is represented. 
This simple recipe covers many important cases, such as $e^A$, $\log A$, and $A^\alpha$, provided $f$ is well defined on the eigenvalues of $A$. 
If $A$ is not diagonalizable, one must use its Jordan normal form, in which case derivatives of $f$ enter, but the principle remains the same. A much more careful analysis is provided in the book by Higham.

Here, we mention a powerful general definition that is obtained when $f$ is analytic on an open domain $\Omega \subset \mathbb{C}$ containing $\Lambda(A)$. 
In this case, one defines
\begin{equation}
   f(A) = \frac{1}{2\pi i}\int_\Gamma f(z)\,(zI - A)^{-1}\,dz,
   \label{eq:resolvent-calculus}
\end{equation}
where $\Gamma$ is a contour in $\Omega$ enclosing the spectrum of $A$. 
This is known as the \emph{Dunford--Taylor functional calculus}\footnote{See N.~Dunford and J.~T.~Schwartz, 
\emph{Linear Operators, Part I}, Interscience (1958).} or \emph{Dunford--Taylor formula}. 
Formula \eqref{eq:resolvent-calculus} shows that the resolvent $(zI-A)^{-1}$ acts as the universal kernel for constructing analytic functions of matrices. 
In later sections, we will see how this connects to the spectral decomposition, to semigroup theory (via $e^{tA}$), and to stability analysis of dynamical systems.

 \chapter{Linear transformations and their inverse maps}
 
\section{Linear transformations between $\mathbb{R}^m \rightarrow \mathbb{R}^n$}
\label{sec:lintrans}

A \emph{linear transformation} is a map $T:\mathbb{R}^m\to \mathbb{R}^n$ of the form
\[
T(\vec x)=A \vec x,
\]
where $A$ is an $n\times m$ matrix. Such transformations include familiar operations such as
rotations, reflections, scalings, and projections.  

\paragraph{Preservation of structure.}
Linear maps preserve the vector space structure: for all $\vec x,\vec y\in\mathbb{R}^m$ and scalars
$\alpha,\beta$, one has
\[
T(\alpha \vec x+\beta \vec y)=\alpha T(\vec x)+\beta T(\vec y).
\]
This means they preserve the origin, lines, and parallelism, though not necessarily distances or angles
(unless $A$ is orthogonal).

\paragraph{Examples.}
\begin{itemize}
\item A rotation in $\mathbb{R}^2$ by angle $\theta$ is given by
\[
R(\theta)=\begin{pmatrix}\cos\theta & -\sin\theta \\ \sin\theta & \cos\theta\end{pmatrix}.
\]
\item A projection onto the $x$-axis is given by
\[
P=\begin{pmatrix}1 & 0 \\ 0 & 0\end{pmatrix}.
\]
\item A scaling by factor $\alpha$ in every direction corresponds to
\[
S(\alpha)=\alpha I.
\]
\end{itemize}

\subsection{Solving linear systems}
\index{key}{linear systems}\index{key}{least squares}\index{key}{pseudoinverse}

One of the most basic but fundamental problems in linear algebra is the solution
of a linear system
\[
A \vec{x} = \vec{b},
\]
where $A \in \mathbb{R}^{n\times m}$ and $\vec{b} \in \mathbb{R}^n$. The behavior of
the solution depends crucially on the shape and rank of $A$:

\begin{itemize}
\item \textbf{Square and nonsingular case:} If $A$ is $n\times n$ and
$\det(A)\neq 0$, then the system admits a unique solution given explicitly by
\[
\vec{x} = A^{-1} \vec{b}.
\]

\item \textbf{Rectangular or rank-deficient case:} If $A$ is not invertible
(either rectangular or singular), one uses the \emph{Moore–Penrose pseudoinverse}
$A^+$ to obtain the least-norm solution:
\[
\vec{x} = A^+ \vec{b}.
\]
This solution minimizes $\|A\vec{x}-\vec{b}\|_2$ among all possible $\vec{x}$, and coincides with
the exact solution when one exists.

\item \textbf{Existence criteria:} In general, consistency can be tested by
constructing the augmented matrix $B=[A\ \vec{b}]$. Then:
\[
\begin{array}{c|c}
\text{Condition} & \text{Solution} \\
\hline
\operatorname{rank}(A) = \operatorname{rank}(B) = m & \text{Unique solution} \\
\operatorname{rank}(A) = \operatorname{rank}(B) < m & \text{Infinitely many solutions} \\
\operatorname{rank}(A) < \operatorname{rank}(B) & \text{No solution (inconsistent)}
\end{array}
\]

\end{itemize}

These cases provide the foundation for more advanced treatments such as least
squares regression, regularized inverses, and iterative methods that we will discuss in a moment. In practice,
explicit inverses are rarely computed directly: factorizations (LU, QR, SVD)
and iterative algorithms provide stable and efficient ways to obtain solutions.

\subsection{General solution of linear systems}
\index{key}{linear systems}\index{key}{least squares}\index{key}{Cramer's rule}\index{key}{pseudoinverse}

A fundamental problem in linear algebra is to solve the system
\[
A \vec{x} = \vec{b},
\]
with $A \in \mathbb{R}^{n\times m}$ and $\vec{b}\in\mathbb{R}^n$. Depending on the rank
and shape of $A$, several distinct cases arise:

\begin{itemize}
\item \textbf{Square and nonsingular.} If $A$ is $n\times n$ and invertible, then
\[
\vec{x} = A^{-1} \vec{b}.
\]
In principle, this is the easiest case from a conceptual standpoint.

\item \textbf{Rank-deficient or rectangular.} If $A$ is not invertible, the system
may have infinitely many or no solutions. In this case the \emph{Moore–Penrose
pseudoinverse} provides the least-norm least-squares solution:
\[
\vec{x} = A^+ \vec{b},
\]
which we will discuss later.

\item \textbf{Existence and uniqueness.} Using the augmented matrix $B=[A\ \vec{b}]$:
\[
\begin{array}{c|c}
\text{Condition} & \text{Solution} \\
\hline
\operatorname{rank}(A)=\operatorname{rank}(B)=m & \text{Unique solution} \\
\operatorname{rank}(A)=\operatorname{rank}(B)<m & \text{Infinitely many solutions} \\
\operatorname{rank}(A)<\operatorname{rank}(B) & \text{No solution}
\end{array}
\]

\item \textbf{Over-determined systems.} For $A\in\mathbb{R}^{n\times m}$ with $n>m$ and
$\operatorname{rank}(A)=m$, the least-squares solution is
\[
\vec{x} = (A^\top A)^{-1} A^\top \vec{b} = A^+ \vec{b},
\]
minimizing $\|A\vec{x}-\vec{b}\|_2^2$. We will also discuss this later.

\item \textbf{Under-determined systems.} For $A\in\mathbb{R}^{n\times m}$ with $n<m$ and
$\operatorname{rank}(A)=n$, the minimum-norm solution is
\[
\vec{x} = A^\top (AA^\top)^{-1} \vec{b},
\]
and in general the full solution set is $\vec{x} = A^+ \vec{b} + (I-A^+ A) \vec{y}$ for arbitrary $\vec{y}$.

\item \textbf{Cramer’s rule.} If $A$ is invertible, then each coordinate can be
written explicitly as
\[
x_i = \frac{\det(A_i)}{\det(A)},
\]
where $A_i$ is $A$ with its $i$-th column replaced by $\vec{b}$ (see also
Section~\ref{sec:matrix_functions_cramer} for a determinantal and spectral
viewpoint).
\end{itemize}

These formulas summarize the basic cases of linear systems and give a general sense of what type of solution one should be looking for.

\subsubsection{General affine transformations}
\label{sec:affine}
\index{key}{affine transformation}

An \emph{affine transformation} is a map of the form
\begin{equation}
\vec x' = M \vec x + \vec b,
\end{equation}
where $M$ is a linear operator (represented by a matrix) and $\vec b$ is a fixed translation vector.  
The linear part $M$ can encode scaling, rotation, shear, reflection, or projection, while $\vec b$ represents
a shift of the origin.  

Affine transformations generalize linear transformations by allowing translations. They are the most
general transformations that preserve straight lines and parallelism, though they do not necessarily
preserve angles or lengths\footnote{See S.~Lang, \emph{Linear Algebra}, Springer (1987).}.

Affine maps can be written as linear maps in homogeneous coordinates:
\[
\tilde{\vec x}=\begin{pmatrix}\vec x\\1\end{pmatrix}, \qquad
\tilde A=\begin{pmatrix} A & \vec b \\ 0 & 1\end{pmatrix}, \qquad
\tilde{\vec x}'=\tilde A\tilde{\vec x}.
\]
The collection of all affine maps forms the \emph{affine group} $\mathrm{Aff}(n)$, which is the semidirect
product $\mathrm{GL}(n,\mathbb{R})\ltimes \mathbb{R}^n$\footnote{See J.~Gallier and J.~Quaintance,
\emph{Notes on Affine and Projective Geometry}, Springer (2020).}.

---

\subsection*{Invertibility of linear maps and connection to matrix inverses}
A central question is whether a linear map $T(\vec x)=A\vec x$ has an \emph{inverse}.  
This requires:
\begin{itemize}
\item $m=n$ (the map must go from $\mathbb{R}^n$ to itself),
\item $A$ must be \emph{invertible}, i.e.\ $\det(A)\neq 0$.
\end{itemize}
In that case, the inverse map $T^{-1}$ is again linear, with
\[
T^{-1}(\vec y)=A^{-1}\vec y.
\]
Thus, the inverse of a linear transformation is represented exactly by the inverse of its matrix.  

\paragraph{Examples.}
\begin{itemize}
\item A rotation matrix $R(\theta)$ is always invertible, with inverse $R(-\theta)=R(\theta)^\top$.
\item A projection matrix $P$ is not invertible, since $\det(P)=0$ and information is lost.
\item A scaling $S(\alpha)$ is invertible if and only if $\alpha\neq 0$, with inverse $S(\alpha)^{-1}=S(1/\alpha)$.
\end{itemize}

\paragraph{Affine case.}
An affine transformation $T(\vec x)=A\vec x+\vec b$ is invertible precisely when $A$ is invertible.
The inverse is then
\[
T^{-1}(\vec y)=A^{-1}(\vec y-\vec b).
\]

The study of linear and affine maps thus naturally leads to the problem of \emph{matrix inversion}:
given $A$, when does $A^{-1}$ exist, how can it be computed, and what properties does it have?
These questions will be the subject of the next section.

---

\section{Matrix inversion}
\label{sec:inverseprob}

Inverting a matrix is a central task in linear algebra, with applications ranging from solving systems of
equations to optimization and statistics. The fundamental problem is:
\begin{equation}
A \vec x=\vec b,
\label{eq:inverse1}
\end{equation}
where $A$ and $\vec b$ are given, and $\vec x$ is unknown.

\paragraph{Equivalent conditions for invertibility.}
For an $n\times n$ matrix $A$, the following are equivalent:
\begin{itemize}
 \item $A$ has an inverse $A^{-1}$,
 \item $\det(A)\neq 0$,
 \item $\ker(A)=\{0\}$,
 \item $A^{\top}$ (or $A^\dagger$ in the complex case) is invertible,
 \item the system $A\vec x=\vec b$ has a unique solution for all $\vec b$,
 \item the left inverse and right inverse coincide.
\end{itemize}

---

\subsection*{Computational approaches}

\paragraph{Iterative inversion.}
In principle, an inverse can be computed iteratively:
\[
X_{k+1}=X_k-X_kAX_k,
\]
starting from $X_0=I$. Such schemes are used in numerical linear algebra for preconditioning.

\paragraph{Spectral decomposition.}
If $A$ is diagonalizable by an orthogonal (or unitary) matrix, $A=O\Sigma O^{\top}$, then
\[
A^{-1}=O \Sigma^{-1} O^{\top},
\]
where $\Sigma^{-1}$ is obtained by inverting each nonzero diagonal element.

\paragraph{Adjugate formula.}
Formally, the inverse can be expressed using cofactors. If $C_{ij}$ is the determinant of the minor
obtained by removing row $i$ and column $j$, then
\begin{equation}
A^{-1}=\frac{1}{\det(A)} C^{\top},
\label{eq:cofinv}
\end{equation}
where $C^{\top}$ is the adjugate matrix\index{key}{adjugate matrix}. This satisfies
\[
A C^{\top}=\det(A) I.
\]

---

\subsection*{Series and polynomial formulas}

The adjugate formula can be related to expansions involving traces and Bell polynomials. For example,
\begin{equation}
A^{-1}=\frac{1}{\det(A)} \sum_{s=1}^{n-1} A^s 
\sum_{k_1,\cdots,k_{n-1}}
\prod_{l=1}^{n-1}\frac{(-1)^{k_l+1}}{l^{k_l} k_l!}
\big(\text{Tr}(A^l)\big)^{k_l},
\label{eq:decdio}
\end{equation}
where the integers $k_l$ satisfy the Diophantine condition
$s+\sum_{l=1}^{n-1} l k_l=n-1$.

Alternatively, using Bell polynomials $B_m$, one may write
\begin{equation}
A^{-1}=\frac{1}{\det(A)} \sum_{s=1}^n A^{s-1}
\frac{(-1)^{n-1}}{(n-s)!} B_{n-s}(t_1,\ldots,t_{n-s}),
\end{equation}
with $t_i=-(i-1)!\,\text{Tr}(A^i)$.

\paragraph{Series expansion.}
If $\|A\|<1$ (for some submultiplicative norm), then
\begin{equation}
(I-A)^{-1}=\sum_{k=0}^\infty A^k,
\end{equation}
a convergent geometric series.

\paragraph{Derivative identity.}
If $A$ depends on a parameter $\alpha$, then
\begin{equation}
\frac{\partial}{\partial \alpha} A^{-1}
=-A^{-1}\,\frac{\partial A}{\partial \alpha}\,A^{-1}.
\end{equation}
This follows from differentiating $A^{-1}A=I$.

---

\subsection*{Block matrix inversion and Schur complement}

If $Q$ has a block structure
\[
Q=\begin{pmatrix}A & B \\ C & D\end{pmatrix},
\]
with $A$ invertible, then
\begin{equation}
Q^{-1}=\begin{pmatrix}
A^{-1}+A^{-1}B(D-CA^{-1}B)^{-1}CA^{-1} & -A^{-1}B(D-CA^{-1}B)^{-1} \\
-(D-CA^{-1}B)^{-1}CA^{-1} & (D-CA^{-1}B)^{-1}
\end{pmatrix},
\end{equation}
where $D-CA^{-1}B$ is the \emph{Schur complement}\index{key}{Schur complement}.  
Such block formulas are heavily used in numerical algorithms and statistics\footnote{See
S.~Boyd and L.~Vandenberghe, \emph{Convex Optimization}, Cambridge University Press (2004).}.

---

\subsection{Laplace and Jacobi identities}

Several useful identities connect determinants, adjugates, and traces.

\paragraph{Jacobi’s differential formula.}
If $A(t)$ depends smoothly on a parameter $t$, then
\begin{equation}
\frac{d}{dt}\det(A(t))=\text{Tr}\!\left(\text{adj}(A(t))\,\frac{dA(t)}{dt}\right).
\end{equation}
Since $\text{adj}(A)=\det(A) A^{-1}$, this becomes
\begin{equation}
\frac{d}{dt}\log\det(A(t))=\text{Tr}\!\left(A(t)^{-1}\frac{dA(t)}{dt}\right).
\end{equation}
For $A(t)=e^{tB}$, this yields the important identity
\begin{equation}
\det(e^{tB})=e^{t\,\text{Tr}(B)}.
\end{equation}

\paragraph{Application to resolvents.}
Let $A(t)=tI-B=R(t,B)^{-1}$ be the resolvent. Then
\[
\frac{d}{dt}\det(A(t))=\det(tI-B)\,\text{Tr}\,(tI-B)^{-1}.
\]

\paragraph{Connection with characteristic polynomials.}
The characteristic polynomial is
\[
p_A(\lambda)=\det(\lambda I-A).
\]
Jacobi’s formula shows that
\[
-\frac{d}{d\lambda}p_A(\lambda)=\text{Tr}\,\text{adj}(\lambda I-A).
\]
Evaluated at $\lambda=0$, this relates the trace of the adjugate to the derivative of the characteristic
polynomial.

\paragraph{Block Jacobi identities.}
For a block matrix
\[
A=\begin{pmatrix}B & D \\ E & C\end{pmatrix}, \qquad
A^{-1}=\begin{pmatrix}W & X \\ Y & Z\end{pmatrix},
\]
Jacobi showed that
\[
\det(A)\det(Z)=\det(B).
\]

Another identity concerns subdeterminants: if $A$ is partitioned into corners
\begin{equation}
A_{tl}, A_{tr}, A_{bl}, A_{br}, A_c,
\end{equation}
then
\begin{equation}
\det(A)\det(A_c)=\det(A_{tl})\det(A_{br})-\det(A_{tr})\det(A_{bl}).
\end{equation}
Such relations are useful in combinatorics and statistical mechanics\footnote{See
K.~Okamoto, ``Determinantal identities in statistical physics,'' \emph{Condensed Matter Physics}
\textbf{3}, 4250 (2000).}.

\subsubsection{Inversion as a minimization problem}
\label{sec:invmin}

The solution of the linear system
\begin{equation}
A\vec x=\vec b
\label{eq:inverse2}
\end{equation}
can also be viewed through the lens of optimization.  
Instead of solving directly, one can look for $\vec x$ that minimizes the functional
\begin{equation}
L(\vec x)=F(A\vec x-\vec b)+R(\vec x),
\label{eq:reginv}
\end{equation}
where $F$ is a vector norm (measuring residual error) and $R$ is a regularization term.  

\begin{itemize}
    \item If $F(\vec y)=\|\vec y\|_2^2$ and $R(\vec x)=0$, then minimizing $L$ gives the least-squares
    solution.
    \item If $R(\vec x)=\lambda\|\vec x\|_2^2$ with $\lambda>0$, one obtains the Tikhonov regularized
    solution, or \emph{ridge regression}.
    \item More general choices of $R$ lead to sparsity-promoting or constrained solutions, widely used in
    statistics and machine learning\footnote{See S.~Boyd and L.~Vandenberghe, \emph{Convex Optimization},
    Cambridge University Press (2004).}.
\end{itemize}

This optimization viewpoint provides stability and flexibility in problems where $A$ is ill-conditioned or
nearly singular. In fact, many iterative solvers (gradient descent, conjugate gradients, etc.) can be derived
directly from minimization principles.

---

\subsection{Matrix invertibility and condition number}
\label{sec:condnum}

We now return to the question of matrix invertibility from a numerical perspective.  
In exact algebra, invertibility is equivalent to $\det(A)\neq 0$. But in computation, the relevant question is
how \emph{sensitive} the solution $\vec x=A^{-1}\vec b$ is to perturbations in $\vec b$ or $A$.

\paragraph{Perturbation in $\vec b$.}
Suppose $\vec b$ changes to $\vec b+\Delta \vec b$. Then the solution changes to
\[
\Delta \vec x=A^{-1}\Delta \vec b.
\]
Using the induced $L^2$ norm (see Sec.~\ref{sec:matrixnorm}), we have
\begin{equation}
\|\Delta \vec x\|\leq \|A^{-1}\|\,\|\Delta \vec b\|.
\end{equation}
Meanwhile,
\begin{equation}
\|\vec b\|=\|A\vec x\|\leq \|A\|\,\|\vec x\|.
\end{equation}
Combining these gives
\begin{equation}
\frac{\|\Delta \vec x\|}{\|\vec x\|}
\;\leq\;\|A\|\,\|A^{-1}\|\;\frac{\|\Delta \vec b\|}{\|\vec b\|}.
\end{equation}

\paragraph{Condition number.}
The product
\begin{equation}
k(A)=\|A\|\,\|A^{-1}\|
\end{equation}
is called the \emph{condition number}\index{key}{condition number}.  
It measures how errors in the input propagate to the output. If $k(A)$ is large, small perturbations in $\vec b$
lead to large relative errors in $\vec x$.  

\paragraph{Perturbation in $A$.}
If $A$ is perturbed to $A+\delta A$, then
\[
\Delta \vec x=A^{-1}\big(\Delta \vec b-\delta A(\vec x+\Delta \vec x)\big).
\]
Following the same steps as above leads to
\begin{equation}
\frac{\|\Delta \vec x\|}{\|\vec x\|}
\;\leq\;\frac{k(A)}{1-k(A)\tfrac{\|\delta A\|}{\|A\|}}
\left(\frac{\|\Delta \vec b\|}{\|\vec b\|}-\frac{\|\delta A\|}{\|A\|}\right).
\end{equation}
If $\|A^{-1}\|\,\|\delta A\|\ll 1$, this simplifies to
\begin{equation}
\frac{\|\Delta \vec x\|}{\|\vec x\|}
\;\lesssim\; k(A)\left(\frac{\|\Delta \vec b\|}{\|\vec b\|}-\frac{\|\delta A\|}{\|A\|}\right).
\end{equation}
Thus the condition number governs sensitivity to perturbations in both $\vec b$ and $A$.

\paragraph{Properties.}
The condition number satisfies:
\begin{itemize}
\item $k(I)=1$,
\item for diagonal $D$, $k(D)=\max_i |D_{ii}|/\min_i |D_{ii}|$,
\item $k(A)\geq 1$ for any nonsingular $A$,
\item $k(UAU^{-1})=k(A)$ for any unitary $U$,
\item $k(A^n)=k(A)^n$,
\item if $A$ is singular, $k(A)=\infty$,
\item $k(cA)=k(A)$ for any nonzero scalar $c$.
\end{itemize}

\paragraph{Remarks.}
The condition number is fundamental in numerical linear algebra\footnote{See
G.~Golub and C.~Van Loan, \emph{Matrix Computations}, Johns Hopkins University Press (1996).}.  
Poorly conditioned matrices ($k(A)\gg 1$) can lead to catastrophic numerical errors — a fact dramatically
illustrated by the Ariane 5 rocket failure, which was triggered by a floating-point conversion error in a
poorly conditioned system\footnote{See D.~P.~Gollmann, ``Why Ariane 5 exploded,'' \emph{IEEE Spectrum}
\textbf{33}, 98–100 (1996).}.

\subsection{Inversion via Singular Value Decomposition (SVD)}
\label{sec:svdinv}
\index{key}{singular value decomposition}

A powerful tool for understanding matrix inversion, particularly when $A$ is rectangular or ill-conditioned,
is the \emph{singular value decomposition} (SVD). Any $m\times n$ matrix $A$ can be written as
\begin{equation}
A=U \Sigma V^\dagger,
\end{equation}
where
\begin{itemize}
    \item $U\in\mathbb{C}^{m\times m}$ and $V\in\mathbb{C}^{n\times n}$ are unitary,
    \item $\Sigma\in\mathbb{R}^{m\times n}$ is diagonal, with entries $\sigma_1\geq \sigma_2\geq\cdots\geq 0$,
    the \emph{singular values}.
\end{itemize}

If $A$ is square and nonsingular, the inverse is simply
\[
A^{-1}=V \Sigma^{-1} U^\dagger,
\]
where $\Sigma^{-1}$ inverts each nonzero diagonal entry.  
If some singular values vanish, then $A$ is not invertible — but this motivates the
\emph{pseudoinverse}.

The SVD highlights numerical stability: small singular values make $A$ nearly singular, and the ratio
$\sigma_{\max}/\sigma_{\min}$ is exactly the 2-norm condition number $k_2(A)$
(see Sec.~\ref{sec:condnum})\footnote{See G.~Golub and C.~Van Loan,
\emph{Matrix Computations}, Johns Hopkins University Press (1996).}.

---
\subsection{Inverse and generalized inverses: general comments}

The ordinary inverse of a square matrix $A\in\mathbb{C}^{n\times n}$, if it exists, is the unique matrix $X$ such that
\[
AX = XA = I.
\]
Invertibility requires that $\det(A)\neq 0$, i.e.\ that $A$ has full rank. In that case the inverse $A^{-1}$ inherits many familiar properties: it is unique, linear in the sense $(AB)^{-1}=B^{-1}A^{-1}$, and its eigenvalues are reciprocals of those of $A$.

But in many contexts, particularly when matrices are singular, or when we want to isolate long-time dynamics of Markov processes, \emph{generalized inverses} are more appropriate. A generalized inverse relaxes some of the conditions of the usual inverse. For example, one often demands only that
\[
AXA = A, \qquad XAX = X,
\]
sometimes also requiring commutation $AX=XA$ or adjoint compatibility. 

Different choices of these defining conditions yield different generalized inverses:
\begin{itemize}
\item the \emph{Moore–Penrose inverse}, defined by the four Penrose equations, ubiquitous in statistics and least squares;
\item the \emph{group inverse}, defined when the matrix has index one, useful in perturbation theory and stochastic models;
\item the \emph{Drazin inverse}, which extends the group inverse to arbitrary index $k$.
\end{itemize}

We will discuss them in detail in what follows.

\subsection{Moore--Penrose inverse}
\label{sec:moorepenrose}
\index{key}{Moore-Penrose inverse} 
\index{key}{pseudo-inverse}

When a matrix $A$ is rectangular or singular, the ordinary inverse does not exist.  
In this case, one uses the \emph{Moore--Penrose pseudoinverse} $A^+$, introduced by Penrose in 1955\footnote{R.~Penrose,
``A generalized inverse for matrices,'' \emph{Proceedings of the Cambridge Philosophical Society} \textbf{51}, 406–413 (1955).}.  
It is the unique matrix satisfying the four \emph{Penrose conditions}:
\begin{enumerate}
\item $A A^+ A=A$,
\item $A^+ A A^+=A^+$,
\item $(A A^+)^\dagger=A A^+$,
\item $(A^+ A)^\dagger=A^+ A$.
\end{enumerate}

Thus $A^+$ generalizes the inverse to all matrices, whether square or rectangular.

\paragraph{Explicit forms.}
If $A\in\mathbb{R}^{m\times n}$ has full row rank ($r(A)=m$), then
\begin{equation}
A^+=A^{\top} (A A^{\top})^{-1}.
\end{equation}
If $A$ has full column rank ($r(A)=n$), then
\begin{equation}
A^+=(A^{\top} A)^{-1} A^{\top}.
\end{equation}
In the general case, the pseudoinverse is best computed using the SVD:
\begin{equation}
A^+=V \Sigma^+ U^\dagger,
\end{equation}
where $\Sigma^+$ is obtained by inverting all nonzero singular values in $\Sigma$ and leaving zeros in
place.

\paragraph{Scalar and vector intuition.}
For a scalar $\alpha$, $\alpha^+=\alpha^{-1}$ if $\alpha\neq 0$, and $0$ otherwise.
For a nonzero vector $\vec x$, regarded as a column matrix, one has
\[
\vec x^+=\frac{\vec x^{\top}}{\vec x\cdot\vec x}.
\]

\paragraph{Block structure.}
If $A=(A_1\; A_2)$ with full rank, one can construct pseudoinverses using projectors:
\begin{eqnarray}
P_2&=&I-A_1 (A_1^{\top} A_1)^{-1} A_1^{\top}, \\
P_1&=&I-A_2 (A_2^{\top} A_2)^{-1} A_2^{\top},
\end{eqnarray}
leading to
\begin{equation}
A^+=\left((P_2 A_1)^+ \;\; (P_1 A_2)^+\right)^{\top}.
\end{equation}
Such formulas are useful in structured least-squares problems.

---

\subsubsection{Greville’s theorem}
\label{sec:greville}
\index{key}{Greville's theorem}

Several algorithms exist for computing the pseudoinverse. One elegant method is
\emph{Greville’s theorem}, a recursive scheme that updates $A^+$ column by column.  
Suppose
\[
A_k=\big(A_{k-1}\ \vline\ \vec a_k\big),
\]
where $\vec a_k$ is the new column. If $A_{k-1}^+$ is known, then
\begin{equation}
A_k^+=\begin{pmatrix}
A_{k-1}^+(I-\vec a_k \vec p_k^{\top}) \\

\vec p_k^{\top}
\end{pmatrix},
\end{equation}
where
\begin{equation}
\vec p_k=\begin{cases}
\frac{(I-A_{k-1} A_{k-1}^+)\vec a_k}{\|(I-A_{k-1} A_{k-1}^+)\vec a_k\|^2}, & \text{if the numerator $\neq 0$}, \\
\frac{A_{k-1}^+ (A_{k-1}^+)^{\top} \vec a_k}{1+\|A_{k-1}^+ \vec a_k\|^2}, & \text{otherwise}.
\end{cases}
\end{equation}
Starting from $A_1=\vec a_1$, with $A_1^+=\vec a_1^{\top}/(\vec a_1\cdot \vec a_1)$ (if $\vec a_1\neq 0$), this
yields $A^+$ after $n$ steps\footnote{See T.N.E.~Greville, ``Some applications of the pseudoinverse of a
matrix,'' \emph{SIAM Review} \textbf{2}, 15–22 (1960).}.

---

\subsubsection{Tikhonov regularization}
\label{sec:tychonov}
\index{key}{Tikhonov regularization}
\index{key}{Tikhonov matrix}

The pseudoinverse can also be defined as a limit of regularized inverses. For any matrix $A$,
\begin{equation}
A^+=\lim_{\gamma\to 0} (A^{\top} A+\gamma^2 I)^{-1} A^{\top}
=\lim_{\gamma\to 0} A^{\top} (A A^{\top}+\gamma^2 I)^{-1}.
\label{eq:tychbas}
\end{equation}
This corresponds to the special case of \emph{Tikhonov regularization} (also known as ridge regression).  

More generally, one can consider the minimization problem (cf.\ Sec.~\ref{sec:invmin}):
\begin{equation}
\mathcal L(\vec x)=\|A\vec x-\vec b\|_P^2+\|\Gamma \vec x-\vec x^0\|_Q^2,
\label{eq:tycreg}
\end{equation}
where $P,Q$ are positive definite weight matrices, $\Gamma$ is a regularization operator, and $\vec x^0$ is a
prior estimate. The minimizer is
\begin{equation}
\vec x=(A^{\top} P A+Q)^{-1}(A^{\top} P \vec b+Q \vec x^0).
\label{eq:gensolinv}
\end{equation}
Choosing $P=I$, $Q=\gamma^2 I$, and $\vec x^0=0$ recovers (\ref{eq:tychbas}).  
If $Q=\Gamma^{\top} \Gamma$, one obtains
\begin{equation}
\vec x=(A^{\top} A+\Gamma^{\top} \Gamma)^{-1} A^{\top} \vec b,
\label{eq:gensolinv2}
\end{equation}
the general Tikhonov regularized solution\footnote{See A.N.~Tikhonov and V.Y.~Arsenin,
\emph{Solutions of Ill-posed Problems}, Winston (1977).}.  
Here $\Gamma$ is called the \emph{Tikhonov matrix}.

---

The pseudoinverse provides a consistent extension of the matrix inverse:
\begin{itemize}
    \item It always exists and is unique.
    \item It yields least-squares solutions when no exact solution exists.
    \item It is stable under regularization, connecting to optimization.
    \item It can be computed via SVD, recursive updates (Greville), or regularization (Tikhonov).
\end{itemize}

\subsubsection{Bott--Duffin inverse}
\label{sec:bottduffin}
\index{key}{Bott-Duffin inverse}

The \emph{Bott--Duffin inverse} was introduced by R.~Bott and R.~J.~Duffin in their 1953 paper on
generalized inverses\footnote{R.~Bott and R.~J.~Duffin, ``On the algebra of networks,'' 
\emph{Transactions of the American Mathematical Society} \textbf{74}, 99–109 (1953).}.  
It is an inverse defined under constraints given by projectors.

\paragraph{Setup.}
Let $G$ be a square matrix acting on a vector space $V$.  
Introduce a projector $P$ and its complement $P'=I-P$.  
Define the \emph{discriminant}
\[
D=\det(GP+P').
\]
If $D\neq 0$, then $(GP+P')$ is invertible, and the Bott--Duffin inverse of $G$ with respect to $P$ is
\begin{equation}
T=P\,(GP+P')^{-1}.
\label{eq:BottDuffin}
\end{equation}

\paragraph{First Bott--Duffin theorem.}
Given a vector $\vec h$, decompose $\vec u=\vec v+\vec v'$ with $\vec v=P\vec u$ and $\vec v'=P'\vec u$.  
Consider the equation
\begin{equation}
G\vec v+\vec v'=\vec h.
\label{eq:fbdt}
\end{equation}
If $D\neq 0$, then there is a unique solution:
\begin{eqnarray}
\vec v &=& T\vec h, \\
\vec v'&=&(I-GT)\vec h.
\end{eqnarray}
The proof follows directly from writing $\vec u=(GP+P')^{-1}\vec h$, and projecting with $P$ and $P'$.

\paragraph{Properties.}
The Bott--Duffin inverse satisfies relations such as
\[
P=TGP=PGT, \qquad T=PT=TP.
\]
Differentiating the potential $\psi=\log D$ with respect to entries of $G$ yields
\[
\partial_{G_{ij}}\psi=T_{ij}, \qquad 
\partial_{G_{ij}} T_{ab}=-T_{ai}T_{bj}.
\]
These identities generalize familiar formulas for matrix derivatives.

\paragraph{Relation to ordinary inverse.}
If $G$ is invertible, define $\tilde D=\det(G^{-1}P'+P)$ and
\[
\tilde T=P'(G^{-1}P'+P)^{-1}.
\]
Then one finds $\tilde D=D/\det(G)$ and $\tilde T=G-GTG$, showing the consistency of the construction.

\paragraph{Variational interpretation.}
If $G$ is symmetric, consider the quadratic form
\[
Q(\vec v)=\tfrac{1}{2}(\vec v-\vec e)^{\top} G (\vec v-\vec e)-\vec e^{\top} \vec v.
\]
The stationary point is
\[
\vec v=T(G\vec e+\vec e'),
\]
illustrating how the Bott--Duffin inverse arises naturally in constrained optimization.

---
\subsection{Group and Drazin inverses}\label{sec:drazin-inverse}

\paragraph{Index.}
The \emph{index} of $A$, $\mathrm{ind}(A)$, is the smallest nonnegative integer $k$ such that
\[f
\text{rank}(A^{k+1})=\text{rank}(A^k).
\]
This parameter determines which generalized inverse is applicable. If $\mathrm{ind}(A)=0$ then $A$ is invertible, if $\mathrm{ind}(A)=1$ then the \emph{group inverse} exists, and for $\mathrm{ind}(A)>1$ we need the \emph{Drazin inverse}.

\paragraph{Group inverse.}
For $\mathrm{ind}(A)=1$, the group inverse $A^\#$ is uniquely defined by
\[
A A^\# A = A,\quad A^\# A A^\# = A^\#,\quad A A^\# = A^\# A.
\]
Basic facts:
\begin{itemize}
\item $A^\#=A^{-1}$ if $A$ is invertible.
\item For a projector $P$, $P^\#=P$.
\item If $A$ is diagonalizable with nonzero eigenvalues $\lambda_i$ and spectral projectors $P_{\lambda_i}$, then
\[
A^\#=\sum_{\lambda_i\neq 0}\frac{1}{\lambda_i}P_{\lambda_i}.
\]
\item The positive powers $\{A^k,(A^\#)^k\}$ form an Abelian group with identity $AA^\#$.
\end{itemize}

\paragraph{Drazin inverse.}
For general $\mathrm{ind}(A)=k$, the Drazin inverse $A^D$ is uniquely defined by
\[
A^k A^D A=A^k,\quad A^D A A^D=A^D,\quad A^DA=AA^D.
\]
It reduces to the group inverse if $k=1$ and to the ordinary inverse if $k=0$.

\paragraph{Spectral representation.}
If $A=S(N\oplus C)S^{-1}$ with $N$ nilpotent and $C$ invertible, then
\[
A^D = S(0\oplus C^{-1})S^{-1}.
\]
Thus the Drazin inverse “inverts what can be inverted” and nullifies the nilpotent sector.

\paragraph{Projection operators.}
The operators
\[
P_0 = I - AA^D, \qquad P_1 = AA^D
\]
are complementary projections onto $\ker(A)$ and $\mathrm{Im}(A)$ respectively. These play a central role in applications to dynamics and Markov processes.

\paragraph{Diagonalizable case.}
If $A$ is diagonalizable with eigenvalues $\{\lambda_i\}$ and projectors $P_{\lambda_i}$, then
\[
A^D=\sum_{\lambda_i\neq 0}\frac{1}{\lambda_i}P_{\lambda_i}.
\]
Hence the Drazin inverse is simply the ordinary inverse on nonzero eigenspaces and vanishes on the zero-eigenvalue subspace.

\paragraph{Integral and limit representations.}
Important constructive formulas:
\begin{align}
A^D &= \int_0^\infty e^{-tA}\,dt \qquad\text{(restricted to nonzero modes)},\\
A^D &= \lim_{t\to 0^+} (tI+A^{k+1})^{-1}A^k.\footnote{H.~Wei and S.~Qiao, \emph{Applied Mathematics and Computation} \textbf{138}, 77–89 (2003).}
\end{align}

\paragraph{Additional properties.}
\begin{itemize}
\item Similarity invariance: $(XAX^{-1})^D = X A^D X^{-1}$.
\item Stability under conjugation: $(A^*)^D = (A^D)^*$.
\item Powers: $(A^\ell)^D = (A^D)^\ell$.
\item Projectors: $(AA^D)^2=AA^D$, $(I-AA^D)^2=I-AA^D$.
\item Commutativity: $A^DA=AA^D$ by definition.
\end{itemize}

\paragraph{Cline’s theorem.}
Among the most elegant closed-form expressions for generalized inverses is due to Cline. 
Suppose $A \in \mathbb{R}^{n \times n}$ admits a \emph{full-rank factorization}, i.e.
\[
A = X Y^\top,
\]
with $X, Y \in \mathbb{R}^{n \times k}$ of full column rank $k$.  
This factorization always exists when $\operatorname{rank}(A) = k$, and it provides a convenient way 
to separate the range and the nullspace structure of $A$ into two tall matrices.

Cline’s result shows that in this setting the \emph{group inverse} $A^\#$ has an explicit form in terms 
of the $k \times k$ Gram matrix $Y^\top X$, namely:
\begin{equation}
A^\# = X \,(Y^\top X)^{-2} Y^\top.
\end{equation}
This formula highlights two important features: first, that the group inverse is completely determined 
by the geometry of the column spaces spanned by $X$ and $Y$; second, that the central object is 
the square $(Y^\top X)^{-2}$, which is invertible since $Y^\top X$ has full rank.  

In practice, Cline’s theorem provides a constructive way to compute group inverses for matrices of 
nontrivial rank, avoiding more abstract operator definitions. It remains a standard reference point 
for applications in matrix theory, statistics, and systems theory where such factorizations are natural.\footnote{R.E.~Cline, ``Inverses of rank invariant powers of a matrix,'' \emph{SIAM Review} \textbf{7}, 43–46 (1965).}
\index{key}{Cline's theorem}

\paragraph{Limit formula for the Drazin inverse.}
Under suitable spectral and index conditions, the Drazin inverse \(A^D\) of a square matrix \(A\) with index \(k\) satisfies the following limit expression:\footnote{H.~Wei and S.~Qiao, “The representation and approximation of the Drazin inverse of a linear operator,” \emph{Applied Mathematics and Computation} 138 (2003) 77–89.}
\begin{equation}
A^D \;=\; \lim_{t \to 0^+} \bigl(t I + A^{k+1}\bigr)^{-1} A^k.
\end{equation}

This formula not only provides a practical computational scheme—regularizing via a small parameter \(t\)—but also highlights deep ties to analytic matrix functions and regularization methods.

A related continuous limit formula, often encountered in computational contexts, takes the form:
\[
A^D = \lim_{\alpha \to 0} \bigl(\alpha^2 I + A^{k+1}\bigr)^{-1} A^k.
\]
This highlights the connection of the Drazin inverse to regularization techniques, by approximating the pseudoinverse via a small regularizing parameter \(\alpha\), and to matrix functions of the form \(A^k ( \alpha^2 I + A^{k+1} )^{-1}\).

\subsubsection{Drazin inverse from matrix differential equations}
\index{key}{Drazin inverse, differential equation representation}

Another perspective on the Drazin inverse comes from the theory of
matrix differential equations. 
Consider the continuous-time evolution
\begin{equation}
\dot R(t) = -\alpha A R(t) + \alpha I,
\label{eq:drazin-diff-eq}
\end{equation}
with initial condition $R(0)$ and $\alpha > 0$. The solution is
\begin{equation}
R(t) = e^{-\alpha t A} R(0) + \alpha \int_0^{\top} e^{-\alpha (t-s)A} I \, ds.
\end{equation}
As $t \to \infty$, provided $A$ has a finite index, the transient
term vanishes and one obtains
\begin{equation}
\lim_{t\to\infty} R(t) = \alpha \int_0^\infty e^{-\alpha s A} \, ds = A^D,
\end{equation}
i.e.\ the Drazin inverse of $A$. In the special case that $A$ is
invertible, $A^D = A^{-1}$.

This representation highlights the connection of the Drazin inverse
to \emph{dynamical systems}: the inverse emerges as the fixed point of
a stable linear differential equation. Such formulations are not only
of theoretical interest but also form the basis of analog algorithms
implemented in memristive crossbar arrays for efficient matrix
inversion.\footnote{J.~Lin, F.~Barrows, F.~Caravelli, ``Memristive Linear Algebra,'' Phys. Rev. Research 7, 023241 (2025).}

\paragraph{Remarks.}
The Drazin inverse plays an important role in differential-algebraic equations and Markov chains,
where singular matrices naturally appear\footnote{M.P.~Drazin, ``Pseudo-inverses in associative rings and
semigroups,'' \emph{American Mathematical Monthly} \textbf{65}, 506–514 (1958).}.

\subsection{Matrix Inverse Identities}
\index{key}{matrix inverses}
Matrix inversion is central in both theory and applications. Here we collect
a number of exact and approximate formulas.

\paragraph{Basic relations.}
\begin{align}
(AB)^{-1} &= B^{-1}A^{-1}, \\
(A^\top)^{-1} &= (A^{-1})^\top, \\
(A^\ast)^{-1} &= (A^{-1})^\ast, \\
(A+B)^\top &= A^\top + B^\top.
\end{align}

\paragraph{Block inversion formulas (Schur complements).}
If 
$\displaystyle M = \begin{bmatrix} A & B \\ C & D \end{bmatrix}$ 
with $A$ invertible, then
\begin{equation}
M^{-1} =
\begin{bmatrix}
A^{-1} + A^{-1}B S^{-1} C A^{-1} & -A^{-1} B S^{-1} \\
- S^{-1} C A^{-1} & S^{-1}
\end{bmatrix},
\qquad
S = D - C A^{-1} B.
\end{equation}
Analogously, if $D$ invertible one uses $A - BD^{-1}C$.

\paragraph{Woodbury and Sherman--Morrison formulas.}
For conformable $U,V,B$:
\begin{align}
(A+U B V)^{-1} &= A^{-1} - A^{-1} U \left(B^{-1} + V A^{-1} U\right)^{-1} V A^{-1}, \\
(A+uv^\top)^{-1} &= A^{-1} - \frac{A^{-1}uv^\top A^{-1}}{1+ v^\top A^{-1} u}.
\end{align}
The first above is a generalization of the Kailath's variant \footnote{cf. Petersen, Pedersen, The Matrix Cookbook.}.
These are invaluable for rank-$k$ updates.

\paragraph{Searle identities (special cases).}
\begin{align}
(I+AB)^{-1} &= I - A(I+BA)^{-1}B, \\
(I+AB)^{-1}A &= A(I+BA)^{-1}, \\
(A^{-1}+B^{-1})^{-1} &= A(A+B)^{-1}B.
\end{align}

\paragraph{Series expansions.}
If $\rho(A)<1$ (spectral radius $<1$),
\begin{align}
(I-A)^{-1} &= \sum_{k=0}^{\infty} A^k, \\
(I+A)^{-1} &= \sum_{k=0}^{\infty} (-1)^k A^k.
\end{align}

\paragraph{Rank-1 update approximations}
In many network and complex-systems contexts one needs approximate update rules
for inverses under small rank-1 perturbations.  
Let $G=A^{-1}$ and consider $A' = A+uv^\top$ with $u,v$ ``small.”  
Then one may approximate
\begin{equation}
(A+uv^\top)^{-1} \approx G - \frac{G u v^\top G}{1+ v^\top G u},
\label{eq:bcvc}
\end{equation}
which is the exact Sherman–Morrison formula.  
\paragraph{Approximation schemes for matrix inverses.}
Besides exact formulas (Woodbury, Sherman--Morrison, block inversion), there
exist useful approximation schemes when direct inversion is impractical.

\subparagraph{Neumann series expansions.}
If the spectral radius $\rho(A)<1$, then
\begin{align}
(I - A)^{-1} &= \sum_{n=0}^{\infty} A^n, \label{eq:neumann-series-1}\\
(I + A)^{-1} &= \sum_{n=0}^{\infty} (-1)^n A^n. \label{eq:neumann-series-2}
\end{align}
Truncations provide simple approximations, e.g.
\[
(I - A)^{-1} \approx I + A + A^2,\qquad
(I + A)^{-1} \approx I - A + A^2.
\]

\subparagraph{Approximation for large symmetric $A$.}
For symmetric $A$ of large norm,
\begin{equation}
A - A(I + A)^{-1}A \;\approx\; I - A^{-1}.
\label{eq:approx-large-symm}
\end{equation}
This avoids explicitly inverting $A$ by instead computing $(I+A)^{-1}$.

\subparagraph{Perturbative expansion around $Q$.}
If $Q$ is invertible and $\sigma^2 \ll \|Q\|$, then
\begin{equation}
(Q + \sigma^2 M)^{-1} \;\approx\; Q^{-1} - \sigma^2 Q^{-1} M Q^{-1}.
\label{eq:approx-perturb}
\end{equation}
More generally,
\begin{equation}
(Q + \sigma^2 M)^{-1}
= Q^{-1}(I + \sigma^2 M Q^{-1})^{-1}
\approx Q^{-1} - \sigma^2 Q^{-1} M Q^{-1} + O(\sigma^4).
\label{eq:approx-perturb-series}
\end{equation}

\paragraph{Rank-1 resolvent approximation}
Let $A\in\mathbb{R}^{N\times N}$ be (typically) nonnegative and sub-stochastic, with
row-sums $r_i=\sum_j A_{ij}$, column-sums $c_j=\sum_i A_{ij}$, and
$m=\frac{1}{N}\sum_{ij}A_{ij}=\frac{1}{N}\sum_i r_i=\frac{1}{N}\sum_j c_j$.
Define the unique rank-1 matrix sharing these marginals
\[
\widehat A \;=\; \frac{1}{mN}\; r\,c^\top .
\]
Using Sherman--Morrison on $\,(I-\widehat A)^{-1}\,$ yields the closed form
\begin{equation}
(I-A)^{-1}\;\approx\;(I-\widehat A)^{-1}
\;=\; I \;+\; \frac{\widehat A}{\,1-\frac{1}{mN}\sum_j r_j c_j\,}.
\label{eq:bcvc-resolvent}
\end{equation}
For instance, when applied to a vector $R=(I-A)^{-1}\mathbf{1}$ one obtains the
double-constraint estimate
\begin{equation}
R_i \;\approx\; 1 \;+\; \frac{r_i}{\,1-\frac{1}{mN}\sum_j r_j c_j\,},
\qquad i=1,\dots,N,
\label{eq:bcvc-influence}
\end{equation}
and, if only row-sums are known (single-constraint case),
\begin{equation}
I_i \;\approx\; 1 \;+\; \frac{r_i}{\,1-\frac{1}{N}\sum_j r_j\,}.
\label{eq:bcvc-influence-rows}
\end{equation}
From a practical perspective, the approximation is most accurate when the interaction matrix is not too
sparse (large spectral gap). It empirically tracks Katz centrality,
PageRank-type scores, trophic levels, and other complex network measures, as we will see later
\footnote{S. Bartolucci, F. Caccioli, F. Caravelli, P. Vivo, \textit{Ranking influential nodes in networks from partial information}, Phys. Rev. Research \textbf{5}, 033123 (2023).}.

\paragraph{Special cases.}
\begin{itemize}
\item For diagonal matrices: $(\mathrm{diag}(d_1,\dots,d_n))^{-1}=\mathrm{diag}(d_1^{-1},\dots,d_n^{-1})$.
\item For orthogonal/unitary $U$: $U^{-1}=U^\top$ or $U^\ast$.
\item For projection $P$: if $P^2=P$, then $P^+=P$ is its Moore–Penrose pseudoinverse.
\end{itemize}

\paragraph{Derivatives of inverses.}
Inverses are continuous functions of their matrices if the inverse exists. To see this, these are formulae that one can use:
\begin{align}
\frac{\partial X^{-1}}{\partial X} &= -X^{-1} (\partial X) X^{-1}, \\
\frac{\partial}{\partial X}\text{Tr}(AX^{-1}B) &= -(X^{-1} B A X^{-1})^\top.
\end{align}

\subsection{Integral representations for inverses}
\label{sec:integralrep}
\index{key}{Integral representations inverse}

Matrix inverses, pseudoinverses, and generalized inverses often admit integral representations. These
expressions connect linear algebra to analysis and are widely used in operator theory and control
theory.

\paragraph{Ordinary inverse.}
If all eigenvalues of a square matrix $A$ have positive real part (i.e.\ $\Re \lambda_i(A)>0$), then
\begin{equation}
A^{-1}=\int_0^\infty e^{-tA}\,dt.
\end{equation}
This is a matrix analogue of the scalar Laplace transform identity
\[
\frac{1}{\alpha}=\int_0^\infty e^{-\alpha t}\,dt, \qquad \Re\alpha>0.
\]

\paragraph{Moore--Penrose pseudoinverse.}
The Moore--Penrose pseudoinverse $A^+$ also admits an integral representation
due to Showalter\footnote{D.~Showalter, ``Representation and computation of the pseudoinverse,''
\emph{Proceedings of the American Mathematical Society} \textbf{18}, 584–586 (1967).}:
\begin{equation}
A^+=\int_0^\infty e^{-t A^\dagger A}\,A^\dagger\,dt.
\end{equation}
This converges for all $A$, since $A^\dagger A$ is positive semidefinite.

\paragraph{Drazin inverse.}
For singular matrices, the \emph{Drazin inverse} (Sec.~\ref{sec:drazin-inverse}) can also be written in integral
form. Let $\Omega$ be the projector onto $\ker(A)$, then
\begin{equation}
A^D=\int_0^\infty e^{-tA}(I-\Omega)\,dt.
\end{equation}
Equivalently, for any integer $m\geq k=\text{ind}(A)$,
\begin{equation}
A^D=\int_0^\infty e^{-t A^{m+1}} A^m\,dt,
\end{equation}
provided $\Re \lambda_i(A)\geq 0$.  
More general formulas hold even when the spectrum extends outside the right half-plane, e.g.
\begin{equation}
A^D=\int_0^\infty 
e^{-t\,(A^k (A^{2k+1})^\dagger A^{k+1})}\,A^k (A^{2k+1})^\dagger\,dt.
\end{equation}
Such representations connect generalized inverses to semigroup theory and functional calculus.

---
\chapter{Matrix factorizations}
\section{Matrix decompositions beyond diagonalization}

\subsection{Schur form}
\label{sec:schur}
\index{key}{Schur decomposition}

Not all matrices are diagonalizable. For example,
\[
A=\begin{pmatrix}
1 & 1\\
0 & 1
\end{pmatrix}
\]
has only one eigenvector and hence cannot be diagonalized.  
However, every square matrix can be reduced to an \emph{upper triangular} form, known as the
\emph{Schur decomposition}.

\paragraph{Complex case.}
Over the complex field, the following theorem holds:

\begin{theorem}[Schur decomposition]
Let $A\in \mathbb{C}^{n\times n}$. Then there exists a unitary matrix $U$ such that
\begin{equation}
A=UTU^\dagger,
\end{equation}
where $T$ is upper triangular with the eigenvalues of $A$ on its diagonal.
\end{theorem}

This is a consequence of the spectral theorem for normal matrices and the fundamental theorem of
algebra, which guarantees that the characteristic polynomial has a complex root\footnote{See R.~Horn
and C.~Johnson, \emph{Matrix Analysis}, Cambridge University Press (1985).}.

\paragraph{Real case.}
Over $\mathbb{R}$, one cannot always diagonalize with real eigenvalues, but the Schur form still holds
with an orthogonal transformation:
\begin{equation}
A=Q T Q^{\top},
\end{equation}
where $Q$ is orthogonal and $T$ is quasi-upper-triangular, with either $1\times 1$ real blocks or $2\times 2$
blocks corresponding to complex conjugate eigenvalues.

\paragraph{Properties.}
\begin{itemize}
    \item The Schur form is numerically stable and forms the basis of many eigenvalue algorithms (e.g.\ QR
    algorithm).
    \item The decomposition implies $A^n=U T^n U^\dagger$, making it useful for computing matrix powers
    and exponentials.
    \item If $T=D+N$ with $D$ diagonal and $N$ strictly upper triangular (nilpotent), then
    \[
    (D+N)^z=\sum_{i=0}^z \binom{z}{i} D^{z-i} N^i,
    \]
    and the series truncates since $N^k=0$ for some $k$.
\end{itemize}

\paragraph{Triangular matrix facts.}
Upper (or lower) triangular matrices are closed under:
\begin{itemize}
    \item addition,
    \item multiplication,
    \item inversion (if all diagonal entries are nonzero),
    \item transpose (which switches upper to lower).
\end{itemize}
Thus the Schur form is preserved under powers: if $A=UTU^\dagger$, then $A^n=UT^nU^\dagger$.

---

The Schur decomposition provides a universal triangularization:
\begin{itemize}
    \item Every complex matrix is unitarily similar to an upper triangular one.
    \item Every real matrix is orthogonally similar to a quasi-upper-triangular one.
    \item It generalizes diagonalization and is fundamental in numerical algorithms.
\end{itemize}

\subsection{Jordan normal form}
\label{sec:jordanform}
\index{key}{Jordan normal form}

We now recall one of the most important results in matrix theory: the \emph{Jordan normal form}.  
It provides the canonical decomposition of any square matrix up to similarity.

\paragraph{Block-diagonal form.}
A matrix is said to be \emph{block diagonal}\index{key}{block diagonal} if it can be written as
\[
A=\begin{pmatrix}
A_1 & 0 & \cdots & 0 \\
0 & A_2 & \ddots & \vdots \\
\vdots & \ddots & \ddots & 0 \\
0 & \cdots & 0 & A_n
\end{pmatrix},
\]
where the $A_i$ are square blocks.

\paragraph{Jordan blocks.}
A \emph{Jordan block} associated with eigenvalue $\lambda_i$ is an $m_i\times m_i$ matrix
\[
J(\lambda_i)=\begin{pmatrix}
\lambda_i & 1 & 0 & \cdots & 0 \\
0 & \lambda_i & 1 & \ddots & \vdots \\
\vdots & \ddots & \lambda_i & \ddots & 0 \\
\vdots & & \ddots & \ddots & 1 \\
0 & \cdots & \cdots & 0 & \lambda_i
\end{pmatrix}.
\]
Equivalently, $J(\lambda_i)=\lambda_i I+N$, where $N$ is nilpotent with $N^{m_i}=0$.

\paragraph{Jordan form.}
A matrix is in \emph{Jordan form} if it is block-diagonal with Jordan blocks on the diagonal.

---
\label{sec:solsys2}
\paragraph{Theorems.}

\begin{theorem}[Jordan decomposition theorem]
Every complex square matrix $A\in\mathbb{C}^{p\times p}$ is similar to a Jordan form matrix $J$, i.e.\
\[
A=QJQ^{-1}.
\]
\end{theorem}

\begin{theorem}[Structure of Jordan form]
Let $A\in\mathbb{C}^{p\times p}$ with distinct eigenvalues $\lambda_1,\ldots,\lambda_s$.
Let $m_i$ be the algebraic multiplicity of $\lambda_i$, and $\mu_i$ the geometric multiplicity.
Then $A$ is similar to a Jordan form matrix $J$ such that:
\begin{enumerate}
\item For each $\lambda_i$, the number of Jordan blocks equals $\mu_i$.
\item $\lambda_i$ appears on the diagonal of $J$ exactly $m_i$ times.
\item The decomposition is unique up to permutation of Jordan blocks.
\end{enumerate}
\end{theorem}

\noindent
These results show that although not every matrix is diagonalizable, every matrix is similar to a
block-triangular matrix with eigenvalues on the diagonal\footnote{See R.~Horn and C.~Johnson,
\emph{Matrix Analysis}, Cambridge University Press (1985).}.

---

\paragraph{Constructive procedure.}
Operationally, the Jordan form can be built from generalized eigenspaces:
\[
E_\lambda^k=\{\vec x : (A-\lambda I)^k \vec x=0\}.
\]
The dimensions of these spaces determine the sizes of Jordan blocks.  
One standard method is the \emph{box diagram algorithm}:
\begin{enumerate}
\item Compute eigenvalues $\lambda_i$, together with algebraic and geometric multiplicities.
\item For each $\lambda$, compute dimensions $\dim E_\lambda^k$. Define increments
\[
d_k=\dim E_\lambda^k-\dim E_\lambda^{k-1}.
\]
\item Construct a Young diagram with $d_1$ boxes in the first row, $d_2$ in the second, and so on.
\item Fill boxes from bottom to top with independent vectors from $E_\lambda^k$ not in $E_\lambda^{k-1}$.
\item Each box stacked above another corresponds to applying $(A-\lambda I)$ to the lower vector.
\item Repeat for each eigenvalue; collect the vectors as columns of $Q$. Then $J=Q^{-1}AQ$.
\end{enumerate}

---

\paragraph{Powers of Jordan blocks.}
If $A=QJQ^{-1}$, then $A^k=QJ^kQ^{-1}$. For a Jordan block $J(\lambda)=\lambda I+N$ with nilpotent $N$,
\[
(\lambda I+N)^k=\sum_{t=0}^k \binom{k}{t} \lambda^{k-t} N^{\top}.
\]
If $N$ is nilpotent of order $\tilde t$, then $N^{\tilde t}=0$, so the sum truncates at $t=\tilde t$.
Thus $J(\lambda)^k$ has:
\begin{itemize}
\item $\lambda^k$ on the diagonal,
\item $\binom{k}{1}\lambda^{k-1}$ on the first superdiagonal,
\item $\binom{k}{2}\lambda^{k-2}$ on the second superdiagonal,
\item and so on, until $N^{\tilde t-1}$.
\end{itemize}
This makes Jordan form invaluable for computing powers and exponentials of matrices.

---

\paragraph{Relation to Kato decomposition.}
The Jordan decomposition is closely related to the \emph{Kato decomposition}\index{key}{Kato decomposition}.
Given eigenvalues $\lambda_s$ of $A$, one can write
\[
A=\sum_{s=1}^r\big(\lambda_s P_s+D_s\big),
\]
where $P_s$ are projectors onto $\ker(A-\lambda_s I)^{m_s}$, and $D_s$ are nilpotent operators.
They satisfy
\[
P_s P_{s'}=\delta_{ss'}P_s,\quad
D_s D_{s'}=\delta_{ss'}D_s,\quad
P_s D_{s'}=D_{s'}P_s=0.
\]
The Kato form thus provides a decomposition with projectors and nilpotents that commute in a simple way,
making it particularly useful for matrix powers and functional calculus\footnote{See T.~Kato,
\emph{Perturbation Theory for Linear Operators}, Springer (1966).}.

\subsubsection{Drazin inverse via Jordan form}
\label{sec:drazinjordan}
\index{key}{Drazin inverse}

The Jordan normal form provides a direct way to compute the Drazin inverse.

\begin{theorem}
Let $A\in\mathbb{C}^{n\times n}$, with Jordan decomposition
\[
A=QJQ^{-1}, \qquad
J=\begin{pmatrix}
J_{nz} & 0 \\
0 & J_{z}
\end{pmatrix},
\]
where $J_{nz}$ collects the Jordan blocks for nonzero eigenvalues, and $J_z$ collects the Jordan blocks for
the eigenvalue $0$. Then the Drazin inverse of $A$ is
\begin{equation}
A^D=Q\begin{pmatrix}
J_{nz}^{-1} & 0 \\
0 & 0
\end{pmatrix}Q^{-1}.
\end{equation}
\end{theorem}

\noindent
\emph{Sketch of proof.} The result follows from the Penrose conditions for the Drazin inverse and the fact
that if $\text{ind}(A)=k$, then $J_z$ consists of nilpotent blocks of size $k$.  

---

\subsection{Singular value decomposition}
\label{sec:svd}
\index{key}{singular value decomposition}

The \emph{singular value decomposition} (SVD) is one of the most important factorizations in linear algebra.
It applies to any matrix, square or rectangular.

\begin{theorem}[Singular Value Decomposition]
For any $A\in\mathbb{C}^{m\times n}$ there exist unitary matrices $U\in\mathbb{C}^{m\times m}$ and
$V\in\mathbb{C}^{n\times n}$ such that
\begin{equation}
A=U\Sigma V^\dagger,
\label{eq:svd}
\end{equation}
where $\Sigma\in\mathbb{R}^{m\times n}$ is diagonal with nonnegative entries
\[
\Sigma=\text{diag}(\sigma_1,\ldots,\sigma_r,0,\ldots,0),
\]
with $\sigma_1\geq \sigma_2\geq \cdots \geq \sigma_r>0$. The $\sigma_i$ are the \emph{singular values} of $A$,
and $r=\text{rank}(A)$.
\end{theorem}

\paragraph{Connection with eigenvalues.}
The singular values are the square roots of the eigenvalues of $A^\dagger A$ (or $AA^\dagger$), both of
which are Hermitian and positive semidefinite. If $\vec v_i$ is an eigenvector of $A^\dagger A$ with
eigenvalue $\sigma_i^2$, then $\vec u_i=A\vec v_i/\sigma_i$ is the corresponding column of $U$.

\paragraph{Rank-1 decomposition.}
The SVD can be written as
\[
A=\sum_{i=1}^r \sigma_i \vec u_i \vec v_i^\dagger,
\]
a decomposition into rank-1 matrices. The number of nonzero singular values equals the rank of $A$.

---

\paragraph{Properties of positive semidefinite matrices.}
If $A$ is positive semidefinite of rank 1, then $A=\vec x \vec x^\dagger$ for some vector $\vec x$.
More generally, $A$ is positive semidefinite if and only if it admits a decomposition
\[
A=\sum_i \lambda_i \vec x_i \vec x_i^\dagger, \qquad \lambda_i\geq 0.
\]

---

\paragraph{Eckart--Young theorem.}
The SVD provides optimal low-rank approximations.  

\begin{theorem}[Eckart--Young--Mirsky]
Let $A\in\mathbb{C}^{m\times n}$ with SVD $A=U\Sigma V^\dagger$.  
For any $1\leq r < \text{rank}(A)$, the best rank-$r$ approximation to $A$ in the spectral or Frobenius norm is
\[
\tilde A=\sum_{i=1}^r \sigma_i \vec u_i \vec v_i^\dagger.
\]
\end{theorem}

This result is fundamental in data analysis, principal component analysis (PCA), and model reduction\footnote{C.~Eckart and G.~Young, ``The approximation of one matrix by another of lower rank,'' \emph{Psychometrika} \textbf{1}, 211–218 (1936).}.

---

\paragraph{Bounds on singular values.}
Singular values satisfy inequalities such as the Schur bound:
\[
\sigma_{\max}^2(A)\leq \max_i \sum_j |a_{ij}|c_j,
\]
where $r_i=\sum_j |a_{ij}|$ are row sums and $c_j=\sum_i |a_{ij}|$ are column sums.  
Such inequalities are used in numerical analysis to estimate $\|A\|_2=\sigma_{\max}(A)$.

---

\subsubsection{Moore--Penrose inverse from SVD}
\label{sec:pseudoinvsvd}

The SVD provides a simple and robust way to compute the Moore--Penrose pseudoinverse.  
If $A=U\Sigma V^\dagger$, then
\begin{equation}
A^+=V \Sigma^+ U^\dagger,
\end{equation}
where $\Sigma^+$ is obtained from $\Sigma$ by inverting all nonzero singular values and leaving zeros in
place.  

It follows that
\[
A A^+=U \begin{pmatrix} I_r & 0 \\ 0 & 0 \end{pmatrix} U^\dagger, \qquad
A^+ A=V \begin{pmatrix} I_r & 0 \\ 0 & 0 \end{pmatrix} V^\dagger,
\]
showing that $A A^+$ and $A^+ A$ are orthogonal projectors onto the column space and row space of $A$,
respectively.  

This construction was first characterized by Penrose in his seminal 1955 paper\footnote{R.~Penrose,
``A generalized inverse for matrices,'' \emph{Proceedings of the Cambridge Philosophical Society}
\textbf{51}, 406–413 (1955).}.

---

\paragraph{Applications.}
The pseudoinverse via SVD is heavily used in statistics, physics, and computer science:
\begin{itemize}
    \item Inverse problems and regularization,
    \item Principal component analysis,
    \item Spectral methods for graphs and bipartite networks\footnote{See F.~Chung, \emph{Spectral Graph Theory}, AMS (1997).}.
\end{itemize}

\subsection{Rank-1 update of the Moore--Penrose inverse}
\index{key}{Moore--Penrose, rank-1 update}

Let $A \in \mathbb{R}^{m\times n}$, and consider a rank-1 update
\[
(A + \vec{c}\,\vec{d}^\top)^+ = A^+ + G,
\]
with $\vec{c} \in \mathbb{R}^m$, $\vec{d} \in \mathbb{R}^n$. Define
\begin{align}
\beta &= 1 + \vec{d}^\top A^+ \vec{c}, \\
\vec{v} &= A^+ \vec{c}, \\
\vec{n} &= (A^+)^\top \vec{d}, \\
\vec{w} &= (I - AA^+)\vec{c}, \\
\vec{m} &= (I - A^+ A)^\top \vec{d}.
\end{align}
The correction $G$ depends on $\|\vec{w}\|$, $\|\vec{m}\|$, and $\beta$, and can be classified into six cases:\footnote{K.B. Petersen and M.S. Pedersen, \emph{The Matrix Cookbook}, Version Nov.~15, 2012.}

\paragraph{Case 1.} If $\|\vec{w}\|\neq 0$ and $\|\vec{m}\|\neq 0$:
\begin{align}
G &= -\vec{v}\,\vec{w}^+ - (\vec{m}^+)^\top \vec{n}^\top + \beta (\vec{m}^+)^\top \vec{w}^+ \\
  &= -\tfrac{1}{\|\vec{w}\|^2} \vec{v}\vec{w}^\top - \tfrac{1}{\|\vec{m}\|^2} \vec{m}\vec{n}^\top
     + \tfrac{\beta}{\|\vec{m}\|^2 \|\vec{w}\|^2} \vec{m}\vec{w}^\top.
\end{align}

\paragraph{Case 2.} If $\|\vec{w}\|=0$, $\|\vec{m}\|\neq 0$, and $\beta=0$:
\begin{align}
G &= -\vec{v}\vec{v}^+ A^+ - (\vec{m}^+)^\top \vec{n}^\top \\
  &= -\tfrac{1}{\|\vec{v}\|^2} \vec{v}\vec{v}^\top A^+ + \tfrac{1}{\|\vec{m}\|^2} \vec{m}\vec{n}^\top.
\end{align}

\paragraph{Case 3.} If $\|\vec{w}\|=0$ and $\beta \neq 0$:
\begin{equation}
G = \tfrac{1}{\beta} \vec{m}\vec{v}^\top A^+ 
    - \tfrac{\beta}{\|\vec{v}\|^2 \|\vec{m}\|^2 + |\beta|^2}
      \Big( \tfrac{\|\vec{v}\|^2}{\beta} \vec{m} + \vec{v} \Big)
      \Big( \tfrac{\|\vec{m}\|^2}{\beta} (A^+)^\top \vec{v} + \vec{n} \Big)^\top.
\end{equation}

\paragraph{Case 4.} If $\|\vec{w}\|\neq 0$, $\|\vec{m}\|=0$, and $\beta=0$:
\begin{align}
G &= -A^+ \vec{n}\vec{n}^+ - \vec{v}\vec{w}^+ \\
  &= -\tfrac{1}{\|\vec{n}\|^2} A^+ \vec{n}\vec{n}^\top - \tfrac{1}{\|\vec{w}\|^2} \vec{v}\vec{w}^\top.
\end{align}

\paragraph{Case 5.} If $\|\vec{m}\|=0$ and $\beta \neq 0$:
\begin{equation}
G = \tfrac{1}{\beta} A^+ \vec{n}\vec{w}^\top 
    - \tfrac{\beta}{\|\vec{n}\|^2 \|\vec{w}\|^2 + |\beta|^2}
      \Big( \tfrac{\|\vec{w}\|^2}{\beta} A^+ \vec{n} + \vec{v} \Big)
      \Big( \tfrac{\|\vec{n}\|^2}{\beta} \vec{w} + \vec{n} \Big)^\top.
\end{equation}

\paragraph{Case 6.} If $\|\vec{w}\|=0$, $\|\vec{m}\|=0$, and $\beta=0$:
\begin{align}
G &= -\vec{v}\vec{v}^+ A^+ - A^+ \vec{n}\vec{n}^+ + \vec{v}\vec{v}^+ A^+ \vec{n}\vec{n}^+ \\
  &= -\tfrac{1}{\|\vec{v}\|^2} \vec{v}\vec{v}^\top A^+ - \tfrac{1}{\|\vec{n}\|^2} A^+ \vec{n}\vec{n}^\top
     + \tfrac{\vec{v}^\top A^+ \vec{n}}{\|\vec{v}\|^2 \|\vec{n}\|^2} \vec{v}\vec{n}^\top.
\end{align}

---

\subsubsection{Additional approximation schemes for matrix inverses}
In addition to exact formulas (Woodbury, Sherman--Morrison, block inversion), 
several iterative and analytical approximation schemes are widely used:

\paragraph{Newton--Schulz iteration.}
Starting from an initial guess $X_0$, iterate
\[
X_{k+1} = 2X_k - X_k A X_k.
\]
This converges quadratically to $A^{-1}$ if $\|I - AX_0\| < 1$.\footnote{See N.~J.~Higham, \emph{Functions of Matrices}, SIAM (2008).}

\paragraph{Chebyshev and polynomial expansions.}
If $\Lambda(A)\subset [a,b]$ with $0<a<b$, one can approximate
\[
A^{-1} \approx \sum_{k=0}^m \alpha_k T_k(A),
\]
where $T_k$ are Chebyshev polynomials, and coefficients $\alpha_k$ are chosen to minimize the uniform error.\footnote{See Y.~Saad, \emph{Iterative Methods for Sparse Linear Systems}, SIAM (2003).}

\paragraph{Low-rank approximations.}
Generalizing Sherman--Morrison, for $U \in \mathbb{R}^{n\times k}, V \in \mathbb{R}^{n\times k}$,
\[
(A + UV^\top)^{-1} = A^{-1} - A^{-1} U (I + V^\top A^{-1} U)^{-1} V^\top A^{-1}.
\]
Truncating when $k\ll n$ yields efficient approximations.\footnote{See G.~Golub and C.~Van Loan, \emph{Matrix Computations}, Johns Hopkins Univ.~Press (1996).}

\paragraph{Randomized SVD approximations.}
For large matrices, randomized sketching methods approximate the dominant singular values/vectors, and thus the pseudoinverse $A^+$.\footnote{See N.~Halko, P.~Martinsson, J.~Tropp, ``Finding structure with randomness,'' \emph{SIAM Review} 53 (2011).}

\paragraph{Tikhonov regularization.}
For ill-conditioned or singular matrices,
\[
A^+ \approx (A^\top A + \lambda I)^{-1} A^\top, \quad \lambda>0,
\]
which stabilizes inversion and connects to ridge regression.\footnote{See S.~Boyd and L.~Vandenberghe, \emph{Convex Optimization}, Cambridge Univ.~Press (2004).}

\paragraph{Harmonic mean approximation.}
For two matrices $A,B$ with similar spectra,
\[
(A+B)^{-1} \approx 2\,(A^{-1}+B^{-1})^{-1}.
\]
Used in covariance estimation and preconditioning.\footnote{See T.~Ando, ``Concavity of certain maps on positive definite matrices,'' \emph{Linear Algebra Appl.} 26 (1979).}

\paragraph{Integral representations.}
For $\Re(\lambda_i(A))>0$,
\[
A^{-1} = \int_0^\infty e^{-tA}\,dt,
\]
allowing quadrature approximations.\footnote{See N.~Dunford and J.~T.~Schwartz, \emph{Linear Operators}, Interscience (1958).}

\subsubsection{Smith form}
\index{key}{Smith form}

The \emph{Smith form} (or \emph{rank-revealing form}) expresses a matrix $A\in\mathbb{R}^{n\times m}$ of
rank $r$ as
\begin{equation}
A=L\begin{pmatrix} I_r & 0 \\ 0 & 0 \end{pmatrix}R,
\end{equation}
where $L\in\mathbb{R}^{n\times n}$ and $R\in\mathbb{R}^{m\times m}$ are invertible, and $I_r$ is the
$r\times r$ identity.  

Such a form exists for every matrix and can be derived from the singular value decomposition:
multiplying rows of $U$ and $V$ by $\sqrt{\sigma_i}$ effectively reduces the SVD to a Smith form\footnote{See R.~Horn and C.~Johnson, \emph{Matrix Analysis}, Cambridge University Press (1985).}.

---

\subsubsection{Polar decomposition}
\index{key}{Polar decomposition}

Every square matrix $A\in\mathbb{C}^{n\times n}$ admits a \emph{polar decomposition}
\begin{equation}
A=QP,
\end{equation}
where $Q$ is unitary (orthogonal if $A$ is real), and $P$ is Hermitian positive semidefinite.  
This decomposition is unique: $Q$ and $P$ are determined uniquely by $A$.

From the SVD $A=U\Sigma V^\dagger$, one has explicitly
\[
Q=UV^\dagger, \qquad P=V\Sigma V^\dagger.
\]

\paragraph{Determinant interpretation.}
Since $\det(Q)$ has modulus 1 and $\det(P)\geq 0$, we obtain
\[
\det(A)=\det(Q)\det(P)=r e^{i\theta},
\]
analogous to the polar representation of complex numbers.

---

\subsubsection{Low-rank approximations}
\index{key}{low-rank approximation}

The SVD yields the best low-rank approximations of a matrix.

\begin{theorem}[Eckart--Young--Mirsky]
Let $A\in\mathbb{R}^{m\times n}$ have SVD $A=\sum_{j=1}^r\sigma_j \vec u_j \vec v_j^{\top}$ with singular
values $\sigma_1\geq \cdots \geq \sigma_r>0$. For $1\leq r'<r$, the minimizer of
\[
\min_{\text{rank}(X)=r'}\|A-X\|_2
\]
is given by
\[
A^{(r')}=\sum_{j=1}^{r'} \sigma_j \vec u_j \vec v_j^{\top},
\]
and
\[
\|A-A^{(r')}\|_2=\sigma_{r'+1}.
\]
\end{theorem}

Thus truncating the SVD provides the optimal approximation in both the spectral and Frobenius norms\footnote{C.~Eckart and G.~Young, \emph{Psychometrika} \textbf{1}, 211–218 (1936).}.

---

\subsubsection{SVD, norms, and condition numbers}

From the SVD, the spectral (operator $2$-) norm is
\begin{equation}
\|A\|_2=\max_{\|\vec x\|_2=1}\|A\vec x\|_2=\sigma_1,
\end{equation}
the largest singular value. Similarly, the smallest singular value $\sigma_r$ gives
\[
\|A^{-1}\|_2=\sigma_r^{-1}, \quad \text{if $A$ is invertible}.
\]
Thus the 2-norm condition number is
\begin{equation}
k_2(A)=\|A\|_2\|A^{-1}\|_2=\frac{\sigma_1}{\sigma_r}.
\end{equation}
This generalizes the spectral radius bound for normal matrices (Sec.~\ref{sec:specradius}) to arbitrary
matrices.

---

\subsection{QR decomposition}
\label{sec:qrdec}
\index{key}{QR decomposition}

Another important factorization is the \emph{QR decomposition}. For any $A\in\mathbb{R}^{n\times m}$,
\begin{equation}
A=QR,
\label{eq:qrdec}
\end{equation}
where $Q\in\mathbb{R}^{n\times n}$ is orthogonal ($Q^{\top}Q=I$), and $R=Q^{\top}A$ is upper triangular. If $A$
is square and nonsingular, the factorization is unique.

---

\paragraph{Triangular systems and substitution.}
Triangular matrices are particularly convenient for solving linear systems.  
If $A$ is lower triangular, then $A\vec x=\vec b$ can be solved by \emph{forward substitution}:
\begin{equation}
x_k=\frac{b_k-\sum_{i=1}^{k-1}A_{ki}x_i}{A_{kk}}, \qquad k=1,\ldots,n.
\end{equation}
If $A$ is upper triangular, the analogous scheme is \emph{back substitution}, starting from $k=n$ and
working upwards.

---

\paragraph{Numerical computation of QR.}
In practice, QR is computed stably via:
\begin{itemize}
    \item \textbf{Householder reflections:} each step introduces zeros below the diagonal in a column.
    \item \textbf{Givens rotations:} used when sparsity must be preserved.
    \item \textbf{Modified Gram--Schmidt:} conceptually simple, but less stable numerically.
\end{itemize}
These algorithms form the backbone of the QR method for eigenvalue computations.

---

\paragraph{Worked example.}
As an illustration, consider:
\begin{verbatim}
A = [1  1;
     1 -1;
     1  1]
\end{verbatim}

Applying Gram--Schmidt orthogonalization to the columns yields
\[
Q=\begin{pmatrix}
1/\sqrt{3} &  1/\sqrt{2} \\
1/\sqrt{3} & -1/\sqrt{2} \\
1/\sqrt{3} &  0
\end{pmatrix},
\qquad
R=\begin{pmatrix}
\sqrt{3} & 0 \\
0 & \sqrt{2}
\end{pmatrix}.
\]
Indeed $A=QR$.

---

\paragraph{Properties.}
\begin{itemize}
    \item $\det(A)=\det(R)$, since $\det(Q)=\pm 1$.
    \item The QR decomposition is stable and numerically efficient.
    \item It underlies many iterative eigenvalue algorithms, notably the QR algorithm.
\end{itemize}

\subsubsection{QR decomposition through Gram--Schmidt}
\label{sec:qrred}
\index{key}{Gram-Schmidt decomposition}

A QR decomposition can be obtained directly by the Gram--Schmidt process
(Sec.~\ref{sec:gramschmidt}).  
Let $A=(\vec a_1\ \vec a_2\ \cdots\ \vec a_n)$, with columns $\vec a_j$.  
Applying Gram--Schmidt produces orthonormal vectors $\vec e_j$, and one finds
\begin{equation}
A=\begin{pmatrix} \vec e_1 & \cdots & \vec e_n \end{pmatrix}
\begin{pmatrix}
\vec a_1\cdot \vec e_1 & \vec a_2\cdot \vec e_1 & \cdots & \vec a_n\cdot \vec e_1\\
0 & \vec a_2\cdot \vec e_2 & \cdots & \vec a_n\cdot \vec e_2\\
\vdots & \vdots & \ddots & \vdots\\
0 & 0 & \cdots & \vec a_n\cdot \vec e_n
\end{pmatrix}.
\end{equation}
This identifies $Q=(\vec e_1\cdots\vec e_n)$ and $R=(\vec e_i\cdot \vec a_j)_{i\leq j}$.

\paragraph{Rank-revealing QR.}
A refinement is the \emph{rank-revealing QR decomposition}. If $A$ has rank $r$, then with column
pivoting one can write
\begin{equation}
AP=QR,
\end{equation}
where $P$ is a permutation matrix, $Q$ is orthogonal, and $R$ is upper triangular with the first $r$
diagonal entries well-conditioned and the rest small.  
This is especially useful in numerical rank determination and least-squares problems\footnote{See G.~Golub and C.~Van Loan, \emph{Matrix Computations}, Johns Hopkins University Press (1996).}.

---

\subsubsection{Householder reduction}
\label{sec:householderred}
\index{key}{Householder reduction}

Gram--Schmidt is conceptually simple but numerically unstable for nearly linearly dependent columns.
A more stable alternative is the \emph{Householder QR decomposition}, which uses reflections.

\paragraph{Householder reflections.}
A Householder reflector is
\[
H=I-2\vec v \vec v^\dagger,
\]
where $\vec v$ is a unit vector. $H$ is unitary and reflects across the hyperplane orthogonal to $\vec v$.

\paragraph{QR via Householder.}
One constructs a sequence of reflectors $H_1,\ldots,H_n$ such that
\[
H_n\cdots H_1 A=\tilde R,
\]
where $\tilde R=\begin{pmatrix}R\\0\end{pmatrix}$ with $R$ upper triangular.  
Since each $H_i$ is unitary,
\[
A=(H_1\cdots H_n)^\dagger \tilde R=Q R,
\]
where $Q=H_1\cdots H_n$ is orthogonal.  

\paragraph{Dimensions.}
If $A$ is $m\times n$, then $Q$ is $m\times m$, $\tilde R$ is $m\times n$, and $R$ is $n\times n$.
The bottom $(m-n)\times n$ block of $\tilde R$ is zero.

\paragraph{Numerical advantages.}
Householder QR is stable and efficient, requiring $O(mn^2)$ operations.  
It is the standard method in numerical linear algebra libraries (e.g.\ LAPACK).

---

\subsubsection{Givens rotations}
\index{key}{Givens rotations}

Another stable approach is via \emph{Givens rotations}, which act only on two coordinates at a time.

\paragraph{Definition.}
A Givens rotation $G(i,j,\theta)$ acts on the $(i,j)$-plane:
\[
G=\begin{pmatrix}
1 & & & & \\
& \ddots & & & \\
& & \cos\theta & \sin\theta & \\
& & -\sin\theta & \cos\theta & \\
& & & & \ddots
\end{pmatrix}.
\]
Applying $G$ to $A$ zeroes out a specific subdiagonal entry.

\paragraph{Algorithm.}
By successive rotations, one introduces zeros below the diagonal, producing an upper triangular $R$.  
The product of all rotations forms an orthogonal matrix $Q$, and we obtain $A=QR$.

\paragraph{Comparison.}
\begin{itemize}
\item Householder QR eliminates an entire column below the diagonal in one step.
\item Givens QR eliminates one entry at a time, but allows selective control.
\end{itemize}
Givens rotations are preferred when $A$ is sparse, since they preserve sparsity better than reflectors.

---

\paragraph{Applications of QR decompositions.}
\begin{itemize}
\item Solving least-squares problems via $QR\vec x=Q^{\top} \vec b$.
\item Eigenvalue computations: the iterative \emph{QR algorithm} applies QR repeatedly to converge to
Schur form (Sec.~\ref{sec:schur}).
\item Rank determination and numerical stability analysis.
\end{itemize}

\subsubsection{Schur decomposition using QR factorization}
\label{sec:qrschur}
\index{key}{QR algorithm}
\index{key}{Schur decomposition}

The Schur form (Sec.~\ref{sec:schur}) can be obtained numerically by the \emph{QR algorithm}, one of the
most important algorithms in numerical linear algebra.  

\paragraph{Algorithm.}
Given a square matrix $A$, one constructs a sequence
\[
A_0=A, \qquad A_{k+1}=R_k Q_k,
\]
where $A_k=Q_k R_k$ is the QR decomposition of $A_k$.  

\paragraph{Theorem.}
If $A$ is diagonalizable, the sequence $A_k$ converges to an upper triangular matrix $T$ whose diagonal
entries are the eigenvalues of $A$. More generally, for any $A$, the limit is a Schur form:
\[
A=UTU^\dagger,
\]
with $U$ unitary and $T$ upper triangular.  

\paragraph{Remarks.}
The QR algorithm is numerically stable and forms the backbone of practical eigenvalue computations in
scientific computing\footnote{See L.~Trefethen and D.~Bau, \emph{Numerical Linear Algebra}, SIAM (1997);
G.~Golub and C.~Van Loan, \emph{Matrix Computations}, Johns Hopkins University Press (1996).}.  

---

\subsubsection{Pseudo-inverse in terms of QR factorization}
\label{sec:qrpinv}

The QR decomposition provides a direct way to compute least-squares solutions and the pseudoinverse.

\paragraph{Full column rank case.}
Let $A\in\mathbb{R}^{m\times n}$ with $m\geq n$ and full column rank. Write $A=QR$, where $Q$ has
orthonormal columns and $R$ is upper triangular and invertible. Then the Moore--Penrose pseudoinverse is
\[
A^+=R^{-1}Q^\top.
\]

\paragraph{General case with rank deficiency.}
If $A$ is rank-deficient, one uses the rank-revealing QR factorization (Sec.~\ref{sec:qrred}), i.e.
\[
AP=Q\begin{pmatrix}R_1 \\ 0\end{pmatrix},
\]
where $R_1$ is invertible and $P$ is a permutation matrix. Then
\[
A^+=P\begin{pmatrix}R_1^{-1}Q^\top & 0\end{pmatrix}.
\]

\paragraph{Applications.}
This construction is used extensively in solving least-squares problems:
\[
\min_{\vec x}\|A\vec x-\vec b\|_2,
\]
whose solution is $\vec x=A^+\vec b$.

---

\subsection{$(P)LU$ decomposition}
\label{sec:ludec}
\index{key}{LU decomposition}

Another classical factorization is the \emph{LU decomposition}, closely related to Gaussian elimination.

\paragraph{Definition.}
A matrix $A\in\mathbb{R}^{n\times n}$ admits an LU decomposition
\[
A=LU,
\]
where $L$ is lower triangular (with unit diagonal entries by convention) and $U$ is upper triangular.

\paragraph{Pivoting.}
To ensure numerical stability, one typically uses row pivoting:
\[
A=PLU,
\]
where $P$ is a permutation matrix. This is known as the \emph{PLU decomposition}, or LU decomposition
with partial pivoting\footnote{The systematic study of LU factorization with pivoting was introduced by
A.~M.~Turing, ``Rounding-off errors in matrix processes,'' \emph{Quarterly Journal of Mechanics and Applied
Mathematics} \textbf{1}, 287–308 (1948).}.

\paragraph{Algorithm.}
The PLU decomposition can be computed via Gaussian elimination:
\begin{enumerate}
\item Initialize $U=A$, $L=I$, $P=I$.
\item For $i=1,\ldots,n-1$:
\begin{itemize}
 \item Choose pivot row $j\geq i$ with largest $|U_{ji}|$. Swap rows $i$ and $j$ in $U$, $L$, and $P$.
 \item For $k=i+1,\ldots,n$, eliminate entry $U_{ki}$ by setting $L_{ki}=U_{ki}/U_{ii}$ and updating row $k$
 of $U$.
\end{itemize}
\end{enumerate}

\paragraph{Solving linear systems.}
Given $A=PLU$, solving $A\vec x=\vec b$ proceeds in three steps:
\[
PA\vec x=P\vec b=LU\vec x,
\]
then forward substitution with $L$, followed by back substitution with $U$.  

---

\subsubsection{Cholesky decomposition}
\label{sec:cholesky}
\index{key}{Cholesky decomposition}

For symmetric positive definite matrices, LU factorization simplifies.

\begin{theorem}[Cholesky]
If $A\in\mathbb{R}^{n\times n}$ is symmetric and positive definite, then there exists a unique lower
triangular matrix $L$ with positive diagonal entries such that
\[
A=LL^\top.
\]
\end{theorem}

If $A$ is Hermitian positive definite, then $A=LL^\dagger$.  

\paragraph{Algorithm.}
Cholesky factorization is computed by:
\begin{equation}
L_{ii}=\sqrt{A_{ii}-\sum_{k=1}^{i-1}L_{ik}^2}, \qquad
L_{ij}=\frac{1}{L_{jj}}\left(A_{ij}-\sum_{k=1}^{j-1}L_{ik}L_{jk}\right), \quad i>j.
\end{equation}

\paragraph{Applications.}
Cholesky decomposition is numerically stable and about twice as fast as LU.  
It is the standard method for solving systems with positive definite matrices, such as those arising in
least-squares problems, covariance matrices in statistics, and kernel matrices in machine learning\footnote{See
N.~Higham, \emph{Accuracy and Stability of Numerical Algorithms}, SIAM (2002).}.

\subsubsection{Polar decomposition and closest orthogonal matrix}
\label{sec:polar}
\index{key}{Polar decomposition}

Given a real square matrix $A$, we may view it as a linear transformation mapping $\vec x$ to $A\vec x$.
Geometrically, such a transformation can always be decomposed into a rotation/reflection and a stretching
transformation. This intuition is formalized in the \emph{polar decomposition}.

\paragraph{Definition.}
The polar decomposition of $A\in\mathbb{R}^{n\times n}$ is
\begin{equation}
A=RP,
\end{equation}
where $R$ is orthogonal ($R^\top R=I$) and $P$ is symmetric positive semidefinite.  
If $A$ is complex, the decomposition takes the form
\[
A=UP,
\]
with $U$ unitary and $P$ Hermitian positive semidefinite.

\paragraph{Existence and uniqueness.}
The decomposition always exists and is unique. Indeed, $A^\top A$ is positive semidefinite, and we define
\[
P=(A^\top A)^{1/2}, \qquad R=A P^{-1}.
\]
Then $R$ is orthogonal, since
\[
R^\top R=P^{-1}A^\top A P^{-1}=P^{-1}P^2 P^{-1}=I.
\]
Thus
\begin{equation}
A=(A^\top A)^{1/2}\,( (A^\top A)^{-1/2}A ),
\end{equation}
with the first factor positive semidefinite and the second orthogonal.  
In the complex case, replace $\top$ with $\dagger$.  

\paragraph{SVD interpretation.}
If $A=U\Sigma V^\dagger$ is the singular value decomposition, then the polar decomposition is
\[
A=(UV^\dagger)(V\Sigma V^\dagger).
\]
Thus the unitary factor is $UV^\dagger$, and the positive factor is $V\Sigma V^\dagger$.

\paragraph{Closest orthogonal matrix.}
An important property of the polar decomposition is that the orthogonal/unitary factor is the
\emph{closest orthogonal matrix} to $A$ in both the spectral and Frobenius norms.  

\begin{theorem}[Orthogonal Procrustes problem]
Let $A\in\mathbb{R}^{n\times n}$. Then
\[
\min_{R\in O(n)}\|A-R\|_F=\|A-R^\ast\|_F,
\]
where
\[
R^\ast=(A^\top A)^{-1/2}A
\]
is the orthogonal factor of the polar decomposition of $A$.  
\end{theorem}

\noindent
The proof uses the invariance of the Frobenius norm under unitary transformations and the SVD of $A$.
This result is fundamental in numerical linear algebra and statistics (e.g.\ in Procrustes analysis,
computer vision, and mechanics)\footnote{See N.~Higham, \emph{Functions of Matrices}, SIAM (2008).}.

---
\chapter{Perturbation of eigenvalues and extensions}
\section{Perturbation theory for eigenvalues}

The perturbation of eigenvalues under additive or multiplicative changes of a matrix 
has been discussed earlier. For example, the Wielandt--Hoffman theorem\index{key}{Wielandt-Hoffman theorem} 
bounds the variation of eigenvalues in terms of the Frobenius norm of the perturbation.  
Here we take a more constructive approach, in the spirit of perturbation theory 
as developed in quantum mechanics.\footnote{See, e.g., J.~Sakurai, 
\emph{Modern Quantum Mechanics}, Addison-Wesley (1994).}

\subsection{Non-degenerate perturbation theory}
\index{key}{Eigenvalues Perturbation theory (non-degenerate)}

Suppose we wish to study how eigenvalues and eigenvectors vary under a small perturbation.  
Let
\begin{equation}
   H(\epsilon)=H_0+\epsilon V,
\end{equation}
where $H_0$ and $V$ are $N\times N$ matrices, and $\epsilon$ is a small parameter.  
The matrices need not be symmetric.  

Let $\lambda^i(\epsilon)$ denote the perturbed eigenvalues and 
$\vec v^i_{L/R}(\epsilon)$ the left and right eigenvectors:
\begin{equation}
   H(\epsilon)\,\vec v^i_{R}(\epsilon)=\lambda^i(\epsilon)\,\vec v^i_{R}(\epsilon),
   \qquad
   \vec v^i_{L}(\epsilon)^{\top} H(\epsilon)=\lambda^i(\epsilon)\,\vec v^i_{L}(\epsilon)^{\top}.
\end{equation}
We assume the normalization
\[
   \vec v^i_{L}(\epsilon)\cdot \vec v^i_{R}(\epsilon)=1,
   \qquad \forall i.
\]

We now expand in powers of $\epsilon$:
\begin{eqnarray}
   \lambda^i(\epsilon) &=& \lambda^i(0) + \epsilon \mu^i + \tfrac{1}{2}\epsilon^2 \nu^i + O(\epsilon^3), \nonumber \\
   \vec v^i_{R}(\epsilon) &=& \vec v^i_{R}(0) + \epsilon \vec e^i_{R} + \tfrac{1}{2}\epsilon^2 \vec g^i + O(\epsilon^3).
\end{eqnarray}
The normalization condition implies 
\[
   \vec v^i_{L}(0)\cdot \vec e^i_{R} = \vec v^i_{R}(0)\cdot \vec e^i_{L}=0.
\]

At first order, substituting into the eigenvalue equation and keeping terms of order $\epsilon$, we obtain
\begin{equation}
   (H_0-\lambda^i(0)I)\,\vec e^i_{R}
   = -\big(V-\mu^i I\big)\,\vec v^i_{R}(0).
\end{equation}
Multiplying on the left by $\vec v^j_{L}(0)$ yields
\begin{equation}
   \vec v^j_{L}(0)\cdot V \vec v^i_{R}(0)
   = \mu^i \delta_{ij} + (\lambda^i(0)-\lambda^j(0))\,\vec v^j_{L}(0)\cdot \vec e^i_{R}.
\end{equation}

Thus:
- For $j=i$:
\[
   \mu^i = \vec v^i_{L}(0)\cdot V \vec v^i_{R}(0).
\]
- For $j\neq i$:
\[
   \vec v^j_{L}(0)\cdot \vec e^i_{R}
   = \frac{\vec v^j_{L}(0)\cdot V \vec v^i_{R}(0)}{\lambda^i(0)-\lambda^j(0)}.
\]

Expanding $\vec e^i_R$ in the basis of $\{\vec v^k_{R}(0)\}$, we obtain
\begin{equation}
   \vec e^i_{R} = \sum_{k\neq i} 
   \frac{\vec v^k_{L}(0)\cdot V \vec v^i_{R}(0)}{\lambda^i(0)-\lambda^k(0)}\,
   \vec v^k_{R}(0).
\end{equation}

At second order, one finds
\begin{equation}
   \nu^i = 2 \sum_{k\neq i} 
   \frac{\big(\vec v^i_{L}(0)\cdot V \vec v^k_{R}(0)\big)
         \big(\vec v^k_{L}(0)\cdot V \vec v^i_{R}(0)\big)}
        {\lambda^i(0)-\lambda^k(0)}.
\end{equation}
The correction to the eigenvector at order $\epsilon^2$ can be derived similarly, 
but the key structure is already captured at first and second order.

We see that if two eigenvalues coincide, denominators vanish and 
the expansion breaks down. In this case one must resort to 
degenerate perturbation theory.

---

\subsection{Degenerate perturbation theory}
\index{key}{Eigenvalues Perturbation theory (degenerate)}

When two or more eigenvalues of $H_0$ are equal, the non-degenerate expansion fails.  
In this case, perturbation theory must be reformulated within the degenerate subspace.  

Suppose $H_0\vec v^i(0)=\lambda \vec v^i(0)$ for $i=1,\dots,M$, with multiplicity $M$.  
Choose an orthonormal basis of this subspace (e.g. via Gram--Schmidt\index{key}{Gram-Schmidt decomposition}) so that
\[
   \vec v^i(0)\cdot \vec v^j(0)=\delta_{ij}.
\]

We seek perturbed eigenvectors of the form
\[
   \vec v(\epsilon)=\sum_{j=1}^M c_j \vec v^j(0).
\]
The eigenvalue equation
\[
   (H_0+\epsilon V)\vec v(\epsilon)=\lambda(\epsilon)\vec v(\epsilon)
\]
reduces, at first order in $\epsilon$, to
\begin{equation}
   \epsilon \sum_{j=1}^M c_j V \vec v^j(0) = (\lambda(\epsilon)-\lambda)\sum_{j=1}^M c_j \vec v^j(0).
\end{equation}
Projecting onto $\vec v^k(0)$ gives
\begin{equation}
   \epsilon \sum_{j=1}^M c_j \, V_{kj} 
   = (\lambda(\epsilon)-\lambda)\,c_k,
\end{equation}
where $V_{kj}=\vec v^k(0)\cdot V \vec v^j(0)$.  

Thus, to first order, the corrections to the eigenvalues are the eigenvalues 
of the $M\times M$ matrix $V_{kj}$ restricted to the degenerate subspace:
\begin{equation}
   \det(V_{ij}-\Delta\lambda\,\delta_{ij})=0.
\end{equation}
The associated eigenvectors give the linear combinations of degenerate states 
that diagonalize $V$.  

This analysis shows that degeneracy lifts under perturbation: the original 
eigenvalue $\lambda$ typically splits into distinct values 
$\lambda+\epsilon \Delta\lambda_i$.  

As a special case, if $[H_0,V]=0$, then $H_0$ and $V$ share eigenvectors, 
and the perturbation theory is exact: no splitting occurs and 
the corrections are simply given by diagonal entries of $V$.

\subsection{Exceptional points and eigenvalue crossings}
\index{key}{Exceptional points (eigenvalues)}\index{key}{Eigenvalue crossing}

A natural question is whether a perturbation can cause two eigenvalues to 
coincide or cross as parameters are varied.  
This plays an important role in quantum phase transitions, 
where so-called \emph{exceptional points} appear.  

\paragraph{Two-dimensional example.}
Consider first the case of a $2\times 2$ Hermitian matrix
\begin{equation}
   H = H_0+P
   = \begin{pmatrix} a_1 & 0 \\ 0 & a_2 \end{pmatrix}
   + \begin{pmatrix} p_1 & p_3 \\ p_3 & p_2 \end{pmatrix},
\end{equation}
where $P$ is the perturbation, depending smoothly on parameters 
$\alpha_1,\ldots,\alpha_k$.  
The eigenvalues are
\begin{eqnarray}
   \lambda_\pm &=& \tfrac{1}{2}\Big(r_1 \pm \sqrt{r_2}\Big), \nonumber \\
   r_1 &=& a_1+a_2+p_1+p_2, \nonumber \\
   r_2 &=& (a_1-a_2+p_1-p_2)^2+4p_3^2.
\end{eqnarray}
Thus, for the two eigenvalues to coincide we must have \emph{simultaneously} $r_1=0$ and $r_2=0$.  

This is the simplest manifestation of the phenomenon of \emph{level repulsion}\index{key}{level repulsion}: 
in generic Hermitian systems, levels avoid crossing unless fine-tuned.  
This was first formalized in the celebrated paper of von Neumann and Wigner (1929).\footnote{J.~von Neumann and E.~Wigner, ``Über merkwürdige diskrete Eigenwerte'', 
\emph{Phys. Z.} \textbf{30}, 465 (1929).}

\begin{figure}[h]
   \centering
   \includegraphics[scale=0.45]{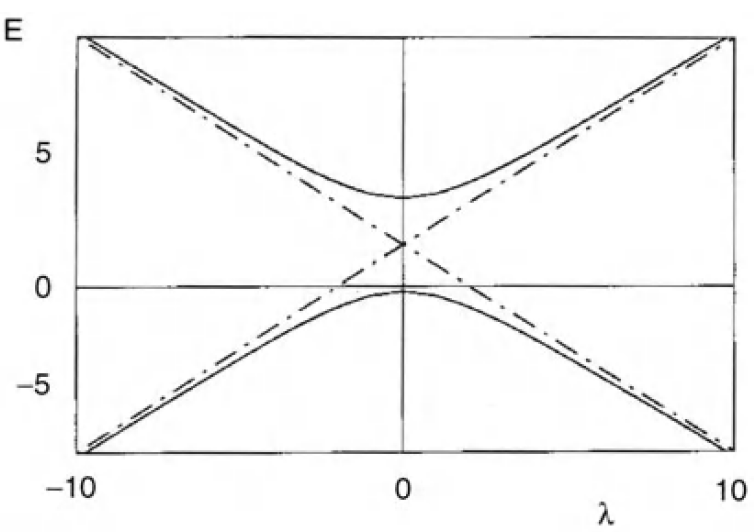}
   \caption{Typical hyperbolic form of level repulsion for two levels.}
   \label{fig:hypf}
\end{figure}

If $P$ is not symmetric,
\[
   P=\begin{pmatrix} p_1 & p_3 \\ p_4 & p_2 \end{pmatrix},
\]
then the eigenvalues become
\begin{equation}
   \lambda_\pm=\tfrac{1}{2}\Big(r_1 \pm \sqrt{\tilde r_2}\Big),
   \qquad
   \tilde r_2=(a_1-a_2+p_1-p_2)^2+4p_3p_4.
\end{equation}
Here, degeneracies can occur in different parameter regimes, 
often leading to complex-valued crossings (exceptional points).

\paragraph{General argument.}
For a Hermitian matrix $H$, the characteristic polynomial
\begin{equation}
   p(\lambda)=\det(\lambda I-H)=c\prod_{i=1}^n(\lambda-\lambda_i),
\end{equation}
has only real roots $\lambda_i\in\mathbb R$.  
A crossing requires both
\begin{eqnarray}
   p(\lambda^*)&=&0, \nonumber \\
   \partial_\lambda p(\lambda^*)&=&0.
   \label{eq:crossing}
\end{eqnarray}
If the perturbation depends on $k$ parameters 
$\alpha_1,\ldots,\alpha_k$, then $\lambda_i(\alpha_1,\ldots,\alpha_k)$ 
varies smoothly with the parameters.\footnote{For a rigorous account, 
see T.~Kato, \emph{Perturbation Theory for Linear Operators}, Springer (1966).}  
Equations~(\ref{eq:crossing}) impose two conditions.  
Thus, generically a manifold of dimension $k-2$ exists where level crossings occur.  

In particular, if $k<2$, crossings are not possible except under 
special symmetries.  
This is the content of the \emph{no-crossing theorem}\index{key}{no-crossing theorem}: 
in a single-parameter family of Hermitian matrices, 
two levels cannot cross unless enforced by symmetry.  

\begin{figure}[h]
   \centering
   \includegraphics[scale=1.3]{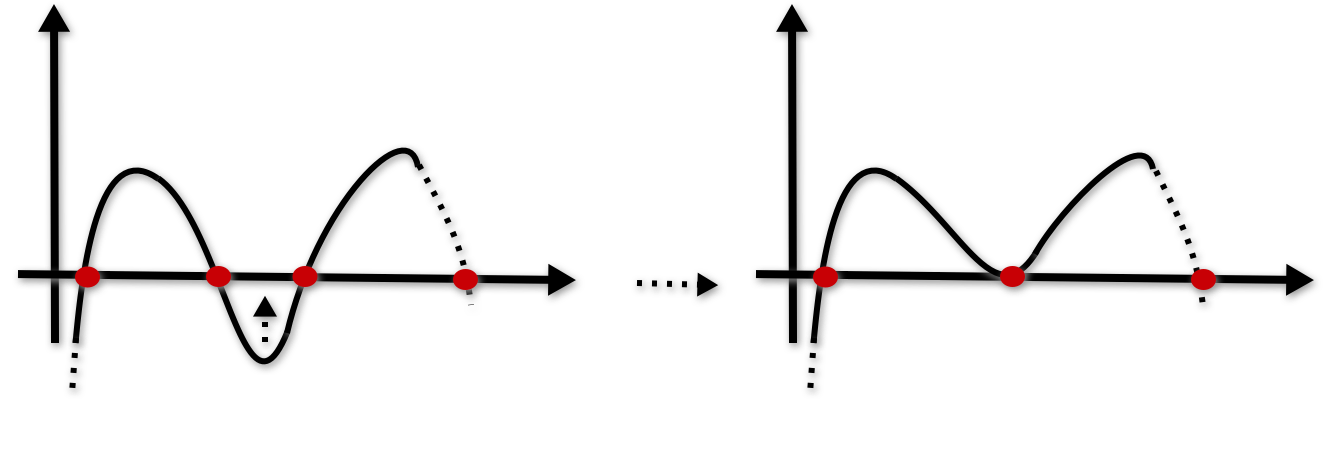}
   \caption{Characteristic polynomial for a self-adjoint matrix.  
   Level crossing requires both $p(\lambda)=0$ and $p'(\lambda)=0$.}
   \label{fig:polrep}
\end{figure}

---

\section{Lie groups and algebras}
\index{key}{Lie groups}
\index{key}{Lie algebras}

For what follows, it is useful to introduce the notions of \emph{Lie groups} and \emph{Lie algebras}, which
generalize the algebraic structures we have seen so far.  
Lie groups provide the language of continuous symmetries, while Lie algebras capture their infinitesimal
generators. Both appear naturally in mathematics, physics, and engineering (e.g.\ control theory).

\paragraph{Definition.}
A Lie group $\mathcal{G}$ is a group whose elements form a smooth manifold, such that the group
operations (multiplication and inversion) are smooth maps.  
In contrast to discrete groups, Lie groups are continuous: their elements depend on continuously varying
parameters.

\paragraph{Examples.}
\begin{itemize}
\item $(\mathbb{R}^n,+)$, vectors with addition.
\item $GL(n,\mathbb{R})$, the general linear group of invertible real $n\times n$ matrices.
\item $O(n,\mathbb{R})$, the orthogonal group; $SO(n,\mathbb{R})$, the special orthogonal group
(determinant $1$).
\item $U(n)$, the unitary group; $SU(n)$, the special unitary group (unitary with determinant $1$).
\item $Sp(2n,\mathbb{R})$, the real symplectic group.
\end{itemize}

These groups arise as transformation groups preserving geometric structures: orthogonal transformations
preserve inner products, symplectic transformations preserve canonical forms, etc.

---

\subsubsection{Actions and adjoint representation}
Let $G$ be a Lie group. Define:
\begin{itemize}
\item Left action: $L_g(h)=gh$.
\item Right action: $R_g(h)=hg^{-1}$.
\item Adjoint action: $\mathrm{Ad}_g(h)=ghg^{-1}$.
\end{itemize}

These satisfy useful identities:
\begin{enumerate}
\item $L_g\circ R_g=R_g\circ L_g$,
\item $\mathrm{Ad}_g=L_g\circ R_g$,
\item $\mathrm{Ad}_g(I)=I$,
\item $\mathrm{Ad}_{g^n}=\mathrm{Ad}_g^n$.
\end{enumerate}

The infinitesimal version of $\mathrm{Ad}_g$ defines the \emph{adjoint representation} of the Lie algebra,
\[
\mathrm{ad}_A(B)=[A,B],
\]
which encodes the commutator structure.

---

\subsubsection{Exponential map}
A crucial fact about Lie groups is that elements near the identity can be expressed as exponentials of Lie
algebra elements:
\[
g=\exp(X), \qquad X\in \mathfrak{g}.
\]
The Lie algebra $\mathfrak{g}$ is the tangent space at the identity, equipped with the Lie bracket
(commutator).

For matrix groups (subgroups of $GL(n)$), this map is particularly simple: $\exp(X)$ is the usual matrix
exponential. The exponential map ensures that continuous symmetries are generated by infinitesimal
ones.

---

\subsubsection{Linear groups}
\index{key}{linear groups}

The general linear group $GL(n,\mathcal{K})$ over a field $\mathcal{K}$ is the archetype of a Lie group.  
Its important subgroups include:
\begin{itemize}
\item $SL(n,\mathcal{K})$: special linear group, $\det(A)=1$.
\item $O(n,\mathcal{K})$: orthogonal group, $A^\top A=I$.
\item $SO(n,\mathcal{K})$: special orthogonal group, $\det(A)=1$.
\item $U(n)$: unitary group, $A^\dagger A=I$.
\item $SU(n)$: special unitary group, $\det(A)=1$.
\item $Sp(2n,\mathbb{R})$: symplectic group.
\end{itemize}

These groups describe familiar transformations: rotations, reflections, unitary rotations in complex
spaces, etc.  

---

\subsubsection{Example: $SO(3)$}
The group $SO(3)$ of real $3\times 3$ rotations is fundamental in geometry and physics. Its Lie algebra
$\mathfrak{so}(3)$ consists of real antisymmetric $3\times 3$ matrices. A basis is
\[
J_x=\begin{pmatrix}0&0&0\\0&0&-1\\0&1&0\end{pmatrix},\quad
J_y=\begin{pmatrix}0&0&1\\0&0&0\\-1&0&0\end{pmatrix},\quad
J_z=\begin{pmatrix}0&-1&0\\1&0&0\\0&0&0\end{pmatrix}.
\]
These are the generators of rotations about the coordinate axes.  

Any element of $SO(3)$ can be written as
\[
R(\vec\theta)=\exp(\theta_x J_x+\theta_y J_y+\theta_z J_z).
\]
The commutation relations are
\[
[J_x,J_y]=J_z,\quad [J_y,J_z]=J_x,\quad [J_z,J_x]=J_y,
\]
the defining relations of $\mathfrak{so}(3)$.

---

\subsubsection{Example: $SU(2)$}
The group $SU(2)$ consists of $2\times 2$ complex unitary matrices of determinant $1$. Its Lie algebra
$\mathfrak{su}(2)$ is spanned by the Pauli matrices:
\[
\sigma_1=\begin{pmatrix}0&1\\1&0\end{pmatrix},\quad
\sigma_2=\begin{pmatrix}0&-i\\i&0\end{pmatrix},\quad
\sigma_3=\begin{pmatrix}1&0\\0&-1\end{pmatrix}.
\]
The commutation relations are
\[
[\sigma_i,\sigma_j]=2i\,\epsilon_{ijk}\,\sigma_k,
\]
where $\epsilon_{ijk}$ is the Levi--Civita symbol.  

$SU(2)$ is a double cover of $SO(3)$: every rotation in three dimensions corresponds to two opposite
elements in $SU(2)$. This relation is fundamental in quantum mechanics, where spin-$\tfrac{1}{2}$
particles transform under $SU(2)$, not $SO(3)$.

---

\subsubsection{Lie algebras and structure constants}
Every Lie group $G$ has an associated Lie algebra $\mathfrak{g}$. Given a basis $\{T_i\}$ of $\mathfrak{g}$,
the commutator defines
\[
[T_i,T_j]=c_{ij}^k T_k,
\]
where $c_{ij}^k$ are the \emph{structure constants}. These constants encode the algebraic structure of
the group locally.  

For example:
\begin{itemize}
\item $\mathfrak{so}(3)$ has structure constants $c_{ij}^k=\epsilon_{ijk}$.
\item $\mathfrak{su}(2)$ has structure constants $c_{ij}^k=2\epsilon_{ijk}$.
\end{itemize}

---

\subsubsection{Cosets and quotient groups}
Let $N$ be a subgroup of $G$.  
Left and right cosets are
\[
gN=\{gn: n\in N\}, \qquad Ng=\{ng: n\in N\}.
\]
If $gN=Ng$ for all $g\in G$, then $N$ is a \emph{normal subgroup}.  
The quotient group $G/N$ is the set of cosets.  

For Lie groups, coset spaces $G/H$ (with $H$ a closed subgroup) are smooth manifolds of dimension
$\dim(G)-\dim(H)$. Examples include spheres, projective spaces, and Grassmannians.

---

\paragraph{Retrospective.}
In this chapter we have explored the algebra of matrices: eigenvalues, eigenvectors, and spectral
decompositions (diagonalization, Schur, Jordan, SVD, QR, LU, Cholesky). We then moved to nonlinear
eigenvalue problems and generalized inverses (Moore--Penrose, Drazin, Bott--Duffin).  

The introduction of Lie groups and algebras places these concepts in a broader framework:  
\begin{itemize}
\item Matrix groups unify geometry and algebra: orthogonal and unitary groups preserve norms,
symplectic groups preserve canonical structures.  
\item Lie algebras encode infinitesimal generators of transformations; commutators capture their algebraic
structure.  
\item The exponential map connects infinitesimal generators to global transformations, much as matrix
exponentials connect Jordan forms to dynamics.  
\end{itemize}

Thus, the study of matrix algebra is not only about computation, but also about the geometry of
transformations. The language of Lie groups and Lie algebras will provide a powerful bridge as we move
from linear algebra to the theory of dynamical systems, networks, and complex systems.


\section{Nonlinear eigenvalue problems}
\index{key}{nonlinear eigenvalue problems}
\index{key}{eigenvalue problems, nonlinear}

So far we have considered linear eigenvalue problems of the form
\begin{equation}
(A-\lambda I)\vec x=0,
\end{equation}
and generalized versions $(A-\lambda B)\vec x=0$ (Sec.~\ref{sec:geneigen}).  

We now move beyond linear dependence in $\lambda$.  

\paragraph{Polynomial eigenvalue problems.}
A general nonlinear eigenvalue problem can be written
\begin{equation}
A(\lambda)\vec x=0, \qquad 
A(\lambda)=\sum_{i=0}^r \lambda^i A_i,
\label{eq:genep}
\end{equation}
with coefficient matrices $A_i$.  

\paragraph{Quadratic case.}
For the quadratic problem
\[
(A_2\lambda^2+A_1\lambda+A_0)\vec x=0,
\]
we set $\vec y=\lambda \vec x$ and obtain the linearized block form
\[
\begin{pmatrix}
A_0 & A_1 \\
0 & I
\end{pmatrix}\begin{pmatrix}
\vec x\\
\vec y
\end{pmatrix}
=\lambda
\begin{pmatrix}
0 & -A_2\\
I & 0
\end{pmatrix}
\begin{pmatrix}
\vec x\\
\vec y
\end{pmatrix}.
\]
Thus quadratic problems can be reduced to generalized linear eigenvalue problems.

\paragraph{General polynomial case.}\label{sec:companion}
For the general polynomial problem \eqref{eq:genep}, one constructs an $rn\times rn$ block matrix
(companion matrix) of Frobenius type:
\[
A=\begin{pmatrix}
0 & I & 0 & \cdots & 0 \\
0 & 0 & I & \cdots & 0 \\
\vdots & & \ddots & \ddots & \vdots \\
0 & 0 & \cdots & 0 & I \\
B_0 & B_1 & \cdots & B_{r-2} & B_{r-1}
\end{pmatrix},
\]
where $B_i=-A_r^{-1}A_i$. Then
\[
A\begin{pmatrix}\vec x\\ \lambda \vec x\\ \lambda^2 \vec x\\ \vdots \\ \lambda^{r-1}\vec x\end{pmatrix}
=\lambda \begin{pmatrix}\vec x\\ \lambda \vec x\\ \lambda^2 \vec x\\ \vdots \\ \lambda^{r-1}\vec x\end{pmatrix}.
\]
Thus the polynomial eigenvalue problem is recast as a standard linear eigenproblem of higher dimension\footnote{See T.~Betcke et al., ``NLEVP: A collection of nonlinear eigenvalue problems,'' \emph{ACM Transactions on Mathematical Software} \textbf{39}, 7 (2013).}.

\subsection{Analytic nonlinear eigenvalue problems}
\index{key}{nonlinear eigenvalue problems}

A broad class of nonlinear eigenvalue problems arises when the matrix depends on $\lambda$ through a
power series or analytic function. Consider
\begin{equation}
\sum_{r=0}^\infty A_r \lambda^r \vec x=0,
\end{equation}
with coefficient matrices $A_r$. A special case is when
\[
A_r=c_r A^r,
\]
for some fixed matrix $A$ and scalar coefficients $c_r$. Then
\begin{equation}
\sum_{r=0}^\infty c_r A^r \lambda^r \vec x=f(\lambda A)\vec x=0,
\end{equation}
where $f$ is the scalar generating function $f(z)=\sum_{r=0}^\infty c_r z^r$ applied to the operator $\lambda A$.

\paragraph{Reduction to scalar zeros.}
If $A$ is diagonalizable, $A=Q\operatorname{diag}(\rho_1,\ldots,\rho_N)Q^{-1}$, then the problem reduces to
finding $\lambda$ such that
\[
f(\lambda \rho_i)=0,
\]
for some eigenvalue $\rho_i$ of $A$. Thus the nonlinear spectrum is determined by the zeros of $f$.

\paragraph{Example.}
If $f(z)=e^z$, then no solution exists since $e^z$ has no zeros.  
If $f(z)=\sin z$, then solutions occur at $\lambda\rho_i=k\pi$, giving
\[
\lambda_{ij}=\frac{x_j}{\rho_i},\qquad x_j\in\{k\pi: k\in\mathbb{Z}\},\ \rho_i\neq 0.
\]

\paragraph{Multiplicity.}
If $t=\dim\ker(A)$ and $N$ is the size of $A$, then only the $N-t$ nonzero eigenvalues contribute to the
construction. If $f$ has $M$ distinct zeros, then the nonlinear problem has $M(N-t)$ solutions:
\[
\lambda_{ij}=\frac{x_j}{\rho_i},\quad i=1,\ldots,N-t,\ j=1,\ldots,M.
\]

---

\subsection{$\lambda$-matrices}
\index{key}{$\lambda$-Matrix}

Polynomial eigenvalue problems are often described in terms of \emph{$\lambda$-matrices}, following
Gantmacher\footnote{F.~R.~Gantmacher, \emph{The Theory of Matrices}, Chelsea Publishing (1959).}.  
A $\lambda$-matrix is a matrix whose entries are polynomials in $\lambda$:
\[
Q(\lambda)=A_r\lambda^r+A_{r-1}\lambda^{r-1}+\cdots+A_1\lambda+A_0,
\]
with coefficient matrices $A_i\in\mathbb{C}^{n\times n}$.  

\paragraph{Spectrum.}
The \emph{spectrum} of $Q$ is defined as
\[
\Lambda(Q)=\{\lambda\in\mathbb{C}:\det(Q(\lambda))=0\}.
\]
For example, in the quadratic case
\[
Q(\lambda)=A_2\lambda^2+A_1\lambda+A_0,
\]
$\det(Q(\lambda))$ is a scalar polynomial of degree $2n$ (assuming $A_2$ is nonsingular), so there are
$2n$ solutions counting multiplicity.

---

\paragraph{Quadratic eigenvalue problems.}
Quadratic problems admit several structured approaches:
\begin{itemize}
\item \emph{Linearization:} introduce variables $\vec y=\lambda\vec x$, leading to a generalized eigenvalue
problem of size $2n$ (as described earlier).
\item \emph{Divisors:} introduce $Q(X)=A_2X^2+A_1X+A_0$. Then
\begin{equation}
Q(\lambda)-Q(X)=A_2(\lambda^2 I-X^2)+A_1(\lambda I-X),
\end{equation}
which factorizes as
\[
Q(\lambda)-Q(X)=(\lambda A_2+A_1 X+A_0)(\lambda I-X).
\]
This relation is known as Bézout’s theorem for quadratic matrix polynomials\index{key}{Bézout theorem}.
\end{itemize}

If a matrix $X$ satisfies $Q(X)=0$, then $X$ is called a \emph{solvent} of the quadratic polynomial.
Such an $X$ provides a factorization of $Q$, analogous to factoring a scalar polynomial when a root is
known.

---

\paragraph{General theory of $\lambda$-matrices.}
The theory of $\lambda$-matrices provides a framework for:
\begin{itemize}
\item Studying parameter-dependent matrices, e.g.\ $A(\lambda)$ in structural mechanics, vibrations, or control theory.
\item Defining matrix divisors and common factors of matrix polynomials.
\item Extending spectral theory to matrix-valued analytic functions.
\end{itemize}

Polynomial and rational eigenvalue problems appear in applications such as:
\begin{itemize}
\item vibration analysis of mechanical structures (quadratic eigenvalue problems),
\item control theory (delay and feedback systems),
\item electromagnetics and photonics (nonlinear dispersion relations).
\end{itemize}
\index{key}{polynomial eigenvalue problem}

\chapter{Graphs}\label{sec:graphs}
\section{What is a graph?}

A graph\index{key}{graph} is one of the simplest relational mathematical objects, and yet one of the most powerful in applications. Formally, a graph $G$ is an ordered pair $(V,E)$, where $V$ is a set of \textit{vertices} (or \textit{nodes}) and $E$ is a set of pairs of vertices called \textit{edges} (or \textit{arcs}). This definition allows us to model a wide variety of systems in which entities interact in pairs: from molecules to social networks, from transportation systems to the Internet.\footnote{See for instance D.~West, \emph{Introduction to Graph Theory}, Prentice Hall (1996); R.~Diestel, \emph{Graph Theory}, Springer (2000); B.~Bollobás, \emph{Modern Graph Theory}, Springer (1998).}

\begin{figure}[h]
    \centering
    \includegraphics[scale=0.4]{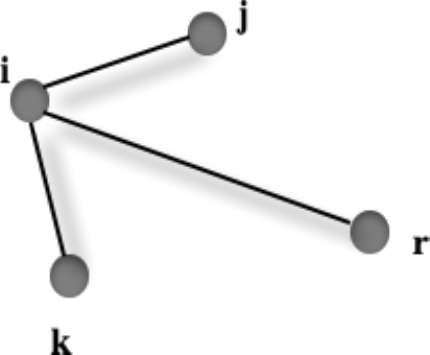}
    \caption{A simple undirected graph with four nodes $i,j,k,r$, and edges $(i,j)$, $(i,r)$ and $(i,k)$.}
    \label{fig:graph0}
\end{figure}

\subsection{Types of graphs}
Graphs come in many variants, depending on the properties we attach to edges:
\begin{itemize}
    \item \textbf{Undirected graphs}: edges are unordered pairs $(i,j)$, meaning that a link has no inherent direction.
    \item \textbf{Directed graphs (digraphs)}: edges are ordered pairs $(i,j)$, representing a relation from $i$ to $j$.
    \item \textbf{Weighted graphs}: each edge carries a weight $w_{ij} \geq 0$, used to model strengths, capacities, or costs.
    \item \textbf{Multigraphs and hypergraphs}: allow multiple edges between the same pair of nodes, or edges that connect more than two nodes.
\end{itemize}

\begin{figure}[h]
    \centering
    \includegraphics[scale=0.4]{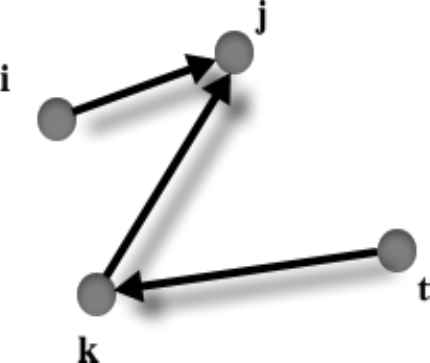}
    \caption{An example of a directed graph on four nodes $i,j,k,t$, with edges $(i,j)$, $(t,k)$ and $(k,r)$.}
    \label{fig:graphdir}
\end{figure}

\subsection{Graph isomorphism}
An important problem in graph theory is whether two graphs are \textit{isomorphic}. 
An \textit{isomorphism} is a bijection between the vertex sets of two graphs $G_1=(V_1,E_1)$ and $G_2=(V_2,E_2)$ such that $(u,v)\in E_1$ if and only if $(f(u),f(v))\in E_2$. 
Put simply, two graphs are isomorphic if, by relabeling the vertices, they become identical. 
Isomorphic graphs share all structural properties, such as degree distribution or connectivity. 
The \emph{Graph Isomorphism Problem}—deciding efficiently whether two graphs are isomorphic—remains one of the few major problems in NP that is neither known to be solvable in polynomial time nor proven NP-complete.\footnote{See M.~Garey and D.~Johnson, \emph{Computers and Intractability}, W.~H.~Freeman (1979). For the breakthrough result showing quasipolynomial-time algorithms, see L.~Babai, \emph{Graph isomorphism in quasipolynomial time}, Proc. 48th ACM Symposium on Theory of Computing (2016).}

\subsection{Matrix representations of graphs}
Graphs can be encoded in matrices, which makes them amenable to linear algebraic methods.
\begin{itemize}
    \item The \textbf{adjacency matrix} $A$ is defined as $A_{ij}=1$ if $(i,j)\in E$, and $0$ otherwise (for weighted graphs, $A_{ij}=w_{ij}$).
    \item The \textbf{incidence matrix} $B$ has rows indexed by vertices and columns by edges, with entries $B_{ve}=1$ if $v$ is incident to edge $e$, and $0$ otherwise.
    \item The \textbf{graph Laplacian} is defined as $L = D - A$, where $D$ is the diagonal matrix of vertex degrees, $D_{ii}=\sum_j A_{ij}$. The Laplacian plays a central role in spectral graph theory and dynamics on networks.\footnote{See F.~Chung, \emph{Spectral Graph Theory}, AMS (1997).}
\end{itemize}

\subsection{Distances and connectivity}
The \textit{distance} between two nodes $u,v \in V$, denoted $d_G(u,v)$, is the length of the shortest path between them, if such a path exists. If no path exists, $d_G(u,v)=\infty$. The graph is said to be \textit{connected} if for every pair of vertices $u,v$ the distance $d_G(u,v)$ is finite. Otherwise, the graph is \textit{disconnected}. An important topological quantity is the number of connected components $c(G)$. If $c(G)=1$, the graph is connected; if $c(G)>1$, the graph splits into $c(G)$ disjoint connected subgraphs.

Connectivity is a basic structural property: for instance, it determines whether a process (such as diffusion, information spreading, or epidemics) can reach the whole graph or only part of it.\footnote{See M.~Newman, \emph{Networks: An Introduction}, Oxford University Press (2010).}

\subsection{Why graphs?}
The language of graphs provides a universal framework for representing relational data. Their simplicity—just nodes and edges—belies their depth: from pure mathematics to modern applications in physics, biology, computer science, and social sciences. In the following sections, we will use graph representations together with spectral methods to extract deeper structural and dynamical insights.

\subsection{Walks and paths}
\index{key}{graph walks}

One of the most fundamental notions in graph theory is that of \emph{walks} and \emph{paths}, which are essential not only in pure mathematics but also in practical applications such as routing, search algorithms, or wireless communications, where shortest paths play a central role.\footnote{For background see D.~West, \emph{Introduction to Graph Theory}, Prentice Hall (1996).}

A \textbf{walk} of length $n$ in a graph $G=(V,E)$ is a sequence of vertices 
\[
v_1, v_2, \ldots, v_{n+1}
\]
such that each consecutive pair $(v_i,v_{i+1})$ forms an edge in $E$. The edges $e_1,\dots,e_n$ that connect these vertices are traversed in order.  

Walks can be of different types:
\begin{itemize}
    \item A \textbf{closed walk} satisfies $v_1 = v_{n+1}$.
    \item A \textbf{path} is a walk in which all vertices are distinct, $v_i \neq v_j$ for $i\neq j$.
    \item A \textbf{cycle} is a closed walk in which the starting and ending vertices coincide, but all other vertices are distinct.
    \item A \textbf{trivial path} has length zero, consisting of a single vertex.
\end{itemize}

A graph is called \textit{acyclic} if it contains no cycles. Acyclic graphs are of special interest because they are easier to classify and analyze than general graphs. The most important class of acyclic connected graphs is the class of \emph{trees}.

\subsection{Bridges and trees}
\begin{figure}[h]
\centering
\includegraphics[scale=0.3]{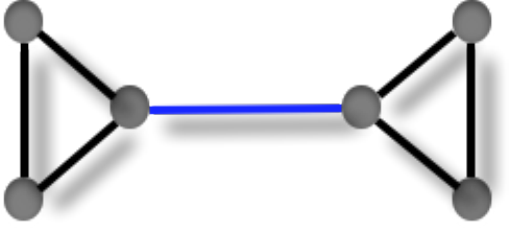}
\caption{An example of a bridge (blue edge) in a graph.}
\label{fig:bridge}
\end{figure}

To introduce trees, we first recall the notion of a \emph{bridge} (also called a \emph{cut edge} or bottleneck). A bridge is an edge $e\in E$ such that removing it increases the number of connected components of the graph, i.e.
\[
c(G-e) > c(G).
\]
For example, in Fig.~\ref{fig:bridge}, if the blue edge is removed, the number of connected components increases from $1$ to $2$.  

A simple but fundamental fact is that an edge $e$ is a bridge if and only if $e$ does not lie in any cycle of $G$.\footnote{See B.~Bollobás, \emph{Modern Graph Theory}, Springer (1998).}  
If $G$ is connected, then removing a bridge always increases the number of connected components by exactly one.  

A \textbf{tree} is a connected graph in which every edge is a bridge; equivalently, it is a connected acyclic graph. Another useful characterization is that in a tree, any two vertices are connected by exactly one simple path.  

The following conditions are equivalent:\footnote{See R.~Diestel, \emph{Graph Theory}, Springer (2000).}
\begin{itemize}
    \item $G$ is a tree.
    \item $G$ is connected and acyclic.
    \item $G$ is connected with $n-1$ edges, where $n=|V|$.
    \item Between any two vertices of $G$ there is exactly one path.
\end{itemize}

\begin{figure}[h]
\centering
\includegraphics[scale=0.3]{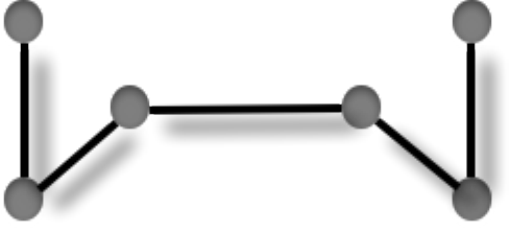}
\caption{A tree with 6 nodes and 5 edges. Every edge is a bridge.}
\label{fig:tree}
\end{figure}

Trees play a central role in graph theory and its applications. Every connected graph $G$ contains a subgraph $T\subseteq G$ that is a tree and includes all the vertices of $G$. Such a subgraph is called a \textbf{spanning tree}. Spanning trees provide minimal skeletons of graphs and form the basis of many algorithms, such as those for network design and optimization (e.g.~Kruskal’s and Prim’s algorithms).\footnote{See R.~Tarjan, \emph{Data Structures and Network Algorithms}, SIAM (1983).}

\subsection{Bipartite graphs}
Another important family of graphs is that of \emph{bipartite graphs}. A graph $G=(V,E)$ is bipartite if the vertex set can be partitioned into two subsets $X$ and $Y$, such that every edge connects a vertex in $X$ with a vertex in $Y$. No edge is allowed to connect two vertices in $X$, nor two in $Y$.  

\begin{figure}[h]
\centering
\includegraphics[scale=0.3]{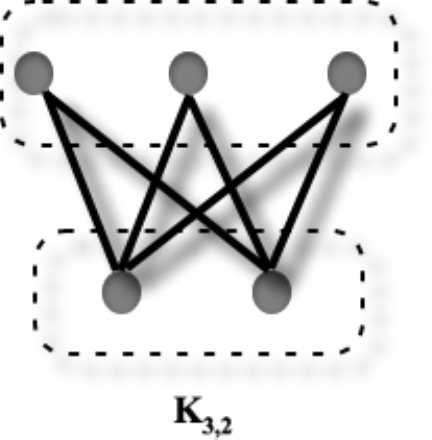}
\caption{An example of the complete bipartite graph $K_{3,2}$.}
\label{fig:bipartite}
\end{figure}

A complete bipartite graph is denoted $K_{m,n}$, with $|X|=m$ and $|Y|=n$. An example of $K_{3,2}$ is shown in Fig.~\ref{fig:bipartite}.  

A fundamental theorem characterizes bipartite graphs:  
\begin{quote}
\textit{A graph is bipartite if and only if it contains no odd cycle as a subgraph.}
\end{quote}
This provides a simple way to check bipartiteness in practice and underlies efficient algorithms for partitioning graphs.\footnote{See D.~West, \emph{Introduction to Graph Theory}, Prentice Hall (1996).}  

Bipartite graphs are ubiquitous in applications: they naturally model systems with two types of entities, such as workers and tasks, customers and products, or authors and papers. Their special structure also makes them amenable to powerful matching and flow algorithms.

\subsection{Graph measures}
\label{graph_meas}

Graphs can be studied at many levels of resolution. Some measures describe the position of \emph{individual vertices} in the network (local measures), while others quantify global structural properties of the whole graph. Such measures, often referred to as \emph{centrality indices} or \emph{network measures}, are usually defined in terms of the adjacency matrix $A$ or other matrix functions derived from it.\footnote{See M.~Newman, \emph{Networks: An Introduction}, Oxford University Press (2010).}

\subsubsection{Local graph measures}

\paragraph{Degree.}
The most basic local measure is the \emph{degree} of a vertex $i$, defined as
\[
k_i = \sum_j A_{ij}.
\]
In directed graphs, one distinguishes the \emph{in-degree} $k_i^{\text{in}} = \sum_j A_{ji}$ and the \emph{out-degree} $k_i^{\text{out}} = \sum_j A_{ij}$.  
For weighted graphs, the \emph{weighted degree} (or \emph{strength}) is given by
\[
k_i^w = \sum_j w_{ij},
\]
where $w_{ij}$ denotes the weight of edge $(i,j)$.  
The sequence $(k_1,k_2,\dots,k_n)$ is often called the \emph{degree sequence} of the graph. Degree distributions play a central role in network science: many real-world systems exhibit highly skewed distributions (such as power laws) rather than the narrow distributions typical of random graphs.\footnote{See A.-L.~Barabási and R.~Albert, \emph{Emergence of scaling in random networks}, Science \textbf{286}, 509 (1999).}

\paragraph{Eigenvector centrality.}
A refinement of degree is \emph{eigenvector centrality}, which accounts not only for the number of connections of a node, but also for the importance of its neighbors.\footnote{See P.~Bonacich, \emph{Power and centrality: A family of measures}, American Journal of Sociology \textbf{92}, 1170–1182 (1987).}  
It is defined as the solution of
\[
c_i = \frac{1}{\lambda}\sum_j A_{ij}c_j ,
\]
that is, the centrality vector $c$ is an eigenvector of $A$, with eigenvalue $\lambda$. Usually one takes the leading eigenvector (associated with the largest eigenvalue) to ensure non-negative entries by the Perron–Frobenius theorem.  
In contrast to simple degree, eigenvector centrality recognizes that “not all connections are equal”: being connected to well-connected nodes yields a higher score.\footnote{For discussion, see M.~Newman, \emph{Networks: An Introduction}, Oxford University Press (2010); J.~Donges et al., \emph{Unified functional network and nonlinear time series analysis for complex systems science: The pyunicorn package}, Chaos \textbf{25}, 113101 (2015).}

\paragraph{Clustering coefficient.}
Another important local measure is the \emph{clustering coefficient}, which quantifies the tendency of neighbors of a node to also be connected with each other.  
For a vertex $v$ of degree $k_v$, the (local) clustering coefficient is defined as
\[
C_v = \frac{2 \cdot \text{number of edges among neighbors of $v$}}{k_v(k_v-1)}.
\]
This expression measures the fraction of possible triangles through $v$ that actually exist.\footnote{See D.~Watts and S.~Strogatz, \emph{Collective dynamics of small-world networks}, Nature \textbf{393}, 440–442 (1998).}  
The clustering coefficient takes values between 0 and 1, with $C_v=1$ if the neighbors of $v$ form a complete subgraph (clique).  

One can also define a \emph{global clustering coefficient}, or \emph{transitivity}, as the ratio
\[
C = \frac{\text{number of closed triplets}}{\text{number of triplets (open and closed)}},
\]
which expresses the overall density of triangles in the graph. High clustering is a hallmark of many real-world networks, especially social networks, where groups of individuals tend to form tightly-knit communities.

\paragraph{Other centralities.}
Additional local measures include \emph{closeness centrality} (inverse of average distance from a vertex to all others), \emph{betweenness centrality} (fraction of shortest paths that pass through a vertex), and many others. Since these are more global in nature, we defer their detailed discussion to Section~\ref{central_meas}.  

\medskip

Local graph measures quantify how “important,” “well-connected,” or “clustered” a single node is relative to its neighbors. They provide the first lens through which we can analyze complex networks, before moving to more global and spectral properties.

 \subsubsection{Global graph measures}

While local measures quantify the role of individual nodes, \emph{global graph measures} provide single-valued characteristics of the entire network. They are often obtained by averaging local measures or by considering graph-wide structures.  

\paragraph{Average degree and degree distribution.}
The simplest global quantity is the \emph{average degree}, defined as
\[
\langle k \rangle = \frac{1}{n}\sum_{i=1}^n k_i,
\]
where $n=|V|$ is the number of vertices. Closely related is the \emph{degree distribution} $P(k)$, the fraction of nodes with degree $k$. The distribution’s shape (e.g.~Poissonian, power-law, or exponential) distinguishes between classical random graphs and many real-world complex networks.\footnote{See M.~Newman, \emph{Networks: An Introduction}, Oxford University Press (2010).}  

\paragraph{Density.}
Another simple global measure is the \emph{edge density}, defined as
\[
\rho = \frac{|E|}{\binom{n}{2}}
\]
for undirected simple graphs. It gives the ratio of actual edges to the maximum possible number of edges. Sparse graphs ($\rho \ll 1$) are the norm in most applications.

\paragraph{Average path length and diameter.}
The \emph{average path length} is the average over all shortest path distances:
\[
\ell = \frac{1}{n(n-1)} \sum_{i\neq j} d_G(i,j),
\]
while the \emph{diameter} of a graph is the maximum of these distances. These measures quantify the “small-world” property: many real networks have surprisingly short path lengths relative to their size.\footnote{See D.~Watts and S.~Strogatz, \emph{Collective dynamics of small-world networks}, Nature \textbf{393}, 440–442 (1998).}  

\paragraph{Global clustering.}
The \emph{global clustering coefficient} (or \emph{transitivity}) is defined as
\[
C = \frac{\text{number of closed triplets}}{\text{number of all triplets (open and closed)}}.
\]
It measures the prevalence of triangles across the entire graph and complements the local clustering coefficient described earlier.

\paragraph{Entropy of degree distribution.}
Entropy-based measures capture network heterogeneity. The Shannon entropy of the degree distribution is defined as
\[
H = - \sum_k P(k) \log P(k),
\]
which quantifies the uncertainty in the degree of a randomly chosen vertex.\footnote{See G.~Bianconi, \emph{Entropy of network ensembles}, Physical Review E \textbf{79}, 036114 (2009).}  
High entropy reflects structural diversity, while low entropy indicates homogeneity.

\paragraph{Spectral measures.}
The spectrum of the adjacency or Laplacian matrix encodes important global properties. For example, the second-smallest Laplacian eigenvalue $\lambda_2$, known as the \emph{algebraic connectivity} or \emph{Fiedler value}, measures how well-connected the graph is. A higher $\lambda_2$ indicates stronger connectivity.\footnote{See M.~Fiedler, \emph{Algebraic connectivity of graphs}, Czechoslovak Mathematical Journal \textbf{23}, 298–305 (1973).}  

Other spectral measures include the spectral radius (largest eigenvalue of $A$), which influences epidemic thresholds and dynamical processes on networks.\footnote{See R.~Pastor-Satorras and A.~Vespignani, \emph{Epidemic spreading in scale-free networks}, Physical Review Letters \textbf{86}, 3200–3203 (2001).}

\paragraph{Assortativity.}
Another global measure is \emph{assortativity}, which quantifies the correlation between the degrees of adjacent vertices. Social networks are often assortative (high-degree nodes connect preferentially to other high-degree nodes), whereas technological and biological networks tend to be disassortative.\footnote{See M.~Newman, \emph{Mixing patterns in networks}, Physical Review E \textbf{67}, 026126 (2003).}  

\paragraph{Random walk measures.}
Global measures can also be defined through random walks on graphs. A central concept is the \emph{mean first passage time} (MFPT) $\tau_{ij}$, the expected number of steps for a random walker starting at $i$ to reach $j$ for the first time. MFPTs capture how efficiently information or particles spread in networks and depend strongly on graph structure.\footnote{See C.~Meyer, \emph{Matrix Analysis and Applied Linear Algebra}, SIAM (2000). For applications to networks, see D.~Brockmann and D.~Helbing, \emph{The hidden geometry of complex, network-driven contagion phenomena}, Science \textbf{342}, 1337–1342 (2013).}  

Recent works have derived MFPTs analytically for specific graph classes (such as trees, lattices, or random graphs).\footnote{See S.~Lee, S.~Khang, and D.~Kim, \emph{Mean first-passage time for random walks on trees}, Physical Review E \textbf{87}, 062810 (2013).}  
MFPTs are intimately related to the eigenvalues of the Laplacian matrix and thus connect random walk theory to spectral graph theory (see Section~\ref{sec:eigenvalues}).

\medskip

Global graph measures condense complex structural information into single indices. They are indispensable for comparing networks, identifying universal features such as small-worldness or scale-freeness, and linking structure to dynamical processes.

\chapter{Connectedness in graphs}
\section{Connectivity in graph theory}

\subsection{Connected and strongly connected graphs}

Connectivity is one of the most basic but also most subtle notions in graph theory.  
For undirected graphs, the concept is straightforward: an undirected graph is said to be \emph{connected} if there exists a path between every pair of vertices. If this is not the case, the graph splits into several connected components, as discussed in Section~\ref{graph_meas}.

For directed graphs (digraphs), the situation is richer. A directed graph $G=(V,E)$ is \emph{strongly connected} if for every ordered pair of vertices $(i,j)$ there exists a directed path from $i$ to $j$.\footnote{See R.~Diestel, \emph{Graph Theory}, Springer (2000).} If this condition holds only when ignoring edge orientations, then $G$ is called \emph{weakly connected}.  

It is often useful to associate a directed graph $G(A)$ with a non-negative matrix $A=(A_{ij})$, where an edge from vertex $i$ to $j$ is present if and only if $A_{ij}\neq 0$. More generally, the graph $G(A^k)$ has an edge from $i$ to $j$ whenever there exists a directed path of length $k$ from $i$ to $j$ in $G(A)$. This correspondence is fundamental when studying the interplay between connectivity and matrix powers: spectral methods often weight paths of different lengths in different ways, and connectivity can be characterized by whether such paths exist.

\subsection{Matrix reducibility and types of connectedness}

The graph-theoretic notion of strong connectivity corresponds precisely to the linear-algebraic concept of irreducibility of a non-negative matrix.\footnote{See C.~Meyer, \emph{Matrix Analysis and Applied Linear Algebra}, SIAM (2000); E.~Seneta, \emph{Non-negative Matrices and Markov Chains}, Springer (1981).}  

\begin{definition}[Reducibility]
A square matrix $A\in \mathbb{R}^{n\times n}$ is called \emph{reducible} if there exists a permutation matrix $P$ such that
\[
C = PAP^T = 
\begin{pmatrix}
A_{11} & A_{12} \\
0 & A_{22}
\end{pmatrix},
\]
where $A_{11}$ and $A_{22}$ are square submatrices of sizes $r$ and $n-r$ with $0<r<n$. Otherwise, $A$ is called \emph{irreducible}.
\end{definition}

The block-triangular form reveals that the vertex set of $G(A)$ can be partitioned into two non-empty subsets with no edges leading back from the second set to the first. In this case, the graph is not strongly connected.  

\begin{theorem}
A non-negative matrix $A$ is irreducible if and only if the directed graph $G(A)$ is strongly connected.
\end{theorem}

\begin{proof}[Sketch]
If $A$ is reducible, then by permuting vertices one finds a block upper-triangular form with a zero block below the diagonal, which implies that some vertices are not reachable from others—hence $G(A)$ is not strongly connected. Conversely, if $G(A)$ is not strongly connected, then its vertices can be ordered so that no edge goes from one component back to another, producing the block form above.  
\end{proof}

This equivalence is a cornerstone of the Perron–Frobenius theory for non-negative matrices. It implies that algebraic properties of $A$ mirror structural properties of the graph $G(A)$.

\paragraph{Nilpotent matrices.}  
If a non-negative square matrix $A$ is nilpotent (i.e. $A^k=0$ for some finite $k$), then $A$ must be reducible. Indeed, if $A$ were irreducible, then by definition there would exist a power $t$ such that every entry of $A^t$ is positive. But this contradicts the vanishing of high powers implied by nilpotency.

\paragraph{Primitive matrices.}  
A non-negative irreducible matrix $A$ is called \emph{primitive} if there exists some power $t$ such that $A^t$ has strictly positive entries.\footnote{See R.~Horn and C.~Johnson, \emph{Matrix Analysis}, Cambridge University Press (1985).}  
Primitivity implies irreducibility, but not conversely: an irreducible matrix may fail to be primitive if, for example, its associated graph is periodic (every cycle length is a multiple of some integer $d>1$).  

A useful observation is that if $A$ is irreducible, then $I+A$ is primitive, since 
\[
(I+A)^n = \sum_{k=0}^n \binom{n}{k} A^k
\]
contains all powers of $A$ and hence eventually all entries become positive. This fact ties together connectivity, paths of arbitrary length, and spectral positivity.

---

The bridge between connectivity in graphs and reducibility in matrices allows us to use linear algebra to classify structural properties of networks, and conversely to interpret algebraic notions like primitivity and nilpotency in purely combinatorial terms.

\section{Special matrices in graph theory}

\subsection{Adjacency matrices}\label{sec:adjmat}

The \emph{adjacency matrix}\index{key}{adjacency matrix} of a finite graph $G=(V,E)$ with $|V|=n$ is the $n\times n$ matrix $A=(a_{ij})$ defined by
\[
a_{ij} =
\begin{cases}
\text{number of edges from $i$ to $j$}, & i\neq j, \\
\text{number of self-loops at $i$}, & i=j.
\end{cases}
\]
For simple undirected graphs without self-loops, $A$ is a symmetric $\{0,1\}$–matrix with zero diagonal.  

If $G$ is undirected, $A=A^T$; if $G$ is directed, $A$ need not be symmetric.  
Graph isomorphisms correspond to permutations of vertices, which act on $A$ by simultaneous row–column permutations.  

The \emph{degree} (or valency) of a vertex $i$ is given by
\[
d_i = \sum_j a_{ij}.
\]
Equivalently, if $\mathcal N(i)$ denotes the neighborhood of $i$, then $d(i)=|\mathcal N(i)|$.  

\paragraph{Handshaking lemma.}  
The sum of all degrees equals twice the number of edges:
\[
\sum_{i=1}^n d_i = 2|E|,
\]
so the number of odd-degree vertices is always even.\footnote{This classical result goes back at least to Euler’s work on the Königsberg bridges problem. See B.~Bollobás, \emph{Modern Graph Theory}, Springer (1998).}  

\paragraph{Powers of $A$.}  
Entries of powers of $A$ have a combinatorial meaning:
\[
(A^k)_{ij} = \text{number of walks of length $k$ from $i$ to $j$}.
\]
This is one of the fundamental connections between graph theory and linear algebra.

\begin{figure}[h]
\centering
\includegraphics[scale=0.3]{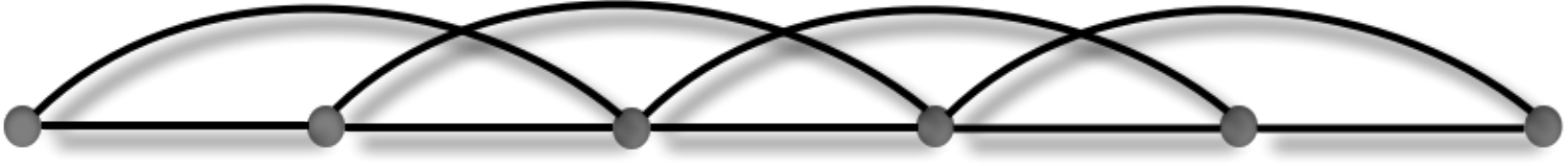}
\caption{A graph whose adjacency matrix is banded.}
\label{fig:banded}
\end{figure}

\paragraph{Banded adjacency matrices.}  
If $A$ can be put, by a permutation of vertices, into a banded form, the graph has a local structure, with edges restricted to nearest or next-to-nearest neighbors. An example is shown in Fig.~\ref{fig:banded}.

\paragraph{Nilpotency.}  
If some power $A^r=0$, then no paths of length $\ge r$ exist, which implies that the graph is acyclic. In particular, strictly upper-triangular adjacency matrices correspond to directed acyclic graphs (DAGs).

\paragraph{Bipartite structure.}  
The adjacency matrix of a bipartite graph can always be put into block form
\[
P^{-1}AP =
\begin{pmatrix}
0 & B \\
B^T & 0
\end{pmatrix},
\]
where $B$ is an $r\times k$ matrix, with $r$ and $k$ the sizes of the two partitions. This reflects the absence of edges within each part.

---

\subsubsection{Incidence matrices}\label{sec:incmat}

The \emph{incidence matrix}\index{key}{incidence matrix} of a graph $G=(V,E)$ with $n=|V|$ vertices and $m=|E|$ edges is the $n\times m$ matrix $B=(b_{i\beta})$ defined by
\[
b_{i\beta} =
\begin{cases}
1 & \text{if vertex $i$ is incident to edge $\beta$ (undirected)}, \\
\pm 1 & \text{if $i$ is the head/tail of $\beta$ (directed)}, \\
0 & \text{otherwise}.
\end{cases}
\]
Thus each column of $B$ corresponds to an edge, and has one or two nonzero entries depending on orientation.

The incidence matrix is related to the adjacency matrix of the \emph{line graph} $\mathscr L(G)$, whose vertices correspond to edges of $G$ and are adjacent when the corresponding edges share a vertex. One has the relation\footnote{See N.~Biggs, \emph{Algebraic Graph Theory}, Cambridge University Press (1993).}
\[
A(\mathscr L(G)) = B^T B - 2 I_m,
\]
where $I_m$ is the $m\times m$ identity.

---

\subsubsection{Laplacian matrices}\label{sec:lapmat}

The \emph{graph Laplacian}\index{key}{Laplacian} is defined by
\[
L = D - A,
\]
where $D=\mathrm{diag}(d_1,\dots,d_n)$ is the diagonal matrix of vertex degrees. Alternatively, using the incidence matrix,
\[
L = B B^T.
\]

\paragraph{Normalized and random-walk Laplacians.}  
Other common variants are:
\[
L_{\mathrm{norm}} = D^{-1/2} L D^{-1/2} = I - D^{-1/2} A D^{-1/2},
\]
\[
L_M = D^{-1} L = I - D^{-1}A,
\]
the latter being the generator of a random walk on the graph.

\paragraph{Properties.}  
For any vector $x\in\mathbb{R}^n$,
\[
x^T L x = \sum_{(i,j)\in E} (x_i - x_j)^2 \ge 0,
\]
so $L$ is positive semidefinite.\footnote{See F.~Chung, \emph{Spectral Graph Theory}, AMS (1997).}  
For weighted graphs with edge weights $a_{ij}>0$, the quadratic form generalizes to
\[
x^T L x = \sum_{(i,j)\in E} a_{ij}(x_i - x_j)^2.
\]

Moreover, $L\vec 1 = 0$, so the smallest eigenvalue is always $\lambda_0=0$. The multiplicity of the zero eigenvalue equals the number of connected components of the graph.  

\paragraph{Analogy with the continuum.}  
The Laplacian on graphs is designed to mimic the continuum Laplacian operator
\[
\Delta u = \sum_{i=1}^n \partial^2_{x_i} u.
\]
Like its continuum counterpart, the graph Laplacian is self-adjoint and encodes diffusion processes. The spectrum of $L$ thus controls random walks, consensus dynamics, and diffusion on networks.\footnote{See J.~Merris, \emph{Laplacian matrices of graphs: A survey}, Linear Algebra Appl. \textbf{197–198}, 143–176 (1994).}  

---

In short, the adjacency matrix encodes direct connections, the incidence matrix encodes node–edge relations, and the Laplacian encodes flows or diffusion. These three matrices form the algebraic foundation of spectral graph theory.

\section{Properties of the Laplacian}

Very often, problems in spectral graph theory reduce to eigenvalue equations of the form
\begin{equation}
L u = \lambda u,
\end{equation}
where $L$ is the Laplacian of a graph. Understanding how the eigenvalues $\{\lambda_i\}$ depend on the structure of the graph, and how they change under the addition or removal of edges, is one of the central tasks of spectral graph theory.\footnote{See F.~Chung, \emph{Spectral Graph Theory}, AMS (1997).}  

\paragraph{Characteristic polynomial and Rayleigh principle.}
Eigenvalues can be studied through the characteristic polynomial
\[
T(\lambda) = \det(L - \lambda I).
\]
An alternative variational description is given by the Rayleigh–Ritz principle: for a symmetric positive semidefinite matrix such as $L$, the smallest nonzero eigenvalue is
\[
\lambda_2 = \min_{\substack{x \perp \vec 1 \\ x \neq 0}} \frac{x^T L x}{x^T x}.
\]
This formulation highlights the dependence of $\lambda_2$ on connectivity and plays a key role in spectral partitioning methods.\footnote{See M.~Fiedler, \emph{Algebraic connectivity of graphs}, Czechoslovak Mathematical Journal \textbf{23}, 298–305 (1973).}  

\subsection{Incidence matrices and rank}

The incidence matrix is particularly useful for studying how Laplacians change when edges are added.  
If $G$ is a graph with Laplacian $L_G$, and a new edge $e$ (represented by the vector $z$) is added, then
\[
L_{G+\{e\}} = L_G + z z^T.
\]
Using the matrix determinant lemma, one obtains
\[
\det(L - \mu I) = \det(L_G - \mu I)\,\det\!\left(I + (L_G - \mu I)^{-1} z z^T\right).
\]
This shows explicitly how the spectrum is perturbed by the addition of a single edge.\footnote{See C.~Meyer, \emph{Matrix Analysis and Applied Linear Algebra}, SIAM (2000).}  

The rank of the incidence matrix also encodes connectivity: if the graph has $n$ vertices and $c(G)$ connected components, then $\mathrm{rank}(B) = n - c(G)$. In particular, for a connected graph, $\mathrm{rank}(B) = n-1$.

\subsection{Laplacian and degree matrices}

For an undirected graph $G$ with adjacency matrix $A$ and degree matrix $D$, the Laplacian is
\begin{equation}
L = D - A.
\label{eq:Ldef}
\end{equation}
The trace of $L$ counts the number of edges:
\begin{equation}
\mathrm{Tr}(L) = \sum_i d_i = 2m,
\end{equation}
where $m=|E(G)|$.  
More generally,
\begin{equation}
\mathrm{Tr}(L^2) = \sum_i d_i^2 + 2m,
\end{equation}
relating the squared eigenvalues of $L$ to vertex degrees and edge count.

\paragraph{Spectral properties.}
The Laplacian is positive semidefinite, since for any vector $x$,
\[
x^T L x = \sum_{(i,j)\in E} (x_i - x_j)^2 \geq 0.
\]
Thus all eigenvalues are nonnegative. Moreover, $L\vec 1 = 0$, so $\lambda_1=0$. The multiplicity of the zero eigenvalue equals the number of connected components of $G$. In particular, $G$ is connected if and only if $0$ is a simple eigenvalue.

\paragraph{Algebraic connectivity and graph partitioning.}
The second smallest eigenvalue $\lambda_2$, called the \emph{algebraic connectivity}, measures how well the graph is connected. Its corresponding eigenvector, the \emph{Fiedler vector}, can be used for graph partitioning: vertices can be divided into groups according to the signs of their entries in this eigenvector.\footnote{See U.~von Luxburg, \emph{A tutorial on spectral clustering}, Statistics and Computing \textbf{17}, 395–416 (2007).}  
This method underlies spectral clustering and many community detection algorithms.

\subsection{Directed graphs}

For directed graphs, the definition of Laplacian is less canonical, since $A$ is not symmetric. A natural generalization is:
\[
L^+ = D^{\text{in}} - A, \qquad L^- = D^{\text{out}} - A^T,
\]
with $D^{\text{in}}_{ii} = \sum_j A_{ij}$ and $D^{\text{out}}_{ii} = \sum_j A_{ji}$.  
From these, one can form symmetrized versions such as
\[
L_1 = \tfrac{1}{2}(L^+ + L^-), \qquad
L_2 = \sqrt{L^+L^- + L^-L^+}.
\]

The spectral properties of these matrices differ, but when the Laplacian is \emph{normal} ($LL^T = L^T L$), it is unitarily diagonalizable. Normality is closely related to the notion of balanced directed graphs, where each vertex has equal in-degree and out-degree.\footnote{See P.~Van Mieghem, \emph{Graph Spectra for Complex Networks}, Cambridge University Press (2011).}  
In fact, it has been shown that if the Laplacian of a directed graph is normal, then the graph is balanced.\footnote{See C.~Wu and L.~Chua, \emph{Synchronization in an array of linearly coupled dynamical systems}, IEEE Trans. Circuits and Systems I \textbf{42}, 430–447 (1995).}  

---

The Laplacian thus sits at the heart of spectral graph theory: its eigenvalues encode connectivity, robustness, and diffusion, while its eigenvectors form the basis for modern methods of clustering and community detection.

\section{Paths, Walks and Spectra}

\subsection{Paths and walks}
The eigenvalues of the adjacency matrix $A$ encode information about walks and paths in a graph.  
If $A^k$ is the $k$-th power of $A$, then
\[
N_k = \sum_{i,j} (A^k)_{ij}
\]
counts the total number of walks of length $k$ in the graph. More generally, $(A^k)_{ij}$ counts the number of walks of length $k$ from $i$ to $j$.  

An upper bound relates these quantities to degrees:
\[
N_k \leq \sum_j d_j^k,
\]
where $d_j$ is the degree of vertex $j$. This reflects the fact that walks proliferate more quickly from high-degree nodes.

\paragraph{Generating functions.}
The generating function for walks is
\[
N_G(z) = \sum_{k=0}^\infty N_k z^k.
\]
Introducing the all-ones vector $\vec u$, one finds
\[
N_G(z) = \vec u^T (I - zA)^{-1} \vec u,
\]
so the generating function is directly related to the \emph{resolvent} of $A$.\footnote{See F.~Chung, \emph{Spectral Graph Theory}, AMS (1997).}  

Using the matrix determinant lemma (Sherman–Morrison identity), one can derive an elegant closed form:
\[
N_G(z) = \frac{1}{z}\left(\frac{\det(I + z(J - A))}{\det(I - zA)} - 1\right),
\]
where $J = \vec u \vec u^T$ is the all-ones matrix.

\subsection{Matrix–tree theorem}
The \emph{matrix–tree theorem} states that the number of spanning trees of a connected graph $G$ is
\[
\tau(G) = \frac{1}{n} \prod_{i=2}^n \mu_i,
\]
where $\mu_2,\dots,\mu_n$ are the nonzero eigenvalues of the Laplacian $L$.\footnote{See N.~Biggs, \emph{Algebraic Graph Theory}, Cambridge University Press (1993).}  
Equivalently, $\tau(G)$ is any cofactor of $L$, i.e. the determinant of $L$ with one row and column removed.  

This formula extends to directed graphs via the notion of \emph{arborescences}. A reverse arborescence is a directed spanning tree oriented toward a root (sink). The number of such arborescences rooted at a vertex $v$ is given by the determinant of the sink-reduced outdegree Laplacian.\footnote{See R.~A.~Brualdi and D.~Cvetković, \emph{A Combinatorial Approach to Matrix Theory and Its Applications}, CRC Press (2009).}

\subsection{Spectral gap and Perron–Frobenius}
Let $\lambda_1 \geq \lambda_2 \geq \cdots \geq \lambda_n$ be the eigenvalues of $A$. The \emph{spectral gap} is
\[
\sigma(A) = \lambda_1 - \lambda_2.
\]
By Perron–Frobenius, for a connected graph with nonnegative adjacency matrix, $\lambda_1$ is simple and its eigenvector can be chosen strictly positive.\footnote{See R.~Horn and C.~Johnson, \emph{Matrix Analysis}, Cambridge University Press (1985).}  

Bounds relate $\lambda_1$ to closed walks:
\[
\lambda_1 \geq \max_{m\geq 1} \max_j \left( (A^m)_{jj} \right)^{1/m}.
\]

The number of closed walks of length $k$ is
\[
W_k = \mathrm{Tr}(A^k) = \sum_i \lambda_i^k.
\]
Its generating function is
\[
W_G(z;j) = \sum_{k=0}^\infty (A^k)_{jj} z^k = \sum_{r=1}^n \frac{(\rho_j^r)^2}{1 - z\lambda_r},
\]
where $\rho^r$ are the eigenvectors of $A$. Thus closed walks are directly tied to the resolvent of $A$.

\section{Subgraph eigenvalues}

If $G_1,G_2$ are vertex-disjoint subgraphs of $G$, then $A_G = A_{G_1} + A_{G_2}$. Inequalities link their spectra:\footnote{See C.~Godsil and G.~Royle, \emph{Algebraic Graph Theory}, Springer (2001).}
\[
\lambda_n(G_1) + \lambda_k(G_2) \leq \lambda_k(G) \leq \lambda_k(G_1) + \lambda_1(G_2).
\]

\subsubsection{Examples}
\begin{itemize}
\item \textbf{Regular graphs:} For a $k$-regular graph, $A\vec 1 = k\vec 1$, so $k$ is an eigenvalue of $A$. The Laplacian eigenvalues are related by $\mu_i = k - \theta_i$, where $\theta_i$ are adjacency eigenvalues.  
\item \textbf{Line graphs:} The line graph $\mathscr L(G)$ has adjacency matrix $A(\mathscr L(G)) = B^T B - 2I$, where $B$ is the incidence matrix of $G$.  
\item \textbf{Complements:} If $\bar G$ is the complement of $G$, then $A_{\bar G} = J - I - A$, and $L_{\bar G} = nI - J - L$. The eigenvectors of $L$ are also eigenvectors of $J$, allowing eigenvalue relations between $G$ and $\bar G$.\footnote{See B.~Bollobás, \emph{Modern Graph Theory}, Springer (1998).}
\end{itemize}

\subsection{Switching methods and isospectrality}
Switching operations can produce non-isomorphic graphs with the same spectrum (isospectral graphs).  

\paragraph{Seidel switching.}  
The Seidel matrix is $S = J - I - 2A$. Seidel switching with respect to a vertex subset changes adjacencies across the cut. Spectra are preserved up to sign, since $S_{\bar G} = -S_G$.

\paragraph{Godsil–McKay switching.}  
Another construction produces cospectral graphs via local modifications that balance eigenvalue multiplicities.\footnote{See C.~Godsil and B.~McKay, \emph{Constructing cospectral graphs}, Aequationes Math. \textbf{25}, 257–268 (1982).}

\section{Graph cohomology}\index{key}{cohomology}

Graphs also admit a cohomological description, particularly useful in network theory and circuit analysis.\footnote{See L.~O.~Chua and P.-M.~Lin, \emph{Computer-Aided Analysis of Electronic Circuits}, Prentice Hall (1975). For a physics perspective see G.~Polettini, \emph{Cycle/cocycle decompositions for the master equation}, J. Stat. Phys. \textbf{161}, 336–352 (2015).}

Consider a connected graph $G$ with $N$ nodes and $M$ edges. Assign an orientation $\mathcal O$ to each edge. Two matrices are fundamental:
\begin{itemize}
\item The \emph{incidence matrix} $B^{\mathcal O}_{\alpha k}$ ($N \times M$), with entries $-1,0,+1$ depending on whether an edge leaves or enters a vertex.
\item The \emph{cycle matrix} $A^{\mathcal O}_{\tilde \xi k}$ ($C \times M$), where $C$ is the number of independent cycles. Entries encode orientation relative to chosen cycles.
\end{itemize}

These correspond to the boundary ($\partial$) and coboundary ($d$) operators in discrete cohomology.\footnote{See A.~Hatcher, \emph{Algebraic Topology}, Cambridge University Press (2002).}

\paragraph{Kirchhoff’s laws.}  
- The \emph{current law} (KCL) is $B \vec i = 0$, expressing conservation of current at each node.  
- The \emph{voltage law} (KVL) is $A \vec v = 0$, expressing that the sum of voltages around any cycle vanishes.  

One usually works with reduced matrices $\tilde B,\tilde A$ to remove linear dependencies. The duality relation $BA^T = A B^T = 0$ reflects Tellegen’s theorem of circuit theory.

\paragraph{Spanning trees and co-chords.}
Choosing a spanning tree $\mathcal T$ allows decomposition of edges into tree edges (co-chords) and non-tree edges (chords). Each chord defines a fundamental cycle. The number of independent cycles is
\[
L = M - N + 1,
\]
a purely topological invariant.  

---

Thus, graph cohomology provides a rigorous language to express conservation laws and dualities, linking algebraic topology with network dynamics and circuit theory.
\chapter{Non-negative matrices}
\section{Motivations}
\label{sec_per_frob}

Matrices with non-negative entries arise naturally in many branches of mathematics and applications.  
In graph theory, adjacency matrices are non-negative by construction, and Laplacians are built from such structures.  
In probability, the transition matrices of Markov chains are non-negative and often stochastic.  
In economics, Leontief input–output models are described by non-negative production matrices, while in demography, Leslie matrices encode non-negative reproduction and survival rates.  
In network science, eigenvector-based centrality measures also derive from non-negative adjacency operators.  
The spectral analysis of these matrices reveals universal organizing principles: regardless of the particular application, non-negativity imposes strong constraints on the location and multiplicity of eigenvalues and on the positivity of eigenvectors.\footnote{For classical treatments see E.~Seneta, \emph{Non-negative Matrices and Markov Chains}, Springer (1981); A.~Berman and R.~Plemmons, \emph{Nonnegative Matrices in the Mathematical Sciences}, SIAM (1994). For modern accounts in linear algebra see R.~Horn and C.~Johnson, \emph{Matrix Analysis}, Cambridge University Press (1985).}

The theory developed by Perron (1907) and Frobenius (1912) formalizes these ideas, proving that irreducible non-negative matrices possess a unique ``dominant'' eigenvalue with a strictly positive eigenvector, and, in the primitive case, a spectral gap separating this eigenvalue from all others.  
This result is not only of theoretical importance but also provides the backbone for probabilistic convergence (Markov chains), growth and stability (population dynamics), and ranking and centrality algorithms (network theory).  

\subsection{Definitions}

A vector $x \in \mathbb{R}^n$ is called \emph{positive} if $x_i > 0$ for all $i$, and \emph{non-negative} if $x_i \geq 0$ for all $i$.  
Similarly, a matrix $A\in\mathbb{R}^{m\times n}$ is \emph{positive} (resp.~non-negative) if all its entries are positive (resp.~non-negative). This should not be confused with symmetric positive definite (SPD) matrices.  

For $A \in \mathbb{C}^{n\times n}$ with eigenvalues $\lambda_1,\dots,\lambda_n$, the \emph{spectral radius} is defined as
\[
\rho(A) = \max_i |\lambda_i|.
\]

\subsection{Perron–Frobenius theorem}\label{sec:perron}
The Perron–Frobenius theorem is one of the cornerstones of spectral theory
with far-reaching implications across mathematics, probability, and the study
of complex systems. It provides the spectral foundation for understanding the
long-time behavior of dynamical processes on nonnegative matrices, including
Markov chains, population growth models, input–output economics, and network
centrality measures.  

Intuitively, Perron–Frobenius guarantees that an irreducible nonnegative system
has a single ``dominant'' eigenvalue, associated with a strictly positive
eigenvector. This eigenpair captures the global steady-state mode of the system:
for Markov chains it corresponds to the stationary distribution, in population
models to the stable growth rate and reproductive distribution, and in network
analysis to the eigenvector centrality that ranks nodes by their long-term
influence.  

The theorem is thus not only an elegant result in linear algebra but also a
conceptual tool: it ensures that positive feedback structures in a connected
system give rise to a unique direction of growth or equilibrium, separating it
cleanly from fluctuations and secondary modes.

\begin{theorem}[Perron–Frobenius]\footnote{See E.~Seneta, \emph{Non-negative Matrices and Markov Chains}, Springer (1981); R.~Horn and C.~Johnson, \emph{Matrix Analysis}, Cambridge University Press (1985); A.~Berman and R.~Plemmons, \emph{Nonnegative Matrices in the Mathematical Sciences}, SIAM (1994).}
Let $A \in \mathbb{R}^{n\times n}$ be an irreducible non-negative matrix. Then:
\begin{enumerate}
\item $A$ has a real eigenvalue $\rho(A)>0$ equal to its spectral radius.  
\item $\rho(A)$ has algebraic and geometric multiplicity one.  
\item There exists a positive eigenvector $x>0$ with $Ax=\rho(A)x$.  
\item If $0\le B\le A$ entrywise and $B\neq A$, then $\rho(B)<\rho(A)$.  
\item There is no other non-negative eigenvector except positive multiples of $x$.  
\item If $A$ is also \emph{primitive} (aperiodic), then all other eigenvalues $\lambda$ satisfy $|\lambda|<\rho(A)$.  
\end{enumerate}
\end{theorem}

Thus, irreducibility guarantees the existence of a unique ``dominant” eigenpair with strictly positive eigenvector, while primitivity strengthens the separation from other eigenvalues.

\textit{Proof sketch}. The proof can be sketched using the compactness of the non-negative unit sphere and the function
\begin{equation}
f(z) = \min_{1\le i \le n,\, z_i\neq 0} \frac{(Az)_i}{z_i}, \qquad z\ge 0,\ z\neq 0.
\label{eq:fz}
\end{equation}
This function has the properties:
\begin{itemize}
\item $f(rz)=f(z)$ for any $r>0$.  
\item If $Az=\lambda z$, then $f(z)=\lambda$.  
\item In general, $f(z)\le f((I+A)^n z)$, with strict inequality unless $z$ is an eigenvector.  
\end{itemize}

The key idea is to maximize $f(z)$ over the non-negative unit sphere. The maximizer $x$ is then an eigenvector corresponding to $f_{\max}$. Compactness ensures existence, positivity of $(I+A)^n$ ensures $x>0$, and one shows $f_{\max}=\rho(A)$.  

\subsection{Monotonicity of the spectral radius}

If $0\le B\le A$ with $B\neq A$, then $\rho(B)<\rho(A)$. Intuitively, increasing any entry of a non-negative matrix can only increase its growth rate. In particular, any principal submatrix (obtained by deleting a row and column) has spectral radius strictly less than $\rho(A)$ when $A$ is irreducible.  

\subsection{Asymptotics for primitive matrices}

If $A$ is primitive, then not only is $\rho(A)$ simple, but the matrix powers converge (after normalization):
\[
\lim_{t\to\infty} \left(\tfrac{1}{\rho(A)} A\right)^t = xy^T,
\]
where $x>0$ and $y>0$ are right and left Perron eigenvectors, normalized so that $y^Tx=1$.  
This result underlies the convergence of Markov chains to a stationary distribution and the long-term dominance of the Perron eigenvector in network centrality measures.\footnote{See C.~Meyer, \emph{Matrix Analysis and Applied Linear Algebra}, SIAM (2000).}

\subsection{Perron root and projections}\label{sec:projectpr}
An elegant application of Perron--Frobenius is the limit
\[
P=\lim_{k\to\infty}\frac{A^k}{\lambda_{pf}^k}.
\]
This operator $P$ is a \emph{projection}: $P^2=P$ and $[P,A]=0$. Indeed, decomposing
\[
\frac{A^r}{\lambda_{pf}^r}=\sum_i \left(\frac{\lambda_i}{\lambda_{pf}}\right)^r
\vec l_i\otimes\vec r_i,
\]
all terms vanish as $r\to\infty$ except the Perron component, yielding
\[
P=\vec l\otimes\vec r,
\]
a rank-1 projector onto the Perron eigenspace.

---

\paragraph{Bounds via Gershgorin disks.}
By the Gershgorin circle theorem (Sec.~\ref{sec:matrixnorm}), any eigenvalue $\lambda$ of $A$
satisfies $|\lambda|\leq \max_i\sum_j |A_{ij}|$. For $A\geq 0$, this gives
\[
\min_i \sum_j A_{ij}\ \leq\ \lambda_{pf}\ \leq\ \max_i \sum_j A_{ij}.
\]
This inequality is sharp for regular graphs and provides useful estimates of growth rates in
population models and random walks.

\subsection{Algebraic and geometric multiplicities}

We now show that the Perron root $\lambda_{\max}=\rho_A$ of an irreducible non-negative matrix $A$ has algebraic and geometric multiplicity equal to $1$. Recall that the geometric multiplicity of an eigenvalue is always less than or equal to its algebraic multiplicity, and both are at least one. Thus, it suffices to prove that the algebraic multiplicity is one.

\paragraph{Notation.} For a square matrix $A$:
\begin{itemize}
\item $A_{(i)}$ denotes the principal submatrix obtained by eliminating the $i$-th row and column.
\item $A_i$ denotes the matrix obtained by zeroing out the $i$-th row and column, without reducing the size of $A$.
\end{itemize}

\paragraph{Lemma.}  
Let $A$ be an $n\times n$ matrix, and let $\Lambda=\mathrm{diag}(\lambda_1,\dots,\lambda_n)$. Then
\[
\frac{\partial}{\partial \lambda_i}\det(\Lambda - A) = \det(\Lambda_{(i)} - A_{(i)}).
\]
This follows by expanding the determinant along the $i$-th row.

\paragraph{Application.}  
Setting $\lambda_i=\lambda$ for all $i$, the chain rule gives
\[
\frac{d}{d\lambda}\det(\lambda I - A) = \sum_{i=1}^n \det(\lambda I - A_{(i)}).
\]
In addition,
\[
\det(\lambda I - A_i) = \lambda \det(\lambda I - A_{(i)}).
\]

Now, by the monotonicity property of the spectral radius (Section~\ref{sec_per_frob}), we know that
\[
\det(\rho_A I - A_{(i)}) > 0 .
\]
Thus, the derivative of the characteristic polynomial of $A$ at $\lambda=\rho_A$ is nonzero. Therefore, $\rho_A$ is a simple root of the characteristic polynomial, i.e.~its algebraic multiplicity equals one. Since geometric multiplicity $\leq$ algebraic multiplicity, both equal one.\footnote{See R.~Horn and C.~Johnson, \emph{Matrix Analysis}, Cambridge University Press (1985).}

---

\subsection{Non-negative eigenvectors}

We next show that the Perron eigenvector is the unique non-negative eigenvector up to scaling.  

\paragraph{Claim.} If $0\leq B \leq A$, then $f_{\max}(B) \leq f_{\max}(A)$, where $f$ is the function defined in Eq.~\eqref{eq:fz}.  

\emph{Proof.} If $z\in Q$ satisfies $sz\leq Bz$, then clearly $sz \leq Az$, so $f_B(z)\leq f_A(z)$.  

Applying this to $A^T$, it follows that $A^T$ has a positive eigenvalue $\eta$ with eigenvector $w>0$, i.e.
\[
w^T A = \eta w^T.
\]
Similarly, $x>0$ is the Perron eigenvector of $A$, $Ax=\lambda_{\max}x$. Then
\[
w^T A x = \eta w^T x = \lambda_{\max} w^T x .
\]
Since $w^Tx>0$, we conclude $\eta=\lambda_{\max}$.  

Now suppose $y\in Q$ satisfies $Ay \leq \mu y$. Then
\[
\lambda_{\max} w^T y = w^T A y \leq \mu w^T y .
\]
Since $w^Ty>0$, this implies $\mu \geq \lambda_{\max}$. If equality holds and $Ay=\mu y$, then $y$ is an eigenvector corresponding to $\lambda_{\max}$. But since $\lambda_{\max}$ has multiplicity one, $y$ must be a scalar multiple of $x$.  

Thus, there is no other non-negative eigenvector except for positive multiples of $x$.\footnote{See E.~Seneta, \emph{Non-negative Matrices and Markov Chains}, Springer (1981).}

---

\subsection{Strict inequality for primitive matrices}

We now show that if $A$ is \emph{primitive} (i.e.~irreducible and aperiodic), then every other eigenvalue has modulus strictly smaller than $\rho_A$.  

\paragraph{Proof.}  
Let $Az=\lambda z$ with $z\in\mathbb{C}^n$, $|\lambda|=\rho_A$. Then
\[
\rho_A |z| = |Az| \leq A|z|.
\]
Hence $f(|z|) \geq \rho_A$, but also $f(|z|)\leq \rho_A$, so equality holds. Thus $|z|$ is a Perron eigenvector of $A$ with eigenvalue $\rho_A$. Moreover, $|Az|=A|z|$ implies that all entries of $z$ have the same complex phase $u$ with $|u|=1$, i.e.~$z_i = u |z_i|$. Therefore $z$ is collinear with $|z|$, and hence corresponds to the Perron root itself. No other eigenvalue can have modulus $\rho_A$.

---

\subsection{Limit for primitive matrices}

Finally, we establish the convergence of normalized powers of $A$.  

Since $A^T$ shares the spectrum of $A$, there exists a positive left Perron eigenvector $y>0$ such that $y^T A = \rho_A y^T$. Choose the scaling $x^T y = 1$, where $x$ is the right Perron eigenvector.  

Define the rank-one matrix $H=xy^T$. Then $H^2=H$, so $H$ is a projection onto the span of $x$. Moreover, $AH=\rho_A H = HA$, so both $R=\mathrm{Im}(H)$ and $N=\ker(H)$ are invariant subspaces of $A$.  

The restriction of $A$ to $R$ is multiplication by $\rho_A$, while the restriction to $N$ has all eigenvalues $<\rho_A$ in modulus.  

Setting $P = \tfrac{1}{\rho_A} A$, we thus have that $P|_N$ has spectral radius $<1$, hence $(P|_N)^t \to 0$ as $t\to\infty$, while $P|_R$ acts as the identity.  

Therefore,
\[
\lim_{t\to\infty} \left(\frac{1}{\rho_A} A \right)^t = xy^T,
\]
the rank-one projection onto the Perron eigenvector direction.\footnote{See C.~Meyer, \emph{Matrix Analysis and Applied Linear Algebra}, SIAM (2000).}

\subsection{Aperiodicity, cyclicity and primitiveness}

So far we have focused on irreducible and primitive matrices. We now refine these notions by introducing periodicity and cyclicity. These are algebraic characterizations that apply to general non-negative matrices and are less intuitive than the simple graph-theoretic notions of connectedness, but are crucial for understanding the spectral structure.  

\paragraph{Cyclicity.}  
Let $A$ be an irreducible non-negative $n\times n$ matrix. For each pair $(i,j)$, define
\[
\mathbb{N}_{ij} = \{ t \in \mathbb{N} : (A^t)_{ij} > 0 \},
\]
the set of walk lengths for which there is a path from $i$ to $j$.  
In particular, $\mathbb{N}_{ii}$ is the set of lengths of closed walks starting and ending at $i$.  
Define
\[
\gamma_i = \gcd(\mathbb{N}_{ii}), \qquad \gamma = \gcd\{ \gamma_i : i \in V \}.
\]
The integer $\gamma$ is called the \emph{cyclicity} (or \emph{period}) of $A$.\footnote{See E.~Seneta, \emph{Non-negative Matrices and Markov Chains}, Springer (1981).}  

Each $\mathbb{N}_{ii}$ is closed under addition and hence forms a numerical semigroup. A classical lemma from number theory states:  

\begin{lemma}
If a set $\mathbb{N}$ of positive integers is closed under addition, then it contains all but finitely many multiples of its $\gcd$. 
\end{lemma}

From this it follows that for all $i$, $\gamma_i=\gamma$. Thus the cyclicity is well-defined.

\paragraph{Primitivity.}  
An irreducible non-negative matrix $A$ is called \emph{primitive} if $\gamma=1$, i.e.~the gcd of all closed walk lengths is one. Equivalently, $A$ is primitive if some power $A^t$ is strictly positive.  

---

\paragraph{Characterizations.}  
The following theorem gives several equivalent conditions for primitivity.

\begin{theorem}
Let $A$ be an irreducible non-negative matrix. Then the following are equivalent:
\begin{enumerate}
\item $A$ is primitive.  
\item All eigenvalues $\lambda\neq \rho(A)$ satisfy $|\lambda|<\rho(A)$.  
\item $\left(\tfrac{1}{\rho(A)}A\right)^t \to xy^T$ as $t\to\infty$, where $x,y>0$ are the Perron right and left eigenvectors.  
\item There exists $i$ such that $\gamma_i=1$.  
\item The cyclicity $\gamma$ equals $1$.  
\end{enumerate}
\end{theorem}

Thus, primitivity is equivalent to aperiodicity in the graph: the system has no “resonant” periodicity.  

\paragraph{Periodicity.}  
The \emph{period} of an irreducible non-negative matrix $A$ is defined as
\[
\mathrm{per}(A) = \gcd\{ m \in \mathbb{N} : (A^m)_{ii}>0 \text{ for some } i \},
\]
i.e.~the gcd of the lengths of all closed directed paths in the associated digraph $G(A)$.  

---

\paragraph{Generalization of Perron–Frobenius.}  
All of the statements of the Perron–Frobenius theorem for positive matrices remain valid for irreducible aperiodic (primitive) matrices. For periodic matrices with period $h>1$, the situation changes: there are $h$ eigenvalues of modulus $\rho(A)$, evenly spaced on the circle of radius $\rho(A)$.  

\begin{example}
Let $A$ be an irreducible non-negative matrix with period $h$ and spectral radius $\rho(A)=r$. Then:
\begin{enumerate}
\item $r>0$ and $r$ is an eigenvalue of $A$.  
\item $r$ is simple, with one-dimensional left and right eigenspaces.  
\item There exist positive left and right eigenvectors $v,w>0$ with $Av = r v$, $w^TA = r w^T$.  
\item $A$ has exactly $h$ simple eigenvalues of modulus $r$, namely $r \zeta_h^k$ where $\zeta_h = e^{2\pi i/h}$ and $k=0,\dots,h-1$.  
\item There exists a permutation matrix $P$ such that
\begin{equation}
PAP^T =
\begin{pmatrix}
0 & A_{12} & 0 & \cdots & 0 \\
0 & 0 & A_{23} & \ddots & \vdots \\
\vdots & & \ddots & \ddots & 0 \\
0 & \cdots & & 0 & A_{h-1,h} \\
A_{h1} & 0 & \cdots & 0 & 0
\end{pmatrix},
\label{eqn:block-matrix-periodic}
\end{equation}
i.e.~$A$ can be permuted into Frobenius normal form, with cyclic block structure of length $h$.
\end{enumerate}
\end{example}

This block form illustrates the decomposition of a periodic irreducible digraph into $h$ classes, each mapping cyclically to the next. When $h=1$, the block structure collapses, and $A$ is primitive.  

\subsection{Nonnegative matrices and $M$-matrices}
\index{key}{M-matrices}

Beyond PF theory, nonnegative matrices play a role in defining \emph{$M$-matrices}, a cornerstone
in numerical linear algebra and dynamical systems\footnote{See R.~Plemmons,
``M-matrix characterizations. I—nonsingular M-matrices,'' \emph{Linear Algebra Appl.} \textbf{18}
(1977), 175–188.}.  

A (nonsingular) $M$-matrix $A$ can be characterized in several equivalent ways:
\begin{itemize}
\item $A=zI-B$ with $B\geq 0$ and $z>\rho(B)$;
\item $A_{ii}>0$, $A_{ij}\leq 0$ for $i\neq j$, and $A^{-1}\geq 0$;
\item All principal minors of $A$ are positive;
\item The real part of every eigenvalue is positive.
\end{itemize}
If $A$ is singular, only weaker conditions apply (nonnegative principal minors, spectral radius
criteria).

\paragraph{Properties.}
\begin{itemize}
\item If $A$ is an $M$-matrix and $A\leq B$ with $B$ having nonpositive off-diagonal entries,
then $B$ is also an $M$-matrix.
\item If $0\leq A\leq B$, then $\rho(A)\leq \rho(B)$ (monotonicity of spectral radius).
\item If $\rho(A)<1$, then $(I-A)^{-1}$ exists and is nonnegative.
\end{itemize}

\paragraph{Kantorovich inequality.}
For symmetric positive definite matrices $A$, one has the bound
\[
\frac{\langle A\vec x,\vec x\rangle\langle A^{-1}\vec x,\vec x\rangle}{\|\vec x\|^4}
\leq \frac{(\lambda_{\max}+\lambda_{\min})^2}{4\lambda_{\max}\lambda_{\min}},
\]
which plays an important role in analyzing convergence of iterative methods.

\paragraph{Ky Fan inequalities.}
Ky Fan proved several extensions relating eigenvalues of nonnegative matrices under partial ordering,
generalizing the inequalities above. In particular, if $0\leq A\leq B$, then interlacing and majorization
results hold for their spectra.\footnote{See K.~Fan, ``Inequalities for eigenvalues of Hermitian
matrices,'' \emph{J.~Math.~Anal.~Appl.} \textbf{2} (1961), 428–434.}

---

Nonnegative and $M$-matrices form a backbone of modern matrix analysis. They unify ideas from
spectral theory (PF root, projection operators), probability (stochastic matrices, Markov chains),
and numerical analysis (stability of iterative methods). Their special structure makes them
simultaneously restrictive and extremely powerful.

\subsubsection{Hawkins--Simon theorem}\label{sec:hstheor}
\index{key}{Hawkins-Simon theorem}

One of the most important applications of Perron--Frobenius theory is the characterization of when the
operator $(I-A)$ is \emph{positively invertible}, i.e.\ when $(I-A)^{-1}\geq 0$.  
This result, known as the \emph{Hawkins--Simon theorem}, originates in the analysis of economic
input--output systems\footnote{D.~Hawkins and H.~A.~Simon,
``Some conditions of macroeconomic stability,'' \emph{Econometrica} \textbf{17} (1949), 245–248.}.
It is also fundamental in the general theory of $M$-matrices and stability analysis.

---

\paragraph{Context.}
We will later discuss the economic interpretation of the model below, but  we 
consider the Leontief input--output model in economics:
\[
\vec x = A\vec x + \vec d,
\]
where $A$ is a nonnegative input matrix representing the economy, $\vec x$ is the vector of gross outputs, and $\vec d$ is
the final demand. Solving for $\vec x$ requires
\[
(I-A)\vec x = \vec d, \qquad \vec x = (I-A)^{-1}\vec d.
\]
For this to be meaningful in economics, $(I-A)^{-1}$ must exist and must be nonnegative, so that
a positive demand vector $\vec d$ produces a positive output vector $\vec x$.  
The Hawkins--Simon theorem provides necessary and sufficient conditions for this.

---

\paragraph{Theorem (Hawkins--Simon).}
Let $A\geq 0$ be an $n\times n$ matrix. Then the following are equivalent:
\begin{enumerate}
\item $(I-A)$ is invertible and $(I-A)^{-1}\geq 0$ (positively invertible);
\item All principal minors of $(I-A)$ are positive;
\item The Perron--Frobenius eigenvalue of $A$ satisfies $\rho(A)<1$.
\end{enumerate}

---

\paragraph{Proof sketch.}
\begin{itemize}
\item[$\bullet$] If $\rho(A)<1$, then the Neumann series
\[
(I-A)^{-1} = I + A + A^2 + \cdots
\]
converges, with each partial sum nonnegative. Hence $(I-A)^{-1}\geq 0$.  
\item[$\bullet$] If $(I-A)^{-1}\geq 0$, then by the Perron--Frobenius theorem,
$\rho(A)<1$; otherwise, the series diverges.  
\item[$\bullet$] The equivalence with the positivity of principal minors follows from the
general characterization of nonsingular $M$-matrices.
\end{itemize}

---

\paragraph{Spectral interpretation.}
The condition $\rho(A)<1$ is crucial: it guarantees that all eigenvalues of $A$ lie inside the unit
disk, so that $(I-A)$ has eigenvalues with positive real part. Indeed, if $\mu$ is an eigenvalue of
$A$, then $(I-A)$ has eigenvalue $1-\mu$, and positivity requires $\mathrm{Re}(1-\mu)>0$.  

---

\paragraph{Importance.}
The Hawkins--Simon theorem connects algebraic, spectral, and combinatorial criteria:
\begin{itemize}
\item In economics, it guarantees that an input--output system admits a unique positive equilibrium
output for any positive demand vector.
\item In matrix theory, it provides a bridge between Perron--Frobenius theory, positive minors,
and the structure of $M$-matrices.
\item In dynamical systems, it ensures stability of linear recursions $\vec x_{t+1}=A\vec x_t$,
since $\rho(A)<1$ implies asymptotic decay.
\end{itemize}

---

\paragraph{Corollary.}
A necessary and sufficient condition for $(I-A)$ to be positively invertible is that
all eigenvalues of $A$ satisfy $\mathrm{Re}(\lambda)<1$. Equivalently, $A$ must be a
nonnegative matrix with spectral radius strictly less than $1$.
\subsection{Symmetries, compressions, and quasi-symmetries: interlacing}\label{sec:matrixsym}
\index{key}{matrix symmetry}
\index{key}{commuting matrices}
\index{key}{interlacing}
\index{key}{matrix compression}

\paragraph{Matrix symmetries and commutators.}
A \emph{symmetry} of a matrix $A$ is another matrix $P$ of the same size such that
\[
[P,A]=PA-AP=0.
\]
The operator $[\cdot,\cdot]$ is called the \emph{commutator}. If $[P,A]=0$, then $A$ and $P$ commute,
which implies they can be simultaneously triangularized, and in the case of normal (in particular,
symmetric or Hermitian) matrices, they can be simultaneously diagonalized. In other words, they
share a common basis of eigenvectors.  

This principle is crucial in physics (e.g.\ quantum mechanics, where commuting observables can be
measured simultaneously), but it also plays an important role in spectral graph theory and matrix
analysis.

---

\paragraph{Interlacing of eigenvalues.}
Two sequences $\{\lambda_i\}$ and $\{\mu_j\}$ are said to \emph{interlace} if
\[
\mu_1 \leq \lambda_1 \leq \mu_2 \leq \lambda_2 \leq \cdots
\]
with alternating inequalities.  

This phenomenon occurs when one compares the eigenvalues of a matrix and those of its compression.

\begin{definition}[Compression]
Let $A$ be a symmetric $n\times n$ matrix. An $m\times m$ matrix $B$ with $m\leq n$ is a
\emph{compression} of $A$ if there exists an orthogonal projection $O$ onto an $m$-dimensional
subspace such that
\[
B=O^\top A O.
\]
\end{definition}

The central result is the Cauchy interlacing theorem (also called Poincaré separation theorem):

\begin{theorem}[Cauchy interlacing]
Let $A$ be a symmetric $n\times n$ matrix with ordered eigenvalues
\[
\lambda_1 \leq \lambda_2 \leq \cdots \leq \lambda_n.
\]
Let $B$ be an $m\times m$ compression of $A$ with ordered eigenvalues
\[
\mu_1 \leq \mu_2 \leq \cdots \leq \mu_m.
\]
Then the eigenvalues of $B$ interlace those of $A$:
\[
\lambda_i \leq \mu_i \leq \lambda_{i+n-m}, \qquad i=1,\ldots,m.
\]
\end{theorem}

This theorem is fundamental in spectral graph theory (e.g.\ eigenvalue bounds for induced subgraphs),
and in perturbation theory, where it controls the spectrum under restriction or projection.

---

\paragraph{Quasi-symmetries.}
The notion of symmetry can be relaxed. A \emph{quasi-symmetry} of $A$ is a matrix $P$ that does not
strictly commute with $A$, but preserves its eigenspaces in a weaker sense. More precisely, if
$\vec x$ is a right eigenvector of $A$ with eigenvalue $\lambda$, and $\vec y$ is a corresponding
left eigenvector, then the operator
\[
P=\vec x \otimes \vec y^\top
\]
satisfies
\[
AP=\lambda P=PA.
\]
Thus, $P$ commutes with $A$ \emph{on the eigenspace associated with $\lambda$}, even if not globally.
Such rank-one operators are sometimes called \emph{spectral projectors} onto the eigenspaces of $A$.

More generally, if $A$ admits a decomposition
\[
A=\sum_i \lambda_i \,\vec y_i \otimes \vec x_i^\top,
\]
then each spectral projector $P_i=\vec x_i \otimes \vec y_i^\top$ satisfies $AP_i=P_iA=\lambda_i P_i$.
These objects are not true symmetries (since they do not commute with $A$ on the whole space), but
they act as \emph{quasi-symmetries} on individual eigenspaces.  

Quasi-symmetries are useful in spectral perturbation theory and in the study of non-Hermitian
matrices, where left and right eigenvectors differ.

\subsection{Directness and periodicity}

So far, we have developed Perron–Frobenius theory algebraically, emphasizing non-negative matrices.  
Although many statements can be made without explicit reference to graphs, it is natural to ask: what do these results say about adjacency matrices of graphs, especially undirected ones?  

We now compare this to the typical block forms of adjacency matrices.

\paragraph{Block structures for undirected graphs.}
Since adjacency matrices of undirected graphs are symmetric, they can appear in several forms (up to permutation similarity):  

\begin{itemize}
\item \textbf{Connected graph:}
\begin{equation}
A =
\begin{pmatrix}
A_{11} & A_{12} \\
A_{12}^T & A_{22}
\end{pmatrix},
\label{eqn:block-matrix-vanilla}
\end{equation}
with each block $A_{ij}$ containing at least one nonzero entry. This corresponds to a typical connected graph.

\item \textbf{Disconnected graph:}
\begin{equation}
A =
\begin{pmatrix}
A_{11} & 0 \\
0 & A_{22}
\end{pmatrix},
\label{eqn:block-matrix-disconnected}
\end{equation}
where the graph splits into two disjoint components.  

\item \textbf{Bipartite graph:}
\begin{equation}
A =
\begin{pmatrix}
0 & A_{12} \\
A_{12}^T & 0
\end{pmatrix}.
\label{eqn:block-matrix-bipartite}
\end{equation}
Here the vertex set is partitioned into two disjoint subsets, with edges only between (never within) the subsets.  

\item \textbf{Line or path-like graph:}  
These lead to tridiagonal banded adjacency matrices (not written explicitly here), reflecting nearest-neighbor connectivity.

\item \textbf{Forbidden symmetric forms:}  
The following matrices are \emph{not} symmetric:
\begin{equation}
A =
\begin{pmatrix}
A_{11} & A_{12} \\
0 & A_{22}
\end{pmatrix},
\label{eqn:block-matrix-reducible1}
\end{equation}
and
\begin{equation}
A =
\begin{pmatrix}
0 & A_{12} \\
0 & 0
\end{pmatrix}.
\label{eqn:block-matrix-reducible2}
\end{equation}
Such forms arise only for directed graphs.
\end{itemize}

\paragraph{Connections to periodicity.}
Comparing Eqn.~(\ref{eqn:block-matrix-periodic}) with these forms yields several insights:
\begin{itemize}
\item One might think that periodicity is only relevant for directed graphs.  
Indeed, if the period $h>2$, the associated adjacency matrix must be nonsymmetric.  
However, the case $h=2$ includes undirected bipartite graphs, whose adjacency matrices have the form \eqref{eqn:block-matrix-bipartite}.  
In this case, the eigenvalues are symmetric about zero, corresponding to the fact that $\pm\rho(A)$ appear as eigenvalues. This matches the second roots of unity.

\item The block form \eqref{eqn:block-matrix-disconnected} can be seen as a degenerate case of \eqref{eqn:block-matrix-reducible1}, with $A_{12}=0$.  
Yet the spectral behavior is quite different: a disconnected graph is reducible, while \eqref{eqn:block-matrix-reducible1} with $A_{12}\neq 0$ represents an irreducible directed structure.  

\item For directed graphs, block upper-triangular forms such as \eqref{eqn:block-matrix-reducible1} and \eqref{eqn:block-matrix-reducible2} occur naturally.  
Unfortunately, spectral theory is much less developed here.\footnote{See P.~Van Mieghem, \emph{Graph Spectra for Complex Networks}, Cambridge University Press (2011).}  
Most practical methods deal with this by symmetrizing the adjacency matrix (e.g.~$A+A^T$ or $(A+A^T)/2$) and then applying undirected spectral techniques.

\item Form \eqref{eqn:block-matrix-reducible1} corresponds to the so-called \emph{bow-tie} structure, famously associated with models of the World Wide Web.  
Although reducible, it can be made irreducible by adding a rank-one perturbation, e.g.
\[
A \longrightarrow A + \varepsilon \vec 1 \vec 1^T.
\]
This has a natural probabilistic interpretation: it corresponds to allowing a random walker to “teleport” uniformly at random.  
This is the key idea behind spectral ranking methods such as Google’s PageRank.\footnote{See S.~Brin and L.~Page, \emph{The anatomy of a large-scale hypertextual Web search engine}, Computer Networks and ISDN Systems \textbf{30}, 107–117 (1998).}  
Such rank-one updates restore irreducibility and primitivity, ensuring convergence to a unique positive stationary distribution.
\end{itemize}

\section{Electrical network approach to graphs}

So far, we have adopted the standard approach to spectral graph theory: 
analyzing graphs through eigenvalues and eigenvectors of associated matrices.  Let us consider a particular way of modelling positive matrices representing graphs (with non-negative entries).
For example, given a graph $G=(V,E)$, we introduced the adjacency matrix $A$ and considered eigenpairs $Av=\lambda v$, as well as the Laplacian $L=D-A$ and the quadratic form
\[
x^T L x = \sum_{(i,j)\in E} (x_i - x_j)^2 .
\]

There are, however, complementary perspectives.  
Instead of starting with eigenvalues, one can bring in \emph{physical intuition} and model a graph as a physical system, where edges behave like springs or resistors.  
Alternatively, one can think directly in terms of vectors obtained from diffusions or random walks on $G$.  
Both perspectives lead to robust and interpretable tools.  
We start here with the physical viewpoint.

---

\subsection{A physical model for a graph}

Many physical systems relax to an equilibrium after perturbation, with forces proportional to displacement.  
This motivates thinking of edges as springs, each with spring constant $k$.  
A spring stretched by displacement $x$ exerts force $F(x)=kx$ and stores potential energy $U(x)=\tfrac{1}{2}kx^2$.  

If each edge $(i,j)$ of $G$ is a spring, the total potential energy is
\[
\sum_{(i,j)\in E} (x_i - x_j)^2 = x^T L x ,
\]
subject to boundary conditions on “nailed-down” vertices.  
The minimum-energy configuration satisfies
\[
x_i = \frac{1}{d_i} \sum_{j:(i,j)\in E} x_j ,
\]
i.e., each vertex takes the average of its neighbors’ values.  
This is the \emph{harmonic property}, central to harmonic analysis on graphs.\footnote{See F.~Chung, \emph{Spectral Graph Theory}, AMS (1997).}

This “spring model” provides intuition, but a more powerful analogy comes from electrical circuits.

---

\subsection{Graphs as electrical networks}

We now model $G=(V,E)$ as an electrical circuit with resistors.  
This analogy allows us to define the notion of \emph{effective resistance} between vertices, which leads to a new metric on graphs closely tied to random walks and spectral properties.\footnote{See P.~Doyle and J.~Snell, \emph{Random Walks and Electric Networks}, Carus Math. Monographs, MAA (1984).}

\paragraph{Resistors and basic laws.}
\begin{itemize}
\item Each edge $(i,j)\in E$ is assigned a resistance $R_{ij}>0$ (or equivalently a conductance $c_{ij}=1/R_{ij}$).  
\item A battery is represented by a source–sink pair $\{a,b\}$, with net current $Y$ injected at $a$ and extracted at $b$.  
\item Ohm’s law states that for potentials $V_i$ on vertices,
\[
Y_{ij} R_{ij} = V_i - V_j ,
\]
where $Y_{ij}$ is the current on edge $(i,j)$ (with $Y_{ij}=-Y_{ji}$).
\item Kirchhoff’s current law: for each node $i$,
\[
\sum_{j \in N(i)} Y_{ij} =
\begin{cases}
 Y, & i=a, \\
 -Y, & i=b, \\
 0, & \text{otherwise}.
\end{cases}
\]
\item Kirchhoff’s potential law: for every cycle $C$, 
\[
\sum_{(i,j)\in C} Y_{ij} R_{ij} = 0.
\]
\end{itemize}

These conditions imply that a well-defined potential function exists on $V$, unique up to an additive constant.  

\paragraph{Effective resistance.}
The \emph{effective resistance} between two vertices $a,b$ is
\[
R_{ab} = \frac{V_a - V_b}{Y},
\]
independent of the chosen net current $Y$.  
Two simple reduction rules hold:
\begin{itemize}
\item \textbf{Series:} $R_{ab}=R_{ac}+R_{cb}$.  
\item \textbf{Parallel:} $R_{ab}=(R_1^{-1}+R_2^{-1})^{-1}$.  
\end{itemize}
Thus effective resistances depend not only on shortest paths but also on the multiplicity of paths.  
In particular, the first additional edge between two nodes greatly reduces $R_{ab}$, while further edges decrease it progressively less — a behavior reminiscent of diffusion and random walks.

---

\subsection{Properties of resistor networks}

The electrical viewpoint connects directly to spectral graph theory.  
Let $L$ be the weighted Laplacian of $G$, with edge weights $w_{ij}=1/R_{ij}$.  
Define the \emph{Laplacian pseudoinverse} $L^+$ as the unique matrix satisfying:
\begin{enumerate}
\item $L^+\vec 1=0$;  
\item for all $w \perp \vec 1$, $L^+w=v$ such that $Lv=w$ and $v\perp \vec 1$.  
\end{enumerate}

Then the effective resistance between vertices $a$ and $b$ is
\[
R_{ab} = (e_a-e_b)^T L^+ (e_a-e_b) = L^+_{aa} - 2L^+_{ab} + L^+_{bb}.
\]
This follows from the existence and uniqueness of solutions to Kirchhoff’s laws.\footnote{See C.~Meyer, \emph{Matrix Analysis and Applied Linear Algebra}, SIAM (2000); S.~Redner, \emph{A Guide to First-Passage Processes}, Cambridge University Press (2001).}

\paragraph{Total effective resistance.}
Summing effective resistances over all pairs gives the \emph{Kirchhoff index}:
\[
R^{\mathrm{tot}} = \sum_{i<j} R_{ij}.
\]
Since $R_{ij}$ defines a metric on $V$, this is the total “size” of the graph in the effective resistance geometry.  
It is closely related to average commute times of random walks and is used to quantify the capacity or robustness of a network.\footnote{See D.~Klein and M.~Randic, \emph{Resistance distance}, J. Math. Chem. \textbf{12}, 81–95 (1993).}

\subsection{Resistance distance and total effective resistance}

The total effective resistance $R^{tot}$ of a graph can be expressed spectrally.  
Let $\lambda_1=0 < \lambda_2 \le \cdots \le \lambda_n$ be the Laplacian eigenvalues. Then:
\[
R^{tot} = n \sum_{i=2}^n \frac{1}{\lambda_i}.
\]
This follows from the spectral representation of $L^+$ and the formula $R_{ab} = (e_a-e_b)^T L^+ (e_a-e_b)$.\footnote{See D.~Klein and M.~Randic, \emph{Resistance distance}, J. Math. Chem. \textbf{12}, 81–95 (1993).}  

As a consequence, $R^{tot}$ can be bounded in terms of $\lambda_2$:
\[
\frac{n}{\lambda_2} \;\le\; R^{tot} \;\le\; \frac{n(n-1)}{\lambda_2}.
\]

---

The effective resistance $R_{ab}$ defines a genuine metric on the vertex set $V$:

\begin{enumerate}
\item \emph{Non-negativity and identity:} $R_{ab}\ge 0$, with $R_{ab}=0 \iff a=b$.  
\item \emph{Symmetry:} $R_{ab}=R_{ba}$.  
\item \emph{Triangle inequality:} $R_{ac} \le R_{ab} + R_{bc}$.
\end{enumerate}

The first two properties are immediate from the properties of $L^+$, which is symmetric positive semidefinite. The triangle inequality requires more work.

\paragraph{Claim 1.}  
Let $Y_{ab}=e_a-e_b$ and $V_{ab}=L^+ Y_{ab}$. Then
\[
V_{ab}(a) \ge V_{ab}(c) \ge V_{ab}(b), \quad \forall c\in V.
\]
That is, the induced potential when $1$ Amp flows from $a$ to $b$ decreases monotonically from $a$ to $b$.  

\emph{Proof (sketch).} For each $c\notin\{a,b\}$, Kirchhoff’s current law implies
\[
V_{ab}(c)=\frac{\sum_{x\sim c} C_{xc}V_{ab}(x)}{\sum_{x\sim c} C_{xc}},
\]
a weighted average of neighbors. Thus $V_{ab}(c)$ cannot exceed the maximum (nor fall below the minimum) of its neighbors. This rules out $V_{ab}(c) > V_{ab}(a)$ or $V_{ab}(c)<V_{ab}(b)$.  

\paragraph{Triangle inequality.}  
Now let $Y_{ab}$, $Y_{bc}$ be unit currents from $a\to b$ and $b\to c$, and $Y_{ac}=Y_{ab}+Y_{bc}$. Corresponding voltages satisfy $V_{ac}=V_{ab}+V_{bc}$. Then
\[
R_{ac} = Y_{ac}^T V_{ac} = Y_{ac}^T V_{ab} + Y_{ac}^T V_{bc}.
\]
By Claim 1, $Y_{ac}^T V_{ab}\le R_{ab}$ and $Y_{ac}^T V_{bc}\le R_{bc}$. Hence $R_{ac}\le R_{ab}+R_{bc}$.  

Thus $R$ is a metric, sometimes called the \emph{resistance distance}.\footnote{See P.~Doyle and J.~Snell, \emph{Random Walks and Electric Networks}, MAA (1984).}  

---

\subsection{Relation to geodesic distance}

Resistance distance and geodesic distance are related but distinct:  
\begin{enumerate}
\item $R_{ab} = d(a,b)$ if and only if there is a unique path between $a$ and $b$.  
\item $R_{ab} < d(a,b)$ otherwise.
\end{enumerate}

\emph{Proof idea.} If there is only one path $P$ from $a$ to $b$, then by Kirchhoff’s laws all current flows along $P$, and $R_{ab}=\sum_{(i,j)\in P}R_{ij}=d(a,b)$. If additional paths exist, currents split, lowering the effective resistance strictly below the geodesic distance.  

Hence on trees (which have unique paths), $R_{ab}=d(a,b)$ for all pairs.

---

\subsection{Examples}

\begin{itemize}
\item \textbf{Complete graph $K_n$:} Minimizes total resistance. One finds
\[
R^{tot}(K_n)=n-1.
\]
\item \textbf{Path graph $P_n$:} Maximizes total resistance among connected graphs:
\[
R^{tot}(P_n) = \frac{1}{6}(n-1)n(n+1).
\]
\item \textbf{Star graph $S_n$:} Minimizes total resistance among trees:
\[
R^{tot}(S_n) = (n-1)^2.
\]
\end{itemize}

These examples show how $R^{tot}$ captures graph “spread” or compactness.  

---
\subsection{Projection operators and Kirchhoff's laws}
\label{sec:proj_kirchhoff}
\index{key}{projection operator}
\index{key}{Kirchhoff's laws}

A natural way to formalize Kirchhoff’s laws in network theory is through the language of
projection operators acting on the space of edge variables.  
Let $G=(V,E)$ be a connected graph with $|V|=n$ nodes and $|E|=m$ edges, and denote by 
$B \in \mathbb{R}^{n \times m}$ its incidence matrix. Each edge $e \in E$ has an arbitrary orientation, and 
$B_{\alpha e} = \pm 1$ if edge $e$ is incident on node $\alpha$ with incoming/outgoing orientation, and $0$ otherwise.\footnote{For a modern overview of the incidence matrix, cut and cycle spaces, see 
Godsil and Royle, \emph{Algebraic Graph Theory}, Springer (2001).}

\paragraph{Kirchhoff’s laws.}
Currents $i \in \mathbb{R}^m$ and node potentials $p \in \mathbb{R}^n$ satisfy the two Kirchhoff laws:
\begin{align}
    B i &= 0 \qquad \text{(KCL)}, \\
    B^{\top} p &= v \qquad \text{(KVL)},
\end{align}
where $v \in \mathbb{R}^m$ denotes edge voltages.  
Thus, the space of admissible currents is $\ker(B)$, also known as the \emph{cycle space}, while the space of admissible voltages is $\operatorname{Im}(B^{\top})$, the \emph{cut space}.  
These two are complementary subspaces of $\mathbb{R}^m$.

\paragraph{Projectors.}\label{sec:projectors}
We introduce two complementary projection operators acting on $\mathbb{R}^m$:\footnote{See F. Caravelli, F. Traversa, M. Di Ventra, ``The complex dynamics of memristive circuits: analytical results and universal slow relaxation
", Phys. Rev. E 95, 022140 (2017);Backhaus and Touchette, 
``Projection operators in network theory,'' \emph{J. Phys. A: Math. Theor.} \textbf{52} (2019), and 
Devriendt, Lambiotte and Touchette, 
``Splitting the Kirchhoff subspaces of a graph,'' \emph{Linear Algebra Appl.} \textbf{675} (2024).}
\begin{align}
    \Omega_A &= A(A^{\top} A)^{-1} A^{\top}, \\
    \Omega_B &= B(B^{\top} B)^{-1} B^{\top},
\end{align}
where $A \in \mathbb{R}^{c \times m}$ is the cycle matrix (rows span the cycle space) and 
$B \in \mathbb{R}^{(n-1)\times m}$ is the reduced incidence matrix.  
It follows that
\[
    \Omega_A + \Omega_B = I_m,
\]
so that $\Omega_A$ projects onto the cycle space, and $\Omega_B$ projects onto the cut space.\footnote{ A. Zegarac, F. Caravelli, ``Memristive Networks: from Graph Theory to Statistical Physics", EPL Perspective 125, 10001 (2019); F. Barrows, F. C. Sheldon, F. Caravelli, ``Network analysis of memristive device circuits: dynamics, stability and correlations", J. Phys. Complex. 6 025003 (2025)}

\paragraph{Spectra of the projectors.}
Since $\Omega_A$ and $\Omega_B$ are orthogonal projectors, their spectra are trivial: eigenvalues belong to $\{0,1\}$.  
The multiplicities are determined by the dimensions of the corresponding spaces:
\begin{align}
    \operatorname{rank}(\Omega_A) &= m-n+1, \\
    \operatorname{rank}(\Omega_B) &= n-1.
\end{align}
Thus, the spectrum of $\Omega_A$ consists of $(m-n+1)$ ones and $(n-1)$ zeros, while the spectrum of $\Omega_B$ consists of $(n-1)$ ones and $(m-n+1)$ zeros.
This directly encodes the dimensions of the cycle and cut spaces.

\paragraph{Euler’s relation.}
A simple yet powerful consequence is obtained by taking the trace of $\Omega_A+\Omega_B=I_m$:
\[
    \operatorname{Tr}(\Omega_A) + \operatorname{Tr}(\Omega_B) = m.
\]
But $\operatorname{Tr}(\Omega_A)=m-n+1$ and $\operatorname{Tr}(\Omega_B)=n-1$, which reproduces Euler’s relation for connected graphs:
\[
    (m-n+1) + (n-1) = m.
\]
This provides a purely algebraic proof of Euler’s formula, highlighting the convenience of the projector framework.

\paragraph{Remarks.}
The decomposition $\mathbb{R}^m = \operatorname{Im}(\Omega_A) \oplus \operatorname{Im}(\Omega_B)$ illustrates the orthogonality of cycle and cut spaces.  
Moreover, the algebraic structure $\Omega_A^2=\Omega_A$, $\Omega_B^2=\Omega_B$, and $\Omega_A\Omega_B=0$ encodes Kirchhoff’s current and voltage laws in a compact operator form.  
This operator-theoretic view shows that spectral analysis of $\Omega_A$ and $\Omega_B$ yields structural invariants of the network, and connects to fundamental physical principles like conservation of flow and potential differences.

\subsection{Extensions to infinite graphs}

Thus far we considered only finite graphs. For infinite graphs, resistance concepts connect to recurrence of random walks.\footnote{See B.~Mohar and W.~Woess, \emph{A survey on spectra of infinite graphs}, Bull. London Math. Soc. \textbf{21}, 209–234 (1989).}  

A random walk is \emph{recurrent} if it returns to the starting point with probability $1$ (equivalently, visits every node infinitely often with probability $1$). Otherwise it is \emph{transient}. For irreducible, aperiodic walks on finite graphs, recurrence is trivial (the walk eventually visits every vertex), but on infinite graphs it is subtle.  

Classical results:  
\begin{itemize}
\item On $\mathbb{Z}$ and $\mathbb{Z}^2$, random walks are recurrent.  
\item On $\mathbb{Z}^d$ for $d\ge 3$, random walks are transient.  
\end{itemize}

This dichotomy illustrates how dimension (i.e.~the degree of branching) influences return probability: higher-dimensional spaces provide “more room to get lost.”\footnote{See L.~Lovász, \emph{Random walks on graphs: A survey}, Combinatorics, Paul Erdős is Eighty, Vol. 2 (1996).}  

These ideas also provide intuition for random-walk based spectral methods on large finite graphs.

The key tool underlying recurrence/transience analysis is \emph{Rayleigh's Monotonicity Law}.  
This states that the effective resistance $R_{ab}$ between two nodes $a,b$ varies monotonically with individual edge resistances: increasing any edge resistance increases $R_{ab}$, while decreasing it lowers $R_{ab}$.\footnote{See P.~Doyle and J.~Snell, \emph{Random Walks and Electric Networks}, Carus Mathematical Monographs, MAA (1984).}  

Two basic graph operations follow from this principle:
\begin{itemize}
\item \emph{Shorting} vertices $u$ and $v$ (identifying them electrically).  
\item \emph{Cutting} edges between $u$ and $v$ (deleting them electrically).
\end{itemize}
Both produce a new graph $G'$ from $G$ and obey:
\begin{itemize}
\item Shorting can only \emph{decrease} $R_{ab}$.  
\item Cutting can only \emph{increase} $R_{ab}$.  
\end{itemize}

These operations allow one to compare effective resistances by bounding them between simpler networks. For example:
- On $\mathbb{Z}^2$, if one “shorts” vertices in Manhattan circles around the origin, then $R_{eff}=\infty$ on the simplified network. Hence $R_{eff}=\infty$ on $\mathbb{Z}^2$, implying recurrence.  
- On $\mathbb{Z}^3$, by carefully “cutting” edges one shows $R_{eff}<\infty$ on the simplified network, so $R_{eff}<\infty$ on $\mathbb{Z}^3$, implying transience.  

A fundamental theorem connects these ideas:  
\begin{quote}
A network is recurrent if and only if $R_{eff}=\infty$.  
\end{quote}
Thus random walks on $\mathbb{Z}$ and $\mathbb{Z}^2$ are recurrent, while those on $\mathbb{Z}^3$ are transient.  
A full proof is left as an exercise: the $\mathbb{Z}$ case is straightforward, $\mathbb{Z}^2$ requires more work, and $\mathbb{Z}^3$ is tricky but classical.\footnote{See B.~Mohar and W.~Woess, \emph{A survey on spectra of infinite graphs}, Bull. London Math. Soc. \textbf{21}, 209–234 (1989); L.~Lovász, \emph{Random walks on graphs: A survey}, Combinatorics, Paul Erdős is Eighty, Vol.~2 (1996).}

---

\section{Types of random walks on networks}

\subsection{Continuous-time random walks (CTRWs)}
\index{key}{continuous time random walks}
\index{key}{CTRW}

So far we mainly considered random walks in discrete time. Interestingly, the mathematical theory of discrete- vs continuous-time random walks differs substantially.\footnote{See I.~Sokolov and J.~Klafter, \emph{First Steps in Random Walks}, Oxford University Press (2011).}  

\paragraph{Settings.}
\begin{itemize}
\item \textbf{Discrete space, discrete time (DTRW):} The usual Markov chain model on graphs.  
\item \textbf{Discrete space, continuous time:} Continuous-time Markov chains on graphs.  
\item \textbf{Continuous space:} Diffusion processes, widely used in physics, biology, and transport science.\footnote{See D.~Grebenkov, R.~Metzler, and N.~Masuda, \emph{Continuous-time random walks, diffusion, and anomalous transport}, Rev. Mod. Phys. (2023).}
\end{itemize}

\paragraph{Discrete-time case.}  
On $\mathbb{Z}$, a walker moves one step left or right with given probabilities at each discrete time step. The length and direction are random variables.

\paragraph{Continuous-time case.}  
In a continuous-time random walk, waiting times between jumps are random, drawn from a distribution $\psi(t)$. The walker remains on discrete vertices, with transitions governed by a stochastic matrix $Q$. This model was introduced by Montroll and Weiss.\footnote{E.~W.~Montroll and G.~H.~Weiss, \emph{Random walks on lattices. II}, J. Math. Phys. \textbf{6}, 167 (1965).}

Formally, the time-dependent transition probabilities are
\[
Q_{x'x}(t) = Q_{x'x}\,\psi(t),
\]
where $\psi(t)$ is the waiting-time density. The propagator (probability of being at $x$ at time $t$ given initial node $x_0$) has Laplace transform
\begin{align}
\tilde P_{x_0x}(s) = \frac{1-\tilde\psi(s)}{s}\,\left[(I-Q\tilde\psi(s))^{-1}\right]_{x_0x}.
\label{eq:ctrw-propagator}
\end{align}
This is the celebrated Montroll–Weiss formula.

\paragraph{Fourier–Laplace domain.}  
Taking a discrete Fourier transform yields
\begin{align}
\tilde P_{x_0x}(s) &= e^{ikx_0}\,\frac{1-\tilde\psi(s)}{s}\,
\frac{1}{1-\tilde\psi(s)(1-\lambda_{0k})},
\end{align}
where $1-\lambda_{0k}$ is the characteristic function of the jump distribution.  

\paragraph{Generalizations.}
\begin{itemize}
\item On arbitrary graphs, $Q$ is any stochastic matrix.  
\item \emph{Heterogeneous CTRW:} If waiting times depend on both nodes and edges (temporal heterogeneity), spatial and temporal parts couple. General expressions are given in Grebenkov (2017).\footnote{See D.~Grebenkov, \emph{Continuous-time random walks on networks with heterogeneous dynamics}, Phys. Rev. E \textbf{95}, 012146 (2017).}  
\item CTRWs can also be seen as random walks on \emph{temporal graphs}, where edges appear intermittently in time.\footnote{See P.~Holme and J.~Saramäki, \emph{Temporal networks}, Phys. Rep. \textbf{519}, 97–125 (2012).}
\end{itemize}

---

\subsection{First- and second-order Markov dynamics}

So far we have described random walks as \emph{first-order Markov processes}, where the next step depends only on the current node.  
In many real-world systems, however, the walk exhibits \emph{memory}: the next step may depend also on the previous node.  
This leads to \emph{second-order Markov dynamics}, which have been studied in empirical networks.\footnote{See V.~Salnikov and R.~Lambiotte, \emph{Effect of memory on the dynamics of random walks on networks}, J. Stat. Mech. (2016).}

\paragraph{Memory networks.}  
To capture such dynamics, one constructs a \emph{memory network}.  
The idea is to replace the original node set $\{1,\dots,N\}$ by an expanded set of \emph{memory nodes}. In the second-order case, memory nodes are pairs $(ij)$ representing a walker currently at $j$ having come from $i$.  

Transitions are then defined as
\[
P\big((ij)\to(jk)\big) = \frac{W((ij)\to(jk))}{\sum_\ell W((ij)\to(j\ell))},
\]
where $W$ are observed transition counts. This normalisation ensures that $\sum_\ell P((ij)\to(j\ell))=1$.  

Thus, a second-order Markov walk can be represented as a \emph{first-order} Markov chain on the memory network, with usual transition matrix.

\paragraph{Comparison with first-order case.}  
For an ordinary walk with transition matrix $T_M$, the probability distribution evolves as
\[
P(i,t+1) = \sum_j P(j,t)\,p(j\to i).
\]

For a memory network, one instead tracks probabilities of memory nodes:
\[
P(i,t+1) = \sum_k P((k,i),t+1) = \sum_{j,k} P((j,k),t)\,p((j,k)\to(k,i)).
\]
Thus memory enlarges the state space but preserves the Markov property once lifted.  

A small perturbation framework can also be defined: let
\[
T = T_M + \Delta T,
\]
where $T_M$ is the transition matrix of the first-order walk and $\Delta T$ encodes memory effects. Deviations in dynamics can then be quantified via $\Delta T$.

---

\section{Spanning trees and the Laplacian resolvent}

A central theorem of algebraic graph theory relates spanning trees to determinants of Laplacian minors — the \emph{matrix–tree theorem}. We now connect this to the Laplacian resolvent.  

Let $L$ be the Laplacian of $G$. Its characteristic polynomial is
\[
C_L(x) = \det(L - xI) = \sum_{k=0}^N C_k(L) x^k.
\]
The coefficients $C_m(L)$ admit a combinatorial interpretation:
\[
C_m(L) = (-1)^m \sum_{F_m} \gamma(F_m),
\]
where the sum runs over all ways of identifying $m$ pairs of nodes, and $\gamma(G)$ denotes the number of spanning trees of the resulting graph $G$.  

In particular, for $m=N-1$ this reduces to the classical matrix–tree theorem:
\[
\tau(G) = \frac{1}{N}\prod_{i=2}^N \lambda_i,
\]
where $\lambda_i$ are the nonzero Laplacian eigenvalues and $\tau(G)$ is the number of spanning trees.\footnote{See N.~Biggs, \emph{Algebraic Graph Theory}, Cambridge University Press (1993).}  
\label{sec:laplacian_spectrum}
Thus the Laplacian resolvent and characteristic polynomial encode spanning-tree counts and their generalisations to identified-node graphs.

\section{Spanning trees, Laplacian resolvent, and Cauchy--Binet}
\index{key}{matrix--tree theorem}\index{key}{spanning trees}\index{key}{Laplacian resolvent}

A central result of algebraic graph theory relates spanning trees to determinants
of Laplacian minors: the celebrated \emph{matrix–tree theorem}. This result can
be seen through the lens of the Laplacian resolvent and the expansion of the
characteristic polynomial.

\subsection{Characteristic polynomial of the Laplacian}
Let $L$ be the Laplacian of a connected graph $G$ on $N$ vertices. Its
characteristic polynomial is
\begin{equation}
C_L(x) = \det(L - xI) = \sum_{m=0}^N C_m(L)\, x^m.
\end{equation}
The coefficients $C_m(L)$ admit a combinatorial interpretation:
\begin{equation}
C_m(L) = (-1)^m \sum_{F_m} \gamma(F_m),
\end{equation}
where the sum runs over all ways of identifying $m$ pairs of nodes, and
$\gamma(G)$ denotes the number of spanning trees of the resulting graph $G$.

\subsection{The matrix--tree theorem}
In particular, for $m=N-1$ one recovers the classical Kirchhoff--Cayley
matrix–tree theorem:
\begin{equation}
\tau(G) = \frac{1}{N}\prod_{i=2}^N \lambda_i,
\end{equation}
where $\lambda_2,\dots,\lambda_N$ are the nonzero Laplacian eigenvalues and
$\tau(G)$ is the number of spanning trees of $G$.\footnote{N.~Biggs,
\emph{Algebraic Graph Theory}, Cambridge University Press (1993).}

Equivalently, $\tau(G)$ is given by the determinant of any $(N-1)\times(N-1)$
principal minor of $L$, i.e.
\begin{equation}
\tau(G) = \det L_{[i]},
\end{equation}
where $L_{[i]}$ is the matrix obtained by deleting the $i$th row and column.

\subsection{Connection to Cauchy--Binet}
The appearance of minors in spanning-tree counts can be understood through
the \emph{Cauchy--Binet formula}, which expresses the determinant of a product
of rectangular matrices. For $A\in\mathbb{R}^{n\times m}$ and
$B\in\mathbb{R}^{m\times n}$ with $m\geq n$,
\begin{equation}
\det(AB) = \sum_{S\subseteq[m],\,|S|=n} \det(A_{[:,S]})\det(B_{[S,:]}).
\end{equation}
Applying this to the Laplacian factorization $L = B B^\top$, where
$B$ is the oriented incidence matrix of $G$, gives
\begin{equation}
\det(L_{[i]}) = \sum_{T\in\mathcal{T}} \big(\det(B_T)\big)^2,
\end{equation}
where the sum runs over all spanning trees $T$, and $B_T$ is the submatrix
of $B$ formed by the edges of $T$. Since $\det(B_T)=\pm 1$, each term
contributes $1$, and the sum counts exactly the number of spanning trees.

\subsection{Resolvent interpretation}\index{key}{resolvent, Laplacian}
Because the Laplacian eigenvalues $\{\lambda_i\}$ enter both the resolvent
\[
R(z) = (L - zI)^{-1} = \sum_{i=2}^N \frac{1}{\lambda_i-z} u_i u_i^\top,
\]
and the characteristic polynomial
\[
C_L(x)=\prod_{i=1}^N (\lambda_i-x),
\]
one can interpret spanning-tree counts and their generalizations as encoded
in the analytic structure of the resolvent. In this sense, the Cauchy–Binet
expansion of determinants provides the combinatorial bridge between spectral
quantities (eigenvalues of $L$) and combinatorial invariants (spanning trees).

\subsection{Resolvent properties}
\index{key}{resolvent}

The \emph{resolvent} of a square matrix $A$ is defined as
\[
R(A,x) = (xI - A)^{-1},
\]
for complex $x$ not in the spectrum of $A$.  
It satisfies
\[
\det R(A,x) = \frac{1}{\det(xI - A)}.
\]

This object is fundamental in spectral analysis because derivatives of $\det(xI-A)$ and $\log \det(xI-A)$ yield information about eigenvalues and eigenvectors.  

\paragraph{Basic identities.}
\begin{itemize}
\item Derivative of the determinant:
\[
\frac{d}{dx}\det(xI - A) = \sum_j \det(xI - A_{\{j\}}),
\]
where $A_{\{j\}}$ denotes the matrix with the $j$th row/column (or link) removed.  

\item Trace identity:
\[
\sum_j \big( (xI - A)^{-1} \big)_{jj} = \sum_k \frac{1}{x - \lambda_k}
= \frac{d}{dx}\log \det(xI - A),
\]
where $\{\lambda_k\}$ are eigenvalues of $A$.
\end{itemize}

Thus,
\[
\partial_\lambda \log p(\lambda) = \mathrm{Tr}\,R(A,\lambda),
\]
where $p(\lambda)=\det(\lambda I-A)$ is the characteristic polynomial.  

---

\paragraph{Perturbations of the Perron–Frobenius eigenvalue.}

Let $A$ be a nonnegative irreducible matrix with Perron root $\lambda_1$ and Perron vector $v_1$.  
One can study sensitivity of $\lambda_1$ and $v_1$ under perturbations using the resolvent and pseudoinverse.\footnote{See Horn and Johnson, \emph{Matrix Analysis}, Cambridge University Press (1985).}  

For $\rho>\lambda_1$, $v_1$ is also the Perron vector of the resolvent $R(A,\rho)=(\rho I-A)^{-1}$.  

Deutch and Neumann proved that
\[
\frac{\partial \lambda_1}{\partial a_{ij}} = (I - R R^{-\sim 1})_{ji},
\]
where $R^{-\sim 1}$ denotes the Moore–Penrose pseudoinverse of $R$.  

Similarly, the second derivative is
\[
\frac{\partial^2 \lambda_1}{\partial a_{ij}\partial a_{kl}}
= (I - RR^{-\sim 1})_{li}\,R^{-\sim 1}_{jk} 
+ (I - RR^{-\sim 1})_{jk}\,R^{-\sim 1}_{li}.
\]

For the Perron vector, suppose $A=A(t)$ with derivative $A'=\partial_t A(t)$.  
If $v_1$ is normalized so that $z^T v_1=1$ for a dual vector $z$, then
\[
v_1' = R^{-\sim 1}A' v_1 
- (z'^T v_1) v_1 
- (z^T R^{-\sim 1}A' v_1) v_1.
\]

These formulas describe the sensitivity of dominant eigenpairs, useful in applications such as ranking and input–output economics.

---

\subsection{Adjacency resolvent and walks}

Consider the matrix inequality
\begin{equation}
[(I-W)^{-1}]_{ij} < 0 \quad \forall i,j,
\label{eq:leontief}
\end{equation}
which arises in input–output economics (the \emph{Leontief inverse}).  
The Hawkins–Simon condition requires nonnegativity of the inverse.\footnote{See Hawkins and Simon, \emph{Note on the Leontief system of equations}, Rev. Econ. Stud. \textbf{15} (1947).}  

Why is the Perron–Frobenius eigenvalue important here? Because it governs the convergence of the Neumann (von Neumann) series expansion:
\[
(I-W)^{-1} = I + W + W^2 + W^3 + \cdots .
\]

\paragraph{Graph interpretation.}  
The $(i,j)$ entry of $W^k$ counts weighted walks of length $k$ from $i$ to $j$.  
If $A$ is the adjacency matrix and $W_{ij}=w_{ij} A_{ij}$, then
\[
W^k_{ij} = \sum_{i_1,\dots,i_{k-1}} w_{ii_1}\cdots w_{i_{k-1}j}\,
A_{ii_1}\cdots A_{i_{k-1}j}.
\]

Bounding by the maximal weight $w^{\max}=\max_{ij}w_{ij}$ gives
\[
W^k_{ij}\le (w^{\max})^k\,A^k_{ij}.
\]

Let $T^k_{\max}=\max_{ij} A^k_{ij}$. Then the series is bounded by
\[
\sum_k (w^{\max})^k T^k_{\max}.
\]

\paragraph{Sign paradox.}  
Why do negative entries sometimes appear in $(I-W)^{-1}$?  
It is a subtle property of divergent series. For instance,
\[
S = \sum_{k=0}^\infty c^k = \frac{1}{1-c}, \quad |c|<1.
\]
But if $c>1$, the series diverges, yet analytic continuation gives $S = -1/(c-1)$, a negative number.  

Similarly, if weights $w_{ij}$ are “too large” relative to the graph structure, the Neumann expansion diverges and negative terms appear in the Leontief inverse.  

\paragraph{Interpretation.}  
Thus, to ensure nonnegative Leontief inverses, one must bound edge weights relative to the spectral radius: the requirement is essentially $\rho(W)<1$. This connects back to Perron–Frobenius: the dominant eigenvalue controls convergence and positivity of the resolvent expansion.

\section{Special graphs}
\subsection{Lattices}
\index{key}{lattice}

Lattices are regular arrangements of points in Euclidean space, and serve as canonical examples of infinite graphs. They are fundamental both in physics (crystal structures, statistical mechanics) and in computer science (coding theory, cryptography).\footnote{See J.~Conway and N.~Sloane, \emph{Sphere Packings, Lattices and Groups}, Springer (1999).}  

---

\paragraph{Definition.}  
Let $B\in\mathbb{R}^{d\times k}$ be a matrix with linearly independent columns $\{\vec b_1,\dots,\vec b_k\}$.  
The lattice generated by $B$ is
\[
L(B)=\{B\vec x : \vec x\in\mathbb{Z}^k\} = \Big\{\sum_{i=1}^k x_i \vec b_i : x_i\in\mathbb{Z}\Big\}.
\]

---

\paragraph{Equivalence of bases.}  
Two bases $B,C$ generate the same lattice iff there exists a \emph{unimodular} integer matrix $U\in\mathbb{Z}^{k\times k}$ with $\det U=\pm 1$ such that
\[
B = C U.
\]
Unimodularity ensures that $U^{-1}$ is also integral, so $L(B)=L(C)$. Thus the choice of basis is not unique, but the lattice itself is well defined.

---

\paragraph{Determinant of a lattice.}  
The \emph{determinant} of $L(B)$, denoted $\det L(B)$, is the volume of the fundamental parallelepiped
\[
P(B) = \Big\{\sum_{i=1}^k x_i \vec b_i : 0\le x_i < 1 \Big\}.
\]
It is given by
\[
\det L(B) = \mathrm{Vol}(P(B)) = \sqrt{\det(B^T B)}.
\]
Geometrically, this quantity is the inverse of the density of lattice points in $\mathbb{R}^d$.

Hadamard’s inequality implies
\[
\det L(B) \;\leq\; \prod_{i=1}^k \|\vec b_i\|,
\]
with equality iff the basis vectors are orthogonal.

---

\paragraph{Shortest vector and Minkowski’s theorem.}  
Let $\mathcal{D}(L(B))$ denote the length of the shortest nonzero lattice vector.  
From Gram–Schmidt orthogonalization, one obtains the bound
\[
\mathcal{D}(L(B)) \;\geq\; \min_i \|\tilde b_i\|,
\]
where $\{\tilde b_i\}$ are the orthogonalized vectors.  

For an upper bound, Minkowski’s theorem applies:  
\begin{quote}
If $L(B)$ is a full-rank lattice in $\mathbb{R}^d$, then there exists a nonzero $\vec x\in L(B)$ with
\[
\|\vec x\| \;\leq\; \sqrt{d}\,\det(L(B))^{1/d}.
\]
\end{quote}
This follows from Blichfeldt’s theorem, which ensures that sufficiently large convex bodies contain two distinct lattice points whose difference lies in $L(B)$.\footnote{See C.~D.~Olds, A.~Lax, and G.~P.~Davidoff, \emph{The Geometry of Numbers}, Springer (2000).}  

The constant
\[
\gamma_d = \sup_B \frac{\mathcal{D}(L(B))}{\det(L(B))^{1/d}}
\]
is the \emph{Hermite constant} in dimension $d$, central to geometry of numbers.

---

\paragraph{Dual lattices.}  
The \emph{dual lattice} (or reciprocal lattice) of $L(B)$ is
\[
L^*(B) = \{\vec y \in \mathrm{span}(B) : \langle \vec y, \vec x \rangle \in \mathbb{Z}, \;\forall \vec x\in L(B)\}.
\]

If $B$ has full rank, then a basis of the dual is
\[
D = B (B^T B)^{-1}.
\]
This satisfies $B^T D=I$. The dual is unique up to unimodular transformations, and satisfies
\[
\det L^*(B) = \frac{1}{\det L(B)}.
\]
Moreover, the dual of the dual lattice is the original lattice.  

\begin{figure}[h]
    \centering
    \includegraphics[scale=0.4]{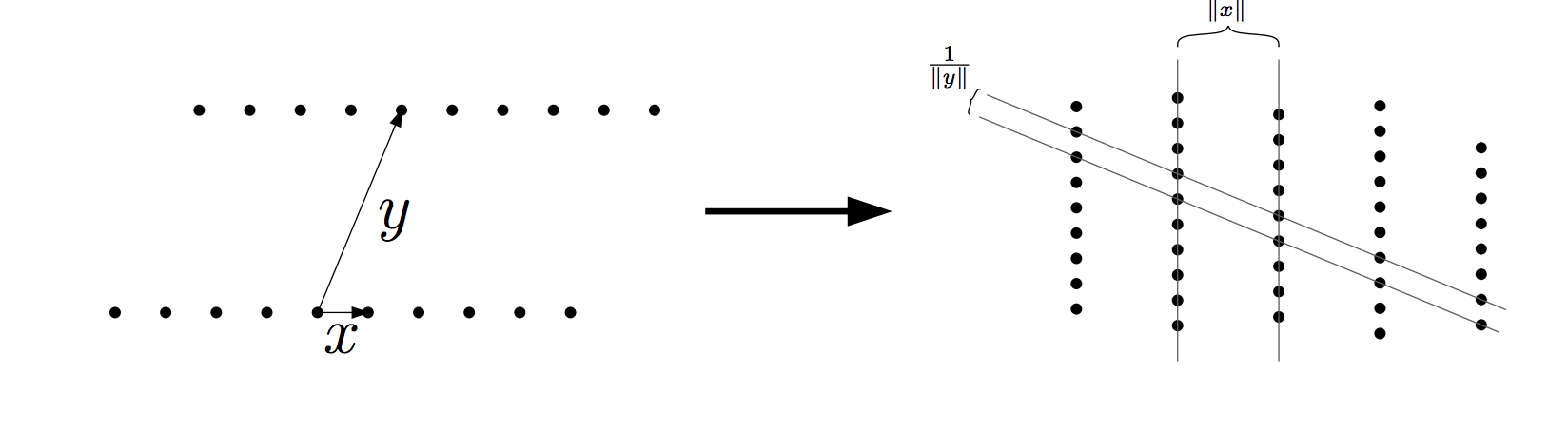}
    \caption{A lattice and its dual (reciprocal) lattice.}
    \label{fig:lattdual}
\end{figure}

Figure~\ref{fig:lattdual} illustrates this duality.

---

\paragraph{Connection to graphs.}  
Lattices provide canonical examples of infinite graphs: the integer lattice $\mathbb{Z}^d$ is the graph with vertex set $\mathbb{Z}^d$ and edges between nearest neighbors.  
Spectral analysis of $\mathbb{Z}^d$ connects to recurrence and transience of random walks (as discussed earlier), while dual lattices arise naturally in Fourier analysis of periodic graphs.

\part{Advanced topics in linear algebra}
\chapter{Advanced methods}
\section{Preamble to Advanced Topics}

Having introduced the fundamental concepts of \emph{matrices} and \emph{graphs}, 
we now shift to more advanced methods. The motivation for this transition is not 
merely pedagogical: in the study of \emph{complex systems}, basic tools such as 
eigenvalues, adjacency matrices, Laplacians, or graph connectivity are often 
insufficient on their own. What gives these notions true power is how they extend 
to more sophisticated frameworks---perturbation theory, matrix functions, spectral 
transforms, or asymptotic analysis. These advanced methods allow us to explore 
phenomena that go beyond static structures: stability of dynamical systems, 
controllability, diffusion on networks, fluctuations, and the embedding of data 
in high-dimensional spaces.

Why was it necessary to begin with the basics? Linear algebra and graph theory 
provide the common language through which different fields---physics, ecology, 
computer science, economics---communicate. For instance, the concept of the 
\emph{resolvent operator}, already seen in both matrix inversion and graph Laplacians, 
reappears across disciplines: in economics as the \emph{Leontief inverse}, in sociology 
as \emph{Katz centrality}, in ecology as the definition of \emph{trophic levels}, and in 
dynamical systems as a tool for studying stability. 
By grounding ourselves in these definitions, we ensure that the reader can navigate 
smoothly between such applications.

The upcoming chapters delve deeper into the \emph{spectral perspective}. We will 
encounter tools such as \emph{Kronecker and tensor products}, \emph{matrix functions 
and their integral representations}, \emph{perturbation theory for eigenvalues}, 
\emph{resolvents and pseudospectra}, and \emph{spectral transforms} (Fourier, Hadamard, 
Haar, Radon). Each of these extends the reach of the simple eigenvalue problem, showing 
how spectra encode rich information about structure and dynamics. The emphasis remains 
on intuition and applicability: rather than exhaustive proofs, we aim to highlight 
connections that illuminate why spectral methods are so broadly useful in complex systems.

In short, the ``basic'' notions of matrices and graphs are the scaffolding. What follows 
is the architecture built upon them: a collection of techniques that allow us to 
\emph{move from structure to dynamics, from static relations to evolving processes, 
and from isolated models to interdisciplinary applications}. It is this shift---from 
foundations to extensions---that defines the advanced part of these notes.

\section{Linear regression, the Markov--Gauss theorem and Gram--Schmidt decompositions}

Let us now introduce the connection between 
linear regression\index{key}{linear regression}, 
the Gram--Schmidt process\index{key}{Gram--Schmidt decomposition}, 
and the QR decomposition\index{key}{QR decomposition}.

A linear regression consists in solving the problem
\begin{equation}
    \vec y = f(\vec x),
\end{equation}
where $f(\vec x)$ is expanded as a linear combination
\[
   f(\vec x) = \sum_{i=1}^k \beta_i f_i(\vec x),
\]
with $f_i(\vec x)$ generic functions of the variables 
$x_1,\ldots,x_k$.  
This is commonly formulated as the minimization of the 
\emph{residual sum of squares} (RSS):
\begin{equation}
    RSS(\vec\beta)
    =\sum_{i=1}^N \big(y_i - \beta_i f_i(\vec x)\big)^2
    =(\vec y- \tilde f \vec \beta)^{\top}(\vec y- \tilde f \vec \beta),
\end{equation}
where the unknown parameters are $\vec \beta=(\beta_0,\cdots,\beta_p)$ 
and $\tilde f=[\vec 1 \ \vec f_1 \cdots \vec f_k]$. 
For simplicity, let us assume $f_i(\vec x)=x_i$, in which case 
we define the design matrix $X=[\vec x_1,\cdots,\vec x_k]$.

The minimization condition is obtained by differentiating:
\begin{equation}
    \partial_{\vec \beta} RSS(\vec \beta)
    = -2 X^{\top}(\vec y - X \vec \beta)=0,
\end{equation}
which yields the standard normal equation
\begin{equation}
    \vec \beta = (X^{\top} X)^{-1} X^{\top} \vec y,
\end{equation}
assuming $X^{\top} X$ is invertible.\footnote{If $X^{\top}X$ is singular, 
the regression problem is ill-posed and one must resort 
to pseudo-inverses or regularization.} 
The regression vector of predictions is then
\begin{equation}
   \hat y = X \vec \beta
   = X(X^{\top} X)^{-1} X^{\top} \vec y,
\end{equation}
which is the orthogonal projection of $\vec y$ 
onto the subspace spanned by the columns of $X$ 
(its ``features''), since $P=X(X^{\top} X)^{-1} X^{\top}$ 
is an orthogonal projector.\footnote{See, e.g., 
the discussion in Horn and Johnson, 
\emph{Matrix Analysis} (Cambridge University Press, 1985).}

The Gram--Schmidt process allows one to connect this 
projection to the QR decomposition. Writing
\[
   X = Z\Gamma=[\vec z_0,\cdots,\vec z_k] \Gamma,
\]
with $\Gamma$ upper triangular and $\vec z_i$ mutually orthogonal, 
we define $D^2 = Z^{\top} Z$, a diagonal matrix whose entries 
are the squared norms of the $\vec z_i$.  
This yields a QR factorization
\[
   X = (Z D^{-1})(D\Gamma) = QR,
\]
where $Q=Z D^{-1}$ is orthogonal and $R=D\Gamma$ is 
upper triangular. Consequently,
\begin{equation}
   X^{\top}X = R^{\top} Q^{\top} Q R = R^{\top} I R,
\end{equation}
so that
\[
   (X^{\top} X)^{-1} X^{\top} = R^{-1} Q^{\top}.
\]
Thus the regression coefficients can be computed directly 
once the QR decomposition of $X$ is available, 
without forming $(X^{\top}X)^{-1}$ explicitly.

Finally, we note that linear regression enforces the condition
\begin{equation}
    X^{\top}(\vec y - \hat y) = X^{\top}(I-\Omega) \vec y = 0,
\end{equation}
meaning that the residual $\vec y-\hat y$ lies in the kernel of $X^{\top}$, 
i.e. it is orthogonal to the feature space $V=\text{Span}(X^{\top})$.
\footnote{The connection between least squares regression and orthogonal 
projections is sometimes referred to as the \emph{Gauss--Markov theorem}, 
which states that under classical assumptions, the least squares estimator 
is the best linear unbiased estimator (BLUE).}

\section{Products: vectors, matrices, tensors}

\subsection{Direct sum}
The simplest operation on vector spaces is the \emph{direct sum}.\index{key}{direct sum}
If $V$ and $W$ are two spaces, we define
\begin{equation}
    V = W_1 \oplus W_2.
\end{equation}
More general sums follow from repeated binary sums. 
For example, if $W_1$ is a linear space over $F^{n_1}$ 
and $W_2$ over $F^{n_2}$, then $V$ is a linear space over 
$F^{n_1+n_2}$ and vectors in $V$ have length $n_1+n_2$.

The direct sum of two matrices $A_1\oplus A_2$ is the block diagonal matrix
\begin{equation}
   A_1\oplus A_2=\begin{pmatrix} 
      A_1 & 0 \\
      0 & A_2 
   \end{pmatrix},
\end{equation}
so that
\[
   (A_1\oplus A_2)(B_1\oplus B_2) = A_1B_1\oplus A_2B_2.
\]
It follows that
\[
   \text{Tr}(A_1\oplus A_2)=\text{Tr}(A_1)+\text{Tr}(A_2),
   \quad
   \det(A_1\oplus A_2)=\det(A_1)\det(A_2).
\]

Similarly, under the Cholesky factorization\index{key}{Cholesky factorization (Kronecker)} 
for positive semidefinite matrices\index{key}{positive semidefinite matrices} we have
\begin{equation}
A \otimes B = (L_A L_A^{\top}) \otimes ( L_B L_B^{\top})
= (L_A\otimes L_B)( L_A^{\top} \otimes L_B^{\top}).
\end{equation}

Likewise, under the QR decomposition\index{key}{QR decomposition (Kronecker)} one finds
\begin{equation}
A \otimes B = (Q_A R_A) \otimes ( Q_B R_B)
= (Q_A\otimes Q_B)( R_A \otimes R_B).
\end{equation}

For the Singular Value Decomposition (SVD)\index{key}{SVD (Kronecker)} we obtain
\begin{eqnarray}
A \otimes B &=& (U_A \Sigma_A V_A^*) \otimes ( U_B \Sigma_B V_B^*) \nonumber \\
&=& (U_A \otimes U_B)(\Sigma_A \otimes \Sigma_B)(V_A^*\otimes V_B^*).
\end{eqnarray}

From this expression it follows immediately that
\[
   \text{rank}(A \otimes B)=\text{rank}(A)\,\text{rank}(B).
\]
It is also interesting to note that if $U$ and $U^\prime$ are orthogonal matrices, 
then $U\otimes U^\prime$ is orthogonal as well.\footnote{This property is inherited 
from the fact that $(U\otimes U^\prime)(U\otimes U^\prime)^{\top} = 
(UU^{\top})\otimes(U^\prime {U^\prime}^{\top}) = I\otimes I = I$.}  

A connection with the Hadamard product also exists.  
If $A = B\otimes C$, with $B$ and $C$ of size $m\times n$, 
then the Hadamard product is contained in $A$ at regular index intervals.  
For example, using index notation $1:k+1:k^2$ to denote the sequence 
$1,k+2,2k+3,\ldots,k^2$, one has
\[
   B\circ C = A_{1:m+1:m^2,\;1:n+1:n^2}.
\]

As a simple example:
\begin{equation}
    \begin{pmatrix}
         a_1 & b_1  \\
         c_1 & d_1
    \end{pmatrix} \otimes     
    \begin{pmatrix}
         a_2 & b_2  \\
         c_2 & d_2 
    \end{pmatrix}
    =
    \begin{pmatrix}
     \mathbf{a_1 a_2} & a_1 b_2 & b_1 a_2 & \mathbf{b_1 b_2} \\
     a_1 c_2 & a_1 d_2 & b_1 c_2 & b_1 d_2 \\
     c_1 a_2 & c_1 b_2 & d_1 a_2 & d_1 b_2 \\
     \mathbf{c_1 c_2} & c_1 d_2 & d_1 c_2 & \mathbf{d_1 d_2} \\
    \end{pmatrix}.
\end{equation}

\subsubsection*{Rank--1 decompositions}
A rank--1 decomposition can also be written using Kronecker products:
\begin{equation}
   M=\sum_i \lambda_i \,\vec x_i \vec x_i^{\top}
   \equiv\sum_i \lambda_i \,\vec x_i \otimes \vec x_i.
\end{equation}

\subsubsection*{Kronecker sum}
The Kronecker product induces the \emph{Kronecker sum}\index{key}{Kronecker sum}.  
For matrices $A$ of size $m$ and $B$ of size $n$, we define
\begin{equation}
    A \oplus B= I_n \otimes A + B \otimes I_m.
\end{equation}
The Kronecker sum has better spectral properties than generic matrix sums.  
In fact, if $\vec x_i$ and $\vec y_j$ are eigenvectors of $A$ and $B$ with eigenvalues 
$\lambda_i$ and $\nu_j$, then $\vec y_j \otimes \vec x_i$ is an eigenvector of 
$A \oplus B$ with eigenvalue $\lambda_i+\nu_j$.

In general the distributive property does not hold:
\begin{eqnarray}
    (A\oplus B)\otimes C &\neq& (A\otimes C)\oplus(B\otimes C), \nonumber \\
    A \otimes (B \oplus C) &\neq& (A \otimes B)\oplus (A \otimes C).
\end{eqnarray}
However, the matrix exponential enjoys a very useful identity:
\begin{equation}
    \exp(A \oplus B)=\exp(A)\otimes \exp(B).
\end{equation}

\subsubsection*{Physical interpretation}
Kronecker sums and products generalize naturally to more than two matrices.  
In quantum physics, they appear in the description of non-interacting systems.  
If a collection of independent systems is described by Hamiltonians $H_i$, then 
the Hamiltonian of the composite system is
\begin{equation}
    H_{\text{whole}}=\bigoplus_i H_i
    =\sum_i I\otimes \cdots \otimes I \otimes 
      \underbrace{H_i}_{\text{$i$-th element}} \otimes I \otimes \cdots  \otimes I,
\end{equation}
which is one of the axioms of both the Koopman--von Neumann 
formulation of classical mechanics and quantum theory.\footnote{See, e.g., 
Koopman (1931), von Neumann (1932).}

\subsubsection*{Kronecker power}
The \emph{Kronecker power}\index{key}{Kronecker power} is a convenient way of expressing 
all possible products of the elements of a vector up to a given order.  
It is denoted as
\begin{equation}
    \vec x^{[i]}= \underbrace{\vec x \otimes \cdots \otimes \vec x}_{i\ \text{times}},
\end{equation}
and is a vector of dimension $n^i$, where $\vec x \in \mathbb{R}^n$.  
For instance, for $n=2$ we have 
$\vec x^{[1]}=\vec x$, 
\[
   \vec x^{[2]}=(x_1^2,\;x_1x_2,\;x_2x_1,\;x_2^2),
\]
and
\[
   \vec x^{[3]}=(x_1^3,\;x_1^2x_2,\;x_1x_2^2,\;x_2^3,\;\ldots).
\]
The Kronecker power is particularly useful in Taylor expansions of multivariate functions.

Finally, Kronecker products are linear.  
If $M$ is a linear operator acting on $V$, then on 
$V^{\otimes n}= V \otimes \cdots \otimes V$ we define
\[
   M(V^{\otimes n}) = M(V) \otimes V \otimes \cdots \otimes V
   + V \otimes M(V) \otimes \cdots \otimes V
   + \cdots + V \otimes \cdots \otimes M(V).
\]

\subsection{Tensor products, outer products, and Kronecker products}

We now discuss in detail another form of product among linear vector spaces, 
which often causes confusion. Very frequently, three different objects are denoted 
by the same symbol $\otimes$: the \emph{Kronecker product} between matrices, 
the \emph{tensor product}, and the \emph{outer product}. 
It is important to clarify their differences, as the notation is often abused 
and can be misleading without proper context.

To begin, recall the familiar scalar (inner) product between vectors:
\begin{equation}
    \vec a \cdot \vec b = \sum_i a_i b_i.
\end{equation}
This operation, which takes two vectors and produces a scalar, can also be written 
as $\vec a^{\top} \vec b$, or, in Einstein summation convention, $a^i b_i$ 
(repeated indices are summed). Here, an upper index $a^i$ represents a contravariant 
(row) vector, while a lower index $b_i$ denotes a covariant (column) vector.  

To make this more explicit, let us define row and column unit vectors:
\begin{equation}
    e_i=\delta_{ij}, \qquad e^i=\delta_{ji},
\end{equation}
so that the identity matrix can be written either as
\[
   I=(e_1\, e_2\, \cdots e_n)
   =\begin{pmatrix} e^1 \\ e^2 \\ \vdots \\ e^n \end{pmatrix}.
\]
Thus, transpose vectors are expressed as $\vec a^{\top}=\sum_i a_i e_i$ 
and $\vec a=\sum_i a_i e^i$.  
In this sense, transposition corresponds to raising or lowering an index.  

The tensor (covariant) product of two vectors is then
\begin{eqnarray}
    \vec a \otimes \vec b &=& \sum_{ij} a_i b_j\, e_i \otimes e_j
     = \begin{pmatrix} a_1 b_1 \\ \vdots \\ a_1 b_n \\ a_2 b_1 \\ \vdots \end{pmatrix}, \nonumber \\
    \vec a^{\top} \otimes \vec b^{\top} &=& \sum_{ij} a_i b_j\, e^i \otimes e^j
     = (a_1 b_1,\ldots,a_1 b_m,a_2 b_1,\ldots).
\end{eqnarray}
Covariant tensor products of vectors are thus Kronecker products, and if the 
original vectors are of length $m$ and $n$, the result has dimension $mn$.  

Note that covariant tensor products are not symmetric, i.e. 
$\vec a\otimes \vec b \neq \vec b \otimes \vec a$.  
Contravariant tensor products, on the other hand, produce matrices. For example,
\begin{equation}
    \vec a \otimes \vec b^{\top} 
    = \sum_{ij} a_i b_j\, e_i \otimes e^j
    = \begin{pmatrix} 
        a_1 b_1 & a_1 b_2 & \cdots & a_1 b_m \\ 
        a_2 b_1 & a_2 b_2 & \cdots & a_2 b_m \\ 
        \vdots  & \vdots  & \ddots & \vdots \\
        a_n b_1 & a_n b_2 & \cdots & a_n b_m  
      \end{pmatrix}.
\end{equation}
Clearly $(\vec a \otimes \vec b^{\top})^{\top} = \vec a^{\top} \otimes \vec b$.  
Such matrices are decomposable (rank--1). In general, one can write
\begin{equation}
    X=\sum_{ij} x_{ij}\, e_i \otimes e^j,
\end{equation}
which need not be of rank--1.  

Furthermore,
\begin{eqnarray}
    \text{vec}(X) &=& \sum_{ij} x_{ij}\, e_i \otimes e_j, \nonumber \\
    \text{vec}(X)^{\top} &=& \sum_{ij} x_{ij}\, e^i \otimes e^j,
\end{eqnarray}
while the transpose operation satisfies $(e_i \otimes e^j)^{\top} = e^i \otimes e_j$.  
In this sense, the tensor product generalizes the Kronecker product.  

Tensor products can be extended further. For brevity of notation, one often writes
\begin{equation}
    e^{i_1} \otimes e^{i_2} \otimes \cdots \otimes e^{i_k}
    \otimes e_{j_1} \otimes \cdots \otimes e_{j_m} 
    = e^{i_1 \cdots i_k}_{j_1 \cdots j_m}.
\end{equation}
Here, the order of indices matters separately for covariant (lower) 
and contravariant (upper) indices, but not between the two types. For instance,
\begin{equation}
    e^{i_1 \cdots i_n}_{j_1 \cdots j_m}
    = \big( {e^{i_1 \cdots i_p}}_{j_1 \cdots j_{p'}} \big)^{i_{p+1}\cdots i_k}{}_{j_{p'+1}\cdots j_m}
\end{equation}
is a valid representation, as long as the relative order of the upper indices 
and the lower indices is maintained.

The basis above represents a generic tensor basis.  
In general, scalar products and the raising or lowering of indices are 
implemented via a metric. In the Euclidean case, the metric is the identity matrix, 
so that
\begin{equation}
    \sum_i e^i \delta_{ij}=e_j, 
    \qquad 
    \sum_j e_i \delta^{ij}=e^j.
\end{equation}
Contraction of a covariant and a contravariant index simply yields a Kronecker delta.  
For example,
\[
    [0,\ldots,\underbrace{1}_i,0,\ldots,0]
    \begin{pmatrix}
    1 & 0 & \cdots & 0 \\
    0 & 1 & \cdots & 0 \\
    \vdots & \vdots & \ddots & \vdots \\
    0 & 0 & \cdots & 1
    \end{pmatrix}
    =
    [0,\ldots,\underbrace{1}_i,0,\ldots,0]^{\top},
\]
with an analogous relation holding for the transpose.  

Thus, outer products can be written as
\begin{equation}
   \vec a\cdot \vec b
   = \sum_{ij} a_i b_j\, e_i \otimes e^j \,\delta_{ij}
   = \sum_i a_i b_i,
\end{equation}
as expected.

\subsection{Tensor products and Schmidt decomposition}

One of the most important decompositions of Hilbert spaces that relies on 
the tensor product is the \emph{Schmidt decomposition theorem}.\index{key}{Schmidt decomposition}  
Let $\mathcal H=\mathcal H_1 \otimes \mathcal H_2$ be the tensor product of two 
Hilbert spaces $\mathcal H_1$ and $\mathcal H_2$. If both are endowed with scalar 
products, $\mathcal H$ is itself a Hilbert space.  

Let $M_1:\mathcal H_1\to\mathcal H_1^\prime$ and 
$M_2:\mathcal H_2\to\mathcal H_2^\prime$ be linear maps.  
Their tensor product acts as
\begin{equation}
   (M_1 \otimes M_2)(\vec v_1 \otimes \vec v_2)
   = M_1 \vec v_1 \otimes M_2 \vec v_2,
\end{equation}
and satisfies
\begin{equation}
   (M_1 \otimes M_2)(M_1^\prime \otimes M_2^\prime)
   = (M_1 M_1^\prime) \otimes (M_2 M_2^\prime).
\end{equation}

Now consider a generic vector $h \in \mathcal H$ of the form
\begin{equation}
   h=\sum_{ij} c_{ij}\, \vec v_i \otimes \vec r_j
   =\sum_i \vec v_i \otimes \Big(\sum_j c_{ij}\,\vec r_j\Big).
\end{equation}
Defining $\vec r_i^\prime = \sum_j c_{ij}\,\vec r_j$, we can rewrite
\[
   h=\sum_i \vec v_i \otimes \vec r_i^\prime.
\]

The Schmidt decomposition theorem states the following:  
\begin{theorem}[Schmidt decomposition]
Let $h \in \mathcal{H}_1 \otimes \mathcal{H}_2$ with $\mathcal{H}_1,\mathcal{H}_2$ Hilbert spaces.  
Then there exist orthonormal bases $\{\vec v_j\}$ for $\mathcal{H}_1$, $\{\vec r_j\}$ for $\mathcal{H}_2$, and non-negative real numbers $c_j$ such that
\begin{equation}
  h=\sum_j c_j\, \vec v_j \otimes \vec r_j.
\end{equation}
The coefficients $c_j$ are called the \emph{Schmidt coefficients}. If $\mathcal{H}_1,\mathcal{H}_2$ are finite-dimensional, only finitely many $c_j$ are non-zero.\footnote{See, e.g., R.~Bhatia, \emph{Matrix Analysis}, Springer (1997); also Nielsen and Chuang, \emph{Quantum Computation and Quantum Information}, Cambridge University Press (2000).}
\end{theorem}

---

\paragraph{First proof (via SVD).}  
Write $h=\sum_{ij} c_{ij}\,\vec v_i^\prime\otimes \vec r_j^\prime$, 
where $\{\vec v_i^\prime\}$ and $\{\vec r_j^\prime\}$ are fixed orthonormal bases.  
Let $C=(c_{ij})$ be the $n\times m$ coefficient matrix.  
Applying the singular value decomposition (SVD),
\begin{equation}
   C=U \Sigma V^*,
\end{equation}
with $U$ unitary of size $n\times n$, $V$ unitary of size $m\times m$, 
and $\Sigma$ a rectangular diagonal matrix with non-negative entries (the singular values).  

Then
\begin{equation}
   h=\sum_i \sigma_i \,(U\vec v_i^\prime)\otimes (V^*\vec r_i^\prime),
\end{equation}
where the vectors $U\vec v_i^\prime$ and $V^*\vec r_i^\prime$ are orthonormal.  
Thus we obtain the Schmidt form, with Schmidt coefficients equal to the singular values $\sigma_i$.\footnote{Some of the $\sigma_i$ may be zero, so only $\min(m,n)$ terms appear in the decomposition.}

---

\paragraph{Second proof (via density matrices and partial trace).}  
Another approach is based on density matrices and partial traces.  

Given $h=\sum_{ij} c_{ij}\,\vec v_i\otimes \vec r_j$,  
define the density operator $\rho$ as
\begin{equation}
   \rho = |h\rangle\langle h|
   = \sum_{ij,i'j'} c_{ij} c_{i'j'}^* \,
   |\vec v_i\rangle\langle \vec v_{i'}|
   \otimes
   |\vec r_j\rangle\langle \vec r_{j'}|.
\end{equation}
The reduced density operators are obtained by taking partial traces:
\begin{eqnarray}
   \rho_1 &=& \mathrm{Tr}_2(\rho)
   = \sum_{jj'} \langle \vec r_{j'}|\vec r_j\rangle
   \sum_{ii'} c_{ij} c_{i'j'}^*\, |\vec v_i\rangle\langle \vec v_{i'}|, \\
   \rho_2 &=& \mathrm{Tr}_1(\rho)
   = \sum_{ii'} \langle \vec v_{i'}|\vec v_i\rangle
   \sum_{jj'} c_{ij} c_{i'j'}^*\, |\vec r_j\rangle\langle \vec r_{j'}|.
\end{eqnarray}
Because the bases are orthonormal, the inner products simplify, giving
\[
   \rho_1=\sum_{i,i',j} c_{ij}c_{i'j}^*\, |\vec v_i\rangle\langle \vec v_{i'}|,
   \qquad
   \rho_2=\sum_{j,j',i} c_{ij}c_{ij'}^*\, |\vec r_j\rangle\langle \vec r_{j'}|.
\]

Both $\rho_1$ and $\rho_2$ are positive semidefinite and Hermitian.  
A fundamental property is that their non-zero eigenvalues coincide.  
These eigenvalues are precisely the squares of the Schmidt coefficients $|c_j|^2$.  

Thus, diagonalizing $\rho_1$ or $\rho_2$ yields the same spectrum of Schmidt coefficients, 
and the corresponding eigenvectors form the orthonormal sets $\{\vec v_j\}$ and $\{\vec r_j\}$ 
appearing in the Schmidt decomposition.  

---

The Schmidt decomposition is especially useful in quantum information theory, 
where it characterizes entanglement. The number of non-zero Schmidt coefficients 
is called the \emph{Schmidt rank} and serves as a measure of the “entanglement dimension” 
of the bipartite state.

\subsection{Kronecker product for matrix equations}

The Kronecker product and the related Kronecker sum have important applications 
in the study of matrix equations.\index{key}{matrix equation} 
A \emph{matrix equation} is an equation in which the unknown is itself a matrix, 
rather than a vector. Such equations arise in a wide range of systems, 
from control theory to signal processing, and it is useful to reduce them to 
linear systems via vectorization.  

The key operation here is the \emph{vectorization} of a matrix.  
For a matrix $V$, the vector $\text{vec}(V)$ is obtained by stacking its columns 
on top of each other to form a single column vector.  
The following identity is central:\footnote{See R.~Horn and C.~Johnson, 
\emph{Matrix Analysis}, Cambridge University Press (1985).}
\begin{equation}
   \text{vec}(U A V) = (V^{\top} \otimes U)\,\text{vec}(A).
\end{equation}
This allows matrix equations to be rewritten as standard linear equations.  

For example:
\begin{itemize}
   \item Multiplication: 
   \[
      AB \quad \mapsto \quad (I \otimes A)\,\text{vec}(B)
      = (B^{\top} \otimes I)\,\text{vec}(A).
   \]
   \item Sylvester-type equations:  
   \[
      \alpha A X + \beta X B
      \quad \mapsto \quad
      (\alpha I \otimes A + \beta B^{\top} \otimes I)\,\text{vec}(X).
   \]
   \item More general bilinear forms:  
   \[
      A X B + C X D
      \quad \mapsto \quad
      (B^{\top} \otimes A + D^{\top} \otimes C)\,\text{vec}(X).
   \]
\end{itemize}

This correspondence means that many matrix equations can be solved using 
linear algebraic techniques once vectorized.  

A particularly important case is the Sylvester equation
\[
   A X + X B = C.
\]
Using the Kronecker product, this becomes
\[
   (I \otimes A + B^{\top} \otimes I)\,\text{vec}(X) = \text{vec}(C).
\]
The solvability condition is clear: the coefficient matrix is invertible if and only if
$\lambda_i(A) + \lambda_j(B) \neq 0$ for all eigenvalues $\lambda_i(A)\in \Lambda(A)$ 
and $\lambda_j(B)\in \Lambda(B)$.  

To see this, let $A=U_A T_A U_A^\dagger$ and $B=U_B T_B U_B^\dagger$ be 
the Schur decompositions. Then
\begin{equation}
   \alpha I \otimes A + \beta B \otimes I
   = (U_B \otimes U_A)\,
     \big(\alpha I \otimes T_A + \beta T_B \otimes I\big)\,
     (U_B^\dagger \otimes U_A^\dagger).
\end{equation}
Since $T_A$ and $T_B$ are triangular, the block in the middle is triangular with 
diagonal entries of the form
\[
   \alpha \lambda_a + \beta \lambda_b,
   \qquad \lambda_a\in \Lambda(A),\ \lambda_b\in \Lambda(B).
\]
Thus invertibility requires that no such combination vanish.  

Thus, the Kronecker product allows one to treat matrix equations 
in exactly the same way as vector equations, a tool that will be exploited 
further in the discussion of matrix functions.

---
\subsection{Kronecker products and the vec operator}
\index{key}{Kronecker product}\index{key}{vec operator}
We now report a summary and few identities. 
As mentioned, the \emph{Kronecker product} of $A \in \mathbb{F}^{m\times n}$ and 
$B \in \mathbb{F}^{p\times q}$ is defined as
\begin{equation}
A \otimes B =
\begin{bmatrix}
a_{11} B & a_{12} B & \cdots & a_{1n} B \\
a_{21} B & a_{22} B & \cdots & a_{2n} B \\
\vdots   & \vdots   & \ddots & \vdots   \\
a_{m1} B & a_{m2} B & \cdots & a_{mn} B
\end{bmatrix}
\in \mathbb{F}^{mp \times nq}.
\end{equation}
It is a block-matrix generalization of the tensor product of vectors.  

The \emph{vec operator}, $\textit{vec}(\cdot)$, stacks the columns of a matrix on top of
each other. For $X = [\mathbf{x}_1,\dots,\mathbf{x}_n]\in \mathbb{F}^{m\times n}$,
\[
\text{vec}(X) =
\begin{bmatrix}
\mathbf{x}_1 \\ \vdots \\ \mathbf{x}_n
\end{bmatrix}
\in \mathbb{F}^{mn}.
\]

\paragraph{Basic properties.}
For conformable matrices $A,B,C,D$:
\begin{align}
(A \otimes B)(C \otimes D) &= (AC) \otimes (BD), \\
(A \otimes B)^\top &= A^\top \otimes B^\top, \\
(A \otimes B)^{-1} &= A^{-1} \otimes B^{-1}, \\
\det(A \otimes B) &= \det(A)^m \det(B)^n, \quad A\in\mathbb{F}^{n\times n}, B\in\mathbb{F}^{m\times m}, \\
\text{Tr}(A \otimes B) &= \text{Tr}(A)\text{Tr}(B), \\
\operatorname{rank}(A \otimes B) &= \operatorname{rank}(A)\,\operatorname{rank}(B).
\end{align}

\paragraph{Vec identities.}
The \textit{vec} operator converts matrix equations into vector equations, enabling the
use of Kronecker products:
\begin{align}
\text{vec}(AXB) &= (B^\top \otimes A)\,\text{vec}(X), \\
\text{vec}(A^\top) &= K_{m,n}\,\text{vec}(A), \quad K_{m,n} \text{ the commutation matrix}, \\
\text{vec}(X^\top) &= K_{m,n}\,\text{vec}(X), \quad K_{m,n} = \sum_{i,j} E_{ij}\otimes E_{ji}.
\end{align}
\paragraph{Commutation and permutation.}
The \emph{commutation matrix} $K_{m,n}$ of size $mn\times mn$ satisfies
\[
K_{m,n}\,\text{vec}(A) = \text{vec}(A^\top).
\]
More generally, permutation matrices can reorder blocks in Kronecker products
systematically.

\paragraph{Applications.}
\begin{itemize}
\item \textbf{Lyapunov equation:} For $AX+XB=C$, we have
\begin{equation}
\text{vec}(X) = (I\otimes A + B^\top\otimes I)^{-1}\,\text{vec}(C).
\end{equation}

\item \textbf{Sylvester equation:} More generally,
\[
\sum_{k} A_k X B_k = C \quad\implies\quad
\text{vec}(X) = \left(\sum_k B_k^\top \otimes A_k\right)^{-1}\,\text{vec}(C).
\]

\item \textbf{Quadratic forms:} $\text{Tr}(A^\top X B X^\top) = \text{vec}(X)^\top (B \otimes A)\,\text{vec}(X)$.

\item \textbf{Least squares:} If $\|AX-B\|_F^2$, then
\[
\|AX-B\|_F^2 = \|\,(I\otimes A)\,\text{vec}(X) - \text{vec}(B)\|_2^2.
\]
\end{itemize}

\paragraph{Spectral properties.}
If $A\in\mathbb{F}^{m\times m}$ has eigenpairs $(\lambda_i, u_i)$ and 
$B\in\mathbb{F}^{n\times n}$ has $(\mu_j, v_j)$, then
\[
A\otimes B \;\;(u_i\otimes v_j) = (\lambda_i \mu_j)(u_i\otimes v_j),
\]
so the spectrum of $A\otimes B$ is the product set $\{\lambda_i\mu_j\}$.

\paragraph{Differential identities.}
\begin{align}
\frac{\partial}{\partial X}\,\text{vec}(AXB) &= (B^\top\otimes A), \\
\frac{\partial}{\partial X}\,\|\text{vec}(X)\|_2^2 &= 2\,\text{vec}(X), \\
\frac{\partial}{\partial X}\,\text{vec}(X)^\top (B\otimes A)\,\text{vec}(X) &= 
A^\top X(B+B^\top)^\top.
\end{align}

\paragraph{Block-structured systems.}
Kronecker products allow compact notation for structured linear systems:
\[
(A\otimes I + I\otimes B)\,\text{vec}(X) = \text{vec}(AX+XB).
\]
Such representations underlie numerical algorithms for differential equations,
signal processing, and control.
\paragraph{Additional vec identities.}
A few further identities are often useful in analysis and derivations:  

\begin{align}
\text{vec}(X+Y) &= \text{vec}(X) + \text{vec}(Y), \\[6pt]
\text{vec}(\alpha X) &= \alpha\,\text{vec}(X), \quad \alpha \in \mathbb{F}, \\[6pt]
\text{vec}(X^\top) &= K_{m,n}\,\text{vec}(X), 
\quad X\in\mathbb{F}^{m\times n}, \\[6pt]
\text{vec}(X^\ast) &= \overline{\text{vec}(X)}, 
\quad \text{(complex conjugation)}, \\[6pt]
\text{vec}(ABC) &= (C^\top \otimes A)\,\text{vec}(B), \\[6pt]
\text{vec}(X)^\top \text{vec}(Y) &= \text{Tr}(X^\top Y), \\[6pt]
\text{vec}(X)^\ast \text{vec}(Y) &= \langle X, Y \rangle_F
= \text{Tr}(X^\ast Y), \\[6pt]
\|X\|_F^2 &= \|\text{vec}(X)\|_2^2.
\end{align}

These identities highlight how the vec operator bridges matrix operations with 
vector and inner product formulations, simplifying derivations in optimization, 
statistics, and numerical linear algebra.

\section{Asymptotic behavior: matrix sequences}

We now discuss sequences of $n\times n$ matrices that approximate one another 
asymptotically in norm, as $n\to\infty$.\index{key}{matrix sequences}  
Two notions of boundedness are relevant:  

\begin{itemize}
   \item \textbf{Strong norm:}  
   A sequence $\{A_n\}$ is bounded in strong norm if
   \[
      \|A_n\|\leq M <\infty \quad \forall n.
   \]
   \item \textbf{Weak norm:}  
   Two sequences $\{A_n\}, \{B_n\}$ satisfy
   \[
      \lim_{n\to\infty} \|A_n-B_n\|=0.
   \]
\end{itemize}

Two sequences are said to be \emph{asymptotically equivalent} 
if both are bounded and $\lim_{n\to\infty}\|A_n-B_n\|=0$, 
which we write as $A_n \sim B_n$.\index{key}{asymptotic equivalence}  

---

\paragraph{Basic properties.}
If $A_n \sim B_n$, then
\[
   \lim_{n\to\infty}\|A_n\| = \lim_{n\to\infty}\|B_n\|,
\]
by the triangle inequality.  
Equivalence is transitive: if $A_n\sim B_n$ and $B_n\sim C_n$, 
then $A_n\sim C_n$.  

If $A_n\sim B_n$ and $C_n\sim D_n$, then 
\[
   A_n C_n \sim B_n D_n,
\]
since $\|AB\|\leq \|A\|\|B\|$.  
If $A_n^{-1}$ and $B_n^{-1}$ exist with bounded norm, then 
\[
   A_n^{-1}\sim B_n^{-1}.
\]

---

\paragraph{Spectral consequences.}
Trace inequalities connect asymptotic equivalence with eigenvalue behavior.  
For example,
\[
   \mathrm{Tr}(A_n-B_n)\leq \|A_n-B_n\|,
\]
implying that the average of the eigenvalues converges.  
More generally, polynomial functions of $A_n$ and $B_n$ share the same asymptotic behavior.  
If $f$ is a polynomial, then
\[
   \lim_{n\to\infty}\frac{1}{n}\Bigg(\sum_i f(\alpha_{n,i})
   -\sum_i f(\beta_{n,i})\Bigg)=0,
\]
where $\{\alpha_{n,i}\}$ and $\{\beta_{n,i}\}$ denote eigenvalues of $A_n$ and $B_n$ respectively.  

By the Weierstrass approximation theorem,\footnote{See, e.g., W.~Rudin, 
\emph{Principles of Mathematical Analysis}, McGraw-Hill (1976).} 
this extends from polynomials to continuous functions $f$.  

In particular, since the determinant is the product of eigenvalues, 
for invertible $A_n$ and $B_n$ one has
\begin{equation}
   \lim_{n\to\infty} (\det A_n)^{1/n}
   = \lim_{n\to\infty} (\det B_n)^{1/n}.
\end{equation}

---

\paragraph{Eigenvalue stability.}
The Wielandt–Hoffman theorem\footnote{See H.~Wielandt (1950); 
K.~Hoffman and R.~Kunze, \emph{Linear Algebra}, Prentice Hall (1971).} 
provides a quantitative estimate:
\begin{equation}
   \sum_{k=1}^n |\alpha_{n,k}-\beta_{n,k}|
   \leq \|A_n-B_n\|_F,
\end{equation}
where $\|\cdot\|_F$ is the Frobenius norm.  
This shows that the eigenvalues of asymptotically equivalent sequences 
remain close in aggregate, guaranteeing stability of spectral properties 
under asymptotic equivalence.

---

\noindent
To conclude, we note that asymptotic equivalence provides a powerful framework: 
if two matrix sequences are asymptotically equivalent, then not only 
their norms but also their eigenvalue distributions and continuous 
spectral functions coincide in the limit $n\to\infty$.

\section{Matrix functions}\label{sec:matrixfunc}
\index{key}{Matrix function}

Matrix functions are a natural generalization of scalar functions.  
One standard definition is via the Taylor expansion:
\begin{equation}
   f(A)=\sum_{j=0}^\infty a_j A^j,
\end{equation}
when it converges.  

A more elegant construction uses the Jordan normal form.\index{key}{Jordan normal form}  
If $A=QJQ^{-1}$, then
\begin{equation}
   f(A)=Q f(J) Q^{-1}.
\end{equation}
Since $J$ is block diagonal, it suffices to define $f(J_{\lambda,r})$ 
for each Jordan block $J_{\lambda,r}$.  
For a block of size $r$, one obtains
\begin{equation}
   f(J_{\lambda,r})=
   \begin{pmatrix}
     f(\lambda) & f'(\lambda) & \tfrac{f''(\lambda)}{2} & \cdots & \tfrac{f^{(r-1)}(\lambda)}{(r-1)!} \\
     0 & f(\lambda) & f'(\lambda) & \cdots & \tfrac{f^{(r-2)}(\lambda)}{(r-2)!} \\
     \vdots & \vdots & \ddots & \ddots & \vdots \\
     0 & 0 & \cdots & f(\lambda) & f'(\lambda) \\
     0 & 0 & \cdots & 0 & f(\lambda)
   \end{pmatrix}.
\end{equation}
Thus, if the largest Jordan block has index $r^*$, 
it suffices that $f$ is $(r^*-1)$-times differentiable.

In particular,
\begin{equation}
   \det f(A)=\prod_{i=1}^\mu f(\lambda_i)^{k_i},
\end{equation}
where $\mu$ is the number of Jordan blocks 
and $k_i$ the size of each block.  

If $A$ is diagonalizable, the definition simplifies to
\begin{equation}
   f(A)=Q f(\Lambda) Q^{-1},
   \qquad f(\Lambda)=\operatorname{diag}\big(f(\lambda_i)\big).
\end{equation}

Another powerful representation uses contour integrals:\index{key}{resolvent}
\begin{equation}
   f(A)=\frac{1}{2\pi i}\int_\Gamma f(z)\,(zI-A)^{-1}\,dz,
\end{equation}
where $\Gamma$ encloses the spectrum $\Lambda(A)$.  
Here, the resolvent
\[
   R(A,z)=(zI-A)^{-1}
\]
plays a central role.  

\paragraph{Sylvester’s theorem.}
If $A$ is diagonalizable, the Sylvester formula\index{key}{Sylvester matrix theorem} states
\begin{equation}
   f(A)=\sum_{i=1}^n f(\lambda_i) A_i,
   \label{eq:spectrdec}
\end{equation}
where $A_i$ are the \emph{Frobenius covariants}\index{key}{Frobenius covariants}:
\begin{equation}
   A_i=\prod_{j\neq i}\frac{A-\lambda_j I}{\lambda_i-\lambda_j}.
\end{equation}
These $A_i$ are projectors onto the eigenspaces.  

---

\paragraph{Fantappiè conditions.}
The formal analogy between scalar and matrix functions was studied by 
Fantappiè, who listed the following conditions:\index{key}{Fantappi\'e conditions}
\begin{itemize}
   \item $f(z)=k \ \Rightarrow \ f(A)=kI$,
   \item $f(z)=z \ \Rightarrow \ f(A)=A$,
   \item $f(z)=g(z)+h(z) \ \Rightarrow \ f(A)=g(A)+h(A)$,
   \item $f(z)=g(z)h(z) \ \Rightarrow \ f(A)=g(A)h(A)$,
   \item $f(z)=g(h(z)) \ \Rightarrow \ f(A)=g(h(A))$.
\end{itemize}

---

\paragraph{Time-dependent functions.}
A particularly important case is when $f$ depends on $tA$, 
where $t$ is a parameter (often interpreted as “time”).  
For example, the matrix exponential
\[
   e^{tA}=\sum_{j=0}^\infty \frac{t^j}{j!}A^j
\]
plays a central role in the study of differential equations and dynamical systems.
\subsection{Matrix exponential}
\index{key}{Matrix exponential}

The matrix exponential plays a central role in linear differential equations, 
control theory, and quantum mechanics. It is defined by the convergent series
\[
   e^{At}=\sum_{k=0}^\infty \frac{t^k}{k!} A^k.
\]

\paragraph{Diagonalizable case.}
If $A$ is diagonalizable, the exponential is straightforward.  
Suppose $A=M \Lambda M^{-1}$ with $\Lambda=\operatorname{diag}(\lambda_1,\ldots,\lambda_n)$.  
Then
\begin{equation}
   e^{At}=M e^{\Lambda t} M^{-1}
   = M\,\operatorname{diag}(e^{\lambda_1 t},\ldots,e^{\lambda_n t})\,M^{-1}.
\end{equation}
Thus, the exponential reduces to exponentiating the eigenvalues.

\paragraph{Non-diagonalizable case: Cayley--Hamilton approach.}
If $A$ is not diagonalizable, we cannot simply diagonalize it.  
Nevertheless, by the Cayley--Hamilton theorem\index{key}{Cayley--Hamilton theorem}, 
$A$ satisfies its characteristic polynomial $p(A)=0$.  
This ensures that $e^{At}$ can be expressed as a finite linear combination of powers of $A$:
\begin{equation}
   e^{At}=\sum_{i=0}^{n-1} y_i(t) A^i,
\end{equation}
for suitable coefficient functions $y_i(t)$.  
These functions can be determined, for example, by the method of 
confluent Vandermonde matrices (see below).

\paragraph{Jordan normal form.}
Using the Jordan decomposition\index{key}{Jordan normal form}, 
if $A=QJQ^{-1}$, then
\[
   e^{At}=Q e^{Jt} Q^{-1}.
\]
For a Jordan block $J_{\lambda,r}$ of size $r$ with eigenvalue $\lambda$, one finds
\begin{equation}
   e^{J_{\lambda,r}t}
   = e^{\lambda t}
   \begin{pmatrix}
     1 & t & \tfrac{t^2}{2!} & \cdots & \tfrac{t^{r-1}}{(r-1)!} \\
     0 & 1 & t & \cdots & \tfrac{t^{r-2}}{(r-2)!} \\
     \vdots & \vdots & \ddots & \ddots & \vdots \\
     0 & 0 & \cdots & 1 & t \\
     0 & 0 & \cdots & 0 & 1
   \end{pmatrix}.
   \label{eq:jordant}
\end{equation}
This provides a constructive way to compute $e^{At}$ even when $A$ is defective.

\paragraph{Kato decomposition.}
Another useful representation is due to Kato.\index{key}{Kato decomposition}  
Suppose $A$ can be decomposed as
\[
   A=\sum_s (\lambda_s P_s+D_s),
\]
where $P_s$ are spectral projectors and $D_s$ nilpotent operators 
with $P_s D_s=D_s P_s=0$.  
Then
\begin{equation}
   e^{At}=\sum_s e^{\lambda_s t}\Big(P_s+\sum_{j<m_s}\frac{t^j}{j!}D_s^j\Big),
   \label{eq:katoexp}
\end{equation}
where $m_s$ is the nilpotency index of $D_s$.  
This formula generalizes the Jordan block expansion.

\paragraph{Confluent Vandermonde approach.}
The exponential can also be expressed using confluent Vandermonde matrices.\index{key}{confluent Vandermonde matrix}  
Consider the ODE
\[
   \frac{d}{dt}\vec x(t)=A\vec x(t),\qquad \vec x(0)=\vec x_0.
\]
The general solution is $\vec x(t)=e^{At}\vec x_0$.  
Let the characteristic polynomial be
\[
   p(\lambda)=(\lambda-\lambda_1)^{\nu_1}\cdots (\lambda-\lambda_m)^{\nu_m}.
\]
Then the fundamental solutions are
\[
   \{t^k e^{\lambda_i t}\mid i=1,\ldots,m,\; k=0,\ldots,\nu_i-1\}.
\]
Using these, one constructs the confluent Vandermonde matrix $V$, 
which allows explicit formulas for the coefficients $y_i(t)$ in the expansion
\[
   e^{At}=\sum_{i=0}^{n-1} y_i(t) A^i.
\]
This representation is closely related to Hermite interpolation of the scalar function $e^{t\lambda}$.\footnote{See N.~Higham, 
\emph{Functions of Matrices}, SIAM (2008).}

\paragraph{Resolvent representation.}
The exponential also admits a Cauchy integral formula:\index{key}{resolvent}
\begin{equation}
   e^{At}=\frac{1}{2\pi i}\int_\Gamma e^{\lambda t}(\lambda I-A)^{-1}\,d\lambda,
\end{equation}
where $\Gamma$ is a contour enclosing the spectrum of $A$.

\paragraph{Properties.}
The matrix exponential enjoys several useful derivative identities.  
If $A(t)$ depends smoothly on $t$, then
\begin{itemize}
   \item $\displaystyle \frac{d}{dt}e^{A(t)}
   =\Big(de^{A'(t)}_{A(t)}\Big)\,e^{A(t)}$,
   \item $\displaystyle \frac{d}{dt}e^{A(t)}
   =e^{A(t)}\Big(de^{A'(t)}_{-A(t)}\Big)$,
   \item $\displaystyle \frac{d}{dt}e^{A(t)}
   =\int_0^1 e^{xA(t)}\,A'(t)\,e^{(1-x)A(t)}\,dx,$
\end{itemize}
where 
\[
   de^X_B = \sum_{k=0}^\infty \frac{1}{(k+1)!}\,\operatorname{ad}_B^k(X),
   \qquad
   \operatorname{ad}_B(X)=[B,X].
\]
These identities are particularly useful for sensitivity analysis 
and perturbation theory of matrix exponentials.

\subsection{Matrix logarithm}
\index{key}{matrix logarithm}

Unlike the exponential, the \emph{matrix logarithm} presents subtleties: 
not every matrix admits a logarithm.  
A matrix logarithm of $A$ is any matrix $X$ such that
\[
   e^X=A.
\]

\paragraph{Properties.}
Some basic properties are:
\begin{itemize}
   \item If $A$ and $B$ commute, $\log(AB)=\log(A)+\log(B)$, and therefore $\log(A^{-1})=-\log(A)$.
   \item In general, $\operatorname{Tr}\log(AB)=\operatorname{Tr}\log(A)+\operatorname{Tr}\log(B)$ 
   whenever the logarithms are defined.\footnote{For precise conditions, 
   see N.~Higham, \emph{Functions of Matrices}, SIAM (2008).}
\end{itemize}
The logarithm is always a multi-valued function, since $\log(\lambda)=\log|\lambda|+i(\theta+2\pi h)$ 
depends on the choice of branch $h\in\mathbb Z$.  
Thus, the logarithm of a real matrix may itself be complex.

\paragraph{Jordan block representation.}
Let $J_n(\lambda)$ be a Jordan block with eigenvalue $\lambda$.  
It can be written as
\[
   J_n(\lambda)=\lambda I+N=\lambda(I+H),
\]
where $N$ (or $H$) is nilpotent.  
Then
\[
   \log J_n(\lambda)=\log(\lambda) I + \log(I+H).
\]
Since $H$ is nilpotent, the series
\[
   \log(I+H)=-\sum_{i=1}^\infty \frac{(-1)^i}{i} H^i
\]
terminates after finitely many terms.  
Therefore the logarithm exists whenever $A$ is invertible ($\lambda\neq 0$ for all eigenvalues).

For example,
\[
   \log \begin{pmatrix} 1 & 1 \\ 0 & 1 \end{pmatrix}
   = \begin{pmatrix} 0 & 1 \\ 0 & 0 \end{pmatrix}.
\]

\paragraph{Applications.}
The matrix logarithm appears in the Baker--Campbell--Hausdorff (BCH) formula:\index{key}{BCH formula}
\[
   C=\log(e^A e^B),
\]
which expands $C$ as a series in commutators of $A$ and $B$.  

---

\subsection{Resolvent formula}
\index{key}{resolvent}

The resolvent is defined by
\[
   R(A,z)=(zI-A)^{-1}.
\]
It is a key tool for defining matrix functions.  
For a Jordan block $J_{\lambda}^k$,
\begin{eqnarray}
R(J_{\lambda}^k,z)&=&
\begin{pmatrix}
z-\lambda & -1 &  &  \\
 & z-\lambda & -1 &  \\
 &  & \ddots & -1 \\
 &  &  & z-\lambda
\end{pmatrix}^{-1} \nonumber \\
&=&
\begin{pmatrix}
(z-\lambda)^{-1} & (z-\lambda)^{-2} & \cdots & (z-\lambda)^{-k} \\
 & (z-\lambda)^{-1} & (z-\lambda)^{-2} & \cdots \\
 & & \ddots & \vdots \\
 & & & (z-\lambda)^{-1}
\end{pmatrix}. \nonumber
\end{eqnarray}
This can be written as
\begin{equation}
   R(J_{\lambda}^k,z)=(z-\lambda)^{-1}I+(z-\lambda)^{-2}D_1+\cdots+(z-\lambda)^{-k}D_{k-1},
   \label{eq:resolventinverse}
\end{equation}
where $D_j$ has ones on the $j$th superdiagonal.  
In general,
\[
   R(A,z)=\bigoplus_{j=1}^p R(J_{\lambda_j}^{k_j},z).
\]

---

\subsection{Matrix square roots and fractional powers}
\index{key}{matrix square root}

A matrix square root of $M$ is a matrix $\sqrt{M}$ such that $M=(\sqrt{M})^2$.  
Square roots and fractional powers of matrices are useful in numerical analysis 
and quantum mechanics.

\paragraph{Exact diagonalization.}
If $A=Q \Sigma Q^{-1}$ with $\Sigma$ diagonal and invertible, then
\[
   \sqrt{A}=Q \sqrt{\Sigma} Q^{-1},
\]
where $\sqrt{\Sigma}$ is diagonal with entries $\sqrt{\lambda_i}$.

\paragraph{Iterative methods.}
Several iterative algorithms exist:
\begin{itemize}
   \item \textbf{Babylonian method:}  
   $X_{k+1}=\tfrac{1}{2}(X_k+B X_k^{-1})$, which converges to $\sqrt{B}$ if $B$ is positive definite.
   \item \textbf{Denman--Beavers iteration:}  
   Initialize $Y_0=B$, $Z_0=I$, then iterate
   \[
      Y_{k+1}=\tfrac{1}{2}(Y_k+Z_k^{-1}), \qquad
      Z_{k+1}=\tfrac{1}{2}(Z_k+Y_k^{-1}).
   \]
   If convergent, $Y_\infty=\sqrt{B}$.
\end{itemize}

\paragraph{Higher roots and fractional powers.}
A $p$th root of $A$ satisfies $X^p=A$.  
Rice’s iterative method is
\[
   X_{k+1}=X_k+\tfrac{1}{p}(A-X_k^p), \qquad X_0=0,
\]
which converges under suitable conditions.  
In general,
\[
   A^\xi = e^{\xi \log(A)},
\]
for $\xi\in\mathbb R$, provided a logarithm of $A$ exists.  

Thus rational powers can be defined by
\[
   A^{p/q}=\big(A^p\big)^{1/q}.
\]

---

\subsection{Functions of idempotent matrices}
\index{key}{idempotent matrices}
If $\Omega$ is a projector, i.e.\ $\Omega^2=\Omega$, then for scalars $s,t \in \mathbb{R}$ one has the identity
\begin{equation}
   f(tI+s\Omega) = (I-\Omega)\,f(s) + \Omega\, f(t+s),
\end{equation}
valid for any scalar function $f$ for which the right-hand side is defined.\footnote{See, e.g., R.~Horn and C.~Johnson, \emph{Matrix Analysis}, Cambridge University Press (1985).}  
This makes functions of projectors particularly tractable and is often used in applications involving spectral decompositions.

\paragraph{Example.}  
Consider the inverse function $f(x)=1/x$ applied to $I-s\Omega$, with $\Omega^2=\Omega$. Using the above rule, we find
\begin{equation}
(I-s\Omega)^{-1} = (I-\Omega)\, f(-s) + \Omega\, f(1-s).
\end{equation}
Since $f(x)=1/x$, this gives
\begin{equation}
(I-s\Omega)^{-1} = (I-\Omega)\cdot 1 + \Omega \cdot \frac{1}{1-s},
\end{equation}
or equivalently,
\begin{equation}
(I-s\Omega)^{-1} = I + \frac{s}{1-s}\,\Omega.
\end{equation}
Thus the inverse is simply the identity plus a scaled projector, showing how functions of projectors reduce to elementary expressions.

---

\subsection{Drazin inverse}
\index{key}{Drazin inverse}

For singular matrices, the inverse does not exist.  
The \emph{Drazin inverse} $A^D$ generalizes the inverse to matrices of nonzero index.  
If $\nu(\lambda)$ denotes the index of $A-\lambda I$, then Drazin proved uniqueness and gave the representation
\begin{equation}
   A^D=\sum_{\lambda(A)\neq 0}P_{\lambda}
   \sum_{k=0}^{\nu(\lambda)-1}\frac{(-1)^k}{\lambda^{k+1}}(A-\lambda I)^k,
\end{equation}
where $P_\lambda$ is the spectral projector onto the eigenspace of $\lambda$.  
This corresponds to the matrix function $f(x)=x^{-1}$, 
but restricted to the nonzero part of the spectrum.
\subsection{Isospectrality}\label{sec:isospectrality}

Two matrices $A$ and $B$ are \emph{isospectral} if they have the same multiset of eigenvalues.  
A sufficient (and in finite dimensions, necessary) condition is \emph{similarity}:
\[
   A = C\,B\,C^{-1},
\]
for some invertible $C$. In this case, $p_A(\lambda)=p_B(\lambda)$ and $\Lambda(A)=\Lambda(B)$.%
\footnote{Isospectrality also arises from isospectral flows $\dot A=[A,\Omega(A)]$ (Lax theory), which conserve the spectrum while evolving $A$.}

\paragraph{Symmetrizing a graph Markov operator via isospectrality.}
Let $A$ be the (weighted) adjacency of an undirected graph and $D=\mathrm{diag}(d_i)$ the degree matrix with $d_i=\sum_j A_{ij}$.  
The row-stochastic Markov operator is $P=D^{-1}A$.  
Although $P$ is generally non-symmetric, it is \emph{similar} to the symmetric matrix
\[
   \tilde P \;=\; D^{-1/2} A D^{-1/2}.
\]
Indeed, with $C=D^{1/2}$ we have $C^{-1} P\, C = D^{-1/2}(D^{-1}A)D^{1/2} = D^{-1/2} A D^{-1/2} = \tilde P$.  
Hence $P$ and $\tilde P$ are isospectral, and $\tilde P$ inherits symmetry (and real spectrum) when $A$ is symmetric.  
This similarity is the standard “symmetrization” used for random walks and normalized Laplacians.

\paragraph{Isospectrality of singular values (two-sided orthogonal/unitary equivalence).}
If two real (resp.\ complex) matrices $A,B\in\mathbb{F}^{m\times n}$ have the same singular values (with multiplicity), then there exist orthogonal (resp.\ unitary) matrices $O_1,O_2$ such that
\[
   A \;=\; O_1\, B\, O_2.
\]
\emph{Proof sketch.} Take SVDs $A=U_A \Sigma V_A^{\!*}$ and $B=U_B \Sigma V_B^{\!*}$ with the \emph{same} diagonal $\Sigma$. Then
$A = (U_A U_B^{\!*})\, B\, (V_B^{\!*} V_A) = O_1\, B\, O_2$, where $O_1,O_2$ are orthogonal/unitary.

\subsection{Cramer's rule for linear systems}\label{sec:matrix_functions_cramer}
We have briefly mentioned this rule earlier in this book.
\index{key}{Cramer's rule}

\textbf{Theorem (Cramer's rule).} Let $A\vec x=\vec b$ with $A\in\mathbb{F}^{n\times n}$ invertible.  
For $k=1,\dots,n$, let $A_k$ be $A$ with its $k$th column replaced by $\vec b$. Then
\[
   x_k \;=\; \frac{\det(A_k)}{\det(A)}.
\]

\paragraph{Consequences and a geometric corollary.}
Let $\vec a_i$ denote the columns of $A$.  
Fix an index set $\mathcal{J}=\{j_1,\dots,j_k\}$ with $k<n$ and write $J=\mathrm{Span}\{\vec a_{j}: j\in\mathcal{J}\}$, \ 
$\overline{\mathcal{J}}=\{1,\dots,n\}\setminus \mathcal{J}$, \ 
$\overline J=\mathrm{Span}\{\vec a_{j}: j\in\overline{\mathcal{J}}\}$.  
Since $\det(A)\neq 0$, the columns are independent and $\mathbb{F}^n=J\oplus \overline J$.

If $\vec b\in J$, then for any $t\in\overline{\mathcal{J}}$, the matrix $A_t$ (replacing the $t$th column by $\vec b$) has two columns in the same subspace $J$, hence its columns are linearly dependent and $\det(A_t)=0$.  
By Cramer’s rule, $x_t=0$ for $t\in\overline{\mathcal{J}}$, i.e. the solution $\vec x$ has support contained in $\mathcal{J}$.  
In particular, if $\overline J=\{0\}$ (i.e.\ $J=\mathbb{F}^n$), then generically all $x_k\neq 0$.

\paragraph{Expression via eigenvalues (determinantal viewpoint).}
Since $\det(A)$ is the product of the eigenvalues of $A$ (counted with multiplicity in an algebraic closure), Cramer’s rule can be written as
\[
   x_k \;=\; \frac{\det(A_k)}{\det(A)} \;=\; 
   \frac{\prod_{i=1}^n \lambda_i^{(k)}}{\prod_{i=1}^n \lambda_i},
\]
where $\{\lambda_i\}$ are the eigenvalues of $A$ and $\{\lambda_i^{(k)}\}$ those of $A_k$.  
This identity is purely determinantal (no spectral assumptions beyond those implicit in $\det$); it is correct provided determinants are taken over the same field and eigenvalues are counted with multiplicity.\footnote{Over $\mathbb{R}$ one may need to pass to $\mathbb{C}$ to factor polynomials; the determinant equality remains valid.}
\subsection{Minimax theorem}
\index{key}{minimax theorem}

The \emph{minimax theorem} provides one of the most powerful 
characterizations of eigenvalues of Hermitian (or symmetric) matrices.  
It links spectral properties directly to optimization principles.  

\paragraph{Statement.}
Let $A\in\mathbb{R}^{n\times n}$ be symmetric with ordered eigenvalues
\[
   \lambda_1 \leq \lambda_2 \leq \cdots \leq \lambda_n.
\]
Then, for $k=1,\ldots,n$, the $k$th eigenvalue can be written as
\begin{equation}
   \lambda_k \;=\; 
   \min_{\dim S=k}\ \max_{\substack{\vec x\in S\\ \vec x\neq 0}} 
   \frac{\vec x^{\top} A \vec x}{\vec x^{\top} \vec x}
   \;=\;
   \max_{\dim S=n-k+1}\ \min_{\substack{\vec x\in S\\ \vec x\neq 0}} 
   \frac{\vec x^{\top} A \vec x}{\vec x^{\top} \vec x}.
   \label{eq:minimax}
\end{equation}
This result is also known as the \emph{Courant--Fischer theorem}.\footnote{
R.~Courant and D.~Hilbert, \emph{Methoden der mathematischen Physik}, Springer (1931).}

\paragraph{Interpretation.}
The theorem states that each eigenvalue arises as the optimum of a 
nested variational problem involving the \emph{Rayleigh quotient}
\[
   R_A(\vec x)=\frac{\vec x^{\top} A \vec x}{\vec x^{\top} \vec x}.
\]
In particular:
- $\lambda_1=\min_{\vec x\neq 0} R_A(\vec x)$, i.e. the ground state minimizes the Rayleigh quotient.
- $\lambda_n=\max_{\vec x\neq 0} R_A(\vec x)$.
- Intermediate eigenvalues are obtained by restricting to $k$-dimensional subspaces.

\paragraph{Spectral relevance.}
This variational view is central to spectral methods.  
It underlies the convergence of iterative algorithms (power method, Lanczos),  
the derivation of eigenvalue bounds (e.g. interlacing),  
and applications to graphs (Cheeger inequalities, spectral clustering).  
It also establishes a direct bridge between optimization and spectral theory: 
the eigenproblem can be formulated as a constrained extremization problem.

\subsection{Fredholm alternatives}\label{sec:fredholm}
\index{key}{Fredholm alternative}

One of the fundamental tools in linear algebra and operator theory 
is the \emph{Fredholm alternative}.  
It characterizes the solvability of systems of linear equations 
and has direct implications for eigenvalue problems.

\paragraph{Finite-dimensional version.}
Let $A$ be a square matrix and $\lambda\in\mathbb{C}$, $\lambda\neq 0$.  
Exactly one of the following holds:
\begin{enumerate}
   \item The homogeneous system $(A-\lambda I)\vec x=0$ has a nontrivial solution $\vec x\neq 0$ 
   (i.e.\ $\lambda$ is an eigenvalue of $A$), or
   \item The inhomogeneous system $(A-\lambda I)\vec v=\vec d$ has a unique solution $\vec v$ 
   for every $\vec d\in\mathbb{C}^n$.
\end{enumerate}
In the second case, the solution $\vec v$ depends continuously on $\vec d$.  
Equivalently, the resolvent operator $(A-\lambda I)^{-1}$ exists and is bounded 
precisely when $\lambda\notin\Lambda(A)$, the spectrum of $A$.

\paragraph{Inequality form.}
The Fredholm alternative also admits a version for inequalities.  
For a matrix $A$ and a vector $\vec b$, exactly one of the following is true:
\begin{enumerate}
   \item The inequality $A\vec x \leq \vec b$ has a solution $\vec x$, or
   \item There exists a vector $\vec z\geq 0$ such that $A^{\top} \vec z=0$ and $\vec b\cdot \vec z < 0$.
\end{enumerate}
These two cases are mutually exclusive.  

\paragraph{Farkas’s lemma.}
A closely related result is Farkas’s lemma, fundamental in linear programming.\index{key}{Farkas's lemma}
\begin{theorem}[Farkas]
For a matrix $A$ and vector $\vec b$, exactly one of the following holds:
\begin{enumerate}
   \item There exists $\vec x\geq 0$ such that $A\vec x=\vec b$, or
   \item There exists $\vec y\geq 0$ such that $A^{\top} \vec y\geq 0$ and $\vec b\cdot \vec y < 0$.
\end{enumerate}
\end{theorem}
This lemma provides the basis for duality in optimization.\footnote{See R.~Rockafellar, \emph{Convex Analysis}, Princeton University Press (1970).}

\begin{figure}[h]
\centering
\includegraphics[scale=0.5]{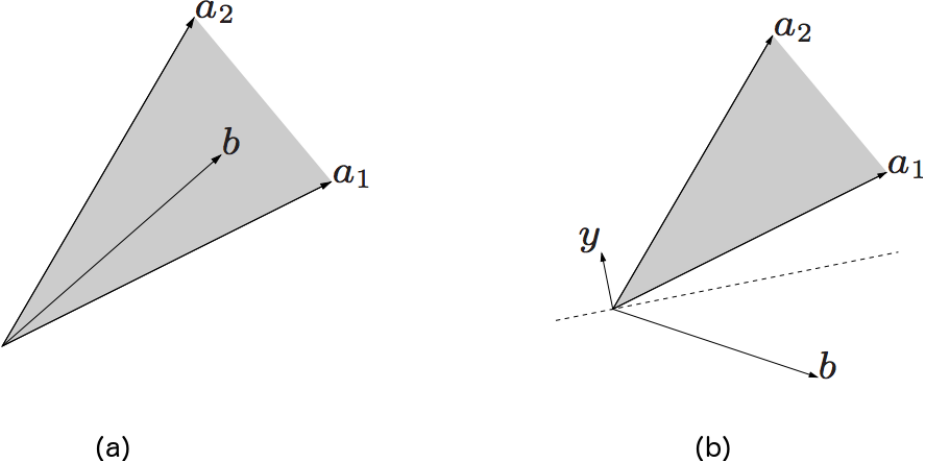}
\caption{Geometric representation of Farkas’s lemma: either (a) a feasible nonnegative solution exists, or (b) a separating hyperplane exists.}
\label{fig:farka}
\end{figure}

---




\section{Characteristic polynomials}
\index{key}{characteristic polynomial}

In the previous chapters we have emphasized how spectral methods allow us to
study the behavior of matrices through their eigenvalues and eigenvectors.
A central object which encodes this spectral information is the
\emph{characteristic polynomial} of a matrix $A$:
\begin{equation}
   p_A(\lambda)=\det(A-\lambda I).
\end{equation}
Its roots $\{\lambda_i\}$ are precisely the eigenvalues of $A$.
Thus, the study of the location and properties of the roots of $p_A(\lambda)$
is essential in understanding stability, asymptotics, and perturbations
of linear systems.

\paragraph{Spectral significance.}
As an example, recall the exponential
\[
   e^{At},
\]
which governs the solution of linear differential equations $\dot x=Ax$.
For large $t$, its growth or decay depends entirely on the spectral values
$\lambda_i$ of $A$:
if $\mathrm{Re}(\lambda_i)<0$ for all $i$, then $e^{At}$ decays,
while eigenvalues with positive real part produce exponential growth.
Thus, stability analysis reduces to the localization of the roots of
$p_A(\lambda)$.

For special classes of matrices, the spectrum is constrained:
\begin{itemize}
   \item Orthogonal and unitary matrices have eigenvalues on the unit circle.
   \item Hermitian matrices have real eigenvalues.
   \item Projection operators have eigenvalues only in $\{0,1\}$.
\end{itemize}
For general matrices, eigenvalues may lie anywhere in the complex plane,
but a rich set of inequalities and inclusion theorems allow one to bound
their location.

---

\subsection{Analytic dependence of eigenvalues}
As discussed in the section on perturbation theory,
simple (non-repeated) roots of polynomials are analytic functions
of the coefficients.  
Consequently, if $A$ has a simple eigenvalue $\lambda(\epsilon)$ depending
on a parameter $\epsilon$, then $\lambda(\epsilon)$ varies analytically
with $\epsilon$.  
However, when eigenvalues have multiplicity greater than one, this property
fails: non-simple eigenvalues can bifurcate and display non-analytic dependence.

---

\subsection{Properties of characteristic roots}

Several classical results give precise information about the distribution
of eigenvalues.

\paragraph{Hermitian matrices.}
If $A$ is Hermitian with eigenvalues
$\lambda_1\geq \lambda_2 \geq \cdots \geq \lambda_n$
and $\{\vec x_i\}$ an orthonormal basis,
then inequalities involving partial sums of the $\lambda_i$ 
bound quadratic forms $\langle A\vec x_j,\vec x_j\rangle$.
These form the starting point of majorization results
(Courant--Fischer, Lidskii, Ky Fan inequalities).%
\footnote{See Horn and Johnson, \emph{Matrix Analysis}, Cambridge Univ. Press (1985).}

\subsubsection{Hermitian interlacing}\index{key}{interlacing}
An important phenomenon is \emph{eigenvalue interlacing}.
If $A$ and $B$ are Hermitian, then by Weyl’s theorem the eigenvalues of
$C=A+B$ satisfy inequalities of the form
\[
   \alpha_j+\beta_{i-j+1} \;\leq\; \gamma_i \;\leq\; \alpha_j+\beta_{i-j+n},
\]
where $\{\alpha\},\{\beta\},\{\gamma\}$ are the ordered eigenvalues of
$A,B,C$.  
Similarly, the eigenvalues of a principal submatrix interlace those of $A$.
This principle is central in graph spectral theory (e.g. Laplacians of induced subgraphs).

\subsubsection{Spectral spread}
The \emph{spectral spread} of a matrix is
\begin{equation}
   s(A)=\max_{i,j}|\lambda_i-\lambda_j|.
\end{equation}
For Hermitian matrices, $s(A)=\lambda_{\max}-\lambda_{\min}$.
In general one has the bound
\[
   s(A)\leq \sqrt{2\|A\|^2-\tfrac{2}{n}|\mathrm{Tr}(A)|^2}
   \;\leq\; \sqrt{2}\,\|A\|,
\]
showing that the spread is controlled by the operator norm.

\subsubsection{Singular values and spectra}
Every matrix $M$ admits a singular value decomposition
$M=U\Sigma V^{\top}$, with $\Sigma=\mathrm{diag}(\sigma_1,\dots,\sigma_n)$,
$\sigma_1\geq\cdots\geq\sigma_n\geq 0$.
Singular values provide universal bounds on eigenvalues:
\begin{align*}
   \prod_i |\lambda_i(A)| &\leq \prod_i \sigma_i(A), \\
   \sum_i |\lambda_i(A)|^p &\leq \sum_i \sigma_i(A)^p, \\
   \lim_{k\to\infty} \sigma_i(A^k)^{1/k} &= |\lambda_i(A)|.
\end{align*}
In particular, $\rho(A)\leq \|A\|$ for any operator norm, with equality
if $A$ is normal (unitarily diagonalizable).

\subsubsection{Diagonally dominant matrices}
\label{sec:diagonallydom}
A classical sufficient condition for invertibility is strict diagonal dominance.
Define
\[
   \xi_i(A)=\sum_{j\neq i}|A_{ij}|.
\]
If $|A_{ii}|>\xi_i(A)$ for all $i$, then $A$ is nonsingular
(Levy--Desplanques theorem).  
More generally, Cassini oval conditions allow invertibility criteria
involving pairs of diagonal entries.

\subsubsection{Ostrowski’s theorem}\index{key}{Ostrowski's theorem}
Ostrowski provided a generalization: defining
$R_i=\sum_j |A_{ij}|$ and $C_i=\sum_j |A_{ji}|$, then if
\[
   |A_{ii}|> R_i^\alpha C_i^{1-\alpha},\qquad 0\leq\alpha\leq 1,
\]
for all $i$, the matrix is invertible.
Further generalizations involve pairs of indices and yield Cassini-type regions
for eigenvalue inclusion.

\subsubsection{Gershgorin domains}\label{sec:gerschgorin}
\index{key}{Gershgorin disks}
One of the most useful localization results is the Gershgorin circle theorem:
if $r_i=\sum_{j\neq i}|A_{ij}|$, then every eigenvalue $\lambda$ of $A$ lies within
at least one disk
\[
   G_i=\{z\in\mathbb{C}:\ |z-A_{ii}|\leq r_i\}.
\]
Thus the spectrum lies in the union $\bigcup_i G_i$.  
This provides immediate bounds on eigenvalues from matrix entries
and is frequently applied in numerical analysis and graph Laplacians.

\begin{tcolorbox}
\begin{verbatim}
% gerschdisc.m
% 
% This function plots the Gershgorin Discs for the matrix
% A passed as an argument.
% It will also plot the centers of such discs, 
% and the actual eigenvalues
% of the matrix.

function gerschdisc(A)

error(nargchk(nargin,1,1));
if size(A,1) ~= size(A,2)
    error('Matrix should be square');
    return;
end

% For each row, we say:
for i=1:size(A,1)
    % The circle has center in 
    % (h,k) where h is the real part of A(i,i) and
    % k is the imaginary part of A(i,i) :
    h=real(A(i,i)); k=imag(A(i,i)); 
    

    % Now we try to compute the radius of the circle, 
    % which is nothing more
    % than the sum of norm of the elements in the row 
    % where i != j
    r=0;
    for j=1:size(A,1)
       if i ~= j 
           r=r+(abs(A(i,j)));
       end    
    end 
    
    % We try to make a vector of points for the circle:
    N=256;
    t=(0:N)*2*pi/N;
    
    % Now we're able to map each of the elements of 
    % this vector into a
    % circle:
    plot( r*cos(t)+h, r*sin(t)+k ,'-');

    % We also plot the center of the circle 
    % for better undesrtanding:
    hold on;
    plot( h, k,'+');
end

% For the circles to be better graphed, we would like 
% to have equal axis:
axis equal;

% Now we plot the actual eigenvalues of the matrix:
ev=eig(full(A));
for i=1:size(ev)
    rev=plot(real(ev(i)),imag(ev(i)),'ro');
end
hold off;
legend(rev,'Actual Eigenvalues');

end
\end{verbatim}
\end{tcolorbox}
\subsection{Spectral inclusion theorems}

In the previous section we introduced Gershgorin’s theorem as a first tool to 
localize the spectrum of a matrix using only its entries.  
Let us illustrate with a concrete example.

\begin{figure}[h]
\centering
\includegraphics[scale=0.3]{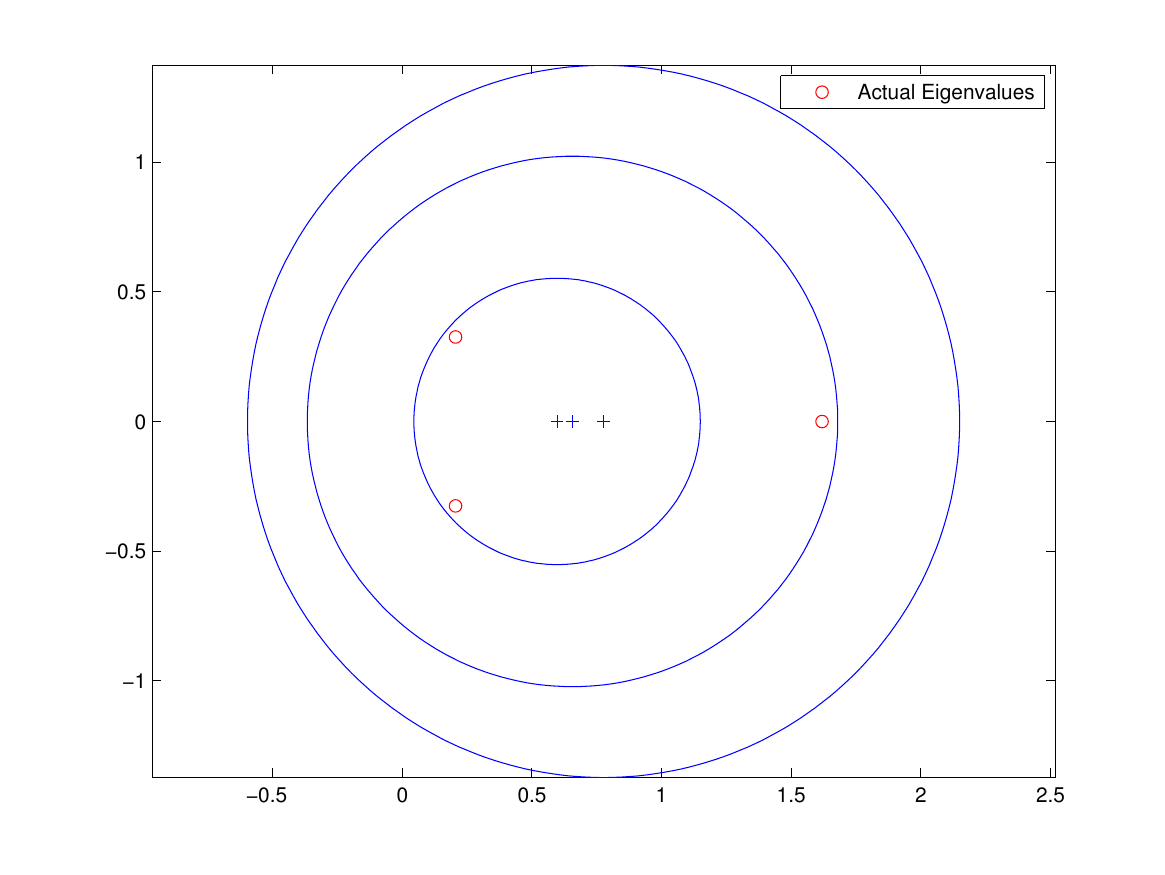}
\caption{The Gershgorin domains of the matrix in eqn.~(\ref{eq:randmat}).
Blue circles: Gershgorin disks. Red dots: eigenvalues of $A$.}
\label{fig:gershgorin}
\end{figure}

\paragraph{Example.}
Consider the random $3\times 3$ matrix
\begin{equation}
A=\begin{pmatrix}
0.5972  &  0.3912 &   0.1610 \\
0.4185  &  0.7764  &  0.9546 \\
0.7417  &  0.2811  &  0.6567
\end{pmatrix}.
\label{eq:randmat}
\end{equation}
The Gershgorin disks are centered at the diagonal entries with radii equal 
to the sum of off-diagonal magnitudes in each row.  
In this case, the disk corresponding to the second row, centered at $0.7764$ 
with radius $1.373$, contains the others, so all eigenvalues of $A$ must lie within it 
(Fig.~\ref{fig:gershgorin}).

---

\paragraph{Brauer--Cassini ovals.}
Gershgorin’s theorem can be sharpened using the \emph{ovals of Cassini}.  
For $i\neq j$, define
\begin{equation}
   K_{ij}=\Big\{z\in\mathbb{C}:\ |z-A_{ii}||z-A_{jj}|\leq \xi_i(A)\,\xi_j(A)\Big\},
\end{equation}
where $\xi_i(A)$ is the sum of absolute values of the off-diagonal entries of row $i$.  
The union $K(A)=\bigcup_{i\neq j}K_{ij}$ yields tighter inclusion sets.

\begin{theorem}[Brauer]
For any $A\in\mathbb{C}^{n\times n}$, $\Lambda(A)\subseteq K(A)$.
\end{theorem}

In general $G(A)\subseteq K(A)$, so Cassini ovals improve Gershgorin’s disks.

---

\paragraph{Irreducible matrices.}
More can be said in the case of irreducible matrices.  
Recall that $A$ is irreducible if no permutation can transform it into 
block upper-triangular form.

\begin{theorem}
Let $A$ be irreducible.  
If an eigenvalue $\lambda$ lies on the boundary of the union of Gershgorin disks, 
then $\lambda$ lies on the boundary of each individual Gershgorin disk.
\end{theorem}

\emph{Proof (sketch).}  
The proof refines the standard Gershgorin argument.  
If $\lambda$ satisfies
\[
   |\lambda-A_{ii}|\leq \sum_{j\neq i}|A_{ij}||x_j|,
\]
with equality for some component of the eigenvector $\vec x$, then irreducibility 
implies that equality must propagate along all nonzero entries of $A$.  
Iterating yields that equality holds for every row, hence $\lambda$ lies on the 
boundary of every disk. \qed

---

\paragraph{Diagonal dominance and nonsingularity.}
From Gershgorin’s theorem it follows that a strictly diagonally dominant matrix
\[
   |A_{ii}|>\sum_{j\neq i}|A_{ij}|\quad \forall i,
\]
cannot have $0$ in any Gershgorin disk, and is therefore nonsingular.  
For weak diagonal dominance ($\geq$), the result may fail.  
However, if $A$ is irreducible and at least one row is strictly diagonally dominant, 
then $A$ is still nonsingular (\emph{irreducible diagonal dominance}).  
If $0$ were an eigenvalue, it would lie on the boundary of every Gershgorin disk, 
contradicting the strict inequality for that row.

---

\subsubsection{Fan’s theorem}\index{key}{Fan's eigenvalue bound}
Bounds can also be derived by comparing with a positive matrix.  
If $|A_{ij}|\leq B_{ij}$ entrywise and $\mu$ is the maximum eigenvalue of $B$, then
\begin{equation}
   |\lambda-A_{ii}|\leq \mu-B_{ii},
\end{equation}
for every eigenvalue $\lambda$ of $A$.  
This result, due to Fan, provides eigenvalue bounds using comparison matrices.

---

\subsubsection{Schur’s inequality}\index{key}{Schur's theorem}
For any $A\in\mathbb{C}^{n\times n}$ with eigenvalues $\{\lambda_i\}$,
\begin{equation}
   \sum_{i=1}^n|\lambda_i|^2 \;\leq\; \sum_{i,j=1}^n |A_{ij}|^2 = \|A\|_F^2,
\end{equation}
where $\|\cdot\|_F$ is the Frobenius norm.  
This follows from the Schur decomposition $A=UTU^{-1}$ with $T$ upper triangular,
since $\|A\|_F^2=\|T\|_F^2=\sum_i|\lambda_i|^2+\sum_{i\neq j}|T_{ij}|^2$.

---

\subsubsection{Bendixson’s theorem}\index{key}{Bendixson's theorem}
For further control of eigenvalue locations, one can bound their real and imaginary parts.  
Let
\[
   B=\tfrac{1}{2}(A+A^\dagger),\quad 
   C=\tfrac{1}{2}(A-A^\dagger),
\]
so that $B$ is Hermitian and $C$ skew-Hermitian.  
If $\{\lambda_i\}$ are the eigenvalues of $A$, then
\begin{align}
   \mu_n \leq \mathrm{Re}(\lambda_i) \leq \mu_1, \\
   |\mathrm{Im}(\lambda_i)|\leq c \sqrt{\tfrac{n(n-1)}{2}},
\end{align}
where $\mu_1,\mu_n$ are the extreme eigenvalues of $B$, and 
$c=\max_{i,j}|C_{ij}|$.  
These inequalities constrain the spectrum to a vertical strip in the complex plane.
\subsubsection{Hirsch’s theorem}\index{key}{Hirsch's theorem}
Bendixson’s bounds on the real and imaginary parts of eigenvalues 
can be generalized to provide uniform control of their magnitude.
Let $a=\max_{i,j}|A_{ij}|$, 
$b=\max_{i,j}|(A+A^\dagger)_{ij}|/2$, 
$c=\max_{i,j}|(A-A^\dagger)_{ij}|/2$.  
Then, for any eigenvalue $\lambda_i$ of $A$,
\begin{eqnarray}
   |\lambda_i| &\leq& n a, \\
   \mathrm{Re}(\lambda_i) &\leq& n b, \\
   \mathrm{Im}(\lambda_i) &\leq& n c.
\end{eqnarray}
This extends Bendixson’s theorem to global spectral bounds.

---

\subsubsection{Browne’s theorem}\index{key}{Browne's theorem}
Eigenvalue magnitudes can also be bounded in terms of singular values.
Let $\sigma_1\geq\sigma_2\geq\cdots\geq\sigma_n\geq 0$ be the singular values of $A$.
Then for every eigenvalue $\lambda_i(A)$,
\begin{equation}
   \sigma_n \;\leq\; |\lambda_i(A)| \;\leq\; \sigma_1.
\end{equation}
Thus, the spectrum is contained in the annulus determined by the largest and 
smallest singular values.

---

\subsection{Additive perturbations of eigenvalues}
\index{key}{Eigenvalues Perturbations (Bounds)}

Perturbation theory is crucial for understanding the stability of spectra
under small changes.  
We now collect two classical results for Hermitian matrices.

\paragraph{Wielandt--Hoffman theorem.}
If $A$ and $A+E$ are Hermitian $n\times n$ matrices with eigenvalues
$\{\lambda_i(A)\}$ and $\{\lambda_i(A+E)\}$ (ordered increasingly), then
\begin{equation}
   \sum_{i=1}^n \big(\lambda_i(A+E)-\lambda_i(A)\big)^2 \;\leq\; \|E\|_F^2.
\end{equation}
In particular, eigenvalues are Lipschitz-continuous with respect to the Frobenius norm.

Another classical bound states that
\begin{equation}
   \lambda_n(E)\;\leq\; \lambda_k(A+E)-\lambda_k(A) \;\leq\; \lambda_1(E),
\end{equation}
providing entrywise control from the extreme eigenvalues of the perturbation $E$.

\paragraph{Sketch of proof.}
For Hermitian $A,B$, one can show
\begin{equation}
   \frac{1}{n}\sum_{k=1}^n |\lambda_k(A)-\lambda_k(B)|^2 \;\leq\; \|A-B\|^2,
\end{equation}
where $\|\cdot\|$ is any unitarily invariant norm.  
Using unitary diagonalizations $A=U\Lambda U^\ast$, $B=W\Gamma W^\ast$, 
and setting $Q=U^\ast W$, one rewrites
\[
   \|A-B\|^2=\frac{1}{n}\sum_{i,j} |\lambda_i(A)-\lambda_j(B)|^2 |q_{ij}|^2.
\]
Since $\{p_{ij}=|q_{ij}|^2/n\}$ forms a doubly stochastic matrix, 
the expression is a convex combination over permutations 
(Birkhoff–von Neumann theorem).  
A convexity argument then shows that the minimum is attained at the identity permutation, 
yielding the stated bound.\footnote{See Wilkinson, \emph{The Algebraic Eigenvalue Problem}, Oxford (1965).}

---

\subsection{Multiplicative perturbations}
\index{key}{Eigenvalues Perturbations (Multiplicative)}

So far we have focused on additive perturbations.  
In applications such as control theory, multiplicative perturbations are equally important.  
Consider stability of $(I-M)^{-1}$ compared to $(I-M\Delta)^{-1}$, 
where $\Delta$ is a structured perturbation.

The \emph{small-gain theorem} provides the relevant criterion.  
Define
\[
   Q_1 = \Delta (I-M\Delta)^{-1},\qquad
   Q_2 = (I-M\Delta)^{-1}.
\]
If $M$ is stable, i.e.\ $\|M\|_\infty<1$, then both $Q_1$ and $Q_2$ remain stable 
whenever $\|\Delta\|_\infty<1$.  
In other words, stability is preserved under multiplicative perturbations 
of sufficiently small norm.

---

Gershgorin disks, Cassini ovals, irreducible dominance, and the inequalities of Fan, Schur, and Bendixson 
form a toolbox for bounding spectra using only matrix entries or norms.  
These results complement the variational (minimax) and perturbative approaches:
while the latter describe eigenvalues through optimization and stability,
spectral inclusion theorems provide explicit geometric regions in the complex plane
containing the spectrum of $A$.

\subsection{Special matrices}
\index{key}{special matrices}

In many areas of applied mathematics and physics, certain classes of matrices arise repeatedly because
their algebraic structure encodes specific properties of the systems they describe. These \emph{special
matrices} often have restricted spectra, stability properties, or symmetry constraints that simplify
analysis. Examples include lattice matrices (Sec.~\ref{eq:latticegreenfun}), stability matrices in
dynamical systems, stochastic and Markov matrices, Laplacian matrices in graph theory, and many others.

We now discuss a few of the most relevant classes.

---

\subsubsection{Stability matrices}
\index{key}{stability matrices}

A \emph{stability matrix} (also called \emph{Jacobian matrix} in dynamical systems) arises in the
linearization of nonlinear dynamical equations.  
Consider a dynamical system
\[
\dot{\vec x}=F(\vec x), \qquad \vec x\in\mathbb{R}^n.
\]
At a fixed point $\vec x^\ast$ with $F(\vec x^\ast)=0$, the linearization is
\[
\dot{\vec y}=A \vec y,\qquad A=\left.\frac{\partial F}{\partial \vec x}\right|_{\vec x=\vec x^\ast}.
\]
The matrix $A$ is called the \emph{stability matrix} at $\vec x^\ast$.

\paragraph{Spectral properties.}
The eigenvalues of $A$ determine the local stability of the fixed point:
\begin{itemize}
\item If $\mathrm{Re}(\lambda_i)<0$ for all $i$, the fixed point is asymptotically stable
(solutions decay to $\vec x^\ast$).
\item If $\mathrm{Re}(\lambda_i)>0$ for some $i$, the fixed point is unstable.
\item If $\mathrm{Re}(\lambda_i)=0$ for some $i$, the system is marginal and higher-order terms
determine stability.
\end{itemize}
Thus, the spectrum $\Lambda(A)$ plays the role of Lyapunov exponents in continuous-time dynamics
(Sec.~\ref{sec:rayleigh}).

\paragraph{Connections with norms.}
From the definition of the spectral radius $\rho(A)$ (Sec.~\ref{sec:matrixnorm}), one has
\[
\max_i \mathrm{Re}(\lambda_i) \leq \rho(A) \leq \|A\|,
\]
so matrix norms provide useful bounds on growth rates. This is consistent with the Rayleigh principle
and variational characterizations of eigenvalues.

\paragraph{Examples.}
\begin{itemize}
\item For $A=-I$, all eigenvalues are $-1$, hence the origin is a globally stable fixed point.
\item For the rotation matrix
$R(\theta)=\begin{pmatrix}0 & -\theta\\ \theta & 0\end{pmatrix}$,
eigenvalues are purely imaginary $\pm i\theta$, corresponding to neutral (center) stability.
\item For upper-triangular matrices, stability is determined directly from the diagonal entries
(Sec.~\ref{sec:jordanform}).
\end{itemize}

\paragraph{Applications.}
Stability matrices appear in:
\begin{itemize}
\item control theory (linear system stability),
\item statistical mechanics (Hessian of free energy landscapes),
\item optimization (Hessian matrix of cost functions),
\item population dynamics and epidemiology (Jacobian of growth equations).
\end{itemize}

---

\paragraph{Remark.}
The study of stability matrices connects spectral analysis of matrices with the qualitative behavior of
nonlinear systems. Later chapters (e.g.\ on dynamical systems and networks) will revisit these ideas in
greater depth.

\subsubsection{Stochastic matrices}
\index{key}{stochastic matrices}

A \emph{stochastic matrix} is a square matrix $P\in\mathbb{R}^{n\times n}$ whose entries are
nonnegative and whose rows sum to one:
\begin{equation}
P_{ij}\geq 0,\qquad \sum_{j=1}^n P_{ij}=1 \quad \forall i.
\end{equation}
Such matrices naturally describe transitions of discrete-time Markov chains, where $P_{ij}$ is the
probability of moving from state $i$ to state $j$ in one step.  
By construction, multiplying a probability vector $\vec p$ (with nonnegative entries summing to $1$)
by $P$ yields another probability vector, so stochastic matrices preserve the probability simplex.

\paragraph{Variants.}
\begin{itemize}
\item \emph{Row-stochastic} matrices: rows sum to one (the convention used here).
\item \emph{Column-stochastic} matrices: columns sum to one (arise in PageRank and network analysis).
\item \emph{Doubly-stochastic} matrices: both rows and columns sum to one. By Birkhoff’s theorem,
doubly-stochastic matrices are convex combinations of permutation matrices.
\end{itemize}

---

\paragraph{Spectral properties.}
Stochastic matrices satisfy several important eigenvalue constraints:
\begin{itemize}
\item The spectral radius satisfies $\rho(P)=1$.
\item The largest eigenvalue is $\lambda_1=1$, with a nonnegative eigenvector (the stationary
distribution). This follows from the Perron--Frobenius theorem\footnote{See R.~Horn and C.~Johnson,
\emph{Matrix Analysis}, Cambridge University Press (1985).}.
\item All eigenvalues satisfy $|\lambda_i|\leq 1$.
\item If $P$ is irreducible and aperiodic, the Perron--Frobenius eigenvalue $1$ is simple, and the
corresponding eigenvector is the unique stationary distribution.
\item Eigenvalues on the unit circle determine periodicity: e.g.\ if $-1$ is an eigenvalue, the chain
is bipartite and alternates between two classes of states.
\end{itemize}

---

\paragraph{Connections with norms and convexity.}
Since $P$ preserves the $\ell^1$-norm of probability vectors, we have
\[
\|P\vec p\|_1=\|\vec p\|_1
\]
for any probability vector $\vec p$. This reflects conservation of total probability.  
Moreover, convexity implies that if $\vec p$ and $\vec q$ are probability distributions, then
\[
P(\theta \vec p+(1-\theta)\vec q)=\theta P\vec p+(1-\theta)P\vec q,
\]
so stochastic matrices act as linear maps that respect probabilistic mixtures.

---

\paragraph{Examples.}
\begin{itemize}
\item The \emph{identity matrix} $I$ is stochastic: the process remains in the same state.
\item The \emph{uniform random walk} on $n$ states is described by
$P_{ij}=\tfrac{1}{n}$, which is doubly-stochastic with stationary distribution uniform on all states.
\item On a graph $G$, the random walk transition matrix is $P_{ij}=A_{ij}/d_i$, where $A$ is the adjacency
matrix and $d_i$ the degree of vertex $i$. This links stochastic matrices to graph Laplacians.
\end{itemize}

---

\paragraph{Applications.}
Stochastic matrices appear throughout probability theory, physics, and data science:
\begin{itemize}
\item Markov chains and random walks on graphs,
\item ergodic theory and mixing times,
\item PageRank algorithm (column-stochastic),
\item statistical physics (transfer matrices, Glauber dynamics),
\item population dynamics and evolutionary models,
\item queuing theory and stochastic processes.
\end{itemize}

---

Stochastic matrices form a convex set: convex combinations of stochastic matrices are stochastic.
They thus encode mixtures of random processes. Their spectral analysis via Perron–Frobenius
theory provides the fundamental link between algebra and probability: the long-term behavior of a
Markov chain is determined by the leading eigenvector of its transition matrix.
\subsubsection{Stochastic matrices}
\index{key}{stochastic matrices}

A \emph{stochastic matrix} is a square matrix $P\in\mathbb{R}^{n\times n}$ whose entries are
nonnegative and whose rows sum to one:
\begin{equation}
P_{ij}\geq 0,\qquad \sum_{j=1}^n P_{ij}=1 \quad \forall i.
\end{equation}
Such matrices naturally describe transitions of discrete-time Markov chains, where $P_{ij}$ is the
probability of moving from state $i$ to state $j$ in one step.  
By construction, multiplying a probability vector $\vec p$ (with nonnegative entries summing to $1$)
by $P$ yields another probability vector, so stochastic matrices preserve the probability simplex.

\paragraph{Variants.}
\begin{itemize}
\item \emph{Row-stochastic} matrices: rows sum to one (the convention used here).
\item \emph{Column-stochastic} matrices: columns sum to one (arise in PageRank and network analysis).
\item \emph{Doubly-stochastic} matrices: both rows and columns sum to one. By Birkhoff’s theorem,
doubly-stochastic matrices are convex combinations of permutation matrices.
\end{itemize}

---

\paragraph{Spectral properties.}
Stochastic matrices satisfy several important eigenvalue constraints:
\begin{itemize}
\item The spectral radius satisfies $\rho(P)=1$.
\item The largest eigenvalue is $\lambda_1=1$, with a nonnegative eigenvector (the stationary
distribution). This follows from the Perron--Frobenius theorem\footnote{See R.~Horn and C.~Johnson,
\emph{Matrix Analysis}, Cambridge University Press (1985).}.
\item All eigenvalues satisfy $|\lambda_i|\leq 1$.
\item If $P$ is irreducible and aperiodic, the Perron--Frobenius eigenvalue $1$ is simple, and the
corresponding eigenvector is the unique stationary distribution.
\item Eigenvalues on the unit circle determine periodicity: e.g.\ if $-1$ is an eigenvalue, the chain
is bipartite and alternates between two classes of states.
\end{itemize}

---

\paragraph{Connections with norms and convexity.}
Since $P$ preserves the $\ell^1$-norm of probability vectors, we have
\[
\|P\vec p\|_1=\|\vec p\|_1
\]
for any probability vector $\vec p$. This reflects conservation of total probability.  
Moreover, convexity implies that if $\vec p$ and $\vec q$ are probability distributions, then
\[
P(\theta \vec p+(1-\theta)\vec q)=\theta P\vec p+(1-\theta)P\vec q,
\]
so stochastic matrices act as linear maps that respect probabilistic mixtures.

---

\paragraph{Examples.}
\begin{itemize}
\item The \emph{identity matrix} $I$ is stochastic: the process remains in the same state.
\item The \emph{uniform random walk} on $n$ states is described by
$P_{ij}=\tfrac{1}{n}$, which is doubly-stochastic with stationary distribution uniform on all states.
\item On a graph $G$, the random walk transition matrix is $P_{ij}=A_{ij}/d_i$, where $A$ is the adjacency
matrix and $d_i$ the degree of vertex $i$. This links stochastic matrices to graph Laplacians.
\end{itemize}

---

\paragraph{Applications.}
Stochastic matrices appear throughout probability theory, physics, and data science: Markov chains and random walks on graphs, ergodic theory and mixing times, PageRank algorithm (column-stochastic),  statistical physics (transfer matrices, Monte Carlo methods),
 population dynamics and evolutionary models, queuing theory and stochastic processes.

---

Stochastic matrices form a convex set: convex combinations of stochastic matrices are stochastic.
They thus encode mixtures of random processes. Their spectral analysis via Perron–Frobenius
theory provides the fundamental link between algebra and probability: the long-term behavior of a
Markov chain is determined by the leading eigenvector of its transition matrix.
\subsubsection{Hermitian matrices}

Hermitian matrices are the complex analogue of real symmetric matrices.
They are exactly the matrices that equal their own adjoint: $A^*=A$. This single symmetry
quietly explains all the pleasant features we use in practice. For instance, the quadratic
form $x^*Ax$ is always real, so $A$ acts like a genuine ``energy'' operator; and the action
of $A$ on orthonormal vectors respects orthogonality after an appropriate change of basis.

\medskip
\textbf{Spectral picture.}  
Every Hermitian matrix admits an orthonormal basis of eigenvectors. Concretely, there
exists a unitary $U$ and a real diagonal matrix $\Lambda$ with $A=U\Lambda U^*$.
The entries of $\Lambda$ are the eigenvalues, and they are \emph{real}. In this basis
the matrix simply scales each coordinate by a real number. As an immediate corollary,
the operator $2$-norm of $A$ is just the largest absolute eigenvalue:
$\|A\|_2=\max_i|\lambda_i|$.

\medskip
\textbf{Quadratic forms and positivity.}  
Because $x^*Ax\in\mathbb{R}$ for all $x$, one can read positivity directly from this form:
\[
A\succeq 0 \quad\Longleftrightarrow\quad x^*Ax\ge 0\ \text{for all }x
\quad\Longleftrightarrow\quad \lambda_i\ge 0\ \text{for all }i.
\]
Positive definite ($\succ 0$) means strict inequality for $x\neq 0$ (equivalently,
all eigenvalues $>0$), and is the regime of Cholesky factorizations $A=R^*R$.
This viewpoint also gives the Löwner order: we write $A\preceq B$ when $B-A\succeq 0$,
i.e.\ when $x^*Ax\le x^*Bx$ for every $x$.

\medskip
\textbf{Variational viewpoint.}  
The Rayleigh quotient $R_A(x)=\frac{x^*Ax}{x^*x}$ ranges exactly between the smallest and
largest eigenvalues:
\[
\lambda_{\min}\ \le\ R_A(x)\ \le\ \lambda_{\max}\qquad(x\neq 0),
\]
and it is maximized/minimized by the corresponding eigenvectors. This gives a practical,
basis-free way to estimate or even compute eigenvalues.

\medskip
\textbf{Functional calculus.}  
Diagonalization allows us to apply real functions to $A$ as if we were applying them to
numbers: $f(A)=Uf(\Lambda)U^*$. In particular $|A|=\sqrt{A^2}$ is Hermitian and $\exp(A)\succ 0$.
When two Hermitian matrices commute, they can be diagonalized by the same unitary, so any
polynomial in them is again Hermitian.

\medskip
\textbf{Examples.}  
The matrix $\begin{bmatrix}2 & i\\ -i & 3\end{bmatrix}$ is Hermitian; its eigenvalues are real.
Over $\mathbb{R}$, Hermitian just means symmetric: $\begin{bmatrix}1&2\\2&1\end{bmatrix}$ is
Hermitian with eigenvalues $3$ and $-1$. In contrast, a non-diagonal upper triangular matrix
(say $\begin{bmatrix}0&1\\0&0\end{bmatrix}$) is not Hermitian and cannot have an orthonormal
eigenbasis.

\medskip
\textbf{Takeaway.}  
Hermitian is the precise algebraic condition ($A^*=A$) that guarantees the geometric and
numerical niceties we want: real spectrum, orthonormal eigenvectors, stable quadratic forms,
and a calculus that treats matrices like real numbers after a unitary change of basis.

\subsubsection{Banded matrices}

Many matrices that arise from discretizing differential equations or fitting local models
(splines, Markov chains, nearest-neighbor couplings) are \emph{banded}: all the action sits
near the diagonal and everything far away is zero. Formally, $A\in\mathbb{C}^{n\times n}$
is banded if there exist integers $\ell,u\ge 0$ such that
\[
A_{ij}=0\quad\text{whenever}\quad i-j>\ell\ \ \text{or}\ \ j-i>u.
\]
We call $\ell$ the lower bandwidth and $u$ the upper bandwidth; the \emph{total bandwidth}
is $\ell+u+1$. In the symmetric case it is common to speak of a single (half-)bandwidth
$b$ with $A_{ij}=0$ for $|i-j|>b$.

\medskip
\textbf{Pictures and examples.}  
Tridiagonal matrices ($\ell=u=1$) are the textbook case; they encode nearest-neighbor
couplings along a line. Pentadiagonal matrices ($\ell=u=2$) appear when a second-order
stencil widens, or when you couple next-nearest neighbors. A small pentadiagonal example:
\[
\begin{bmatrix}
\ast & \ast & \ast &        &        &        &   \\
\ast & \ast & \ast & \ast   &        &        &   \\
\ast & \ast & \ast & \ast   & \ast   &        &   \\
     & \ast & \ast & \ast   & \ast   & \ast   &   \\
     &      & \ast & \ast   & \ast   & \ast   & \ast \\
     &      &      & \ast   & \ast   & \ast   & \ast \\
     &      &      &        & \ast   & \ast   & \ast
\end{bmatrix}
\]
Only entries within two steps of the diagonal are possibly nonzero.

\medskip
\textbf{Why we care (storage and ops).}  
If only $\ell+u+1$ diagonals can be nonzero, we should store only those.
In \emph{compressed diagonal storage} the memory scales like $O\!\left(n(\ell+u+1)\right)$,
rather than $O(n^2)$. The matrix–vector product inherits this structure:
\[
\text{matvec } y=Ax \ \ \text{costs}\ \ O\!\left(n(\ell+u+1)\right),
\]
since each row touches at most $\ell+u+1$ entries. This is the main practical win.

\medskip
\textbf{Solving linear systems.}  
The band structure persists under elimination in important cases. For a tridiagonal
system $Ax=b$ with nonzero sub/superdiagonal, the Thomas algorithm (specialized Gaussian
elimination without pivoting) solves the system in $O(n)$ time. More generally, banded
Gaussian elimination and Cholesky factorization (for Hermitian positive definite $A$)
run in
\[
O\!\left(n(\ell+u)^2\right)\quad\text{time and}\quad O\!\left(n(\ell+u)\right)\ \text{storage}.
\]
In the SPD case no pivoting is needed and the Cholesky factor has the \emph{same} bandwidth.
With partial pivoting (nonsymmetric problems), a small amount of \emph{fill} can grow the
band, so diagonally dominant or well-conditioned problems are preferable if you want to
preserve the structure.

\medskip
\textbf{Algebra of bandwidth.}  
Bandwidth behaves predictably under basic operations. Scaling does nothing. Addition
keeps the larger band: if $A$ has $(\ell_1,u_1)$ and $B$ has $(\ell_2,u_2)$, then
$A+B$ has bandwidth $(\max\{\ell_1,\ell_2\},\max\{u_1,u_2\})$. Products widen bands at most
additively: $AB$ has lower bandwidth $\le \ell_1+\ell_2$ and upper bandwidth $\le u_1+u_2$.
These estimates are tight in typical stencil constructions.

\medskip
\textbf{Graph and ordering viewpoint.}  
The sparsity pattern of a (symmetric) banded matrix is the adjacency of a path graph
after some vertex ordering. Reordering the unknowns can shrink the band (and reduce fill)
without changing the problem itself. Algorithms such as (reverse) Cuthill–McKee exploit
this idea; they are standard preprocessing steps before sparse factorization.

\medskip
\textbf{Block structure.}  
Many applications are \emph{block-banded}: each nonzero along the band is itself a small
$r\times r$ block. All the statements above remain true if you count bandwidth in blocks:
storage and matvec costs gain a factor of $r^2$, and factorizations scale with the block size
accordingly.

\medskip
\textbf{Takeaway.}  
“Banded’’ means locality: interactions decay with distance from the diagonal. That locality
buys you linear (or nearly linear) storage, fast matvecs, and direct solvers whose cost
depends on the band, not on the full dimension. When you can see or enforce a band, you
should—your algorithms and memory footprint will thank you.
\subsubsection{Lattice matrices}
\index{key}{Lattice matrices}

A useful class of structured matrices arises from discrete lattices with Euclidean symmetry, widely used
in statistical mechanics and condensed matter physics\footnote{See C.~Kittel,
\emph{Introduction to Solid State Physics}, Wiley (2005); L.~P.~Kadanoff, \emph{Statistical Physics:
Statics, Dynamics and Renormalization}, World Scientific (2000); G.~Parisi,
\emph{Statistical Field Theory}, Addison--Wesley (1988).}.  

Let $\mathcal{H}_d$ be a $d$-dimensional hypercubic lattice with unit spacing, and associate a node
index $i$ to each lattice site with coordinates
\[
\vec r_i=(n_i^1,\ldots,n_i^d),\qquad n_i^\nu\in\mathbb{N}.
\]
A \emph{lattice matrix} $A_{ij}$ is then defined by
\begin{equation}
A_{ij}=\begin{cases}
\alpha & \text{if } i=j,\\
\beta  & \text{if $i,j$ are nearest neighbors in $\mathcal{H}_d$},\\
0 & \text{otherwise},
\end{cases}
\end{equation}
i.e.\ it contains on the diagonal a constant $\alpha$, and on each edge of the hypercubic lattice a
constant $\beta$. This structure encodes the translational symmetry
\[
A_{\vec r_i,\vec r_j}=A_{\vec r_i+\vec q,\vec r_j+\vec q},
\]
for any lattice translation $\vec q\in\mathbb{Z}^d$. If the lattice spacing is $a\neq 1$, we simply
rescale $\vec r\mapsto a\vec r$.

---

\paragraph{Fourier representation.}
Because of translational invariance, $A$ can be diagonalized in Fourier space. One finds the integral
representation
\begin{equation}
A_{km}=\left(\frac{a}{2\pi}\right)^d \int_{-\pi/a}^{\pi/a} d\vec p\,
A(\vec p)\, e^{i(\vec r_k-\vec r_m)\cdot\vec p},
\end{equation}
with $\vec p$ restricted to the first Brillouin zone
$B=[-\pi/a,\pi/a]^d$\index{key}{Brillouin zone}.  
The Fourier symbol of $A$ is
\begin{equation}
A(\vec p)=\alpha+2\beta\sum_{\nu=1}^d \cos(p_\nu).
\end{equation}
This is consistent with the general spectral decomposition of translation-invariant operators discussed
earlier.

The inverse matrix is obtained by inverting $A(\vec p)$ pointwise:
\begin{eqnarray}
(A^{-1})_{km}&=&\left(\frac{a}{2\pi}\right)^d
\int_{-\pi/a}^{\pi/a} d\vec p\,
\frac{e^{i(\vec r_k-\vec r_m)\cdot\vec p}}
{\alpha+2\beta\sum_{\nu=1}^d \cos(p_\nu)}.
\label{eq:latticegreenfun}
\end{eqnarray}
This integral kernel is called the \emph{lattice Green function}\index{key}{lattice Green function} and
plays a central role in lattice models of statistical physics and random walks\footnote{See J.~Montroll,
``Random walks on lattices,'' \emph{Proc. Symp. Appl. Math.} \textbf{16}, 193–220 (1964);
G.~S.~Joyce, ``On the simple cubic lattice Green function,'' \emph{J.~Phys.~A} \textbf{5}, L65 (1972).}.

---

\paragraph{One-dimensional example.}
Consider a one-dimensional chain with
\[
A_{km}=\beta_1 \delta_{k,m+1}+\beta_2 \delta_{k+1,m}+\alpha \delta_{k,m}.
\]
Using Fourier basis vectors $M_{qk}=e^{iqk}$ (plane waves), one finds that $A$ is diagonal in Fourier
space:
\[
A(q)=\alpha+\beta_1 e^{iq}+\beta_2 e^{-iq}.
\]
If $A$ is symmetric, $\beta_1=\beta_2=\beta$ and
$A(q)=\alpha+2\beta\cos q$.  
If $A$ is skew-symmetric, $\beta_1=-\beta_2$ and $\alpha=0$, giving
$A(q)=2i\beta\sin q$.  
This is consistent with the spectral theorem (Sec.~\ref{sec:eigenvalues}): symmetric matrices have real
eigenvalues, while skew-symmetric ones have purely imaginary eigenvalues occurring in $\pm$ pairs.

---

\paragraph{Continuum and large-dimension limits.}
In the continuum limit $a\to 0$, one expands
\[
A(\vec q)\approx \alpha+2\beta-\beta\|\vec q\|^2+O(\|\vec q\|^4),
\]
so the lattice operator approaches a Laplacian-like form.  

In the large-$d$ limit, integral representations such as
\begin{equation}
\frac{1}{x}=\int_0^\infty dq\, e^{-qx}
\end{equation}
combined with Bessel function identities yield asymptotic estimates of $(A^{-1})_{km}$ in terms of
modified Bessel functions $I_n(x)$, connecting to correlation lengths in statistical mechanics.

---

\subsubsection{Matrices representing bipartite and ordered graphs}
\index{key}{bipartite matrices}
\index{key}{consistently ordered matrices}

The lattice matrices above can be understood as adjacency-like matrices of regular graphs. More generally,
symmetric matrices associated with bipartite or consistently ordered graphs also display distinctive
spectral properties\footnote{See R.~Diestel, \emph{Graph Theory}, Springer (2017).}.

\paragraph{Bipartite matrices.}
If a symmetric matrix corresponds to a bipartite graph, it can be permuted into block form
\[
A'=\begin{pmatrix}0 & B\\ B^\top & 0\end{pmatrix}.
\]
Such matrices always have eigenvalues occurring in $\pm$ pairs: if $\vec v=(\vec x,\vec y)$ is an
eigenvector with eigenvalue $\lambda$, then $(\vec x,-\vec y)$ is an eigenvector with eigenvalue
$-\lambda$.  

\paragraph{Consistently ordered (T-)matrices.}
More generally, matrices associated with linearly ordered partitions of vertices can be permuted into
block tridiagonal form (a T-matrix):
\begin{equation}
T=\begin{pmatrix}
A_1 & B_1 & 0 & \cdots & 0 \\
B_1' & A_2 & B_2 & \ddots & 0 \\
0 & B_2' & A_3 & \ddots & \vdots \\
\vdots & \ddots & \ddots & \ddots & B_{p-1} \\
0 & 0 & \cdots & B_{p-1}' & A_p
\end{pmatrix}.
\label{eq:tmatrix}
\end{equation}
Such matrices inherit ordering properties: strictly lower and upper triangular parts remain invariant
under the permutation.  

A key property, generalizing the bipartite case, is that the eigenvalues of
\[
A(\alpha)=\alpha L+\tfrac{1}{\alpha}U,
\]
with $L$ and $U$ the strictly lower and upper triangular parts of $A$, are independent of $\alpha$.
This follows from a similarity transformation by a diagonal scaling matrix $X$, analogous to the
argument in the bipartite case.

---

\paragraph{Connections.}
These examples of lattice, bipartite, and ordered matrices illustrate how algebraic structure
(symmetry, sparsity, block form) constrains the spectrum.  
They connect naturally to topics discussed earlier:  
\begin{itemize}
\item Fourier diagonalization,  
\item eigenvalue symmetry of Hermitian and skew-Hermitian matrices (Sec.~\ref{sec:eigenvalues}),  
\item and matrix inverses via Green functions.  
\end{itemize}
Such structured matrices play a central role in both pure mathematics (group representations,
combinatorics) and physics (lattice models, spin systems, statistical mechanics).
 
\chapter{Spectral and discrete transforms}\label{sec:spectral-transforms}

This chapter is devoted to the analysis of certain linear transformations of vectors, which have
continuous counterparts that are well known in analysis and physics. We emphasize both their
properties and applications. Most of the discrete transforms can be interpreted as sampled versions
of continuous transforms.
\section{Fourier transform}
\subsection{Fourier: discrete transformations}
The Fourier transform is one of the most important integral transforms. For a function
$f(x)\in L^2(\mathbb{R})$, its Fourier transform is defined as
\begin{equation}
    \hat f(k)=\frac{1}{\sqrt{2\pi}} \int_{-\infty}^\infty f(x) e^{-2 \pi i x k}\, dx,
\end{equation}
with inverse
\begin{equation}
    f(x)=\frac{1}{\sqrt{2\pi}} \int_{-\infty}^\infty \hat f(k) e^{2 \pi i x k}\, dk.
\end{equation}
Different conventions exist, e.g.\ with normalization factor $1$ instead of $1/\sqrt{2\pi}$.

---

\paragraph{Properties.}
The Fourier transform has several fundamental properties:
\begin{itemize}
    \item \textbf{Linearity:} $a g(x)+b q(x) \mapsto a \hat g(k)+b \hat q(k)$.  
    \item \textbf{Translation:} $f(x-x_0)\mapsto e^{-2\pi i x_0 k}\hat f(k)$.  
    \item \textbf{Scaling:} $f(ax)\mapsto \frac{1}{|a|}\hat f(k/a)$.  
    \item \textbf{Conjugation:} $f^*(x)\mapsto \hat f^*(-k)$.  
    \item \textbf{Derivatives:} $\partial_x f(x)\mapsto 2\pi i k \hat f(k)$.  
\end{itemize}

Two key results are Parseval’s identity and Plancherel’s theorem:
\begin{align}
    \int_{-\infty}^{\infty} f(x) g^*(x)\, dx &= \int_{-\infty}^{\infty} \hat f(k) \hat g^*(k)\, dk, \\
    \int_{-\infty}^{\infty} |f(x)|^2\, dx &= \int_{-\infty}^{\infty} |\hat f(k)|^2\, dk.
\end{align}

---

\paragraph{Fourier series.}
If $f(x)$ is periodic on $[-T/2,T/2]$, then it can be expanded in a Fourier series:
\begin{equation}
    f(x)=a_0+\sum_{n=1}^{\infty} \Big[a_n \cos\!\left(\tfrac{2 \pi n x}{T}\right)
    + b_n \sin\!\left(\tfrac{2 \pi n x}{T}\right)\Big],
\end{equation}
with coefficients
\begin{align}
a_n&=\frac{2}{T} \int_{-T/2}^{T/2} f(x) \cos\!\left(\tfrac{2 \pi n x}{T}\right)\, dx, \\
b_n&=\frac{2}{T} \int_{-T/2}^{T/2} f(x) \sin\!\left(\tfrac{2 \pi n x}{T}\right)\, dx.
\end{align}
This is the discrete counterpart of the continuous Fourier transform, restricted to integer
multiples of the fundamental frequency.

---

\paragraph{Application: differential equations.}
For linear differential equations with constant coefficients,
\begin{equation}
    \sum_{j=0}^n b_j \partial_t^j y(t)=\sum_{j=0}^n a_j \partial_t^j x(t),
\end{equation}
taking the Fourier transform converts differentiation into multiplication by $(-i\omega)$, yielding
\begin{equation}
     \hat y(\omega)=\frac{\sum_{j=0}^n a_j (-i \omega)^j}{\sum_{j=0}^n b_j (-i \omega)^j}\,\hat x(\omega)
     =\frac{P(\omega)}{Q(\omega)} \hat x(\omega).
\end{equation}
This diagonalization in frequency space is the basis of spectral methods in differential equations.

---

\paragraph{Discrete Fourier transform (DFT).}
In practice we deal with sampled signals. Let $\vec x=(x_0,\ldots,x_{N-1})^{\top}$ be a vector of $N$
samples. The \emph{discrete Fourier transform} (DFT) is defined by the linear map
\[
\vec k = W \vec x,
\]
where $W$ is the $N\times N$ Fourier matrix with entries
\begin{equation}
    W_{mn} = \frac{1}{\sqrt{N}}\, \omega^{mn}, \qquad \omega=e^{-2\pi i/N},
\end{equation}
with indices $m,n=0,\ldots,N-1$. Explicitly,
\[
W=\frac{1}{\sqrt{N}}
\begin{pmatrix}
1 & 1 & 1 & \cdots & 1 \\
1 & \omega & \omega^2 & \cdots & \omega^{N-1} \\
1 & \omega^2 & \omega^4 & \cdots & \omega^{2(N-1)} \\
\vdots & \vdots & \vdots & \ddots & \vdots \\
1 & \omega^{N-1} & \omega^{2(N-1)} & \cdots & \omega^{(N-1)^2}
\end{pmatrix}.
\]

This is a \emph{Vandermonde matrix} built from the $N$-th roots of unity $\alpha_m=\omega^m$. Matrices whose spectrum is defined among the roots of unity have a variety of applications.\footnote{The Caesar cipher — one of the earliest substitution ciphers, shifting letters by a fixed amount on the circular alphabet — may seem far from spectral methods. Yet the connection lies in the spectrum of the shift (or permutation) operator on the alphabet: its eigenvalues are roots of unity, and diagonalization makes the periodicity of the cipher transparent. In this sense, spectral analysis already appeared in cryptography long before its modern role in complex systems. Certain cyphers can also be defined with site dependent shifts  $f(k)$ that are dependent on a certain series $f(k)$.} Its
determinant is
\begin{equation}
    \det(W)=\frac{1}{N^{N/2}} \prod_{1\leq m<n\leq N} (\alpha_m-\alpha_n).
\end{equation}
The DFT matrix is unitary, i.e.\ $W^\dagger W=I$, ensuring that the transform preserves the $L^2$
norm of the signal (discrete Plancherel theorem).

---

\paragraph{Remarks.}
\begin{itemize}
    \item The DFT is invertible with inverse given by $W^{-1}=W^\dagger$.
    \item Computationally, the DFT is implemented via the \emph{Fast Fourier Transform} (FFT),
    which reduces complexity from $O(N^2)$ to $O(N\log N)$.
    \item In signal processing, the DFT is used to analyze frequencies of discrete signals, filter
    noise, and compress information.
\end{itemize}

We now verify the unitarity of the Fourier matrix $W$. Consider
\[
M_{ij}=\sum_{k} W_{ik} W^*_{kj}.
\]
Explicitly,
\[
M_{ij}=\frac{1}{N}\sum_{k=0}^{N-1} \omega^{ik}\omega^{-kj}
=\frac{1}{N}\sum_{k=0}^{N-1} \omega^{(i-j)k}.
\]
If $i=j$, the sum gives $M_{ii}=1$. If $i\neq j$, then $\omega^{i-j}$ is a nontrivial $N$th root of unity and hence
\[
\sum_{k=0}^{N-1}\omega^{(i-j)k}=\frac{1-(\omega^{i-j})^N}{1-\omega^{i-j}}=0.
\]
Thus $M=I$ and
\begin{equation}
WW^*=I,
\end{equation}
so $W$ is indeed a unitary matrix.

---

\paragraph{Determinant and Vandermonde structure.}
A second proof, often found in the literature, uses the determinant of Vandermonde matrices. The norm of the determinant can be expressed as
\begin{eqnarray}
   |\det(W)|
   &=&\frac{1}{N^{N/2}} \prod_{1\leq m< n\leq N} \big|1-e^{-\tfrac{2 \pi i}{N}(n-m)}\big| \nonumber \\
   &=&\frac{1}{N^{N/2}} \prod_{k=1}^{N-1} \big|2\sin(\tfrac{\pi k}{N})\big|^{N-k}.
\end{eqnarray}
Using the classical trigonometric identity
\[
\prod_{k=1}^{N-1} \sin\left(\tfrac{\pi k}{N}\right)=\frac{N}{2^{N-1}},
\]
one shows after simplification that
\[
|\det(W)|=1,
\]
which is consistent with $W$ being unitary.

---

\subsection{Polynomial and trigonometric interpolation viewpoint}
The DFT can also be understood as a special case of interpolation. Given $n+1$ points $(x_i,y_i)$, the polynomial interpolant
\[
P_n(x)=\sum_{i=0}^n a_i x^i
\]
is determined by the Vandermonde system
\begin{equation}
\begin{pmatrix}
1 & x_0 & \cdots & x_0^n \\
1 & x_1 & \cdots & x_1^n \\
\vdots & \vdots & \ddots & \vdots \\
1 & x_n & \cdots & x_n^n
\end{pmatrix}
\begin{pmatrix}
a_0 \\ a_1 \\ \vdots \\ a_n
\end{pmatrix}
=
\begin{pmatrix}
y_0 \\ y_1 \\ \vdots \\ y_n
\end{pmatrix}.
\end{equation}
The DFT arises when one considers interpolation not on real monomials $x^j$ but on complex exponentials $e^{i x}$ evaluated at the $N$th roots of unity. This corresponds to \emph{trigonometric interpolation}, in which the basis functions are periodic.

Thus the DFT can be seen as evaluating a trigonometric interpolant of the form
\[
p(x)=\sum_{k=0}^{N-1} Y_k e^{i 2\pi k x/N},
\]
with data sampled at $x_n=2\pi n/N$. This perspective connects the DFT to the general theory of interpolation, and explains why its matrix form is a Vandermonde built from roots of unity.

---

\subsection{Symmetry properties}
The DFT inherits symmetry properties that mirror those of the continuous Fourier transform. For a real vector $\vec x\in\mathbb{R}^N$ with transform $\vec k=W\vec x$:
\begin{itemize}
    \item $\Re(\vec k)$ is symmetric, $\Im(\vec k)$ is antisymmetric.
    \item If $\vec x$ is symmetric, then $\vec k$ is real and symmetric.
    \item If $\vec x$ is antisymmetric, then $\vec k$ is purely imaginary and antisymmetric.
\end{itemize}
Analogous statements hold for purely imaginary $\vec x$ with roles of real/imaginary parts swapped.

---

\subsection{Convolution and circulant matrices}
One of the most important properties is the convolution theorem: the DFT converts convolution into multiplication. To see this, define the circulant matrix generated by $\vec x=(x_0,\ldots,x_{N-1})$:
\[
C(\vec x)=
\begin{pmatrix}
x_0 & x_1 & \cdots & x_{N-1}\\
x_{N-1} & x_0 & \cdots & x_{N-2}\\
\vdots & \vdots & \ddots & \vdots\\
x_1 & x_2 & \cdots & x_0
\end{pmatrix}.
\]
Then convolution with another vector $\vec y$ can be written as
\[
\vec z=C(\vec x)\vec y.
\]
The DFT diagonalizes circulant matrices: the eigenvectors of $C(\vec x)$ are the Fourier modes, and the eigenvalues are the DFT of $\vec x$. Thus,
\[
W C(\vec x) W^\dagger = \mathrm{diag}(\hat x_0,\ldots,\hat x_{N-1}),
\]
and
\[
\widehat{\vec x\ast \vec y}=\hat x \circ \hat y,
\]
where $\circ$ denotes the Hadamard product. This gives an elegant proof of the convolution theorem.

---

\subsection{Difference equations}
Just as the Fourier transform simplifies differential equations with constant coefficients, the DFT simplifies linear difference equations. For example, the second-order difference scheme
\[
f_{k+1}+(b h^2+a h-2) f_k +(1-ah) f_{k-1}=h^2 q_k
\]
arises from discretizing $f''(x)+af'(x)+bf(x)=q(x)$ by finite differences. Applying the DFT,
\[
\hat f_j = \frac{h^2}{\omega^j+\alpha+\beta \omega^{-j}} \hat q_j,
\]
which provides the solution in frequency space. Inverse transforming yields $f_k$.

---

This operator viewpoint highlights the unity between continuous Fourier analysis, polynomial and trigonometric interpolation, circulant matrices, and spectral methods for solving differential and difference equations.

\subsubsection{Discrete Laplace and $z$-transforms}
The discrete Fourier transform is closely related to another classical tool in analysis and signal processing: the \emph{Laplace transform}\index{key}{Laplace transform} and its discrete analogue, the \emph{$z$-transform}\index{key}{$z$-transform}.

For a continuous function $x(t)$, the Laplace transform is defined as
\begin{equation}
    \tilde X(s)=\int_0^\infty x(t)\, e^{-s t}\, dt,
\end{equation}
where $s=\sigma+i \omega$ is a complex variable. Restricting to the imaginary axis ($\sigma=0$) gives the \emph{unilateral Fourier transform}.  

For a discrete-time signal $(x_i)_{i\ge 0}$, the analogue is the $z$-transform:
\begin{equation}
    X(z)=\sum_{i=0}^\infty x_i z^{-i},
\end{equation}
with $z=e^{s \Delta T}$, where $\Delta T$ is the sampling interval. Thus $X(z)$ is a complex analytic function on an annulus of the complex plane, rather than a finite vector. As $\Delta T\to 0$, the $z$-transform converges to the Laplace transform.

The region of convergence (ROC)\index{key}{region of convergence} of $X(z)$ is the set of complex $z$ where the series converges. Just as for Fourier transforms, the ROC encodes important stability and causality properties.

\paragraph{Properties.} 
If $x_i \mapsto X(z)$, then:
\begin{itemize}
    \item Time shift: $x_{i-m}u(i-m) \mapsto z^{-m} X(z)$.
    \item Advance: $x_{i+m}u(i) \mapsto z^m \left(X(z)-\sum_{n=0}^{m-1} x_n z^{-n}\right)$.
    \item Convolution: if $x^1_i \mapsto X^1(z)$ and $x^2_i \mapsto X^2(z)$, then 
    \[
    (x^1\star x^2)_i = \sum_{m} x^1_m x^2_{i-m} \quad\mapsto\quad X^1(z)X^2(z).
    \]
    \item Multiplication by $i$: $i x_i \mapsto -z \tfrac{d}{dz}X(z)$.
    \item Exponential scaling: $c^i x_i \mapsto X\!\left(\tfrac{z}{c}\right)$.
    \item Time reversal: $x_{-i}\mapsto X(1/z)$.
\end{itemize}

Two further facts connect coefficients with analytic properties:
\begin{equation}
    x_0 = \lim_{z\to\infty} X(z), \qquad 
    \lim_{i\to\infty} x_i = \lim_{z\to 1} X(z),
\end{equation}
so the long-time behaviour of the signal is encoded in the poles of $X(z)$ inside the unit circle.

\paragraph{Application to linear difference equations.}
The $z$-transform is particularly effective for solving constant-coefficient difference equations. For example,
\begin{equation}
    y_i+\sum_{j=1}^n a_j y_{i-j}=\sum_{j=0}^n b_j x_{i-j},
\end{equation}
with initial conditions $y_{-k}=0$ for $k\geq 0$, becomes in the $z$-domain
\begin{equation}
    Y(z)=\frac{\sum_{j=0}^n b_j z^{-j}}{1+\sum_{j=1}^n a_j z^{-j}}\, X(z)
    =\frac{P(z)}{Q(z)} X(z).
\end{equation}
The rational function $\tfrac{P(z)}{Q(z)}$ is the \emph{transfer function}\index{key}{transfer function}, whose poles determine the stability of the system. If any pole lies outside the unit circle, the system is unstable.

---
\section{Hadamard transformations}
\subsection{Hadamard and Walsh transforms}
Another family of discrete transforms arises from \emph{Hadamard matrices}\index{key}{Hadamard matrix}.  
A Hadamard matrix $H_n$ of order $n$ is an $n\times n$ matrix with entries $\pm 1$ such that columns (and rows) are mutually orthogonal. For example:
\[
H_1=(1), \qquad 
H_2=\begin{pmatrix}1&1\\1&-1\end{pmatrix}.
\]

In general,
\[
H^\top H = n I, \qquad \det(H)=\pm n^{n/2},
\]
the latter being \emph{Hadamard’s determinant bound}\index{key}{Hadamard’s determinant bound}.  

Hadamard matrices exist for orders $n=1,2$ and for all multiples of $4$ conjecturally (the \emph{Hadamard conjecture}), and can be constructed recursively by \emph{Sylvester’s construction}\index{key}{Sylvester construction}:
\[
H_{2n}=\begin{pmatrix} H_n & H_n \\ H_n & -H_n \end{pmatrix}.
\]
Scaling by $1/\sqrt{n}$ yields an orthogonal transform. For instance,
\[
\hat H_2=\frac{1}{\sqrt{2}}H_2
\]
is exactly the $2$-point DFT.

\paragraph{Walsh matrices.}
A \emph{Walsh matrix}\index{key}{Walsh matrix} is a Hadamard matrix whose rows are ordered by increasing number of sign changes. This ordering, due to Walsh, produces the so-called \emph{Walsh functions}, which take only values $\pm 1$ and form an orthogonal basis of square waves.  

For example:
\[
H_4=\begin{pmatrix}
1&1&1&1\\
1&-1&1&-1\\
1&1&-1&-1\\
1&-1&-1&1
\end{pmatrix}
\quad\longrightarrow\quad
W_4=\begin{pmatrix}
1&1&1&1\\
1&1&-1&-1\\
1&-1&-1&1\\
1&-1&1&-1
\end{pmatrix}.
\]

Walsh functions are piecewise-constant, orthogonal functions on $[0,1]$, expressible as products of \emph{Rademacher functions}\index{key}{Rademacher functions}:
\begin{equation}
    R_n(t)=\mathrm{sign}\!\left(\sin(2^{n+1}\pi t)\right).
\end{equation}
Explicitly,
\begin{equation}
    W_k(t)=\prod_{j=0}^\infty R_j(t)^{k_j},
\end{equation}
where $k_j$ are binary digits of $k$.  

\paragraph{Connection to Haar functions.}
The Rademacher functions can in turn be expanded in terms of \emph{Haar wavelets}\index{key}{Haar wavelets}:
\[
r_n(t)=\frac{1}{2^{n/2}}\sum_{k=0}^{2^n-1}\psi_{n,k}(t),
\]
where $\psi_{n,k}(t)$ are the Haar basis functions. This shows how Walsh and Hadamard transforms are linked to wavelet analysis, and how binary structure naturally appears in discrete orthogonal transforms.

---

These alternative transforms (Hadamard, Walsh, Haar) are widely used in coding theory, digital communications, and image compression, where their simplicity (entries are just $\pm1$) makes them computationally efficient compared to the Fourier transform.

\subsection{Haar transform}
For completeness, we now introduce the \emph{Haar transform}, which is historically the first example of a wavelet transform, dating back to Alfréd Haar (1909). It plays a central role in data compression, image analysis, and denoising.\index{key}{Haar transform}\index{key}{wavelets}

The Haar system is a family of square-integrable functions on $[0,1]$, called \emph{Haar wavelets}\index{key}{Haar functions}, defined by
\begin{equation}
    e^k_n(x)=
    \begin{cases}
    2^{n/2}, & \frac{k-1}{2^n}<x<\frac{k-\tfrac{1}{2}}{2^n}, \\[6pt]
    -2^{n/2}, & \frac{k-\tfrac{1}{2}}{2^n}\leq x<\frac{k}{2^n}, \\[6pt]
    0 & \text{otherwise},
    \end{cases}
\end{equation}
where $n\geq 0$ is the \emph{scale} and $k=1,\dots,2^n$ is the \emph{translation index}.  
These functions satisfy
\[
\int_0^1 e^k_n(x) e^{k'}_{n'}(x)\,dx = \delta_{n,n'}\delta_{k,k'},
\]
so they form an orthonormal basis of $L^2([0,1])$.

\paragraph{Discrete Haar transform.}
The discrete Haar transform (DHT)\index{key}{discrete Haar transform} is obtained by sampling the Haar wavelets at $2^n$ points. It can be written in matrix form:
\begin{eqnarray}
    M_1&=&\frac{1}{\sqrt{2}}H_2, \\
    M_n&=&\frac{1}{\sqrt{2^n}}
    \begin{pmatrix}
      H_{n-1}\otimes [1\ 1]\\[3pt]
      2^{n/2} I_{2^{n-1}} \otimes [1\ -1]
    \end{pmatrix},
\end{eqnarray}
where $H_2$ is the Hadamard matrix, $I_m$ the $m\times m$ identity, and $\otimes$ denotes the Kronecker product. Each $M_n$ is an orthogonal matrix: $M_n^\top M_n=I_{2^n}$.

For instance:
\[
M_1=\frac{1}{\sqrt{2}}\begin{pmatrix}1&1\\1&-1\end{pmatrix}, \quad
M_2=\frac{1}{2}\begin{pmatrix}
1&1&1&1\\
1&1&-1&-1\\
\sqrt{2}&-\sqrt{2}&0&0\\
0&0&\sqrt{2}&-\sqrt{2}
\end{pmatrix}.
\]

\paragraph{Properties and applications.}
Unlike Hadamard or Fourier transforms, Haar matrices are highly \emph{sparse}: many entries are zero. This makes them computationally efficient, and their localized support means that they capture \emph{local features} of data (edges, discontinuities). The inverse Haar transform is simply the transpose, since $M_n$ is orthogonal.

Applications include Image compression (e.g. JPEG2000 uses wavelet-based methods),  signal denoising by thresholding Haar coefficients and multiresolution analysis, e.g. decomposing signals into coarse and fine scales.

---

\subsection{Radon transform}
We now turn to the \emph{Radon transform}\index{key}{Radon transform}, a fundamental tool in image analysis, tomography, and inverse problems. Introduced by Johann Radon (1917), it is the mathematical basis of computed tomography (CT) and other imaging modalities.\index{key}{tomography}

\paragraph{Definition.}
For a 2D function $f(x,y)$, the Radon transform is the collection of its line integrals:
\begin{equation}
    \hat f(\theta,s)=\int_{\mathbb{R}^2} f(x,y)\, \delta(x\cos\theta+y\sin\theta-s)\, dx\,dy,
\end{equation}
where $\theta$ is the angle of the line normal, and $s$ is the signed distance from the origin. The function $\hat f(\theta,s)$ is called a \emph{sinogram}\index{key}{sinogram}, as a single point in the plane maps to a sinusoidal curve in $(\theta,s)$-space.

\paragraph{Projection-slice theorem.}
A key fact is the \emph{projection-slice theorem}\index{key}{projection-slice theorem}: the 1D Fourier transform of $\hat f(\theta,s)$ in the $s$ variable equals the restriction of the 2D Fourier transform of $f(x,y)$ to the line at angle $\theta$:
\begin{equation}
    \mathcal{F}_s\{\hat f(\theta,s)\}(k)=
    \mathcal{F}_{x,y}\{f(x,y)\}(k\cos\theta,k\sin\theta).
\end{equation}
Thus, by collecting projections $\hat f(\theta,s)$ for all $\theta\in[0,\pi)$, one recovers the full Fourier transform of $f$, and hence $f$ itself via inverse Fourier transform. This is the theoretical foundation of CT reconstruction.

\paragraph{Geometric interpretation.}
- A single point $(x_0,y_0)$ in the plane maps to a sinusoidal curve $\hat f(\theta,s)=\delta(s-x_0\cos\theta-y_0\sin\theta)$ in the sinogram.  
- A line in the plane maps to a delta ridge in $(\theta,s)$-space.  

This duality is essential for understanding how spatial features appear in projection data.

\paragraph{Discrete Radon transform.}
In practice, we discretize $f(x,y)$ into pixels and approximate integrals by sums. The discrete Radon transform can be represented as a linear mapping
\[
\vec y = A \vec x,
\]
where $\vec x$ is the vectorized image and $\vec y$ the projection data. The system matrix $A$ encodes line integrals through pixels, and is typically sparse but very ill-conditioned. Reconstruction therefore requires regularization (e.g. Tikhonov regularization\index{key}{Tikhonov regularization}):
\[
\vec x = \arg\min_{\vec x} \left( \|A\vec x-\vec y\|_2^2 + \mu \|\Gamma(\vec x-\vec x^*)\|_2^2 \right),
\]
where $\Gamma$ is a regularization operator (often a derivative matrix), and $\vec x^*$ a prior estimate.

The Radon transform is used in computed tomography (medical CT, industrial CT), geophysical imaging (e.g. seismic tomography) and pattern recognition in machine learning (Radon transforms detect oriented structures such as lines).

---

\noindent
Thus, the Haar and Radon transforms illustrate two complementary aspects of matrix transforms:  
- Haar provides \emph{localized multiscale analysis} (good for compression and denoising).  
- Radon provides a \emph{global projection framework} (good for reconstruction and inverse problems).  
Both fit naturally within the broader theory of linear operators and spectral analysis.

\section{Spectral transforms}
\subsection{Mellin transform and spectral zeta functions}\label{sec:mellin}
\index{key}{Mellin transform}\index{key}{spectral zeta function}\index{key}{heat kernel}

Spectral analysis of matrices (and operators) is often enriched by tools from analytic number theory.  
In particular, the \emph{spectral zeta function} and the \emph{Mellin transform} connect eigenvalues of matrices to quantities such as determinants, heat kernels, and combinatorial invariants.

\paragraph{Spectral zeta function.}
Given a matrix (or graph Laplacian) with eigenvalues $0=\lambda_0 \leq \lambda_1 \leq \cdots \leq \lambda_{n-1}$, we define the associated \emph{zeta function} as
\begin{equation}
    \zeta(s)=\sum_{i\neq 0}\frac{1}{\lambda_i^s}.
\end{equation}
This is a direct analogue of the Riemann zeta function, but built from the spectrum of the operator. Differentiating with respect to $s$ gives
\begin{equation}
    \zeta^\prime(s) = -\sum_{i\neq 0}\frac{\log \lambda_i}{\lambda_i^s}, \qquad 
    \zeta^\prime(0)=\log\Big(\prod_{i\neq 0} \lambda_i\Big).
\end{equation}
Thus $\zeta^\prime(0)$ encodes the product of all nonzero eigenvalues.

\paragraph{Application to spanning trees.}
For a graph $G$, Kirchhoff’s matrix-tree theorem relates the Laplacian eigenvalues to the number of spanning trees $\tau(G)$. One can express it as
\begin{equation}
    \tau(G)=\frac{\prod_v d_v}{\sum_v d_v}\, e^{\zeta^\prime(0)},
\end{equation}
where $d_v$ are the vertex degrees. This shows how $\zeta^\prime(0)$ naturally appears in combinatorial quantities.

\paragraph{Heat kernel representation.}
The spectral zeta function is connected to the heat kernel via a Mellin transform.  
Recall the Laplace transform identity
\[
    \rho^{-s} = \frac{1}{\Gamma(s)} \int_0^\infty e^{-\rho t} t^{s-1}\, dt.
\]
Applying this to each eigenvalue,
\begin{eqnarray}
\frac{1}{\Gamma(s)}\int_0^\infty t^{s-1}\left(\mathrm{Tr}(e^{-tA})-1\right)dt
&=&\frac{1}{\Gamma(s)}\int_0^\infty t^{s-1}\sum_{i\neq 0}e^{-\lambda_i t}\,dt \\
&=&\sum_{i\neq 0}\lambda_i^{-s} \\
&=&\zeta(s).
\end{eqnarray}
Thus the spectral zeta function is the Mellin transform of the (reduced) heat trace.

\paragraph{Spectral density.}
For a diagonalizable operator $A$ with eigenvalues $\{\lambda_m\}$ we define the empirical spectral density
\begin{equation}
    f_\lambda(t)=\frac{1}{N}\sum_{m=1}^N \delta(t-\lambda_m).
\end{equation}
Using the Dirac representation, one can connect it to the Laplace transform
\[
f_\lambda(t)=\frac{1}{2\pi i}\int_{c-i\infty}^{c+i\infty} e^{zt}\, \phi_\lambda(z)\,dz, 
\qquad 
\phi_\lambda(z)=\frac{1}{N}\sum_m e^{-z\lambda_m}=\frac{1}{N}\mathrm{Tr}(e^{-zA}).
\]
By Mellin transforming the eigenvalues,
\[
\lambda_m^{-s}=\frac{1}{\Gamma(s)}\int_0^\infty x^{s-1} e^{-\lambda_m x}\,dx,
\]
we recover the Dirichlet series for the zeta function:
\[
\zeta(s)=\sum_m \lambda_m^{-s} = \frac{1}{\Gamma(s)} \int_0^\infty x^{s-1}\,\mathrm{Tr}(e^{-xA})\,dx.
\]
Hence the eigenvalue density can itself be written as a Mellin transform:
\begin{equation}
f_\lambda(t)=\frac{1}{2\pi i N}\int_{c^\prime-i\infty}^{c^\prime+i\infty} t^{s-1}\,\zeta(s)\,ds.
\end{equation}

\paragraph{Heat kernel and Mellin transform.}
The heat kernel associated with a graph Laplacian $\Delta$ is the operator
\[
F(t,x,y)=[e^{-t\Delta}]_{xy}=\langle x|e^{-t\Delta}|y\rangle,
\]
which satisfies the heat equation
\[
\partial_t F(t,x,y)=-\Delta F(t,x,y).
\]
Its Mellin transform is
\begin{equation}
    K(s,x,y)=\frac{1}{\Gamma(-s)}\int_0^\infty t^{-s-1} F(t,x,y)\,dt,
\end{equation}
which encodes fractional (negative) powers of $\Delta$. Conversely,
\begin{equation}
    F(t,x,y)=\frac{1}{2\pi i}\int_{c-i\infty}^{c+i\infty}\Gamma(-s)\, t^s\,K(s,x,y)\,ds.
\end{equation}
This duality between heat kernels and spectral zeta functions is a central tool in spectral geometry and mathematical physics (e.g. in quantum field theory, index theorems, and quantum gravity).

---

\subsection{Stieltjes transform}
\index{key}{Stieltjes transform}
The \emph{Stieltjes transform} is a powerful analytic tool to study the spectral distribution of a matrix or operator.  
Given a probability measure $\mu$ supported on $\mathbb{R}$ (e.g.\ the empirical spectral distribution of a matrix), its Stieltjes transform is defined as
\begin{equation}
    s_\mu(z) = \int_{-\infty}^\infty \frac{1}{x-z}\,d\mu(x), \qquad z \in \mathbb{C}\setminus \mathrm{supp}(\mu).
\end{equation}
For a finite $n\times n$ Hermitian matrix $A$ with eigenvalues $\{\lambda_i\}_{i=1}^n$, the empirical measure is
\[
\mu_A = \frac{1}{n}\sum_{i=1}^n \delta(x-\lambda_i),
\]
so that
\begin{equation}
    s_A(z) = \frac{1}{n}\sum_{i=1}^n \frac{1}{\lambda_i-z}.
\end{equation}
Equivalently, the Stieltjes transform can be written as a normalized trace of the resolvent\index{key}{resolvent}:
\begin{equation}
    s_A(z) = \frac{1}{n}\mathrm{Tr}\,(A-zI)^{-1}.
\end{equation}

\paragraph{Properties.}
\begin{itemize}
    \item $s_\mu(z)$ is analytic in $\mathbb{C}\setminus\mathbb{R}$ and satisfies $\Im(s_\mu(z)) \cdot \Im(z)>0$ (Nevanlinna–Pick property).
    \item The measure $\mu$ can be recovered from $s_\mu$ via the Stieltjes inversion formula:
    \begin{equation}
        \mu([a,b])=\lim_{\eta\downarrow 0}\frac{1}{\pi}\int_a^b \Im\, s_\mu(x+i\eta)\, dx.
    \end{equation}
    Thus, the Stieltjes transform fully characterizes the spectral measure.
    \item If $A$ is Hermitian and bounded, then as $|z|\to \infty$, 
    \begin{equation}
        s_A(z) \sim -\frac{1}{z} - \frac{1}{n}\frac{\mathrm{Tr}(A)}{z^2} - \cdots,
    \end{equation}
    providing a moment expansion (moments of $\mu_A$).
\end{itemize}

\paragraph{Connection to resolvents.}
The resolvent of a matrix $A$ is
\begin{equation}
    R(z) = (A-zI)^{-1}.
\end{equation}
The entries of $R(z)$ encode information about the Green’s function of $A$, and its normalized trace is exactly the Stieltjes transform:
\[
s_A(z) = \frac{1}{n}\mathrm{Tr}\,R(z).
\]
In physics and probability, $s_A(z)$ is often referred to as the \emph{Green’s function}.

In general, the Stieltjes transform is used in \textbf{spectral distribution recovery}, e.g. using the inversion formula, one can reconstruct the eigenvalue density of a matrix from its Stieltjes transform.It is also widely used in \textbf{random matrix theory}\footnote{See, e.g., T.~Tao, \emph{Topics in Random Matrix Theory}, AMS (2012); G.~Anderson, A.~Guionnet, and O.~Zeitouni, \emph{An Introduction to Random Matrices}, Cambridge University Press (2010); L.~Pastur and M.~Shcherbina, \emph{Eigenvalue Distribution of Large Random Matrices}, AMS (2011); P.~Vivo, G.~Livan, and M.~Novaes, \emph{Introduction to Random Matrices: Theory and Practice}, Springer (2020).}: the Stieltjes transform is the central tool to prove convergence of empirical spectral distributions. For example, Wigner’s semicircle law can be derived by showing that $s_A(z)$ converges almost surely to the Stieltjes transform of the semicircle density. In free probability, the $R$-transform and free convolutions are often formulated in terms of the Stieltjes transform.
 In numerical analysis instead, the Stieltjes transform is related to continued fractions, Padé approximants, and iterative methods for spectral density estimation.

\paragraph{Example.}
For the Wigner semicircle distribution $\mu_{sc}$ with density
\[
\rho_{sc}(x)=\frac{1}{2\pi}\sqrt{4-x^2}\,\mathbf{1}_{[-2,2]}(x),
\]
its Stieltjes transform is
\begin{equation}
    s_{\mu_{sc}}(z) = \frac{-z+\sqrt{z^2-4}}{2},
\end{equation}
where the branch of the square root is chosen with $\Im\sqrt{z^2-4}>0$ when $\Im z>0$. This explicit formula is often the starting point for random matrix limit laws.

\section{Embeddings}
\index{key}{embeddings}

In many applications — from data science to compressed sensing and randomized numerical linear algebra —
we are interested in embedding a high-dimensional vector space into a lower-dimensional one,
while approximately preserving geometric quantities such as norms or pairwise distances.
This idea is central to dimensionality reduction, randomized algorithms, and approximation theory.

\subsection{Johnson–Lindenstrauss lemma}\label{sec:JL}
\index{key}{Johnson-Lindenstrauss lemma}

One of the most celebrated results in this direction is the \emph{Johnson–Lindenstrauss lemma},
which shows that high-dimensional points can be embedded into a surprisingly low-dimensional space
while preserving pairwise distances up to small distortion.

\textbf{Theorem (Johnson–Lindenstrauss).}  
Let $0<\varepsilon<1$ and let $X=\{\vec x_1,\dots,\vec x_m\}$ be a set of $m$ points in $\mathbb{R}^n$.  
There exists a linear map $f:\mathbb{R}^n \to \mathbb{R}^k$ with $k=O(\varepsilon^{-2}\log m)$ such that
for all $i,j$,
\begin{equation}
(1-\varepsilon)\|\vec x_i-\vec x_j\|_2^2 \leq \|f(\vec x_i)-f(\vec x_j)\|_2^2 \leq (1+\varepsilon)\|\vec x_i-\vec x_j\|_2^2.
\end{equation}

Equivalently, a set of $m$ points in $\mathbb{R}^n$ can be embedded into $O(\log m)$ dimensions with only small distortion in Euclidean distances.\footnote{W.B.~Johnson and J.~Lindenstrauss, ``Extensions of Lipschitz mappings into a Hilbert space,'' \emph{Contemporary Mathematics}, 26 (1984): 189–206.}  

\paragraph{Proof idea.}  
A random projection suffices: let $R$ be a $k\times n$ matrix with i.i.d.\ Gaussian entries $N(0,1)$ (or $\pm 1$ entries).
Then with high probability,
\[
f(\vec x)=\frac{1}{\sqrt{k}}R\vec x
\]
satisfies the inequality above, due to concentration of measure and the Chernoff bound on $\chi^2$ random variables.

The lemma is widely used in: dimensionality reduction for clustering and nearest-neighbor search, compressed sensing and sketching algorithms, speeding up kernel methods and regression, and embedding finite metric spaces into $\ell_2$.\footnote{See S.~Dasgupta and A.~Gupta, ``An elementary proof of the Johnson–Lindenstrauss lemma,'' \emph{Random Structures \& Algorithms} 22(1), 60–65 (2003); W.~Johnson and J.~Lindenstrauss, ``Extensions of Lipschitz mappings into a Hilbert space,'' \emph{Contemporary Mathematics} 26, 189–206 (1984); S.~Vempala, \emph{The Random Projection Method}, AMS (2004); D.~Achlioptas, ``Database-friendly random projections,'' \emph{Journal of Computer and System Sciences} 66(4), 671–687 (2003).}

\subsection{Gordon's theorem}\label{sec:gordon}
\index{key}{Gordon's theorem}
\index{key}{escape through the mesh theorem}

A related and deeper result in high-dimensional geometry is \emph{Gordon’s escape through a mesh theorem},
which controls how random subspaces intersect with convex sets in high dimensions.

\textbf{Theorem (Gordon, Escape through a Mesh).}  
Let $K\subseteq \mathbb{S}^{n-1}$ (the unit sphere) be a subset, and let $E$ be a random subspace of codimension $m$
(chosen uniformly at random). Then, with high probability,  
\begin{equation}
E \cap K = \emptyset \quad \text{if } m \gtrsim w(K)^2,
\end{equation}
where $w(K)$ is the Gaussian mean width of $K$, defined as
\[
w(K) = \mathbb{E}_{\vec g\sim N(0,I)}\left[\sup_{\vec x\in K}\langle \vec g, \vec x\rangle\right].
\]

The theorem states that if $K$ is a ``thin'' set on the sphere (measured by its Gaussian width),
then a random subspace of dimension $n-m$ will typically avoid it once $m$ exceeds $w(K)^2$.  
This is a sharp threshold phenomenon.

Gordon’s theorem plays a central role across several domains. In \emph{compressed sensing}, it provides the probabilistic foundation for the fact that random linear measurements suffice to recover sparse signals.\footnote{See E.~Candès and T.~Tao, ``Decoding by linear programming,'' \emph{IEEE Transactions on Information Theory} 51(12), 4203–4215 (2005); D.~Donoho, ``Compressed sensing,'' \emph{IEEE Transactions on Information Theory} 52(4), 1289–1306 (2006).}  
In \emph{convex geometry}, it quantifies the behavior of intersections of convex bodies with random subspaces, extending classical results of Dvoretzky and Milman.\footnote{See V.~Milman and G.~Schechtman, \emph{Asymptotic Theory of Finite Dimensional Normed Spaces}, Springer (1986).}  
In \emph{randomized numerical linear algebra}, it underlies guarantees for sketching methods and randomized embeddings used in fast regression and matrix approximation.\footnote{See P.~Drineas and M.~Mahoney, ``RandNLA: Randomized Numerical Linear Algebra,'' \emph{Communications of the ACM} 59(6), 80–90 (2016).}  
Finally, in \emph{machine learning}, it has been used to analyze the robustness of random feature maps and to provide generalization bounds for high-dimensional classifiers.\footnote{See S.~Mendelson, ``Learning without concentration,'' \emph{Journal of the ACM} 62(3), 1–25 (2015).}

The result is due to Gordon (1988)\footnote{Y.~Gordon, ``On Milman’s inequality and random subspaces which escape through a mesh in $\mathbb{R}^n$,'' in \emph{Geometric Aspects of Functional Analysis}, Springer Lecture Notes in Mathematics, vol.~1317 (1988), pp.~84–106.}, 
and is closely related to Dvoretzky’s theorem on almost-spherical sections of convex bodies\footnote{Dvoretzkys theorem states that in every high-dimensional normed space, there exist almost-Euclidean sections of large dimension. More precisely:
For every $\epsilon > 0$, there exists a dimension $k = k(\epsilon)$ such that every Banach space of sufficiently large dimension n contains a k-dimensional subspace that is $(1 + \epsilon)$-isomorphic to the Euclidean space of $k$ dimensions.
Equivalently: high-dimensional convex bodies always have low-dimensional sections that are nearly spherical; see A. Dvoretzky, ``Some results on convex bodies and Banach spaces," Proc. Int. Symp. Linear Spaces, Jerusalem 1960, pp. 123160, Pergamon Press (1961).}.

\chapter{Spectra and quadratic forms}
\section{Quadratic forms}
\subsection{Congruence and Sylvester's inertia theorem for quadratic forms}\label{sec:sylvesterinertia}

A \emph{quadratic form} in real variables $\vec x=(x_1,\dots,x_n)$ is an expression of the type
\begin{equation}
    Q(\vec x)=\sum_{i,j=1}^n a_{ij} x_i x_j,
\end{equation}
which can be written compactly as
\begin{equation}
    Q(\vec x) = {\vec x}^{\top} A \vec x,
\end{equation}
with $A$ an $n\times n$ real matrix. Note that this matrix representation is not unique: for instance,
\[
Q(\vec x) = {\vec x}^{\top} A \vec x = {\vec x}^{\top}\frac{A+A^{\top}}{2}\vec x,
\]
so that the symmetric part of $A$ always defines the same quadratic form. Thus, without loss of generality, quadratic forms are studied via \emph{symmetric matrices}.

\paragraph{Congruence.}  
Two quadratic forms $Q(\vec x)={\vec x}^{\top} A \vec x$ and $Q'(\vec x)={\vec x}^{\top} B \vec x$ are said to be \emph{congruent} if there exists an invertible matrix $P$ such that
\begin{equation}
    B = P^{\top} A P.
\end{equation}
This differs from similarity ($B=P^{-1}AP$), but congruence preserves the essential properties of quadratic forms such as the number of positive, negative, and zero squares.

\paragraph{Sylvester's law of inertia.}  
For a real symmetric matrix $A$, Sylvester’s theorem (also called the \emph{law of inertia}) states that there exists an invertible $P$ such that
\begin{equation}
    P^{\top} A P = D = \mathrm{diag}(1,\dots,1,\,-1,\dots,-1,\,0,\dots,0),
\end{equation}
with $p$ positive entries, $n$ negative entries, and $z$ zero entries on the diagonal. The triple
\[
\mathrm{In}(A)=(p,n,z),
\]
called the \emph{inertia} of $A$, is an invariant under congruence transformations. In particular:
- The rank $r=p+n$ is preserved,
- The nullity $z$ is preserved,
- The difference $s=p-n$, called the \emph{signature}\index{key}{matrix signature}, is preserved.

\textbf{Theorem (Sylvester’s Law of Inertia).} Two real symmetric matrices $A,B\in \mathbb{R}^{n\times n}$ are congruent if and only if they have the same inertia, i.e.\ the same numbers of positive, negative, and zero eigenvalues.\footnote{See F.R.~Gantmacher, \emph{The Theory of Matrices}, AMS Chelsea (1959); R.A.~Horn and C.R.~Johnson, \emph{Matrix Analysis}, Cambridge University Press (1985).}

\paragraph{Principal minors and inertia.}  
An equivalent formulation of inertia uses the signs of the leading principal minors $\Delta_k=\det(A^k)$, where $A^k$ denotes the $k\times k$ principal submatrix. The number of sign changes in the sequence $(\Delta_0,\Delta_1,\dots,\Delta_n)$ gives the negative index of inertia (Jacobi’s criterion).\footnote{Classic treatments can be found in Gantmacher, \emph{The Theory of Matrices}.}

\paragraph{Haynsworth’s inertia additivity formula.}  
For block-partitioned Hermitian matrices
\begin{equation}
A=\begin{pmatrix}
A_{11} & A_{12} \\
A_{21} & A_{22}
\end{pmatrix}, \qquad A_{21}=A_{12}^\dagger,
\end{equation}
Haynsworth’s theorem states that
\begin{equation}
    \mathrm{In}(A) = \mathrm{In}(A_{11}) + \mathrm{In}(A/A_{11}),
\end{equation}
where $A/A_{11}=A_{22}-A_{21}A_{11}^{-1}A_{12}$ is the \emph{Schur complement}\index{key}{Schur complement}.\footnote{E.V.~Haynsworth, ``Determination of the inertia of a partitioned Hermitian matrix,'' \emph{Linear Algebra Appl.}, 1(1):73–81 (1968).}  
This leads immediately to the determinant identity
\[
\det(A)=\det(A_{11})\,\det(A/A_{11}),
\]
generalizing the $2\times 2$ block determinant formula.

\paragraph{Special block cases.}  
- If $A_{11}=a\in\mathbb{R}$ and $A_{22}=b$, one recovers $\det(A)=ab- A_{12}A_{21}$.  
- If $A_{11}=A_{22}=M$ and $A_{12}=A_{21}=N$, then
\begin{equation}
    \det(A)=\det(M+N)\,\det(M-N).
\end{equation}
This factorization appears in circuit theory as \emph{Kron reduction}.\footnote{See F. Dörfler and F. Bullo, ``Kron reduction of graphs with applications to electrical networks,'' \emph{IEEE Trans. Circuits Syst. I}, 60.1 (2013): 150–163.}

\paragraph{Eigenvalues and Schur complement.}  
The eigenvalues of a block matrix $A$ satisfy
\begin{equation}
    0=\det(A-\lambda I) = \det(A_{11}-\lambda I)\,\det(A_{22}-A_{21}(A_{11}-\lambda I)^{-1} A_{12}-\lambda I).
\end{equation}
Thus, if $A_{12}$ or $A_{21}$ vanish, the spectrum of $A$ is simply the union of the spectra of $A_{11}$ and $A_{22}$.  
More generally, for special structures (e.g.\ $A_{11}=0$), this leads to \emph{conformal eigenvalue problems} of the type
\[
A_{22}\vec v - \lambda \vec v + \frac{A_{21}A_{12}}{\lambda}\vec v=0,
\]
whose eigenvalues are roots of quadratic polynomials $\lambda^2-\alpha_i\lambda-\beta_i=0$ in each invariant subspace.

\subsubsection{Properties of quadratic forms and degeneracy}
The positivity of a quadratic form is closely tied to the eigenvalues of symmetric matrices. For a general (not necessarily symmetric) $A$, positivity of $Q(\vec x)={\vec x}^{\top} A \vec x$ cannot be read from the eigenvalues of $A$ directly. However,
\[
Q(\vec x)={\vec x}^{\top} \frac{A+A^{\top}}{2}\vec x,
\]
so positivity depends only on the symmetric part $\frac{A+A^{\top}}{2}$, which is Hermitian and has real eigenvalues. This is consistent with the definition of positive (semi)definiteness in operator theory.\footnote{See Horn and Johnson, \emph{Matrix Analysis}, Ch.~7.}  

\begin{itemize}
    \item If all eigenvalues of $\frac{A+A^{\top}}{2}$ are strictly positive, then $Q(\vec x)>0$ for all $\vec x\neq 0$ (positive definite).
    \item If they are non-negative, then $Q(\vec x)\geq 0$ (positive semidefinite).
    \item The number of positive and negative eigenvalues of $\frac{A+A^{\top}}{2}$ determines the inertia of $Q$.
\end{itemize}

As an illustration, consider
\[
A=\begin{pmatrix}0.0838 & 0.9133 \\ 0.2290 & 0.1524\end{pmatrix}.
\]
The eigenvalues of $A$ differ from those of its symmetrized version $(A+A^{\top})/2$, confirming that positivity should be understood through the symmetric part.
\subsection{Quadratic forms and optimization}
\index{key}{quadratic forms}\index{key}{optimization}

Quadratic forms naturally arise in optimization, statistics, and machine learning.
Recall that any quadratic form in $x\in\mathbb{R}^n$ can be written as
\begin{equation}
f(x) = \vec x^\top A \vec x + \vec b^\top \vec x + c,
\end{equation}
with $A\in\mathbb{R}^{n\times n}$ symmetric (w.l.o.g.\ since 
$\vec x^\top A \vec x = \vec x^\top \tfrac{1}{2}(A+A^\top) \vec x$), $\vec b\in\mathbb{R}^n$, and $c\in\mathbb{R}$.
\paragraph{Definiteness and Sylvester's law.}
The nature of the quadratic form is determined by the eigenvalues of $A$.
By Sylvester’s law of inertia, any real symmetric $A$ can be brought to a diagonal form 
with the same number of positive, negative, and zero entries under congruence. 
Hence the inertia of $A$ (the counts of positive, negative, zero eigenvalues) fully determines
whether the quadratic form is positive definite, semidefinite, or indefinite.\footnote{See R.~Horn and C.~Johnson, \emph{Matrix Analysis}, Cambridge University Press (2013).}

\begin{itemize}
\item If $A \succ 0$ (positive definite), then $f$ is strictly convex and has a unique global minimizer.
\item If $A \succeq 0$ (positive semidefinite), then $f$ is convex; minimizers may not be unique.
\item If $A$ has both positive and negative eigenvalues, the form is indefinite and $f$ is not convex.
\end{itemize}

\paragraph{Gradient and Hessian.}
From matrix calculus:\footnote{See Petersen \& Pedersen, \emph{The Matrix Cookbook}, Version Nov. 15, 2012.}
\begin{align}
\nabla_{\vec{x}} f(\vec{x}) &= (A+A^\top)\vec{x} + b = 2A\vec{x}+b \quad (\text{if $A$ symmetric}), \\
\nabla^2_{\vec{x}} f(\vec{x}) &= A+A^\top = 2A.
\end{align}
Thus the Hessian coincides with $2A$, and convexity is guaranteed when $A\succeq 0$.

\paragraph{Unconstrained minimization.}
The stationary point satisfies
\[
\nabla_{\vec{x}} f(\vec{x}^\star)=0 \quad \implies \quad A\vec{x}^\star = -\tfrac{1}{2}b.
\]
If $A$ is invertible and positive definite, then the unique minimizer is
\[
\vec{x}^\star = -\tfrac{1}{2}A^{-1}b.
\]

\paragraph{Constrained case.}
For quadratic minimization under linear constraints,
\[
\min_{\vec{x}} \; \vec{x}^\top A \vec{x} \quad \text{s.t.\ } C\vec{x}=d,
\]
the Karush--Kuhn--Tucker system reads
\[
\begin{bmatrix} 2A & C^\top \\ C & 0\end{bmatrix}
\begin{bmatrix} \vec{x} \\ \lambda \end{bmatrix}
=\begin{bmatrix}0\\ d\end{bmatrix}.
\]

\paragraph{Useful derivatives.}
Matrix calculus provides the following identities, often needed in optimization:\footnote{See Petersen \& Pedersen, \emph{The Matrix Cookbook}.}
\begin{align}
\frac{\partial}{\partial \vec{x}}\,\vec{x}^\top A \vec{x} &= (A+A^\top)\vec{x}, \\
\frac{\partial}{\partial A}\,\vec{x}^\top A \vec{x} &= \vec{x}\vec{x}^\top, \\
\frac{\partial}{\partial \vec{x}}\,\frac{\vec{x}^\top A \vec{x}}{\vec{x}^\top B \vec{x}} &= 
\frac{2}{(\vec{x}^\top B \vec{x})^2}\Big((\vec{x}^\top B \vec{x})A\vec{x}-(\vec{x}^\top A \vec{x})B\vec{x}\Big).
\end{align}

\paragraph{Quadratic forms in statistics and ML.}
Quadratic objectives appear ubiquitously:
\begin{itemize}
\item Regression: $\|A\vec{x}-b\|^2 = \vec{x}^\top A^\top A \vec{x} - 2b^\top A \vec{x} + \|b\|^2$.
\item PCA: maximize $\vec{x}^\top \Sigma \vec{x}$ subject to $\|\vec{x}\|=1$, yielding eigenvalue problems.
\item Ridge regression: $\|A\vec{x}-b\|^2+\lambda\|\vec{x}\|^2=\vec{x}^\top(A^\top A+\lambda I)\vec{x}-2b^\top A \vec{x}+\|b\|^2$.
\end{itemize}

\paragraph{Rayleigh quotient.}
The Rayleigh quotient
\[
R(\vec{x})=\frac{\vec{x}^\top A \vec{x}}{\vec{x}^\top \vec{x}}
\]
satisfies $\min_i \lambda_i(A)\leq R(\vec{x})\leq \max_i \lambda_i(A)$.
Thus extremizing quadratic forms on the unit sphere reduces to eigenvalue problems.

\paragraph{Probabilistic quadratic forms.}
If $\vec{x}\sim \mathcal{N}(m,\Sigma)$, then:\footnote{See Petersen \& Pedersen, \emph{The Matrix Cookbook}.}
\begin{align}
\mathbb{E}[\vec{x}^\top A \vec{x}] &= \mathrm{Tr}(A\Sigma)+m^\top A m, \\
\mathrm{Var}(\vec{x}^\top A \vec{x}) &= 2\,\mathrm{Tr}(A\Sigma A\Sigma)+4m^\top A\Sigma A m.
\end{align}
Such results appear in random quadratic optimization, stochastic control, and random matrix theory.

\paragraph{Quadratic optimization and convex sets.}
These results connect naturally to convex analysis. Recall that a convex set $\mathcal S$ satisfies:
for any $\vec{x},\vec{y}\in\mathcal S$, the line segment $\theta \vec{x}+(1-\theta)\vec{y}\in\mathcal S$ for all $\theta\in[0,1]$.
Thus, when $f$ is convex (e.g.\ $A\succeq 0$), any local minimizer is a global minimizer.\footnote{See S.~Boyd and L.~Vandenberghe, \emph{Convex Optimization}, Cambridge University Press (2004).}

Convex sets admit extremal points, faces, and polyhedral descriptions. For example,
\[
\mathcal P=\{\vec{x}\in\mathbb{R}^n \mid A \vec{x} \preceq \vec{b},\; C\vec{x}=d\}
\]
defines a polyhedron. Balls and ellipsoids $B(\vec{x}_c,r)=\{\vec{x}:\|\vec{x}-\vec{x}_c\|\leq r\}$ and
$E(\vec{x}_c,P)=\{\vec{x}:(\vec{x}-\vec{x}_c)^\top P^{-1}(\vec{x}-\vec{x}_c)\leq 1\}$ with $P\succ 0$ are convex as well.
Thus quadratic optimization problems are best understood in the setting of convex sets and convex functions.

In general, the intersection of any number of convex sets is convex.
\subsection{Convexity, convex hulls, and duality}
\index{key}{convex sets}\index{key}{convex optimization}

Convex geometry plays a central role in optimization. Faces of convex sets 
always lie on the topological boundary of the set. The fundamental 
operation is the \emph{convex hull} of a collection of points. 

\paragraph{Convex hull.}
Given points $\vec{x}_1,\dots,\vec{x}_n\in\mathbb{R}^d$, their convex hull is
\begin{equation}
\operatorname{ch}(\vec{x}_1,\dots,\vec{x}_n)
=\left\{\sum_{i=1}^n \theta_i \vec{x}_i \;\Big|\; \theta_i\ge 0, \ \sum_{i=1}^n \theta_i=1\right\}.
\label{eq:convexhull}
\end{equation}
Every convex hull has nonempty interior provided the generating points span
$\mathbb{R}^d$.\footnote{See R.~Rockafellar, \emph{Convex Analysis}, Princeton Univ.~Press (1970).}

A classical result of Minkowski--Carathéodory states that if $\mathcal{S}$ is a compact convex
subset of $\mathbb{R}^d$, then any $\vec{x}\in\mathcal{S}$ can be expressed as a convex combination
of at most $d+1$ extreme points.\footnote{See S.~Boyd and L.~Vandenberghe,
\emph{Convex Optimization}, Cambridge Univ.~Press (2004).}

\paragraph{Differential conditions for convexity.}
For smooth functions $f:\mathbb{R}^n\to\mathbb{R}$:
\begin{itemize}
\item First-order condition: $f$ is convex iff
\[
f(\vec{y})\;\geq\; f(\vec{x})+\nabla f(\vec{x})^\top(\vec{y}-\vec{x}),\quad \forall \vec{x},\vec{y}\in\mathbb{R}^n.
\]
\item Second-order condition: if $f$ is twice differentiable, then convexity is equivalent to
\[
\nabla^2 f(\vec{x})\succeq 0\quad \forall \vec{x}.
\]
\end{itemize}
For example:
\begin{itemize}
\item $f(\vec{x})=\tfrac{1}{2}\vec{x}^\top P \vec{x}+q^\top \vec{x}+r$ has $\nabla f(\vec{x})=P\vec{x}+q$, $\nabla^2 f(\vec{x})=P$.
\item $f(\vec{x})=\|A\vec{x}-\vec{b}\|^2$ has $\nabla f(\vec{x})=2A^\top(A\vec{x}-\vec{b})$, $\nabla^2 f(\vec{x})=2A^\top A\succeq 0$.
\end{itemize}

\paragraph{Jensen’s inequality.}
Another key characterization: if $f$ is convex, then
\begin{equation}
f(\theta \vec{x}+(1-\theta)\vec{y}) \leq \theta f(\vec{x})+(1-\theta)f(\vec{y}),\quad 0\leq \theta\leq 1,
\end{equation}
or equivalently,
\begin{equation}
f(\mathbb{E}[\vec{X}]) \;\leq\; \mathbb{E}[f(\vec{X})].
\end{equation}

\paragraph{Operations preserving convexity.}
Convexity is preserved under: 
\begin{itemize}
\item nonnegative weighted sums, 
\item composition with affine maps, 
\item pointwise maxima and suprema, 
\item minimization over some variables,
\item perspective transformations. 
\end{itemize}
For instance, if $f$ is convex, then so is $f(A\vec{x}+\vec{b})$. Likewise, 
\[
g(\vec{x})=\max\{g_1(\vec{x}),\dots,g_m(\vec{x})\}
\]
is convex whenever each $g_i$ is convex. As an example, the spectral function
\begin{equation}
g(X)=\lambda_{\max}(X)=\max_{\|\vec{z}\|=1} \vec{z}^\top X \vec{z}
\end{equation}
is convex in $X$. This connects with Rayleigh’s principle: the inequality
\[
\lambda_{\max}(X)\le r
\]
is equivalent to $rI-X\succeq 0$.

\paragraph{Duality.}
Consider the constrained minimization problem
\begin{align}
\text{minimize} \quad & f(\vec{x}) \nonumber \\
\text{subject to} \quad & f_i(\vec{x})\leq 0,\ \ h_j(\vec{x})=0.
\label{eq:minprob}
\end{align}
The associated Lagrangian is
\begin{equation}
L(\vec{x},\lambda,\nu) = f(\vec{x})+\sum_i \lambda_i f_i(\vec{x}) + \sum_j \nu_j h_j(\vec{x}),
\label{eq:lagdual}
\end{equation}
with multipliers $\lambda\succeq 0$. The Lagrangian dual function is defined as
\[
g(\lambda,\nu) = \inf_{\vec{x}} L(\vec{x},\lambda,\nu),
\]
which is always concave in $(\lambda,\nu)$, even if $f$ is not convex.
Maximizing $g(\lambda,\nu)$ subject to $\lambda\succeq 0$ gives the 
dual problem, itself a convex optimization problem.

\paragraph{Weak and strong duality.}
Weak duality always holds:
\[
f(\tilde{\vec{x}})\;\geq\; L(\tilde{\vec{x}},\lambda,\nu)\;\geq\; g(\lambda,\nu).
\]
Strong duality (equality of primal and dual optimal values) holds under 
regularity conditions such as Slater’s constraint qualification (existence of a strictly feasible point).\footnote{See Boyd \& Vandenberghe, \emph{Convex Optimization}, Ch.~5.}

\paragraph{Example: two-way partitioning.}
Duality methods are powerful even for nonconvex problems. Consider
\begin{equation}
\min_{\vec{x}\in\{\pm 1\}^n} \vec{x}^\top W \vec{x}, 
\end{equation}
which is equivalent to minimizing over quadratic forms with constraints $x_i^2=1$
(ground states of Ising models). The Lagrangian dual becomes
\begin{align}
g(\nu) &= \inf_{\vec{x}} \left(\vec{x}^\top W \vec{x} + \sum_i \nu_i(x_i^2-1)\right) \nonumber \\
&= \inf_{\vec{x}} \left(\vec{x}^\top (W+\operatorname{diag}(\nu))\vec{x}\right) - \mathbf{1}^\top \nu \nonumber \\
&=\begin{cases}
-\,\mathbf{1}^\top \nu, & W+\operatorname{diag}(\nu)\succeq 0,\\
-\infty, & \text{otherwise}.
\end{cases}
\end{align}
This relaxation gives lower bounds on the original combinatorial optimization problem.
\subsection{Convexity-preserving transformations and advanced optimization}
\index{key}{convex optimization}\index{key}{convexity-preserving transformations}

Beyond first- and second-order conditions, there are several powerful tools
for proving convexity of matrix-valued functions and for transforming
nonconvex problems into convex ones.

\paragraph{Convexity via line restrictions.}
A fundamental theorem states that a function $f:\mathbb{R}^n\to\mathbb{R}$ is convex 
if and only if, for all $M_1,M_2$, the univariate function
\begin{equation}
g(t)=f(M_1+t M_2),\quad t\in\mathbb{R},
\end{equation}
is convex. As an example, consider
\begin{equation}
f(M)=\log\det(M),\quad M\succ 0.
\end{equation}
Then
\begin{align}
g(t)&=\log\det(M_1+tM_2) \\
&=\log\det(M_1)+\log\det\!\left(I+t\,M_1^{-1/2}M_2M_1^{-1/2}\right)\nonumber\\
&=\log\det(M_1)+\sum_i \log(1+t\lambda_i), \nonumber
\end{align}
where $\lambda_i$ are the eigenvalues of $M_1^{-1/2}M_2M_1^{-1/2}$. Since each
$\log(1+t\lambda_i)$ is convex, $g$ is convex, hence $f(M)=\log\det(M)$ is convex.\footnote{See S.~Boyd and L.~Vandenberghe, \emph{Convex Optimization}, Cambridge Univ.~Press (2004).}\paragraph{Barrier functions.}
Functions such as
\begin{equation}
f(\vec{x})=-\sum_i \log(b_i-a_i^\top \vec{x}),
\end{equation}
are convex because $-\log(\cdot)$ is convex and convexity is preserved under affine composition and positive summation. Such functions are used as \emph{barriers} in interior-point methods.

\paragraph{Perspective transformation.}
If $g$ is convex, then its perspective
\begin{equation}
g(\vec{x},t)=t\,g(\vec{x}/t),\quad t>0,
\end{equation}
is convex in $(\vec{x},t)$. This device connects entropy and information measures:
for example, applying it to Shannon entropy shows that the Kullback--Leibler divergence is convex.

\paragraph{Conjugate functions.}
The Fenchel conjugate of $f$,
\begin{equation}
f^*(\vec{y})=\sup_{\vec{x}} \big(\vec{y}^\top \vec{x}-f(\vec{x})\big),
\end{equation}
is always convex, regardless of $f$.\footnote{See R.~Rockafellar, \emph{Convex Analysis}, Princeton Univ.~Press (1970).}  
For quadratic $f(\vec{x})=\tfrac{1}{2}\vec{x}^\top Q\vec{x}$, one finds
\begin{equation}
f^*(\vec{y})=\tfrac{1}{2} \vec{y}^\top Q^{-1}\vec{y}.
\end{equation}

\paragraph{Geometric programming.}
Certain nonconvex problems can be transformed into convex form.  
Consider the program
\begin{align}
\text{minimize}\quad & \sum_k c_k \prod_j x_j^{\alpha_{j,k}} \\
\text{subject to}\quad & f_i(\vec{x})\le 1,\quad h_i(\vec{x})=1,\nonumber
\end{align}
with $c_k>0$, $\alpha_{j,k}\in\mathbb{R}$. Substituting $x_j=e^{y_j}$ yields
\begin{align}
\text{minimize}\quad & \log\!\left(\sum_k \exp\!\left(\sum_j \alpha_{j,k} y_j + \log c_k\right)\right) \\
\text{subject to}\quad & \log f_i(e^{\vec{y}})\le 0,\quad \log h_i(e^{\vec{y}})=0, \nonumber
\end{align}
which is convex.

\paragraph{Linear programming.}
A central result of convex analysis: minimizing a linear functional over a convex polytope attains its optimum at an extreme point. A linear program has the form
\begin{align}
\text{minimize}\quad & c^\top \vec{x}+d \\
\text{subject to}\quad & G\vec{x}\preceq h,\quad A\vec{x}=b. \label{eq:linearprogram}
\end{align}
This principle underlies the geometry of linear programming and will later be used in the proof of the Wielandt--Hoffman theorem.

\paragraph{Semi-definite programming (SDP).}
Linear programs can be generalized to matrix inequalities:
\begin{align}
\text{minimize}\quad & c^\top \vec{x}+d \nonumber \\
\text{subject to}\quad & F(\vec{x})=\sum_{i}x_i F_i+G \preceq 0,\quad A\vec{x}=b,
\end{align}
where $F_i$ are symmetric matrices. The matrix inequality constraint $F(\vec{x})\preceq 0$ is a \emph{linear matrix inequality} (LMI).  
Linear programs are a special case, since $G\vec{x}\preceq h$ can be written as $\operatorname{diag}(G\vec{x}-h)\preceq 0$.  

Eigenvalue optimization is naturally embedded in SDP form. For example:
\begin{equation}
\min \lambda_{\max}(A(\vec{x})),\quad A(\vec{x})=A_0+\sum_i x_i A_i
\end{equation}
is equivalent to
\begin{align}
\text{minimize}\quad & t \\
\text{subject to}\quad & A(\vec{x})-tI\preceq 0.
\end{align}
Similarly, spectral norm minimization
\[
\min \|A(\vec{x})\|_2=\sqrt{\lambda_{\max}(A(\vec{x})A(\vec{x})^\top)}
\]
has an SDP relaxation
\begin{align}
\text{minimize}\quad & t \\
\text{subject to}\quad & 
\begin{pmatrix}
tI & A(\vec{x}) \\ A(\vec{x})^\top & tI
\end{pmatrix}
\succeq 0,
\end{align}
using the Schur complement.

\paragraph{Example: D-optimal design.}
Consider the problem
\begin{align}
\text{minimize}\quad & \log\det\!\left(\sum_{i=1}^p x_i v_i v_i^\top\right)^{-1} \\
\text{subject to}\quad & x_i\ge 0,\quad \sum_i x_i=1,
\end{align}
which minimizes the volume of an ellipsoid determined by vectors $\{v_i\}$.  
Its dual can be written
\begin{align}
\text{maximize}\quad & \log\det W+n\log n \\
\text{subject to}\quad & v_i^\top W v_i \le 1,\quad i=1,\dots,p,
\end{align}
corresponding to the minimum-volume ellipsoid centered at the origin.  
The Lagrangian derivation connects determinant minimization to convex duality and ellipsoidal approximations.
\section{Matrix iterations and fixed points}\label{sec:matrixiterations}
\index{key}{iterative methods}
\index{key}{fixed points}

This section introduces iterative methods used to solve either the minimization of a quadratic functional
or, equivalently, to compute the solution of a linear system of equations.  
While these methods are described in some detail, we do not provide a full convergence-rate analysis,
which belongs to numerical analysis. Instead, we focus on the underlying ideas and why they work.

\subsection{Linear iterative methods}
We begin with iterations of the form
\begin{equation}
\vec{x}^{(k+1)} = A\,\vec{x}^{(k)} + \vec{b},
\label{eq:itmethod}
\end{equation}
and ask: does \(\vec{x}^{(k)}\) converge to a fixed point \(\vec{x}^\ast\) as \(k\to\infty\)?
Formally, we say that \(\vec{x}^{(k)}\to \vec{x}^\ast\) if
\[
\lim_{k\to\infty}\big\|\vec{x}^{(k)}-\vec{x}^\ast\big\|=0,
\]
for some vector norm \(\|\cdot\|\).

\paragraph{Consistency.}
An iteration is \emph{consistent}\index{key}{consistent iterations} if there exists \(\vec{x}^\ast\) such that
\begin{equation}
\vec{x}^\ast = A\,\vec{x}^\ast + \vec{b}.
\label{eq:itmethodas}
\end{equation}
In this case, \(\vec{x}^\ast = (I-A)^{-1}\vec{b}\).
Thus, consistency requires that \(I-A\) be invertible, i.e.\ \(1\notin \Lambda(A)\).
(Convergence of the iteration additionally requires \(\rho(A)<1\); see below.)

\paragraph{Error propagation.}
Let \(\vec{q}^{(k)}=\vec{x}^{(k)}-\vec{x}^\ast\) denote the error at step \(k\).
Then
\begin{equation}
\vec{q}^{(k+1)} = A\,\vec{q}^{(k)},
\end{equation}
so that
\[
\vec{q}^{(k)} = A^{k}\vec{q}^{(0)}, \qquad 
\big\|\vec{q}^{(k)}\big\| \le \|A^{k}\|\,\big\|\vec{q}^{(0)}\big\|.
\]
Hence \(\vec{q}^{(k)}\to 0\) provided \(\rho(A)<1\).
This simple scheme is the \emph{Richardson iteration}\index{key}{Richardson iteration}.

\paragraph{Convergence factors.}
The quantity \(\|A^{k}\|\) is called the \emph{convergence factor after \(k\) iterations}.
The asymptotic average convergence factor is
\[
q_k=\|A^{k}\|^{1/k}, \qquad \lim_{k\to\infty} q_k = \rho(A),
\]
while the \emph{average convergence rate} is
\[
R_k(A)=-\frac{1}{k}\log\|A^{k}\|.
\]

\subsubsection{Jacobi and Gauss–Seidel methods}
Suppose we want to solve
\begin{equation}
A\vec{x}=\vec{b}.
\label{eq:jacobi}
\end{equation}
Split \(A=D+N\), where \(D\) is easily invertible (typically the diagonal of \(A\)).
Then the iteration
\begin{equation}
\vec{x}^{(k+1)} = D^{-1}\!\left(\vec{b}-N\,\vec{x}^{(k)}\right)
\end{equation}
is the \emph{Jacobi method}\index{key}{Jacobi method}.
Convergence requires \(\rho(-D^{-1}N)<1\), which holds, for instance, if \(A\) is (strictly) diagonally dominant.

The \emph{Gauss–Seidel method}\index{key}{Gauss-Seidel method} uses the splitting \(A=L+U\),
where \(L\) is lower triangular (including the diagonal) and \(U\) is strictly upper triangular:
\begin{equation}
\vec{x}^{(k+1)} = L^{-1}\!\left(\vec{b}-U\,\vec{x}^{(k)}\right).
\end{equation}
Again, convergence requires \(\rho(I-L^{-1}A)<1\).
In practice, Gauss–Seidel often converges faster than Jacobi.

More generally, for any splitting \(A=M+N\) with \(M\) invertible, define the
\emph{stationary iterative method}\index{key}{stationary iterative methods}
\begin{equation}
\vec{x}^{(k+1)} = -\,M^{-1}N\,\vec{x}^{(k)} + M^{-1}\vec{b}.
\end{equation}

\subsubsection{Projection methods}
Another important family are projection methods, including Galerkin and Petrov–Galerkin schemes.
The idea is to approximate the solution of \eqref{eq:jacobi} in a subspace,
while imposing orthogonality of the residual against another subspace.

Let \(\vec{x}^{(0)}\) be an initial guess. Seek \(\vec{x}=\vec{x}^{(0)}+V\vec{y}\) with \(\vec{y}\in\mathbb{R}^m\),
where \(V\in\mathbb{R}^{n\times m}\) spans a trial subspace \(\mathcal{V}\).
Impose the Petrov–Galerkin condition
\begin{equation}
A\vec{x}-\vec{b} \ \perp\ \mathcal{K},
\end{equation}
where \(\mathcal{K}\) is a test subspace with basis \(K\).
Equivalently, for the residual \(\vec{r}_0=\vec{b}-A\vec{x}^{(0)}\), the coefficients satisfy
\begin{equation}
K^{\top} A V \,\vec{y}=K^{\top} \vec{r}_0.
\end{equation}
Thus the updated approximation is
\begin{equation}
\vec{x}^{(1)} = \vec{x}^{(0)} + V\,(K^{\top} A V)^{-1}K^{\top} \vec{r}_0,
\end{equation}
and, more generally, the iteration
\begin{equation}
\vec{x}^{(k+1)}=\vec{x}^{(k)} + V\,(K^{\top} A V)^{-1}K^{\top}\big(\vec{b}-A\vec{x}^{(k)}\big).
\end{equation}

\paragraph{When does it work?}
If \(A\) is symmetric positive definite and \(\mathcal{V}=\mathcal{K}\), then \(K^{\top}AV\) is invertible
(the classical Galerkin case).
If \(\mathcal{K}=A\mathcal{V}\) (the \emph{oblique projection} case), the system matrix is again invertible
provided \(A\) is nonsingular.

\paragraph{Minimization interpretation.}
Projection methods can be seen as solving
\[
\min_{\vec{x}\in \vec{x}^{(0)}+\mathcal{V}} \ \|A\vec{x}-\vec{b}\|_2,
\]
or, more generally, minimizing quadratic functionals over subspaces.
This connects them directly to iterative methods for optimization.

For details, see classical numerical analysis texts\footnote{Y.~Saad, \emph{Iterative Methods for Sparse Linear Systems}, SIAM (2003). \newline
R.~Varga, \emph{Matrix Iterative Analysis}, Springer (2000).}.
A similar argument applies if, instead of minimizing \(\|A\vec{x}-\vec{b}\|_2\) over \(\vec{x}\in \vec{x}^{(0)}+\mathcal{V}\),
we ask for a solution in \(\mathcal{K}\) such that the residual is orthogonal to \(\mathcal{V}=A\mathcal{K}\).  
In that case one must have
\begin{equation}
\langle \vec{b}-A\vec{y}, \vec{k} \rangle = 0 \quad \forall \vec{k} \in A\mathcal{K},
\end{equation}
which is another form of Petrov–Galerkin projection.

\paragraph{Comparison with gradient methods.}
To compare with unconstrained methods, recall the \emph{steepest descent algorithm}\index{key}{steepest descent}, 
which minimizes the functional
\[
F(\vec{x})=\|A\vec{x}-\vec{b}\|^2,
\]
for a symmetric positive definite \(A\).
Starting from \(\vec{x}^{(0)}\), one iterates
\begin{equation}
\vec{x}^{(k+1)}=\vec{x}^{(k)}-\gamma\, \nabla F(\vec{x}^{(k)}),
\end{equation}
with \(\gamma>0\) a step size. Since
\[
\nabla F(\vec{x})=2A^{\top}\big(A\vec{x}-\vec{b}\big),
\]
this scheme uses the full gradient direction.  
By contrast, projection methods (Galerkin, Petrov–Galerkin) use projected search directions
via operators such as \(V\,(K^{\top}AV)^{-1}\!K^{\top}\), and hence can accelerate convergence by restricting
to subspaces where the residual is controlled.

\subsubsection{Krylov subspace methods}\label{sec:krylov}
Projection methods become especially powerful when combined with \emph{Krylov subspaces}, which
naturally arise in linear algebra and control theory.  

Given a nonsingular $A\in\mathbb{R}^{n\times n}$ and $\vec b\in\mathbb{R}^n$, the solution to $A\vec x=\vec b$
lies in a Krylov subspace whose dimension is the degree of the minimal polynomial of $A$.
Indeed, by the Cayley–Hamilton theorem\index{key}{Cayley-Hamilton theorem}, $A$ satisfies a polynomial
$q(A)=0$ of degree $m$, so that
\begin{equation}
A^{-1}=-\sum_{i=0}^{m-1}\alpha_{i+1}A^i,
\end{equation}
and hence
\begin{equation}
\vec x=A^{-1}\vec b\in \text{span}\{\vec b, A\vec b,\ldots,A^{m-1}\vec b\}.
\end{equation}

\paragraph{Definition.} The $m$-th Krylov subspace generated by $A$ and $\vec b$ is
\begin{equation}
\mathcal K_m(A,\vec b)=\text{span}\{\vec b, A\vec b, \ldots, A^{m-1}\vec b\}.
\end{equation}
Its dimension is at most $n$, and stabilizes once $A^m\vec b$ can be expressed as a linear
combination of previous powers, yielding the \emph{minimal polynomial} of $(A,\vec b)$.

\paragraph{Companion matrices.}
If $\mathcal K_m(A,\vec b)$ has dimension $m$, then
\begin{equation}
A^m\vec b=\sum_{j=0}^{m-1}\alpha_j A^j\vec b,
\end{equation}
which corresponds to the Frobenius companion matrix\index{key}{Frobenius companion} $F_m(\alpha)$,
showing that the restriction of $A$ to $\mathcal K_m$ has Hessenberg form.

\paragraph{Arnoldi iteration.}
This motivates projecting $A$ onto $\mathcal K_m(A,\vec b)$ via an orthonormal basis
$\{\vec q_1,\ldots,\vec q_m\}$ obtained from $\{A^j\vec b\}$ by Gram–Schmidt orthogonalization.  
The Arnoldi iteration\index{key}{Arnoldi iteration} constructs $Q_m=(\vec q_1,\ldots,\vec q_m)$ such that
\begin{equation}
AQ_m=Q_{m+1}H_m,
\end{equation}
where $H_m$ is an $(m+1)\times m$ upper Hessenberg matrix.  
Thus, $Q_m^{\top}AQ_m=H_m$, reducing $A$ to Hessenberg form.  
This procedure is numerically stabilized by the modified Gram–Schmidt method.

\paragraph{Lanczos method.}
If $A$ is Hermitian, Arnoldi iteration simplifies to the \emph{Lanczos method}\index{key}{Lanczos method},
where $H_m$ becomes tridiagonal. The recurrence
\begin{equation}
A\vec q_j=\beta_{j-1}\vec q_{j-1}+\alpha_j\vec q_j+\beta_j\vec q_{j+1}
\end{equation}
is the \emph{Lanczos relation}\index{key}{Lanczos relation}.  
Diagonalizing the tridiagonal matrix $H_m$ yields Ritz approximations to eigenvalues of $A$.

\paragraph{GMRES.}
Krylov subspaces also underlie iterative solvers for $A\vec x=\vec b$.  
The \emph{Generalized Minimal Residual method} (GMRES)\index{key}{GMRES} seeks $\vec x\in \vec x^0+\mathcal K_m(A,\vec r_0)$
such that the residual $A\vec x-\vec b$ has minimal Euclidean norm.  
Writing $A Q_m=Q_{m+1}H_m$, one reduces the problem to
\[
\min_{\vec y}\|\vec r_0+H_m\vec y\|_2,
\]
which can be solved efficiently since $H_m$ is small.  
Thus GMRES combines projection ideas with Krylov subspace structure.

\paragraph{Connection to orthogonal polynomials.}
The tridiagonal structure of the Lanczos method is identical to the recurrence
satisfied by orthogonal polynomials. In fact, the Lanczos process can be viewed
as generating such polynomials relative to the spectral measure of $A$,
a perspective that connects Krylov methods to approximation theory.

For further details, see standard texts\footnote{Y.~Saad, \emph{Iterative Methods for Sparse Linear Systems}, SIAM (2003). \newline
C.~Trefethen and L.~N.~Trefethen, \emph{Numerical Linear Algebra}, SIAM (1997).}.

\index{key}{Generalized minimum residual method} \index{key}{GMRES}

\subsection{Line search}

We now discuss in detail the \emph{steepest descent} and the \emph{conjugate gradient} methods.  
Both are instances of iterative schemes known as \textit{line search} methods, where one updates
\begin{equation}
    \vec x^{\,k+1}=\vec x^{\,k}+\alpha_k \vec r^{\,k},
    \label{eq:linesearch1}
\end{equation}
with a step length $\alpha_k$ chosen at each iteration and a search direction $\vec r^k$ depending on the method.  

Our goal is to solve $A\vec x=\vec b$. As seen earlier, this problem is equivalent to minimizing the quadratic functional
\begin{equation}
    E(\vec x)=\tfrac{1}{2}\vec x^{\top} A \vec x - \vec x^{\top} \vec b,
\end{equation}
up to an irrelevant additive constant. The gradient is
\begin{equation}
    \nabla E(\vec x)=\tfrac{1}{2}(A^{\top}+A)\vec x-\vec b.
\end{equation}
If $A$ is symmetric, this simplifies to $\nabla E(\vec x)=A\vec x-\vec b$, showing that the solution of $A\vec x=\vec b$ is exactly the critical point of $E(\vec x)$.  
When $A$ is symmetric positive definite, $E(\vec x)$ is strictly convex, ensuring uniqueness of the minimizer.

\subsubsection{Steepest descent}

The method of steepest descent starts from an initial guess $\vec x^0$ and moves in the direction of the negative gradient:
\[
\vec r^k=-\nabla E(\vec x^k)=\vec b-A\vec x^k,
\]
which is exactly the residual of the linear system. The update rule is
\begin{equation}
    \vec x^{\,k+1}=\vec x^{\,k}+\alpha_k \vec r^k.
\end{equation}

The optimal step size $\alpha_k$ is chosen to minimize $E(\vec x^k+\alpha \vec r^k)$.  
Differentiating yields
\begin{equation}
    \alpha_k=\frac{\vec r^k\cdot \vec r^k}{\vec r^k\cdot A\vec r^k}.
\end{equation}
A key property follows: successive gradients are orthogonal,
\[
\nabla E(\vec x^{\,k+1}) \cdot \nabla E(\vec x^{\,k})=0.
\]
Thus, steepest descent follows a zig-zag path towards the solution, often resulting in slow convergence.

\begin{figure}
    \centering
    \includegraphics[scale=0.4]{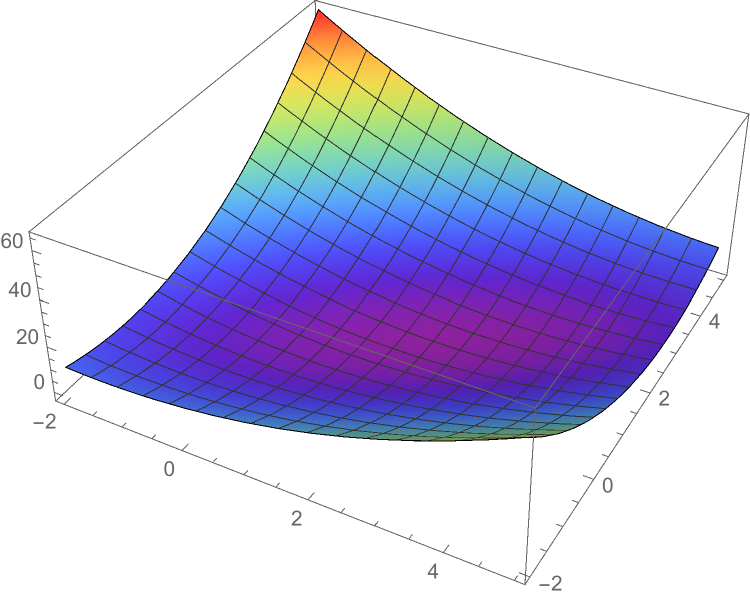} \quad
    \includegraphics[scale=0.3]{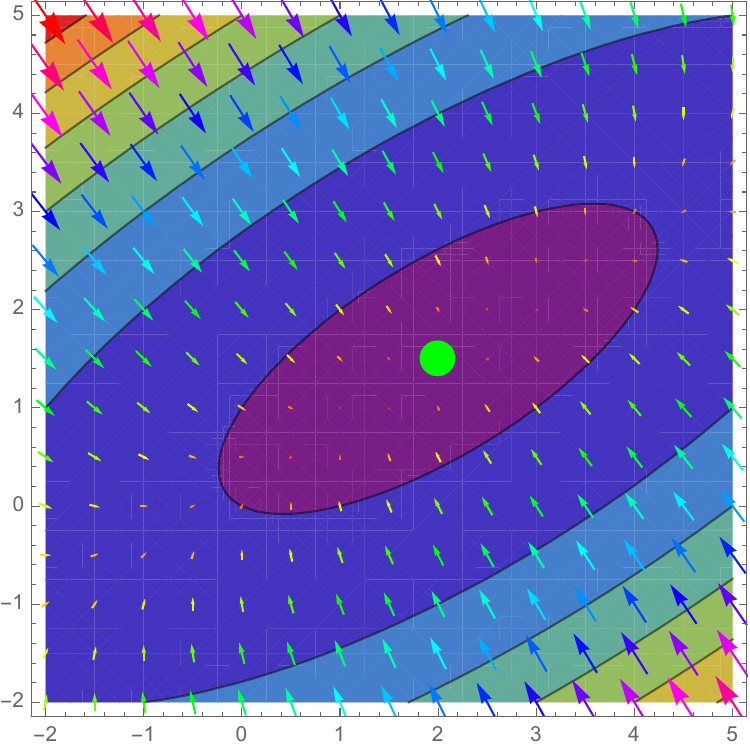}
    \caption{Steepest descent iterations for $E(x,y)=x^2+2y^2-2xy-x-2y$. 
    The gradient direction $-\nabla E(x,y)$ is orthogonal to the contour lines, leading to a characteristic zig-zag trajectory.}
    \label{fig:graddescent}
\end{figure}

\subsubsection{Conjugate gradient}

The \emph{conjugate gradient method}\index{key}{conjugate gradient method} improves upon steepest descent by replacing orthogonality of gradients with \textit{conjugacy with respect to $A$}.  
Two vectors $\vec d^i$ and $\vec d^j$ are said to be $A$-conjugate if
\[
(\vec d^i)^{\top} A \vec d^j=0 \quad (i\neq j).
\]

The iteration is
\begin{equation}
    \vec x^{\,k+1}=\vec x^{\,k}+\alpha_k \vec d^k,
    \label{eq:cd1}
\end{equation}
with $\alpha_k$ chosen so that the residual is orthogonal to $\vec d^k$:
\begin{equation}
    \alpha_k=\frac{(\vec r^k)^{\top} \vec d^k}{(\vec d^k)^{\top} A \vec d^k}.
\end{equation}

\paragraph{Key result.}
Because successive directions are chosen $A$-conjugate, the method converges \emph{exactly in at most $n$ steps} (for an $n\times n$ SPD matrix), in exact arithmetic.  
This is in stark contrast to steepest descent, which may require many more iterations.

\paragraph{Constructing conjugate directions.}
Starting from $\vec d^0=\vec r^0$, the search directions are built by Gram–Schmidt $A$-orthogonalization:
\begin{equation}
    \vec d^i=\vec r^i+\sum_{j=0}^{i-1}\beta_{ij}\vec d^j, 
    \qquad 
    \beta_{ij}=-\frac{(\vec r^i)^{\top} A \vec d^j}{(\vec d^j)^{\top} A \vec d^j}.
\end{equation}
This ensures $A$-conjugacy of all $\vec d^i$.

\paragraph{Practical considerations.}
In finite precision arithmetic, rounding errors may spoil perfect conjugacy, so convergence may deteriorate. Nonetheless, in practice conjugate gradient is one of the most efficient iterative solvers for large, sparse SPD systems. Its popularity comes from combining fast convergence with modest memory requirements.

\begin{figure}
    \centering
    \includegraphics[scale=0.4]{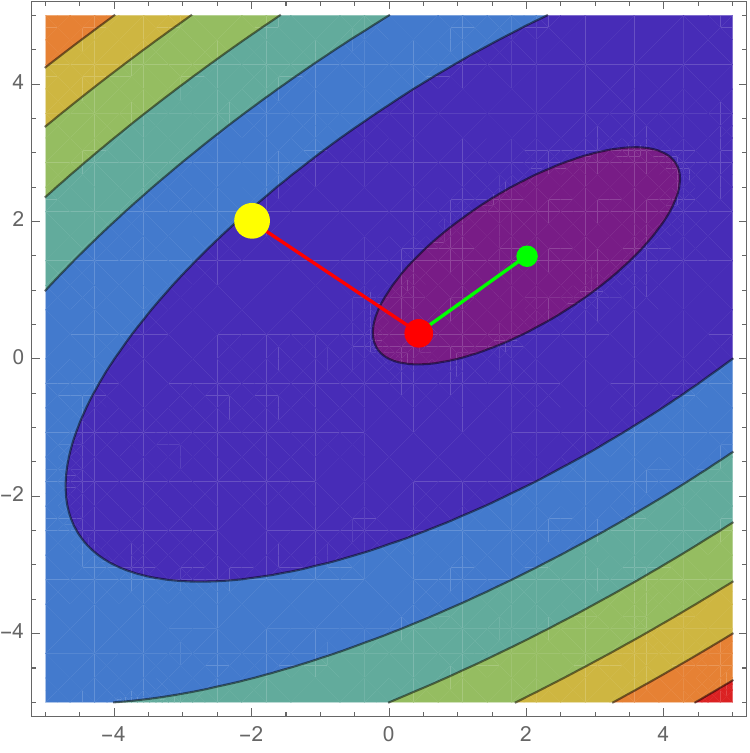}
    \caption{Conjugate gradient applied to $E(x,y)=x^2+2y^2-2xy-x-2y$.  
    Starting from $\vec x=(-2,2)$ (yellow), the method converges to the minimum in two iterations by following $A$-conjugate directions.}
    \label{fig:conjgrad}
\end{figure}

\subsubsection{The von Neumann--Ulam scheme: Monte Carlo sampling} 
\index{key}{von Neumann-Ulam linear solver}

A less well-known but elegant method to solve linear systems was introduced by von Neumann and Ulam, and later popularized by Forsythe and Leibler. The idea is to interpret the Neumann series expansion of the inverse as the expectation of a random walk process.

Consider the linear system
\begin{equation}
    \vec x=A\vec x+\vec b,
\end{equation}
which, provided $(I-A)$ is invertible, has the solution
\begin{equation}
    \vec x=(I-A)^{-1}\vec b.
\end{equation}
If $\rho(A)<1$, the Neumann expansion
\begin{equation}
    (I-A)^{-1}=I+A+A^2+\cdots
\end{equation}
is convergent. The key observation is that each term $A^m$ can be represented as the contribution of random walks of length $m$ on a weighted graph.

---

\paragraph{Random walk interpretation.}
Let $A\in \mathbb R^{n\times n}$ be nonnegative. We introduce a stochastic matrix $P$ with $\sum_j P_{ij}=p_i\leq 1$, and define
\[
A = P\circ V,
\]
where $\circ$ denotes the Hadamard product and $V_{ij}$ is a weight matrix. To complete the random walk description we add an absorbing state $n+1$, with transition matrix
\begin{equation}
    \tilde P=\begin{pmatrix} P & 1-\vec p \\ 0 & 1 \end{pmatrix},
\end{equation}
where $1-\vec p=(1-p_1,\dots,1-p_n)^{\top}$. Thus, a walk continues within $\{1,\dots,n\}$ according to $P$, but may terminate with probability $1-p_i$ in the absorbing state.

A trajectory of the walk starting at $i$ is given by $(i,i_0,i_1,\dots,i_m,k)$, with $k$ the terminal state before absorption. We define the payoff variable
\begin{equation}
    W_{ij}=\begin{cases}
    \dfrac{V_{i i_0}V_{i_0 i_1}\cdots V_{i_m j}}{1-p_j}, & \text{if the trajectory ends in $j$,} \\
    0, & \text{otherwise}.
    \end{cases}
\end{equation}
The absorbing boundary ensures trajectories almost surely terminate.

---

\paragraph{Main result.}
Taking expectations over all possible random walks,
\begin{eqnarray}
    \langle W_{ij}\rangle 
    &=& \delta_{ij}+\sum_{m=1}^\infty \sum_{i_1,\dots,i_m} 
        V_{i i_1}\cdots V_{i_m j}\;
        P_{i i_1}\cdots P_{i_m j} \nonumber \\
    &=& \delta_{ij}+\sum_{m=1}^\infty (A^m)_{ij} \nonumber \\
    &=& (I-A)^{-1}_{ij}.
\end{eqnarray}
Thus the inverse matrix entries are recovered as expectations of the random walk payoff $W_{ij}$.

---

\paragraph{Solving the linear system.}
The solution of $A\vec x+\vec b=\vec x$ can then be written as
\begin{equation}
    x_i=\sum_j W^s_{ij} b_j,
\end{equation}
where $W^s_{ij}$ denotes the Monte Carlo estimate of $(I-A)^{-1}_{ij}$ obtained from a finite number of simulated walks. With sufficient sampling, the law of large numbers guarantees convergence.

---

\paragraph{Generalization to matrix functions.}
The method extends beyond the inverse. If the payoff is modified by including a factor $f_m$ depending on the walk length $m$, namely
\begin{equation}
    W_{ij}=\begin{cases}
    f_m \dfrac{V_{i i_0}\cdots V_{i_m j}}{1-p_j}, & \text{if the trajectory ends in $j$,}\\
    0, & \text{otherwise},
    \end{cases}
\end{equation}
then
\begin{equation}
    \langle W_{ij}\rangle=\delta_{ij}+\sum_{m=1}^\infty f_m (A^m)_{ij}\equiv f(A)_{ij}.
\end{equation}
This provides a general Monte Carlo framework for evaluating functions of matrices.

---

The von Neumann–Ulam scheme offers a probabilistic method for approximating solutions of large linear systems and matrix functions, especially useful when $A$ is sparse and high-dimensional. While the method is not always competitive with deterministic solvers in terms of convergence speed, it is embarrassingly parallel and conceptually important in the development of \emph{probabilistic numerical linear algebra}.

\index{key}{von Neumann-Ulam inverse}
\begin{tcolorbox}
\rule{\textwidth}{0.4pt}
\begin{verbatim}
% We calculate the inverse of I-A
% function InvM=vonNeumannUlam(A,nm)
% nm= Monte Carlo samples
function InvM=vonNeumannUlam(A,nm)
n=size(A,1);
% I choose a trivial transition probability, 
% uniform among the urns and the exit urn
% I do not use the P below but it 
% is equivalent to the process we consider below
% P=ones(n+1);
% P=1/(n+1)*P; 
% P(n+1,1:n)=0;
% P(n+1,n+1)=1;
%%%%%%%%%%
% This is the reward matrix according to the game
W=A*(n+1);
InvM=zeros(n);
% I play the game for every starting element of matrix
for i=1:n
        for nmc=1:nm
            % I initialize the reward
            w=1;
            q=i;
            while (q<n+1)
            % I pick a jump with random prob
            qn=floor(rand*(n+1))+1;
          %  display(qn)
             if (qn<n+1)
                % If not in abs. state, 
                % I update the reward string
                w=w*W(q,qn);
                q=qn;
             else
                    % I am in the right urn, then 
                    % I add the reward to my tentative
                    % inverse, the factor (n+1) comes 
                    % from the exit probability
                    InvM(i,q)=InvM(i,q)+w*(n+1);
                    % this gets me out of the while loop
                    q=qn;
             end
            end
        end
end
% We normalize the matrix
InvM=InvM/nm;
end
\end{verbatim}
\rule{\textwidth}{0.4pt}

\end{tcolorbox}

\section{Nonlinear systems: iterative methods and fixed points}

Up to this point, our discussion of iterative algorithms has focused mainly on 
\emph{linear} problems, such as solving $A\vec x=\vec b$ or minimizing quadratic forms. 
However, in applications across physics, engineering, economics, and computer science, 
one almost always encounters \emph{nonlinear} systems of equations. These can take the form
\[
f_i(\vec x)=0, \qquad i=1,\dots,m,
\]
where the functions $f_i$ are not linear in $\vec x$.  
Nonlinear systems arise naturally in optimization, dynamical systems, equilibrium models, 
nonlinear differential equations, and control theory.  

Unlike linear problems, where powerful direct methods (LU decomposition, QR factorization, Krylov subspaces) 
or iterative solvers (Jacobi, Gauss–Seidel, GMRES) exist, nonlinear systems rarely admit closed-form solutions. 
As a result, one must resort to iterative approximation schemes. The basic philosophy is to 
\emph{linearize} the nonlinear system around a current estimate, solve the resulting linear problem, 
and update the solution iteratively. This naturally connects nonlinear methods to the linear machinery 
developed earlier: Jacobians generalize matrices, and their inverses (or approximations thereof) 
drive convergence.

Several families of methods will be introduced:

\begin{itemize}
    \item \textbf{Newton iterations:} direct generalizations of one-dimensional Newton–Raphson 
    methods, based on local Taylor expansions and the Jacobian of the system.
    \item \textbf{Quasi-Newton methods:} efficient variants (such as Broyden’s method) that update 
    approximations to the Jacobian or its inverse without recomputing it fully at every step.
    \item \textbf{Nonlinear optimization connections:} many nonlinear systems can be recast as 
    minimization of a residual function, leading naturally to line search, steepest descent, and 
    nonlinear conjugate gradient methods.
    \item \textbf{Constrained systems:} projection techniques (such as Rosen’s gradient projection method) 
    adapt descent algorithms to respect constraints on the variables.
\end{itemize}

It is worth stressing that convergence in the nonlinear setting is much more delicate than in 
the linear one: issues such as basins of attraction, local vs. global minima, and ill-conditioning 
of Jacobians play a major role. Nevertheless, the iterative framework remains similar: one defines 
a sequence $\vec x^{k+1}=\Phi(\vec x^k)$ for a suitable update map $\Phi$, and convergence analysis 
often reduces to the study of fixed points of $\Phi$ in the spirit of Banach’s contraction principle.

In the following subsections, we introduce Newton’s method and its variants, explore their relation 
to minimization problems, and discuss generalizations such as nonlinear conjugate gradients and 
gradient projection methods. These techniques are the nonlinear analogue of the linear iterative 
methods presented earlier, and together they form the backbone of modern numerical optimization.

\subsubsection{Newton iterations}\index{key}{Newton iterations}
Many of the ideas developed for linear systems extend to nonlinear ones, with the necessary care.  
We consider a system of nonlinear equations:
\begin{equation}
    f_i(\vec x)=0,
    \label{eq:systemnoneq}
\end{equation}
for which closed-form solutions are rarely available. Expanding $f_i$ in a multivariate Taylor series around $\vec x$ gives
\begin{equation}
    f_i(\vec x+\vec h)=f_i(\vec x)+\vec h\cdot \nabla f_i(\vec x)+O(\|\vec h\|^2).
\end{equation}
If $\vec z$ is a solution, $f(\vec z)=0$, then setting $\vec z=\vec x+\vec h$ implies
\begin{equation}
    J(\vec x)\vec h=-f(\vec x),
\end{equation}
where $J_{ij}(\vec x)=\partial_j f_i(\vec x)$ is the Jacobian.  

This motivates the recursive Newton (Newton–Raphson) method\footnote{See for example Ortega and Rheinboldt, \emph{Iterative Solution of Nonlinear Equations in Several Variables}, Academic Press (1970).}:
\begin{eqnarray}
    J(\vec x^{\,k}) \, \vec h^{\,k+1} &=& -f(\vec x^{\,k}), \nonumber \\
    \vec x^{\,k+1} &=& \vec x^{\,k} + \vec h^{\,k+1}, \nonumber
\end{eqnarray}
or equivalently,
\begin{equation}
    \vec x^{\,k+1}=\vec x^{\,k}-J(\vec x^{\,k})^{-1} f(\vec x^{\,k}).
\end{equation}
At each iteration one must compute the Jacobian, and solve a linear system involving it. The method converges quadratically if the initial guess lies within the correct basin of attraction, but it can be expensive and unstable otherwise.

---

\paragraph{Quasi-Newton methods (Broyden’s update).}
Broyden suggested economizing the evaluation of $J^{-1}$ by updating it rank-one at each step instead of recomputing it entirely\footnote{Broyden, C.G., ``A Class of Methods for Solving Nonlinear Simultaneous Equations'', \emph{Math. Comp.}, 19(92), 577–593 (1965).}.  

If we denote $\delta \vec x^k=\vec x^{\,k+1}-\vec x^{\,k}$ and $\delta \vec f_k=f(\vec x^{\,k+1})-f(\vec x^{\,k})$, then consistency requires
\[
J_{k+1}\,\delta \vec x^k=\delta \vec f_k.
\]
Broyden’s rank-one update satisfying this is
\begin{equation}
    J_{k+1}=J_k+\frac{(\delta \vec f_k-J_k \,\delta \vec x^k)\, (\delta \vec x^k)^{\top}}{\|\delta \vec x^k\|^2}.
\end{equation}
Using the Sherman–Morrison formula, one can update the inverse $J_{k+1}^{-1}$ directly from $J_k^{-1}$ without recomputing it. This defines the family of quasi-Newton methods, which are widely used in large-scale optimization.

---

\subsubsection{Systems as minimization problems}
Another strategy is to reformulate nonlinear equations as an optimization problem. Define
\begin{equation}
    g(\vec x)=\tfrac{1}{2}\|f(\vec x)\|_2^2=\tfrac{1}{2} f(\vec x)\cdot f(\vec x).
\end{equation}
Then $\nabla g(\vec x)=J(\vec x)^{\top} f(\vec x)$, so any zero of $f$ is a minimizer of $g$.  
A simple line search update is
\begin{equation}
    \vec x^{\,k+1}=\vec x^{\,k}-\alpha_k \nabla g(\vec x^{\,k}),
\end{equation}
where the step size $\alpha_k$ is chosen by minimizing
\[
    g(\vec x^{\,k}-\alpha \nabla g(\vec x^{\,k})).
\]

---

\subsubsection{Nonlinear conjugate gradient}
The nonlinear conjugate gradient method combines line search with conjugacy ideas from the linear case. The recursion is
\begin{eqnarray}
    \vec x^{\,k+1} &=& \vec x^{\,k}+\alpha_k \vec d^{\,k}, \nonumber \\
    \vec d^{\,k+1} &=& -\nabla g(\vec x^{\,k+1})+\beta_k \vec d^{\,k}, \qquad \vec d^0=-\nabla g(\vec x^0),
\end{eqnarray}
with $\alpha_k$ determined by line search. Several formulas exist for $\beta_k$; in the Fletcher–Reeves scheme\footnote{Fletcher, R. and Reeves, C.M., ``Function Minimization by Conjugate Gradients'', \emph{Comput. J.} 7 (1964), 149–154.}:
\[
\beta_k=\frac{\|\nabla g(\vec x^{\,k+1})\|^2}{\|\nabla g(\vec x^{\,k})\|^2}.
\]

---

\subsubsection{Rosen projections}
The gradient projection method (due to Rosen\footnote{J.B. Rosen, ``The Gradient Projection Method for Nonlinear Programming'', \emph{J. Soc. Indust. Appl. Math.}, 8(1), 181–217 (1960).}) enforces linear constraints while following a descent direction.  

Suppose we minimize $f(\vec W)$ subject to $B\vec W=0$. An update has the form
\begin{equation}
    \vec W_{t+1}=\vec W_t+\alpha \vec s,
\end{equation}
where $\vec s$ lies in the tangent space of the constraints. The steepest feasible descent direction solves
\begin{eqnarray}
    &\min_{\vec s} & \ \ \vec s\cdot \nabla_{\!W} f(\vec W) \\
    &\text{s.t.} & \ \ B\vec s=0,\quad \|\vec s\|=1.
\end{eqnarray}
Introducing multipliers $\vec \lambda,\mu$ leads to the Lagrangian
\[
\mathcal L(\vec s,\vec \lambda,\mu)=\vec s\cdot \nabla_W f- \vec s\cdot B\vec \lambda -\mu(\|\vec s\|^2-1).
\]
Solving the stationarity conditions yields
\begin{equation}
    \vec s \propto (I-B(B^{\top}B)^{-1}B^{\top})\nabla_W f,
\end{equation}
that is, the projection of the gradient onto the nullspace of $B$.  

The resulting constrained update is
\begin{equation}
    \vec W_{t+1}=\vec W_t+\alpha\, P \nabla_W f, \qquad 
    P=I-B(B^{\top}B)^{-1}B^{\top},
\end{equation}
where $P$ is the orthogonal projector onto the feasible subspace. This provides a natural extension of gradient descent to constrained nonlinear optimization.

\section{Fixed Point Theorems: Existence and Uniqueness Principles}

In the previous sections we introduced iterative methods for solving linear and nonlinear systems.  
A recurring theme was to design an iteration of the form
\[
    \vec x^{k+1} = \Phi(\vec x^k),
\]
and then study whether the sequence $\{\vec x^k\}$ converges to the true solution.  
From this point of view, solving algebraic equations or optimization problems often reduces to the 
existence of a \emph{fixed point} of a map $\Phi:X\to X$, that is, a point $\vec x^\ast$ such that
\[
    \Phi(\vec x^\ast)=\vec x^\ast.
\]

This observation explains why fixed point theory is central to numerical analysis and to many 
applications in mathematics, physics, and economics. For example:
\begin{itemize}
    \item The Jacobi and Gauss–Seidel methods can be seen as fixed point iterations of the form 
    $\vec x^{k+1}=A\vec x^k+\vec b$.
    \item Newton’s method for nonlinear equations updates $\vec x^{k+1}=\Phi(\vec x^k)$, where 
    $\Phi$ depends on the inverse of the Jacobian.
    \item Equilibria in game theory (e.g.~Nash equilibria) or in dynamical systems correspond to fixed 
    points of appropriate maps.
\end{itemize}

The power of fixed point theorems lies in providing \emph{general conditions} under which a fixed point 
must exist (and is often unique), without having to construct it explicitly. These results give a rigorous 
theoretical foundation for the iterative algorithms we have been using: they guarantee that our iterations 
are not merely heuristic, but converge under suitable assumptions.

Throughout this section, we focus on maps $f:X\to X$ in the setting of Banach spaces\index{key}{Banach space}.  
We equip $X$ with a metric $\rho:X\times X\to \mathbb R$, and introduce the notions of contraction and 
non-expansive maps. A contraction ensures that iterates are pulled closer together at each step, 
which is precisely the mechanism that guarantees convergence of fixed point iterations.

---

\subsection{Contraction-type fixed-point theorems}

\paragraph{Banach's Fixed Point Theorem.}
\index{key}{Banach's fixed point theorem}  
Let $X$ be a complete metric space and $f:X\to X$ a contraction. Then $f$ admits a unique fixed point.  

\emph{Sketch of proof.} Suppose $\vec x_{n+1}=f(\vec x_n)$. Then
\[
\rho(\vec x_{n+1},\vec x_n)\leq \lambda^n \rho(f(\vec x_0),\vec x_0).
\]
This implies $\{\vec x_n\}$ is a Cauchy sequence. Completeness gives convergence to some $\vec x^*$, and continuity ensures $f(\vec x^*)=\vec x^*$. Uniqueness follows since if both $\vec x_1,\vec x_2$ are fixed points then
\[
\rho(\vec x_1,\vec x_2)\leq \lambda \rho(\vec x_1,\vec x_2)\implies \vec x_1=\vec x_2.
\]

---

\paragraph{Extensions of Banach’s theorem.}
Several generalizations relax the strict contraction condition:

- \textbf{Boyd–Wong theorem.}\footnote{D.W. Boyd and J.S.W. Wong, \emph{On Nonlinear Contractions}, Proc. Amer. Math. Soc. 20 (1969), 458–464.}  
If $f:X\to X$ satisfies
\[
\rho(f(x_1),f(x_2))\leq \phi(\rho(x_1,x_2))
\]
for some continuous $\phi:\mathbb R^+\to \mathbb R^+$ with $\phi(r)<r$, then $f$ has a unique fixed point.

- \textbf{Caristi’s theorem.}\footnote{J. Caristi, \emph{Fixed Point Theorems for Mappings Satisfying Inwardness Conditions}, Trans. Amer. Math. Soc. 215 (1976), 241–251.}  
Let $X$ be a complete metric space and $T:X\to X$ such that
\[
d(x,T(x))\leq q(x)-q(T(x)),
\]
for some lower semicontinuous $q:X\to\mathbb R$ bounded from below. Then $T$ has a fixed point.

- \v{C}iri\v{c} theorem.  
If $f:X\to X$ satisfies
\[
\rho(f(x_1),f(x_2))\leq \lambda \max\{q_1,q_2,q_3,q_4,q_5\}, \quad \lambda<1,
\]
where $q_1=\rho(x_1,x_2)$, $q_2=\rho(x_1,f(x_2))$, $q_3=\rho(f(x_1),x_2)$, $q_4=\rho(x_1,f(x_1))$, $q_5=\rho(x_2,f(x_2))$, then $f$ has a unique fixed point with convergence rate $O(\lambda^n)$.

---

\paragraph{Weak contractions.}
A weaker notion allows
\[
\rho(f(x_1),f(x_2))\leq \rho(x_1,x_2), \qquad \forall x_1\neq x_2.
\]
Such weak contractions need not be strict contractions, but a theorem guarantees:  

\textbf{Theorem.} If $f$ is a weak contraction on a compact metric space, then $f$ admits a unique fixed point $\vec x^*$. Moreover, $f^n(x_0)\to \vec x^*$ for any $x_0\in X$.

---

\paragraph{Bessaga’s theorem.}
\index{key}{Bessaga's theorem}  
If a map $f:X\to X$ admits a unique fixed point $\vec x^*$, then there exists an equivalent metric $d_\epsilon$ (for any $\epsilon\in(0,1)$) such that $(X,d_\epsilon)$ is complete and $f$ is a contraction with constant $\epsilon$.  

This shows that the existence of a unique fixed point is essentially equivalent to the existence of a suitable contracting metric.

---

\subsection{Fixed points for non-expansive maps}
Banach’s theorem does not apply directly to non-expansive maps. A classical result is:

\paragraph{Browder–Kirk theorem.}\footnote{F.E. Browder, \emph{Fixed-Point Theorems for Nonexpansive Mappings}, Proc. Natl. Acad. Sci. USA 53 (1965), 1272–1276. \\ W.A. Kirk, \emph{A Fixed Point Theorem for Mappings Which Do Not Increase Distances}, Amer. Math. Monthly 72 (1965), 1004–1006.}  
Let $X$ be a uniformly convex Banach space, and $C\subseteq X$ a nonempty, closed, bounded, convex set. If $f:C\to C$ is non-expansive, then $f$ has a fixed point.

---

\subsubsection{Connections with ergodic theory}
Fixed point results are closely related to ergodic theorems.  

Consider an operator $T$ on a Banach space with $\|T\vec x\|\leq \|\vec x\|$. Define the Cesàro averages
\[
P\vec x=\lim_{n\to\infty}\frac{1}{n+1}\sum_{k=0}^n T^k\vec x.
\]
\textbf{Riesz mean ergodic theorem.}\footnote{See E. Hille and R.S. Phillips, \emph{Functional Analysis and Semi-Groups}, AMS (1957).}  
The operator $P$ is a continuous projection onto the closed subspace spanned by $\{T^i\vec x : i\geq 0\}$.  

This shows how fixed-point and averaging arguments connect linear operator theory with ergodic properties.

\subsection{Brouwer fixed point theorem}

\textbf{Brouwer's fixed point theorem.}\index{key}{Brouwer's fixed point theorem}  
Let $S\subset \mathbb{R}^n$ be convex and compact. Let $T:S\rightarrow S$ be a continuous map. Then there exists a fixed point $\vec x^*\in S$ such that
\[
T(\vec x^*)=\vec x^*.
\]
This fundamental result has wide applications, for instance in proving the existence of Nash equilibria in game theory.\footnote{See K. G. Binmore, \emph{Game Theory: A Very Short Introduction}, Oxford University Press (2007).}

---

\paragraph{Application: positivity of operators.}
One classical application is to the Perron–Frobenius theory. Suppose $A$ is a matrix with strictly positive entries. Consider the map
\[
f(\vec x)=\frac{A\vec x}{\|A\vec x\|_1}.
\]
The map $f$ is continuous and sends the simplex (a convex, compact set) into itself. By Brouwer's theorem, there exists $\vec x^*$ such that $f(\vec x^*)=\vec x^*$, i.e.
\[
A\vec x^*=\lambda \vec x^*, \qquad \lambda=\|A\vec x^*\|_1>0.
\]
Thus, $A$ has a positive eigenvalue with a positive eigenvector, which is the content of the Perron–Frobenius theorem.

---

\paragraph{Schauder–Tychonoff theorem.}
\index{key}{Schauder–Tychonoff theorem}  
A generalization of Brouwer’s result applies to infinite-dimensional spaces.  

Let $S$ be a convex, compact subset of a locally convex topological vector space. If $T:S\to S$ is continuous, then $T$ has a fixed point.\footnote{See F.E. Browder, \emph{Fixed Point Theory and Nonlinear Problems}, Bull. Amer. Math. Soc. 9 (1983).}

---

\paragraph{Schaefer’s theorem.}
\index{key}{Schaefer's theorem}  
Let $X$ be a Banach space and $f:X\to X$ a continuous map. Define
\[
Q=\{x\in X : x=\lambda f(x)\ \text{for some}\ \lambda\in [0,1]\}.
\]
If $Q$ is bounded, then $f$ has a fixed point.\footnote{See K. Schaefer, \emph{Über die Methode der a-priori-Schranken}, Math. Ann. 129 (1955).}  
This result is widely used in nonlinear functional analysis.

---

\paragraph{Krasnosel’skii’s theorem.}
\index{key}{Krasnosel’skii's theorem}  
Let $X$ be a Banach space and $C\subset X$ a nonempty, closed, convex set. Suppose $f,g:C\to X$ with $f$ continuous, $g$ a contraction. Then there exists $\bar x\in C$ such that
\[
f(\bar x)+g(\bar x)=\bar x.
\]
This fixed-point principle is particularly useful for nonlinear differential and integral equations.\footnote{See M.A. Krasnosel’skii, \emph{Topological Methods in the Theory of Nonlinear Integral Equations}, Pergamon (1964).}

---

\paragraph{Markov–Kakutani theorem.}
\index{key}{Markov–Kakutani theorem}  
Let $K$ be a nonempty compact convex subset of a locally convex space $X$, and let $\mathcal G$ be a commuting family of continuous affine maps of $K$ into itself. Then there exists $\bar x\in K$ fixed by all maps in $\mathcal G$.\footnote{A.A. Markov, S. Kakutani, \emph{Two Fixed Point Theorems}, Proc. Imp. Acad. Tokyo 16 (1940).}  
This result underlies ergodic theory and invariant measures.

---

\paragraph{Kakutani and Ky Fan generalizations.}
\index{key}{Kakutani's fixed point theorem}  
Kakutani’s theorem generalizes Brouwer’s result to \emph{multivalued} functions.  

Let $K\subset\mathbb R^n$ be convex, compact and nonempty. If $\Gamma:K\to 2^K$ is an upper semicontinuous correspondence with nonempty, convex values, then $\Gamma$ has a fixed point.

A further generalization by Ky Fan extends this to locally convex spaces:  

\textbf{Kakutani–Ky Fan theorem.}\index{key}{Kakutani–Ky Fan theorem}  
Let $K$ be a nonempty, convex, compact subset of a locally convex space $X$. If $f:K\to 2^K$ is upper semicontinuous, with nonempty, convex and closed values, then $f$ admits a fixed point.  
This version plays a central role in economics and game theory.

---

\subsection{Knaster–Tarski theorem}
Order-theoretic fixed point results complement the topological ones.  

\textbf{Knaster–Tarski theorem.}\index{key}{Knaster–Tarski theorem}  
Let $(X,\leq)$ be a complete lattice, and let $T:X\to X$ be an order-preserving map. Then the set of fixed points of $T$ is nonempty and forms a complete lattice itself.\footnote{See A. Tarski, \emph{A Lattice-Theoretical Fixpoint Theorem and Its Applications}, Pacific J. Math. 5 (1955).}  

Applications appear in lattice theory, formal semantics, and computer science (e.g. semantics of recursive programs).

---

\paragraph{Bourbaki–Witt theorem.}
\index{key}{Bourbaki–Witt theorem}  
If $T$ is an order-preserving self-map on a chain-complete poset such that $T(x)\geq x$ for all $x$, then $T$ admits a fixed point.  
This is often used in constructive mathematics and domain theory.

---

\subsection{Hutchinson theorem and fractals}
\index{key}{Hutchinson theorem}  
A striking application of contraction-type fixed point theorems is in fractal geometry.  

\textbf{Hutchinson’s theorem.}  
Let $(X,d)$ be a complete metric space, and let $\{f_i\}_{i=1}^m$ be a finite set of contractions $f_i:X\to X$. Then there exists a unique nonempty compact set $K\subset X$ such that
\[
K=\bigcup_{i=1}^m f_i(K).
\]
This set $K$ is called the \emph{attractor} of the iterated function system (IFS).\footnote{See J. Hutchinson, \emph{Fractals and Self-Similarity}, Indiana Univ. Math. J. 30 (1981).}  

The theorem provides the rigorous mathematical foundation for fractals: Cantor sets, Sierpiński gaskets, and many others arise as fixed points of iterated contractions.

\chapter{Pseudo-spectra and non-normality}\label{sec:pseudospectra}

A fundamental distinction in linear algebra is between \emph{normal} and \emph{non-normal} matrices.  For us, this is particularly important in view that some iterative schemes or dynamical systems might deviate substantially from the asymptotic fixed point. The non-normality of the matrices, as we see here, is the culprit.

A matrix $A\in \mathbb{C}^{n\times n}$ is called \textit{normal}\index{key}{normal matrix} if
\begin{equation}
    A^\dagger A = A A^\dagger,
    \label{eq:normality}
\end{equation}
where $A^\dagger$ is the conjugate transpose of $A$.  
Equivalently, a matrix is normal if and only if it can be diagonalized by a unitary transformation,
\begin{equation}
    A = U D U^\dagger,
\end{equation}
with $U$ unitary and $D$ diagonal.\footnote{See R.~Horn and C.~Johnson, \emph{Matrix Analysis}, Cambridge University Press (1985).}

Indeed, if $A=UDU^\dagger$ then $A^\dagger A=UD^\dagger D U^\dagger$ and $A A^\dagger = U D D^\dagger U^\dagger$, which coincide since $D$ is diagonal.  

Important subclasses of normal matrices include:  
\begin{itemize}
    \item Hermitian (or symmetric, if real) matrices, $A=A^\dagger$,
    \item skew-Hermitian matrices, $A=-A^\dagger$,
    \item orthogonal matrices, $A^\top A=I$,
    \item unitary matrices, $A^\dagger A=I$.
\end{itemize}

Normal matrices enjoy several important properties:  
\begin{itemize}
    \item They admit an orthonormal basis of eigenvectors.  
    \item Their right and left eigenvectors coincide up to conjugation.  
    \item The kernel and range are preserved: $\ker(A)=\ker(A^\dagger)$ and $\operatorname{im}(A)=\operatorname{im}(A^\dagger)$.  
    \item If $A$ is normal, then so is $A^{-1}$ (when it exists).  
    \item If $A$ and $B$ are normal and commute, then they are simultaneously diagonalizable by a unitary matrix, hence $A+B$ is also normal.\footnote{See R.~Bhatia, \emph{Matrix Analysis}, Springer (1997).}
\end{itemize}

By contrast, a \textit{non-normal}\index{key}{non-normal matrix} matrix is one for which $AA^\dagger\neq A^\dagger A$. Such matrices cannot be unitarily diagonalized, but only triangularized (via the Schur decomposition). This lack of unitary diagonalization results in significant differences in their spectral and dynamical behavior.  

In particular, functions of non-normal matrices can behave very differently from the normal case. Two key examples are
\begin{equation}
    e^{At}, \qquad A^k,
\end{equation}
which arise naturally in the solution of linear systems:
\begin{equation}
    \frac{d}{dt}\vec x = A \vec x \quad \implies \quad \vec x(t)=e^{At}\vec x_0,
\end{equation}
and
\begin{equation}
    \vec x_{n} = A \vec x_{n-1} \quad \implies \quad \vec x_n=A^n \vec x_0.
\end{equation}
For normal $A$, the asymptotic growth or decay of solutions is governed entirely by the spectral radius. For non-normal matrices, however, transient growth can occur even if all eigenvalues lie within the stable region of the complex plane.\footnote{See L.~N.~Trefethen and M.~Embree, \emph{Spectra and Pseudospectra: The Behavior of Nonnormal Matrices and Operators}, Princeton University Press (2005).}  

This motivates the study of the \emph{pseudospectrum}.  
Recall that the \textit{spectrum} $\Lambda(A)$ of a matrix $A$ is defined as
\begin{equation}
    \Lambda(A)=\{ z\in\mathbb{C} : (zI-A)\ \text{is not invertible}\},
\end{equation}
which is equivalent to the set of poles of the resolvent operator
\begin{equation}
    R(z,A)=(zI-A)^{-1}.
\end{equation}
That is, the spectrum consists of points $z$ in the complex plane such that
\begin{equation}
    \|R(z,A)\|=\|(zI-A)^{-1}\|=\infty.
\end{equation}

If $A$ is diagonalizable, then near an eigenvalue $\lambda_i$ we have
\[
\|(zI-A)^{-1}\|\sim \frac{1}{|z-\lambda_i|},
\]
so that
\begin{equation}
    \|(zI-A)^{-1}\|_2=\max_{i}\frac{1}{|z-\lambda_i|} = \frac{1}{\operatorname{dist}(z,\Lambda(A))}.
    \label{eq:distnorm}
\end{equation}
This relation, however, breaks down for non-normal matrices, where $\|(zI-A)^{-1}\|$ can become extremely large far away from the spectrum. The study of this phenomenon leads to the notion of pseudospectra, which we introduce next.

It is therefore natural to introduce a generalization that accounts for distances to the set of eigenvalues, namely the \emph{pseudospectrum}.  
For $\epsilon>0$, the $\epsilon$-pseudospectrum of $A$ is defined as
\begin{equation}
    \Lambda_\epsilon(A)=\{z\in \mathbb{C}\ : \|(zI-A)^{-1}\|_2>\tfrac{1}{\epsilon}\}.
\end{equation}
Unlike the spectrum (a discrete set of points), the pseudospectrum forms \emph{regions} in the complex plane, controlled by the tolerance $\epsilon$.

There are several equivalent formulations (for the spectral norm):
\begin{itemize}
    \item $\Lambda_\epsilon(A)=\{z\in \mathbb{C}\ : \|(zI-A)^{-1}\|_2>\tfrac{1}{\epsilon}\}$,
    \item $\Lambda_\epsilon(A)=\{z\in \mathbb{C}\ : \sigma_{\min}(zI-A)<\epsilon\}$, where $\sigma_{\min}$ denotes the smallest singular value,
    \item $\Lambda_\epsilon(A)=\{z\in \mathbb{C}\ : z\in \Lambda(A+\Delta A),\ \|\Delta A\|<\epsilon\}$,
\end{itemize}
i.e. $z$ belongs to the pseudospectrum of $A$ precisely when it is an approximate eigenvalue of some perturbed matrix $A+\Delta A$ with $\|\Delta A\|<\epsilon$.\footnote{For a comprehensive account, see L.~N.~Trefethen and M.~Embree, \emph{Spectra and Pseudospectra}, Princeton University Press (2005).}

If the norm is the $L^2$ (spectral) norm, pseudospectra are invariant under unitary similarity:  
\[
    \Lambda_\epsilon(U A U^\dagger)=\Lambda_\epsilon(A).
\]

\paragraph{Normal vs.~non-normal matrices.}  
For a normal matrix $A$, the resolvent satisfies eqn.~(\ref{eq:distnorm}), and one can show that
\[
    \Lambda_\epsilon(A) = \Lambda(A) + B_\epsilon,
\]
where $B_\epsilon=\{z\in \mathbb{C}: |z|<\epsilon\}$ is the $\epsilon$-ball.  
For non-normal matrices, this equality fails, and only inclusions hold:
\begin{equation}
    \Lambda(A)+B_\epsilon \subseteq \Lambda_\epsilon(A).
\end{equation}

More generally, if $A=V D V^{-1}$ with $D$ diagonal and $V$ invertible, then
\begin{equation}
    \|(zI-A)^{-1}\|\leq \|V\|\,\|V^{-1}\|\cdot \|(zI-D)^{-1}\|
    = \frac{\kappa(V)}{\operatorname{dist}(z,\Lambda(A))},
\end{equation}
where $\kappa(V)=\|V\|\|V^{-1}\|$ is the condition number of the eigenbasis.  
This leads to the Bauer–Fike theorem\index{key}{Bauer-Fike theorem}:
\begin{equation}
    \Lambda(A)+B_\epsilon \subseteq \Lambda_\epsilon(A) \subseteq \Lambda(A)+B_{\kappa(V)\epsilon}.
\end{equation}
Equivalently, the eigenvalues of a perturbed matrix $A+\Delta A$ (with $\|\Delta A\|<\epsilon$) must lie within a $\kappa(V)\epsilon$ neighborhood of $\Lambda(A)$.  
For normal matrices, $\kappa(V)=1$, so the inclusion becomes an equality.

\paragraph{Basic properties.}  
The pseudospectrum satisfies several natural invariance rules:
\begin{itemize}
    \item $\Lambda_\epsilon(A+cI)=c+\Lambda_\epsilon(A)$,
    \item $\Lambda_\epsilon(\alpha A)=\alpha\Lambda_\epsilon(A)$ for $\alpha\in\mathbb{C}$,
    \item $\Lambda_\epsilon(A\oplus B)=\Lambda_\epsilon(A)\cup\Lambda_\epsilon(B)$,
    \item $\Lambda_\epsilon(A^\dagger)=\Lambda_\epsilon(A)^*$.
\end{itemize}

\paragraph{Pseudo-eigenvectors.}  
The pseudospectrum admits a natural vector generalization.  
If
\[
    \|(zI-A)\vec u\|<\epsilon,
\]
then $z$ is an $\epsilon$-\textit{pseudoeigenvalue} and $\vec u$ is a corresponding \textit{pseudoeigenvector}\index{key}{pseudoeigenvector}\index{key}{pseudoeigenvalue}.  
Thus, pseudospectra characterize both approximate eigenvalues and the sensitivity of eigenvectors under perturbations.

\paragraph{Growth of matrix functions.}  
One key motivation for pseudospectra is to analyze the growth of $e^{At}$.  
Define
\begin{equation}
    \alpha(A)=\lim_{t\to\infty}\frac{1}{t}\log\|e^{At}\|, \qquad
    \omega(A)=\lim_{t\to 0^+}\frac{1}{t}\log\|e^{At}\|.
\end{equation}
Here $\alpha(A)$ is called the \emph{spectral abscissa} and $\omega(A)$ the \emph{numerical abscissa}\index{key}{spectral abscissa}\index{key}{numerical abscissa}.  
For normal matrices, $\alpha(A)$ equals the largest real part of an eigenvalue, but for non-normal matrices $\omega(A)$ can be much larger, capturing transient growth.\footnote{See K.~Dowler, ``Pseudospectra and transient growth,'' \emph{SIAM Review} (2017).}  
In fact, $\omega(A)=\sup\sigma\!\left(\tfrac{A+A^\dagger}{2}\right)$, i.e. the largest eigenvalue of the Hermitian part of $A$.\footnote{This follows from the Hille–Yosida theorem on semigroups of operators.}

\paragraph{Schur form estimates.}  
Let $A=UTU^\dagger$ be the Schur decomposition, with $T$ upper triangular and eigenvalues on its diagonal. Then
\[
T=\begin{pmatrix}
\lambda_1 & * & \cdots & * \\
0 & \lambda_2 & \cdots & * \\
\vdots & & \ddots & \vdots \\
0 & \cdots & 0 & \lambda_n
\end{pmatrix}.
\]
Explicit bounds for $\|T^k\|$ can be obtained by combinatorial expansions of off-diagonal terms. If $\rho(T)$ denotes the spectral radius and $M=\max_{i<j}|T_{ij}|$, one has
\begin{equation}
    \|T^k\|_2 \leq \sqrt{\sum_{l=0}^{n-1}\Bigg(\sum_{j=0}^l \binom{l}{j}\binom{k}{j} M^j \rho(T)^{\,k-j}\Bigg)^2}.
\end{equation}
Similarly, the Frobenius norm can be bounded as
\begin{equation}
    \|T^k\|_F \leq \sqrt{n\rho(T)^{2k}+\sum_{i<j}\Bigg|\sum_{z=1}^{j-i}\binom{j-i-1}{z-1}\binom{k}{z}M^z \rho(T)^{\,k-z}\Bigg|^2}.
\end{equation}
These estimates highlight how non-normality (large $M$) can cause transient amplification even when $\rho(T)<1$, i.e. when the spectrum lies in the stable region.

\subsection{The Kreiss Matrix Theorem and Growth Bounds}

The pseudospectrum allows us to quantify how non-normality affects the growth of matrix powers and semigroups. A central result in this direction is the Kreiss matrix theorem, which establishes an equivalence between uniform resolvent bounds and power-boundedness.

\medskip
\noindent
\textbf{Theorem (Kreiss matrix theorem, original version).}\footnote{See Kreiss, H.-O., \emph{Über die Stabilitätsdefinition für Differenzengleichungen die partielle Differentialgleichungen approximieren}, BIT Numerical Mathematics \textbf{2}, 153–181 (1962).}\index{key}{Kreiss matrix theorem}  
For a matrix $A\in\mathbb{C}^{n\times n}$, the following are equivalent:
\begin{enumerate}
    \item There exists $C>0$ such that
    \[
        \|A^n\|\leq C \quad \forall n\geq 0.
    \]
    \item There exists $C>0$ such that the resolvent satisfies
    \[
        \|R(z,A)\|\leq \frac{C}{|z|-1}, \qquad |z|>1.
    \]
    \item There exists $C>0$ such that $A$ can be put in upper-triangular form $A^{ut}=SAS^{-1}$ with $\|S^{-1}\|\leq C$, and
    \[
        |A^{ut}_{ii}|\leq 1, \qquad |A^{ut}_{ij}|\leq C\ \max\!\left(1-|A^{ut}_{ii}|,\ 1-|A^{ut}_{jj}|,\ ||A^{ut}_{ii}|-A^{ut}_{ii}|\right).
    \]
    \item There exists $C>0$ such that for each $A$ there is an Hermitian matrix $H$ with
    \[
        \tfrac{1}{C}I \leq H \leq CI, \qquad A^\dagger H A \leq H.
    \]
\end{enumerate}

\paragraph{Kreiss constant.}  
The theorem above motivates the definition of the \emph{Kreiss constant}
\begin{equation}
    \mathcal K(A)=\sup_{|z|>1} (|z|-1)\,\|(zI-A)^{-1}\|
    = \sup_{\epsilon>0} \frac{\rho_\epsilon(A)-1}{\epsilon},
\end{equation}
where $\rho_\epsilon(A)$ denotes the \emph{$\epsilon$-pseudospectral abscissa},
\begin{equation}
    \rho_\epsilon(A)=\sup\{\operatorname{Re}(z): z\in \Lambda_\epsilon(A)\}.
\end{equation}
Thus $\mathcal K(A)$ measures how far pseudospectral growth can extend beyond the unit disk.

\medskip
\noindent
\textbf{Theorem (Kreiss matrix theorem, modern form).} \footnote{See Trefethen, L.~N., and Embree, M., \emph{Spectra and Pseudospectra}, Princeton University Press (2005).}   
For every $A\in \mathbb{C}^{n\times n}$,
\begin{equation}
    \mathcal K(A)\ \leq\ \sup_{k\geq 0}\|A^k\|\ \leq\ e\,n\,\mathcal K(A).
\end{equation}

Thus, the resolvent norm completely controls the growth of matrix powers, up to an explicit multiplicative factor depending only on the dimension $n$. This is especially striking because even when $\rho(A)<1$ (the spectral radius lies inside the unit circle), $\sup_k\|A^k\|$ may be arbitrarily large if $\mathcal K(A)$ is large — a hallmark of highly non-normal matrices.

\paragraph{Interpretation.}  
For normal matrices, $\mathcal K(A)=1$, and $\sup_k \|A^k\|=1$ if $\rho(A)\leq 1$.  
For non-normal matrices, however, large transient growth may occur before eventual asymptotic decay, with the size of the amplification governed by $\mathcal K(A)$. This explains phenomena such as transient instability in fluid mechanics, control theory, and signal processing.

\paragraph{Joint pseudospectral abscissa and spectral bound.}  
In many applications (e.g., switched linear systems or control theory), one must analyze not a single matrix but a set $\mathcal A=\{A_1,\dots,A_m\}$.  
Two key quantities are:
\begin{itemize}
    \item The \emph{joint spectral radius}
    \[
        \rho(\mathcal A)=\lim_{k\to\infty}\ \sup\ \|A_{i_1}A_{i_2}\cdots A_{i_k}\|^{1/k},
    \]
    which generalizes $\rho(A)$ to families of matrices;
    \item The \emph{joint pseudospectral abscissa}
    \[
        \alpha_\epsilon(\mathcal A)=\sup\{\operatorname{Re}(z): z\in \Lambda_\epsilon(A_i)\ \text{for some}\ A_i\in\mathcal A\},
    \]
    which captures approximate growth rates under perturbations.
\end{itemize}
The Kreiss theorem admits natural generalizations to this setting, where one studies bounds of the type
\[
    \sup_{k}\|A_{i_1}A_{i_2}\cdots A_{i_k}\| \ \sim\ f\big(\sup_i \alpha_\epsilon(A_i)\big),
\]
linking transient amplification to pseudospectral geometry for the whole family.  
\footnote{See Blondel, V.~D., and Nesterov, Y., ``Computationally efficient approximations of the joint spectral radius,'' \emph{SIAM J. Matrix Anal. Appl.} \textbf{27}, 256–272 (2005).}

\chapter{Matrix Equations}\label{sec:matrixequations}

Matrix equations are ubiquitous in applied mathematics, physics, engineering, and computer science.  
They arise naturally in control theory, numerical analysis, optimization, and dynamical systems.  
In this chapter we review several fundamental classes of matrix equations, ranging from simple linear forms to more structured operator equations, such as the Lyapunov and Sylvester equations.  

\section{Simplest Non-trivial Equations: $XA=B$ and $AXB=C$}

\subsection{Equation $XA=B$}
The equation
\begin{equation}
    X A = B
\end{equation}
appears in many applications, including system identification and matrix factorization problems.  

If $A$ is square and invertible, the solution is immediate:
\[
    X = B A^{-1}.
\]

If, however, $A$ is not invertible or the system is inconsistent, the problem must be interpreted in a least-squares sense.  
That is, one seeks $X$ minimizing $\|XA-B\|$ in some matrix norm.  
For the Frobenius norm, one considers
\begin{equation}
    \|XA-B\|_F^2 = \mathrm{Tr}\!\left((XA-B)(XA-B)^{\top}\right).
    \label{eq:traceeqs}
\end{equation}

Using the Moore--Penrose pseudoinverse\index{key}{pseudo-inverse}, introduced by Penrose\footnote{R.~Penrose, ``A generalized inverse for matrices,'' \emph{Proc. Cambridge Phil. Soc.} \textbf{51}, 406–413 (1955).},  
one can show that
\begin{equation}
    \|XA-B\|_F^2 = (BA^+ - X) A A^{\top} (BA^+ - X)^{\top} + B(I - A^+ A)B^{\top},
\end{equation}
where the second term is independent of $X$ and positive semi-definite.  
The minimum is attained at
\[
    X = B A^+,
\]
which is therefore the least-squares solution of $XA=B$.

\subsection{Equation $AXB=C$}
A slightly more general case is the equation
\begin{equation}
    A X B = C,
\end{equation}
with $A,B,C$ given matrices.  

\paragraph{Case 1: $A$ and $B$ invertible.}  
If $A,B$ are square and invertible, the unique solution is
\[
    X = A^{-1} C B^{-1}.
\]

\paragraph{Case 2: General case.}  
If $A$ or $B$ are not invertible, solvability requires a compatibility condition.  
In particular,
\begin{equation}
    A A^+ C B^+ B = C
    \label{eq:matrixreq}
\end{equation}
must hold. This condition ensures that $C$ lies in the appropriate range spaces determined by $A$ and $B$.

When the rows of $A$ are linearly independent and the columns of $B$ are linearly independent, then $AA^+=I$ and $B^+B=I$, and condition \eqref{eq:matrixreq} is automatically satisfied.

\paragraph{General solution.}  
The solution of the homogeneous problem can be written as
\begin{equation}
    X = A^+ C B^+ + Y - A^+ A Y B B^+,
\end{equation}
where $Y$ is arbitrary.  
This expression reduces to $X=A^{-1}CB^{-1}$ if $A$ and $B$ are invertible.  
The formula is again due to Penrose\footnote{R.~Penrose, ibid.}.

\paragraph{Vectorized form.}  
Using the Kronecker product, the equation can be equivalently expressed as
\begin{equation}
    (B^{\top} \otimes A)\,\mathrm{vec}(X) = \mathrm{vec}(C),
\end{equation}
where $\mathrm{vec}(\cdot)$ denotes column-wise vectorization.  
This form is particularly useful in numerical linear algebra.

\section{Lyapunov and Sylvester Equations}

\subsection{Continuous Lyapunov Equation}
The \emph{Lyapunov equation} (or continuous-time Lyapunov equation) is the matrix equation
\begin{equation}
    \mathcal{L}_A(X) \equiv A^* X + X A = Y,
    \label{eq:lyapunov}
\end{equation}
where $A$ and $Y$ are given and $X$ is unknown.  
The operator $\mathcal L_A$ is known as the \emph{Lyapunov operator}.

\paragraph{Applications.}  
This equation plays a central role in the study of dynamical systems and stability theory.  
In fact, \eqref{eq:lyapunov} admits a positive definite solution $X$ if and only if the linear system
\[
    \dot{\vec x} = A \vec x
\]
is asymptotically stable.  

\paragraph{Characterization.}  
The following theorem summarizes the equivalence of several characterizations.

\medskip
\noindent
\textbf{Theorem.} The following statements are equivalent:
\begin{enumerate}
    \item $\exists Y>0$ such that $\exists X>0$ with $\mathcal L_A(X)=-Y$.
    \item $\forall Y>0$, $\exists X>0$ such that $\mathcal L_A(X)=-Y$.
    \item $\mathrm{Re}(\Lambda(A))<0$, i.e. the spectrum of $A$ lies strictly in the left half-plane.
\end{enumerate}

\emph{Sketch of proof.}  
(2)$\implies$(1) is immediate.  
For (3)$\implies$(1), note that if $\mathrm{Re}(\Lambda(A))<0$, then $\|e^{At}\|\leq M e^{-\gamma t}$ for some $M,\gamma>0$. Define
\[
    X=\int_0^\infty e^{A^* t} Y e^{A t}\, dt.
\]
Differentiation under the integral shows $\mathcal L_A(X)=-Y$.  
Finally, (1)$\implies$(3) follows by evaluating $\vec x^\dagger \mathcal L_A(X)\vec x$ on eigenvectors of $A$ and using positivity of $Y$ and $X$.

\paragraph{Spectrum of the Lyapunov operator.}  
It follows that
\[
    \Lambda(\mathcal L_A)=\{\lambda_1+\lambda_2^* : \lambda_1,\lambda_2 \in \Lambda(A)\}.
\]
Indeed, if $\vec v,\vec w$ are eigenvectors of $A$ with eigenvalues $\lambda_1,\lambda_2$, then
\[
    \mathcal L_A(\vec v \vec w^\dagger)=(\lambda_1+\lambda_2^*) \vec v \vec w^\dagger.
\]

If $\mathcal L_A(X)<0$, then $\vec x^\dagger X \vec x$ serves as a Lyapunov function for the system $\dot{\vec x}=A\vec x$.

\subsection{Sylvester Equation}
A closely related equation is the \emph{Sylvester equation}:
\begin{equation}
    A X + X B = Y,
\end{equation}
of which the Lyapunov equation is a special case.  

Using the Kronecker product, this can be vectorized as
\begin{equation}
    (I \otimes A + B^{\top} \otimes I)\,\mathrm{vec}(X) = \mathrm{vec}(Y).
\end{equation}
This form makes clear that uniqueness requires $(I \otimes A + B^{\top} \otimes I)$ to be non-singular.

\paragraph{Solvability.}  
The equation admits a solution (Roth, 1952\footnote{W.E. Roth, ``On direct product matrices,'' \emph{Bull. Amer. Math. Soc.} \textbf{58}, 23–40 (1952).}) if and only if
\[
    \begin{pmatrix} A & 0 \\ 0 & B \end{pmatrix}
    \quad \text{and} \quad
    \begin{pmatrix} A & Y \\ 0 & B \end{pmatrix}
\]
are similar.

\paragraph{Integral solutions.}  
One representation of the solution is
\begin{equation}
    X=-\frac{1}{4\pi^2}\int_{\Gamma_1}\int_{\Gamma_2}\frac{(\lambda I-A)^{-1} Y (\gamma I-B)^{-1}}{\gamma+\lambda}\, d\lambda\, d\gamma,
\end{equation}
where $\Gamma_1,\Gamma_2$ enclose $\Lambda(A)$ and $\Lambda(B)$ respectively.  
Equivalently, the solution can be written in exponential form as
\begin{equation}
    X=-\int_0^\infty e^{At} Y e^{Bt}\, dt,
\end{equation}
valid when the spectra of $A$ and $B$ lie in the left half-plane.

\subsection{Stein Equation}
The \emph{Stein equation} (or discrete-time Lyapunov equation) is given by
\begin{equation}
    X + A X B = Y.
    \label{eq:steineq}
\end{equation}
If $B=-A^{\top}$, then this equation arises as the discrete analogue of the continuous Lyapunov equation, naturally associated with discrete-time dynamical systems.

In the Kronecker product form, it is easy to see that this can be written as:
\begin{equation}
\left(I+B^{\top} \otimes A \right)\text{vec}(X)=\text{vec}(Y)
\end{equation}
The condition for the existence of a solution is again a spectral one:
\begin{equation}
\lambda_i(A) \lambda_j(B)+1 \neq 0\ \forall i,\forall j.
\end{equation}
The solution can be found by means of a Smith iteration, i.e.
\begin{eqnarray}
X_{k+1}=Y+ A X_{k} B
\end{eqnarray}
with $X_0=I$, if it converges.

\subsubsection{A direct generalization}
A natural extension of the Stein/Sylvester equation is the \emph{generalized Sylvester equation}
\begin{equation}
AXD + EXB = Y,
\end{equation}
where $A,E \in \mathbb{R}^{m\times m}$, $D,B \in \mathbb{R}^{n\times n}$, and $X$ is the unknown matrix.  
This formulation encompasses the classical Sylvester equation ($AX+XB=Y$) as the special case $D=I$, $E=I$, and is important in control theory, descriptor systems, and model reduction.\footnote{See P.~Van Dooren, ``The generalized eigenstructure problem in linear system theory,'' \emph{IEEE Transactions on Automatic Control}, 26(1):111–129 (1981); D.~S.~Bernstein, \emph{Matrix Mathematics}, Princeton University Press (2009).}

\subsection{Riccati equation(s)}

The classical Riccati equations appear in continuous- and discrete-time control theory.  
In continuous time, the algebraic Riccati equation is
\begin{equation}
A^\ast X + XA + P - X Q X = 0,
\end{equation}
while in discrete time it takes the form
\begin{equation}
X = A^\ast X A + P - X\,(Q+B^\ast X B)^{-1} X.
\end{equation}

\subsubsection{Generalization I}

A direct extension is the \emph{generalized Riccati equation}
\begin{equation}
A^\ast X + X A + \sum_{j=1}^n A_j^\ast X A_j = Y,
\end{equation}
which arises naturally in robust and stochastic control when additional quadratic terms 
model uncertainty channels or noise directions.  
This equation can be seen as a matrix-valued analogue of Riccati-type identities with multiple coupling terms, 
and it reduces to the standard algebraic Riccati equation when the sum contains a single term.  
Such generalizations play a central role in $H^\infty$-control, linear–quadratic–Gaussian problems, 
and systems with multiplicative noise.\footnote{See V.~Ionescu, C.~Oară, and M.~Weiss, 
\emph{Generalized Riccati Theory and Robust Control}, Wiley (1999);  
D.~S.~Bernstein, \emph{Matrix Mathematics}, Princeton University Press (2009).}

\subsection{Quadratic vector equations}
\index{key}{quadratic vector equation}

A natural nonlinear generalization of the linear system
\begin{equation}
A \vec{x} = \vec{b}, \qquad A \in \mathbb{R}^{n\times n},\ \vec{x},\vec{b}\in \mathbb{R}^n,
\end{equation}
is obtained by allowing quadratic terms in the unknown vector.  
A typical form of such a \emph{quadratic vector equation} is
\begin{equation}
B(\vec{x},\vec{x}) + A\vec{x} = \vec{b},
\label{eq:qve}
\end{equation}
where $A \in \mathbb{R}^{n\times n}$ is a linear operator and $B:\mathbb{R}^n\times \mathbb{R}^n \to \mathbb{R}^n$ is a bilinear form (often represented by a third-order tensor).  
In coordinates, this can be written as
\[
(B(\vec{x},\vec{x}))_i = \sum_{j,k=1}^n b_{ijk}\,x_j x_k, \qquad i=1,\dots,n,
\]
so that equation \eqref{eq:qve} reads componentwise
\[
\sum_{j,k=1}^n b_{ijk}\,x_j x_k + \sum_{j=1}^n a_{ij}x_j = b_i.
\]

\paragraph{Interpretation.}  
The linear system $A\vec{x}=\vec{b}$ is recovered when $B=0$.  
Adding the quadratic term captures nonlinear coupling among the components of $\vec{x}$ and leads to richer solution sets.  
Such systems arise in many applied settings:
\emph{Markov chains and stochastic processes}, where $B$ encodes quadratic transition rates; \emph{Queueing theory}, particularly in the analysis of branching processes and retrial queues;  \emph{Control and optimization}, as algebraic Riccati equations are a special case of quadratic vector equations;  \emph{Physics and biology}, where mean-field models and population dynamics often involve quadratic nonlinearities.

\paragraph{Solution methods.}  
Unlike linear systems, quadratic vector equations may have multiple solutions or none at all.  
Common techniques include:
\begin{itemize}
  \item Fixed-point iterations: rewriting \eqref{eq:qve} as $\vec{x} = A^{-1}(\vec{b}-B(\vec{x},\vec{x}))$ when $A$ is invertible.
  \item Newton-type methods: linearizing around a current guess.
  \item Monotone iterations: applicable when $B$ has positivity properties, ensuring convergence to a minimal nonnegative solution.
\end{itemize}

\paragraph{Example.}  
For $\vec{x}\in\mathbb{R}^n$ and a diagonal quadratic operator $B(\vec{x},\vec{x})=\{\beta_i x_i^2\}_{i=1}^n$, equation \eqref{eq:qve} reduces to
\[
\beta_i x_i^2 + \sum_{j=1}^n a_{ij}x_j = b_i, \qquad i=1,\dots,n,
\]
which decouples into $n$ scalar quadratic equations if $A$ is also diagonal.

\medskip
Quadratic vector equations thus provide a unifying framework bridging linear algebra, nonlinear algebraic systems, and stochastic models.\footnote{See G.~Latouche, V.~Ramaswami, \emph{Introduction to Matrix Analytic Methods in Stochastic Modeling}, SIAM (1999); D.~Bini, G.~Latouche, and B.~Meini, \emph{Numerical Methods for Structured Markov Chains}, Oxford University Press (2005).}

\section{Polynomials}\index{key}{polynomials}
\subsection{Basic facts}

Polynomials are among the most fundamental objects in analysis, algebra, and geometry.  We discuss them as the roots of polynomials are associated to the spectra of matrices via Cayley's polynomial.

One reason for their ubiquity is the Weierstrass approximation theorem\index{key}{Weierstrass theorem}: if $F(x)$ is a continuous function on a closed interval $[a,b]$, then there exists a sequence of polynomials $\{p_n(x)\}$ such that
\begin{equation}
    \lim_{n\to \infty} p_n(x) = F(x),
\end{equation}
uniformly on $[a,b]$.  
Thus, polynomials approximate arbitrary continuous functions to any desired accuracy.

\paragraph{Definitions and stability.}
A univariate polynomial is written
\[
f(x)=a_n x^n+\cdots+a_0,\qquad a_n\neq 0.
\]
If the coefficients are real, then complex roots appear in conjugate pairs.  
A polynomial is called \emph{stable} if all of its roots lie in a prescribed domain $\Omega\subset\mathbb{C}$, for instance, the left half-plane (continuous-time stability) or the unit disk (discrete-time stability).

\paragraph{Vieta’s formulas.}
If the roots of $f(x)=a_n x^n+\cdots+a_0$ are $\{z_1,\dots,z_n\}$, then the coefficients are related to the elementary symmetric functions:
\begin{align}
\frac{a_{n-k}}{a_n} &= (-1)^k e_k(z_1,\dots,z_n), \\
e_1(z_1,\dots,z_n) &= \sum_i z_i, \\
e_2(z_1,\dots,z_n) &= \sum_{i<j} z_i z_j, \quad \dots
\end{align}
These relations are known as \emph{Vieta’s formulas}\index{key}{Vieta's formula}. They connect roots, coefficients, and symmetric sums, and appear naturally in expansions of determinants and characteristic polynomials.

\paragraph{Roots as analytic functions of coefficients.}
If $x_r$ is a simple root of $p(x)=\sum_{i=0}^n a_i x^i$, then $x_r$ depends analytically on the coefficients $\{a_i\}$. This follows from the implicit function theorem\index{key}{implicit function theorem}: at a point where $p(x_r,a_i)=0$ and $\partial_x p(x_r,a_i)\neq 0$, the root $x_r$ varies smoothly (indeed, analytically) with the parameters.

\paragraph{Interlacing.}
A polynomial $f$ is said to \emph{interlace} another polynomial $g$ if their roots alternate on the real line. A classical fact is that the roots of $f'$ interlace those of $f$, a property that can be seen from
\[
\frac{f'(x)}{f(x)} = \sum_{i=1}^n \frac{1}{x-z_i}.
\]
Interlacing plays an important role in the theory of orthogonal polynomials, real stability, and spectral graph theory.\footnote{See S.~Fisk, \emph{Polynomials, roots, and interlacing}, unpublished notes (2006).}

\paragraph{Polynomial recurrences.}
Sequences of polynomials defined by three-term recurrences often inherit interlacing properties. For example, if
\[
f_{n}(x) = (a_n x+b_n)f_{n-1}(x)-c_n f_{n-2}(x),\quad f_{-1}=0,\ f_0=1,
\]
with $a_n,c_n>0$, then $f_{n-1}$ interlaces $f_n$.  
Such properties underlie the theory of orthogonal polynomials, including Chebyshev, Legendre, and Hermite families.

\paragraph{Classes of real-rooted polynomials.}
If a polynomial has only real roots, it belongs to the class $\mathcal{P}$. Subclasses include:
\begin{itemize}
    \item $\mathcal{P}^{pos}$: all roots real and non-positive, all coefficients non-negative.
    \item $\mathcal{P}^{alt}$: all roots real and positive, coefficients alternating in sign.
\end{itemize}
These classes are closed under various transformations. For example:
\begin{itemize}
    \item If $f\in\mathcal{P}$ then $f'\in\mathcal{P}$ (Gauss–Lucas theorem).
    \item If $f\in\mathcal{P}^{pos}$, then $f(-x)\in \mathcal{P}^{alt}$.
    \item If $f(x)=\sum_k a_k x^k$ is real-rooted, Newton’s inequalities imply
    \[
    \left(\frac{a_k}{\binom{n}{k}}\right)^2 \geq 
    \frac{a_{k-1}a_{k+1}}{\binom{n}{k-1}\binom{n}{k+1}},
    \]
    which yields \emph{ultra log-concavity}.
    \item Stability in the left half-plane yields further inequalities among coefficients.
\end{itemize}

\paragraph{Toeplitz and Hankel connections.}
Real-rooted polynomials are closely related to matrix positivity.  
If $f(x)=\sum_k a_k x^k$ is real-rooted with non-negative coefficients, Edrei’s theorem\footnote{T.~Edrei, ``On the generating functions of totally positive sequences. II,'' \emph{J. Analyse Math.} \textbf{2} (1952).} states that the Toeplitz matrix $A_{ij}=a_{i-j}$ is totally positive.  
Similarly, if $f(x)=\prod_i(x-\lambda_i)=\sum_{k=0}^n (-1)^k e_k x^{n-k}$ and $m_k=\sum_i \lambda_i^k$, then the Hankel matrix $H_{ij}=m_{i+j-2}$ is positive semidefinite if and only if $f(x)$ is real-rooted.

\medskip
In summary, polynomials form a bridge between algebra, analysis, and matrix theory. Approximation theorems, interlacing, recurrence relations, and connections to totally positive matrices illustrate their central role in both classical and modern mathematics.

\subsection{Multivariate polynomials}
We now extend to several variables.  
A nonzero multivariate polynomial $f(z_1,\dots,z_n)$ is called \emph{stable} if it has no zeros whenever each $\operatorname{Im}(z_i)>0$.  
Stability in this sense generalizes univariate root location results and plays an important role in combinatorics, probability, and control theory.\footnote{See J.~Borcea and P.~Brändén, ``Multivariate Pólya-Schur classification problems in the Weyl algebra,'' \emph{Proc. Lond. Math. Soc.} \textbf{101} (2010).}

\subsection{From Polynomials to Matrices}
We now return to the central bridge between polynomials and matrices: the \emph{companion matrix}, which also appears in a variety of fields.\footnote{The Fibonacci sequence, defined by the recurrence 
$f(k+1) = f(k) + f(k-1)$, can be expressed compactly using the 
eigenvalues of the companion matrix 
$\begin{pmatrix} 1 & 1 \\ 1 & 0 \end{pmatrix}$. 
Its growth rate is governed by the spectral radius of this matrix, 
which is the golden ratio $\varphi = (1+\sqrt{5})/2$. 
In this way, even the simple Fibonacci series is fundamentally spectral: 
its asymptotics emerge directly from diagonalization.}

As seen earlier, the eigenvalues of a matrix are the roots of its characteristic polynomial. Conversely, given a polynomial we can construct a matrix whose eigenvalues are exactly its roots.

Consider the univariate polynomial
\begin{equation}
    p(x)=\sum_{i=0}^n C_i x^i, \qquad C_n=1.
\end{equation}
A companion matrix associated with $p(x)$ is
\begin{equation}
    M=\begin{pmatrix}
         0 & 1 & \cdots & 0  \\
         \vdots & \vdots & \ddots & \vdots \\
         0 & 0 & \cdots & 1\\
         -C_0 & -C_1 & \cdots & -C_{n-1}
    \end{pmatrix},
\end{equation}
also known as the Frobenius companion matrix.\footnote{See R.A.~Horn and C.R.~Johnson, \emph{Matrix Analysis}, Cambridge Univ. Press (1985).}  
It satisfies $\det(M-x I)=\pm p(x)$, and the eigenvalues of $M$ are the roots of $p(x)$.  

Furthermore, the Vandermonde matrix
\begin{equation}
    V=\begin{pmatrix}
         1 & 1 & \cdots & 1  \\
         \lambda_1 & \lambda_2 & \cdots & \lambda_n \\
         \vdots & \vdots & \ddots & \vdots \\
         \lambda_1^{n-1}& \lambda_2^{n-1} & \cdots & \lambda_n^{n-1}
    \end{pmatrix}
\end{equation}
diagonalizes $M$, provided the $\lambda_i$ are distinct:
\[
    V M V^{-1}=\mathrm{diag}(\lambda_1,\dots,\lambda_n).
\]

\subsection{Resultants and Sylvester Matrices}
For two univariate polynomials
\[
    p(x)=\sum_{i=0}^m p_i x^i, \qquad q(x)=\sum_{i=0}^n q_i x^i,
\]
the \emph{Sylvester matrix} $S_{p,q}$ encodes whether $p$ and $q$ have common roots. Its determinant, the \emph{resultant}, satisfies
\begin{equation}
    R_{p,q}=\det S_{p,q}=p_m^n q_n^m \prod_{i,j}(x_i^p-x_j^q).
\end{equation}
Thus $R_{p,q}=0$ if and only if $p$ and $q$ have a common root.  
The discriminant $\Delta(f)$ of a polynomial $f$ is related: $\Delta(f)=\tfrac{1}{a_n}R_{f,f'}$.  

\subsection{Macaulay Matrices}
A generalization to several variables is provided by the \emph{Macaulay matrix}, which organizes the coefficients of a system of polynomials into a structured linear system.  
If $f_1,\dots,f_k$ are polynomials, the Macaulay matrix $M(d)$ at degree $d$ is constructed by multiplying each $f_i$ with all monomials of degree at most $d-\deg(f_i)$. The rows correspond to these multiplied polynomials, and the columns to all monomials of degree $\leq d$.  
Resultants of multivariate systems can then be extracted from determinants of appropriate submatrices of $M(d)$.\footnote{See I.~Emiris, ``Sparse elimination and applications in kinematics,'' Ph.D. thesis, UC Berkeley (1994).}

\subsection{Stable Polynomials}
Finally, stability.  
For a linear dynamical system $\dot x = A x$, stability requires $\mathrm{Re}(\lambda)<0$ for all eigenvalues of $A$, i.e. the characteristic polynomial of $A$ has all roots in the left half-plane.  

This motivates the definition of \emph{stable polynomials}, whose root locations determine stability criteria in control theory (Lyapunov’s theorem, Routh–Hurwitz conditions, etc.).  
Thus, much of matrix stability analysis reduces to studying properties of the associated polynomials.

Let us now briefly consider a certain class of polynomials with special root and coefficient structures.  

If a polynomial has only real roots, we denote it as belonging to the set $\mathcal P$. If its coefficients are all of the same sign, and all roots are real and negative, then it belongs to $\mathcal P^{pos}$. If instead it has all real roots, all positive, and coefficients of alternating signs, then it belongs to $\mathcal P^{alt}$.  

Some useful closure properties are as follows:
\begin{itemize}
    \item If $g(x)\in \mathcal P$, then $g'(x)\in \mathcal P$ (real-rootedness is preserved under differentiation).
    \item If $g\in \mathcal P$ and all coefficients are positive, then $g\in \mathcal P^{pos}$.
    \item If $g\in \mathcal P$, then there exists $\alpha$ such that $g(x+\alpha)\in \mathcal P$.
    \item $f(x) \in \mathcal P^{pos}$ if and only if $f(-x)\in \mathcal P^{alt}$.
    \item $f(-x^2)$ has only real roots if and only if $f(x)\in \mathcal P^{pos}$.
    \item $f(x^2)$ has only real roots if and only if $f(x)\in \mathcal P^{alt}$.
    \item If $g$ interlaces $f$ and $f,g \in \mathcal P^{pos}$, then $f(-x^2)+x g(-x^2)\in \mathcal P$.
    \item If $f(-x)$ and $f(x)$ interlace, then the signs of the coefficients of $f$ alternate as $++--++\cdots$.
    \item If $f(x)$ has no real roots, then $f(x)>0$ everywhere and its even and odd parts both have only real roots.
    \item Newton’s inequalities: if $f(x)=\sum_k a_k x^k$ is real-rooted, then
    \[
    \left(\frac{a_k}{\binom{n}{k}}\right)^2 \geq 
    \frac{a_{k-1}a_{k+1}}{\binom{n}{k-1}\binom{n}{k+1}}, \qquad 1\leq k \leq n-1,
    \]
    which implies \emph{ultra-log-concavity}\index{key}{ultra log concavity}.
    \item If $f(x)$ is stable (all roots in the left half-plane), then 
    \[
        \frac{a_{i+1}a_{i-1}}{a_{i+2}a_{i-2}}\geq 1.
    \]
\end{itemize}

\paragraph{Toeplitz and Hankel connections.}  
Real-rooted polynomials are closely linked to total positivity. If $g(x)=\sum_k a_k x^k$ is real-rooted, Edrei’s theorem\footnote{T.~Edrei, ``On the generating functions of totally positive sequences. II,'' \emph{J. Analyse Math.} \textbf{2} (1952).} states that the Toeplitz matrix $A_{ij}=a_{i-j}$ is totally positive, i.e., all its minors are positive.  
Similarly, if $g(x)=\prod_i(x-\lambda_i)=\sum_{k=0}^n (-1)^k e_k x^{n-k}$ and $m_k=\sum_i \lambda_i^k$, then the Hankel matrix $H_{ij}=m_{i+j-2}$ is positive semidefinite if and only if $g(x)$ is real-rooted.

\subsection{Multivariate polynomials}
We now extend to several variables.  
A nonzero multivariate polynomial $f(z_1,\dots,z_n)$ is called \emph{stable} if it has no zeros whenever each $\operatorname{Im}(z_i)>0$.  
Stability in this sense generalizes univariate root location results and plays an important role in combinatorics, probability, and control theory.\footnote{See J.~Borcea and P.~Brändén, ``Multivariate Pólya-Schur classification problems in the Weyl algebra,'' \emph{Proc. Lond. Math. Soc.} \textbf{101} (2010).}

\subsection{From Polynomials to Matrices}
We now return to the central bridge between polynomials and matrices: the \emph{companion matrix}.  
As seen earlier, the eigenvalues of a matrix are the roots of its characteristic polynomial. Conversely, given a polynomial we can construct a matrix whose eigenvalues are exactly its roots.

Consider the univariate polynomial
\begin{equation}
    p(x)=\sum_{i=0}^n C_i x^i, \qquad C_n=1.
\end{equation}
A companion matrix associated with $p(x)$ is
\begin{equation}
    M=\begin{pmatrix}
         0 & 1 & \cdots & 0  \\
         \vdots & \vdots & \ddots & \vdots \\
         0 & 0 & \cdots & 1\\
         -C_0 & -C_1 & \cdots & -C_{n-1}
    \end{pmatrix},
\end{equation}
also known as the Frobenius companion matrix.\footnote{See R.A.~Horn and C.R.~Johnson, \emph{Matrix Analysis}, Cambridge Univ. Press (1985).}  
It satisfies $\det(M-x I)=\pm p(x)$, and the eigenvalues of $M$ are the roots of $p(x)$.  

Furthermore, the Vandermonde matrix
\begin{equation}
    V=\begin{pmatrix}
         1 & 1 & \cdots & 1  \\
         \lambda_1 & \lambda_2 & \cdots & \lambda_n \\
         \vdots & \vdots & \ddots & \vdots \\
         \lambda_1^{n-1}& \lambda_2^{n-1} & \cdots & \lambda_n^{n-1}
    \end{pmatrix}
\end{equation}
diagonalizes $M$, provided the $\lambda_i$ are distinct:
\[
    V M V^{-1}=\mathrm{diag}(\lambda_1,\dots,\lambda_n).
\]

\subsection{Resultants and Sylvester Matrices}
For two univariate polynomials
\[
    p(x)=\sum_{i=0}^m p_i x^i, \qquad q(x)=\sum_{i=0}^n q_i x^i,
\]
the \emph{Sylvester matrix} $S_{p,q}$ encodes whether $p$ and $q$ have common roots. Its determinant, the \emph{resultant}, satisfies
\begin{equation}
    R_{p,q}=\det S_{p,q}=p_m^n q_n^m \prod_{i,j}(x_i^p-x_j^q).
\end{equation}
Thus $R_{p,q}=0$ if and only if $p$ and $q$ have a common root.  
The discriminant $\Delta(f)$ of a polynomial $f$ is related: $\Delta(f)=\tfrac{1}{a_n}R_{f,f'}$.  

\subsection{Macaulay Matrices}
A generalization to several variables is provided by the \emph{Macaulay matrix}, which organizes the coefficients of a system of polynomials into a structured linear system.  
If $f_1,\dots,f_k$ are polynomials, the Macaulay matrix $M(d)$ at degree $d$ is constructed by multiplying each $f_i$ with all monomials of degree at most $d-\deg(f_i)$. The rows correspond to these multiplied polynomials, and the columns to all monomials of degree $\leq d$.  
Resultants of multivariate systems can then be extracted from determinants of appropriate submatrices of $M(d)$.\footnote{See I.~Emiris, ``Sparse elimination and applications in kinematics,'' Ph.D. thesis, UC Berkeley (1994).}

\subsection{Stable Polynomials}
Finally, stability.  
For a linear dynamical system $\dot x = A x$, stability requires $\mathrm{Re}(\lambda)<0$ for all eigenvalues of $A$, i.e. the characteristic polynomial of $A$ has all roots in the left half-plane.  

This motivates the definition of \emph{stable polynomials}, whose root locations determine stability criteria in control theory (Lyapunov’s theorem, Routh–Hurwitz conditions, etc.).  
Thus, much of matrix stability analysis reduces to studying properties of the associated polynomials.

\part{Applications in dynamical systems}
\chapter{Dynamical systems}
\section{ODE and Vectorization}

Spectral methods emerge naturally when dynamical laws are linear (or locally linear) and time evolution is governed by matrices.  
In complex systems, this perspective ties stability, transients, and long-time behavior directly to spectral data (eigenvalues, Jordan structure) and matrix functions (exponentials, resolvents).\footnote{See R.~A.~Horn and C.~R.~Johnson, \emph{Matrix Analysis}, Cambridge Univ. Press (1985); N.~J.~Higham, \emph{Functions of Matrices}, SIAM (2008).}

\medskip
\noindent\textbf{From higher-order ODEs to first-order state space.}
We begin by discussing how a linear ODE of higher order can be written in a first order form. A scalar linear ODE with constant coefficients
\begin{equation}
    a_n x^{(n)}(t)+a_{n-1} x^{(n-1)}(t)+\cdots+a_0 x(t)=f(t)
\label{eq:ode}
\end{equation}
can be lifted to a first-order system by the companion construction. \label{sec:companionmatode}
With the state $\vec y(t)=(x,\dot x,\dots,x^{(n-1)})^{\top}$,
\[
    \frac{d}{dt}\,\vec y(t)=A \vec y(t)+\vec d(t),
\]
where $A$ is the (transpose of the) Frobenius companion matrix
\begin{equation}
    A=\begin{pmatrix}
         0 & 1 & 0 & \cdots & 0  \\
         0 & 0 & 1 &  \cdots &0 \\ 
         \vdots &  & \ddots & \ddots & \vdots\\
         0 & \cdots & \cdots & 0 & 1 \\
         -\tfrac{a_0}{a_n} & - \tfrac{a_1}{a_n} & \cdots & \cdots & -\tfrac{a_{n-1}}{a_n}
    \end{pmatrix}.
    \label{eq:companionmatode}
\end{equation}
The eigenvalues of $A$ are the roots of the characteristic polynomial of \eqref{eq:ode}, so stability, oscillation, and resonance reduce to spectral properties of $A$.\footnote{Classic ODE references: E.~A.~Coddington and N.~Levinson, \emph{Theory of Ordinary Differential Equations}, McGraw–Hill (1955); G.~Teschl, \emph{Ordinary Differential Equations and Dynamical Systems}, AMS (2012).}

\subsection{Linear Autonomous Systems}
\label{sec:lindiffeq}
For $\dot{\vec x}(t)=A\vec x(t)$ the solution operator is the matrix exponential,
\[
   \vec x(t)=e^{At}\vec x(0).
\]
Hence asymptotic stability is characterized by the spectral abscissa: $\Re \lambda_i<0$ for all $\lambda_i\in \Lambda(A)$.\footnote{See H.~K.~Khalil, \emph{Nonlinear Systems} (3rd ed.), Prentice Hall (2002), Ch.~2; Higham (2008) for $e^{At}$.}

\paragraph{Two-dimensional case (trace–determinant picture).}
For
\[
\dot{\vec x}(t)=\begin{pmatrix}a & b \\ c & d\end{pmatrix}\vec x(t),
\]
the eigenvalues satisfy $\lambda^2-\lambda\,\mathrm{Tr}(A)+\det(A)=0$.  
Stability regions and bifurcations can be visualized in the trace–determinant plane, a standard phase-portrait tool in low-dimensional complex systems.

\paragraph{Jordan form and transient growth.}
If $A$ is diagonalizable, $e^{At}$ is governed directly by its eigenvalues.  
Non-diagonalizable $A$ introduces Jordan blocks and polynomial prefactors $t^k e^{\lambda t}$.  
This explains how non-normality can yield large transient amplification even when $\Re \lambda<0$, a mechanism relevant to tipping and bursts in complex networks.

\subsection{Discrete Time and Spectral Radius}

Time discretization with step $\delta t$ yields a linear recurrence
\[
   \vec y_{k+1}=M \vec y_k,
\]
with $M$ determined by the numerical scheme (e.g., $M=I+\delta t\,A$ for forward Euler).  
Stability is governed by the spectral radius:
\[
   \rho(M)<1 \ \Rightarrow\  \vec y_k \to 0.
\]
This connects continuous-time spectra to discrete-time contractivity, and underpins stability analysis of iterative algorithms.\footnote{See G.~H.~Golub and C.~F.~Van Loan, \emph{Matrix Computations} (4th ed.), Johns Hopkins (2013), and Dahlquist’s classical results on A-stability.}

\subsection{Linearization Near Equilibria}

For a nonlinear system $\dot x = v(t,x)$, linearization along a reference solution $x^*(t)$ gives the variational equation
\[
   \dot \xi(t)=A(t)\,\xi(t), \qquad A(t)=\frac{\partial v}{\partial x}(t,x^*(t)).
\]
Local stability, growth rates, and sensitivity are thus encoded in the (possibly time-varying) spectrum of $A(t)$; this is the gateway from nonlinear dynamics to spectral analysis.\footnote{See M.~Hirsch, S.~Smale, and R.~Devaney, \emph{Differential Equations, Dynamical Systems, and an Introduction to Chaos} (3rd ed.), Academic Press (2012).}

\subsection{Time-Periodic Systems and Floquet Theory}

When $A(t+T)=A(t)$ is $T$-periodic, a single period encapsulates the dynamics via the monodromy operator $P$ (mapping $x(0)$ to $x(T)$).  
Its eigenvalues—the \emph{Floquet multipliers}—govern stability.  
Floquet’s theorem states that a fundamental solution can be written as
\[
   X(t)=\Phi(t)\,e^{t\Lambda},
\]
with $\Phi(t+T)=\Phi(t)$ and $\Lambda$ constant.\footnote{G.~Floquet, ``Sur les équations différentielles linéaires à coefficients périodiques,'' \emph{Ann. ENS} (1883). Standard modern treatments appear in many dynamical systems texts, e.g., Teschl (2012).}
This factorization separates periodic modulation from exponential growth/decay—an idea used widely in waves, oscillators, and parametric resonance.

\subsection{Lyapunov Exponents}\label{sec:lyapunov}

The characteristic Lyapunov exponent of a trajectory $g(t)$ is
\[
\chi(g)=\lim_{t\to\infty}\sup \frac{1}{t}\ln\|g(t)\|.
\]
For $\dot x=Ax$, the largest Lyapunov exponent equals the spectral abscissa $\max_i \Re(\lambda_i)$, linking asymptotic growth directly to eigenvalues.  
In nonautonomous or random settings, Oseledets’ theorem generalizes this notion to multiplicative cocycles.\footnote{V.~I.~Oseledets, ``A multiplicative ergodic theorem,'' \emph{Trans. Moscow Math. Soc.} 19 (1968).}

\subsection{Transition Matrices}

The state transition matrix $X(t,t_0)$ satisfies $\vec x(t)=X(t,t_0)\vec x(t_0)$.  
For constant $A$, $X(t,t_0)=e^{A(t-t_0)}$; for general $A(t)$, $X$ solves a matrix ODE.  
Via Laplace transform, $X$ connects to the resolvent $(zI-A)^{-1}$, tying time-domain propagation to spectral resolvent analysis used elsewhere in the book.\footnote{Higham (2008), Chs.~1–3.}

\subsection{Remark on Hamiltonian Systems}\label{sec:hamiltonian}

Hamiltonian linearization has the form $\dot x=J\nabla H$, with the symplectic matrix
\[
   J=\begin{pmatrix}0 & -I \\ I & 0\end{pmatrix}, \quad J^2=-I.
\]
Eigenvalues of $J A$ occur in symmetric $\pm i\omega$ pairs, reflecting the oscillatory nature of Hamiltonian flows.\footnote{V.~I.~Arnold, \emph{Mathematical Methods of Classical Mechanics}, Springer (1978).}

\subsubsection{Linear Hamiltonian Systems}

For a quadratic Hamiltonian
\[
   H=\tfrac{1}{2}\sum_{i=1}^n \Omega_i(p_i^2+q_i^2),
\]
in canonical coordinates $\vec x=(\vec q,\vec p)$, the dynamics obey
\begin{equation}
   \dot{\vec x}=J A(t)\vec x,
\end{equation}
with $A(t)=A(t+T)$ symmetric and $J$ symplectic.  
If $X(t)$ is a fundamental matrix, symplecticity yields
\begin{equation}
   X(t)^{\top} J X(t)=J, \qquad \det(X(t))=1.
\end{equation}
For the monodromy matrix $\Pi=X(T)$,
\[
   \Pi^{\top} J \Pi = J, \qquad \Pi^{-T}=J\Pi J^{-1},
\]
so $\Pi$ is similar to its inverse transpose.

\paragraph{Spectral symmetry.}
Floquet multipliers (eigenvalues of $\Pi$) come in reciprocal pairs: if $\rho$ is a multiplier, so is $1/\rho$.  
Hence $\pm 1$ have even algebraic multiplicity, and the characteristic polynomial $\det(\Pi-\rho I)$ is palindromic, i.e. $g(\rho+1/\rho)=0$.  
This reciprocity is a hallmark of symplectic dynamics and underlies stability bands and tongues in many complex oscillatory systems.\footnote{Arnold (1978); for numerical aspects, see E.~Hairer, C.~Lubich, and G.~Wanner, \emph{Geometric Numerical Integration}, Springer (2006).}

\subsection{Lyapunov stability criterion}\label{sec:lyapunovstab}
\index{key}{Lyapunov stability}\index{key}{Lyapunov function}

An equilibrium point $\vec x^\ast$ of $\dot{\vec x}=f(\vec x)$ is said to be
\emph{Lyapunov stable} if for every $\varepsilon>0$ there exists $\delta>0$ such that
$\|\vec x(0)-\vec x^\ast\|<\delta$ implies $\|\vec x(t)-\vec x^\ast\|<\varepsilon$
for all $t\geq 0$.  
It is \emph{asymptotically stable} if it is Lyapunov stable and, in addition,
$\vec x(t)\to \vec x^\ast$ as $t\to\infty$.

\paragraph{Lyapunov functions.}
A standard tool to verify stability is the construction of a \emph{Lyapunov function}:
a continuously differentiable function $V:\mathbb{R}^n\to\mathbb{R}$ such that
\begin{enumerate}
\item $V(\vec x^\ast)=0$ and $V(\vec x)>0$ for $\vec x\neq \vec x^\ast$ (positive definite),
\item $\dot V(\vec x)=\nabla V(\vec x)^\top f(\vec x)\leq 0$ in a neighborhood of $\vec x^\ast$.
\end{enumerate}
If these conditions hold, then $\vec x^\ast$ is Lyapunov stable.
If in addition $\dot V(\vec x)<0$ for all $\vec x\neq \vec x^\ast$, then $\vec x^\ast$
is asymptotically stable.\footnote{See H.~K.~Khalil, \emph{Nonlinear Systems} (3rd ed.), Prentice Hall (2002).}

\paragraph{Linear systems.}
For the linear system $\dot{\vec x}=A\vec x$, the origin $\vec x^\ast=0$ is an equilibrium.
A quadratic Lyapunov function candidate is
\begin{equation}
V(\vec x)=\vec x^\top P \vec x,\qquad P=P^\top\succ 0.
\end{equation}
Then
\begin{equation}
\dot V(\vec x)=\vec x^\top (A^\top P+PA)\vec x.
\end{equation}
Hence, stability reduces to the \emph{Lyapunov inequality}
\begin{equation}
A^\top P+PA \prec 0,
\end{equation}
which has a solution $P\succ 0$ if and only if all eigenvalues of $A$ have negative real parts.  
This provides an algebraic characterization of asymptotic stability for linear ODEs, complementing the spectral abscissa condition discussed in Sec.~\ref{sec:lindiffeq}.\footnote{See R.~A.~Horn and C.~R.~Johnson, \emph{Matrix Analysis}, Cambridge Univ. Press (1985); Khalil (2002).}

\section{Local Stability Analysis of Nonlinear Systems}\label{sec:stability}

We now turn to nonlinear dynamical systems and recall how their local behavior reduces to the spectral theory already discussed in earlier sections.  

Consider
\begin{equation}
   \frac{d\vec Y}{dt}=\vec f(\vec Y),
\end{equation}
with $\vec f:\mathbb{R}^n\to \mathbb{R}^n$ smooth.  
An equilibrium point $\vec Y^*$ satisfies $\vec f(\vec Y^*)=0$.  

\paragraph{Linearization.}
Expanding around $\vec Y^*$ gives the \emph{Jacobian matrix}
\[
   M_{ij}=\frac{\partial f_i}{\partial Y_j}\Big|_{\vec Y=\vec Y^*},
\]
so that small perturbations $\Delta\vec Y$ satisfy
\begin{equation}
   \Delta \vec Y(t)=e^{Mt}\Delta \vec Y(0).
\end{equation}
Thus, the local dynamics are governed by the exponential of $M$, as in the linear case discussed in Sec.~\ref{sec:lindiffeq}.  

\paragraph{Spectral criterion.}
The equilibrium $\vec Y^*$ is asymptotically stable if all eigenvalues of $M$ satisfy $\Re\lambda<0$.  
This is the classical Lyapunov stability criterion\footnote{A.M.~Lyapunov, \emph{The General Problem of the Stability of Motion}, Taylor \& Francis (1992) [English translation].}, and shows once again how local stability reduces to spectral properties of a linear operator.  

\paragraph{Exponential representation.}
The matrix exponential $e^{Mt}$ is defined by its convergent series
\[
   e^{Mt}=\sum_{n=0}^\infty \frac{t^n}{n!}M^n,
\]
which has appeared repeatedly in our analysis (see also the discussion of Jordan blocks in Sec.~\ref{sec:solsys2}).  
Its spectral norm controls transient growth, while its asymptotic behavior is dictated by the spectral abscissa $\alpha(M)=\max_i \Re\lambda_i$.

\paragraph{Applications.}
Linearization and Jacobian spectra are used across applied sciences: in biology to study equilibria of population models\footnote{J.D.~Murray, \emph{Mathematical Biology}, Springer (2002).}, in epidemiology for disease thresholds, and in physics for stability of equilibria in nonlinear oscillators.

\paragraph{Special case of two dimensions.}
The stability of linear systems in the plane provides a vivid illustration
of how spectra control dynamics. Consider the system
\[
\dot{\vec{x}} = A \vec{x}, \qquad A \in \mathbb{R}^{2\times 2}.
\]
The trajectories of the system are governed entirely by the eigenvalues of $A$:

\begin{itemize}
  \item If both eigenvalues have negative real part, the origin is a \emph{stable node}:
  trajectories decay exponentially to the origin.
  \item If both eigenvalues have positive real part, the origin is an \emph{unstable node}:
  trajectories diverge exponentially.
  \item If eigenvalues are real with opposite signs, the origin is a \emph{saddle point}:
  trajectories approach along the stable eigendirection and diverge along the unstable one.
  \item If eigenvalues are complex with negative real part, the origin is a \emph{stable spiral}:
  trajectories spiral inward with exponential decay.
  \item If eigenvalues are complex with positive real part, the origin is an \emph{unstable spiral}:
  trajectories spiral outward.
  \item If eigenvalues are purely imaginary, the origin is a \emph{center}:
  trajectories are closed orbits, neutrally stable.
\end{itemize}

\begin{figure}[h]
\centering
\includegraphics[width=0.8\textwidth]{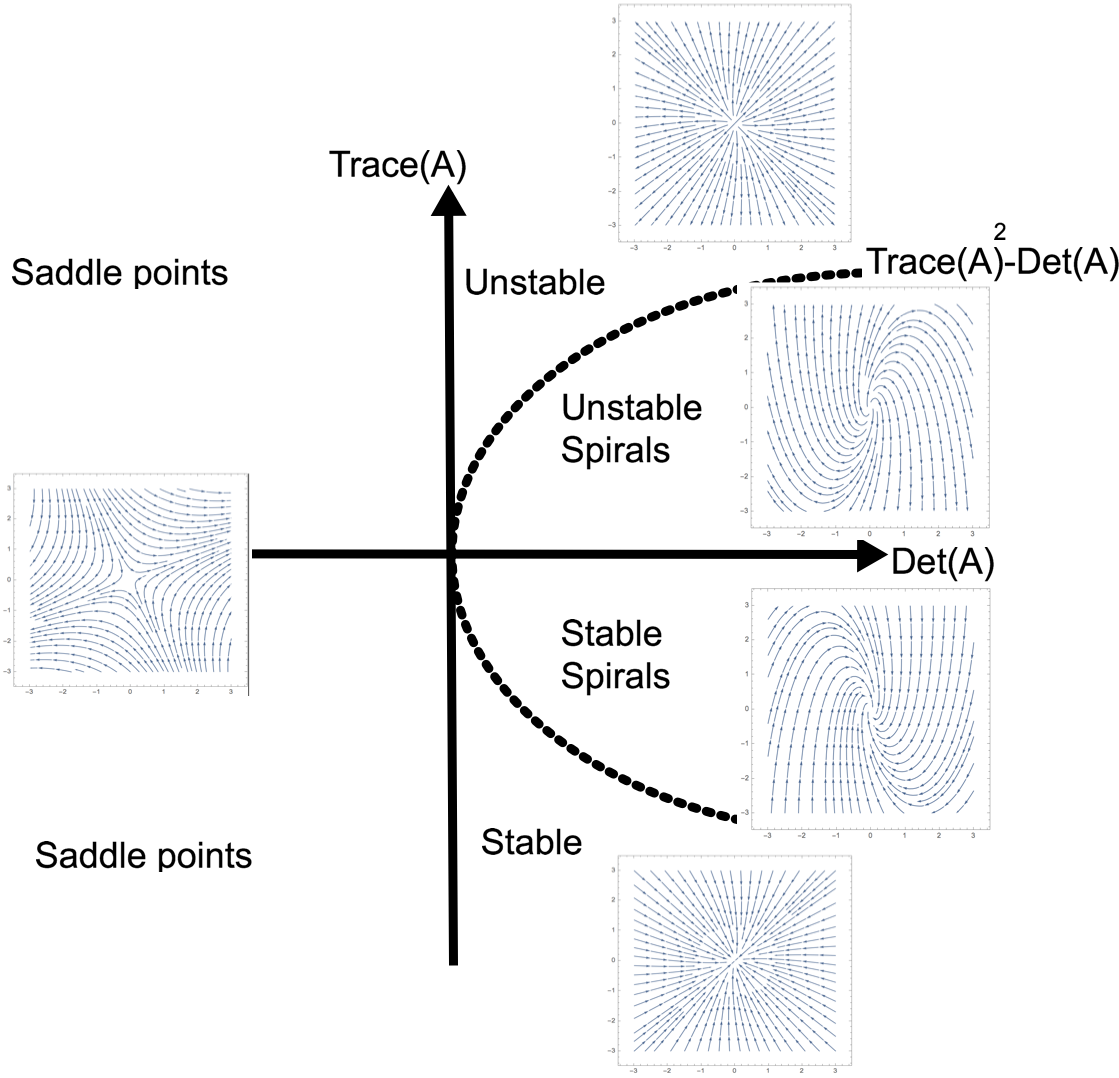}
\caption{Phase portraits of two-dimensional linear systems for different
eigenvalue configurations. The spectral properties of $A$ fully determine
the qualitative dynamics.}
\label{fig:stability2d}
\end{figure}

These phase portraits are shown in Fig.~\ref{fig:stability2d}.  
They demonstrate the intimate relation between stability and spectra: the
\emph{real parts} of the eigenvalues determine growth or decay, while the
\emph{imaginary parts} encode oscillatory behavior. Thus the spectrum of $A$
fully classifies local dynamics in two dimensions.

\subsection{Formal solution using Laplace transforms}\label{sec:lincontr}

We now revisit linear dynamical systems with external forcing,
\begin{equation}
   \dot{\vec Y}(t)=A\vec Y(t)+B\vec u(t),
   \label{eq:dyn}
\end{equation}
where $\vec Y(t)\in\mathbb{R}^n$ is the state vector and $\vec u(t)$ is an external input.  

\paragraph{Laplace domain solution.}
Taking Laplace transforms gives
\[
   \vec Y(s) = (sI-A)^{-1}\vec Y(0)+(sI-A)^{-1}B\vec u(s),
\]
so that the inverse transform yields
\begin{equation}
   \vec Y(t)=\mathcal{L}^{-1}\{(sI-A)^{-1}\vec Y(0)\}+\mathcal{L}^{-1}\{(sI-A)^{-1}B\vec u(s)\}.
\end{equation}
The first term corresponds to the homogeneous solution, while the second describes the forced response.  

\paragraph{Resolvent connection.}
Since
\begin{equation}
   e^{At}=\mathcal{L}^{-1}\{(sI-A)^{-1}\},
\end{equation}
the resolvent $(sI-A)^{-1}$ and the semigroup $e^{At}$ are Laplace duals.\footnote{A.~Pazy, \emph{Semigroups of Linear Operators and Applications to Partial Differential Equations}, Springer (1983).}  
This makes explicit that the poles of the resolvent coincide with the eigenvalues of $A$, controlling growth and decay.  

\paragraph{Integral representation.}
By Mellin’s inversion formula,
\begin{equation}
   e^{At}=\frac{1}{2\pi i}\int_{\gamma-i\infty}^{\gamma+i\infty}(sI-A)^{-1}e^{st}\,ds,
\end{equation}
where $\gamma>\max_{\lambda\in\Lambda(A)}\Re\lambda$.  
This contour integral is evaluated by residues at the eigenvalues of $A$, again highlighting the spectral connection.\footnote{L.N.~Trefethen and M.~Embree, \emph{Spectra and Pseudospectra}, Princeton University Press (2005).}  

\paragraph{Duhamel’s principle.}
In time domain, the solution of (\ref{eq:dyn}) is
\begin{equation}
   \vec Y(t)=e^{At}\vec Y(0)+\int_0^{\top} e^{A(t-s)}B\vec u(s)\,ds,
\end{equation}
known as Duhamel’s formula.  
Equivalently, in terms of the state transition matrix $X(t,t_0)=e^{(t-t_0)A}$,
\begin{equation}
   \vec Y(t)=X(t,t_0)\vec Y(t_0)+\int_{t_0}^{\top} X(t,s)B\vec u(s)\,ds.
\end{equation}

\section{Time-continuous dynamics and non-autonomous systems}\label{sec:differenceeq}

We now generalize to time-dependent operators.  
Let $\vec Y(t)$ evolve according to
\begin{equation}
   \dot{\vec Y}(t)=A(t)\vec Y(t),
   \label{eq:matrixvecevolutionexp}
\end{equation}
with solution operator $U(t,t_0)$ defined by
\begin{equation}
   \vec Y(t)=U(t,t_0)\vec Y(t_0).
   \label{eq:matrixevecvolution}
\end{equation}

\paragraph{Commuting case.}
If $A(t)$ commutes at different times, $[A(t),A(s)]=0$, then
\[
   U(t,t_0)=\exp\!\left(\int_{t_0}^{\top} A(s)\,ds\right).
\]
This is a direct generalization of the scalar exponential solution.  

\paragraph{Non-commuting case.}
In general $A(t)A(s)\neq A(s)A(t)$, so (\ref{eq:matrixvecevolutionexp}) cannot be integrated directly.  
Instead, one introduces the \emph{time-ordered exponential}
\[
   U(t,t_0)=\mathcal{T}\exp\!\left(\int_{t_0}^{\top} A(s)\,ds\right),
\]
where $\mathcal{T}$ denotes chronological ordering.\footnote{E.~Coddington and N.~Levinson, \emph{Theory of Ordinary Differential Equations}, McGraw–Hill (1955).}  
This operator still inherits spectral properties from $A(t)$, but its analysis requires Floquet theory for periodic $A(t)$ or Lyapunov exponents for general $A(t)$ (see Secs.~\ref{sec:stability} and \ref{sec:lindiffeq}).  

\paragraph{Spectral viewpoint.}
Thus, both constant and time-dependent systems reduce to the study of resolvents and semigroups.  
The key distinction is that in the non-autonomous case, the lack of commutativity complicates spectral analysis, and motivates the use of Floquet multipliers, Lyapunov exponents, and pseudospectral bounds.

\subsection{Exponential operator families and the Cauchy operator}

We now extend the analysis from vectorial equations to operator evolution.  
As discussed in Sec.~\ref{sec:lincontr}, the fundamental solution of a linear ODE with constant $A$ can be written as $U(t)=e^{At}$.  
If $A(t)$ depends on time, the corresponding \emph{Cauchy operator} $U(t)$ satisfies
\begin{equation}
   U'(t)=A(t)U(t), \qquad U(0)=I.
   \label{eq:timord1}
\end{equation}
In quantum mechanics, this $U(t)$ is called the \emph{S-matrix} or evolution operator.  

The difficulty is that in general $A(t)$ and $A(s)$ do not commute, so that
\[
   e^{\int_0^{\top} A(s)ds} \neq \prod_{s} e^{A(s)\,ds}.
\]
Thus, new techniques are required to construct $U(t)$. Two complementary approaches are the \emph{Dyson time-ordering expansion} and the \emph{Magnus expansion}.

\subsubsection{Dyson time-ordering expansion}
\label{sec:dysonseries}
Equation \eqref{eq:timord1} can be written in integral form
\begin{equation}
   U(t)=I+\int_0^{\top} A(t_1)U(t_1)\,dt_1.
   \label{eq:timeord2}
\end{equation}
Iterating gives the series
\begin{equation}
   U(t)=I+\int_0^{\top} A(t_1)\,dt_1 + \int_0^{\top} dt_1 \int_0^{t_1} dt_2\,A(t_1)A(t_2)+\cdots,
\end{equation}
which generalizes to the \emph{Dyson series}\footnote{F.J.~Dyson, 
``The radiation theories of Tomonaga, Schwinger, and Feynman,'' \emph{Phys. Rev.} \textbf{75}, 486 (1949).}
\begin{equation}
   U(t)=\sum_{k=0}^\infty \int_0^{\top} dt_1 \int_0^{t_1}dt_2 \cdots \int_0^{t_{k-1}}dt_k\, A(t_1)\cdots A(t_k).
   \label{eq:dysonseries}
\end{equation}
The nested integrals impose the ordering $t_1>t_2>\cdots>t_k$.  
This motivates the \emph{time-ordering operator} $\mathcal T$, defined by
\[
   \mathcal T(A(t_1)A(t_2))=\begin{cases}
   A(t_1)A(t_2), & t_1>t_2,\\
   A(t_2)A(t_1), & t_2>t_1,
   \end{cases}
\]
so that \eqref{eq:dysonseries} can be written compactly as
\begin{equation}
   U(t)=\mathcal T\exp\!\left(\int_0^{\top} A(s)\,ds\right).
   \label{eq:dysonseries2}
\end{equation}
This expression generalizes the matrix exponential and plays a central role in perturbation theory, control theory, and quantum mechanics (where $A\to -iH(t)$ with $H$ the Hamiltonian).  

\subsubsection{Magnus expansion}

An alternative representation was introduced by Magnus\footnote{W.~Magnus, ``On the exponential solution of differential equations for a linear operator,'' \emph{Commun. Pure Appl. Math.} \textbf{7}, 649–673 (1954).}.  
Instead of ordered products, one seeks $U(t)$ in exponential form
\begin{equation}
   U(t)=\exp(\Omega(t)),
\end{equation}
with $\Omega(t)$ expressed as an infinite series
\[
   \Omega(t)=\sum_{k=1}^\infty \Omega_k(t).
\]
The first terms are
\begin{eqnarray}
   \Omega_1(t)&=&\int_0^{\top} A(t_1)\,dt_1, \nonumber \\
   \Omega_2(t)&=&\frac{1}{2}\int_0^{\top} dt_1\int_0^{t_1}dt_2\,[A(t_1),A(t_2)], \nonumber \\
   \Omega_3(t)&=&\frac{1}{6}\int_0^{\top} dt_1\int_0^{t_1}dt_2\int_0^{t_2}dt_3
   \big([A(t_1),[A(t_2),A(t_3)]]\nonumber \\
   &&\hspace{4cm}+[A(t_3),[A(t_2),A(t_1)]]\big), \nonumber
\end{eqnarray}
and so on.  

Unlike the Dyson expansion, which produces series of nested integrals, the Magnus expansion preserves the exponential form at every truncation.  
This ensures structural properties such as symplecticity and conservation laws are respected exactly, which is crucial in Hamiltonian systems.\footnote{S.~Blanes, F.~Casas, J.A.~Oteo, and J.~Ros, 
``The Magnus expansion and some of its applications,'' \emph{Phys. Rep.} \textbf{470}, 151–238 (2009).}  

\paragraph{Convergence.}
The series converges whenever
\[
   \int_0^{\top} \|A(s)\|\,ds < \pi,
\]
providing a practical criterion.  
In practice, truncated Magnus expansions are widely used in numerical analysis of differential equations and in physics (Floquet theory, NMR, quantum control).  

\paragraph{Comparison.}
Dyson’s expansion highlights the role of chronological ordering and makes the connection with resolvent expansions of Sec.~\ref{sec:lincontr}.  
Magnus’ approach emphasizes qualitative preservation: even low-order truncations approximate $U(t)$ while preserving symplectic structure and spectral reciprocity.  
Together they illustrate two complementary philosophies of operator exponentials: ordered integrals versus structured exponentials.
\subsubsection{Fer expansion}
\index{key}{Fer expansion}
The Fer method is another strategy to approximate the solution of
\begin{equation}
   U'(t)=A(t)U(t), \qquad U(0)=I,
   \label{eq:timord1_fer}
\end{equation}
by factorizing the evolution operator as a \emph{product of exponentials}.  
One assumes
\[
   U(t)=\prod_{i=1}^n e^{F_i(t)}\, Y_n(t),
\]
with each $F_i(t)$ determined recursively.  
Starting from $A_0(t)=A(t)$, one defines
\begin{eqnarray}
   F_{n+1}(t)&=&\int_0^{\top} A_n(s)\,ds, \nonumber \\
   A_{n+1}(t)&=&e^{-F_{n+1}(t)} A_n(t) e^{F_{n+1}(t)}
      -\int_0^1 e^{-xF_{n+1}(t)} A_n(t) e^{xF_{n+1}(t)}\,dx.
\end{eqnarray}
This recursion generates a rapidly convergent expansion: if $A(t)=\epsilon \tilde A(t)$ with $\|\tilde A\|=O(1)$, then $A_n(t)=O(\epsilon^{2^n})$.  
Convergence is guaranteed under the condition
\[
   \int_0^{\top} \|A(s)\|\,ds < 0.860\ldots
\]
(see \emph{Moan and Niesen, 2008}).  
The Fer expansion is therefore particularly efficient when the operator norm of $A$ is small, and it is often used in numerical integration of differential equations where preserving the spectral structure of $U(t)$ is critical.

\subsubsection{Baker--Campbell--Hausdorff formula and Hadamard series}
\index{key}{Baker-Campbell-Hausdorff formula}
\index{key}{Hadamard series}

A central difficulty in time-dependent evolution problems is that exponentials of operators do not commute in general: $e^A e^B \neq e^{A+B}$.  
The \emph{Baker--Campbell--Hausdorff} (BCH) formula addresses this by writing
\begin{equation}
   \log(e^A e^B)=A+B+\tfrac{1}{2}[A,B]+\tfrac{1}{12}([A,[A,B]]-[B,[A,B]])-\tfrac{1}{24}[B,[A,[A,B]]]+\cdots
\end{equation}
as an infinite series of nested commutators.  
No closed formula for all coefficients is known, but many useful truncations exist.

A related tool is the \emph{Hadamard series}, defined for
\[
   F(s)=e^{sA}Be^{-sA}=\sum_{n=0}^\infty \frac{s^n}{n!}\underbrace{[A,[A,\cdots[A}_{n \text{ times}},B]\cdots],
\]
which describes the adjoint action of $e^{sA}$ on $B$.
In physics this corresponds to the Heisenberg time evolution of operators.  

From BCH and Hadamard, one can derive compact product formulas such as the so-called \emph{triple product formula}:
\begin{equation}
   e^{tV(t)}=e^{tA}e^{tB}e^{tA},
\end{equation}
with $V(t)=2A+B+t^2[(A+B),[B,A]]+\cdots$, which forms the basis for higher-order splitting methods.

These expansions are useful in spectral analysis because they provide controlled approximations to $\exp(A+B)$, while quantifying how non-commutativity ($[A,B]$) perturbs eigenvalues and stability regions.

\subsubsection{Trotter--Suzuki decomposition}
\index{key}{Trotter-Suzuki decomposition}
The Trotter product formula gives the simplest rigorous way to approximate $e^{A+B}$:
\begin{equation}
   e^{A+B}=\lim_{n\to\infty}\Big(e^{A/n}e^{B/n}\Big)^n.
   \label{eq:suzukitrotter}
\end{equation}
This is a direct consequence of the \emph{Lie--Trotter product formula}.  
A second-order symmetric variant due to Suzuki is
\begin{equation}
   e^{A+B}=\lim_{n\to\infty}\Big(e^{A/(2n)}e^{B/n}e^{A/(2n)}\Big)^n,
   \label{eq:suzukitrotter2}
\end{equation}
which has error $O(1/n^2)$.  

In general, the Suzuki recursive construction yields $(2k)$-th order approximants
\[
   S_{2k+2}(t)=S_{2k}(a_{2k}t)^2 \, S_{2k}((1-4a_{2k})t)\, S_{2k}(a_{2k}t)^2,
\]
with $a_{2k}=1/(4-4^{1/(2k+1)})$, and error $O(t^{2k+1}/n^{2k})$ after $n$ steps.

\paragraph{Spectral significance and applications.}
Trotter--Suzuki formulas preserve important structural properties:  
unitarity for skew-Hermitian operators, symplecticity for Hamiltonian matrices, and stability of the spectral radius.  
These decompositions are ubiquitous in quantum mechanics (time-splitting in many-body Hamiltonians), PDE solvers (operator-splitting methods), and numerical linear algebra (approximation of matrix exponentials).  
Their effectiveness stems from approximating $e^{A+B}$ in terms of simpler exponentials $e^A$ and $e^B$ whose spectral properties are explicitly known.

\medskip

\noindent
In summary, the Fer, BCH/Hadamard, and Trotter--Suzuki methods all address the same spectral difficulty: the non-commutativity of operators.  
They provide systematic expansions or product formulas to approximate evolution operators while preserving key spectral invariants, and are therefore central in both theoretical spectral analysis and computational practice.\footnote{See: W.~Magnus (1954); H.~F.~Trotter (1959); M.~Suzuki (1976); Blanes and Casas (2009).}
\section{Linear systems out of equilibrium}

So far we have considered time-independent systems, where stability and asymptotic behavior are determined by the spectrum $\Lambda(A)$ and the properties of the exponential $e^{At}$ (cf.~Sec.~\ref{sec:stability}).  
We now turn to systems out of equilibrium, where the generator of the dynamics $A(t)$ depends explicitly on time.  
In this regime, eigenvalues, eigenvectors, and the resolvent evolve with $t$, and perturbative tools are required to track how the spectral structure deforms.

\subsection{Adiabatic switching}
\index{key}{Adiabatic switching}

A standard way of introducing time dependence is to split the generator into a reference part $A_0$ (exactly solvable) and a perturbation $A_1(t)$, i.e.
\[
   A(t)=A_0+A_1(t).
\]
If $A_1(t)$ is switched on smoothly from $0$ to a fixed perturbation $A_1$, one speaks of \emph{adiabatic switching}.  
A common model is
\[
   A_1(t)=e^{-\epsilon |t|} A_1, \qquad \epsilon>0,
\]
so that in the distant past the system is governed purely by $A_0$, while for $t\to 0$ the perturbation is fully present.

The adiabatic idea underlies both mathematical and physical results:
\begin{itemize}
   \item In quantum mechanics, it appears in the \emph{adiabatic theorem} and in the
   \emph{Gell--Mann--Low formula}, which expresses perturbed eigenstates of $A_0+A_1$ in terms of the adiabatic limit of the interacting evolution operator\footnote{See M.~Gell-Mann and F.~Low, \emph{Bound states in quantum field theory}, Phys.~Rev.~84 (1951), and Kato’s treatment in \emph{Perturbation Theory for Linear Operators}, Springer (1966).}.
   \item In spectral theory, adiabatic switching provides a controlled way of relating $\Lambda(A_0)$ and $\Lambda(A_0+A_1)$ by tracking how resolvents deform with time (cf.~Sec.~\ref{sec:perron} and Sec.~\ref{sec:matrixsym}).
\end{itemize}

\subsection{Dynamical eigenvalue perturbation theory}
\index{key}{Eigenvalues Perturbation Theory (dynamical)}

Classical perturbation theory studies how eigenvalues of $A+\epsilon B$ expand in powers of $\epsilon$.  
In the dynamical setting, one asks: how do eigenvalues and eigenvectors evolve when $A(t)$ varies smoothly in time?

Kato’s \emph{analytic perturbation theory}\footnote{T.~Kato, \emph{Perturbation Theory for Linear Operators}, Springer (1966).} shows that if $A(t)$ depends analytically on $t$, then eigenvalues can be expanded as analytic functions of $t$ (possibly branching when eigenvalues cross).  
Formally, if $A(t)=A_0+t B+o(t)$, then
\[
   \lambda_i(t)=\lambda_i(0)+t\, v_i^\dagger B v_i + O(t^2),
\]
where $v_i$ is the eigenvector of $A_0$ associated with $\lambda_i(0)$.  
Higher-order terms involve resolvents, cf.~the Rayleigh--Schrödinger series discussed in Sec.~\ref{sec:matrixnorm}.  

In practice, this theory provides estimates of how spectral abscissae, pseudospectra, and stability regions evolve under slowly varying perturbations, which is central in control theory and dynamical stability analysis.

\subsection{Gronwall inequality and growth bounds}
\index{key}{Gronwall inequality}

A key tool for bounding the growth of solutions of time-dependent systems
\[
   \dot x(t)=A(t)x(t),
\]
is the \emph{Gronwall inequality}.  
In its matrix form, it states that if
\[
   \|x(t)\|\leq \alpha + \int_{t_0}^{\top} \beta(s)\|x(s)\|\,ds,
\]
then
\[
   \|x(t)\|\leq \alpha \exp\!\left(\int_{t_0}^{\top} \beta(s)\,ds\right).
\]
This shows that the exponential growth rate of solutions is controlled by the integral of $\|\!A(s)\|\!$.  
Spectrally, this provides an upper bound on the \emph{Lyapunov exponents} of the system (cf.~Sec.~\ref{sec:lyapunovstab}), and hence on the asymptotic stability.

\subsection{Liouville’s formula}
\index{key}{Liouville's formula}

Another classical spectral tool is Liouville’s formula for the determinant of the fundamental solution $X(t)$ of $\dot X=A(t)X$, namely
\begin{equation}
   \det(X(t))=\det(X(t_0))\,\exp\!\left(\int_{t_0}^{\top} \mathrm{Tr}(A(s))\,ds\right).
\end{equation}
Thus the \emph{phase volume} scales with the exponential of the integrated trace of $A(t)$, which is the sum of instantaneous eigenvalues.  
In particular:
\begin{itemize}
   \item If $\mathrm{Tr}(A(t))<0$ on average, trajectories contract in volume, reflecting dissipative dynamics.
   \item If $\mathrm{Tr}(A(t))=0$, the system preserves volume, as in Hamiltonian systems (Sec.~\ref{sec:hamiltonian}).
\end{itemize}

This spectral perspective links local eigenvalue information (through $\mathrm{Tr}(A(t))$) to global dynamical invariants (volume contraction rates, Lyapunov exponents).  

\medskip

As we have seen, out-of-equilibrium linear systems are governed by the same spectral objects as time-independent systems: eigenvalues, resolvents, traces.   Adiabatic switching, dynamical perturbation theory, Gronwall-type inequalities, and Liouville’s formula provide complementary ways to control the evolution of spectra and stability in the time-dependent case.

\subsection{Floquet exponents and Mathieu's equation}
\index{key}{Floquet theory}
\index{key}{Floquet exponent}
\index{key}{Mathieu equation}

Consider a linear time-dependent system of the form
\begin{equation}
   \frac{d}{dt}\vec x(t)=A(t)\vec x(t),
\end{equation}
where $A(t)$ is a $T$-periodic matrix, i.e.\ $A(t+T)=A(t)$.  
The central result of Floquet theory is that although $\vec x(t)$ need not be periodic, solutions can be written in the form
\begin{equation}
   \vec x(t)=e^{\alpha t}\,\vec p(t),
\end{equation}
where $\vec p(t)$ is $T$-periodic and $\alpha\in\mathbb{C}$ is called a \emph{Floquet exponent}.

\paragraph{Monodromy matrix.}
Let $M(t)$ be the fundamental solution defined by
\begin{eqnarray}
   \frac{d}{dt}M(t)&=&A(t)M(t), \\
   M(0)&=&I.
\end{eqnarray}
Then the \emph{monodromy matrix} is $M(T)$, which propagates solutions over one period.  
The eigenvalues of $M(T)$ are called \emph{Floquet multipliers}, and their logarithms (scaled by $1/T$) are the Floquet exponents:
\begin{equation}
   \mu_i \in \Lambda(M(T)), 
   \qquad
   \alpha_i=\frac{1}{T}\log \mu_i.
\end{equation}
Thus Floquet exponents are spectral quantities, directly determined by the eigenvalues of the period map.

\paragraph{Stability.}
The stability of the system depends on the real parts of the exponents:
\begin{itemize}
   \item If $\Re(\alpha_i)<0$ for all $i$, the solution is asymptotically stable.
   \item If some $\Re(\alpha_i)>0$, trajectories grow exponentially.
   \item If all $\Re(\alpha_i)=0$ and the multipliers lie on the unit circle, the system is marginally stable.
\end{itemize}
This criterion is a natural extension of the constant-coefficient case (Sec.~\ref{sec:stability}), where stability was dictated by $\Re(\Lambda(A))<0$.

\paragraph{Mathieu’s equation.}
A canonical example of Floquet theory is the \emph{Mathieu equation}:
\begin{equation}
   \frac{d^2}{dt^2}x(t)+\big(\delta+2\epsilon \cos(2t)\big)x(t)=0,
\end{equation}
which can be rewritten as a first-order system
\begin{equation}
   \frac{d}{dt}\begin{pmatrix}x \\ \dot x\end{pmatrix}
   =\begin{pmatrix}
        0 & 1 \\
        -(\delta+2\epsilon \cos 2t) & 0
     \end{pmatrix}\begin{pmatrix}x \\ \dot x\end{pmatrix}.
\end{equation}
Here $A(t)$ is $T=\pi$-periodic.  
The associated monodromy matrix $M(\pi)$ has eigenvalues (multipliers) that determine stability.  
Regions in the $(\delta,\epsilon)$ plane where $|\mu_i|>1$ correspond to instability bands, while $|\mu_i|=1$ indicates stability.  
This leads to the famous \emph{Ince–Strutt diagram} of stability tongues\footnote{See N.~W.~McLachlan, \emph{Theory and Application of Mathieu Functions}, Dover (1964).}.

\paragraph{Spectral interpretation.}
From a spectral viewpoint:
\begin{itemize}
   \item Floquet multipliers $\mu_i$ play the role of eigenvalues of the period-$T$ evolution operator $M(T)$.
   \item Floquet exponents $\alpha_i$ are logarithmic lifts of $\mu_i$, analogous to Lyapunov exponents (Sec.~\ref{sec:lyapunov}).
   \item The transition from stability to instability corresponds to eigenvalues of $M(T)$ crossing the unit circle in the complex plane.
\end{itemize}

\medskip

Floquet theory thus extends spectral analysis of autonomous systems to periodic non-autonomous systems.  
In particular, Mathieu’s equation illustrates how time-periodic coefficients can destabilize a system even when the average dynamics is stable --- a phenomenon central in mechanics, physics, and control theory.
\subsection{Transients and pseudospectra}
\index{key}{pseudospectrum}
\index{key}{transient growth}

We recall the definition of the $\epsilon$-\emph{pseudospectrum} of a matrix $A$:
\begin{equation}
   \sigma_\epsilon(A)=\{ z\in\mathbb{C}:\exists\,\vec v\in\mathbb{C}^n,\ \|\vec v\|=1,\ \|(zI-A)\vec v\|<\epsilon \}.
\end{equation}
By construction, $\sigma(A)\subseteq \sigma_\epsilon(A)$, and the sets are nested:
$\sigma_{\epsilon_1}(A)\subseteq\sigma_{\epsilon_2}(A)$ if $\epsilon_1\leq\epsilon_2$.  
Equivalently, one can write
\begin{equation}
   \sigma_\epsilon(A)=\{z\in\mathbb{C}:\|(zI-A)^{-1}\|>1/\epsilon\},
\end{equation}
showing the direct connection with the \emph{resolvent} (see Sec.~\ref{sec:resolvent}).

\paragraph{Unitary invariance.}
Since the $L^2$-norm is unitarily invariant,
\[
   \|(zI-U^*AU)^{-1}\|=\|(zI-A)^{-1}\|,
\]
we have $\sigma_\epsilon(A)=\sigma_\epsilon(U^*AU)$.  
This implies that pseudospectra are spectral invariants, though they depend strongly on whether $A$ is normal or non-normal.

\paragraph{Normal vs.~non-normal matrices.}
If $A$ is normal (i.e.\ $AA^\dagger=A^\dagger A$), then $A$ can be unitarily diagonalized and pseudospectra are simply fattened spectra:
\begin{equation}
   \sigma_\epsilon(A)=\sigma(A)+B_\epsilon(0),
\end{equation}
where $B_\epsilon(0)$ is the disk of radius $\epsilon$.  
For non-normal matrices, however, $\sigma_\epsilon(A)$ may be drastically larger: pseudospectral contours can extend far from the spectrum, reflecting the possibility of large \emph{transient growth} even if all eigenvalues are stable\footnote{L.~Trefethen and M.~Embree, \emph{Spectra and Pseudospectra}, Princeton Univ.~Press (2005).}.

\paragraph{Perturbation bounds.}
Let $A=VDV^{-1}$ be diagonalizable with eigenvectors in $V$.  
Defining the condition number $k(V)=\|V\|\|V^{-1}\|$, Bauer--Fike theory (Sec.~\ref{sec:pseudospectra}) gives
\begin{equation}
   \sigma(A)+B_\epsilon(0)\ \subseteq\ \sigma_\epsilon(A)\ \subseteq\ \sigma(A)+B_{k(V)\epsilon}(0).
\end{equation}
Thus, if $V$ is ill-conditioned (large $k(V)$), small perturbations may lead to large spectral shifts.

\paragraph{Transient growth.}
Consider the evolution operator $e^{tA}$, which appears in ODE systems $\dot{\vec x}=A\vec x$.  
Let $\alpha(A)=\max_i \Re(\lambda_i)$ be the \emph{spectral abscissa}.  
Then one can show
\begin{equation}
   e^{t\alpha(A)} \leq \|e^{tA}\| \leq k(V)\,e^{t\alpha(A)}.
\end{equation}
For normal $A$, $k(V)=1$, so growth is governed purely by the eigenvalues.  
For non-normal $A$, transient amplification can occur even when $\alpha(A)<0$, an effect central to hydrodynamic stability and control theory.

\paragraph{Illustrative example.}
A striking demonstration is given by the Jordan-type perturbation
\[
   C=\begin{pmatrix}
     1 & -1 & 0 & \cdots & 0 \\
     0 & 1 & -1 & \cdots & 0 \\
     \vdots & & \ddots & \ddots & \vdots \\
     0 & \cdots & 0 & 1 & -1 \\
     \epsilon & 0 & \cdots & 0 & 1
   \end{pmatrix},
\]
which for $\epsilon=0$ has all eigenvalues equal to $1$.  
Introducing a perturbation $\epsilon\ll 1$ splits the spectrum into $N$ distinct roots $\lambda=1-(-\epsilon)^{1/N}$.
For $N=100$ and $\epsilon=10^{-16}$, the eigenvalues shift by $O(1)$, despite the perturbation being tiny.  
This illustrates the extreme pseudospectral sensitivity of non-normal matrices.

\paragraph{Connection to transients.}
The pseudospectrum is thus directly tied to transient behavior:
\begin{equation}
   \rho(A)^n \leq \|A^n\| \leq k(V)\,\rho(A)^n,
\end{equation}
where $\rho(A)$ is the spectral radius.  
Large values of $\|A^n\|$ for moderate $n$ correspond to non-modal transient growth, even if $\rho(A)<1$.  
This is the foundation of the Kreiss matrix theorem, which bounds transient growth via pseudospectral abscissae.

\paragraph{Residue calculus.}
Finally, the connection to functional calculus is worth emphasizing.  
Given a holomorphic function $f$, one can write
\begin{equation}
   f(A)=\frac{1}{2\pi i}\int_\Gamma (zI-A)^{-1}f(z)\,dz,
\end{equation}
known as the \emph{Dunford--Taylor integral}.  
Here $\Gamma$ encloses the spectrum of $A$.  
This highlights again that the resolvent and pseudospectrum are the central objects in analyzing both spectral and transient properties of linear operators.

\begin{figure}
   \centering
   \includegraphics[scale=0.3]{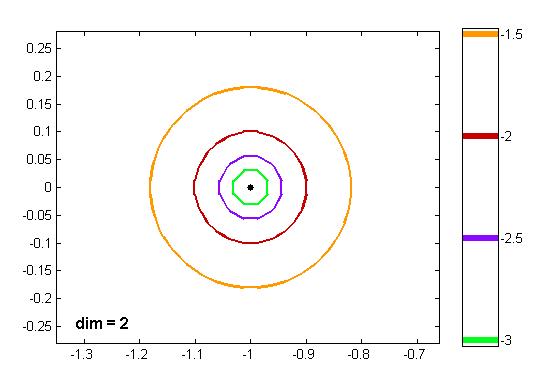}
   \includegraphics[scale=0.3]{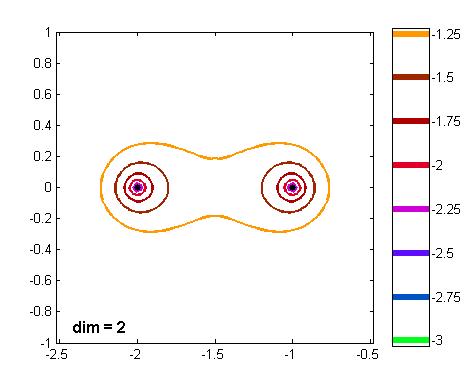}
   \caption{Examples of pseudospectra for two $2\times 2$ matrices. 
   On the left, $M_1=\begin{pmatrix}-1&1\\0&-1\end{pmatrix}$; 
   on the right, $M_2=\begin{pmatrix}-1&5\\0&-2\end{pmatrix}$. 
   Contours indicate $\epsilon$-levels (log scale). 
   The non-normal matrix $M_2$ exhibits far larger pseudospectral regions, signaling potential transient growth.}
\end{figure}
\subsection{More general linear dynamics}
\index{key}{joint spectral radius}
So far we have considered powers of a single matrix $A$, where the asymptotic growth is controlled by its spectral radius $\rho(A)$.  
In many applications --- switched systems, iterative algorithms, wavelet theory, control --- one is interested in products of \emph{different} matrices:
\begin{equation}
   \vec x_n=A_n A_{n-1}\cdots A_1 \vec x_0.
\end{equation}
In this case, the relevant asymptotic growth rate is the \emph{joint spectral radius} (JSR):
\begin{equation}
   \rho(\mathcal M)=\lim_{t\to\infty}\ \sup_{A_i\in\mathcal M}\ \|A_t\cdots A_1\|^{1/t},
\end{equation}
where $\mathcal M$ is the set of admissible matrices\footnote{G.-C. Rota and W. Strang, \emph{A note on the joint spectral radius}, Indag.~Math. 22 (1960).}.  
This generalizes Gelfand's formula for a single operator and provides a stability criterion for switched linear systems: stability is guaranteed iff $\rho(\mathcal M)<1$.

\paragraph{Pseudospectral extension.}
Since the JSR is often hard to compute exactly, one uses norm bounds.  
If $\|A\|\leq \rho_\epsilon(A)/\epsilon$, where $\rho_\epsilon(A)$ is the $\epsilon$-pseudospectral abscissa (see Sec.~\ref{sec:pseudospectra}), then
\begin{equation}
   \|\vec x_n\|\leq \prod_{i=1}^n \|A_i\|\ \|\vec x_0\|
   \ \leq\ \prod_{i=1}^n \frac{\rho_\epsilon(A_i)}{\epsilon}\ \|\vec x_0\|.
\end{equation}
Thus pseudospectra provide a robust way of bounding the growth of products of matrices, and are central in the analysis of robustness and control.

\paragraph{Continuous-time limit.}
The same reasoning carries over to continuous evolution operators.  
For a smooth $A(t)$ one has the time-ordered exponential (see Sec.~\ref{sec:dysonseries}):
\begin{equation}
   e^{\int_0^{\top} A(\xi)\,d\xi}=\lim_{n\to\infty}\prod_{i=1}^n\left(I+A\!\left(\tfrac{it}{n}\right)\tfrac{t}{n}\right).
\end{equation}
By submultiplicativity of the norm, this yields
\begin{equation}
   \big\|e^{\int_0^{\top} A(\xi)\,d\xi}\big\|\leq 
   \exp\!\left(\int_0^{\top} \frac{\rho_\epsilon(A(\xi))}{\epsilon}\,d\xi\right).
\end{equation}
This inequality shows how pseudospectral bounds extend from static to time-dependent systems.

\subsection{Nonlinear systems and linearization}
The flow of a (possibly nonlinear) dynamical system is the map
\begin{equation}
   f^{\top}(\vec x_0)=\vec x(t),
\end{equation}
satisfying the semigroup property $f^{t+s}=f^{\top}\circ f^s$.  
For nonlinear dynamics
\begin{equation}
   \dot x_i(t)=v_i(\vec x,t),
\end{equation}
linearization near a trajectory leads to the Jacobian:
\begin{equation}
   J^{\top}_{ij}(\vec x_0)=\frac{\partial x_i(t)}{\partial x_{0j}},
\end{equation}
so that infinitesimal perturbations evolve as
\[
   \delta \vec x(t)=J^{\top}(\vec x_0)\,\delta \vec x(0).
\]
For small $\delta t$, $J^{\delta t}(\vec x)\approx I+A(\vec x)\,\delta t$, where $A(\vec x)=\nabla v(\vec x)$.  
In the limit, one recovers the time-ordered exponential (cf.~Sec.~\ref{sec:dysonseries}):
\begin{equation}
   J^{\top}(\vec x_0)=\mathcal T\exp\!\left(\int_0^{\top} A(x(\tau))\,d\tau\right).
\end{equation}
Thus, even nonlinear flows can be studied via spectral properties of local Jacobians, extending the linear theory.
\subsection{More general linear dynamics}
\index{key}{joint spectral radius}
So far we have considered powers of a single matrix $A$, where the asymptotic growth is controlled by its spectral radius $\rho(A)$.  
In many applications --- switched systems, iterative algorithms, wavelet theory, control --- one is interested in products of \emph{different} matrices:
\begin{equation}
   \vec x_n=A_n A_{n-1}\cdots A_1 \vec x_0.
\end{equation}
In this case, the relevant asymptotic growth rate is the \emph{joint spectral radius} (JSR):
\begin{equation}
   \rho(\mathcal M)=\lim_{t\to\infty}\ \sup_{A_i\in\mathcal M}\ \|A_t\cdots A_1\|^{1/t},
\end{equation}
where $\mathcal M$ is the set of admissible matrices\footnote{G.-C. Rota and W. Strang, \emph{A note on the joint spectral radius}, Indag.~Math. 22 (1960).}.  
This generalizes Gelfand's formula for a single operator and provides a stability criterion for switched linear systems: stability is guaranteed iff $\rho(\mathcal M)<1$.

\paragraph{Pseudospectral extension.}
Since the JSR is often hard to compute exactly, one uses norm bounds.  
If $\|A\|\leq \rho_\epsilon(A)/\epsilon$, where $\rho_\epsilon(A)$ is the $\epsilon$-pseudospectral abscissa (see Sec.~\ref{sec:pseudospectra}), then
\begin{equation}
   \|\vec x_n\|\leq \prod_{i=1}^n \|A_i\|\ \|\vec x_0\|
   \ \leq\ \prod_{i=1}^n \frac{\rho_\epsilon(A_i)}{\epsilon}\ \|\vec x_0\|.
\end{equation}
Thus pseudospectra provide a robust way of bounding the growth of products of matrices, and are central in the analysis of robustness and control.

\paragraph{Continuous-time limit.}
The same reasoning carries over to continuous evolution operators.  
For a smooth $A(t)$ one has the time-ordered exponential (see Sec.~\ref{sec:dysonseries}):
\begin{equation}
   e^{\int_0^{\top} A(\xi)\,d\xi}=\lim_{n\to\infty}\prod_{i=1}^n\left(I+A\!\left(\tfrac{it}{n}\right)\tfrac{t}{n}\right).
\end{equation}
By submultiplicativity of the norm, this yields
\begin{equation}
   \big\|e^{\int_0^{\top} A(\xi)\,d\xi}\big\|\leq 
   \exp\!\left(\int_0^{\top} \frac{\rho_\epsilon(A(\xi))}{\epsilon}\,d\xi\right).
\end{equation}
This inequality shows how pseudospectral bounds extend from static to time-dependent systems.

\subsection{Nonlinear systems and linearization}
The flow of a (possibly nonlinear) dynamical system is the map
\begin{equation}
   f^{\top}(\vec x_0)=\vec x(t),
\end{equation}
satisfying the semigroup property $f^{t+s}=f^{\top}\circ f^s$.  
For nonlinear dynamics
\begin{equation}
   \dot x_i(t)=v_i(\vec x,t),
\end{equation}
linearization near a trajectory leads to the Jacobian:
\begin{equation}
   J^{\top}_{ij}(\vec x_0)=\frac{\partial x_i(t)}{\partial x_{0j}},
\end{equation}
so that infinitesimal perturbations evolve as
\[
   \delta \vec x(t)=J^{\top}(\vec x_0)\,\delta \vec x(0).
\]
For small $\delta t$, $J^{\delta t}(\vec x)\approx I+A(\vec x)\,\delta t$, where $A(\vec x)=\nabla v(\vec x)$.  
In the limit, one recovers the time-ordered exponential (cf.~Sec.~\ref{sec:dysonseries}):
\begin{equation}
   J^{\top}(\vec x_0)=\mathcal T\exp\!\left(\int_0^{\top} A(x(\tau))\,d\tau\right).
\end{equation}
Thus, even nonlinear flows can be studied via spectral properties of local Jacobians, extending the linear theory.
\subsection{Local stability}
A central tool for understanding dynamical systems near equilibria is the linearization
\begin{equation}
   \frac{d}{dt}\vec x = A\vec x,
\end{equation}
where $A$ is the Jacobian matrix of the vector field, $A_{ij}=\partial_{x_j} f_i(\vec x^*)$ at the fixed point $\vec x^*$.  

The spectral properties of $A$ govern the local behavior:
\begin{itemize}
   \item If all eigenvalues satisfy $\Re(\lambda)<0$, the equilibrium is \emph{asymptotically stable}.
   \item If any eigenvalue has $\Re(\lambda)>0$, the equilibrium is unstable.
   \item If eigenvalues have mixed signs, the fixed point is a \emph{saddle}.
   \item Purely imaginary eigenvalues correspond to neutral stability and possible cycles.
\end{itemize}

More generally, along trajectories the linearized dynamics is captured by the Jacobian flow
\[
   J^{\top}(\vec x_0)=\mathcal T\exp\!\left(\int_0^{\top} A(\vec x(\tau))\,d\tau\right),
\]
whose eigenvalues (the multipliers) classify directions as expanding, contracting, or marginal.  

\paragraph{Beyond hyperbolicity.} 
If all eigenvalues of $A$ have nonzero real part (a \emph{hyperbolic} equilibrium), the local dynamics is topologically equivalent to that of the linear system.  
If some eigenvalues lie on the imaginary axis, nonlinear terms become essential. In this case, results such as the \emph{Center Manifold Theorem} provide a reduction of dynamics to the subspace associated with these eigenvalues\footnote{See L.~Perko, \emph{Differential Equations and Dynamical Systems}, Springer (2001).}. We do not pursue these constructions here, since our focus remains on spectral analysis.  
\subsubsection{Hartman--Grobman theorem}
The Hartman--Grobman theorem formalizes the spectral criterion for local stability.  
If $x^*$ is a \emph{hyperbolic equilibrium} of a nonlinear dynamical system  
(all eigenvalues of the Jacobian $A=\partial f/\partial x|_{x^*}$ satisfy $\Re(\lambda)\neq 0$),  
then the nonlinear flow is topologically conjugate to its linearization
\[
   \dot x = A(x-x^*).
\]
In other words, near $x^*$ the trajectories of the nonlinear system are qualitatively equivalent to those of the linear system $e^{At}$.  
This emphasizes that spectral data (eigenvalues of $A$) completely determine the local phase portrait around hyperbolic fixed points.\footnote{See P. Hartman, \emph{Ordinary Differential Equations}, Wiley (1964).}

\subsubsection{Non-resonant systems and Poincaré linearization}
For equilibria with eigenvalues satisfying certain \emph{non-resonance conditions}, the nonlinear system can even be analytically conjugated to its linear part.  
In particular, if no eigenvalue $\lambda_s$ satisfies a resonance relation of the form
\[
   \lambda_s = \sum_{i=1}^n k_i \lambda_i, \qquad \sum_i k_i \geq 2,
\]
then (under mild assumptions, e.g. Siegel’s condition) there exists a change of variables that reduces the system to $\dot y = Ay$.  
This result, known as the \emph{Poincaré linearization theorem}, again underlines the robustness of spectral properties for classifying local dynamics near equilibria.  

\subsection{Remarks on global stability}
While spectral analysis provides a complete description of local dynamics near hyperbolic fixed points, it does not capture global phenomena such as attractors, limit cycles, or chaotic behavior.  
Global stability often requires Lyapunov functions or geometric/topological tools such as the Poincaré--Bendixson theorem.  
For example, in two dimensions, bounded trajectories that avoid equilibria must approach a periodic orbit.  
Such results go beyond purely spectral methods, but are essential for understanding the global structure of nonlinear flows.  
\subsubsection{Resonant systems}
When the eigenvalues of the Jacobian $A$ satisfy a \emph{resonance relation} of the form
\[
   \lambda_s = \sum_{i=1}^n k_i \lambda_i, 
   \qquad \sum_{i=1}^n k_i \geq 2,
\]
the system is called \emph{resonant}.  
In this case the Poincaré linearization theorem no longer applies: nonlinear terms corresponding to the resonant monomials cannot be removed by analytic changes of coordinates.  

Instead, the dynamics can be simplified using \emph{normal form theory}, in which the system is transformed into a polynomial system that retains the resonant terms:
\[
   \dot y = Ay + R(y),
\]
where $R(y)$ contains only resonant monomials.  
These resonant contributions govern the long-time behavior of the flow and may lead to bifurcations or the appearance of invariant cycles.  

A classical example is the \emph{Hopf bifurcation}: if $A$ has a pair of purely imaginary eigenvalues $\pm i\omega$, resonance with quadratic or cubic nonlinearities can generate a stable or unstable periodic orbit. This connects resonance to the spectral mechanism underlying cycle stability.\footnote{See J. Guckenheimer and P. Holmes, \emph{Nonlinear Oscillations, Dynamical Systems, and Bifurcations of Vector Fields}, Springer (1983).}  

Thus, in resonant cases the eigenvalue spectrum still constrains the dynamics, but nonlinear terms aligned with resonances dictate the qualitative behavior near equilibrium.
\subsection{Carleman linearization}\label{sec:carleman}
\index{key}{Carleman linearization}

One of the most striking results in the study of nonlinear dynamical systems is the \emph{Carleman linearization}:  
any analytic nonlinear system of ordinary differential equations (ODEs) can be embedded into an infinite-dimensional linear system.  
This idea goes back to T. Carleman (1932), and has been rediscovered in many contexts, from functional analysis to modern control theory.\footnote{T. Carleman, ``Application de la théorie des équations intégrales linéaires aux systèmes d'équations différentielles non linéaires,'' \emph{Acta Mathematica} 59, 63–87 (1932). For a modern introduction see A.~Kowalski and H.~Steeb, \emph{Nonlinear Dynamical Systems and Carleman Linearization}, World Scientific (1991).}

\paragraph{From Taylor expansion to Kronecker structure.}
Let $\vec f(\vec x)$ be an analytic vector field on $\mathbb{R}^n$, and consider the nonlinear system
\begin{equation}
    \frac{d}{dt}\vec x = \vec f(\vec x).
    \label{eq:nonlinODE}
\end{equation}
Expanding $\vec f$ in a multivariate Taylor series around a reference point $\vec h$, we may collect terms by total degree.  
Writing $\vec p=\vec x-\vec h$, the $k$-th order contribution can be expressed as a matrix acting on Kronecker powers of $\vec p$:
\begin{equation}
   \vec f(\vec x) = \sum_{k=0}^\infty A_k \vec p^{[k]},
\end{equation}
where $\vec p^{[k]}$ is the column vector of all monomials of degree $k$ (dimension $n^k$), and $A_k$ is an $n\times n^k$ coefficient matrix.  
This uses the Kronecker power notation introduced in Sec.~\ref{sec:matrixequations}.

\paragraph{Lifted infinite-dimensional dynamics.}
Introduce new variables corresponding to all monomials of $\vec x$:
\[
   y=(x_1,\,x_2,\,\ldots,\,x_n,\,x_1^2,\,x_1x_2,\,\ldots)^{\top}.
\]
In this lifted space, the nonlinear system becomes \emph{linear}:
\begin{equation}
    \frac{d}{dt} y = \mathcal{A} y,
    \label{eq:carleman_linear}
\end{equation}
where $\mathcal{A}$ is an infinite block upper-triangular matrix built recursively from the $A_k$.  
Its structure ensures that monomials of degree $k$ only couple to degrees $\geq k$.

A simple example is the scalar ODE $\dot x = x^2$, which leads to
\[
\frac{d}{dt}
\begin{pmatrix}
x \\ x^2 \\ x^3 \\ \vdots
\end{pmatrix}
=
\begin{pmatrix}
0 & 1 & 0 & 0 & \cdots \\
0 & 0 & 2 & 0 & \cdots \\
0 & 0 & 0 & 3 & \cdots \\
\vdots & \vdots & \vdots & \ddots & \ddots
\end{pmatrix}
\begin{pmatrix}
x \\ x^2 \\ x^3 \\ \vdots
\end{pmatrix}.
\]

\paragraph{General formalism.}
Pouly, Bouvrie, and Maggio recently reformulated Carleman linearization in a systematic operator-theoretic way, clarifying its scope and limitations.\footnote{C.~Pouly, J.~Bouvrie, and G.~Maggio, ``Carleman linearization in terms of matrix and tensor operations,'' arXiv:1711.02552 (2017).}  
In their framework, the nonlinear vector field is represented by a collection of \emph{Carleman tensors}, each encoding the multilinear action of $k$th-order terms on Kronecker powers of the state vector.  
The lifted operator $\mathcal{A}$ is then expressed compactly in block form using Kronecker sums and tensor contractions.  
This makes explicit how vectorization and Kronecker calculus provide the natural language for Carleman embeddings.

\paragraph{Spectral viewpoint.}
Equation \eqref{eq:carleman_linear} is solved formally by
\[
   y(t)=e^{\mathcal{A}t}y(0).
\]
Thus, the nonlinear flow of \eqref{eq:nonlinODE} is governed by the \emph{spectrum of $\mathcal{A}$}.  
Stability, growth rates, and resonances of the nonlinear system can be studied using spectral properties of $\mathcal{A}$, in close analogy with linear systems.  

\paragraph{Truncation and Carleman approximation.}
In applications one truncates $\mathcal{A}$ at degree $m$, obtaining the \emph{Carleman approximation}, a finite matrix of size $N(m)=\sum_{k=1}^m n^k$.  
This yields a sequence of finite-dimensional linear systems approximating the true nonlinear flow on the domain of analyticity of $\vec f$.  
Although the dimension grows rapidly with $m$, the approximation allows the application of linear spectral tools (eigenvalues, pseudospectra, resolvent bounds).

\paragraph{Implications.}
Carleman linearization emphasizes that nonlinear phenomena remain accessible to spectral analysis once lifted appropriately.  
In particular:
\begin{itemize}
    \item Long-time behavior is tied to the spectral abscissa of $\mathcal{A}$ (cf. Sec.~\ref{sec:pseudospectra}).
    \item Truncation may induce significant pseudospectral effects, altering stability predictions.
    \item Nonlinear resonances appear as generalized Jordan-type structures in the blocks of $\mathcal{A}$.
\end{itemize}
Beyond being a theoretical bridge, this operator-theoretic view connects to practical algorithms, such as Carleman-based model order reduction, control design, and even modern machine-learning inspired techniques. The properties of dynamical system in the Carleman linearization can also mapped to a combinatorial framework for certain systems \footnote{See for instance F. Caravelli, Y.T. Lin, ``On the Combinatorics of Lotka-Volterra equations", Physica A,  670, 130484 (2025);
M. Forets, A. Pouly,  Explicit Error Bounds for Carleman Linearization, arXiv:1711.02552 }.

\subsection{Mori–Zwanzig formalism from Carleman linearization}
\index{key}{Mori--Zwanzig formalism}

While Carleman linearization embeds a nonlinear ODE into an infinite-dimensional linear system, 
practical use requires reducing this system back to a manageable size.  
A powerful technique from statistical mechanics is the \emph{Mori–Zwanzig formalism}, originally developed for projection methods in irreversible processes.\footnote{H.~Mori, ``Transport, collective motion, and Brownian motion,'' \emph{Prog. Theor. Phys.} 33(3), 423–455 (1965); R.~Zwanzig, ``Nonlinear generalized Langevin equations,'' \emph{J. Stat. Phys.} 9(3), 215–220 (1973). See also A.~Chorin, O.~Hald, and R.~Kupferman, ``Optimal prediction and the Mori–Zwanzig representation of irreversible processes,'' \emph{PNAS} 97(7), 2968–2973 (2000).}  

\paragraph{Linear setting.}
Consider the linear ODE system
\begin{equation}
\frac{d}{dt}\vec r = A \vec r,
\end{equation}
where $\vec r$ is the (infinite-dimensional) Carleman state vector.  
Partition $\vec r=(\vec y,\vec z)$ into \emph{resolved} observables $\vec y$ (low-order monomials of interest) and \emph{unresolved} observables $\vec z$ (higher-order monomials).  
Correspondingly, the matrix $A$ decomposes into blocks
\[
A=\begin{pmatrix}
A_{rr} & A_{ru} \\
A_{ur} & A_{uu}
\end{pmatrix}.
\]

\paragraph{Elimination of unresolved variables.}
Solving formally for $\vec z(t)$ and substituting back yields an exact reduced equation for $\vec y(t)$:
\begin{equation}
\frac{d}{dt}\vec y(t) = A_{rr}\vec y(t) + \int_0^t K(t-s)\,\vec y(s)\,ds + F(t),
\end{equation}
where
\[
K(t)=A_{ru}e^{A_{uu}t}A_{ur}, \qquad F(t)=A_{ru}e^{A_{uu}t}\vec z(0).
\]
This is a generalized Langevin equation: the effect of unresolved modes appears as a \emph{memory kernel} $K(t)$ and a \emph{noise term} $F(t)$ determined by initial conditions.

\paragraph{Interpretation.}
Thus, Carleman with Mori–Zwanzig produces an exact reduced model for the original nonlinear system, consisting of:
\begin{itemize}
  \item a Markovian part governed by $A_{rr}$,
  \item a non-Markovian memory integral encoding feedback of unresolved modes,
  \item and a fluctuating force depending on unresolved initial conditions.
\end{itemize}
This formulation connects nonlinear dynamics to the generalized Langevin formalism of statistical mechanics, showing that nonlinearities induce memory and noise once projected to finite observables.

\paragraph{Combinatorial structure.}
For quadratic systems like Lotka–Volterra the Mori–Zwanzig reduction admits a \emph{lattice-walk interpretation}\footnote{F.~Caravelli, Y.~T.~Lin, (ibid)}  : terms in the expansion of $e^{A_{uu}t}$ correspond to weighted walks on a lattice of monomials, and the generating function of these walks obeys a PDE whose solution encodes the reduced dynamics.
This provides a surprising bridge between nonlinear ODEs, combinatorics, and statistical mechanics.

\paragraph{Implications.}
The Mori–Zwanzig formalism thus complements Carleman linearization:
\begin{itemize}
  \item Carleman makes nonlinear systems linear but infinite-dimensional.
  \item Mori–Zwanzig reduces the infinite system back to a finite subsystem with memory.
\end{itemize}
Together, they yield a principled framework to derive formal exact solutions, construct approximations, or design data-driven reductions via Koopman and operator-theoretic methods.

\subsection{Practical application: Proper Orthogonal Decomposition (POD)}
\index{key}{Proper Orthogonal Decomposition (POD)}

The \emph{Proper Orthogonal Decomposition} (POD) is a widely used method of dimensionality reduction for dynamical systems.  
Its main goal is to approximate a high-dimensional time-dependent signal by a small number of \emph{orthogonal modes}, each carrying maximal variance.  
In spirit, POD is closely related to Principal Component Analysis (PCA), and connects naturally to the spectral theory of correlation operators and to Koopman-type decompositions discussed earlier.  

\paragraph{Setup.}  
Suppose we have a field $\vec v(x,t)\in\mathbb{R}^Z$ sampled at $K$ time instants $t_1,\ldots,t_K$ over a discretized domain of size $N^D$ (e.g., a $D$-dimensional spatial grid with $N$ points in each direction).  
We form the vectorized snapshots
\[
   \vec v_i = \mathrm{vec}_x\big(\vec v(x,t_i)\big)\in \mathbb{R}^{M}, \qquad M=Z\,N^D,
\]
and arrange them into the snapshot matrix
\[
   V = \big[ \vec v_1 \ \vec v_2 \ \cdots \ \vec v_K \big]\in \mathbb{R}^{M\times K}.
\]

\paragraph{Correlation operator and eigenvalue problem.}  
The POD modes are obtained by solving the spectral problem for the correlation matrix
\begin{equation}
   C = \frac{1}{K} VV^{t}\in \mathbb{R}^{M\times M}.
   \label{eq:PODcorr}
\end{equation}
By construction, $C$ is symmetric and positive semidefinite, so its eigenvalues $\lambda_m$ are real and nonnegative, and the eigenvectors $\phi_m$ form an orthonormal basis:
\begin{equation}
   C \phi_m = \lambda_m \phi_m, \qquad \langle \phi_m,\phi_n\rangle = \delta_{mn}.
\end{equation}
The eigenvalues quantify the variance (or energy) captured by each mode.

\paragraph{Spectral interpretation.}  
Equation \eqref{eq:PODcorr} is precisely a spectral decomposition: POD seeks the orthogonal eigenbasis of the correlation operator.  
Equivalently, since $C=VV^{\top}/K$, the POD modes are the left singular vectors of $V$ (cf.\ Sec.~\ref{sec:svd}).  
Thus POD is directly linked to the singular value decomposition (SVD), with $\lambda_m$ the squared singular values.  

\paragraph{Approximation of dynamics.}  
Each snapshot $\vec v_i$ can be expanded as
\begin{equation}
   \vec v_i \approx \sum_{m=1}^r c^i_m \phi_m,
\end{equation}
where $r\ll M$ is the chosen truncation rank, and the coefficients are obtained by projection,
\[
   c^i_m = \langle \vec v_i,\phi_m\rangle.
\]
This gives a reduced-order model that captures the dominant coherent structures in the data.

\paragraph{Applications.}  
POD is ubiquitous in fluid dynamics, signal processing, and control theory.  
In spectral terms, it provides the optimal rank-$r$ approximation of the snapshot ensemble in the $L^2$ sense, by the Eckart–Young theorem (Sec.~\ref{sec:svd}).  
It also connects naturally to the Koopman operator approach, where POD modes approximate Koopman modes under ergodicity assumptions.

POD is a direct application of spectral decomposition: a nonlinear dynamical system is represented by data snapshots, which are compressed via the eigenstructure of the correlation operator.  
This exemplifies how spectral methods bridge dynamical systems, data analysis, and operator theory.

\section{More on Koopman and Perron–Frobenius}\label{sec:koopman}
\subsection{Two averages, one result: ergodicity}
\index{key}{ergodicity}

One of the central notions in dynamical systems is the equivalence between 
\emph{time averages} along trajectories and \emph{ensemble averages} with respect to an invariant measure.  

\paragraph{Time averages.}  
Given an observable $a(x)$ along a trajectory $x(t)$, the time average over a horizon $T$ is
\begin{equation}
    \langle a \rangle_T = \frac{1}{T}\int_{t_0}^{t_0+T} a(x(t'))\,dt'.
    \label{eq:timeavg}
\end{equation}
If the limit $T\to\infty$ exists, it defines the asymptotic time average.  

In the Koopman formalism, where $(U^{\top} a)(x)=a(f^{\top}(x))$ acts on observables, eq.~\eqref{eq:timeavg} becomes
\begin{equation}
    \langle a \rangle_T = \frac{1}{T}\int_{t_0}^{t_0+T} (U^{t'} a)(x_0)\,dt'.
\end{equation}

\paragraph{Ensemble averages.}  
On the other hand, given a probability density $\rho(x)$ on phase space, the ensemble average is
\begin{equation}
    \langle a \rangle_\rho = \int a(x)\,\rho(x)\,dx,
\end{equation}
where $\rho$ evolves in time under the \emph{Perron–Frobenius operator} $\mathcal{L}^{\top}$ as $\rho_t = \mathcal{L}^{\top} \rho_0$.  
If $\rho^*$ is an invariant density, $\mathcal{L}^{\top} \rho^* = \rho^*$, then
\[
   \langle a \rangle_\rho = \int a(x)\,\rho^*(x)\,dx
\]
represents the long-time statistics of the system.  

\paragraph{Ergodicity.}  
The system is called \emph{ergodic} if for almost every initial condition $x_0$
\begin{equation}
   \lim_{T\to\infty} \frac{1}{T}\int_0^{\top} a(f^{\top}(x_0))\,dt
   \ =\ \int a(x)\,\rho^*(x)\,dx,
\end{equation}
i.e. time averages equal ensemble averages.  
Spectrally, this corresponds to the fact that both $U^{\top}$ and $\mathcal{L}^{\top}$ admit a \emph{unique} invariant mode associated with the eigenvalue $1$ (the constant function for Koopman, the invariant density for Perron–Frobenius).  
All other spectral components must decay, which is equivalent to a spectral gap separating $\lambda=1$ from the rest of the spectrum.\footnote{See M.~Dellnitz and O.~Junge, \emph{On the approximation of complicated dynamical behavior}, SIAM J. Numer. Anal. (1999).}

\paragraph{Physical meaning.}  
In the Koopman picture, ergodicity means that the only fixed observables are the constants:
\[
   U^{\top} g = g \quad \Longrightarrow \quad g(x)\equiv \text{const}.
\]
In the Perron–Frobenius picture, ergodicity means that any initial density converges to the unique invariant density:
\[
   \lim_{t\to\infty} \mathcal{L}^{\top} \rho = \rho^*.
\]

This duality reflects the microscopic origin of macroscopic irreversibility: while trajectories are reversible, the spectrum of $\mathcal{L}$ typically exhibits decay of non-invariant modes, leading to the \emph{arrow of time}.  
Ergodicity thus provides the rigorous foundation for replacing time averages with statistical ensembles in statistical mechanics.
\subsection{Hamiltonian case}
We now turn to the important special case of \emph{Hamiltonian flows}, where phase-space volume is preserved and spectral methods take on a symplectic structure.

\paragraph{Liouville dynamics.}
Recall that for a density $\rho(x,t)$ transported by a vector field $\vec v(x)$,
\[
   \partial_t \rho + \nabla \cdot (\rho \vec v) = 0.
\]
If $\nabla \cdot \vec v = 0$ (incompressible flow), this reduces to
\[
   \partial_t \rho + \vec v \cdot \nabla \rho = 0,
\]
which is precisely Liouville’s equation. This expresses that $\rho$ is constant along trajectories: probability density is transported without compression. Equivalently, the Perron–Frobenius operator $\mathcal L^{\top}$ is \emph{unitary}, and so is the Koopman operator $U^{\top}$.\footnote{See V.~I. Arnold, \emph{Mathematical Methods of Classical Mechanics}, Springer (1989).}

\paragraph{Hamilton’s equations in symplectic form.}
For a Hamiltonian $H(p,q)$, the canonical equations are
\[
   \dot p = -\partial_q H, \qquad \dot q = \partial_p H.
\]
Defining $x=(p,q)\in \mathbb{R}^{2d}$ and the symplectic matrix
\[
   Q=\begin{pmatrix}
        0 & -I_d \\
        I_d & \ \ 0
   \end{pmatrix},
\]
we can write the dynamics compactly as
\[
   \dot x = Q \nabla_x H(x).
\]
The Jacobian of the flow is $A(x)=Q\, \nabla^2 H(x)$, where $\nabla^2 H$ is the Hessian of $H$.

\paragraph{Symplectic invariance.}
Let $J^{\top}(x)$ denote the Jacobian of the flow at time $t$. Then
\[
   \frac{d}{dt} J^{\top}(x) = A(x(t))\, J^{\top}(x), \qquad J^0=I.
\]
A standard calculation shows that
\[
   (J^{\top})^\top Q\, J^{\top} = Q,
\]
which means $J^{\top} \in Sp(2d,\mathbb{R})$, the real symplectic group.  
As consequences:
\begin{itemize}
   \item $\det(J^{\top})=1$, i.e. phase-space volume is preserved (Liouville’s theorem).
   \item The eigenvalues of $J^{\top}$ occur in reciprocal pairs: if $\mu$ is an eigenvalue, so is $1/\mu$.
   \item The Perron–Frobenius and Koopman operators are \emph{unitary}, hence their spectra lie on the unit circle.
\end{itemize}
These spectral features are crucial in Hamiltonian systems: stability is not decided by contraction, but by oscillatory or resonant modes.

\subsection{Spectral characterization of dynamical systems}
\index{key}{Koopman eigenfunctions}
We now illustrate how spectral methods, via Koopman operators, characterize two qualitatively different dynamical regimes.

\paragraph{Limit cycles.}
Consider the simple planar system
\[
   r'(t)=r(1-r^2), \qquad \theta'(t)=1,
\]
whose trajectories converge to the unit circle $r=1$ with uniform angular velocity.  
An observable detecting crossings of the $x$-axis can be modeled by a smoothed delta function $a(\theta) \approx \delta_\epsilon(\theta)$.  
Since $\theta$ is $2\pi$-periodic, the action of the Koopman operator $U^{\top}$ is diagonalized by Fourier modes:
\[
   a(\theta) = \sum_{k\in\mathbb{Z}} g_k e^{i k \theta}, \qquad
   U^{\top} e^{i k \theta} = e^{i k t}\, e^{i k \theta}.
\]
Thus the Koopman spectrum on limit cycles consists of purely imaginary eigenvalues $i k$, corresponding to harmonics of the fundamental frequency.\footnote{M. Budisić, R. Mohr, I. Mezić, \emph{Applied Koopmanism}, Chaos (2012).}

\paragraph{Hyperbolic fixed points (Hartman–Grobman).}
For systems with a hyperbolic equilibrium $x^*$, the Hartman–Grobman theorem states that the nonlinear flow is topologically conjugate to its linearization $\dot y = Ay$.  
Let $\{\lambda_i,\vec v_i\}$ be eigenpairs of $A$. The Koopman eigenfunctions are
\[
   \phi_i(y) = \vec v_i^\top y, \qquad U^{\top} \phi_i = e^{\lambda_i t}\,\phi_i.
\]
General observables can be expanded in monomials of $\phi_i$, producing eigenvalues of the form $e^{(k_1 \lambda_1+\cdots+k_n \lambda_n) t}$.  
Thus the Koopman spectrum encodes the linear stability structure of the equilibrium, but applies in a nonlinear neighborhood via the conjugacy.

In both the oscillatory (limit-cycle) and hyperbolic cases, the Koopman operator diagonalizes the dynamics in a basis of eigenfunctions.  
This reveals why spectral methods are so powerful: invariant sets (cycles, fixed points, tori) correspond to pure point spectra, while chaotic dynamics typically correspond to continuous spectra.

\paragraph{Case of Poincaré linearization.}
Another important situation is given by \emph{non-resonant} systems, where the nonlinear flow can be analytically conjugated to its linearization.  
Recall the Poincaré linearization theorem (see Sec.~\ref{sec:stability}): if the Jacobian at a fixed point has eigenvalues
$\{\lambda_1,\dots,\lambda_n\}$ with $\mathrm{Re}(\lambda_i)\neq 0$, and the spectrum is non-resonant (i.e. it satisfies Siegel’s condition),
then there exists an analytic change of coordinates $y=h(x)$ such that
\[
   \dot y = Ay, \qquad A=\mathrm{diag}(\lambda_1,\dots,\lambda_n).
\]

In these new coordinates the Koopman operator acts diagonally:
\[
   U^{\top} \phi_i(y) = e^{\lambda_i t}\, \phi_i(y),
\]
with Koopman eigenfunctions given by the eigencoordinates $\phi_i(y)=y_i$.  
For a general observable $a(y)$, one can expand in multivariate monomials
\[
   a(y)=\sum_{k_1,\dots,k_n} \alpha_{k_1,\dots,k_n}\, \phi_1^{k_1}\cdots \phi_n^{k_n},
\]
and thus
\[
   U^{\top} a(y)=\sum_{k_1,\dots,k_n} \alpha_{k_1,\dots,k_n}\, 
   e^{(k_1 \lambda_1+\cdots+k_n \lambda_n)t} \phi_1^{k_1}\cdots \phi_n^{k_n}.
\]

Hence, under Poincaré linearization, the Koopman spectrum is completely determined by integer linear combinations of the Jacobian eigenvalues.  
This provides an \emph{analytic spectral characterization} of the nonlinear dynamics near the fixed point, sharper than the purely topological one given by the Hartman–Grobman theorem.  

Physically, this means that when resonances are absent the dynamics is not only qualitatively equivalent to its linearization, but also spectrally equivalent: the Koopman eigenfunctions form a complete analytic basis adapted to the linearized flow.\footnote{See S.~Wiggins, \emph{Introduction to Applied Nonlinear Dynamical Systems and Chaos}, Springer (2003).}

\subsection{Practical application: Dynamic Mode Decomposition}
\index{key}{Dynamic Mode Decomposition (DMD)}

Dynamic Mode Decomposition (DMD) is a modern data-driven algorithm that extracts spectral information directly from time-resolved measurements of a dynamical system. 
It is best understood as a finite-dimensional approximation of the Koopman operator\footnote{See P.~J.~Schmid, \emph{Dynamic mode decomposition of numerical and experimental data}, J.~Fluid Mech.~\textbf{656} (2010), 5–28.}, which acts linearly on observables even for nonlinear systems (see Sec.~\ref{sec:koopman}). 
From a numerical linear algebra viewpoint, DMD is closely related to Krylov subspace methods (Sec.~\ref{sec:krylov}), Carleman linearization (Sec.~\ref{sec:carleman}), and Ritz eigenvalue approximations.

\paragraph{Snapshot matrices.} 
Suppose we collect a sequence of ``snapshots'' of the system at equally spaced sampling times, e.g.\ $\vec v_1,\dots,\vec v_n \in \mathbb{R}^m$, where each $\vec v_i$ represents the discretized state of the system at time $t_i$. 
We form the data matrices
\begin{equation}
    V_1^n=\{ \vec v_1,\dots,\vec v_n\}, \qquad 
    V_L^{n-1}=\{\vec v_1,\dots,\vec v_{n-1}\}, \qquad
    V_R^{n-1}=\{\vec v_2,\dots,\vec v_n\}.
\end{equation}
The central assumption of DMD is that there exists a linear operator $A$ such that
\begin{equation}
    \vec v_{i+1}\approx A \vec v_i,
\end{equation}
so that
\begin{equation}
    A V_L^{n-1}\approx V_R^{n-1}.
\end{equation}
In this way, the snapshot sequence $\{ \vec v_1,A\vec v_1,\dots,A^{n-1}\vec v_1\}$ forms a Krylov sequence (Sec.~\ref{sec:krylov}).

\paragraph{Residual and companion matrix.}
Because the last snapshot cannot, in general, be represented exactly as a Krylov iterate, we write
\begin{equation}
    \vec v_n=V_L^{n-1}\vec a+\vec r,
\end{equation}
where $\vec r$ is a residual. This leads to the companion-matrix relation
\begin{equation}
    V_R^{n-1}=V_L^{n-1} S+\vec r \vec e_{n-1}^{\top},
\end{equation}
with $S$ a Frobenius companion matrix (cf.~Sec.~\ref{sec:companion}).

\paragraph{Projected operator via SVD.}
To approximate $S$, we compute the singular value decomposition
\begin{equation}
    V_L^{n-1}=U \Sigma V^*,
\end{equation}
and define the reduced operator
\begin{equation}
    \tilde S=U^* V_R^{n-1} V \Sigma^{-1}.
\end{equation}
This $\tilde S$ is a low-dimensional representation of $A$ restricted to the Krylov subspace spanned by the snapshots.

\paragraph{Dynamic modes and Ritz eigenvalues.}
Finally, the spectral decomposition of $\tilde S$ provides the \emph{dynamic modes}. If $\vec y_i$ is an eigenvector of $\tilde S$ with eigenvalue $\mu_i$, then
\begin{equation}
    \vec \Phi_i = U \vec y_i
\end{equation}
is the corresponding DMD mode. The eigenvalues $\mu_i$ are Ritz eigenvalues, and define approximate growth/decay rates and frequencies:
\begin{equation}
    \lambda_i = \frac{\log(\mu_i)}{\Delta t},
\end{equation}
where $\Delta t$ is the snapshot interval. If $\mathrm{Re}(\lambda_i)<0$, the mode decays; if $\mathrm{Re}(\lambda_i)>0$, it grows.

\paragraph{Interpretation.}
Thus, DMD provides a purely spectral decomposition of empirical data into modes that evolve exponentially in time, closely mirroring Koopman eigenfunctions. 
It has been successfully applied in fluid dynamics (e.g.~airfoil stall control\footnote{A.~Mohan, \emph{Analysis of airfoil stall control using dynamic mode decomposition}, private communication (2019).}), neuroscience, finance, and many other complex systems (see also Kutz et al.\footnote{J.~N.~Kutz et al., \emph{Dynamic Mode Decomposition: Data-Driven Modeling of Complex Systems}, SIAM (2016).}).

In summary, DMD is a practical algorithmic realization of the spectral ideas developed in this book: the Koopman operator, Krylov decompositions, and companion matrices appear naturally in its formulation.

\begin{tcolorbox}
\begin{verbatim}
function [modes,mode_ampl,floqvalsOrder,
   ... freqs,dOrder] = ...
dmd_function(inputMatrix,dt,filename)
%DMD analysis on input data matrix
% Arvind T. Mohan
% PhD-Aerospace Engineering
% High Fidelity Computational Multiphysics Laboratory
% The Ohio State University


% Data prep
Nsnaps = size(inputMatrix,2);

%% DMD algorithm
gridSize = size(inputMatrix,1);

disp('Starting DMD computation...');

tic
% Compute Mean
Umean = mean(inputMatrix,2);

% Step 1: Building Matrix V1^n-1

V1 = inputMatrix(:,1:end-1);

% Computation Starts

% Step 3: Singular Value Decomposition of the V1 matrix 
disp('SVD in process...')
[U,E,W] = svd(V1,0);
disp('Operation Successful.');

% Step 2: Building V2 Matrix

V2 = inputMatrix(:,2:end);

% Step 4: Computing S matrix

S = U'*V2*W/E;

clear V2 W E
% Step 5: Eigendecomposition of Matrix S
disp('Computing Eigendecomposition of S Matrix...')
[eigenvecs,eigenvals] = eig(S);
disp('Operation Successful.');
clear S

ritz = diag(eigenvals); % saving the Ritz values
clear eigenvals


%% Post Processing of DMD data

% Computing Modes

disp('Computing DMD Modes...')
modes = U*eigenvecs;
clear U eigenvecs
[~,nummodes] = size(modes);
disp('Operation Successful.'); 

% Sorting Ritz values in descending order
[~,index] = sort(real(ritz),'descend');
ritz = ritz(index);
modes = modes(:,index);

% Computing Floquet values
%dt = timestep*delta; % Actual time interval

floqvals = log(ritz)/dt;

%% DMD Part 2
% Reconstruction and post-processing

%% Reconstruction

%Computing Scaling Coeffs for modes using 1st snapshot
disp('Computing Scaling Coeffs');
u1 = inputMatrix(:,1);
d = modes\u1;
clear u1

% Computing Scaled modes

modes_scaled = zeros(gridSize,nummodes);

parfor i=1:nummodes
   modes_scaled(:,i) =  modes(:,i)*d(i);
end

% Compute Scaled Mode Norms

mode_norms = zeros(nummodes,1);
parfor i=1:nummodes
   mode_norms(i) = norm(modes_scaled(:,i));
end

mode_ampl_unordered = mode_norms/(max(mode_norms));

% Order Modes and other vars by amplitude

[~,idx] = sort(mode_ampl_unordered,'descend');
mode_ampl = mode_ampl_unordered(idx);
clear mode_ampl_unordered
modes_scaled = modes_scaled(:,idx);
modes = modes(:,idx);
ritzOrder = ritz(idx);
floqvalsOrder = floqvals(idx);
dOrder = d(idx);
freqs = imag(floqvalsOrder)/(2*pi);
% % Build Vandermonde Matrix using Ritz values
% 
% T = fliplr(vander(ritz));
% 
% T = T(idx,:);
% 
% % Computing Reconstructed Velocity
% disp('Computing Reconstructed Data . . .');
% 
% dmdrecon = modes_scaled(:,1:reconmodes) ....
%     *T(1:reconmodes,:);

time1 = toc;
time1 = time1/60;
%cd('../')
disp('saving results')
filename = [filename 'DMD' '.mat'];
save(filename,'modes','floqvalsOrder','freqs', ...
    'ritzOrder','dOrder','mode_ampl','time1','-v7.3');
sprintf('Results saved in %s !!!',filename)
end

    
\end{verbatim}
\end{tcolorbox}
\section{Stable polynomials}
\index{key}{stable polynomial}
\index{key}{Hurwitz polynomial}

The concept of polynomial stability arises naturally when analyzing the spectral properties of linear differential equations.  
Consider the $n$th-order constant-coefficient ODE
\begin{equation}
    a_0 x^{(n)}+a_1 x^{(n-1)}+\cdots+a_n x(t)=z(t).
\end{equation}
As shown earlier (cf.~Sec.~\ref{sec:companionmatode}), this ODE can be recast as a first-order vectorial system
\begin{equation}
    \dot{\vec x}=A \vec x+\vec z,
\end{equation}
with $A$ the Frobenius companion matrix
\begin{equation}
    A=\begin{pmatrix}
    0 & 1 & 0 & \cdots & 0 \\
    0 & 0 & 1 & \cdots & 0 \\
    \vdots & \vdots & \ddots & \ddots & \vdots \\
    0 & 0 & \cdots & 0 & 1 \\
    - \tfrac{a_n}{a_0} & - \tfrac{a_{n-1}}{a_0} & \cdots & \cdots & -\tfrac{a_1}{a_0}
    \end{pmatrix},
    \label{eq:hurwitzmat}
\end{equation}
and forcing $\vec z=(0,\dots,0,z(t))^{\top}$.  
The stability of the system depends solely on the eigenvalues of $A$; equivalently, on the zeros of the characteristic polynomial
\begin{equation}
    f(s)=a_0 s^n+a_1 s^{n-1}+\cdots+a_n.
\end{equation}
A polynomial is called \emph{Hurwitz stable} if all its roots lie strictly in the open left half-plane $\mathrm{Re}(s)<0$.  
Thus, spectral stability of the ODE is equivalent to geometric stability of the associated polynomial.

\subsection{Geometric criteria: Leonhard–Mikhailov}
\index{key}{Leonhard–Mikhailov criteria}
One classical geometric approach is the Leonhard–Mikhailov criterion, which interprets the polynomial $f(s)$ along the imaginary axis $s=i\omega$.  
If $f(i\omega)=R(\omega)e^{i\theta(\omega)}$, then as $\omega$ runs from $0$ to $\infty$, the argument $\theta(\omega)$ winds around the complex plane.  
A Hurwitz polynomial is characterized by the fact that $f(i\omega)$ changes quadrants exactly $n$ times\footnote{See Mikhailov, \emph{Theory of Stability of Motion}, Gostekhizdat (1938).}.  
This provides a spectral–geometric visualization: stability corresponds to the absence of roots crossing into $\mathrm{Re}(s)\geq 0$.  

\begin{figure}[h]
    \centering
    \includegraphics[scale=0.3]{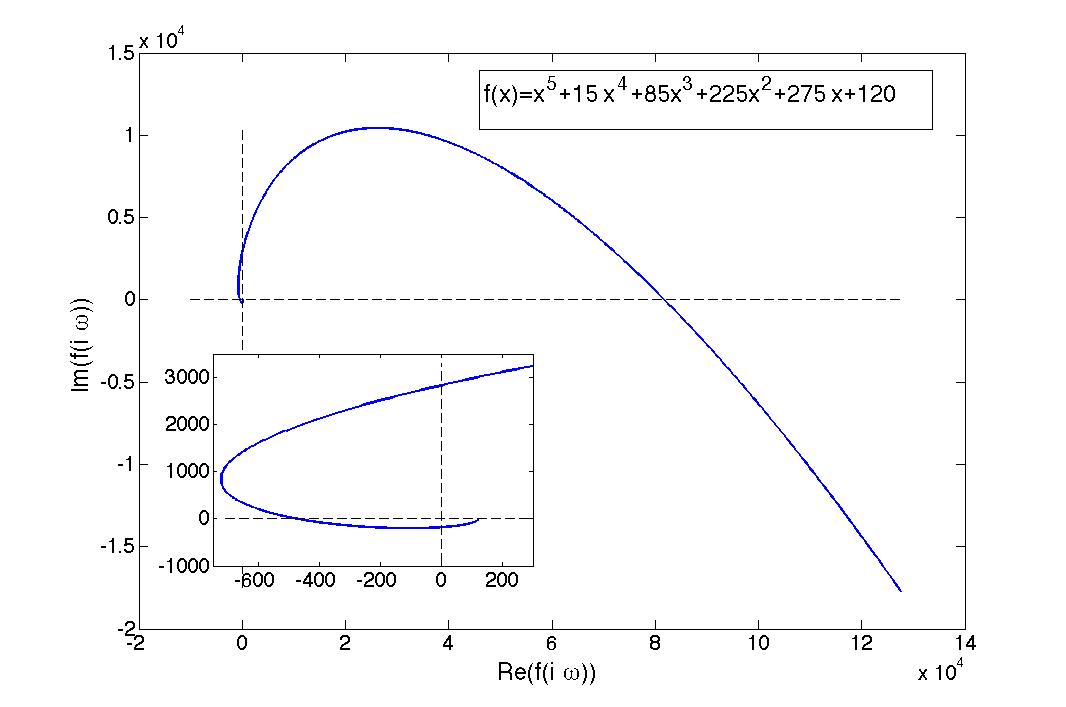}
    \caption{Leonhard–Mikhailov criterion applied to $f(x)=(1+x)(2+x)(3+x)(4+x)(5+x)$, which has five negative real roots.}
    \label{fig:Lcrit}
\end{figure}

\subsection{Routh–Hurwitz and Liénard–Chipart}
\index{key}{Routh-Hurwitz theorem}
Another powerful characterization is the Routh–Hurwitz theorem.  
Given the characteristic polynomial of a matrix $A$,
\begin{equation}
    \det(\lambda I-A)=\lambda^n+b_1\lambda^{n-1}+\cdots+b_{n-1}\lambda+b_n,
\end{equation}
all eigenvalues $\lambda_i$ satisfy $\mathrm{Re}(\lambda_i)<0$ if and only if a sequence of Hurwitz determinants is positive:
\begin{equation}
    \Delta_1>0,\ \Delta_2>0,\ \dots,\ \Delta_n>0,
\end{equation}
where
\begin{equation}
\Delta_k=\det\begin{pmatrix}
b_1 & 1 & 0 & \cdots & 0 \\
b_3 & b_2 & b_1 & \cdots & 0 \\
b_5 & b_4 & b_3 & \cdots & 0 \\
\vdots & \vdots & \vdots & \ddots & \vdots \\
b_{2k-1} & b_{2k-2} & b_{2k-3} & \cdots & b_k
\end{pmatrix}.
\end{equation}
The Liénard–Chipart criterion refines this by requiring only half of the determinants to be checked\footnote{See Gantmacher, \emph{The Theory of Matrices}, Vol.~II, Chelsea (1959).}.  
These determinantal tests are spectral in nature: they indirectly verify the negativity of the real parts of all roots without computing them explicitly.

\subsection{Discrete-time stability: Schur polynomials}
\index{key}{Schur stability}
In discrete-time dynamics, stability is governed by the location of polynomial roots inside the unit disk $|z|<1$.  
A \emph{Schur polynomial} is defined as one whose roots all lie strictly within the unit disk.  
For example, $P(s)=4s^3+3s^2+2s+1$ is Schur stable.  

Hurwitz and Schur stability are connected via a Möbius transformation\footnote{See Jury, \emph{Inners and Stability of Dynamical Systems}, Wiley (1974).}:  
if $M(s)$ is Schur stable of degree $n$, then
\begin{equation}
    P(s)=(s-1)^n M\!\left(\frac{s+1}{s-1}\right)
\end{equation}
is Hurwitz stable.  
This maps the unit disk conformally to the left half-plane, highlighting the spectral equivalence between continuous- and discrete-time stability.  

\medskip
\noindent
In summary, the study of stable polynomials provides multiple complementary lenses on spectral stability:  
geometric (Leonhard–Mikhailov), determinantal (Routh–Hurwitz, Liénard–Chipart), and conformal/discrete (Schur).  
Each connects the roots of the characteristic polynomial—i.e.\ the spectrum of the associated companion matrix—to asymptotic properties of dynamical systems (cf.~Secs.~\ref{sec:lyapunov}, \ref{sec:pseudospectra}).
\subsubsection{Bistritz stability criterion}
\index{key}{Bistritz stability criterion}
The Bistritz stability criterion is a recursive algorithm for testing whether a polynomial is \emph{Schur stable}, i.e.~all its roots lie strictly within the unit disk $|z|<1$.  
Given a real-coefficient polynomial
\[
P(z)=a_0 z^n+a_1 z^{n-1}+\cdots+a_n, \qquad a_0>0,
\]
the method constructs two auxiliary sequences of polynomials recursively, and evaluates their signs at $z=1$.  
If the number of sign changes equals the degree $n$, the polynomial is Schur stable.  

Spectrally, this means that the eigenvalues of the Frobenius companion matrix (cf.~Sec.~\ref{eq:hurwitzmat}) all satisfy $|\lambda_i|<1$.  
The Bistritz criterion is particularly useful in digital signal processing and control theory where discrete-time stability is essential\footnote{See H. Bistritz, ``Zero location of polynomials with applications,'' \emph{Proc. IEEE}, 72(9):1131–1142 (1984).}.

\subsubsection{Jury stability criterion}
\index{key}{Jury stability criterion}
The Jury test is another discrete-time analogue of the Routh–Hurwitz criterion.  
It provides necessary and sufficient conditions for all roots of a polynomial
\[
P(z)=a_0 z^n+a_1 z^{n-1}+\cdots+a_n
\]
to lie inside the unit disk.  

The procedure constructs a so-called \emph{Jury table}, analogous to the Routh array, and checks positivity conditions row by row.  
Concretely, one tests:
\begin{itemize}
    \item $|a_n|<a_0$,
    \item $P(1)>0$, $(-1)^n P(-1)>0$,
    \item all leading minors of the Jury table positive.
\end{itemize}
If these hold, then all roots satisfy $|\lambda_i|<1$.  
This criterion is widely used in digital filter design and discrete control systems\footnote{See E.I. Jury, \emph{Theory and Application of the Z-Transform Method}, Wiley (1964).}.  
From a spectral perspective, the Jury test is equivalent to bounding the spectral radius $\rho(A)<1$ for the companion matrix $A$ of $P(z)$.

\subsubsection{Kharitonov's theorem}
\index{key}{Kharitonov's theorem}
Kharitonov's theorem addresses the stability of \emph{interval polynomials}, i.e.~polynomials with coefficients varying independently within prescribed intervals:
\[
P(s)=a_0 s^n+a_1 s^{n-1}+\cdots+a_n,\qquad a_i \in [\underline{a}_i,\overline{a}_i].
\]
The remarkable result is that the entire family is Hurwitz stable (all roots in $\mathrm{Re}(s)<0$) if and only if four specific \emph{Kharitonov polynomials}, constructed by alternating the bounds of the coefficient intervals, are Hurwitz stable.  

This reduces an infinite verification problem to just four tests.  
Spectrally, this corresponds to guaranteeing that the eigenvalues of all possible companion matrices associated with $P(s)$ lie in the left half-plane, by checking only four representatives\footnote{See V.L. Kharitonov, ``Asymptotic stability of an equilibrium position of a family of systems of linear differential equations,'' \emph{Differentsial'nye Uravneniya} 14(11):2086–2088 (1978).}.  
This theorem is fundamental in robust control and uncertain systems analysis.

\subsubsection{Nyquist stability}
\index{key}{Nyquist stability criterion}
The Nyquist criterion is a frequency-domain graphical test used to determine the stability of a closed-loop system from its open-loop transfer function $G(s)H(s)$.  
The Nyquist plot traces $G(i\omega)H(i\omega)$ as $\omega$ ranges over $\mathbb{R}$, and the winding number of the plot around the point $-1$ determines the number of closed-loop poles in the right half-plane.  

Formally, if $P$ is the number of poles of $G(s)H(s)$ in $\mathrm{Re}(s)>0$, and $N$ is the number of clockwise encirclements of $-1$, then the number of unstable closed-loop poles is
\[
Z=N+P.
\]
Stability requires $Z=0$.  

From the spectral perspective, Nyquist analysis is again a test on the location of the roots of the characteristic equation $1+G(s)H(s)=0$, i.e.~on the eigenvalues of the closed-loop system matrix.  
It can thus be seen as a frequency-domain visualization of spectral stability conditions, closely related to the Leonhard–Mikhailov plot (cf.~Sec.~\ref{fig:Lcrit}) and to Routh–Hurwitz determinants\footnote{See Nyquist, H., ``Regeneration theory,'' \emph{Bell System Technical Journal} 11:126–147 (1932).}.

\medskip
\noindent
These criteria (Bistritz, Jury, Kharitonov, Nyquist) complement the Routh–Hurwitz and Schur tests presented earlier.  
All can be reinterpreted as different ways of bounding or characterizing the spectrum of companion matrices or transfer operators:  
geometric (Nyquist, Leonhard–Mikhailov), determinantal (Routh–Hurwitz, Jury), recursive (Bistritz), and robust (Kharitonov).  
Together they illustrate the central role of spectral analysis in stability theory.


\part{Applications in complex systems: some examples}
\chapter{Applications in Complex Systems}

The aim of this chapter is not to provide an exhaustive catalogue of all 
applications of spectral methods in complex systems. Such a task would be 
impossible: the range of disciplines in which these tools appear---from 
ecology to finance, from neuroscience to computer science---is too broad, 
and each field has developed its own specialized techniques, models, and 
terminology. Instead, what follows is a somewhat curated selection of examples that 
highlight recurring themes: how the spectrum of a matrix encodes stability, 
how projection operators reveal conservation laws, or how resolvents and 
Green’s functions provide a unifying language across domains.

The guiding philosophy remains the same as in earlier chapters: rather than 
proving general theorems in their most abstract form, we focus on concrete 
instances where linear algebra illuminates structure and dynamics. The 
applications collected here should therefore be read as illustrations of a 
perspective, not as a definitive account. Many other important directions 
exist -- such as spectral methods in machine learning, epidemiology, or 
quantum systems -- but fall outside the scope of these notes. 

In this sense, the chapter is best understood as an invitation: a set of 
worked examples that show how spectral tools migrate across disciplines, 
and how the same mathematical objects---resolvents, Laplacians, projectors, 
eigenvalues---reappear in unexpected guises throughout the study of complex 
systems.

\section{Gaussian integration and multivariate analysis}
\index{key}{Gaussian integration}\index{key}{multivariate Gaussian}\index{key}{quadratic forms}\index{key}{partition functions}

One of the most fundamental and widely used identities in mathematics is the
\emph{Gaussian integral}.  
It is the prototypical example where linear algebra, determinants, and quadratic
forms converge in a single formula.  
Gaussian integrals lie at the heart of probability theory and statistics, underpinning
partition functions in statistical mechanics, and provide the computational
backbone of modern machine learning.  

For any symmetric positive definite matrix $A \in \mathbb{R}^{n\times n}$ and vector
$\vec c \in \mathbb{R}^n$, one has the closed-form identity
\begin{equation}
\int_{\mathbb{R}^n} 
\exp\!\Big(-\tfrac{1}{2} \vec x^\top A \vec x + \vec c^\top \vec x\Big)\,d\vec x
= (2\pi)^{n/2}\, (\det A)^{-1/2}
\exp\!\Big(\tfrac{1}{2} \vec c^\top A^{-1} \vec c\Big).
\label{eq:gaussianintegral}
\end{equation}
Two structural features stand out:
\begin{itemize}
\item The \emph{determinant} $\det A$ controls the normalization constant,
or equivalently the volume element of the Gaussian measure.
\item The \emph{inverse} $A^{-1}$ encodes correlations, and determines all
conditional and marginal distributions.
\end{itemize}
This single identity already illustrates why determinants and inverses play a
central role throughout this book.

\subsection{Gaussian measures and partition functions}
A random vector $\vec x \sim \mathcal{N}(\vec m,\Sigma)$ has density
\[
p(\vec x) = \frac{1}{\sqrt{(2\pi)^n \det \Sigma}}
\exp\!\Big(-\tfrac{1}{2}(\vec x-\vec m)^\top \Sigma^{-1}(\vec x-\vec m)\Big),
\]
with mean $\vec m$ and covariance $\Sigma$.  
The normalization constant is precisely a Gaussian integral of the form
\eqref{eq:gaussianintegral}, and can be interpreted as the \emph{partition function}
of the quadratic form.

This observation connects several fields at once:
\begin{itemize}
\item In \textbf{statistical mechanics}, the free energy of a quadratic Hamiltonian
is given by $\log\det$ of its covariance operator.
\item In \textbf{Bayesian inference}, marginal likelihoods for linear–Gaussian
models have the same structure, and Gaussian integrals enable closed-form
posterior updates.
\item In \textbf{machine learning}, multivariate Gaussians serve as the foundation
for Gaussian processes, variational approximations, and latent-variable models.
\end{itemize}

Thus the Gaussian integral is not just a convenient identity, but a paradigmatic
example where spectral quantities — determinants, inverses, eigenvalues —
govern the behavior of complex systems across disciplines.

\subsection{Marginals, conditionals, and inference}
Partition $\vec x=(\vec x_a,\vec x_b)$ with covariance
\[
\Sigma = \begin{bmatrix} \Sigma_a & \Sigma_c \\ \Sigma_c^\top & \Sigma_b\end{bmatrix},
\]
we obtain
\begin{align}
p(\vec x_a) &= \mathcal{N}(\vec x_a \mid \vec\mu_a, \Sigma_a), \\
p(\vec x_a \mid \vec x_b) &= 
\mathcal{N}\!\Big(\vec x_a \mid \vec\mu_a + \Sigma_c \Sigma_b^{-1}(\vec x_b-\vec\mu_b),
\;\Sigma_a - \Sigma_c \Sigma_b^{-1} \Sigma_c^\top\Big).
\end{align}
Thus marginals preserve Gaussianity, and conditionals depend explicitly on
Schur complements and matrix inversion.  
These properties underpin Gaussian graphical models, Kalman filtering,
and Gaussian process regression.

\subsection{Quadratic forms and expectations}
For $\vec x\sim \mathcal{N}(0,\Sigma)$ and symmetric $A$,
\[
\mathbb{E}[\vec x^\top A \vec x] = \mathrm{Tr}(A\Sigma), 
\qquad
\mathrm{Var}(\vec x^\top A \vec x) = 2\,\mathrm{Tr}\!\big[(A\Sigma)^2\big],
\]
as detailed earlier in the section on quadratic forms.  
Such identities are ubiquitous in random matrix theory, portfolio risk estimation,
and in the physics of Gaussian fields.

\subsection{Connections to spectral methods}
Because $\det \Sigma = \prod_i \lambda_i$ and $\Sigma^{-1}$ has eigenvalues
$1/\lambda_i$, Gaussian integrals reduce to spectral quantities. In particular:
\begin{itemize}
\item Normalization constants involve $\prod_i \lambda_i^{-1/2}$.
\item Conditional covariances use Schur complements, which are linked to
eigenvalue interlacing.
\item Free energies in Gaussian models reduce to sums over $\log \lambda_i$.
\end{itemize}
Gaussian integration therefore illustrates in one setting how determinants,
inverses, and eigenvalues appear together in closed form.

\medskip
Gaussian integrals thus serve as a unifying theme of this book: they tie together
matrix inverses, determinants, spectral decompositions, quadratic forms, and
applications to inference and physics.  
They are central both in quantum field theory and in statistical mechanics, and
reappear throughout modern machine learning in Gaussian processes, variational
inference, and models of random fields.
\subsection{Transformations and algebra of Gaussian integrals}
\index{key}{Gaussian integrals!transformations}\index{key}{quadratic forms}\index{key}{spectral decomposition}

Beyond normalization and conditionals, Gaussian integration admits a rich
algebra of transformations that are fundamental across applications.  
These identities show how covariance matrices transform under linear mappings,
how means shift under changes of variables, and how quadratic forms can be
rearranged by completing the square.  
All of them follow directly from determinant and inverse identities.

\paragraph{Linear transformations.}
If $\vec x \sim \mathcal{N}(\vec m_x,\Sigma_x)$ and $\vec y \sim \mathcal{N}(\vec m_y,\Sigma_y)$, then
for fixed matrices $A,B$ and a shift vector $\vec c$,
\begin{equation}
A\vec x + B\vec y + \vec c \;\sim\; 
\mathcal{N}\!\big(A\vec m_x+B\vec m_y+\vec c,\;
A\Sigma_x A^\top + B\Sigma_y B^\top\big).
\end{equation}
Thus covariances transform quadratically, a property exploited in linear regression,
Kalman filtering, and signal processing.

\paragraph{Change of variables and Jacobians.}
Under a linear change of variables $\vec x = A \vec y$, densities acquire a Jacobian factor.  
For $\vec x \sim \mathcal{N}(\vec m,\Sigma)$ one has
\begin{equation}
\mathcal{N}_{A\vec x}[\vec m,\Sigma]
= \frac{1}{|\det A|}\;
\mathcal{N}_{\vec x}\!\big(A^{-1}\vec m,\;(A^\top \Sigma^{-1} A)^{-1}\big),
\end{equation}
assuming $A$ is invertible.  
Here $|\det A|^{-1}$ accounts for the volume distortion of $A$, while the
transformed covariance involves the inverse $(A^\top \Sigma^{-1}A)^{-1}$.

\paragraph{Completing the square.}
Quadratic forms can always be recast in “centered” form. For symmetric $A$,
\begin{equation}
-\tfrac{1}{2} \vec x^\top A \vec x + \vec b^\top \vec x
= -\tfrac{1}{2}(\vec x-A^{-1}\vec b)^\top A(\vec x-A^{-1}\vec b)
+ \tfrac{1}{2}\vec b^\top A^{-1}\vec b.
\end{equation}
This simple identity underpins Gaussian integration: it makes explicit how $A^{-1}$
controls mean shifts, while $\det A$ appears in normalization.

\paragraph{Solving linear systems via Gaussian expectations.}
Completing the square also shows that solving a linear system can be recast as
an expectation under a Gaussian measure.  
Consider the distribution
\[
p(\vec x)\;\propto\; \exp\!\Big(-\tfrac{1}{2}\vec x^\top A \vec x + \vec b^\top \vec x\Big),
\]
with $A\succ 0$.  
This is a Gaussian with mean $\langle \vec x \rangle = A^{-1}\vec b$ and covariance $A^{-1}$.  
Thus computing the solution of $A\vec x=\vec b$ is equivalent to evaluating the mean
of $\vec x$ under this Gaussian distribution.  
This perspective links numerical linear algebra to probabilistic inference and
provides the basis for Monte Carlo methods for solving linear systems.
\paragraph{Product of two Gaussians.}
The product of two Gaussian densities is again Gaussian (up to a normalization
constant). Writing
\begin{align}
&-\tfrac{1}{2}(\vec x-\vec m_1)^\top \Sigma_1^{-1}(\vec x-\vec m_1)
-\tfrac{1}{2}(\vec x-\vec m_2)^\top \Sigma_2^{-1}(\vec x-\vec m_2) \nonumber \\
&\qquad = -\tfrac{1}{2}(\vec x-\vec m_c)^\top \Sigma_c^{-1}(\vec x-\vec m_c)+C,
\end{align}
with
\begin{align}
\Sigma_c^{-1} &= \Sigma_1^{-1} + \Sigma_2^{-1}, \\
\vec m_c &= (\Sigma_1^{-1}+\Sigma_2^{-1})^{-1}(\Sigma_1^{-1}\vec m_1 + \Sigma_2^{-1}\vec m_2),
\end{align}
we see that \emph{precisions} (inverses of covariances) add, while the combined mean
$\vec m_c$ is a precision-weighted average of the individual means.  
This principle is fundamental in Bayesian updating: posterior precision =
prior precision + data precision.

\subsection{Spectral connections}
These formulas illustrate recurring themes:
\begin{itemize}
\item Determinants $\det \Sigma$ control normalization constants.
\item Inverses $\Sigma^{-1}$ (precisions) govern conditioning, products,
and updated means.
\item Eigenvalues $\{\lambda_i\}$ of $\Sigma$ determine scaling:
$\det \Sigma = \prod_i \lambda_i$, and $\Sigma^{-1}$ has spectrum $\{1/\lambda_i\}$.
\item Completing the square corresponds to shifting into the eigenbasis of $A$,
where quadratic forms decouple into sums of $\lambda_i y_i^2$.
\end{itemize}

Gaussian algebra is therefore a unifying arena for the concepts of this book:
matrix inversion, determinants, Schur complements, and spectral decompositions.
It connects directly to inference (Bayesian updating, marginalization),
machine learning (Gaussian processes, graphical models), and physics (free
energies, fluctuation–dissipation theorems).

\subsection{Moments of Gaussian distributions}
\index{key}{Gaussian moments}\index{key}{quadratic forms}\index{key}{trace identities}

Another important application of Gaussian integration is the computation of
moments. Because the Gaussian distribution is entirely determined by its mean
and covariance, higher moments can be expressed in closed form using trace and
quadratic identities.

\paragraph{First and second moments.}
For $\vec x \sim \mathcal{N}(\vec m,\Sigma)$,
\begin{align}
\mathbb{E}[\vec x] &= \vec m, \label{eq:gauss-mean}\\
\mathrm{Cov}(\vec x,\vec x) &= \Sigma. \label{eq:gauss-cov}
\end{align}
For linear transformations $A\vec x$,
\begin{align}
\mathbb{E}[A\vec x] &= A\vec m, \\
\mathrm{Var}[A\vec x] &= A \Sigma A^\top, \\
\mathrm{Cov}[A\vec x, B\vec y] &= A\,\mathrm{Cov}(\vec x,\vec y)\,B^\top.
\end{align}
These properties show how covariances propagate through linear maps —
the reason Gaussians are closed under marginalization and conditioning.

\paragraph{Quadratic forms.}
If $\vec x \sim \mathcal{N}(\vec m,\Sigma)$ and $A$ is symmetric,
\begin{align}
\mathbb{E}[\vec x^\top A \vec x] &= \mathrm{Tr}(A\Sigma) + \vec m^\top A \vec m, \\
\mathrm{Var}(\vec x^\top A \vec x) &= \mathrm{Tr}(A\Sigma(A+A^\top)\Sigma)
+ \vec m^\top (A+A^\top)\Sigma(A+A^\top)\vec m + \mathrm{Tr}(A\Sigma).
\end{align}
In the isotropic case $\Sigma=\sigma^2 I$,
\[
\mathrm{Var}(\vec x^\top A \vec x) = 2\sigma^4 \,\mathrm{Tr}(A^2) + 4\sigma^2 \vec m^\top A^2 \vec m.
\]
Thus quadratic statistics of Gaussians reduce to traces of powers of $A\Sigma$,
linking moments directly to eigenvalues and spectral norms.

\paragraph{Cross-covariance of quadratic forms.}
If $\vec x\sim \mathcal{N}(0,\sigma^2 I)$ and $A,B$ are symmetric,
\[
\mathrm{Cov}(\vec x^\top A \vec x,\, \vec x^\top B \vec x) = 2\sigma^4 \,\mathrm{Tr}(AB).
\]
This identity underlies random matrix fluctuations, Wishart distributions, and
applications in statistics (e.g.\ variance component estimation).

\paragraph{Cubic forms.}
Gaussian third moments vanish in the centered case, but more general cubic
expectations can be written explicitly. If $\vec x\sim \mathcal{N}(\vec m,M)$ and
$\vec b$ is fixed,
\begin{equation}
\mathbb{E}[(\vec x^\top \vec b)\, \vec x \vec x^\top]
= \vec m \vec b^\top (M + \vec m \vec m^\top)
+ (M+\vec m \vec m^\top)\vec b \vec m^\top
+ \vec b^\top \vec m \,(M-\vec m \vec m^\top).
\end{equation}
Although less common in practice, such formulas appear in perturbative
inference, Gaussian process expansions, and field-theoretic calculations.

\subsection{Spectral interpretation}
All these Gaussian moment identities admit a natural spectral meaning:
\begin{itemize}
\item $\mathrm{Tr}(A\Sigma)$ sums eigenvalue–weighted variances.
\item $\mathrm{Tr}(A^2)$ captures the spread of quadratic fluctuations.
\item Cross-covariances $\mathrm{Tr}(AB)$ link the spectra of operators $A$ and $B$.
\end{itemize}
In other words, Gaussian moments are nothing but trace identities in disguise,
built from the eigenvalues of $A\Sigma$.  
This makes them indispensable in random matrix theory, statistical mechanics,
and inference problems.

\section{Statistical mechanics of Gaussian variables}
\index{key}{Gaussian variables}\index{key}{partition function}\index{key}{free energy}\index{key}{entropy}

The quadratic forms studied above appear naturally as Hamiltonians in statistical
mechanics.  
Let $\vec x\in\mathbb{R}^n$ be a vector of Gaussian variables, and define the
quadratic Hamiltonian
\begin{equation}
H(\vec x) = \tfrac{1}{2}\,\vec x^\top A \vec x,
\end{equation}
with $A$ symmetric positive definite. Introducing an inverse temperature
$\beta = 1/(k_B T)$, the Boltzmann weight is
\begin{equation}
p(\vec x) = \frac{1}{Z} \exp\!\big(-\beta H(\vec x)\big)
= \frac{1}{Z} \exp\!\Big(-\tfrac{\beta}{2}\,\vec x^\top A \vec x\Big).
\end{equation}

\subsection{Partition function and free energy}
The normalization constant $Z(\beta)$ is the \emph{partition function},
\begin{equation}
Z(\beta) = \int_{\mathbb{R}^n} e^{-\tfrac{\beta}{2} \vec x^\top A \vec x}\, d\vec x.
\end{equation}
By Gaussian integration,
\begin{equation}
Z(\beta) = (2\pi)^{n/2} (\det(\beta A))^{-1/2}
= (2\pi/\beta)^{n/2} (\det A)^{-1/2}.
\end{equation}
The corresponding free energy is
\begin{equation}
F(\beta) = -\frac{1}{\beta}\ln Z(\beta)
= -\frac{n}{2\beta}\ln(2\pi/\beta) + \frac{1}{2\beta}\ln\det A.
\end{equation}
This shows explicitly how determinants of $A$ encode thermodynamic potentials.

\subsection{Moments and fluctuations}
The average energy is
\begin{equation}
\langle H \rangle = -\frac{\partial}{\partial \beta}\ln Z(\beta)
= \frac{n}{2\beta}.
\end{equation}
The variance of the energy follows from the second derivative:
\begin{equation}
\mathrm{Var}(H) = \frac{\partial^2}{\partial \beta^2}\ln Z(\beta)
= \frac{n}{2\beta^2}.
\end{equation}
These reproduce the quadratic–form results derived earlier, now reinterpreted as
thermodynamic averages.

\subsection{Entropy}
The entropy of the Gaussian measure is
\begin{equation}
S = -\langle \ln p(\vec x)\rangle
= \frac{n}{2}\Big(1 + \ln(2\pi)\Big) + \tfrac{1}{2}\ln\det(\beta^{-1} A^{-1}).
\end{equation}
In terms of the eigenvalues $\{\lambda_i\}$ of $A$,
\begin{equation}
S = \frac{n}{2}\Big(1 + \ln(2\pi/\beta)\Big) - \frac{1}{2}\sum_{i=1}^n \ln \lambda_i.
\end{equation}
Thus entropy depends additively on the spectral data of $A$.

\subsection{Spectral decomposition and thermodynamics}
Diagonalizing $A=U\Lambda U^\top$, with $\Lambda=\mathrm{diag}(\lambda_1,\dots,\lambda_n)$,
decouples the Hamiltonian:
\begin{equation}
H(\vec x) = \tfrac{1}{2}\sum_{i=1}^n \lambda_i y_i^2,\qquad \vec y=U^\top \vec x.
\end{equation}
The system reduces to $n$ independent harmonic modes with stiffnesses
$\lambda_i$. Then:
\begin{itemize}
\item The partition function factorizes: $Z=\prod_i (2\pi/(\beta \lambda_i))^{1/2}$.
\item The mean energy per mode is $\langle H_i\rangle = 1/(2\beta)$.
\item Mode fluctuations are $\langle y_i^2\rangle = 1/(\beta \lambda_i)$,
consistent with the covariance $\Sigma = (\beta A)^{-1}$.
\end{itemize}

\subsection{Connections to free energy and inference}
These results mirror the role of Gaussians in inference and information theory:
\begin{itemize}
\item The free energy $F=-\tfrac{1}{\beta}\ln Z$ is the log-partition function,
equivalent to the log normalizer in exponential families.
\item The entropy coincides with the Shannon entropy of a multivariate Gaussian.
\item Fluctuations $\langle \vec x^\top A \vec x\rangle$ encode covariance structure,
directly tied to inversion of $A$.
\item Eigenmodes $\lambda_i$ simultaneously control statistical variability
and thermodynamic stiffness.
\end{itemize}

\medskip
The statistical mechanics of Gaussian Hamiltonians demonstrates vividly how
matrix inversion, determinants, and spectra intertwine.  
Free energies are sums over $\ln \lambda_i$, entropies reflect spectral spread, and
fluctuations are trace identities of $A^{-1}$.  
The Gaussian distribution is thus the unifying bridge between linear algebra and
thermodynamic reasoning, connecting statistical mechanics, probability, and
machine learning.

\section{Leslie model}
\index{key}{Leslie model}

One of the simplest but most influential applications of spectral methods in complex systems is the \emph{Leslie matrix model} of population dynamics.  
Matrix population models have become increasingly important in ecology and demography for predicting the growth and long-term structure of populations. Early work on projection methods dates back to Cannan (1895) and Whelpton (1936), but it was in the 1940s that Bernardelli (1941), Lewis (1942), and especially Patrick H.~Leslie (1945) formalized these methods using linear algebraic tools.\footnote{See P.H.~Leslie, ``On the use of matrices in certain population mathematics,'' \emph{Biometrika}, 33(3):183–212 (1945); and P.H.~Leslie, ``Some further notes on the use of matrices in population mathematics,'' \emph{Biometrika}, 35(3/4):213–245 (1948).}  

Leslie’s insight was to combine fertility and survival information into a single \emph{projection matrix}, now called the \emph{Leslie matrix}.  
The model is discrete in time (generations are advanced step by step) and structured by age classes (each individual belongs to a class based on age interval).  

\subsection*{The model}
Let $\vec n(t)$ be the population vector at time $t$, with entries $n_i(t)$ representing the number of individuals in age class $i$. Then
\begin{equation}
    \vec n(t+1)=A \vec n(t),
\end{equation}
where $A$ is the \emph{Leslie matrix}, typically written as
\begin{equation}
A=\begin{bmatrix}
F_1 & F_2 & \cdots & F_{w-1} & F_w \\
P_1 & 0   & \cdots & 0       & 0 \\
0   & P_2 & \cdots & 0       & 0 \\
\vdots & \ddots & \ddots & \vdots & \vdots \\
0 & \cdots & 0 & P_{w-1} & 0
\end{bmatrix},
\end{equation}
with:
\begin{itemize}
    \item $F_i$ = fertility rate of individuals in age class $i$, contributing newborns to age class $1$ in the next generation;
    \item $P_i$ = survival probability of individuals in age class $i$ to survive and progress to class $i+1$.
\end{itemize}

\subsection*{Spectral properties}
The power of the Leslie model lies in the fact that its long-term dynamics are governed by the \emph{spectrum} of $A$.  

Since all entries of $A$ are non-negative, the Perron–Frobenius theorem (see Chapter~\ref{sec:perron}) applies:
\begin{enumerate}
    \item There exists a real, positive dominant eigenvalue $\lambda_{\mathrm{PF}}$ of $A$.
    \item The associated right eigenvector $\vec v_{\mathrm{PF}}$ has non-negative components and describes the \emph{stable age distribution}.
    \item The associated left eigenvector $\vec w_{\mathrm{PF}}$ can be normalized to represent the \emph{reproductive value} of each age class.
\end{enumerate}

Thus:
\begin{itemize}
    \item If $\lambda_{\mathrm{PF}}>1$, the population grows asymptotically at exponential rate $\lambda_{\mathrm{PF}}$.
    \item If $\lambda_{\mathrm{PF}}=1$, the population stabilizes.
    \item If $\lambda_{\mathrm{PF}}<1$, the population eventually declines to extinction.
\end{itemize}

This spectral interpretation transforms demographic questions into purely linear-algebraic computations. In fact, the asymptotic solution takes the form
\begin{equation}
    \vec n(t) \;\sim\; c \, \lambda_{\mathrm{PF}}^{\top} \vec v_{\mathrm{PF}}, \qquad t\to\infty,
\end{equation}
where $c$ depends on the initial condition $\vec n(0)$ but the growth rate and shape are determined only by the Perron eigenpair.

\subsection*{Connections to earlier spectral methods}
Several tools discussed in earlier chapters naturally extend to Leslie matrices:
\begin{itemize}
    \item The \textbf{resolvent} $(zI-A)^{-1}$ describes the generating function of population trajectories (cf.~Sec.~\ref{sec:resolvent}).
    \item \textbf{Pseudospectral analysis} (Sec.~\ref{sec:matrixsym}) helps assess robustness of growth rates to perturbations in fertility or survival probabilities, which is important in ecology where parameter values are uncertain.
    \item The \textbf{Lyapunov stability} viewpoint (Sec.~\ref{sec:lyapunovstab}) connects $\lambda_{\mathrm{PF}}$ to exponential stability of the trivial fixed point $\vec n=0$.
\end{itemize}

\subsection*{Extensions and stochastic generalizations}
Leslie himself extended the model in 1959 to include density-dependent effects, allowing survival or fertility to depend on total population size.  
Later, Pollard (1966) developed stochastic Leslie models where fertility and survival are treated as random variables, making the system equivalent to a stochastic matrix evolution problem closely related to random matrix theory.\footnote{J.~H.~Pollard, ``On the Use of the Direct Matrix Product in Analysing Stochastic Leslie Models,'' \emph{Biometrika} 53(3/4), 397–415 (1966).}

\subsection*{Applications}
The Leslie model is widely used in:
\begin{itemize}
    \item \textbf{Ecology and conservation}: projecting endangered species populations, evaluating which age classes contribute most to growth.
    \item \textbf{Epidemiology}: modeling progression through disease stages, where age classes correspond to states of infection.
    \item \textbf{Demography and policy}: forecasting human populations under different fertility/mortality assumptions.
\end{itemize}

\medskip
\noindent
The Leslie matrix illustrates the central message of this book: \emph{spectral properties of linear operators govern the fate of complex systems}.  
Here, Perron–Frobenius links the leading eigenvalue to population growth, and the corresponding eigenvectors to stable structure and reproductive value.  
This idea will reappear in network analysis, epidemiology, and beyond.

\section{Input-output economic theory}\label{sec:leontief}
{Leontief economy}
{Input-output analysis}
{Leontief inverse}

\subsection{Historical background}
Wassily Leontief (1906–1999), awarded the Nobel Prize in Economics in 1973, developed the celebrated \emph{input–output model} as a way to systematically describe the interdependencies of industrial sectors in an economy. His first applications in the 1930s and 40s used U.S.\ data to forecast demand for steel, coal, and labor during wartime planning. The model, now standard in both economics and policy analysis, is one of the clearest examples of how linear algebra and spectral methods reveal the structure and dynamics of complex systems.

\subsection{The basic model}
Input–output analysis begins by partitioning the economy into $n$ productive
sectors (agriculture, manufacturing, services, etc.).  
Each sector produces a homogeneous good, some of which is used as input
by other sectors, and some of which goes to final demand (households,
government, exports).  

Let $z_{ij}$ denote the physical flow of good $i$ used by sector $j$ as an
intermediate input in a given accounting period, and let $x_j$ denote the total
gross output of sector $j$.  
The \emph{technical coefficient} $a_{ij}$ is defined as
\begin{equation}
a_{ij} = \frac{z_{ij}}{x_j},
\end{equation}
i.e.\ the quantity of good $i$ required as input per unit of gross output of sector $j$.  
Collecting all coefficients $a_{ij}$ into the matrix
\[
A = (a_{ij}) \in \mathbb{R}^{n\times n},
\]
we obtain the \emph{technical coefficient matrix} (or input–output coefficient
matrix).  

By construction, each column $j$ of $A$ is \emph{normalized by output} $x_j$,
so that $a_{ij}$ measures an input requirement relative to one unit of sector $j$’s output.  
The coefficients are typically estimated from national accounts and input–output
tables, where rows correspond to industries supplying goods, and columns
to industries demanding them.\footnote{See W.~Leontief, \emph{The Structure of American Economy, 1919–1929}, Harvard University Press (1941); W.~Leontief, \emph{Input–Output Economics}, 2nd ed., Oxford University Press (1986).  
For theoretical foundations see P.~Sraffa, \emph{Production of Commodities by Means of Commodities}, Cambridge University Press (1960).  
On vertically integrated sectors and dynamic aspects see L.~Pasinetti, \emph{Lectures on the Theory of Production}, Columbia University Press (1977).  
For modern data applications see R.~E.~Miller and P.~D.~Blair, \emph{Input–Output Analysis: Foundations and Extensions}, 2nd ed., Cambridge University Press (2009).}

Let $\vec x$ denote the vector of gross outputs, and $\vec f$ the vector of final
demands (consumption, investment, exports).  
The accounting identity of input–output analysis reads
\begin{equation}
\vec x = A \vec x + \vec f,
\end{equation}
meaning that each sector’s output is absorbed by intermediate input
requirements $A\vec x$ plus final demand $\vec f$.  

\paragraph{The Leontief inverse.}
Rearranging yields
\begin{equation}
\vec x = (I - A)^{-1}\vec f,
\end{equation}
where $(I-A)^{-1}$ is the \emph{Leontief inverse}.  
The $(i,j)$ entry of $(I-A)^{-1}$ gives the total (direct and indirect) amount
of good $i$ required to deliver one additional unit of final demand in sector $j$.  

The matrix $A$ captures the \emph{technology of production}: each column describes the recipe by which a sector combines intermediate inputs to produce one unit of output. Because each entry $a_{ij}$ is normalized by the gross output $x_j$, the coefficients are dimensionless and scale-invariant, which makes them comparable across industries of very different sizes. A crucial mathematical requirement is that the spectral radius $\rho(A)$ is less than one, ensuring that $(I-A)^{-1}$ exists and that production can satisfy final demand without requiring infinite intermediate inputs. The Leontief inverse then plays a dynamic role: it traces how a change in final demand in one sector propagates, directly and indirectly, throughout the entire web of inter-industry linkages.

Thus, the technical coefficient matrix $A$ serves as the cornerstone of classical
and modern input–output analysis, linking technology, interdependence, and
equilibrium production in a unified linear-algebraic framework.

\subsection{Spectral conditions for productivity}
From a spectral point of view, the system is well-defined if and only if $I-A$ is invertible. Since $A$ is nonnegative, the Perron–Frobenius theorem (see Chapter~\ref{sec:perron}) applies. The condition can be stated in several equivalent ways:
\begin{itemize}
    \item \textbf{Spectral radius condition:}
    \begin{equation}
    \rho(A) < 1,
    \end{equation}
    where $\rho(A)$ is the spectral radius. This ensures that the Neumann series expansion
    \begin{equation}
    (I-A)^{-1} = I + A + A^2 + \cdots
    \end{equation}
    converges.
    \item \textbf{Hawkins–Simon condition:} all principal minors of $I-A$ must be strictly positive. This is equivalent to requiring that $I-A$ is an \emph{M-matrix} (see Section~\ref{sec:hstheor}). Intuitively, no subset of industries can require more input than it collectively produces.
    \item \textbf{Perron–Frobenius eigenvalue condition:} if $\lambda_{pf}$ is the PF eigenvalue of $A$, then:
    \begin{equation}
    \lambda_{pf} < 1 \quad \Rightarrow \quad \text{productive economy.}
    \end{equation}
    See again Sec. ~\ref{sec_per_frob}.
\end{itemize}

\subsection{Worked example: a two-sector economy}
Consider the economy with two industries, agriculture and manufacturing. The technical coefficient matrix is
\begin{equation}
A=\begin{bmatrix}
0.4 & 0.2 \\
0.3 & 0.1
\end{bmatrix}.
\end{equation}
The spectral radius is the maximum eigenvalue of $A$:
\begin{equation}
\Lambda(A)=\{0.5,\,0.0\}, \quad \rho(A)=0.5<1.
\end{equation}
Thus the system is productive. Alternatively, the Hawkins–Simon condition requires positivity of the leading principal minors of $I-A$:
\begin{equation}
I-A=\begin{bmatrix}
0.6 & -0.2 \\
-0.3 & 0.9
\end{bmatrix},\quad
\det(0.6)=0.6>0,\quad
\det(I-A)=0.48>0.
\end{equation}
Both checks confirm feasibility. The Leontief inverse is then
\begin{equation}
(I-A)^{-1} = \begin{bmatrix}
1.67 & 0.37 \\
0.56 & 1.11
\end{bmatrix}.
\end{equation}
A one-unit increase in final demand for manufacturing thus requires $0.37$ units of agriculture and $1.11$ units of manufacturing output, including indirect effects.

\subsection{Extensions and applications}
The Leontief framework extends naturally to:
\begin{itemize}
    \item \textbf{Dynamic models:} introducing time-dependence or capital accumulation leads to Leontief-type differential equations.
    \item \textbf{Trade and global value chains:} measures such as \emph{upstreamness} and \emph{downstreamness} (see Section~\ref{sec:resolvent}) are built from powers of $A$ and its resolvent.
    \item \textbf{Network resilience:} interpreting $A$ as the adjacency matrix of a weighted directed graph connects input–output economics to percolation and contagion dynamics (cf. Chapter~\ref{sec:graphs}).
\end{itemize}

In all these cases, the spectral radius of $A$ and the Perron–Frobenius eigenvector govern the long-term sustainability and distribution of activity across sectors, showing again how economic complexity is encoded in spectral properties of nonnegative matrices.
\subsection{Production networks, growth amplification, and ecological analogies}
{production networks}
{economic growth}
{output multipliers}

So far we have emphasized Leontief’s input–output model as a tool for analyzing short-term multipliers of shocks in an economy.  
However, recent work shows that the same spectral structures also control \emph{long-term growth}.  
In particular, the notion of \emph{output multipliers} provides a bridge between production networks, economic growth, and even ecological food webs.\footnote{J.~McNerney, C.~Savoie, F.~Caravelli, V.~M.~Carvalho, and J.~D.~Farmer, ``How production networks amplify economic growth,'' \emph{Proc. Natl. Acad. Sci. USA} \textbf{119}, e2106031118 (2022).}  

\paragraph{Output multipliers.}
Recall that the Leontief system is given by
\[
   \vec x = A^{\top} \vec x + \vec f,
\]
where $A$ is the matrix of input coefficients ($a_{ij}$ = fraction of $j$ used in production of $i$), $\vec x$ is gross output, and $\vec f$ is final demand.  
The Leontief inverse,
\[
   H = (I - A^{\top})^{-1},
\]
encodes the total requirements (direct and indirect) of producing goods.  

The \emph{output multiplier} of industry $i$ is then defined by
\begin{equation}
   L_i = \left[(I - A^{\top})^{-1} \vec 1\right]_i,
   \label{eq:output_multiplier}
\end{equation}
where $\vec 1$ is the vector of ones.  
Intuitively, $L_i$ is the average length of production chains leading to industry $i$, weighting each link by expenditure shares.\footnote{See also R.~E.~Miller and P.~D.~Blair, \emph{Input–Output Analysis: Foundations and Extensions}, Cambridge University Press (2009).}  
It measures how many steps, on average, a unit of final demand for $i$ traces back through upstream suppliers before reaching households.

\paragraph{Recursive form and ecological analogy.}
Equation \eqref{eq:output_multiplier} can be written recursively:
\begin{equation}
   L_i = 1 + \sum_j a_{ji} L_j,
   \label{eq:recursive_multiplier}
\end{equation}
where $a_{ji}$ is the fraction of inputs to $i$ purchased from $j$.  
This formula is identical in form to the definition of trophic levels in ecology, where the trophic level of a species is one plus the average trophic level of its prey.\footnote{D.~M.~Post, ``The long and short of food-chain length,'' \emph{Trends Ecol. Evol.} \textbf{17}, 269–277 (2002).}  
Thus, industries in an economy have \emph{trophic levels} in the production network: agriculture and services often occupy short chains (low $L_i$), while manufacturing industries sit at higher levels (long $L_i$).

\paragraph{Price dynamics and inherited improvement.}
Suppose productivity in industry $i$ improves at rate $\gamma_i$, meaning less input is required per unit output.  
The rate of real price change $r_i$ of good $i$ can then be decomposed as
\begin{equation}
   r_i = -\gamma_i + \sum_j a_{ji} r_j,
   \label{eq:pricechange}
\end{equation}
which states: prices fall due to local productivity ($-\gamma_i$) and due to inherited cost reductions from suppliers.  
In vector form,
\[
   \vec r = -H^{\top} \vec \gamma.
\]
Thus, productivity improvements accumulate along supply chains: industries with higher $L_i$ experience faster expected price declines, because they inherit improvements from many upstream industries.\footnote{McNerney et al., PNAS (2022), Eq.~(3).}

\paragraph{Aggregate growth.}
Aggregating across industries yields a simple formula for GDP growth:
\begin{equation}
   g = \tilde \gamma \, \bar L,
   \label{eq:growth}
\end{equation}
where $\tilde \gamma$ is the average productivity improvement rate across industries, and $\bar L$ is the average output multiplier of the economy (weighted by final demand shares).\footnote{McNerney et al., PNAS (2022), Eq.~(5).}  
Equation \eqref{eq:growth} shows that growth depends both on \emph{how fast technology improves} and on \emph{how deep the production network is}.  
Countries with longer supply chains ($\bar L$ large) exhibit systematically faster growth given the same productivity improvement.

\paragraph{Interpretation.}
Equation \eqref{eq:growth} provides a spectral reinterpretation of growth theory: the Leontief inverse $(I-A^{\top})^{-1}$, which already appears in input–output multipliers, also governs the long-run accumulation of technological change.  
The analogy with ecology is instructive: in ecosystems, longer trophic chains amplify fragility (Sec.~\ref{sec:maytheorem}), while in economies, longer production chains amplify growth.  
Both cases are manifestations of the same recursive spectral structure.

In summary, production networks demonstrate how resolvent operators $(I-A)^{-1}$ mediate not only shock propagation but also long-term amplification of productivity improvements.  
The unifying role of spectral methods thus extends from economics to finance (Sec.~\ref{sec:debtrank}) and ecology (Sec.~\ref{sec:maytheorem}), linking stability, contagion, and growth through the common language of eigenvalues and resolvents.

\section{PageRank and Ranking}
{PageRank}
{centrality measures}
{Google matrix}
{Markov chain}
{spectral gap}

\subsection{Centrality measures}
\label{central_meas}
The importance of nodes in a network can often be quantified by notions of \emph{centrality}.  
One of the earliest spectral measures is the \emph{eigenvector centrality}\footnote{P.~Bonacich, ``Factoring and weighting approaches to status scores and clique identification,'' \emph{Journal of Mathematical Sociology} \textbf{2}, 113--120 (1972).}
, which associates to each node $i$ a score $c_i$ proportional to the sum of the scores of its neighbors.  
Formally, for a graph with adjacency matrix $A$,
\begin{equation}
c_i = \frac{1}{\lambda}\sum_{j} A_{ij} c_j,
\end{equation}
or in vector notation,
\begin{equation}
A\vec c=\lambda\vec c.
\end{equation}
Thus, centralities correspond to the components of an eigenvector of $A$.  
By the Perron–Frobenius theorem (see Section~\ref{sec_per_frob}), the eigenvector associated with the largest eigenvalue $\lambda_{\max}$ can be chosen to have nonnegative entries, making it a natural choice for ranking nodes in networks.

Eigenvector centrality has limitations: in graphs with sinks, dangling nodes, or reducible structures, it may not be well-defined or unique.  
This motivates the construction of the \emph{Google matrix} and the celebrated PageRank algorithm.

\subsection{The Google matrix}
The PageRank algorithm, introduced by Brin and Page in 1998, was originally designed to rank web pages on the basis of the hyperlink structure of the World Wide Web.  
The web can be represented by a directed graph with adjacency matrix $A$, where $A_{ij}=1$ if page $j$ links to page $i$. The problem is to determine a global ranking of importance that reflects both direct and indirect link structure.

To this end, one first defines a column-stochastic matrix $S$ by normalizing each column of $A$ by its out-degree $k_j$:
\begin{equation}
S_{ij}=\frac{A_{ij}}{k_j}, \qquad k_j=\sum_i A_{ij}.
\end{equation}
$S$ is the transition matrix of a Markov chain describing a random surfer moving along hyperlinks.  
However, $S$ is not always irreducible (due to dangling nodes with $k_j=0$) and may be periodic, which would prevent the existence of a unique stationary distribution.

To solve this, Brin and Page introduced the \emph{Google matrix},\footnote{S. Brin and L. Page, ``The anatomy of a large-scale hypertextual Web search engine,'' \emph{Computer Networks and ISDN Systems} \textbf{30}, 107--117 (1998).}

\begin{equation}
G=\alpha S+(1-\alpha)\frac{1}{N}\vec 1\vec 1^{\top},
\end{equation}
where $N$ is the number of nodes, $\vec 1$ the all-ones vector, and $\alpha \in (0,1)$ the \emph{damping factor} (typically $\alpha=0.85$ in practice).  
This corresponds to a random walk in which, with probability $\alpha$, the surfer follows a hyperlink, and with probability $1-\alpha$, jumps to a uniformly random page.  

\subsection{Spectral properties}
The matrix $G$ is stochastic, irreducible, and aperiodic. By the Perron–Frobenius theorem, its largest eigenvalue is $\lambda_1=1$, and the corresponding eigenvector $v^{pr}$ has strictly positive components.  
This eigenvector is called the \emph{PageRank vector}, and its entries define the stationary distribution of the random walk.  

The rest of the spectrum lies strictly inside the unit disk: $|\lambda_i|\leq \alpha$ for $i>1$.  
The gap $1-\alpha$ between the dominant eigenvalue and the second largest eigenvalue in modulus is called the \emph{spectral gap}, and it controls the convergence rate of the \emph{power method}.  
The larger the gap, the faster the convergence to the PageRank vector.  

\subsection{Computation and convergence}
In practice, PageRank is computed by iterating
\begin{equation}
\vec v^{(k+1)}=G\vec v^{(k)},
\end{equation}
starting from an arbitrary probability vector $v^{(0)}$.  
Convergence to $v^{pr}$ is guaranteed, and the number of iterations required is proportional to the inverse spectral gap.  
This is an application of the general theory of linear iterative methods (see Chapter~\ref{sec:matrixiterations}).

\subsection{Extensions and interpretations}
The PageRank construction admits many variations:
\begin{itemize}
    \item \textbf{Personalized PageRank:} replace the uniform teleportation vector $\frac{1}{N}\vec 1$ with a preference vector $p$, biasing the random walk.
    \item \textbf{Block PageRank:} applied to bipartite or multilayer networks.
    \item \textbf{Topic-sensitive PageRank:} multiple personalized vectors, useful in search engines.
\end{itemize}

From the perspective of spectral methods, PageRank is a natural generalization of eigenvector centrality: it guarantees uniqueness and positivity of the ranking vector even on reducible or irregular networks.  
It can also be interpreted as finding the invariant measure of a Markov chain with teleportation, a link to the Perron–Frobenius operator introduced in Section~\ref{sec:perron}.

Beyond search engines, PageRank has been applied to citation networks, biological networks, transportation systems, and even international trade flows.  
Its robustness and reliance on spectral properties make it a paradigmatic example of how local interactions in a complex system (hyperlinks, citations, or interactions) give rise to global emergent order.

\subsection{Relation of centrality measures to the resolvent}
\label{sec:resolvent_centrality}

We have already introduced several centrality measures in networks
--- eigenvector centrality, Katz centrality, PageRank, communicability ---
all of which can be interpreted through the unifying lens of \emph{matrix functions}.  
At the heart of this unification lies the \emph{resolvent operator},
\[
   R(A,z) = (zI - A)^{-1},
\]
which we studied in Section~\ref{sec:resolvent}.  
The resolvent encodes information about walks of all lengths in the network: 
its Laurent expansion shows that the coefficient of $z^{-k}$ corresponds to paths of length $k$.  
Thus, the resolvent naturally acts as a generating object for ranking schemes.

\paragraph{Katz centrality.}
Katz centrality, introduced in the 1950s,\footnote{L.~Katz, ``A new status index derived from sociometric analysis,'' \emph{Psychometrika} \textbf{18}, 39--43 (1953).}  
was one of the first attempts to move beyond simple degree centrality by incorporating paths of \emph{all lengths}, but with a decay factor so that distant nodes contribute less.  
Formally,
\begin{equation}
   \vec K(A) = (I - \alpha A)^{-1}\vec  - \vec 1,
   \qquad 0 < \alpha < \frac{1}{\rho(A)},
\end{equation}
where $\rho(A)$ is the spectral radius of $A$ and $\vec 1$ is the all-ones vector.  
The subtraction of $\vec 1$ removes trivial length-zero contributions.  

Expanding the inverse yields
\[
\vec   K(A) = \sum_{k=1}^\infty \alpha^k A^k \vec 1,
\]
which makes the interpretation transparent: $K(A)_i$ counts the number of walks of all lengths emanating from node $i$, weighted by $\alpha^k$.  
Thus Katz centrality interpolates between \emph{local} (short paths) and \emph{global} (long paths) influence, with $\alpha$ tuning the scale.  
In modern terms, Katz centrality is simply the resolvent $R(A,1/\alpha)$ applied to $\vec 1$.

\paragraph{Exponential centrality.}
A parallel family of measures is based on the matrix exponential,\footnote{See A.~Aprahamian, D.~J.~Higham, and N.~J.~Higham, ``Matching exponential-based and resolvent-based centrality measures,'' \emph{Journal of Complex Networks} \textbf{4}(2), 157--176 (2016).}
\begin{equation}
   \vec C(A) = \exp(\beta A)\vec 1 = \sum_{k=0}^\infty \frac{\beta^k}{k!} A^k \vec 1,
\end{equation}
where $\beta>0$ is a scaling parameter.  
Compared with Katz centrality, exponential centrality weights long walks more strongly at short lengths but discounts them faster for large $k$, due to the factorial decay $1/k!$.  
Both are walk-based measures, but with different decay kernels: Katz emphasizes \emph{geometric} weighting ($\alpha^k$), exponential emphasizes \emph{factorial} weighting ($1/k!$).

\paragraph{Dunford--Taylor functional calculus.}
The connection between these families can be made precise using the Dunford--Taylor representation of analytic matrix functions (Sec.~\ref{sec:resolvent}):
\begin{equation}
   f(A) = \frac{1}{2\pi i} \int_\Gamma (zI - A)^{-1} f(z)\,dz,
\end{equation}
where $\Gamma$ is a contour enclosing the spectrum of $A$.  
This identity shows that \emph{any} analytic centrality function (exponential, Katz, logarithmic, fractional power) can be expressed in terms of the resolvent.  
For instance,
\[
   f(z) = e^{\beta z} \quad \text{(exponential centrality)}, \qquad 
   f(z) = \frac{1}{1-\alpha z} \quad \text{(Katz centrality)}.
\]
From this perspective, Katz, exponential, communicability, and PageRank are not separate constructs but different resolvent-based weightings.

\paragraph{Approximations and spectral gap.}
Direct computation of $(I - \alpha A)^{-1}$ can be expensive for large, dense networks, and ill-conditioned near $\alpha \rho(A) = 1$.  
However, many networks are effectively low-rank, and the leading eigenvalue dominates when the spectral gap
\[
   \lambda_1 - \max\{|\lambda_2|,\dots,|\lambda_n|\}
\]
is large.  
In such cases, low-rank approximations of the resolvent suffice.  
Recent work\footnote{A.~Bartolucci, F.~Caccioli,F.~Caravelli, P. ~Vivo, ``
Ranking influential nodes in networks from aggregate local information,'' Phys. Rev. Research 5, 033123(2023).}  
shows that even rank-one approximations (using degree sequences) can accurately approximate Katz and PageRank rankings in many empirical systems.  
This explains why simple heuristics based on degrees often perform surprisingly well.

Resolvent-based centrality measures provide a unifying framework:
\begin{itemize}
   \item \textbf{Katz centrality}: resolvent weighting with geometric decay.
   \item \textbf{Exponential/communicability}: resolvent weighting with factorial decay.
   \item \textbf{PageRank}: resolvent with teleportation correction.
   \item \textbf{Spectral approximations}: dominated by the Perron root and spectral gap.
\end{itemize}
Thus, apparently diverse notions of influence, from sociometric status (Katz), to web search (PageRank), to communicability in physics, are all variations on a single theme: the resolvent as a generator of weighted walks.

\section{Ecologies}
In the field of population ecology, a great deal of effort has been devoted to understanding the stability of large, interacting systems of species.  
Empirical observations show that only relatively small groups of species form persistent and 
structured ecological networks, such as food webs or mutualistic communities.  
A longstanding question has been whether this empirical regularity is a consequence of 
\emph{size-induced instability}: does a system with many species, and many potential 
interactions, become unstable purely as a result of its complexity?  

Spectral methods provide a natural mathematical framework for this problem.  
The key object is the \emph{diet matrix} $A$ {diet matrix},  
encoding how the abundance of each species changes in response to the others.  
The entries of $A$ capture interaction types:
\begin{itemize}
    \item $A_{ij}>0$ if species $i$ benefits from species $j$ (e.g.\ prey consumes predator biomass);
    \item $A_{ij}<0$ if species $i$ suffers from the presence of $j$ (e.g.\ being consumed);
    \item $A_{ij},A_{ji}>0$ corresponds to mutualism;
    \item $A_{ij},A_{ji}<0$ corresponds to competition;
    \item $A_{ij}>0, A_{ji}=0$ corresponds to parasitism.
\end{itemize}
The population dynamics are then written in the linearized form
\begin{equation}
\frac{d}{dt} \vec p = A \vec p,
\end{equation}
where $\vec p$ is the vector of species abundances.  
As in Section~\ref{sec:stability}, the system is stable if and only if all eigenvalues of $A$ 
lie in the left half-plane (i.e.\ have negative real part).  
Thus, stability is entirely governed by the spectrum $\Lambda(A)$.

\subsection{Instability: May's theorem}
{May's theorem}
\label{sec:maytheorem}
The modern study of ecological stability begins with Robert May’s celebrated work in the early 1970s.\footnote{R.~M.~May, ``Will a large complex system be stable?'' \emph{Nature} \textbf{238}, 413--414 (1972).}\footnote{R.~M.~May, \emph{Stability and Complexity in Model Ecosystems}, Princeton University Press (1973).}  
May asked a provocative question: \emph{what happens when the interaction matrix of an ecosystem is random}?  
In other words, rather than analyzing a carefully structured food web, suppose that interactions among species are assigned at random.  

He made the following assumptions:
\begin{itemize}
    \item The community contains $S$ species.
    \item Each potential pair $(i,j)$ interacts with probability $C$, called the \emph{connectance}.
    \item Nonzero off-diagonal entries $A_{ij}$ are drawn from a distribution with mean $0$ and variance $\sigma^2$.
    \item Diagonal entries represent self-regulation, fixed to a stabilizing constant $A_{ii}=-d$ with $d>0$.
\end{itemize}

\subsection{Random matrix spectrum and stability}

Random matrix theory implies that, in the limit of large $S$, the eigenvalues of $A$ concentrate in a disk of radius $\sqrt{SC}\,\sigma$ centered at $-d$ in the complex plane.\footnote{For a mathematical background see M.~L.~Mehta, \emph{Random Matrices}, Academic Press (2004).}  
Therefore, the system is stable (all eigenvalues have negative real parts) if the disk lies entirely in the left half-plane, i.e.
\begin{equation}
   \sqrt{SC}\,\sigma < d.
   \label{eq:maytheorem}
\end{equation}
This inequality is the essence of \textbf{May’s theorem}: in large, random ecosystems, stability becomes harder to achieve as the number of species $S$, the connectance $C$, or the interaction strength $\sigma$ increase.  
The threshold $\sqrt{SC}\,\sigma = d$ marks a sharp spectral transition, reminiscent of phase transitions in statistical mechanics.

\paragraph{Spectral intuition.}
A heuristic picture follows from Gershgorin’s circle theorem (Sec.~\ref{sec:gerschgorin}).  
Each eigenvalue of $A$ lies within a disk centered at $A_{ii}$ with radius $\sum_{j\neq i} |A_{ij}|$.  
On average, this radius is of order $\sqrt{SC}\,\sigma$, while the diagonal stabilizing term is $-d$.  
Thus, stability requires that the stabilizing self-regulation outweighs the destabilizing fluctuations, leading to the criterion \eqref{eq:maytheorem}.

\subsection{Beyond random interactions}

While May’s theorem was striking, it relies on the strong assumption that all interaction types are equally likely and symmetrically distributed.  
In practice, ecological interactions are highly structured:
\begin{itemize}
    \item \emph{Predator–prey} interactions ($A_{ij}>0$, $A_{ji}<0$);
    \item \emph{Mutualistic} interactions ($A_{ij},A_{ji}>0$);
    \item \emph{Competitive} interactions ($A_{ij},A_{ji}<0$).
\end{itemize}
Allesina and Tang later extended May’s framework to these cases.\footnote{S.~Allesina and S.~Tang, ``Stability criteria for complex ecosystems,'' \emph{Nature} \textbf{483}, 205--208 (2012).}  
They showed that predator–prey networks are far more stable than random expectation, mutualistic networks are highly destabilizing, and competitive networks are intermediate.  
From a spectral perspective, these distinctions arise because symmetries in $A$ determine whether eigenvalues are shifted predominantly along the imaginary axis (predator–prey) or along the real axis (mutualism and competition).

\subsection{Connection to resolvents and pseudospectra}

The stability criterion can also be recast in terms of the resolvent norm (Sec.~\ref{sec:resolvent_centrality}).  
Large resolvent norms indicate sensitivity of eigenvalues to perturbations, a hallmark of ecological fragility.  
Thus, pseudospectral analysis\footnote{L.~N.~Trefethen and M.~Embree, \emph{Spectra and Pseudospectra}, Princeton University Press (2005).} provides a natural refinement of May’s theorem: even if eigenvalues lie safely in the left half-plane, large pseudospectral sets $\sigma_\epsilon(A)$ reveal the potential for transient growth and instability under small perturbations.\footnote{See for instance F. Caravelli, P. Staniczenko, Bounds on transient instability for complex ecosystems, PLoS ONE 11(6): e0157876 (2016) }

\subsection{Historical note and legacy}

May’s theorem was initially controversial, since it appeared to imply that large, complex ecosystems could never be stable.  
This ``complexity–stability paradox'' triggered decades of research on the role of \emph{structure} in stabilizing ecosystems: modularity, nestedness, trophic hierarchies, and degree heterogeneity can all counteract the destabilizing effects predicted by random matrix theory.\footnote{For reviews see J.~Grilli, T.~Rogers, and S.~Allesina, ``Modularity and stability in ecological communities,'' \emph{Nature Communications} \textbf{7}, 12031 (2016).}  
Thus, spectral methods provided both the ``doom theorem’’ for complexity and, at the same time, the mathematical tools to explain how real ecosystems circumvent it.

\section{Correlation cleaning in portfolio theory}
\index{key}{random matrix theory}\index{key}{Markowitz portfolio}\index{key}{Marčenko--Pastur distribution}\index{key}{correlation matrix cleaning}

\subsection{Markowitz portfolio theory}
In classical finance, Markowitz introduced the mean--variance framework to
balance return and risk.\footnote{H.~Markowitz, ``Portfolio Selection,''
\emph{Journal of Finance} 7(1), 77–91 (1952).}  
Given $N$ assets with expected returns $\mu$ and covariance matrix $\Sigma$, 
The variance of a portfolio with weights $w$ is
\[
\sigma^2(w) = w^\top \Sigma w.
\]
The optimal portfolio under a budget constraint $w^\top \vec 1=1$ involves
the inverse covariance:
\[
w^\ast \propto \Sigma^{-1}\vec 1.
\]
Thus, reliable estimation of $\Sigma$ and its inverse is central to risk
management. However, when the number of observations $T$ is not much larger
than the number of assets $N$, the empirical covariance
\[
\hat \Sigma = \frac{1}{T} XX^\top
\]
(where $X$ is the $N\times T$ matrix of demeaned returns) is dominated by
noise. This leads to unstable and unrealistic portfolios.

\subsection{The Marčenko--Pastur distribution}
Random matrix theory provides a null model for the spectrum of empirical
covariance matrices in the absence of true correlations.  
If $X$ has i.i.d.~Gaussian entries, the eigenvalue density of
$\hat \Sigma$ converges, as $N,T\to\infty$ with $q=N/T$ fixed, to the
Marčenko--Pastur law:\footnote{V.A.~Marčenko and L.A.~Pastur, 
``Distribution of eigenvalues for some sets of random matrices,''
\emph{Mathematics of the USSR-Sbornik} 1(4), 457–483 (1967).}
\[
\rho(\lambda) = \frac{\sqrt{(\lambda_+ - \lambda)(\lambda - \lambda_-)}}{2\pi q \lambda},
\qquad \lambda_\pm = (1\pm\sqrt{q})^2.
\]
All eigenvalues of $\hat \Sigma$ are expected to fall within $[\lambda_-,\lambda_+]$,
except for genuine correlations which appear as outliers.  
This provides a natural baseline to distinguish signal from noise.

\subsection{Empirical discovery: financial correlations}
The empirical spectrum of stock return correlations was first compared to the
Marčenko--Pastur distribution in the late 90s \footnote{L.~Laloux, P.~Cizeau, J.-P.~Bouchaud, M.~Potters, 
``Noise dressing of financial correlation matrices,'' 
\emph{Physical Review Letters} 83, 1467–1470 (1999).\ This was the
first application of random matrix theory to financial data, showing that most
empirical eigenvalues fall within the MP bulk. Also, V.~Plerou, P.~Gopikrishnan, B.~Rosenow, L.A.N.~Amaral, H.E.~Stanley, 
``Universal and nonuniversal properties of cross-correlations in financial time
series,'' \emph{Physical Review Letters} 83, 1471–1474 (1999).}
Both studies found that the bulk of eigenvalues in financial correlation
matrices are consistent with the MP distribution, indicating noise, while only
a few large eigenvalues carry genuine market and sector information.
\subsection{Cleaning techniques}
The raw empirical covariance $\hat \Sigma = \frac{1}{T}XX^\top$ is unstable in
the high-dimensional regime. Bouchaud and collaborators proposed several
systematic methods to regularize or ``clean'' it.\footnote{See L.~Laloux,
P.~Cizeau, J.-P.~Bouchaud, M.~Potters, ``Noise dressing of financial correlation
matrices,'' \emph{Phys.~Rev.~Lett.} 83, 1467 (1999); J.-P.~Bouchaud and
M.~Potters, \emph{Theory of Financial Risk and Derivative Pricing},
Cambridge University Press (2003).}

\paragraph{Eigenvalue clipping.}
Diagonalize the empirical covariance $\hat \Sigma = U \Lambda U^\top$, with
eigenvalues $\Lambda = \mathrm{diag}(\lambda_1,\dots,\lambda_N)$.  
Random matrix theory predicts that eigenvalues in the interval
$[\lambda_-,\lambda_+]$ are noise-dominated, where $\lambda_\pm=(1\pm\sqrt{q})^2$
for $q=N/T$. Clipping replaces these bulk eigenvalues by their average
$\bar\lambda$:
\[
\tilde \Lambda_{ii} = 
\begin{cases}
\lambda_i, & \lambda_i > \lambda_+, \\
\bar\lambda, & \lambda_i \in [\lambda_-,\lambda_+].
\end{cases}
\]
The cleaned covariance is then reconstructed as $\tilde \Sigma = U \tilde \Lambda U^\top$.
This preserves informative outliers (market and sector modes) while flattening noise.

\paragraph{Shrinkage.}  
Ledoit and Wolf (2004) introduced an asymptotically optimal linear shrinkage estimator for high-dimensional covariance matrices. It takes the convex form:
\[
\tilde \Sigma = (1-\alpha)\, \hat\Sigma \;+\; \alpha\, \mu I,
\]
where the shrinkage target is a scaled identity matrix, with \(\mu = \tfrac1N\mathrm{Tr}(\hat\Sigma)\), and the shrinkage intensity \(\alpha\in[0,1]\) is chosen to minimize the expected Frobenius-norm loss relative to the true covariance. This estimator is shown to be well-conditioned and superior to the sample covariance matrix when \(N/T\) is large.\footnote{O.~Ledoit and M.~Wolf, ``A well-conditioned estimator for large-dimensional covariance matrices,'' \emph{Journal of Multivariate Analysis}, \textbf{88}, 365–411 (2004).}

\paragraph{Rotationally invariant estimators (RIE).}
A more sophisticated approach assumes that in the null model eigenvectors are
random (Haar-distributed). Thus the only reliable information lies in the
eigenvalue spectrum. Given the empirical eigenvalues $\{\lambda_i\}$, one
constructs a cleaned spectrum $\{\hat \xi_i\}$ by solving
\[
\hat \xi_i = \frac{\lambda_i}{|1 - q + q \lambda_i g(\lambda_i - i0^+)|^2},
\]
where $g(z) = \frac{1}{N}\sum_{j=1}^N \frac{1}{z-\lambda_j}$ is the empirical
resolvent. The cleaned covariance is $\tilde \Sigma = U \hat \Xi U^\top$, with
$\hat \Xi=\mathrm{diag}(\hat \xi_1,\dots,\hat \xi_N)$. This procedure, though more
involved, yields asymptotically optimal estimators for large $N,T$.

All three methods act on the spectrum of $\hat \Sigma$: eigenvalue clipping is
a coarse projection, shrinkage is a convex combination, and RIE is a nonlinear
spectral transform. Their common goal is to stabilize the inverse covariance
$\hat \Sigma^{-1}$, which is highly sensitive to noise but essential for risk
management in the Markowitz framework.

These cleaning methods stabilize the inverse covariance, producing more robust
portfolios in the Markowitz sense. Beyond finance, correlation cleaning has
been applied in neuroscience, genomics, and climate data, wherever the
high-dimensional regime $N/T$ is large. The central lesson is that spectral
methods from random matrix theory provide the statistical mechanics of noisy
correlations: the Marčenko--Pastur law gives the baseline, and deviations
identify structure.

\section{Transfer matrix spectrum for Ising models}
{transfer matrix}
Statistical mechanics is far from a \emph{spectral theory}, but there are cases in which this is the case. The example we want to provide here is the Ising model in one dimension.  
The central object is the \textit{Gibbs measure}, defined for a system with Hamiltonian $H$ at 
inverse temperature $\beta=1/T$ as
\begin{equation}
   \mathbb P(\sigma) = \frac{e^{-\beta H(\sigma)}}{Z(\beta)},
\end{equation}
where the normalizing constant
\begin{equation}
   Z(\beta)=\sum_{\{\sigma\}} e^{-\beta H(\sigma)}
\end{equation}
is the \emph{partition function}.  
Thermodynamic quantities, such as the free energy $F=-\tfrac{1}{\beta}\log Z$, internal energy, 
and susceptibilities, are all encoded in $Z(\beta)$.  
The transfer matrix method, pioneered by Kramers and Wannier in the 1940s\footnote{Kramers \& Wannier, 
\emph{Statistics of the Two-Dimensional Ferromagnet. Part I}, Phys. Rev. (1941).}, 
reveals a deep connection between $Z(\beta)$ and the spectral properties of an appropriate matrix.

\paragraph{1D Ising model.}
Consider the ferromagnetic Ising chain with Hamiltonian
\begin{equation}
   H(\sigma) = -J \sum_{i=1}^N \sigma_i \sigma_{i+1} - h \sum_{i=1}^N \sigma_i,
   \qquad \sigma_i \in \{\pm 1\},
\end{equation}
with periodic boundary conditions $\sigma_{N+1}=\sigma_1$.  
The Gibbs weight can be factorized as a product of local Boltzmann factors,
\begin{equation}
   Z = \sum_{\{\sigma\}} \prod_{j=1}^N 
   \exp\!\left(\beta J \sigma_j\sigma_{j+1} + \tfrac{\beta h}{2}(\sigma_j+\sigma_{j+1})\right).
\end{equation}
This factorization allows us to write $Z$ as the trace of a product of $2\times 2$ matrices,
\begin{equation}
   Z = \mathrm{Tr}(\mathcal T^N),
\end{equation}
where
\begin{equation}
   \mathcal T =
   \begin{bmatrix}
      e^{\beta(J+h)} & e^{-\beta J} \\
      e^{-\beta J}   & e^{\beta(J-h)}
   \end{bmatrix}
   \label{eq:transfermatrix}
\end{equation}
is the \textit{transfer matrix}.  

Since $\mathcal T$ is real symmetric, it is diagonalizable with real eigenvalues.  
Thus,
\begin{equation}
   Z = \lambda_1^N + \lambda_2^N, \qquad \lambda_1 \geq \lambda_2 \geq 0.
\end{equation}
In the thermodynamic limit $N\to\infty$, the free energy density is controlled by the largest eigenvalue:
\begin{equation}
   f = -\frac{1}{\beta}\lim_{N\to\infty}\frac{1}{N}\log Z
     = -\frac{1}{\beta}\log \lambda_1.
\end{equation}
Thus the \emph{thermodynamics is spectrally determined}.  

\paragraph{Correlation length and eigenvalue gaps.}
The ratio $\lambda_2/\lambda_1$ determines correlation decay.  
Indeed, the two-point function behaves as
\begin{equation}
   g(r) \sim \left(\frac{\lambda_2}{\lambda_1}\right)^r = e^{-r/\xi},
\end{equation}
so that the correlation length $\xi$ is given by
\begin{equation}
   \xi^{-1} = -\log\!\left(\frac{\lambda_2}{\lambda_1}\right).
\end{equation}
In this way, the spectral gap of the transfer matrix controls physical observables.  

\paragraph{Explicit eigenvalues.}
The eigenvalues of \eqref{eq:transfermatrix} are
\begin{equation}
   \lambda_{1,2} = e^{\beta J}\cosh(\beta h) 
   \pm \sqrt{e^{2\beta J}\sinh^2(\beta h)+e^{-2\beta J}}.
\end{equation}
At zero field $h=0$, $\lambda_1=2\cosh(\beta J)$ and $\lambda_2=2\sinh(\beta J)$, so that
\begin{equation}
   \frac{\lambda_2}{\lambda_1} = \tanh(\beta J).
\end{equation}
As expected, $\xi$ diverges at $T=0$ and vanishes at $T\to\infty$.

\paragraph{From Ising to Potts models.}
The method extends naturally to the $q$-state Potts model, where $\sigma_i\in\{1,\ldots,q\}$, with
\begin{equation}
   \mathcal T = (e^{\beta J}-1)I_q + J_q,
   \label{eq:pottstransfer}
\end{equation}
where $J_q$ is the $q\times q$ all-ones matrix.  
The eigenvalues are $\lambda_1=e^{\beta J}+q-1$ and $\lambda_2=e^{\beta J}-1$, with multiplicity $q-1$.  
Again, the free energy and correlation length follow directly from $\lambda_1$ and $\lambda_2$.

\paragraph{Spectral growth and phase transitions.}
For 1D models, $\lambda_1>\lambda_2$ strictly at finite $T$, so correlations decay exponentially and no phase transition occurs.  
In 2D, however, the transfer matrix becomes exponentially larger ($2^N\times 2^N$ for Ising on an $N\times N$ torus), and the thermodynamic limit can produce eigenvalue degeneracies.  
Onsager’s famous solution of the 2D Ising model\footnote{L. Onsager, \emph{Crystal Statistics. I. A Two-Dimensional Model with an Order–Disorder Transition}, Phys. Rev. (1944).} 
used precisely this transfer matrix framework.  
Here, the nontrivial spectral structure of $\mathcal T$ encodes the phase transition at critical temperature $T_c$.  

The transfer matrix method highlights the deep link between statistical mechanics and spectral theory: 
partition functions are traces of matrix powers, free energies come from logarithms of leading eigenvalues, 
and correlation lengths arise from eigenvalue gaps.

\section{Heat Kernels and Diffusion}
{heat kernel}
{diffusion processes}

Heat kernels provide one of the most powerful tools for connecting linear algebra, spectral 
theory, and dynamics on networks.  
In essence, the heat kernel describes how an initial distribution of ``heat'' (or probability) 
diffuses over time under the action of a Laplacian operator.  
This idea is central not only in differential geometry and partial differential equations, 
but also in graph theory, complex networks, and data science.

\subsection{Heat kernels of graph Laplacians}
Let $L=D-A$ be the (combinatorial) graph Laplacian, where $A$ is the adjacency matrix and 
$D$ is the diagonal degree matrix.  
The \emph{heat equation} on the graph is
\begin{equation}
   \frac{d}{dt} \vec f(t) = -L \vec f(t),
   \label{eq:heat-graph}
\end{equation}
with initial condition $\vec f(0)=\vec f_0$.  
The solution is given in terms of the \emph{heat kernel}
\begin{equation}
   \vec f(t) = e^{-tL}\vec f_0, \qquad
   H_t = e^{-tL}.
\end{equation}
Here $H_t$ is an $n\times n$ symmetric, positive semidefinite matrix.  
Its spectral decomposition is immediate: if $L=U\Lambda U^\top$ with eigenvalues $\lambda_i$, then
\begin{equation}
   H_t = U \, \mathrm{diag}(e^{-t\lambda_1},\ldots,e^{-t\lambda_n}) \, U^\top.
\end{equation}
Thus, diffusion attenuates modes according to their eigenvalues: high-frequency eigenvectors 
(decaying fast) correspond to local fluctuations, while low-frequency modes persist longer.  
This makes the heat kernel a natural tool for multi-scale analysis on graphs, used for clustering, 
community detection, and manifold learning.\footnote{See Chung, \emph{Spectral Graph Theory}, AMS (1997).}

\paragraph{Series expansion.}
Expanding $e^{-tL}$ yields
\begin{equation}
   H_t = e^{-t}\sum_{k=0}^\infty \frac{t^k}{k!} W^k,
\end{equation}
where $W=D^{-1}A$ is the random walk transition matrix.  
Thus, $H_t$ can be seen as averaging over random walks of all lengths, weighted by a 
Poisson distribution with mean $t$.

\subsection{Heat kernel PageRank}
Standard PageRank (Section~\ref{central_meas}) is defined as
\begin{equation}
   \vec \pi^\top = \alpha \vec f^\top \big(I - (1-\alpha)W\big)^{-1}
   = \alpha \vec f^\top \sum_{k=0}^\infty (1-\alpha)^k W^k,
\end{equation}
where $W=D^{-1}A$ and $\alpha$ is the teleportation parameter.  
When $\alpha\to 1$, this Neumann series diverges, reflecting the fact that $W$ has 
Perron-Frobenius eigenvalue $1$.  

To address this, Chung and colleagues proposed the \emph{heat kernel PageRank}~\footnote{Fan Chung, 
\emph{The Heat Kernel as the PageRank of a Graph}, PNAS (2007).}.  
It replaces the geometric weighting by a Poissonian one:
\begin{equation}
   \vec h_t^\top = \vec f^\top e^{-t}\sum_{k=0}^\infty \frac{t^k}{k!}W^k
   = \vec f^\top e^{-tL'},
\end{equation}
where $L'=I-W$ is the random walk Laplacian (also called the tilted Laplacian).  
This approach has strong convergence guarantees and interprets PageRank as a diffusion process.

\subsection{Heat kernels, Mellin transforms, and zeta functions}
Heat kernels are intimately connected to zeta functions and Mellin transforms 
(Section~\ref{sec:mellin}).  
Recall that the spectral zeta function of an operator $L$ is
\begin{equation}
   \zeta_L(s) = \sum_{\lambda_i>0} \lambda_i^{-s}.
\end{equation}
By the Mellin transform identity,
\begin{equation}
   \lambda^{-s} = \frac{1}{\Gamma(s)} \int_0^\infty t^{s-1} e^{-t\lambda} \, dt,
\end{equation}
we find that
\begin{equation}
   \zeta_L(s) = \frac{1}{\Gamma(s)} \int_0^\infty t^{s-1} \mathrm{Tr}(e^{-tL}) \, dt.
   \label{eq:zeta-heat}
\end{equation}
In other words, the zeta function is the Mellin transform of the heat trace $\mathrm{Tr}(H_t)$.  
This is a cornerstone of spectral geometry, with deep implications in number theory, quantum 
field theory, and dynamical systems.\footnote{See Gilkey, \emph{Invariance Theory, the Heat Equation and the Atiyah–Singer Index Theorem}, CRC Press (1995).}

\paragraph{Connection to the resolvent.}
Using Dunford–Taylor calculus (Section~\ref{sec:resolvent}), one may also write
\begin{equation}
   (zI-L)^{-1} = \int_0^\infty e^{-t(zI)} e^{-tL} \, dt,
\end{equation}
so the resolvent is the Laplace transform of the heat kernel.  
This duality highlights the unifying role of functional calculus: the resolvent, exponential, 
heat kernel, and zeta function are different faces of the same spectral object.

\subsection{Diffusion as a spectral probe}
Diffusion processes on graphs are therefore natural \emph{spectral filters}:  
they amplify or suppress eigenmodes depending on their eigenvalues.  
For small $t$, the expansion of $H_t$ probes local structure (short paths);  
for large $t$, it reveals global structure (low-frequency eigenvectors).  
This is why heat kernel methods have found applications in:
\begin{itemize}
   \item graph clustering and community detection,
   \item semi-supervised learning on graphs,
   \item defining diffusion distances and graph kernels in machine learning,
   \item ranking and recommendation systems (via heat kernel PageRank).
\end{itemize}

\medskip
Heat kernels act as a bridge between diffusion dynamics, random walks, 
spectral decompositions, and analytic continuations via Mellin and zeta transforms.  
They provide a unifying framework that links the discrete world of graphs to the continuous 
methods of PDEs and spectral geometry.

\section{Bloch theorem a.k.a.\ Floquet's theorem in space}
{Bloch's theorem}
{Floquet's theorem}

One of the most important results in quantum solid-state physics is \emph{Bloch’s theorem}, which describes the structure of wavefunctions in a periodic medium.\footnote{F.~Bloch, ``Über die Quantenmechanik der Elektronen in Kristallgittern,'' \emph{Zeitschrift für Physik} \textbf{52}, 555--600 (1929).}  
It provides the spectral cornerstone behind the modern theory of electronic bands, band gaps, and conductivity in crystalline materials.  
Mathematically, Bloch’s theorem is the spatial analogue of Floquet’s theorem in dynamical systems: just as a time-periodic system admits solutions of the form 
$\phi(t)=e^{\alpha t}p(t)$ with $p(t)$ periodic (Sec.~\ref{sec:stability}), a quantum particle in a spatially periodic potential admits wavefunctions of the form
\begin{equation}
   \psi(\vec r) = e^{i \vec k\cdot \vec r}\,u_{\vec k}(\vec r),
   \qquad u_{\vec k}(\vec r+a\hat e_j)=u_{\vec k}(\vec r),
   \label{eq:bloch}
\end{equation}
where $u_{\vec k}$ is periodic with the same periodicity as the lattice.  
The vector $\vec k$ is called the \emph{quasi-momentum}, and its domain of definition is the \emph{Brillouin zone}, the fundamental cell of the reciprocal lattice.

\subsection{Periodic Hamiltonians and Floquet analogy}

Consider the one-dimensional Schrödinger operator
\begin{equation}
   H = -\frac{\hbar^2}{2m}\frac{d^2}{dx^2} + V(x),
\end{equation}
with $V(x+a)=V(x)$ periodic of lattice spacing $a$.  
A prototypical example is a tight array of potential wells
\begin{equation}
   V(x) = \sum_{j\in \mathbb Z} V_{\mathrm{well}}(x-ja).
\end{equation}
By Floquet theory applied to the second-order differential operator,\footnote{See W.~Magnus and S.~Winkler, \emph{Hill’s Equation}, Dover (1979).} the eigenfunctions of $H$ can be taken of the Bloch form \eqref{eq:bloch}.  
Thus, spectral analysis of $H$ reduces to spectral analysis of a family of operators $H(k)$ parametrized by quasi-momentum $k\in[-\pi/a,\pi/a]$, i.e. over the first Brillouin zone.

The analogy with Floquet theory is direct:
\begin{itemize}
   \item For ODEs with $T$-periodic coefficients, Floquet’s theorem guarantees solutions $\phi(t)=e^{\alpha t}p(t)$ with $p(t+T)=p(t)$.\footnote{G.~Floquet, ``Sur les équations différentielles linéaires à coefficients périodiques,'' \emph{Ann. de l’École Normale Supérieure} \textbf{12}, 47--88 (1883).}
   \item For Schrödinger operators with $a$-periodic potentials, Bloch’s theorem guarantees solutions $\psi(x)=e^{ikx}u(x)$ with $u(x+a)=u(x)$.
\end{itemize}
In both cases, the dynamics separates into an exponential factor (growth rate $\alpha$ or plane wave $k$) and a periodic factor.

\subsection{Fourier (plane-wave) expansion}

To make the structure explicit, expand $\psi(x)$ in Fourier modes,
\begin{equation}
   \psi(x)=\sum_{q}\psi_q e^{iqx}, \qquad q=\frac{2\pi j}{Na},
\end{equation}
and similarly the potential $V(x)=\sum_{q} V_q e^{iqx}$.  
The Schrödinger equation then becomes
\begin{equation}
   \frac{\hbar^2 q^2}{2m}\,\psi_q + \sum_{q'} V_{q-q'}\psi_{q'} = E \psi_q.
   \label{eq:fourierbloch}
\end{equation}
Equation \eqref{eq:fourierbloch} is an infinite-dimensional eigenvalue problem with Toeplitz structure: the Hamiltonian in Fourier space is block-banded with couplings determined by Fourier coefficients of $V(x)$.\footnote{For a pedagogical introduction, see C.~Kittel, \emph{Introduction to Solid State Physics}, Wiley (2005).}  
Truncating to finitely many modes produces a matrix $M$, whose eigenvalues $E(k)$ define the approximate band structure.

\subsection{Example: cosine potential}

\begin{figure}
    \centering
    \includegraphics[scale=0.4]{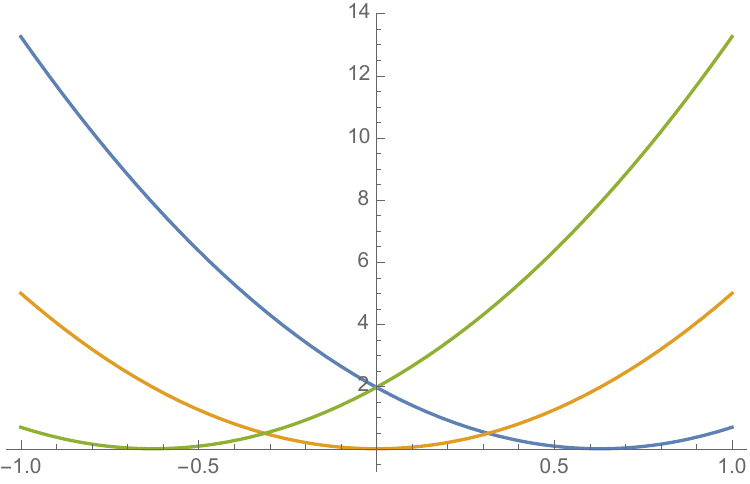}\quad
    \includegraphics[scale=0.4]{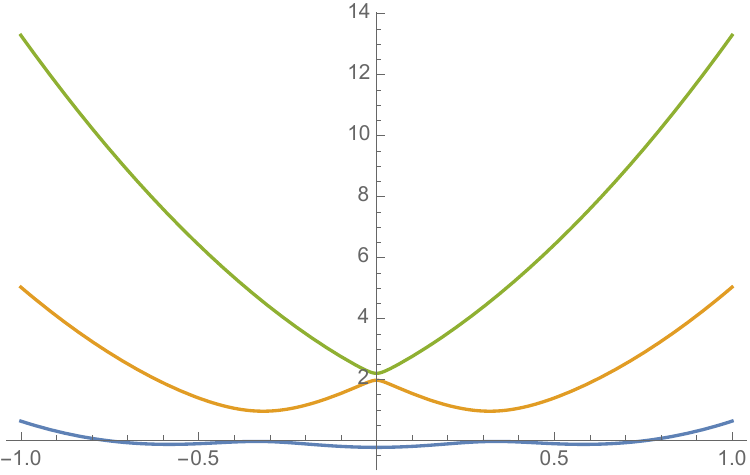}
    \caption{Energy bands as functions of quasi-momentum $k$. Left: free particle ($U=0$). Right: periodic cosine potential ($U\neq 0$) shows opening of spectral gaps at the Brillouin zone boundaries.}
    \label{fig:energybands}
\end{figure}

For $V(x)=U\cos(sx)$, only couplings between $q$ and $q\pm s$ survive, so $M$ takes a tridiagonal block form:
\begin{equation}
   M =
   \begin{bmatrix}
      \ddots & \ddots &        &        &     \\
      \ddots & \frac{\hbar^2(q-s)^2}{2m} & U &        &     \\
             & U & \frac{\hbar^2 q^2}{2m} & U &     \\
             &   & U & \frac{\hbar^2(q+s)^2}{2m} & \ddots \\
             &   &   & \ddots & \ddots
   \end{bmatrix}.
\end{equation}
Diagonalizing $M$ yields energy bands $E_\nu(k)$ as functions of quasi-momentum $k$.  
For $U=0$, one recovers free-particle parabolas $E=\hbar^2 k^2/(2m)$ periodically folded into the Brillouin zone.  
For $U\neq 0$, spectral gaps open at zone boundaries, as illustrated in Fig.~\ref{fig:energybands}.  
These gaps are the origin of the distinction between metals, semiconductors, and insulators.\footnote{N.~W.~Ashcroft and N.~D.~Mermin, \emph{Solid State Physics}, Saunders College (1976).}

\subsection{Higher dimensions and Bravais lattices}

In higher dimensions, let $\Lambda$ be a Bravais lattice with basis vectors $\{\vec a_j\}$.  
A potential satisfies $V(\vec r+\vec a_j)=V(\vec r)$ for all $\vec a_j\in\Lambda$.  
Bloch’s theorem then states that eigenfunctions take the form
\begin{equation}
   \psi_{\vec k}(\vec r) = e^{i\vec k\cdot \vec r} u_{\vec k}(\vec r),
\end{equation}
with $u_{\vec k}$ periodic with respect to $\Lambda$.  
The eigenvalues $E_\nu(\vec k)$ define the \emph{band structure}, i.e. a family of dispersion relations indexed by crystal momentum $\vec k$ in the Brillouin zone.

\subsection{Spectral interpretation}

From a spectral viewpoint, Bloch’s theorem reduces the spectral problem of an infinite periodic operator to a direct integral of finite-dimensional problems indexed by $\vec k$.  
That is,
\begin{equation}
   \sigma(H) = \bigcup_{\vec k\in \text{BZ}} \sigma(H(\vec k)),
\end{equation}
where $\sigma$ denotes the spectrum and BZ the Brillouin zone.\footnote{See M.~Reed and B.~Simon, \emph{Methods of Modern Mathematical Physics IV: Analysis of Operators}, Academic Press (1978).}  
Thus the energy spectrum of a crystal is the union of band spectra of finite Toeplitz-type matrices parametrized by $\vec k$.

\bigskip

Bloch’s theorem is therefore a paradigmatic example of how spectral analysis of operators on periodic domains leads to profound physical consequences.  
It not only underlies electronic band theory and the classification of solids, but also provides a direct bridge to Floquet theory in dynamical systems, illustrating once more how periodicity and spectra govern both temporal and spatial dynamics.

\section{Projector operators in memristive networks}

Memristors are two-terminal passive devices whose resistance depends on the history of the current or voltage that has passed or has been applied to them.\footnote{L.~Chua, ``Memristor--the missing circuit element,'' \emph{IEEE Trans. Circuit Theory}, 18(5):507--519 (1971).} In their simplest form, a \emph{current-controlled memristor} can be modeled by
\begin{equation}
   V(t) = R(x) I(t), \qquad \dot x = -\alpha x + \beta I(t),
   \label{eq:memristor}
\end{equation}
where $R(x) = R_{\mathrm{off}} (1-x) + R_{\mathrm{on}} x$ interpolates between two limiting resistances $R_{\mathrm{off}}>R_{\mathrm{on}}$, $x\in[0,1]$ is the internal memory variable, $\alpha>0$ is a decay constant, and $\beta$ sets the response to current input. Such devices exhibit fading memory: in the absence of current, $x$ relaxes back to zero at rate $\alpha$.\footnote{D.~B.~Strukov, G.~Snider, D.~Stewart, and R.~S.~Williams, ``The missing memristor found,'' \emph{Nature}, 453:80--83 (2008).}

Memristors are thus dynamical resistive devices, distinguished from standard resistors by their ability to encode past signals. They naturally arise in nanoscale physics and have been proposed for neuromorphic architectures, unconventional computation, and optimization tasks.\footnote{M.~Di Ventra and Y.~V.~Pershin, ``Memcomputing: a computing paradigm to store and process information on the same physical platform,'' \emph{Nature Phys.} 9:200--202 (2013).}

\subsection{Network representation and projectors}

When many memristors are connected into a circuit defined by a graph $\mathcal{G}$, Kirchhoff’s laws impose constraints on possible voltages and currents. As we saw in Section~\ref{sec:projectors}, these constraints can be compactly expressed using projector operators:
\begin{equation}
   \Omega_A = I - B(B^{\top}B)^{-1}B^{\top}, \qquad 
   \Omega_B = B(B^{\top}B)^{-1}B^{\top},
\end{equation}
where $B$ is the incidence matrix of $\mathcal{G}$. The operator $\Omega_A$ projects edge variables onto the cycle space (enforcing Kirchhoff’s current law), while $\Omega_B$ projects onto the cut space (enforcing Kirchhoff’s voltage law). They satisfy $\Omega_A+\Omega_B=I_m$, $\Omega_A^2=\Omega_A$, $\Omega_B^2=\Omega_B$, and have spectra supported on $\{0,1\}$, with multiplicities determined by the dimensions of the cycle and cut spaces.

The decomposition of edge space into cycles and cuts is central for memristive dynamics: memory variables live on edges, but their admissible evolutions are constrained to respect global conservation laws.

\subsection{Self-consistent memory dynamics}

Using this projector formalism, the dynamics of a memristive network driven by a source vector $\vec S$ can be written in compact form as\footnote{F.~Caravelli, F.~Sheldon, F.~Traversa, ``Global minimization via classical tunneling assisted by collective force field formation,'' \emph{Sci. Adv.}, 7(52):eabf7867 (2021).}
\begin{equation}
   \dot{\vec x} = \frac{1}{\beta} \,(I - \chi \Omega_A X)^{-1}\,\Omega_A \vec S - \alpha \vec x,
   \label{eq:memproj}
\end{equation}
where:
\begin{itemize}
   \item $\vec x=(x_1,\ldots,x_m)$ is the vector of internal memory states of the $m$ edges;
   \item $X=\mathrm{diag}(x_1,\ldots,x_m)$;
   \item $\chi=(R_{\mathrm{off}}-R_{\mathrm{on}})/R_{\mathrm{off}}$ measures the contrast between on/off resistances;
   \item $\Omega_A$ is the projector onto the cycle space, encoding current conservation.
\end{itemize}

Equation~\eqref{eq:memproj} is a nonlinear, self-consistent memory equation: the right-hand side depends on $\vec x$ both directly and through the resolvent-like inverse $(I - \chi \Omega_A X)^{-1}$. This coupling makes each memristor’s evolution dependent on the global state of the network. The structure is reminiscent of nonlocal feedback in spin systems or mean-field models, but here it arises directly from Kirchhoff’s constraints.

\subsection{Spectral and structural properties}

Several properties of Eq.~\eqref{eq:memproj} follow directly from the spectral nature of projector operators:
\begin{enumerate}
   \item \textbf{Stability bounds.} The resolvent structure $(I - \chi \Omega_A X)^{-1}$ ensures stability as long as $\|\chi \Omega_A X\|<1$, which can be understood in terms of Gershgorin bounds on the eigenvalues of $\Omega_A X$.
   \item \textbf{Collective coupling.} Since $\Omega_A$ has nontrivial kernel and image, the dynamics naturally split into cycle and cut modes. The feedback term couples all memory variables along cycles, producing collective effects such as avalanches, synchronization, and metastable relaxation.\footnote{Y.~V.~Pershin and M.~Di Ventra, ``Memory effects in complex materials and nanoscale systems,'' \emph{Adv. Phys.} 60(2):145--227 (2011).}
   \item \textbf{Conservation laws.} Kirchhoff’s current law appears as the projection $\Omega_A\vec S$, which filters external drives into cycle-consistent currents. Kirchhoff’s voltage law is enforced by the complementary projection $\Omega_B$.
   \item \textbf{Attractors and computation.} The nonlinear feedback makes the system relax toward global attractors in state space, which can correspond to approximate solutions of optimization problems. This motivates the study of memristive networks as physical analogues of constraint satisfaction solvers.
\end{enumerate}

\subsection{Physical interpretation}

Equation~\eqref{eq:memproj} shows that projector operators provide the natural language for memristive network dynamics:
\begin{itemize}
   \item \emph{Local memory with global consistency.} Each memristor updates its internal state according to the local current, but global consistency is imposed by projection onto the cycle space.
   \item \emph{Emergent computation.} Because the evolution couples all variables nonlocally, the system can exhibit emergent behaviors such as spontaneous organization, constraint satisfaction, or ``memcomputing'' capabilities.
   \item \emph{Spectral partition.} The spectral decomposition of $\Omega_A$ and $\Omega_B$ partitions the dynamics into cycle and cut modes, clarifying which degrees of freedom govern long-term memory and relaxation.
\end{itemize}

Memristive networks thus exemplify how projector operators, originally a linear-algebraic tool, acquire physical meaning in the analysis of nonlinear dynamical devices.\footnote{F. Caravelli, G. Milano, A. Stieg, C. Ricciardi, S. Brown, Z. Kuncic, Self-organizing memristive networks as physical computing systems, arXiv:2509.00747 (2025)}

\section{Epidemic spreading on networks}
{epidemic threshold}
{SIS model}
{SIR model}

A paradigmatic example of spectral methods in complex systems is the analysis of epidemic processes on networks.\footnote{R.~M.~Anderson and R.~M.~May, \emph{Infectious Diseases of Humans: Dynamics and Control}, Oxford University Press (1992).}  
The central question is: \emph{given a contact network and disease parameters, when does an epidemic outbreak occur?}  
Mathematically, this is encoded in the \emph{epidemic threshold} --- the critical condition separating extinction of infection from sustained spread.

\subsection{SIS and SIR models on graphs}

Consider a population structured as a graph $G=(V,E)$ with adjacency matrix $A$.  
Each node represents an individual, and each edge a possible transmission channel.  
Two classical compartmental models are:
\begin{itemize}
   \item \textbf{SIS model} (susceptible $\to$ infected $\to$ susceptible): infected individuals transmit disease to neighbors at rate $\beta$, and recover at rate $\mu$.
   \item \textbf{SIR model} (susceptible $\to$ infected $\to$ recovered): individuals transmit at rate $\beta$ but, once recovered, remain immune forever.
\end{itemize}

At the mean-field level, the probability $p_i(t)$ that node $i$ is infected evolves according to
\begin{equation}
   \dot p_i(t) = -\mu p_i(t) + \beta (1-p_i(t)) \sum_j A_{ij} p_j(t).
   \label{eq:sis}
\end{equation}
Linearizing near the disease-free equilibrium $p_i=0$ gives
\begin{equation}
   \dot{\vec p}(t) \approx \left(\beta A - \mu I\right)\vec p(t).
\end{equation}

\subsection{Spectral threshold condition}

By Perron–Frobenius theory (Sec.~\ref{sec:perron}), the growth rate is governed by the largest eigenvalue $\lambda_1$ of $A$.  
Thus the epidemic threshold is
\begin{equation}
   \frac{\beta}{\mu} < \frac{1}{\lambda_1}.
   \label{eq:epidemic_threshold}
\end{equation}
If $\beta/\mu$ exceeds $1/\lambda_1$, the disease-free state becomes unstable and the epidemic can spread through the network.\footnote{Y.~Wang, D.~Chakrabarti, C.~Wang, and C.~Faloutsos, ``Epidemic spreading in real networks: an eigenvalue viewpoint,'' \emph{Proc. 22nd International Symposium on Reliable Distributed Systems}, 2003.}  
This result is striking: the epidemic threshold depends only on the \emph{spectral radius} of the adjacency matrix, not on local details.

\paragraph{Gershgorin intuition.}
A heuristic bound follows from Gershgorin’s theorem (Sec.~\ref{sec:gerschgorin}): each eigenvalue lies within a disk centered at zero with radius equal to the degree of a node.  
Thus $\lambda_1 \leq \max_i d_i$, where $d_i$ is the degree of node $i$.  
This recovers the intuitive statement that highly connected ``hubs'' make epidemics easier to sustain.

\paragraph{Resolvent interpretation.}
The resolvent expansion (Sec.~\ref{sec:resolvent}) provides another viewpoint.  
Near the threshold, the expected number of infections generated by one initial case can be expressed as a walk sum:
\begin{equation}
   \sum_{k=0}^\infty \left(\frac{\beta}{\mu}\right)^k A^k = (I - \tfrac{\beta}{\mu} A)^{-1},
\end{equation}
valid when $\beta/\mu < 1/\lambda_1$.  
Here the resolvent acts as the generating function of infection paths, and divergence at $\beta/\mu=1/\lambda_1$ marks the epidemic transition.

\subsection{Extensions and applications}

The spectral threshold condition has been extended to weighted, directed, and temporal networks.  
For directed graphs, the relevant $\lambda_1$ is that of the nonnegative adjacency or transmission matrix.  
For temporal networks, the effective adjacency is replaced by time-aggregated or supra-Laplacian operators.\footnote{A.~Barrat, M.~Barthélemy, and A.~Vespignani, \emph{Dynamical Processes on Complex Networks}, Cambridge University Press (2008).}  

Epidemic thresholds play a key role in public health and cyber-security.  
For example, in immunization strategies, removing nodes that maximize the reduction in $\lambda_1$ (e.g. hubs) is more efficient than random vaccination.\footnote{R.~Pastor-Satorras and A.~Vespignani, ``Epidemic spreading in scale-free networks,'' \emph{Phys. Rev. Lett.} \textbf{86}, 3200 (2001).}  
Thus, the spectral radius serves as a unifying control parameter for contagion in biological, social, and technological systems.

Epidemic spreading on networks illustrates how Perron–Frobenius eigenvalues, Gershgorin bounds, and resolvent expansions converge to explain a single phenomenon: the existence of a sharp epidemic threshold controlled by the spectral radius of the contact network.

\section{Synchronization phenomena and the Master Stability Function}
{synchronization}
{master stability function}
{Pecora--Carroll}

Synchronization --- the tendency of coupled dynamical units to evolve in unison --- is one of the most striking emergent behaviors in complex systems, from fireflies flashing together to power grids operating in phase.  
The modern mathematical analysis of synchronization rests on spectral methods, in particular the eigenvalues of the Laplacian matrix.  
A unifying framework is provided by the \emph{master stability function} (MSF), introduced by Pecora and Carroll in the 1990s.\footnote{L.~M.~Pecora and T.~L.~Carroll, ``Master Stability Functions for Synchronized Coupled Systems,'' \emph{Phys. Rev. Lett.} \textbf{80}, 2109 (1998).}

\subsection{Coupled oscillators on networks}

Consider $n$ identical dynamical systems (oscillators) with state $\vec x_i(t)\in\mathbb{R}^d$, coupled along the edges of a graph with Laplacian $L$.  
The equations of motion are
\begin{equation}
   \dot{\vec x}_i = F(\vec x_i) - \sigma \sum_{j} L_{ij} H(\vec x_j),
   \label{eq:coupled_oscillators}
\end{equation}
where $F$ describes the local dynamics, $H$ the coupling function, and $\sigma>0$ the coupling strength.  
Examples include Kuramoto phase oscillators, chaotic Rössler systems, or neural oscillators.

The fully synchronized state is defined by $\vec x_i(t)=\vec s(t)$ for all $i$, where $\vec s(t)$ satisfies the isolated system $\dot{\vec s}=F(\vec s)$.

\subsection{Linear stability and Laplacian spectrum}

To analyze stability of synchronization, linearize Eq.~\eqref{eq:coupled_oscillators} about the synchronized state.  
Perturbations $\delta \vec x_i = \vec x_i - \vec s$ evolve according to
\begin{equation}
   \dot{\delta \vec X} = \left[ I_n \otimes DF(\vec s) - \sigma L \otimes DH(\vec s) \right] \delta \vec X,
\end{equation}
where $\otimes$ denotes the Kronecker product, $DF$ and $DH$ are Jacobians of $F$ and $H$, and $\delta \vec X$ stacks all perturbations.

Because $L$ is symmetric (for undirected networks), it admits an orthogonal eigendecomposition $L = V \Lambda V^{\top}$ with eigenvalues $0=\lambda_1 \leq \lambda_2 \leq \cdots \leq \lambda_n$.  
Transforming to Laplacian eigenmodes decouples the system into $n$ blocks:
\begin{equation}
   \dot{\eta}_k = \left[ DF(\vec s) - \sigma \lambda_k DH(\vec s) \right] \eta_k,
   \qquad k=1,\dots,n.
   \label{eq:variational_block}
\end{equation}

\subsection{The Master Stability Function}

Equation \eqref{eq:variational_block} shows that each Laplacian mode evolves independently, with effective coupling parameter $\alpha = \sigma \lambda_k$.  
The \emph{master stability function} $\Lambda(\alpha)$ is defined as the largest Lyapunov exponent of
\begin{equation}
   \dot{\eta} = \left[ DF(\vec s) - \alpha DH(\vec s) \right] \eta.
   \label{eq:msf}
\end{equation}
Thus $\Lambda(\alpha)$ depends only on the node dynamics $(F,H)$, not on network structure.  
Network effects enter solely through the Laplacian eigenvalues $\{\lambda_k\}$, which rescale $\sigma$.

\paragraph{Stability criterion.}
Synchronization is stable if $\Lambda(\sigma \lambda_k)<0$ for all nonzero eigenmodes $k=2,\dots,n$.  
The zero eigenvalue $\lambda_1=0$ corresponds to perturbations along the synchronization manifold and is neutral by construction.  
Therefore, stability requires the rescaled eigenvalues $\sigma \lambda_k$ to lie entirely within the interval of $\alpha$ where $\Lambda(\alpha)<0$.

\subsection{Spectral implications}

The MSF framework reveals a direct link between Laplacian spectra and synchronization:
\begin{itemize}
   \item If $\Lambda(\alpha)<0$ for all $\alpha>\alpha_c$, then synchronization requires $\sigma \lambda_2 > \alpha_c$.  
   Hence $\lambda_2$ (the algebraic connectivity; see Sec.~\ref{sec:laplacian_spectrum}) sets the onset of synchronizability.
   \item If $\Lambda(\alpha)<0$ only within a finite interval $(\alpha_{\min},\alpha_{\max})$, then stability requires
   \[
      \frac{\lambda_n}{\lambda_2} < \frac{\alpha_{\max}}{\alpha_{\min}},
   \]
   i.e. a bound on the Laplacian eigenratio.  
   This explains why heterogeneous networks (large $\lambda_n/\lambda_2$) are often harder to synchronize.
\end{itemize}

\paragraph{Connection to Floquet theory.}
The MSF is computed from the variational equation \eqref{eq:msf}, which has time-dependent coefficients if $\vec s(t)$ is periodic or chaotic.  
Thus its Lyapunov exponents are determined using Floquet theory (Sec.~\ref{sec:stability}).  
In this way, synchronization theory is a direct application of Floquet methods to high-dimensional, networked systems.

\subsection{Applications}

The master stability framework has been applied broadly:
\begin{itemize}
   \item Power grid stability, where generators must remain synchronized across a transmission network.\footnote{P.~J.~Menck, J.~Heitzig, J.~Kurths, and H.~J.~Schewe, ``How dead ends undermine power grid stability,'' \emph{Nature Communications} \textbf{5}, 3969 (2014).}
   \item Neural dynamics, where synchronization and desynchronization underlie cognitive function and disorders.\footnote{S.~Boccaletti, V.~Latora, Y.~Moreno, M.~Chavez, and D.-U.~Hwang, ``Complex networks: Structure and dynamics,'' \emph{Physics Reports} \textbf{424}, 175--308 (2006).}
   \item Coupled lasers and electronic circuits, where synchronization enables secure communications.\footnote{L.~M.~Pecora, T.~L.~Carroll, G.~A.~Johnson, D.~J.~Mar, and J.~F.~Heagy, ``Fundamentals of synchronization in chaotic systems, concepts, and applications,'' \emph{Chaos} \textbf{7}, 520--543 (1997).}
\end{itemize}

\bigskip
The master stability function demonstrates how Laplacian eigenvalues control synchronization stability in networks.  
It generalizes the role of spectra in linear stability analysis (Floquet theory) to nonlinear, high-dimensional, and networked settings, providing a universal spectral criterion for collective behavior.

\section{Financial stability and systemic risk}
{financial stability}
{systemic risk}
{DebtRank}
\label{sec:debtrank}
Financial networks provide another arena where spectral methods are central.  
Banks and financial institutions are linked by interbank loans and derivative exposures, forming a weighted, directed network of liabilities.  
The collapse of a single node can propagate losses to its creditors, triggering cascades of defaults.  
Understanding the \emph{stability} of such systems is therefore a core problem in systemic risk analysis.\footnote{See D.~Garlaschelli, S.~Battiston, M.~Riccaboni, and F.~Vega-Redondo, ``The architecture of complex weighted networks,'' \emph{Proc. Natl. Acad. Sci. USA} \textbf{102}, 7794--7799 (2005).}

\subsection{Interbank liability matrices}

Let $L$ denote the \emph{liability matrix}, where $L_{ij}$ is the exposure of bank $i$ to bank $j$ (i.e. how much $i$ owes to $j$).  
A shock that reduces the equity of bank $i$ may propagate to its creditors proportional to $L_{ij}$.  
Normalizing by total assets produces a matrix $W$ of relative exposures:
\begin{equation}
   W_{ij} = \frac{L_{ij}}{E_j},
\end{equation}
where $E_j$ is the equity of institution $j$.  
Thus $W_{ij}$ measures the fraction of $j$'s capital that would be wiped out if $i$ defaults.

\paragraph{Spectral condition.}
Contagion dynamics in its simplest linear form can be expressed as
\begin{equation}
   \vec h^{(t+1)} = W^{\top} \vec h^{(t)},
\end{equation}
where $h_i^{(t)}$ is the cumulative distress of bank $i$ at time $t$.  
By Perron–Frobenius theory (Sec.~\ref{sec:perron}), the long-term amplification of shocks is governed by the spectral radius $\rho(W)$.  
If $\rho(W)<1$, shocks eventually dissipate; if $\rho(W)>1$, small shocks can be amplified into systemic crises.\footnote{S.~Battiston, D.~Delli Gatti, M.~Gallegati, B.~Greenwald, and J.~E.~Stiglitz, ``Liaisons dangereuses: Increasing connectivity, risk sharing, and systemic risk,'' \emph{Journal of Economic Dynamics and Control} \textbf{36}, 1121--1141 (2012).}

This condition is mathematically parallel to the Leontief model of economics (Sec.~\ref{sec:leontief}): in both cases the resolvent $(I-W)^{-1}$ determines whether activity converges (stable economy) or diverges (crisis).

\subsection{DebtRank: systemic impact measure}

To go beyond the simple threshold $\rho(W)=1$, Battiston \emph{et al.} proposed \emph{DebtRank}, a recursive measure of systemic importance.\footnote{S.~Battiston, M.~Puliga, R.~Kaushik, P.~Tasca, and G.~Caldarelli, ``DebtRank: Too Central to Fail? Financial Networks, the FED and Systemic Risk,'' \emph{Scientific Reports} \textbf{2}, 541 (2012).}

Define the state of each bank $i$ by a variable $h_i \in [0,1]$, representing the fraction of its equity that is distressed.  
Starting from an initial shock $\vec h^{(0)}$, distress propagates iteratively according to
\begin{equation}
   h_j^{(t+1)} = \min\left\{1,\, h_j^{(t)} + \sum_i W_{ij} \Delta h_i^{(t)}\right\},
\end{equation}
where $\Delta h_i^{(t)} = h_i^{(t)} - h_i^{(t-1)}$ is the incremental distress transmitted at step $t$.  
The process continues until no new distress propagates.

The \emph{DebtRank} of a set of initially shocked nodes $S$ is defined as the total fraction of equity in the system that eventually becomes distressed:
\begin{equation}
   R(S) = \sum_j h_j^{(\infty)} v_j,
\end{equation}
where $v_j$ is the economic value (e.g. assets) of institution $j$.  
Thus $R(S)$ quantifies the systemic importance of a bank or group of banks, not only their own size but their ability to transmit shocks.

\subsection{Spectral interpretation and parallels}

From a spectral point of view, DebtRank is a nonlinear extension of resolvent-based contagion.  
Linearizing around small shocks, the recursion reduces to
\[
   \vec h^{(\infty)} \approx (I - W^{\top})^{-1} \vec h^{(0)},
\]
so that the resolvent determines amplification, and $\rho(W)$ is the critical parameter.  
In practice, nonlinear saturation (the $\min\{1,\cdot\}$) ensures that losses are capped at 100\% of equity, preventing unbounded growth.

\paragraph{Parallel with Leontief economics.}
This picture is the financial dual of Leontief input–output analysis: in economics, $(I-A)^{-1}$ encodes production multipliers, while in finance, $(I-W)^{-1}$ encodes distress multipliers.  
In both cases, the spectral radius decides whether multipliers converge.  
The difference is interpretive: in one case amplification is productive (economic growth), in the other destructive (systemic collapse).

\subsection{Applications}

DebtRank has been applied to real financial networks, such as the US Federal Reserve’s large-value payment system, to identify institutions that are ``too central to fail.''  
It has influenced discussions of macroprudential regulation, capital buffers, and systemic risk monitoring.\footnote{S. Battiston, D. Delli Gatti, M. Gallegati, B. Greenwald, J. E. Stiglitz,, ``Default cascades: When does risk diversification increase stability?'' \emph{Journal of Financial Stability} \textbf{8}, 138--149 (2012).}

Financial stability can be understood as a spectral problem on interbank liability matrices: the spectral radius governs solvency thresholds, and DebtRank provides a nonlinear extension that quantifies systemic impact.  
This is the financial analogue of the Leontief framework, with the resolvent acting as a contagion operator for crises.

\section{Reservoir Computing: Dynamics Meets Spectra}
{Reservoir computing}
{Echo state networks}
{Spectral radius and stability}

Reservoir computing (RC) is a paradigm in machine learning and nonlinear signal processing that exploits the dynamics of large recurrent networks.  
Unlike standard neural networks, only the output layer is trained, while the internal recurrent network---called the \emph{reservoir}---is left fixed after initialization.  
This leads to an extremely efficient training procedure, and, perhaps more importantly, highlights the role of spectral properties of the reservoir in shaping its computational power.
\subsection{Basic idea: Echo State Networks}
{Echo state networks}
{Reservoir computing}

Reservoir computing was originally introduced in the early 2000s through two closely related models: 
the \emph{Echo State Network} (ESN) of Jaeger\footnote{H. Jaeger, ``The echo state approach to analysing and training recurrent neural networks,'' GMD Report 148, 2001.} 
and the \emph{Liquid State Machine} (LSM) of Maass et al.\footnote{W. Maass, T. Natschläger, and H. Markram, ``Real-time computing without stable states: A new framework for neural computation based on perturbations,'' Neural Computation 14 (2002): 2531–2560.}.  
Both models share the same central idea: rather than training all the parameters of a recurrent neural network, one fixes the internal recurrent network (the \emph{reservoir}) at random, and trains only the linear readout weights.  

\paragraph{Architecture.}  
An ESN is composed of three layers:
\begin{enumerate}
  \item \textbf{Input layer.}  
  An external signal $u(t) \in \mathbb{R}^{d_{in}}$ is injected into the network via an input weight matrix $W_{\mathrm{in}} \in \mathbb{R}^{N \times d_{in}}$, which determines how the input is mapped into the high-dimensional reservoir state space.  
  In practice, $W_{\mathrm{in}}$ is usually sparse and randomly initialized, with weights scaled by an input scaling parameter.
  
  \item \textbf{Reservoir (recurrent layer).}  
  The heart of the ESN is the reservoir, a recurrent dynamical system of size $N$, described by its state vector $\vec x(t) \in \mathbb{R}^N$.  
  Its evolution is given by
  \begin{equation}
     \vec x(t+1) = f\!\left(W \vec x(t) + W_{\mathrm{in}} u(t) + \vec b\right),
     \label{eq:esndyn}
  \end{equation}
  where $W \in \mathbb{R}^{N \times N}$ is the reservoir weight matrix (fixed after initialization), $f(\cdot)$ is a nonlinear activation function (typically $\tanh$ or sigmoid), and $\vec b$ is an optional bias term.  
  The recurrent weights $W$ are generally sparse and initialized randomly, but their spectral radius $\rho(W)$ (the maximum modulus of their eigenvalues) is a crucial hyperparameter that controls stability and memory (see below).
  
  \item \textbf{Readout layer.}  
  The output is a linear function of the reservoir state,
  \begin{equation}
     \hat y(t) = W_{\mathrm{out}} \vec x(t),
  \end{equation}
  where $W_{\mathrm{out}} \in \mathbb{R}^{d_{out} \times N}$ are the only trainable parameters of the network.  
  Training reduces to solving a linear regression (often with ridge regularization), which can be performed efficiently by closed-form least-squares or gradient descent methods.
\end{enumerate}

\paragraph{Echo state property.}  
For the ESN to work reliably, the reservoir dynamics must satisfy the \emph{echo state property (ESP)}: the effect of any initial condition $\vec x(0)$ must vanish asymptotically, so that $\vec x(t)$ is uniquely determined by the input history $\{u(\tau)\}_{\tau \leq t}$.  
Formally, this requires that the homogeneous system $\vec x(t+1) = f(W \vec x(t))$ be asymptotically stable.  
A sufficient (but not necessary) condition is that the spectral radius of $W$ satisfies
\[
   \rho(W) < 1,
\]
which ensures that perturbations decay over time.  
In practice, $\rho(W)$ is often tuned to be close to but slightly less than $1$, to balance long memory with stability.  

\paragraph{Memory and expressivity.}  
Because the reservoir is high-dimensional and recurrent, its states encode a nonlinear fading memory of past inputs.  
Inputs injected at time $t-k$ are stored implicitly in the coordinates of $\vec x(t)$, with the rate of decay controlled by $\rho(W)$.  
This makes ESNs particularly well-suited for tasks involving temporal dependencies, such as speech recognition, chaotic time series prediction, or control problems.  

\paragraph{Training.}  
Unlike traditional recurrent neural networks, where training involves backpropagation through time, the ESN requires training only in the output layer:
\begin{equation}
   W_{\mathrm{out}} = \arg \min_{W} \sum_t \| y(t) - W \vec x(t) \|^2 + \lambda \|W\|^2,
\end{equation}
where $\lambda$ is a regularization parameter.  
This leads to extremely fast training, since the problem reduces to standard linear regression.

\paragraph{Spectral connection.}  
From a spectral perspective, the ESN can be seen as an input-driven dynamical system filtered through the powers of $W$.  
Expanding the recurrence iteratively, and linearizing $f$ near the origin, one finds
\[
   \vec x(t) \approx \sum_{k=0}^\infty W^k W_{\mathrm{in}} u(t-k),
\]
which resembles the Neumann expansion of the resolvent $(I - W)^{-1}$ (see Sec.~\ref{sec:resolvent}).  
Hence the eigenvalues of $W$ determine the reservoir’s memory depth, richness, and stability. Non-normal reservoirs (Sec.~\ref{sec:pseudospectra}) can also generate transient amplification, which has been shown to enhance computational performance in practice.

The Echo State Network is a paradigm where learning is shifted entirely to a simple spectral filtering problem: a fixed recurrent matrix $W$ transforms inputs into a rich dynamical basis, and spectral analysis (e.g., eigenvalues, resolvent expansions, pseudospectra) predicts and explains the network’s computational capabilities.

\subsection{Spectral viewpoint}
The reservoir can be understood as a high-dimensional linear system perturbed by nonlinearities.  
If $f$ is approximately linear around the origin, the update rule is
\begin{equation}
   \vec x(t+1) \approx W \vec x(t) + W_{\mathrm{in}} u(t).
   \label{eq:reservoirlin}
\end{equation}
Equation \eqref{eq:reservoirlin} shows that the expressive power and stability of the reservoir are governed by the eigenvalues of $W$:
\begin{itemize}
  \item The \emph{echo state property}---which ensures that the reservoir forgets initial conditions and is driven uniquely by inputs---holds if the spectral radius $\rho(W)<1$.  
  This condition guarantees asymptotic stability of the homogeneous system $\vec x(t+1)=W\vec x(t)$.
  \item The \emph{memory capacity} of the reservoir depends on how close $\rho(W)$ is to unity: values of $\rho(W)$ just below $1$ allow information to persist over long times, while smaller values quickly erase memory.
  \item Non-normality of $W$ (Sec.~\ref{sec:matrixsym}) plays a key role: even if $\rho(W)<1$, strong transient growth may occur due to pseudospectral effects, enriching the dynamics and improving computational capability.
\end{itemize}

From a spectral perspective, RC can be interpreted as follows:
\begin{enumerate}
  \item The reservoir state $\vec x(t)$ is obtained by filtering the input signal $u(t)$ through the resolvent of $W$:
  \begin{equation}
     \vec x(t) = \sum_{k=0}^{\infty} W^k W_{\mathrm{in}} u(t-k),
  \end{equation}
  provided $\rho(W)<1$. This resembles the Neumann expansion of the resolvent (Sec.~\ref{sec:resolvent}).
  \item The spectrum of $W$ thus controls the effective memory kernel of the reservoir. Eigenvalues near the unit circle correspond to slow decays and long memory, while small eigenvalues correspond to fast forgetting.
  \item Echo state networks can therefore be seen as stochastic approximations to \emph{Koopman-type expansions} (Sec.~\ref{sec:koopman}), where nonlinear dynamics are embedded into a linear high-dimensional spectral representation.
\end{enumerate}

\subsection{Why spectra matter in practice}
The spectral approach provides both intuition and design principles:
\begin{itemize}
  \item Reservoirs with $\rho(W)\approx 1$ achieve a balance between stability and richness, often leading to optimal predictive performance.\footnote{See H.~Jaeger, ``The echo state approach to analysing and training recurrent neural networks,'' GMD Report 148 (2001).}
  \item Spectral shaping of $W$ (e.g.\ using orthogonal, unitary, or structured random matrices) has been shown to enhance efficiency and reduce training variance.\footnote{M.~Lukoševičius and H.~Jaeger, ``Reservoir computing approaches to recurrent neural network training,'' Computer Science Review 3 (2009).}
  \item Non-normal spectra allow the reservoir to exploit transient amplification (Sec.~\ref{sec:pseudospectra}), an effect parallel to those observed in fluid dynamics and control theory.
\end{itemize}

\medskip
Reservoir computing is therefore an excellent modern illustration of how spectral properties govern nonlinear computation: by analyzing the eigenvalues, pseudospectra, and resolvents of the reservoir weight matrix, one can predict and tune the memory, stability, and computational richness of the system.

\section{Rate matrices, thermodynamics, and generalized inverses}
{thermodynamics}{Markov chains}{Drazin inverse}

The language of linear algebra and spectral methods finds a natural and powerful application in the study of
\emph{complex systems} --- ranging from biochemical reaction networks and ecological systems to economic
dynamics and information processing. In such settings, one often deals with stochastic dynamics, typically
formulated in terms of \emph{rate matrices} (or generators) of continuous-time Markov processes. These rate
matrices are generally singular, which makes the ordinary inverse inapplicable. Instead, one is naturally led
to the use of generalized inverses, particularly the \emph{Drazin inverse}, which isolates stationary modes
and encodes relaxation dynamics.\footnote{See G.~E.~Crooks, \emph{On the Drazin inverse of the rate matrix},
Tech.~Note 011 (2018).}

\paragraph{Rate matrices and Markov dynamics.}
Consider a finite-state continuous-time Markov process. Its dynamics are governed by a rate matrix
$R\in\mathbb{R}^{n\times n}$ with entries
\[
R_{ij}\geq 0 \quad (i\neq j), \qquad \sum_j R_{ij}=0.
\]
The time evolution of the probability vector $p(t)$ is given by the master equation
\begin{equation}
\frac{d}{dt}p(t) = R p(t).
\end{equation}
Because each column of $R$ sums to zero, $R$ has at least one zero eigenvalue. For ergodic processes, there
is exactly one stationary distribution $\pi$, i.e.
\[
R\pi = 0.
\]
Thus $R$ is singular, and cannot be inverted in the usual sense.

\paragraph{The role of the Drazin inverse.}
The Drazin inverse $R^D$ of the rate matrix provides precisely the object needed to handle such cases. As
we have seen in Section~\ref{sec:drazin-inverse}, the Drazin inverse inverts $R$ on its non-null eigenspaces,
while annihilating the stationary mode:
\[
R^D R = R R^D = I - \pi \vec 1^\top.
\]
This property allows us to express deviations from stationarity in terms of $R^D$, since for any probability
distribution $p$, we have
\[
R^D R p = p - \pi.
\]
In other words, the operator $R^D R$ projects $p$ onto its deviation from equilibrium.\footnote{Crooks (2018),
Sec.~2.1.} This property already illustrates why the Drazin inverse is the natural generalized inverse for
stochastic generators.

\paragraph{Relaxation times and spectral gaps.}
The eigenvalues of $R$ all have nonpositive real part, with one zero eigenvalue. The spectral gap
$\Delta = -\max\{\Re \lambda_i : \lambda_i\neq 0\}$ controls the asymptotic rate of relaxation back to
equilibrium. The Drazin inverse encodes this relaxation quantitatively: for example, if $\delta p(0)=p(0)-\pi$,
then
\[
\int_0^\infty \delta p(t)\,dt = R^D \delta p(0).
\]
Thus $R^D$ maps deviations from stationarity to their \emph{integrated relaxation}, providing effective
relaxation times for each mode.\footnote{See also Mandal \& Jarzynski, \emph{J.~Stat.~Mech.}, 063204 (2016).}

\paragraph{Thermodynamic response and friction.}
One of the most striking applications of $R^D$ is in nonequilibrium thermodynamics. When a system is driven
slowly in time by external parameters $\lambda(t)$, the mean excess work can be written as
\[
\langle \beta W_{\mathrm{ex}}\rangle = \dot\lambda^\top \zeta \,\dot\lambda,
\]
where the \emph{friction tensor} $\zeta$ is given by
\[
\zeta = (d_\lambda \ln \pi)^\top (-R^D) (d_\lambda \ln \pi)\,\pi.
\]
This formula shows that dissipated work is governed by the interplay of stationary fluctuations
($d_\lambda \ln \pi$) and dynamical relaxation ($R^D$).\footnote{Crooks (2018), Sec.~3.} In this way,
generalized inverses connect to information geometry, since the Fisher information matrix of the stationary
distribution also appears in the expression for $\zeta$.

\paragraph{Currents and fluctuations.}
In nonequilibrium steady states, one is often interested in probability currents $J$, which can be defined by
a skew-symmetric incidence matrix $F$ as $J=R\odot F$, where $\odot$ denotes the Hadamard product.
The Drazin inverse provides a compact expression for current autocorrelations:
\[
C_{JJ} = -\vec 1^\top (R\odot F) R^D (R\odot F)\pi.
\]
Thus correlations of fluctuating currents are expressed directly in terms of $R^D$ and the stationary
distribution. This mirrors the structure of Green–Kubo relations in nonequilibrium statistical mechanics,
with the Drazin inverse replacing ordinary inverses.

\paragraph{Projection operators and thermodynamic geometry.}
As shown in Section~\ref{sec:drazin-inverse}, the operator $P_0 = I - R R^D$ projects onto the stationary
distribution, while $P_1 = R R^D$ projects onto the transient subspace. This decomposition allows one to
separate thermodynamic quantities into stationary and transient parts. For example, the relative entropy
between instantaneous and stationary distributions can be written compactly using $P_1$ and $R^D$,
highlighting the geometric structure of thermodynamics near equilibrium.\footnote{See G. Crooks, (2018), Sec.~3.1.}

\paragraph{Broader perspective.}
The appearance of the Drazin inverse in thermodynamics and stochastic processes underscores the unity
of linear algebra and statistical physics. Just as eigenvalues and spectra determine the stability and dynamics
of linear systems, generalized inverses describe what happens when singularity is inevitable: the presence of
conserved quantities, stationary states, or irreversible modes. By projecting out these modes, the Drazin
inverse provides the correct algebraic framework for describing relaxation, fluctuations, and dissipation in
complex systems.

To conclude, the Drazin inverse is not only a mathematical curiosity, but also a central tool in the physics
of complex systems: it governs how systems relax, how they respond, and how they dissipate energy. It is
precisely in these contexts that the abstract theory of generalized inverses finds some of its most vivid
applications.
\section{Principal Component Analysis (PCA) and Spectral Methods}
\index{keu}{Principal Component Analysis (PCA)}

Principal Component Analysis (PCA) is one of the central techniques in data
analysis and machine learning. It provides a way to reduce the dimensionality
of a dataset while retaining as much of the variance (information) as possible. 
From the viewpoint of spectral methods, PCA is a direct application of the
eigendecomposition of covariance matrices.

\subsection{Setup and formulation}
Let $X \in \mathbb{R}^{n \times p}$ be a data matrix, with $n$ samples and $p$
features, assumed to be centered so that $\frac{1}{n}\sum_{i=1}^n x_i = 0$. The
sample covariance matrix is
\begin{equation}
    \Sigma = \frac{1}{n} X^\top X.
\end{equation}
PCA seeks orthogonal directions $u_1, \dots, u_k \in \mathbb{R}^p$ (the
principal components) that maximize the projected variance:
\begin{equation}
    u_1 = \arg\max_{\|u\|=1} u^\top \Sigma u.
\end{equation}
By the Rayleigh--Ritz theorem, the solution is given by the leading eigenvector
of $\Sigma$, with eigenvalue $\lambda_1$ equal to the maximum variance along
that direction. Iterating this procedure yields the subsequent components.

\subsection{Spectral perspective}
From a spectral point of view:
\begin{itemize}
    \item The eigenvalues $\lambda_1 \geq \lambda_2 \geq \cdots \geq \lambda_p$
    represent the variances captured by successive principal components.
    \item The cumulative variance ratio
    \[
    R_k = \frac{\sum_{i=1}^k \lambda_i}{\sum_{j=1}^p \lambda_j}
    \]
    quantifies how much of the total variability is retained by the first $k$
    components.
    \item The low-dimensional representation of the data is obtained as
    $Y = X U_k$, where $U_k = (u_1,\dots,u_k)$.
\end{itemize}
Thus, PCA reduces to an eigenvalue problem on the covariance matrix, making
it a quintessential spectral method.

\subsection{Connections and generalizations}
Several related techniques build directly on PCA:
\begin{enumerate}
    \item \textbf{Singular Value Decomposition (SVD).} Applying the SVD
    $X = U \Sigma V^\top$ directly provides the PCA decomposition. The right
    singular vectors $V$ are the principal components, and the singular values
    squared correspond to eigenvalues of the covariance matrix.
    \item \textbf{Factor Analysis.} A probabilistic generalization, in which the
    observed data is modeled as a linear combination of latent factors plus
    noise. PCA can be seen as the maximum likelihood estimator in the noiseless
    Gaussian case.
    \item \textbf{Independent Component Analysis (ICA).} Goes beyond PCA by
    looking for statistically independent components (not just uncorrelated),
    using higher-order cumulants rather than covariance.
    \item \textbf{Kernel PCA.} Nonlinear extension: given a feature map
    $\phi(x)$ into a Hilbert space, PCA is performed on the Gram matrix
    $K_{ij} = \langle \phi(x_i), \phi(x_j)\rangle$. By the spectral theorem, this
    yields nonlinear embeddings of the data.
    \item \textbf{Nonnegative Matrix Factorization (NMF).} Provides additive
    decompositions $X \approx WH$ with nonnegativity constraints, often more
    interpretable in applications such as topic modeling or image analysis.
\end{enumerate}

\subsection{Applications in inference and complex systems}
PCA and its spectral generalizations enter a variety of inference problems:
\begin{itemize}
    \item \emph{Noise reduction and signal denoising.} By retaining only the top
    eigenmodes, one effectively filters noise dominated by low-variance
    directions.
    \item \emph{Feature extraction and visualization.} PCA projects high
    dimensional data into two or three dimensions for visualization.
    \item \emph{Complex systems.} In ecology and finance, PCA is used to extract
    dominant modes of variability in correlation matrices of species or assets.
    \item \emph{Network science.} Spectral embeddings (e.g. Laplacian
    eigenmaps) are closely related to PCA, with applications to community
    detection and dimensionality reduction of graphs.
\end{itemize}

PCA epitomizes how spectral methods solve inference problems: by reducing
complex high-dimensional data to a low-rank approximation, interpretable
through eigenvalue spectra. Its connection to the SVD shows that inference can
often be recast as a problem of matrix factorization. Generalizations such as
kernel PCA or ICA highlight how the same spectral principles extend to
nonlinear and higher-order settings, expanding the toolbox of inference in
machine learning and complex systems.

\section{Matrix inversion in statistical inference}
{inference}{Gaussian models}{precision matrix}{graphical models}{generalized inverses}

Matrix inversion is ubiquitous in statistical inference. In Gaussian models, the \emph{precision}
matrix $\Sigma$ determines the geometry of fluctuations, while its inverse $C=\Sigma^{-1}$ is the
covariance. Estimators, posteriors, conditionals, regularizers, and information metrics all reduce
to solving linear systems or inverting structured matrices.\footnote{C.M.~Bishop, \emph{Pattern
Recognition and Machine Learning}, Springer (2006), Chs.~2--3.}  

\paragraph{Gaussian likelihood and the role of the precision.}
For zero-mean $x\in\mathbb{R}^n$ with precision $\Sigma\succ0$,
\[
p_\Sigma(x)=\frac{1}{(2\pi)^{n/2}(\det\Sigma)^{1/2}}\exp\!\Big(-\tfrac12 x^\top \Sigma x\Big),
\]
the log-likelihood for $M$ samples with empirical covariance $S=\tfrac1M\sum_k x^{(k)}x^{(k)\top}$ is
\[
\ell(\Sigma)=\frac{M}{2}\Big(\log\det\Sigma-\operatorname{Tr}(S\Sigma)\Big)+\text{const}.
\]
Setting $\nabla_\Sigma \ell=0$ gives the MLE condition $\Sigma^{-1}=S$, i.e.\ \emph{estimating the covariance
is matrix inversion}.\footnote{K.P.~Murphy, \emph{Machine Learning: A Probabilistic Perspective},
MIT Press (2012), \S4.3.} In high dimensions ($n\gg M$), $S$ is singular and one replaces the
inverse with a generalized inverse or a regularized inverse (see below).

\paragraph{Conditionals of a joint Gaussian.}
For a joint Gaussian 
\[
\begin{bmatrix}x\\y\end{bmatrix}\sim\mathcal{N}\!\left(
\begin{bmatrix}\mu_x\\\mu_y\end{bmatrix},\;
\begin{bmatrix}\Sigma_{xx}&\Sigma_{xy}\\\Sigma_{yx}&\Sigma_{yy}\end{bmatrix}\right),
\]
the conditional is $p(x\mid y)=\mathcal{N}(\mu_{x\mid y},\,\Sigma_{x\mid y})$ with
\begin{align}
\mu_{x\mid y}&=\mu_x+\Sigma_{xy}\Sigma_{yy}^{-1}(y-\mu_y),\\
\Sigma_{x\mid y}&=\Sigma_{xx}-\Sigma_{xy}\Sigma_{yy}^{-1}\Sigma_{yx}.
\end{align}
Thus inference reduces to inverting $\Sigma_{yy}$ (or solving linear systems involving it).\footnote{Bishop (2006), Eq.~2.116; Murphy (2012), \S2.3.3.}

\paragraph{Bayesian linear regression.}
With $y=Xw+\varepsilon$, $\varepsilon\sim\mathcal{N}(0,\beta^{-1}I)$, and Gaussian prior
$w\sim\mathcal{N}(m_0,\Lambda_0^{-1})$, the posterior is Gaussian with precision and mean
\[
\Lambda_n=\Lambda_0+\beta X^\top X,\qquad 
m_n=\Lambda_n^{-1}\big(\Lambda_0 m_0+\beta X^\top y\big).
\]
Again the core step is \emph{inverting a shifted Gram matrix}.\footnote{Bishop (2006), \S3.3.}  
The same matrix appears in ridge regression / Tikhonov regularization:
\[
\hat w=(X^\top X+\lambda I)^{-1}X^\top y,\quad \lambda>0,
\]
which stabilizes the inverse when $X^\top X$ is ill-conditioned.\footnote{A.E.~Hoerl and R.W.~Kennard, 
``Ridge Regression: Biased Estimation for Nonorthogonal Problems,'' \emph{Technometrics} 12 (1970) 55–67.}

\paragraph{Gaussian graphical models (sparse precision).}
In an undirected Gaussian graphical model, zeros in the precision $\Theta=\Sigma^{-1}$ encode conditional
independence. Given empirical $S$, the (penalized) MLE solves the convex program
\[
\max_{\Theta\succ0}\ \log\det\Theta-\operatorname{Tr}(S\Theta)-\lambda\|\Theta\|_1,
\]
whose solution is a sparse \emph{inverse covariance}.\footnote{O.~Banerjee, L.E.~Ghaoui, A.~d'Aspremont,
``Model Selection Through Sparse Maximum Likelihood Estimation,'' \emph{JMLR} 9 (2008) 485–516; 
J.~Friedman, T.~Hastie, R.~Tibshirani, ``Sparse inverse covariance estimation with the graphical lasso,''
\emph{Biostatistics} 9 (2008) 432–441.} Efficient solvers rely on Cholesky factorizations and the
matrix determinant lemma rather than forming $\Sigma$ explicitly.\footnote{S.~Boyd and L.~Vandenberghe,
\emph{Convex Optimization}, Cambridge (2004), \S7.5.}

\paragraph{Gaussian processes (GPs).}
Posterior predictions for GPs require solving systems with $(K+\sigma^2 I)$:
\[
m_\ast = k_\ast^\top (K+\sigma^2 I)^{-1}y, \qquad 
\mathrm{cov}_\ast = k_{\ast\ast}-k_\ast^\top (K+\sigma^2 I)^{-1} k_\ast,
\]
with $K_{ij}=k(x_i,x_j)$.\footnote{C.E.~Rasmussen and C.K.I.~Williams, 
\emph{Gaussian Processes for Machine Learning}, MIT Press (2006), Ch.~2.}  
Low-rank/inducing methods and the Woodbury identity are used to avoid dense $O(n^3)$ inversion.\footnote{M.~Seeger, C.K.I.~Williams, N.~Lawrence, 
``Fast Forward Selection to Speed up Sparse Gaussian Process Regression,'' \emph{AISTATS} (2003); 
M.~Titsias, ``Variational Learning of Inducing Variables,'' \emph{AISTATS} (2009).}

\paragraph{Variational inference and natural gradients.}
In mean-field/structured VI, one minimizes a KL or ELBO that contains terms like 
$\tfrac12(\operatorname{Tr}(S\Sigma)-\log\det\Sigma)$; gradients require $\Sigma^{-1}$ via 
$\partial_{\Sigma}\log\det\Sigma=(\Sigma^{-1})^\top$.\footnote{D.M.~Blei, A.~Kucukelbir, J.D.~McAuliffe, 
``Variational Inference: A Review for Statisticians,'' \emph{JASA} 112 (2017) 859–877.}  
Optimization with \emph{natural gradients} preconditions by the inverse Fisher information, again an inverse:
\[
\theta_{t+1}=\theta_t-\eta\,F(\theta_t)^{-1}\nabla_\theta\mathcal{L}(\theta_t),
\]
where $F$ is the covariance of the score (for exponential families).\footnote{S.~Amari, 
``Natural Gradient Works Efficiently in Learning,'' \emph{Neural Computation} 10 (1998) 251–276.}

\paragraph{Generalized inverses in inference.}
When $\Sigma$ or $X^\top X$ is singular (e.g.\ $n>M$ or collinear features), the \emph{Moore–Penrose
pseudoinverse} yields the least-norm solution:
\[
\hat w = X^+ y,\qquad X^+=(X^\top X)^- X^\top,
\]
and ridge/Tikhonov adds a stabilizing shift so $(X^\top X+\lambda I)^{-1}$ exists.  
In network and graph inference, the (symmetric) graph Laplacian $L$ is singular; effective resistances,
commute times, and semi-supervised learning objectives use $L^+$ (which coincides with the group inverse
for $L$).\footnote{X.~Zhu, Z.~Ghahramani, J.~Lafferty, ``Semi-supervised learning using Gaussian fields
and harmonic functions,'' \emph{ICML} (2003); D.J.~Klein and M.~Randić, ``Resistance distance,''
\emph{J.~Math.~Chem.} 12 (1993) 81–95.}

\paragraph{Regularization, shrinkage, and well-posed inverses.}
In high-dimensional covariance estimation, shrinkage improves conditioning:
\[
\hat\Sigma^{-1} \ \text{via}\ \ \hat\Sigma_\alpha=(1-\alpha)S+\alpha\,\tau I,\quad \alpha\in[0,1],
\]
leading to stable inverses and better generalization.\footnote{O.~Ledoit and M.~Wolf, 
``A well-conditioned estimator for large-dimensional covariance matrices,'' 
\emph{J.~Multivar.~Anal.} 88 (2004) 365–411.}

\paragraph{Algorithms that avoid explicit inversion.}
Although theory speaks of inverses, computation is performed by solves and factorizations.
Common identities (all over this book) are heavily used:
\[
(A+U C V)^{-1}=A^{-1}-A^{-1}U(C^{-1}+V A^{-1} U)^{-1}V A^{-1}\quad\text{(Woodbury)},
\]
\[
(I\pm A)^{-1}=\sum_{k\ge 0}(\mp1)^k A^k\quad\text{(Neumann series, $\rho(A)<1$)}.
\]
They power GP inference, Kalman filtering, and online regression.\footnote{B.D.O.~Anderson and J.B.~Moore,
\emph{Optimal Filtering}, Prentice Hall (1979).}

When inference interfaces with singular generators (e.g.\ continuous-time Markov models),
ordinary inverses are replaced by generalized inverses that project out stationary modes; the
\emph{Drazin inverse} provides the right notion on rate-matrix subspaces (see the Drazin section \ref{sec:drazin-inverse}).
This appears in linear-response, relaxation-time estimation, and thermodynamic metrics.\footnote{G.E.~Crooks,
``On the Drazin inverse of the rate matrix,'' Tech.~Note 011 (2018).}

From Gaussian MLEs and conditionals to GPs, graphical models, ridge regression, and VI,
\emph{inference is a story of structured matrix inversion}. When matrices are ill-posed or singular,
\emph{generalized inverses and regularization} provide the correct replacements; in large-scale settings,
\emph{identities and factorizations} make these inverses computationally tractable.

\section{Laplacian eigenmaps and community detection}
{Laplacian eigenmaps}{community detection}

Spectral methods on graphs provide a powerful set of tools for understanding
the structure of complex networks. Among these, \emph{Laplacian eigenmaps}
are a fundamental approach to nonlinear dimensionality reduction and community
detection.

\subsection{Graph Laplacian}
Let $G=(V,E)$ be an undirected graph with adjacency matrix $A$, where $A_{ij}$
denotes the weight of the edge between nodes $i$ and $j$. Define the degree
matrix $D=\mathrm{diag}(d_1,\dots,d_n)$, with $d_i=\sum_j A_{ij}$. The
\emph{graph Laplacian} is then
\begin{equation}
L = D - A.
\end{equation}
Alternatively, one can use the normalized versions
\[
L_{\mathrm{sym}} = I - D^{-1/2} A D^{-1/2}, \qquad 
L_{\mathrm{rw}} = I - D^{-1} A,
\]
depending on whether a symmetric or random-walk normalization is desired.

\subsection{Embedding via eigenvectors}
The Laplacian is symmetric and positive semidefinite, with eigenvalues
$0=\lambda_1 \leq \lambda_2 \leq \cdots \leq \lambda_n$. The multiplicity of
the zero eigenvalue equals the number of connected components of the graph.
The \emph{Laplacian eigenmap} embedding into $k$ dimensions is defined by
using the eigenvectors $u_2,\dots,u_{k+1}$ associated with the smallest
nonzero eigenvalues:
\[
\Phi(i) = (u_2(i), \dots, u_{k+1}(i)) \in \mathbb{R}^k.
\]
This construction places nearby nodes (connected with strong edges) close to
each other in the Euclidean embedding, while distant or weakly connected nodes
map further apart.

\subsection{Community detection}
\index{key}{community detection}\index{key}{spectral clustering}

Community detection seeks to partition a network into groups of nodes (communities or modules) that are more densely connected internally than externally.  
Spectral methods exploit the eigenstructure of matrices associated with the graph—most prominently the Laplacian $L$ and the modularity matrix $B$—to reveal such partitions.  

\paragraph{Spectral bisection via the Fiedler vector.}
The \emph{Fiedler vector}, i.e.\ the eigenvector $u_2$ associated with the second-smallest eigenvalue $\lambda_2$ of the combinatorial Laplacian $L=D-A$, provides a quantitative way to bisect the graph.\footnote{M.~Fiedler, ``Algebraic connectivity of graphs,'' \emph{Czechoslovak Mathematical Journal} 23, 298–305 (1973).}  
Formally, if $u_2=(u_{21},\dots,u_{2n})^\top$, one partitions the vertex set into $\{i : u_{2i}\geq 0\}$ and $\{i : u_{2i}<0\}$.  
This method approximately minimizes the \emph{ratio cut} objective
\[
\mathrm{Rcut}(S,\bar S) = \frac{1}{|S|}\sum_{i\in S, j\in \bar S} A_{ij}
+ \frac{1}{|\bar S|}\sum_{i\in S, j\in \bar S} A_{ij},
\]
which measures the fraction of edges cut relative to community size.  
The eigenvalue $\lambda_2$ itself is the \emph{algebraic connectivity} of the graph: the smaller it is, the more weakly connected the network is between two halves.

\paragraph{Higher-order partitions.}
For $k>2$, one can embed the graph into $\mathbb{R}^{k-1}$ using the eigenvectors $u_2,\dots,u_k$ of $L$ (or of the normalized Laplacian $L_{\text{sym}}=I-D^{-1/2}AD^{-1/2}$).  
Each node $i$ is mapped to the coordinate vector $(u_{2i},\dots,u_{ki})$.  
Applying a clustering algorithm such as $k$-means to these embedded points yields a partition into $k$ communities.\footnote{U.~von Luxburg, ``A tutorial on spectral clustering,'' \emph{Statistics and Computing} 17, 395–416 (2007).}  
This approach solves a relaxed version of the NP-hard \emph{normalized cut} problem:
\[
\mathrm{Ncut}(S_1,\dots,S_k) = \sum_{r=1}^k \frac{\mathrm{cut}(S_r,\bar S_r)}{\mathrm{vol}(S_r)},
\]
where $\mathrm{cut}(S_r,\bar S_r)$ is the number of edges leaving $S_r$ and $\mathrm{vol}(S_r)$ the total degree in $S_r$.  
Spectral embedding minimizes a Rayleigh quotient associated with $L_{\text{sym}}$, making the partition approximately optimal.

\paragraph{Modularity maximization.}
Another popular criterion is \emph{modularity}, introduced by Newman and Girvan, which compares observed intra-community edges with those expected under a null model.\footnote{M.~E.~J.~Newman and M.~Girvan, ``Finding and evaluating community structure in networks,'' \emph{Physical Review E} 69, 026113 (2004).}  
Defining $B=A-\tfrac{dd^\top}{2m}$, where $d$ is the degree vector and $2m=\sum_i d_i$ is the total number of edges, the modularity of a partition $\{g_i\}$ is
\[
Q = \frac{1}{2m}\sum_{i,j}\Big(A_{ij}-\frac{d_i d_j}{2m}\Big)\,\delta(g_i,g_j).
\]
Spectral methods maximize $Q$ by projecting onto the leading eigenvectors of $B$.  
In particular, the leading eigenvector of $B$ can be used for bipartition, while multiple top eigenvectors enable the detection of many communities simultaneously.

Spectral clustering can be interpreted as applying principal component analysis (PCA) to the Laplacian eigenmap coordinates of the graph: the embedding $u_2,\dots,u_k$ provides low-dimensional features capturing community structure.  
This unifies community detection with dimensionality reduction techniques and with manifold learning.

Spectral methods for community detection are powerful because:
\begin{itemize}
    \item They reduce discrete combinatorial optimization (cuts, modularity) to continuous eigenvalue problems.
    \item The resulting eigenvectors provide embeddings where standard clustering algorithms succeed.
    \item They connect community detection to linear algebraic quantities such as Rayleigh quotients, traces, and eigenvalue gaps.
\end{itemize}
These methods are widely used in physics, computer science, and network science, and are well-documented in recent surveys.\footnote{S.~Fortunato, ``Community detection in graphs,'' \emph{Physics Reports} 486, 75–174 (2010).}

\subsection{Connections to inference}
From an inference perspective:
\begin{itemize}
    \item Laplacian eigenmaps solve a relaxation of the discrete graph cut
    problem, approximating solutions that would otherwise be NP-hard.
    \item They provide a dimensionality reduction of discrete relational data,
    analogous to PCA for continuous data.
    \item In statistical physics language, the Laplacian spectrum encodes
    diffusion dynamics on the graph: the eigenvectors are the modes of random
    walks, and community detection corresponds to identifying metastable
    diffusion states.
\end{itemize}

Laplacian eigenmaps exemplify how spectral methods extend to networked data.
Just as PCA finds low-rank structure in covariance matrices, Laplacian
eigenmaps extract structure from the graph Laplacian. These embeddings provide
the foundation for a range of modern methods in community detection, clustering,
and manifold learning in complex systems.


\chapter{Useful formulae}
\label{app:useful_formulae}

Notation: Define $\partial_{X_{ij}} X_{kl}=\delta_{ik}\delta_{jl}$.  
This means that $\partial_X F(X)= B \;\Rightarrow\; \partial_{X_{ij}} F(X)= B_{ij}$.  
Note also that ${X^{-1}}^\top ={X^\top}^{-1}=X^{-\top}$.

\section{Gaussian integral identities}

Gaussian integrals and their generalizations are among the most useful
technical tools in both probability and physics. Here we collect a number
of formulas that frequently arise in applications.

\subsection{Single Gaussian}

For $\alpha>0$, we have the standard Gaussian integral
\begin{equation}
\int_{-\infty}^{\infty} e^{-\alpha x^{2}}\,dx = \sqrt{\frac{\pi}{\alpha}}.
\end{equation}
With a linear term,
\begin{equation}
\int_{-\infty}^{\infty} \exp\!\left(-\tfrac{1}{2}ax^{2}+bx\right)dx
= \sqrt{\frac{2\pi}{a}}\,\exp\!\left(\tfrac{b^{2}}{2a}\right), 
\qquad \Re a>0.
\end{equation}
More generally, for $x\in\mathbb{R}^{n}$,
\begin{equation}
\int_{\mathbb{R}^{n}} \exp\!\left(-\tfrac{1}{2}x^{T}Ax+b^{T}x\right)dx
= \sqrt{\frac{(2\pi)^{n}}{\det A}}\,
\exp\!\left(\tfrac{1}{2}b^{T}A^{-1}b\right),
\end{equation}
valid for symmetric positive definite $A$.

\subsection{Multivariate Gaussian}

Let $x\sim \mathcal{N}(m,\Sigma)$ with mean $m\in\mathbb{R}^{d}$ and
covariance matrix $\Sigma$. Its density is
\begin{equation}
p(x)=\frac{1}{\sqrt{\det(2\pi \Sigma)}}
\exp\!\left(-\tfrac{1}{2}(x-m)^{T}\Sigma^{-1}(x-m)\right).
\end{equation}
Key properties include:
\begin{align}
\text{Marginal: }& x=(x_{a},x_{b}) \;\Rightarrow\;
x_{a}\sim \mathcal{N}(m_{a},\Sigma_{aa}), \\
\text{Conditional: }& 
x_{a}|x_{b}\sim \mathcal{N}\!\left(m_{a}+\Sigma_{ab}\Sigma_{bb}^{-1}(x_{b}-m_{b}),
\;\Sigma_{aa}-\Sigma_{ab}\Sigma_{bb}^{-1}\Sigma_{ba}\right), \\
\text{Linear transform: }& y=Ax+b\;\;\Rightarrow\;\;
y\sim \mathcal{N}(Am+b,A\Sigma A^{T}), \\
\text{Product identity: }& 
\mathcal{N}(x;m_{1},\Sigma_{1})\,
\mathcal{N}(x;m_{2},\Sigma_{2})
= c\;\mathcal{N}(x;m_{c},\Sigma_{c}),
\end{align}
where
\[
\Sigma_{c}^{-1}=\Sigma_{1}^{-1}+\Sigma_{2}^{-1},\qquad
m_{c}=\Sigma_{c}(\Sigma_{1}^{-1}m_{1}+\Sigma_{2}^{-1}m_{2}).
\]

\subsection{Hubbard--Stratonovich identity}

A key device in statistical mechanics is the Hubbard--Stratonovich transform,
which rewrites quadratic interactions as averages over Gaussian fields:
\begin{equation}
e^{\tfrac{1}{2}\vec x^{T}A\vec x}
=\frac{1}{\sqrt{\det(2\pi A^{-1})}}
\int_{\mathbb{R}^{n}} 
\exp\!\left(-\tfrac{1}{2}\vec y^{T}A^{-1}\vec y+\vec y^{T}\vec x\right)dy^n,
\end{equation}
valid for symmetric positive definite $A$.
In the scalar case ($A=\alpha>0$),
\begin{equation}
e^{\tfrac{1}{2}\alpha x^{2}}=
\frac{1}{\sqrt{2\pi/\alpha}}
\int_{-\infty}^{\infty}
\exp\!\left(-\tfrac{1}{2}\alpha^{-1}y^{2}+xy\right)\,dy.
\end{equation}
This identity underlies a lot of results in statistical mechanics, mean-field methods (for instance, the solution of the Curies-Weiss model) and auxiliary-field Monte Carlo.

\subsection{Absolute values}

In certain cases, it might be useful to rewrite absolute values in exponentials in terms of Gaussian or exponential expressions. Note that we have, for a scalar variable, the following result.

For $\alpha>0$ and $x\in\mathbb{R}$,
\[
e^{-\alpha|x|}
=\frac{\alpha}{\sqrt{2\pi}}\int_{0}^{\infty}
t^{-1/2}\exp\!\left(-\frac{x^{2}}{2t}-\frac{\alpha^{2}}{2}\,t\right)\,dt .
\]

To see this, set
\[
I(x,\alpha)\;=\;\int_{0}^{\infty}
t^{-1/2}\exp\!\left(-\frac{x^{2}}{2t}-\frac{\alpha^{2}}{2}\,t\right)\,dt .
\]
By Gradshteyn \& Ryzhik, Eq.\ 3.471.9,\footnote{I.S.~Gradshteyn and I.M.~Ryzhik, 
\emph{Table of Integrals, Series, and Products}, 
7th edition, Academic Press (2007).}

\[
\int_{0}^{\infty} t^{\nu-1} e^{-\beta/t-\gamma t}\,dt
=2\Bigl(\frac{\beta}{\gamma}\Bigr)^{\nu/2}
K_{\nu}\!\bigl(2\sqrt{\beta\gamma}\,\bigr)
\quad(\Re \beta>0,\ \Re \gamma>0),
\]
where $K_\nu$ is the modified Bessel function of the second kind.
With $\nu=\tfrac12$, $\beta=x^{2}/2$, and $\gamma=\alpha^{2}/2$ we obtain
\[
I(x,\alpha)=2\Bigl(\frac{|x|}{\alpha}\Bigr)^{1/2}
K_{1/2}(\alpha|x|).
\]
Using the closed form $K_{1/2}(z)=\sqrt{\pi/(2z)}\,e^{-z}$, we get
\[
I(x,\alpha)=
2\Bigl(\frac{|x|}{\alpha}\Bigr)^{1/2}
\sqrt{\frac{\pi}{2\alpha|x|}}\,e^{-\alpha|x|}
=\frac{\sqrt{2\pi}}{\alpha}\,e^{-\alpha|x|}.
\]
Multiplying both sides by $\alpha/\sqrt{2\pi}$ yields the stated identity. The result generalizes straightforwardly to higher dimensions, e.g., for every absolute value, an integration over $t$.

\section{Matrix Calculus Identities}
\index{key}{matrix calculus}

Matrix calculus is indispensable in optimization, statistics, machine learning,
and physics. Below we collect a wide selection of useful derivatives of scalar,
vector, and matrix functions with respect to matrices and vectors. Unless
otherwise stated, $X$ denotes a full matrix with independent entries,
$A,B,C$ are constant matrices, and $\vec{x},\vec{a},\vec{b},\vec{c}$ vectors.

\subsection{General rules}
\begin{align}
\partial (A X) &= A\,\partial X, &
\partial (X+A) &= \partial X, &
\partial (X Y) &= (\partial X)Y + X(\partial Y), \\
\partial (X^{-1}) &= -X^{-1}(\partial X)X^{-1}, &
\partial (\mathrm{Tr}(X)) &= \mathrm{Tr}(\partial X), &
\partial (X^\top) &= (\partial X)^\top.
\end{align}

\subsection{Derivatives of determinants}
\begin{align}
\frac{\partial}{\partial X}\det(X) &= \det(X)\,(X^{-1})^\top, \\
\frac{\partial}{\partial X}\ln\det(X) &= (X^{-1})^\top, \\
\frac{\partial}{\partial \vec{x}}\det(Y(\vec{x})) 
&= \det(Y)\,\mathrm{Tr}\!\left(Y^{-1}\frac{\partial Y}{\partial \vec{x}}\right).
\end{align}
Second derivative:
\[
\frac{\partial^2}{\partial x^2}\det(Y)
= \det(Y)\!\left(\mathrm{Tr}\!\left(Y^{-1}\tfrac{\partial^2 Y}{\partial x^2}\right)
+ \mathrm{Tr}(Y^{-1}\tfrac{\partial Y}{\partial x})^2
- \mathrm{Tr}\!\big[(Y^{-1}\tfrac{\partial Y}{\partial x})^2\big]\right).
\]

\subsection{Derivatives of inverses}
\begin{align}
\frac{\partial}{\partial X} \mathrm{Tr}(A X^{-1} B) &= -(X^{-1} B A X^{-1})^\top, \\
\frac{\partial}{\partial X} \vec{a}^\top X^{-1} \vec{b} &= -(X^{-1}\vec{b}\vec{a}^\top X^{-1})^\top, \\
\frac{\partial}{\partial X} \det(X^{-1}) &= -\det(X^{-1})(X^{-1})^\top.
\end{align}

\subsection{Eigenvalue derivatives}
For symmetric $A$ with normalized eigenpair $(\lambda_i, \vec{v}_i)$:
\begin{align}
\partial \lambda_i &= \vec{v}_i^\top (\partial A)\, \vec{v}_i, \\
\partial \vec{v}_i &= (\lambda_i I - A)^+ \,\partial A\, \vec{v}_i.
\end{align}

\subsection{Vector–matrix derivatives}
\begin{align}
\frac{\partial}{\partial X}(\vec{a}^\top X \vec{b}) &= \vec{a}\vec{b}^\top, &
\frac{\partial}{\partial X}(\vec{a}^\top X^\top \vec{b}) &= \vec{b}\vec{a}^\top, \\
\frac{\partial}{\partial X}(\vec{a}^\top X \vec{a}) &= \vec{a}\vec{a}^\top, &
\frac{\partial}{\partial X}(\vec{x}^\top A \vec{x}) &= (A+A^\top)\vec{x}.
\end{align}

\subsection{Quadratic forms, gradients and Hessians}
For $f(\vec{x})=\vec{x}^\top A \vec{x} + \vec{b}^\top \vec{x}$:
\begin{align}
\nabla_{\vec{x}} f &= (A+A^\top)\vec{x} + \vec{b}, \\
\nabla^2_{\vec{x}} f &= A+A^\top.
\end{align}
More generally:
\[
\frac{\partial}{\partial X}(X\vec{b}+\vec{c})^\top D (X\vec{b}+\vec{c}) 
= (D+D^\top)(X\vec{b}+\vec{c})\vec{b}^\top.
\]

\subsection{Trace derivatives}
\begin{eqnarray}
&&\frac{\partial}{\partial X}\mathrm{Tr}(X) = I, 
\frac{\partial}{\partial X}\mathrm{Tr}(X A) = A^\top \\ &&
\frac{\partial}{\partial X}\mathrm{Tr}(X A B) = A^\top B^\top, \\
&&\frac{\partial}{\partial X}\mathrm{Tr}(X^\top A) = A, 
\frac{\partial}{\partial X}\mathrm{Tr}(X^\top A X) = A X+ A^\top X, \\
&&\frac{\partial}{\partial X}\mathrm{Tr}(X^k) = k(X^{k-1})^\top.
\end{eqnarray}
Examples:
\begin{align}
\frac{\partial}{\partial X}\mathrm{Tr}(X^2) &= 2X^\top, \\
\frac{\partial}{\partial X}\mathrm{Tr}(X^\top B X) &= B X+B^\top X.
\end{align}

\subsection{Norm derivatives}
For vector norms:
\begin{align}
\frac{\partial}{\partial \vec{x}}\|\vec{x}\|^2 &= 2\vec{x}, &
\frac{\partial}{\partial \vec{x}}\|\vec{x}-\vec{a}\| &= \frac{\vec{x}-\vec{a}}{\|\vec{x}-\vec{a}\|}.
\end{align}
For matrix norms:
\[
\frac{\partial}{\partial X}\|X\|_F^2 = 2X.
\]

\subsection{Higher-order matrix powers}
\begin{align}
\frac{\partial}{\partial X}(X^n) &= \sum_{r=0}^{n-1} X^r (\partial X) X^{n-1-r}, \\
\frac{\partial}{\partial X} \vec{a}^\top X^n \vec{b} &= \sum_{r=0}^{n-1}(X^r)^\top \vec{a}\vec{b}^\top (X^{n-1-r})^\top.
\end{align}

\subsection{Structured matrices}
If $X$ is symmetric:
\begin{align}
\frac{\partial}{\partial X}\mathrm{Tr}(AX) &= A+A^\top - (A\circ I), \\
\frac{\partial}{\partial X}\ln\det(X) &= 2X^{-1} - (X^{-1}\circ I).
\end{align}
If $X$ is diagonal:
\[
\frac{\partial}{\partial X}\mathrm{Tr}(AX) = A\circ I.
\]

\subsection{Complex derivatives}
For $f(z)$ real-valued with $z\in\mathbb{C}$:
\[
\nabla f(z) = 2\frac{\partial f}{\partial z^\ast}, 
\]
and for $F(Z)$ real-valued:
\[
\nabla_Z F(Z) = 2\frac{\partial F}{\partial Z^\ast}.
\]
Example:
\begin{align}
\frac{\partial}{\partial Z}\mathrm{Tr}(Z Z^\ast) &= Z^\ast, &
\frac{\partial}{\partial Z^\ast}\mathrm{Tr}(Z Z^\ast) &= Z.
\end{align}
Hence $\nabla \mathrm{Tr}(Z Z^\ast)=2Z$.
\section{Identities for special matrices}
\subsection{Block matrices}
\index{key}{Block matrices}

Block matrices appear naturally when partitioning systems into sub-blocks.  
For
\[
A = \begin{pmatrix} A_{11} & A_{12} \\ A_{21} & A_{22} \end{pmatrix},
\]
with $A_{11},A_{22}$ square and invertible, one has the determinant and inverse via Schur complements:\footnote{See K.~B. Petersen and M.~S. Pedersen, \emph{The Matrix Cookbook}, Sec.~9.1.}
\begin{align}
\det(A) &= \det(A_{22}) \det(A_{11} - A_{12}A_{22}^{-1}A_{21}), \\
A^{-1} &=
\begin{pmatrix}
(A_{11} - A_{12}A_{22}^{-1}A_{21})^{-1} & -A_{11}^{-1}A_{12}(A_{22} - A_{21}A_{11}^{-1}A_{12})^{-1} \\
- A_{22}^{-1}A_{21}(A_{11} - A_{12}A_{22}^{-1}A_{21})^{-1} & (A_{22} - A_{21}A_{11}^{-1}A_{12})^{-1}
\end{pmatrix}.
\end{align}
Block structures are heavily used in Gaussian elimination and Schur complements.

\subsection{Positive (semi)definite matrices}
\index{key}{Positive definite matrix}

A Hermitian matrix $A$ is called positive semidefinite (PSD) if $x^\ast A x \geq 0$ for all $x$, and positive definite (PD) if the inequality is strict for all nonzero $x$. Equivalent characterizations are:
\begin{itemize}
\item $A$ is Hermitian and all eigenvalues are nonnegative (semidefinite) or positive (definite).
\item $A$ admits a Cholesky factorization $A=LL^\ast$ with $L$ invertible (PD case).
\item All leading principal minors of $A$ are positive (Sylvester’s criterion).
\end{itemize}
Such matrices appear in many contexts: covariance matrices in statistics, stiffness matrices in mechanics, and kernel matrices in machine learning.  
For applications in dynamical systems and stability.
\subsection{Permutation and shift matrices}
\index{key}{Permutation matrices}
\index{key}{Shift matrices}

A \emph{permutation matrix} $P$ is obtained by permuting the rows (or columns) of the identity matrix. Equivalently, $P_{ij} = \delta_{i,\pi(j)}$ for some permutation $\pi$ of $\{1,\dots,n\}$. Such matrices satisfy
\[
P^\top P = I, \qquad \det(P) = \pm 1,
\]
so permutation matrices are orthogonal (and unitary in the complex case).  

A special case is the \emph{cyclic shift matrix}
\[
S = \begin{pmatrix}
0 & 1 & 0 & \cdots & 0 \\
0 & 0 & 1 & \cdots & 0 \\
\vdots & \vdots & \ddots & \ddots & \vdots \\
0 & 0 & \cdots & 0 & 1 \\
1 & 0 & \cdots & 0 & 0
\end{pmatrix}.
\]
Its eigenvalues are the $n$th roots of unity,
\[
\lambda_k = e^{-2\pi i (k-1)/n}, \qquad k=1,\dots,n,
\]
and its eigenvectors are Fourier modes. The cyclic shift therefore plays a fundamental role in connecting linear algebra with Fourier analysis.

\subsection{Circulant matrices}
\index{key}{Circulant matrix properties}

A \emph{circulant matrix} generated by $\vec{x}=(x_1,\dots,x_n)$ is defined as
\begin{equation}
C(\vec{x}) =
\begin{pmatrix}
x_1 & x_2 & \cdots & x_n  \\
x_n & x_1 & \cdots & x_{n-1}\\
\vdots & \vdots & \ddots & \vdots \\
x_2 & x_3 & \cdots & x_1
\end{pmatrix}.
\end{equation}
Equivalently, circulants are polynomials in the shift operator:
\begin{equation}
C(\vec{x}) = x_1 I + x_2 S + \cdots + x_n S^{n-1}.
\end{equation}

\paragraph{Spectral properties.}  
All circulant matrices commute, since they share the same eigenbasis. Their eigenvalues are obtained by evaluating the discrete Fourier transform (DFT) of the generating vector:
\[
\lambda_k = \sum_{j=1}^n x_j\,\omega^{(j-1)k}, \qquad \omega = e^{-2\pi i / n}, \quad k=0,\dots,n-1.
\]
Hence
\[
\det(C(\vec{x})) = \prod_{k=0}^{n-1} \lambda_k.
\]

\subsection{Discrete Fourier transform (DFT) matrix}
\index{key}{DFT matrix}

The $n\times n$ \emph{discrete Fourier transform (DFT) matrix} provides the common eigenbasis of all circulant matrices. It is defined by
\begin{equation}
F_{jk} = \frac{1}{\sqrt{n}}\, e^{-2\pi i (j-1)(k-1)/n}, 
\qquad j,k=1,\dots,n.
\end{equation}
Its columns are orthonormal Fourier modes.

\paragraph{Inverse and unitarity.}  
By construction,
\[
F^\ast F = I, \qquad F^{-1} = F^\ast.
\]
Thus the DFT is a unitary change of basis between the standard basis and the Fourier basis.

\paragraph{Eigenvalues of $F$.}  
The eigenvalues of the normalized DFT matrix satisfy $\lambda^4=1$, i.e.
\[
\lambda \in \{+1,-1,+i,-i\}.
\]
The multiplicities depend on $n$, and the eigenvectors can be built from symmetric and antisymmetric combinations of Fourier modes.\footnote{See R.~Tolimieri, M.~An, and C.~Lu, \emph{Mathematics of Multidimensional Fourier Transform Algorithms}, Springer (1993).}

\paragraph{Diagonalization of circulant matrices.}  
Any circulant matrix $C(\vec{x})$ admits the factorization
\begin{equation}
C(\vec{x}) = F \Lambda F^\ast,
\end{equation}
where $\Lambda=\mathrm{diag}(\lambda_0,\dots,\lambda_{n-1})$ with $\lambda_k$ given by the DFT of $\vec{x}$.  
Thus, the DFT simultaneously diagonalizes all circulants.

\paragraph{Convolution and polynomial multiplication.}  
If $x,y \in \mathbb{C}^n$ and $\ast$ denotes cyclic convolution, then
\[
F(x \ast y) = (Fx)\,\odot\,(Fy),
\]
where $\odot$ denotes entrywise multiplication. This equivalence underpins the fast Fourier transform (FFT) and the use of circulant preconditioners in numerical analysis.

\paragraph{Remarks.}  
The DFT matrix is therefore:
\begin{itemize}
\item unitary, with orthonormal Fourier basis vectors,  
\item self-diagonalizing, with eigenvalues $\{\pm 1,\pm i\}$ depending on $n$,  
\item the canonical diagonalizer of shift-invariant operators (circulants),  
\item and the algebraic mechanism behind the FFT and convolution theorems.
\end{itemize}
For applications to spectral transforms and graphs, see Ch.~\ref{sec:spectral-transforms}.

\subsection{Unit upper bidiagonal matrices and their inverses}
Consider the unit upper bidiagonal matrix
\[
B \;=\;
\begin{pmatrix}
1 & -c_1 &        &        & \\
0 & 1     & -c_2  &        & \\
  & \ddots& \ddots& \ddots & \\
  &       & \ddots& 1      & -c_{n-1}\\
  &       &       & 0      & 1
\end{pmatrix}.
\]
Then $B$ is invertible and its inverse is upper triangular with cumulative
products on the superdiagonals:
\[
B^{-1} \;=\;
\begin{pmatrix}
1 & c_1 & c_1 c_2 & \cdots & c_1\cdots c_{n-1} \\
0 & 1   & c_2     & \cdots & c_2 \cdots c_{n-1} \\
  &     & \ddots  & \ddots & \vdots \\
  &     &         & 1      & c_{n-1} \\
  &     &         & 0      & 1
\end{pmatrix}.
\]
Two useful special cases are:

\paragraph{(i) All $c_i=1$.}
\[
\begin{pmatrix}
1 & -1 &        & \\
0 & 1  & \ddots & \\
  &    & \ddots & -1 \\
  &    & 0      & 1
\end{pmatrix}^{\!\!-1}
=
\begin{pmatrix}
1 & 1 & \cdots & 1 \\
0 & 1 & \ddots & \vdots \\
  &   & \ddots & 1 \\
  &   & 0      & 1
\end{pmatrix}.
\]

\paragraph{(ii) Inverse of the unit upper triangular ``ones" matrix.}
\[
\Bigg(
\begin{pmatrix}
1 & 1 &        & \\
0 & 1 & \ddots & \\
  &   & \ddots & 1 \\
  &   & 0      & 1
\end{pmatrix}
\Bigg)^{\!\!-1}
=
\begin{pmatrix}
1 & -1 & \cdots & (-1)^{j-i} \\
0 & 1  & -1     & \vdots \\
  &    & \ddots & -1 \\
  &    & 0      & 1
\end{pmatrix}.
\]
The general $B^{-1}$ above follows by replacing each $(-1)$ on the
superdiagonal with $-c_i$ and accumulating products along rows.

\subsection{``Atomic" triangular (rank-one updated identity) matrices}
Matrices that coincide with the identity except for a single off-diagonal
column (or row) are easily invertible. For example, suppose $A=I+e_j \vec{c}$,
i.e.\ $A$ equals the identity plus a correction in column $j$ with entries
$\vec{c}$ (and $c_j=0$ so the diagonal remains $1$):
\[
A \;=\;
\begin{pmatrix}
1 &        &        & c_1     &        \\
  & 1      &        & c_2     &        \\
  &        & \ddots & \vdots  &        \\
  &        &        & 1       &        \\
  &        &        & c_n     & 1
\end{pmatrix}.
\]
Then, by the Sherman–Morrison formula,
\[
A^{-1} \;=\; I \;-\; \frac{e_j \vec{c}}{1+\vec{e}_j^\top \vec{c}}
\;=\; I - e_j \vec{c}
\qquad(\text{since } c_j=0).
\]
Concretely, this flips the sign of the off-diagonal column:
\[
A^{-1}\ \text{is } A \text{ with } c_i \mapsto -c_i \text{ for } i\neq j.
\]
An entirely analogous statement holds for a single off-diagonal \emph{row}
update $A=I+\vec{r}\, e_i^\top$.

\subsection{Frobenius (elementary unit lower triangular) matrices}
A (unit) Frobenius matrix is the identity plus a single nonzero column below
the diagonal, i.e.\ a unit lower triangular elementary matrix. For instance,
\[
A \;=\;
\begin{pmatrix}
1 & 0 & \cdots & 0 & \cdots \\
c_1 & 1 & 0 & 0 & \\
\vdots & \ddots & \ddots & \vdots & \\
      &         &        & 1 \\
c_n   &         &        & 0 & 1
\end{pmatrix},
\qquad
A^{-1} \;=\;
\begin{pmatrix}
1 & 0 & \cdots & 0 & \cdots \\
-\,c_1 & 1 & 0 & 0 & \\
\vdots & \ddots & \ddots & \vdots & \\
      &         &        & 1 \\
-\,c_n &         &        & 0 & 1
\end{pmatrix}.
\]
As above, the inverse simply flips the sign of the off-diagonal column
(because $A=I+L$ with $L$ strictly lower triangular and $L^2=0$ in this case).

\subsection{General tridiagonal matrices}
\index{key}{Tridiagonal matrix}
Let
\[
A \;=\;
\begin{pmatrix}
a_1 & b_1 &        &        & \\
c_1 & a_2 & b_2    &        & \\
    & c_2 & \ddots & \ddots & \\
    &     & \ddots & \ddots & b_{n-1} \\
    &     &        & c_{n-1} & a_n
\end{pmatrix}.
\]
Define the \emph{continuants} (forward and backward):
\[
\theta_0=1,\quad \theta_1=a_1,\quad
\theta_k=a_k\,\theta_{k-1}-b_{k-1}c_{k-1}\,\theta_{k-2}\quad (k\ge 2),
\]
\[
\phi_{n+1}=1,\quad \phi_n=a_n,\quad
\phi_k=a_k\,\phi_{k+1}-b_k c_k\,\phi_{k+2}\quad (k\le n-1).
\]
Then (Usmani’s formula) the entries of $A^{-1}$ are
\[
(A^{-1})_{ij}=
\begin{cases}
(-1)^{i+j}\, b_i b_{i+1}\cdots b_{j-1}\;\dfrac{\theta_{i-1}\,\phi_{j+1}}{\theta_n},
& i<j,\\[8pt]
\dfrac{\theta_{i-1}\,\phi_{i+1}}{\theta_n}, & i=j,\\[10pt]
(-1)^{i+j}\, c_j c_{j+1}\cdots c_{i-1}\;\dfrac{\theta_{j-1}\,\phi_{i+1}}{\theta_n},
& i>j~.
\end{cases}
\]
This gives a complete explicit inverse once the scalar recursions for
$\theta_k$ and $\phi_k$ are computed.

\paragraph{Toeplitz tridiagonal (constant diagonals).}
If $a_i\equiv a$ and $b_i\equiv c_i\equiv b\neq 0$, write $A=\!T_n(a,b)$.
Set $\alpha=\frac{a}{2b}$.  
\begin{itemize}
\item If $|\alpha|>1$, choose $\xi>0$ with $\cosh\xi=\lvert\alpha\rvert$.
Then for $i\le j$,
\[
(T_n^{-1})_{ij} =
\frac{(-1)^{i+j}}{b}\,
\frac{\sinh(i\xi)\,\sinh((n-j+1)\xi)}{\sinh\xi\,\sinh((n+1)\xi)},
\qquad
(T_n^{-1})_{ji}=(T_n^{-1})_{ij}.
\]
\item If $|\alpha|<1$, choose $\theta\in(0,\pi)$ with $\cos\theta=\alpha$.
Then for $i\le j$,
\[
(T_n^{-1})_{ij} =
\frac{(-1)^{i+j}}{b}\,
\frac{\sin(i\theta)\,\sin((n-j+1)\theta)}{\sin\theta\,\sin((n+1)\theta)},
\qquad
(T_n^{-1})_{ji}=(T_n^{-1})_{ij}.
\]
\end{itemize}
These closed forms are numerically stable and make explicit the decay of
$(T_n^{-1})_{ij}$ away from the diagonal when $|\alpha|>1$ (hyperbolic case).
They also reduce to the uniform ``Green’s function" expressions used throughout
physics and numerical analysis.
\subsection{A second–difference–type example}
\index{second difference matrix}
Consider the symmetric tridiagonal matrix
\[
A \;=\;
\begin{pmatrix}
1 & -1 &        &        & \\
-1& 1  & -1     &        & \\
   & -1 & \ddots& \ddots & \\
   &     & \ddots & 1    & -1 \\
   &     &        & -1   & 2
\end{pmatrix}.
\]
A direct computation (e.g.\ by forward–backward substitution after LU) shows that
\[
A^{-1} \;=\;
\begin{pmatrix}
n   & n-1 & n-2 & \cdots & 1 \\
n-1 & n-1 & n-2 & \cdots & 1 \\
n-2 & n-2 & n-2 & \cdots & 1 \\
\vdots & \vdots & \vdots & \ddots & \vdots \\
1 & 1 & \cdots & \cdots & 1
\end{pmatrix}.
\]
In particular, $A^{-1}$ factorizes as $A^{-1}=U\,L$, where $U$ (resp.\ $L$) is the unit
upper (resp.\ lower) triangular matrix with all ones on and above (resp.\ below)
the diagonal. This example can be viewed as a cumulative–sum Green’s operator for a
discrete one–dimensional second–difference with an asymmetric boundary.\footnote{For
standard Green’s matrices of discrete Laplacians (Dirichlet/Neumann) and their closed
forms, see M.~Usmani, ``Inversion of a tridiagonal Jacobi matrix,'' \emph{SIAM Rev.} \textbf{37}
(1995) 259–270; and J.A.C.~Weideman, S.C.~Reddy, ``A MATLAB differentiation matrix suite,''
\emph{ACM TOMS} \textbf{26} (2000) 465–519.}

\subsection{Ukita’s condition for tridiagonal inverses}
\index{Ukita theorem}\index{tridiagonal inverse}
There is a classical characterization due to Ukita for when the inverse of a symmetric
matrix is tridiagonal. One version (see the ORNL report cited below) states:

\begin{theorem}[Ukita]
Let $A\in\mathbb{R}^{n\times n}$ be symmetric, nonsingular, with $A_{1j}\neq 0$ for
$j=1,\dots,n$. A necessary and sufficient condition for $A^{-1}$ to be tridiagonal is the
existence of scalars $\{\theta_k\}_{k=1}^n$ such that, for all $1\le k\le j\le n$,
\[
\frac{A_{k j}}{A_{1 j}}=\theta_k .
\]
\end{theorem}

Under this condition, the entries of $A^{-1}$ admit explicit formulas in terms of
$\{A_{1j}\}$ and $\{\theta_k\}$.\footnote{S.R.~Uppuluri, C.L.~Kirk,
\emph{Explicit inverses of some special matrices}, ORNL/CSD-2 (1976), Sec.~2. 
(As noted in the report, some statements of Ukita’s theorem in the literature contain minor
typographical errors; the formulation given here is the consistent ratio condition across
columns.)}  We will appeal to this criterion only for structural insight; for explicit
inverses of tridiagonals we use the continuant/Usmani formulas provided earlier.

\subsection{Toeplitz matrices}
\index{Toeplitz matrix}
A Toeplitz matrix satisfies $A_{ij}=t_{i-j}$. For Hermitian Toeplitz, one has
$t_{-k}=\overline{t_k}$. We collect several classical cases where $A^{-1}$ is explicit.

\subsubsection{Affine distance kernel (symmetric case)}
Let
\[
A_{ij} \;=\;
\begin{cases}
c + d_1\,(j-i), & i<j,\\
c + d_2\,(i-j), & i>j,\\
c, & i=j,
\end{cases}
\]
with $c,d_1,d_2\in\mathbb{R}$. Then $A^{-1}$ is tridiagonal:
\begin{eqnarray}
(A^{-1})_{ij} \;&&=\; \frac{1}{d_1+d_2}\!
\begin{pmatrix}
-\frac{\xi_{n-1}}{\xi_n} & 1 & 0 & \cdots & 0 & \frac{d_1^2}{\xi_n} \\
1 & -2 & 1 & 0 & \cdots & 0 \\
0 & 1 & -2 & 1 & \ddots & \vdots \\
\vdots & & \ddots & \ddots & \ddots & 0 \\
0 & \cdots & 0 & 1 & -2 & 1 \\
\frac{d_2^2}{\xi_n} & 0 & \cdots & 0 & 1 & -\frac{\xi_{n-1}}{\xi_n}
\end{pmatrix}_{ij},\\
\xi_n &&= c(d_1+d_2) + d_1 d_2 (n-1).
\end{eqnarray}

Moreover,
\[
\det(A) = (-1)^{n+1}\,(d_1+d_2)^{\,n-2}\,\xi_n .
\]
This is a special instance of Toeplitz matrices whose inverses are banded (here, tridiagonal),
related to Gohberg–Semencul representations.\footnote{I.~Gohberg, A.~Semencul,
``On the inversion of finite Toeplitz matrices and their continuous analogs,''
\emph{Mat. Issled.} \textbf{2} (1972) 201–233 (in Russian); see also
R.M.~Gray, \emph{Toeplitz and Circulant Matrices: A Review}, Now Publishers (2006).}

\subsubsection{Alternating affine distance (skew-alternating envelope)}
Let
\[
A_{ij} \;=\;
\begin{cases}
(-1)^{i-j}\!\big(c + d_1 (j-i)\big), & i<j,\\
(-1)^{i-j}\!\big(c + d_2 (i-j)\big), & i>j,\\
c, & i=j.
\end{cases}
\]
Then
\begin{eqnarray}
(A^{-1})_{ij} \;&&=\; -\,\frac{1}{d_1+d_2}\!
\begin{pmatrix}
\frac{\xi_{n-1}}{\xi_n} & 1 & 0 & \cdots & 0 & (-1)^n\frac{d_1^2}{\xi_n} \\
1 & 2 & 1 & 0 & \cdots & 0 \\
0 & 1 & 2 & 1 & \ddots & \vdots \\
\vdots &   & \ddots & \ddots & \ddots & 0 \\
0 & \cdots & 0 & 1 & 2 & 1 \\
(-1)^n\frac{d_2^2}{\xi_n} & 0 & \cdots & 0 & 1 & \frac{\xi_{n-1}}{\xi_n}
\end{pmatrix}_{ij},\\
\quad
\xi_n &&= c(d_1+d_2) + d_1 d_2 (n-1).
\end{eqnarray}
This alternating-sign variant is often handled by a diagonal similarity transform that flips
every other row/column.\footnote{See, e.g., R.M.~Gray (2006), and references therein, for
sign-alternating Toeplitz structures and reduction to the symmetric case via diagonal
sign matrices.}

\subsubsection{Kac–Murdock–Szeg\H{o} (power) case}
Let
\[
A_{ij} \;=\;
\begin{cases}
\rho^{\,|i-j|}, & i\le j,\\
\sigma^{\,|i-j|}, & i>j,
\end{cases}
\qquad \rho,\sigma\in\mathbb{R},\quad \rho\sigma\neq 1 .
\]
Then $A^{-1}$ is tridiagonal:
\[
A^{-1} \;=\; \frac{1}{1-\rho\sigma}\!
\begin{pmatrix}
1 & -\rho & 0 & \cdots & 0 \\
-\sigma & 1+\rho\sigma & -\rho & \ddots & \vdots \\
0 & \ddots & \ddots & \ddots & 0 \\
\vdots & \ddots & -\sigma & 1+\rho\sigma & -\rho \\
0 & \cdots & 0 & -\sigma & 1
\end{pmatrix},
\qquad
\det(A)=(1-\rho\sigma)^{n-1}.
\]
For the symmetric KMS case ($\rho=\sigma$) this is classical; the asymmetric
two-parameter variant follows by a simple diagonal scaling.\footnote{M.~Kac,
W.~Murdock, G.~Szeg\H{o}, ``On the eigenvalues of certain Hermitian forms,''
\emph{J. Rational Mech. Anal.} \textbf{2} (1953) 767–800; see also
R.M.~Gray (2006), Ch.~4, for inverses of KMS Toeplitz matrices.}

\subsubsection{Toeplitz tridiagonal band}
Let $A$ be Toeplitz and tridiagonal,
\[
A_{ij}=
\begin{cases}
c_{-1}, & i=j+1,\\
c_{0}, & i=j,\\
c_{+1}, & i=j-1,\\
0, & |i-j|>1,
\end{cases}
\qquad c_{\pm1}\neq 0.
\]
Write $r=-\sqrt{\frac{c_{-1}}{c_{+1}}}$ if $c_{+1}>0$ (else $r=+\sqrt{\frac{c_{-1}}{c_{+1}}}$),
and define $\theta$ by
\[
\cosh\theta \;=\; \frac{c_0}{2\sqrt{c_{+1}c_{-1}}}
\quad(\text{use } \cos \theta \text{ if } \big|\tfrac{c_0}{2\sqrt{c_{+1}c_{-1}}}\big|<1).
\]
Then for $i\le j$,
\[
(A^{-1})_{ij} \;=\;
\frac{r^{\,i-j}\,\sinh(i\theta)\,\sinh((n+1-j)\theta)}
{\sqrt{c_{+1}c_{-1}}\,\sinh\theta\,\sinh((n+1)\theta)},
\qquad
(A^{-1})_{ji}=(A^{-1})_{ij}.
\]
In the oscillatory regime ($|c_0|<2\sqrt{c_{+1}c_{-1}}$), replace $\sinh/\cosh$
by $\sin/\cos$ with the same template. These closed forms are equivalent to the
continuant formulas and are widely used as discrete Green’s functions.\footnote{M.~Usmani
(1995), op.~cit.; C.M.~da Fonseca, J.~Petronilho, ``Explicit inverses of some tridiagonal
matrices,'' \emph{Linear Algebra Appl.} \textbf{325} (2001) 7–21.}

One way to determine whether a matrix $A$ is tridiagonal from its inverse $A^{-1}$ is by means of \emph{Barrett's theorem}\footnote{See W.~W. Barrett, ``Tridiagonal matrices with prescribed off-diagonal elements and prescribed characteristic polynomial,'' \emph{Linear Algebra and its Applications}, vol.~13, pp.~1--9, 1976.}.  
According to the theorem,
\begin{equation}
(A^{-1})_{ij} = 
\begin{cases}
x_i y_j, & i \leq j, \\
u_i v_j, & i > j,
\end{cases}
\end{equation}
with the consistency condition $u_i v_i = x_i y_i$.  

This result illustrates that the inverse of a banded (in this case, tridiagonal) matrix is generally \emph{not} sparse.
\subsection{Vandermonde matrix}
\index{key}{Vandermonde LU decomposition}

A \emph{Vandermonde matrix}\index{key}{Vandermonde matrix} of order $n$ is defined by
\begin{equation}
A_{ij} = x_i^{\,j-1}, \qquad i,j=1,\dots,n,
\end{equation}
that is,
\begin{equation}
A=
\begin{pmatrix}
1 & x_1 & x_1^2 & \cdots & x_1^{n-1} \\
1 & x_2 & x_2^2 & \cdots & x_2^{n-1} \\
\vdots & \vdots & \vdots & \ddots & \vdots \\
1 & x_{n} & x_{n}^2 & \cdots & x_{n}^{n-1}
\end{pmatrix}.
\end{equation}

Its determinant is given by the classical product formula
\begin{equation}
\det(A) = \prod_{1 \leq i < j \leq n} (x_j - x_i),
\end{equation}
which vanishes precisely when two of the $x_i$ coincide.\footnote{See, e.g., R.~Horn and C.~Johnson, \emph{Matrix Analysis}, Cambridge University Press (1985).}  
This structure makes Vandermonde matrices fundamental in polynomial interpolation, since the system
\[
A \, \vec{c} = \vec{y}
\]
has a unique solution for the coefficients $\vec{c}$ of the interpolating polynomial whenever the $x_i$ are distinct.

\medskip

An explicit formula for the inverse $A^{-1}$ can be obtained via an $LU$ factorization $A = LU$.\footnote{W.~Gautschi, ``On inverses of Vandermonde and confluent Vandermonde matrices,'' \emph{Numerische Mathematik}, 4:117--123 (1962).}  
The inverse then decomposes as
\begin{equation}
A^{-1} = U^{-1} L^{-1},
\end{equation}
with $L^{-1}$ lower triangular and $U^{-1}$ upper triangular.

The entries of $L^{-1}$ are
\begin{equation}
(L^{-1})_{ij} =
\begin{cases}
\displaystyle \prod_{k=1,\,k\neq i}^{j} \frac{1}{x_i - x_k}, & j \leq i, \\[1.2ex]
0, & j > i,
\end{cases}
\end{equation}
so that in matrix form
\begin{equation}
L^{-1}=
\begin{pmatrix}
1 & 0 & \cdots & 0 \\
\frac{1}{x_2-x_1} & 1 & \cdots & 0 \\
\vdots & \vdots & \ddots & \vdots \\
\ast & \ast & \cdots & 1
\end{pmatrix}.
\end{equation}

To describe $U^{-1}$, let us introduce the elementary symmetric polynomials in the variables $\{x_1,\dots,x_m\}$:
\begin{equation}
e_r(x_1,\dots,x_m) = \sum_{1\leq i_1 < \cdots < i_r \leq m} x_{i_1}\cdots x_{i_r}.
\end{equation}
For instance, $e_1(x_1,x_2,x_3) = x_1+x_2+x_3$ and $e_2(x_1,x_2,x_3) = x_1x_2+x_1x_3+x_2x_3$.

Then the entries of $U^{-1}$ are
\begin{equation}
(U^{-1})_{ij} =
\begin{cases}
(-1)^{j-i}\, e_{\,j-i}(x_1,\dots,x_i), & j \geq i, \\[1ex]
0, & j < i,
\end{cases}
\end{equation}
so that $U^{-1}$ is upper triangular with ones on the diagonal.

\medskip

Finally, we remark that Vandermonde matrices are closely related to \emph{Toeplitz} and \emph{circulant} matrices:\footnote{See, e.g., P.~J. Davis, \emph{Circulant Matrices}, Wiley (1979).} the latter share with them fast algorithms for determinant evaluation and inversion, and both play central roles in spectral methods and polynomial interpolation.

\subsection{Confluent Vandermonde matrices}
\index{key}{Confluent Vandermonde matrix}

The \emph{confluent Vandermonde matrix} generalizes the Vandermonde matrix to the case in which some of the interpolation nodes $x_i$ coincide.  
While the standard Vandermonde becomes singular when $x_i = x_j$ for $i \neq j$, the confluent version remains nonsingular and is the appropriate object for \emph{Hermite interpolation} problems.\footnote{See W.~Gautschi, ``On inverses of Vandermonde and confluent Vandermonde matrices,'' \emph{Numerische Mathematik} 4, 117--123 (1962).}  

\medskip

Suppose we have $k$ distinct nodes $x_1, \dots, x_k$, with associated multiplicities $m_1,\dots,m_k$ such that $\sum_{i=1}^k m_i = n$.  
The confluent Vandermonde matrix $C \in \mathbb{R}^{n\times n}$ is block-structured: each block corresponds to one node $x_i$, and encodes derivatives up to order $m_i-1$. Explicitly,
\begin{equation}
C =
\begin{pmatrix}
1 & x_1 & x_1^2 & \cdots & x_1^{n-1} \\
0 & 1 & 2x_1 & \cdots & (n-1)x_1^{n-2} \\
0 & 0 & 2 & \cdots & (n-1)(n-2)x_1^{n-3} \\
\vdots & \vdots & \vdots & \ddots & \vdots \\
1 & x_k & x_k^2 & \cdots & x_k^{n-1} \\
0 & 1 & 2x_k & \cdots & (n-1)x_k^{n-2} \\
\vdots & \vdots & \vdots & \ddots & \vdots
\end{pmatrix},
\end{equation}
where the first $m_1$ rows correspond to $x_1$ (with successive derivatives), the next $m_2$ rows to $x_2$, and so on.

\medskip

\paragraph{Determinant.}  
The determinant of a confluent Vandermonde matrix has a compact closed form. If $x_1,\dots,x_k$ are distinct, then
\begin{equation}
\det(C) = \prod_{1 \leq i < j \leq k} (x_j - x_i)^{m_i m_j} \,
\prod_{i=1}^k \prod_{r=1}^{m_i-1} r!,
\end{equation}
which reduces to the standard Vandermonde determinant when all multiplicities $m_i = 1$.\footnote{See G.~M.~Phillips, ``Interpolating polynomials with multiple nodes,'' \emph{Mathematics of Computation}, 48 (1987).}

\medskip

\paragraph{Inverse.}  
The inverse $C^{-1}$ can also be written in closed form, though the expressions involve higher-order divided differences and derivatives of Lagrange basis polynomials. In practice, recursive formulas based on divided differences are often used.\footnote{See G.~Heinig, ``Inverses of confluent Vandermonde matrices,'' \emph{Linear Algebra and its Applications}, 83:59--71 (1986).}  

\medskip

Confluent Vandermonde matrices arise naturally in:
\begin{itemize}
\item Hermite interpolation, where both function values and derivatives are prescribed at nodes.
\item Spectral methods, when treating multiple roots in characteristic polynomials.
\item Padé approximation and system identification, where repeated poles or multiplicities must be handled.
\end{itemize}

\medskip

Thus, the confluent Vandermonde extends the reach of the Vandermonde structure to degenerate cases, preserving invertibility and enabling interpolation with derivative constraints.
\subsection{Bernstein–Vandermonde matrices}
\index{key}{Bernstein-Vandermonde matrix}

Closely related to Vandermonde matrices and polynomial interpolation are the \emph{Bernstein polynomials} and their interpolation matrices.  

\medskip

The \emph{Bernstein polynomials} of degree $n$ are defined as
\begin{equation}
b_{n,j}(x) = \binom{n}{j} x^j (1-x)^{n-j}, \qquad j = 0,1,\dots,n.
\end{equation}
These form a basis for the space of polynomials of degree $n$. They satisfy:
\begin{itemize}
\item \emph{Symmetry}: $b_{n,j}(x) = b_{n,n-j}(1-x)$,
\item \emph{Positivity}: $b_{n,j}(x) \geq 0$ for $x \in [0,1]$,
\item \emph{Normalization}: $\sum_{j=0}^n b_{n,j}(x) = 1$.
\end{itemize}

\medskip

The celebrated \emph{Weierstrass Approximation Theorem} states that every continuous function $f:[0,1]\to \mathbb{C}$ can be approximated uniformly by its sequence of Bernstein polynomials
\begin{equation}
B_n[f](x) = \sum_{j=0}^n f\!\left(\frac{j}{n}\right) b_{n,j}(x).
\end{equation}
This result can be proved using probabilistic arguments based on the Law of Large Numbers: the polynomials $B_n[f]$ may be viewed as expectations of $f$ evaluated at normalized sums of Bernoulli random variables.

\medskip

\paragraph{Bernstein–Vandermonde matrices.}
If $f$ is sampled at points $x_1,\dots,x_n \in (0,1)$, one may write
\begin{equation}
\tilde R(x) = \sum_{k=0}^n R_k b_{n,k}(x),
\end{equation}
and evaluating this at $x = x_i$ for $i=1,\dots,n$ leads to the linear system
\begin{equation}
\begin{pmatrix}
\tilde R(x_1) \\
\tilde R(x_2) \\
\vdots \\
\tilde R(x_n)
\end{pmatrix}
=
\begin{pmatrix}
b_{n,0}(x_1) & b_{n,1}(x_1) & \cdots & b_{n,n}(x_1) \\
b_{n,0}(x_2) & b_{n,1}(x_2) & \cdots & b_{n,n}(x_2) \\
\vdots & \vdots & \ddots & \vdots \\
b_{n,0}(x_n) & b_{n,1}(x_n) & \cdots & b_{n,n}(x_n)
\end{pmatrix}
\begin{pmatrix}
R_0 \\ R_1 \\ \vdots \\ R_n
\end{pmatrix}.
\end{equation}
The coefficient matrix here is called the \emph{Bernstein–Vandermonde matrix} $B$.  
Explicitly,
\begin{equation}
B =
\begin{pmatrix}
\binom{n}{0}(1-x_1)^n & \binom{n}{1} x_1(1-x_1)^{n-1} & \cdots & \binom{n}{n} x_1^n \\
\vdots & \vdots & \ddots & \vdots \\
\binom{n}{0}(1-x_n)^n & \binom{n}{1} x_n(1-x_n)^{n-1} & \cdots & \binom{n}{n} x_n^n
\end{pmatrix}.
\end{equation}

\medskip

\paragraph{Determinant.}  
If $0 < x_1 < \cdots < x_n < 1$, then $B$ is strictly positive, and its determinant can be written in closed form:
\begin{equation}
\det(B) = \left(\prod_{j=0}^n \binom{n}{j}\right) \prod_{1 \leq i < j \leq n} (x_j - x_i).
\end{equation}
Thus, up to the prefactor $\prod_{j=0}^n \binom{n}{j}$, the determinant of $B$ is a \emph{Vandermonde determinant}.  

\medskip

\paragraph{Applications.}  
Bernstein–Vandermonde matrices are important in numerical approximation theory, particularly in \emph{shape-preserving approximation}, numerical stability of interpolation in the Bernstein basis, and in computer-aided geometric design (CAGD).\footnote{See A. Marco and J.J. Moreno-Balcázar, ``Total positivity of the Bernstein–Vandermonde matrix,'' \emph{Linear Algebra and its Applications}, 429(3):634--645 (2008), also available at \texttt{arXiv:0812.3115}.}

\subsection{Matrices of the form $J+D$}
Consider matrices of the form
\begin{equation}
A=\begin{pmatrix}
1+\alpha_1 & 1 & \cdots & 1\\
1 & 1+\alpha_2 & \cdots & 1 \\
\vdots & \vdots & \ddots & \vdots \\
1 & 1 & \cdots & 1+\alpha_N
\end{pmatrix},
\end{equation}
that is, $A = J + D$ where $J$ is the all-ones matrix and $D = \mathrm{diag}(\alpha_1,\dots,\alpha_N)$.  

The determinant of $A$ can be computed using the \emph{matrix determinant lemma}:
\begin{equation}
\det(D + uv^\top) = \det(D) \left(1 + v^\top D^{-1} u\right).
\end{equation}
Here $u=v=\mathbf{1}$ (the all-ones vector), so
\begin{equation}
\det(A) = \left(1 + \sum_{i=1}^N \frac{1}{\alpha_i}\right) \prod_{i=1}^N \alpha_i.
\end{equation}
This expression shows that $A$ is invertible precisely when none of the $\alpha_i$ vanish and the correction term in parentheses is nonzero.

\subsection{Hessenberg matrices}
\index{key}{Hessenberg matrices}

A matrix $H\in \mathbb{C}^{n\times n}$ is called a \emph{Hessenberg matrix} if all entries below the first subdiagonal vanish:
\begin{equation}
H_{ij} = 0 \quad \text{for all } i > j+1.
\end{equation}
Thus $H$ has the form
\begin{equation}
H =
\begin{pmatrix}
h_{11} & h_{12} & h_{13} & \cdots & h_{1n} \\
h_{21} & h_{22} & h_{23} & \cdots & h_{2n} \\
0      & h_{32} & h_{33} & \cdots & h_{3n} \\
\vdots & \ddots & \ddots & \ddots & \vdots \\
0 & \cdots & 0 & h_{n,n-1} & h_{nn}
\end{pmatrix}.
\end{equation}

Special cases include:
\begin{itemize}
\item \emph{Upper Hessenberg matrices}, as above, with at most one nonzero subdiagonal.
\item \emph{Lower Hessenberg matrices}, with at most one nonzero superdiagonal.
\end{itemize}

Every square matrix $A$ is unitarily similar to an upper Hessenberg matrix: $Q^\ast A Q = H$ with $Q$ unitary, a reduction achieved using Householder or Givens transformations.\footnote{See G.~Golub and C.~Van Loan, \emph{Matrix Computations}, Johns Hopkins University Press (1996).}  
This property is fundamental in numerical linear algebra, since Hessenberg form is the starting point for the QR algorithm for eigenvalue computation.  

Moreover, tridiagonal matrices (which are upper Hessenberg with one subdiagonal and one superdiagonal) are a particularly important subclass, used extensively in spectral methods and iterative solvers.

\subsection{Diagonal matrix with one symmetric row/column}
\index{key}{Diagonal with 1 row symmetric}

We consider the inversion of matrices of the form
\begin{equation}
M=\begin{pmatrix}
a_0 & -b & -b & \cdots & -b \\
-b & a_1 & 0   & \cdots & 0 \\
-b & 0   & a_2 & \cdots & 0 \\
\vdots & \vdots & \ddots & \ddots & \vdots \\
-b & 0 & \cdots & 0 & a_n
\end{pmatrix},
\label{eq:matriximp}
\end{equation}
i.e. a diagonal matrix perturbed by a single symmetric row and column with constant entries $-b$.  

\paragraph{Determinant.}  
Using the matrix determinant lemma (or direct cofactor expansion), one finds
\begin{equation}
\det(M) = \left( a_0 - b^2 \sum_{k=1}^n \frac{1}{a_k} \right) \prod_{k=1}^n a_k.
\end{equation}
This shows that $M$ is invertible whenever all $a_k \neq 0$ and the correction term in parentheses is nonzero.  

\paragraph{Cofactor structure.}  
By definition,
\[
(M^{-1})_{ij} = \frac{1}{\det(M)} C_{ji},
\]
where $C_{ij}=(-1)^{i+j}\det(M^{\tilde i \tilde j})$ is the cofactor of $M$.  
Due to the structure of $M$, the cofactor matrix $C$ is itself symmetric. In particular:
\begin{itemize}
\item For $i=j>0$,
\begin{equation}
C_{ii} = \left(\prod_{\substack{k=0 \\ k \neq i}}^n a_k \right) - b^2 \sum_{\substack{k=1 \\ k\neq i}}^n \prod_{\substack{j \neq k,i \\ j>0}} a_j.
\end{equation}
\item For $i=0, j>0$ (and symmetrically $j=0, i>0$),
\begin{equation}
C_{0j} = C_{j0} = -b^2 \prod_{\substack{k=1 \\ k \neq j}}^n a_k.
\end{equation}
\item For $i\neq j$, $i,j>0$,
\begin{equation}
C_{ij} = -b^2 \prod_{\substack{k=1 \\ k \neq i,j}}^n a_k.
\end{equation}
\end{itemize}

\paragraph{Block inversion.}  
Collecting these results, one may express the action of $M^{-1}$ on a vector $\vec S = (S_0,S_1,\dots,S_n)^\top$. If $S_0=0$, the reduced system for the components $(i_1,\dots,i_n)^\top$ reads
\begin{equation}
\begin{pmatrix}
i_1 \\ \vdots \\ i_n
\end{pmatrix}
=
\frac{1}{\det(M)}
\begin{pmatrix}
q_1 & b^2 c_{12} & b^2 c_{13} & \cdots & b^2 c_{1n} \\
b^2 c_{12} & q_2 & b^2 c_{23} & \cdots & \vdots \\
b^2 c_{13} & b^2 c_{23} & \ddots & \ddots & \vdots \\
\vdots & \cdots & \ddots & q_{n-1} & b^2 c_{\,n-1,n} \\
b^2 c_{1n} & \cdots & \cdots & b^2 c_{\,n-1,n} & q_n
\end{pmatrix}
\begin{pmatrix}
S_1 \\ \vdots \\ S_n
\end{pmatrix},
\end{equation}
with coefficients
\begin{align}
c_{ij} &= \prod_{\substack{k=1 \\ k\neq i,j}}^n a_k, \\
q_i &= a_0 \prod_{\substack{k=1 \\ k\neq i}}^n a_k - b^2 \sum_{\substack{k=1 \\ k\neq i}}^n \prod_{\substack{j\neq k,i}}^n a_j.
\end{align}

Matrices of the type \eqref{eq:matriximp} arise naturally in problems where a diagonal system is coupled symmetrically to a single degree of freedom (for example, in impedance network reductions or star–mesh transformations in circuit theory). Their inversion is thus of practical relevance in applied linear algebra and physics.

\subsection{Idempotent matrices and Projection operators}
\index{key}{Idempotent matrix}
\index{key}{Projector matrix}

A matrix $\Omega$ is idempotent if $\Omega^2 = \Omega$. These are precisely the projection operators . Their eigenvalues lie in $\{0,1\}$, with $\operatorname{rank}(\Omega)=\operatorname{Tr}(\Omega)$.  
If $\Omega$ is Hermitian, it is an \emph{orthogonal projector}; otherwise it is an \emph{oblique projector}.\footnote{See R.~Horn and C.~Johnson, \emph{Matrix Analysis}, Cambridge University Press (1985).}

\paragraph{Range and kernel.}  
A projection $\Omega$ naturally splits the vector space into two complementary subspaces:  
\begin{equation}
\mathbb{C}^n = \mathrm{Im}(\Omega) \oplus \ker(\Omega),
\end{equation}
where $\mathrm{Im}(\Omega) = \{ \Omega v : v\in \mathbb{C}^n\}$ is the image (the subspace onto which $\Omega$ projects), and $\ker(\Omega) = \{v : \Omega v = 0\}$ is the kernel (the orthogonal complement in the symmetric case). Every vector $x$ can thus be written uniquely as
\[
\vec x = \Omega \vec x + (I-\Omega)\vec x,
\]
where $\Omega \vec x \in \mathrm{Im}(\Omega)$ and $(I-\Omega)\vec x \in \ker(\Omega)$.  

\paragraph{Orthogonal vs. oblique projectors.}  
A projection $\Omega$ is said to be:
\begin{itemize}
\item \emph{Orthogonal} if $\Omega = \Omega^\ast$ (i.e. Hermitian). In this case, $\mathrm{Im}(\Omega)$ and $\ker(\Omega)$ are orthogonal subspaces. Orthogonal projectors minimize distance: for any $y \in \mathrm{Im}(\Omega)$, 
\[
\|\;\vec x - \Omega \vec x\;\|_2 = \min_{\vec y \in \mathrm{Im}(\Omega)} \|\vec x-\vec y\|_2.
\]
\item \emph{Oblique} if $\Omega \neq \Omega^\ast$. In this case, the decomposition $\mathbb{C}^n = \mathrm{Im}(\Omega) \oplus \ker(\Omega)$ still holds, but the two subspaces are not orthogonal. Oblique projections are useful in constrained optimization and generalized inverses.\footnote{See P.-A.~Absil and A.~Krishna, ``Oblique projectors and their applications,'' \emph{Linear Algebra Appl.} 435: 2021–2039 (2011).}
\end{itemize}

\paragraph{Eigenspectrum.}  
Since $\Omega^2=\Omega$, every eigenvalue $\lambda$ must satisfy $\lambda^2 = \lambda$, hence
\[
\Lambda(\Omega) \subset \{0,1\}.
\]
The algebraic multiplicity of $1$ equals $\mathrm{rank}(\Omega) = \dim(\mathrm{Im}(\Omega))$, while the multiplicity of $0$ equals $\dim(\ker(\Omega))$. Thus the spectrum of a projector directly encodes the dimensions of the range and kernel.  

\paragraph{Matrix representations.}  
Any orthogonal projection can be written in the form
\begin{equation}
\Omega = A(A^\ast A)^{-1}A^\ast,
\end{equation}
where $A$ is any full-column-rank matrix with $\mathrm{Im}(A)=\mathrm{Im}(\Omega)$.\footnote{See R.~Horn and C.~Johnson, \emph{Matrix Analysis}, Cambridge University Press (1985).}  
Non-orthogonal projectors admit similar factorizations but with a second matrix specifying the direction of projection.  

\paragraph{Pseudo-inverse.}  
Projection operators cannot be inverted unless they are trivial ($\Omega=0$ or $\Omega=I$). However, they do possess a simple Moore–Penrose pseudoinverse\index{key}{pseudo-inverse}.  
For an orthogonal projector $\Omega$, the pseudoinverse is itself:
\begin{equation}
\Omega^+ = \Omega.
\end{equation}
Indeed, the Penrose conditions are satisfied:
\[
\Omega \Omega^+ \Omega = \Omega, \quad \Omega^+ \Omega \Omega^+ = \Omega^+, \quad (\Omega \Omega^+)^\ast = \Omega \Omega^+, \quad (\Omega^+ \Omega)^\ast = \Omega^+ \Omega.
\]

\paragraph{Solution of projected systems.}  
Consider the equation $\Omega x = b$. This has solutions iff $b \in \mathrm{Im}(\Omega)$. A general solution is
\begin{equation}
\vec x = \Omega \vec b + (I-\Omega) \vec y, \qquad\vec  y \in \mathbb{C}^n \text{ arbitrary}.
\end{equation}
That is, the solution is determined up to addition of a vector in $\ker(\Omega)$.  

Projectors thus provide a natural language for decomposing vector spaces into complementary subspaces, and they appear throughout applied mathematics: in numerical least-squares problems (normal equations are solved by projection), in statistics (hat matrices in regression), and in physics (constraint operators and observables in quantum mechanics).\footnote{See C.~Meyer, \emph{Matrix Analysis and Applied Linear Algebra}, SIAM (2000).}
\section{Identities}
\label{sec:identities}

This section collects a number of useful matrix identities for determinants, inverses, and eigenvalues.  
Many of these can be found in compact form in the \emph{Matrix Cookbook}\footnote{K.~B. Petersen and M.~S. Pedersen, \emph{The Matrix Cookbook}, available at \texttt{http://matrixcookbook.com}.}.

\subsection{General recap of useful identities we encountered}

\paragraph{Jacobi’s formula.}  
If $A=A(\alpha)$ depends smoothly on a parameter $\alpha$,
\begin{equation}
\frac{\partial}{\partial \alpha}\det(A) = \operatorname{Tr}\!\bigl(\operatorname{adj}(A)\, \tfrac{\partial A}{\partial \alpha}\bigr)
= \det(A)\,\operatorname{Tr}\!\bigl(A^{-1} \tfrac{\partial A}{\partial \alpha}\bigr).
\end{equation}
This identity connects determinants and traces later, and underlies log–det
convexity results.

\paragraph{Exponential trace identity.}
For any square matrix $B$,
\begin{equation}
\det(e^B) = e^{\text{Tr}(B)}.
\end{equation}
This follows by applying Jacobi’s formula to $A(t)=e^{tB}$.

\paragraph{Resolvent identities.}
For $A\in\mathbb{C}^{n\times n}$, the resolvent is $R(A,z)=(zI-A)^{-1}$.  
Two classical identities are:
\begin{align}
R(A,z_1)-R(A,z_2) &= (z_2-z_1)\,R(A,z_1)R(A,z_2), \\
R(A+C,z)-R(A,z) &= R(A+C,z)\,C\,R(A,z).
\end{align}
These play a central role in perturbation theory and spectral analysis.

\paragraph{Woodbury and Sherman–Morrison.}
For invertible $A$ and conformable $U,V,B$,
\begin{align}
(A+UBV)^{-1} &= A^{-1} - A^{-1}U\bigl(B^{-1}+VA^{-1}U\bigr)^{-1}VA^{-1}, \\
(A+uv^\top)^{-1} &= A^{-1} - \frac{A^{-1}uv^\top A^{-1}}{1+v^\top A^{-1}u}.
\end{align}
These identities are invaluable for rank-$k$ updates.

\paragraph{Condition number.}
For any invertible $A$, the spectral condition number is
\begin{equation}
\kappa(A) = \|A\|\cdot \|A^{-1}\| = \frac{\sigma_{\max}(A)}{\sigma_{\min}(A)},
\end{equation}
where $\sigma_{\max},\sigma_{\min}$ are the extreme singular values.  
This measures sensitivity of solutions of $Ax=b$ to perturbations.

\paragraph{Neumann series.}
If $\|A\|<1$, then
\begin{equation}
(I-A)^{-1} = \sum_{k=0}^\infty A^k.
\end{equation}
This gives an explicit expansion for approximate inverses.

\paragraph{Cayley–Hamilton theorem.}
Every square matrix satisfies its own characteristic polynomial:
\begin{equation}
p_A(A)=\det(\lambda I-A)|_{\lambda=A}=0.
\end{equation}
This allows one to express high powers of $A$ as linear combinations
of lower ones.

\subsection{Determinants}

\paragraph{Matrix determinant lemma (Sylvester’s theorem).}  
For $A \in \mathbb{C}^{n\times n}$ invertible and $u,v \in \mathbb{C}^n$,
\begin{equation}
\det(A + u v^\top) = \bigl(1 + v^\top A^{-1} u\bigr)\,\det(A)
= \det(A) + v^\top \operatorname{adj}(A)\, u,
\end{equation}
where $\operatorname{adj}(A) = \det(A)\,A^{-1}$ is the adjugate matrix.

\paragraph{Matrix derivatives.}  
\begin{align}
\frac{\partial}{\partial X} \det(X) &= \det(X)\,(X^{-1})^\top, \\
\frac{\partial}{\partial X} \det(A X B) &= \det(A X B)\,(X^{-1})^\top, \\
\frac{\partial}{\partial x} \det(Y) &= \det(Y)\,\operatorname{Tr}\!\Bigl(Y^{-1}\frac{\partial Y}{\partial x}\Bigr).
\end{align}

\paragraph{Quadratic forms in determinants.}  
If $X$ is square and invertible, then
\begin{equation}
\frac{\partial}{\partial X}\det(X^\top A X) = 2\,\det(X^\top A X)\,(X^{-1})^\top.
\end{equation}
If $X$ is square and $A$ is symmetric,
\begin{equation}
\frac{\partial}{\partial X}\det(X^\top A X) = 2\,\det(X^\top A X)\,A X\,(X^\top A X)^{-1}.
\end{equation}
If $X$ is rectangular and $A$ not necessarily symmetric,
\begin{equation}
\frac{\partial}{\partial X}\det(X^\top A X) =
\det(X^\top A X)\Bigl( A X (X^\top A X)^{-1} + A^\top X (X^\top A^\top X)^{-1}\Bigr).
\end{equation}

\paragraph{Perturbative expansion.}  
For small $\epsilon$,
\begin{equation}
\det(I + \epsilon A) = 1 + \epsilon\,\operatorname{Tr}(A)
+ \tfrac{1}{2}\epsilon^2\bigl( \operatorname{Tr}(A)^2 - \operatorname{Tr}(A^2)\bigr) + O(\epsilon^3).
\end{equation}

\subsection{Inverses}

\paragraph{Derivatives of the inverse.}
\begin{align}
\frac{\partial}{\partial x} Y^{-1} &= -\,Y^{-1}\Bigl(\frac{\partial Y}{\partial x}\Bigr)Y^{-1}, \\
\frac{\partial}{\partial X_{ij}}(X^{-1})_{kl} &= - (X^{-1})_{ki}(X^{-1})_{jl}.
\end{align}

\paragraph{Composed functions.}
\begin{align}
\frac{\partial}{\partial X}\det(A X^{-1} B) &= -\,(X^{-1} B A X^{-1})^\top, \\
\frac{\partial}{\partial X}\operatorname{Tr}\bigl((X+A)^{-1}\bigr) &= -\,(X+A)^{-\top}(X+A)^{-\top}.
\end{align}

\subsection{Eigenvalues}

Since
\[
\operatorname{Tr}(X) = \sum_{i=1}^n \lambda_i, \qquad \det(X) = \prod_{i=1}^n \lambda_i,
\]
where $\{\lambda_i\}$ are the eigenvalues of $X$, we have
\begin{align}
\frac{\partial}{\partial X}\sum_i \lambda_i(X) &= I, \\
\frac{\partial}{\partial X}\prod_i \lambda_i(X) &= \det(X)\,(X^{-1})^\top.
\end{align}

\subsection{Formulas for linear systems}
\index{key}{Cramer's rule}

Consider the linear system
\begin{equation}
A \vec{x} = \vec{b}, \qquad A \in \mathbb{C}^{n\times n}, \ \vec{b} \in \mathbb{C}^n.
\end{equation}
If $A$ is invertible, the solution is $\vec{x} = A^{-1}\vec{b}$.  
An explicit (though computationally impractical) formula for each component is given by \emph{Cramer’s rule}:
\begin{equation}
x_i = \frac{\det(A_i)}{\det(A)},
\end{equation}
where $A_i$ is the matrix obtained from $A$ by replacing its $i$-th column with $\vec{b}$.  

While rarely used in numerical practice (since determinant computation is unstable for large systems), Cramer’s rule is conceptually useful: it shows that each solution component is a rational function of the data. It also underpins sensitivity analysis and connections between linear systems and geometry.\footnote{See R.~Horn and C.~Johnson, \emph{Matrix Analysis}, Cambridge University Press (1985).}

\subsection{Sandwiched equations}
\index{key}{sandwiched equations}

A common situation in applied linear algebra is the evaluation of bilinear forms of the type
\begin{equation}
f(X) = \vec{a}^\top F(X)\, \vec{b},
\end{equation}
where $F$ is some matrix function, $X$ is a matrix variable, and $\vec{a},\vec{b}$ are fixed vectors.  
Such ``sandwiched" expressions appear in many contexts, e.g.:
\begin{itemize}
\item path-counting in graphs via $\vec{a}^\top A^k \vec{b}$,
\item Green’s functions or resolvents $\vec{a}^\top (zI-X)^{-1}\vec{b}$,
\item sensitivity analysis of quadratic forms.
\end{itemize}

The derivatives of these expressions with respect to $X$ follow standard rules.\footnote{See K.~B. Petersen and M.~S. Pedersen, \emph{The Matrix Cookbook}, \texttt{http://matrixcookbook.com}.}

\paragraph{Linear case.}
\begin{equation}
\frac{\partial}{\partial X}\,\vec{a}^\top X \vec{b} = \vec{a}\vec{b}^\top.
\end{equation}

\paragraph{Transpose case.}
\begin{equation}
\frac{\partial}{\partial X}\,\vec{a}^\top X^\top \vec{b} = \vec{b}\vec{a}^\top.
\end{equation}

\paragraph{Inverse case.}
\begin{equation}
\frac{\partial}{\partial X}\,\vec{a}^\top X^{-1}\vec{b} = -\,X^{-\top}\,\vec{a}\vec{b}^\top\,X^{-\top}.
\end{equation}

\medskip

These formulas show how ``sandwiched" bilinear forms can be differentiated efficiently, without explicitly expanding $F(X)$. They are particularly useful when $F(X)$ is the inverse or a resolvent, since these arise in optimization, control theory, and spectral graph analysis.


\section{Matrix Equalities and Inequalities}
\label{sec:matrix-equalities}

\subsection{Matrix equalities}
\index{key}{matrix equalities}

If $A$ and $B$ commute, then the binomial expansion holds:
\begin{equation}
(A+B)^n=\sum_{k=0}^n \binom{n}{k}A^{n-k} B^k.
\end{equation}

Some other useful equalities include
\begin{equation}
\frac{A^k-B^k}{2}(A-B)=\Biggl(\sum_{i=1}^k A^{k-i+1} B^i\Biggr)(A-B)^2,
\end{equation}
and
\begin{equation}
\left(\frac{A+B}{2}\right)^{k+1}=\frac{A^{k+1}+B^{k+1}}{2}-\frac{A^k-B^k}{2}\frac{A-B}{2}.
\end{equation}

For positive operators, these identities lead to \emph{Löwner-type inequalities}:
\begin{equation}
\left(\frac{A+B}{2}\right)^{k+1}\preceq \frac{A^{k+1}+B^{k+1}}{2}-\frac{A^k-B^k}{2}\frac{A-B}{2},
\end{equation}
where $\preceq$ denotes the Löwner partial order.\footnote{See R.~Bhatia, \emph{Matrix Analysis}, Springer (1997).}

\subsubsection{Inversion sums}
\index{key}{matrix sum inversion}

For invertible matrices $A$ and $B$, we have
\begin{align}
A+B &= A(A^{-1}+B^{-1})B \\
     &= B(A^{-1}+B^{-1})A.
\end{align}
From this, one obtains the identities
\begin{align}
A+I &= A(A^{-1}+I)=(A^{-1}+I)A, \\
A^{-1}+I &= A^{-1}(A+I)=(A+I)A^{-1}.
\end{align}

For the inverse,
\begin{align}
(A^{-1}+B^{-1})^{-1} &= A(A+B)^{-1} B \\
&= A - A(A+B)^{-1}A \\
&= B(B+A)^{-1}A \\
&= B - B(A+B)^{-1}B.
\end{align}

In particular,
\begin{equation}
(A^{-1}+I)^{-1} = A(A+I)^{-1} = (A+I)^{-1}A = I - (I+A)^{-1}.
\end{equation}

Other related formulas include
\begin{equation}
(A+B)^{-1}+(A-B)^{-1}=2(A-B A^{-1} B)^{-1},
\end{equation}
and
\begin{equation}
(A+B)^{-1}=(A-BA^{-1}B)^{-1}-(B-AB^{-1}A)^{-1}.
\end{equation}

As an application, if $C=A+iB$ with $A=A^\dagger$ Hermitian and $B=-B^\dagger$ skew-Hermitian, then
\begin{equation}
(A+iB)^{-1}=(A+BA^{-1}B)^{-1}-i(B+AB^{-1}A)^{-1}.
\end{equation}

Finally, if $\|A^{-1}B\|<1$, the Neumann series gives
\begin{equation}
(A+B)^{-1}=\sum_{k=0}^\infty (-1)^k (A^{-1}B)^k A^{-1}.
\end{equation}
This expansion is useful for perturbative inverses.\footnote{See K.~B. Petersen and M.~S. Pedersen, \emph{The Matrix Cookbook}, \texttt{http://matrixcookbook.com}.}

\subsubsection{Projection operators}
\index{key}{projection operator identities}

If $\Omega$ is an idempotent projection ($\Omega^2=\Omega$), then
\begin{align}
(I-\Omega)\Omega &= 0, \\
\Omega(I-\Omega) &= 0, \\
\Omega^\dagger &= \Omega \quad \text{(orthogonal case)}.
\end{align}
For the Moore–Penrose pseudoinverse,
\[
\Omega^+ = \Omega,
\]
so projectors are self-pseudoinverse. Moreover, $\operatorname{Tr}(\Omega)=\operatorname{rank}(\Omega)$, and the eigenvalues lie in $\{0,1\}$.\footnote{See R.~Horn and C.~Johnson, \emph{Matrix Analysis}, Cambridge University Press (1985).}

\subsubsection{Sherman–Morrison–Woodbury formula}
\index{key}{Sherman–Morrison–Woodbury formula}

For invertible $A\in\mathbb{C}^{n\times n}$, $C\in\mathbb{C}^{m\times m}$, and matrices $U\in\mathbb{C}^{n\times m}$, $V\in\mathbb{C}^{m\times n}$,
\begin{equation}
(A+UCV)^{-1}=A^{-1}-A^{-1}U \bigl(C^{-1}+VA^{-1}U\bigr)^{-1}VA^{-1}.
\end{equation}
This generalizes the Sherman–Morrison rank-one update and is widely used in statistics, optimization, and numerical analysis.\footnote{See G.~Golub and C.~Van Loan, \emph{Matrix Computations}, Johns Hopkins University Press (1996).}

\subsubsection{Jensen's inequality}
\index{key}{Jensen's inequality}

Let $\vec z:[a,b]\to \mathbb{R}^n$ be differentiable. Then, for any positive semidefinite matrix $R$,
\begin{equation}
\int_a^b \dot{\vec z}(\xi)^\top R\,\dot{\vec z}(\xi)\,d\xi 
\;\geq\; \frac{\bigl(\vec z(b)-\vec z(a)\bigr)^\top R\bigl(\vec z(b)-\vec z(a)\bigr)}{b-a}.
\end{equation}
This is a matrix-valued version of Jensen’s inequality.\footnote{See T. Ando, ``Concavity of certain maps on positive definite matrices and applications to Hadamard products,'' \emph{Linear Algebra Appl.} 26, 203–241 (1979).}

\subsubsection{Wirtinger's inequalities}
\index{key}{Wirtinger's inequalities}

Let $\vec z:[a,b]\to \mathbb{R}^n$ and $R\succ 0$.  

\textbf{Inequality 1.} If $\vec z(a)=\vec z(b)$,
\begin{equation}
\int_a^b \vec z(\xi)^\top R \vec z(\xi)\,d\xi \;\leq\; \frac{(b-a)^2}{4\pi^2}\int_a^b \dot{\vec z}(\xi)^\top R \dot{\vec z}(\xi)\,d\xi.
\end{equation}

\textbf{Inequality 2.} If $\vec z(a)=\vec z(b)=0$,
\begin{equation}
\int_a^b \vec z(\xi)^\top R \vec z(\xi)\,d\xi \;\leq\; \frac{(b-a)^2}{\pi^2}\int_a^b \dot{\vec z}(\xi)^\top R \dot{\vec z}(\xi)\,d\xi.
\end{equation}

\textbf{Inequality 3.} If $\vec z(a)=0$,
\begin{equation}
\int_a^b \vec z(\xi)^\top R \vec z(\xi)\,d\xi \;\leq\; \frac{4(b-a)^2}{\pi^2}\int_a^b \dot{\vec z}(\xi)^\top R \dot{\vec z}(\xi)\,d\xi.
\end{equation}

These inequalities are widely used in control theory and PDE analysis.\footnote{See E. Fridman, \emph{Introduction to Time-Delay Systems}, Birkhäuser (2014).}
\subsection{Determinantal equalities}
\index{key}{determinantal equalities}

We start with a block determinant identity. For conformable matrices,
\begin{equation}
\det\!\begin{pmatrix}
 A & C \\
 D & B
\end{pmatrix}
= \det(B)\,\det(A - C B^{-1} D),
\label{eq:deteq}
\end{equation}
where $B$ is assumed invertible. This is the \emph{Schur complement formula}, fundamental in block matrix analysis.\footnote{See R.~Horn and C.~Johnson, \emph{Matrix Analysis}, Cambridge University Press (1985).}

\paragraph{Sylvester’s determinant theorem.}  
For $A \in \mathbb{C}^{m\times n}$ and $B \in \mathbb{C}^{n\times m}$,
\begin{equation}
\det(I_m + AB) = \det(I_n + BA).
\end{equation}
This identity (also known as Sylvester’s theorem) is remarkable because it relates two determinants of different dimension.\footnote{See K.~B. Petersen and M.~S. Pedersen, \emph{The Matrix Cookbook}, Sec.~9.1.}

\paragraph{Rank-one updates.}  
If $u,v \in \mathbb{C}^n$, then
\begin{equation}
\det(k I + u v^\top) = k^n + k^{n-1}\, v^\top u.
\end{equation}
This is a special case of the matrix determinant lemma.

\paragraph{Transpose identities.}  
For conformable $A,B$,
\begin{equation}
\det(I + A^\top B) = \det(I + AB^\top) = \det(I + B^\top A).
\end{equation}

\paragraph{Determinant products.}  
For invertible $A,B$,
\begin{equation}
\det(A+B)\,\det(A-B) = \det(B)\,\det(AB^{-1}A - B).
\end{equation}

\paragraph{Log–trace relation.}  
Since $\det(A) = \prod_i \lambda_i$ and $\operatorname{Tr}(A) = \sum_i \lambda_i$, one has in general
\begin{equation}
\log\det(A) = \operatorname{Tr}(\log A),
\end{equation}
where $\log A$ is the matrix logarithm defined by functional calculus.  
This equality underlies the notion of the \emph{Fredholm determinant}, defined as
\begin{equation}
\det(I + A) = \exp\!\bigl(\operatorname{Tr}\log(I+A)\bigr),
\end{equation}
which extends the determinant concept to certain classes of infinite-dimensional operators.\footnote{See R.~Bhatia, \emph{Matrix Analysis}, Springer (1997).}

Expanding the logarithm gives the series
\begin{equation}
\log\det(I + zA) = \operatorname{Tr}\log(I+zA) = \sum_{k=1}^\infty \frac{(-1)^{k+1}}{k}\, \operatorname{Tr}(A^k) z^k,
\end{equation}
valid for sufficiently small $z$.

\paragraph{Perturbative expansion.}  
For small $\epsilon$,
\begin{equation}
\det(I + \epsilon A) = 1 + \epsilon \operatorname{Tr}(A) + \tfrac{1}{2}\epsilon^2\bigl(\operatorname{Tr}(A)^2 - \operatorname{Tr}(A^2)\bigr) + O(\epsilon^3).
\end{equation}

\paragraph{Determinant–trace connection.}  
Determinant identities can be used to derive trace identities. For example, from
\begin{equation}
\det(A+BC+\epsilon T) = \det(A+\epsilon T)\,\det\!\bigl(I + C(A+\epsilon T)^{-1}B\bigr),
\end{equation}
and expanding to first order in $\epsilon$, one obtains
\begin{equation}
\operatorname{Tr}\bigl((A+BC)^{-1}T\bigr)
= \operatorname{Tr}(A^{-1}T) - \operatorname{Tr}\!\Bigl((I+CA^{-1}B)^{-1} C A^{-1} T A^{-1} B\Bigr).
\end{equation}

\subsection{Trace inequalities}
\index{key}{trace inequalities}

We now recall some fundamental inequalities involving the trace.

\paragraph{Functions of Hermitian matrices.}  
If $A$ is Hermitian with spectral decomposition $A=\sum_j \lambda_j P_j$, then for a scalar function $f$,
\begin{equation}
\operatorname{Tr}(f(A)) = \sum_j f(\lambda_j).
\end{equation}
If $f$ is monotone and convex, this yields powerful operator inequalities. For instance, if $A\succeq B$ then $f(A)\succeq f(B)$ for $f$ operator-monotone.\footnote{See R.~Bhatia, \emph{Matrix Analysis}, Springer (1997).}

\paragraph{Simple trace inequalities.}  
If $A,B\succeq 0$, then for any positive integer $m$,
\begin{equation}
\operatorname{Tr}\bigl((AB)^m\bigr) \leq \bigl(\operatorname{Tr}(AB)\bigr)^m.
\end{equation}
More generally, for $m > s > 0$,
\begin{equation}
\operatorname{Tr}\bigl((AB)^m\bigr) \leq \Bigl(\operatorname{Tr}\bigl((AB)^s\bigr)\Bigr)^{m/s}.
\end{equation}
These inequalities follow from Hölder-type inequalities for traces.\footnote{See F.~Zhang, \emph{Matrix Theory: Basic Results and Techniques}, Springer (2011).}
\subsubsection{Gibbs’ inequality}
\index{key}{Gibbs' inequality}

Let $A$ be Hermitian and $\rho \succeq 0$ a density matrix ($\operatorname{Tr}\rho=1$). Then
\begin{equation}
\operatorname{Tr}(\rho A) + \operatorname{Tr}(\rho \log \rho) \;\geq\; -\log \operatorname{Tr}(e^{-A}),
\end{equation}
with equality if and only if
\[
\rho = \frac{e^{-A}}{\operatorname{Tr}(e^{-A})}.
\]
This inequality is fundamental in statistical mechanics and quantum information theory.\footnote{See M.~Nielsen and I.~Chuang, \emph{Quantum Computation and Quantum Information}, Cambridge University Press (2000).}

\subsubsection{Peierls–Bogoliubov inequality}
\index{key}{Peierls-Bogoliubov inequality}

If $R,F$ are Hermitian with $\operatorname{Tr}(e^R)=1$, and if $f=\operatorname{Tr}(F e^R)$, then
\begin{equation}
\operatorname{Tr}(e^F e^R) \;\geq\; \operatorname{Tr}(e^{F+R}) \;\geq\; e^f.
\end{equation}
This inequality provides bounds on the free energy in statistical mechanics.\footnote{See R.~Bhatia, \emph{Matrix Analysis}, Springer (1997).}

\subsubsection{Golden–Thompson and Lieb inequalities}
\index{key}{Golden-Thompson inequality}
\index{key}{Lieb inequality}

If $A,B$ are Hermitian,
\begin{equation}
\operatorname{Tr}(e^{A+B}) \;\leq\; \operatorname{Tr}(e^A e^B),
\end{equation}
the celebrated \emph{Golden–Thompson inequality}.  

Lieb proved powerful extensions. For positive operators $A,B,C$,
\begin{equation}
\operatorname{Tr}\!\left(e^{\log A - \log B + \log C}\right) \leq \int_0^\infty \operatorname{Tr}\!\bigl(A(\xi I+B)^{-1}C(\xi I+B)^{-1}\bigr)\,d\xi.
\end{equation}
These inequalities underlie monotonicity of quantum relative entropy.\footnote{See E.~H.~Lieb, ``Convex trace functions and the Wigner–Yanase–Dyson conjecture,'' \emph{Advances in Math.} 11, 267–288 (1973).}

\subsubsection{Klein’s inequality}
\index{key}{Klein inequality}

If $f$ is differentiable and convex, and $A,B$ are Hermitian,
\begin{equation}
\operatorname{Tr}\bigl(f(A)-f(B)-(A-B)f^\prime(B)\bigr) \;\geq\; 0.
\end{equation}
This generalizes scalar convexity to the operator setting.\footnote{See R.~Bhatia, \emph{Matrix Analysis}, Springer (1997).}

\subsubsection{Araki–Lieb–Thirring inequalities}
\index{key}{Lieb-Thirring inequalities}
\index{key}{Araki inequalities}

The Lieb–Thirring inequality generalizes Bellman’s inequality. For positive $A,B$,
\begin{equation}
\operatorname{Tr}\bigl((AB)^m\bigr) \leq \operatorname{Tr}(A^m B^m).
\end{equation}
Lieb and Thirring also proved that for $r\geq 1$,
\begin{equation}
\operatorname{Tr}\bigl((B^{1/2} A B^{1/2})^r\bigr) \leq \operatorname{Tr}(B^r A^r B^r).
\end{equation}
Araki extended this to the following family of inequalities: for $r\geq 1$ and $q\geq 0$,
\begin{equation}
\operatorname{Tr}\bigl((B^{1/2} A B^{1/2})^{rq}\bigr) \leq \operatorname{Tr}\bigl((B^{r/2} A^r B^{r/2})^q\bigr).
\end{equation}
If $0 \leq r \leq 1$, the inequality is reversed:\footnote{See H.~Araki, ``On an inequality of Lieb and Thirring,'' \emph{Letters in Math. Phys.} 19, 167–170 (1990).}
\begin{equation}
\operatorname{Tr}\bigl((B^{r/2} A^r B^{r/2})^q\bigr) \leq \operatorname{Tr}\bigl((B^{1/2} A B^{1/2})^{rq}\bigr).
\end{equation}

\subsubsection{Schur–Horn theorem}
\index{key}{Schur-Horn theorem}

For any Hermitian $D \in \mathbb{C}^{m\times m}$ with diagonal entries $d_i$ and eigenvalues $\lambda_i$ in non-increasing order,
\begin{equation}
\sum_{i=1}^n d_i \;\leq\; \sum_{i=1}^n \lambda_i, \qquad \forall n,
\end{equation}
and
\begin{equation}
\sum_{i=1}^m d_i = \sum_{i=1}^m \lambda_i.
\end{equation}
This is the classical \emph{Schur–Horn theorem}, relating diagonals to spectra.\footnote{See R.~Horn and C.~Johnson, \emph{Matrix Analysis}, Cambridge University Press (1985).}

\paragraph{von Neumann’s trace inequality.}  
If $A,B$ are Hermitian with singular values $\{\alpha_i\}$, $\{\beta_i\}$ sorted in non-increasing order, then
\begin{equation}
|\operatorname{Tr}(AB)| \;\leq\; \sum_{i=1}^n \alpha_i \beta_i.
\end{equation}
This inequality plays a central role in unitarily invariant norms.\footnote{See R.~Bhatia, \emph{Matrix Analysis}, Springer (1997).}

\subsection{Traces and determinants}
\index{key}{trace and determinant relations}

For any square matrix $A$,
\begin{equation}
\det(e^A) = e^{\operatorname{Tr}(A)}.
\end{equation}
Equivalently, if $A=e^L$, then
\begin{equation}
\log\det(A) = \operatorname{Tr}(L).
\end{equation}

\paragraph{AM–GM inequality.}  
For positive real numbers $x_1,\dots,x_n$,
\[
\frac{1}{n}\sum_{i=1}^n x_i \;\geq\; \Bigl(\prod_{i=1}^n x_i\Bigr)^{1/n}.
\]
Applied to the eigenvalues of a positive definite matrix $A$, this gives
\begin{equation}
\frac{\operatorname{Tr}(A)}{n} \;\geq\; \det(A)^{1/n}.
\end{equation}

\paragraph{Mixed determinant–trace inequalities.}  
If $A,B\succeq 0$, then
\begin{equation}
n\,( \det(A)\det(B))^{m/n} \;\leq\; \operatorname{Tr}(A^m B^m).
\end{equation}
As a corollary, if $\det(B)=1$, then
\begin{equation}
n\,\det(A)^{1/n} \;\leq\; \operatorname{Tr}(AB).
\end{equation}

\paragraph{Bounds involving $\log\det$.}  
For $A\succeq 0$,
\begin{equation}
\operatorname{Tr}(I-A^{-1}) \;\leq\; \log\det(A) \;\leq\; \operatorname{Tr}(A-I),
\end{equation}
with equality at $A=I$.

A chain of inequalities also holds:
\begin{equation}
\frac{n}{\operatorname{Tr}(A^{-1})} \;\leq\; \det(A)^{1/n} \;\leq\; \frac{1}{n}\operatorname{Tr}(A) \;\leq\; \sqrt{\tfrac{1}{n}\operatorname{Tr}(A^2)}.
\end{equation}
Here the central inequality is the matrix AM–GM inequality, while the last step is by Cauchy–Schwarz.\footnote{See R.~Bhatia, \emph{Positive Definite Matrices}, Princeton University Press (2007).}
\subsubsection{Determinantal inequalities}
\index{key}{determinantal inequalities}

For any $A\in\mathbb{C}^{n\times n}$,
\begin{equation}
\label{eq:hadamard-col}
|\det(A)|^2 \;\le\; \prod_{j=1}^n \sum_{i=1}^n |a_{ij}|^2,
\end{equation}
i.e. $|\det(A)|\le \prod_{j=1}^n \|\vec a_j\|_2$, where $\vec a_j$ are the columns of $A$ (Hadamard’s bound).\footnote{See R.~Horn and C.~Johnson, \emph{Matrix Analysis}, Cambridge University Press (1985).}  
If $A\succeq 0$ is Hermitian positive semidefinite, then the \emph{Hadamard inequality} gives
\begin{equation}
\label{eq:hadamard-diag}
\det(A) \;\le\; \prod_{j=1}^n a_{jj},
\end{equation}
with equality iff $A$ is diagonal in the standard basis.\footnote{Horn--Johnson (1985).}  
As a corollary of \eqref{eq:hadamard-col}, if $|a_{ij}|\le B$ then
\begin{equation}
\label{eq:hadamard-B}
|\det(A)| \;\le\; B^n\, n^{\,n/2}.
\end{equation}

\medskip

\paragraph{Minkowski determinant theorem and log-concavity.}
For $A,B\succ 0$ (Hermitian positive definite),
\begin{equation}
\label{eq:minkowski}
\det(A+B)^{1/n} \;\ge\; \det(A)^{1/n} + \det(B)^{1/n}.
\end{equation}
Equivalently, the map $A\mapsto \det(A)^{1/n}$ is \emph{concave} on the cone of positive definite matrices.\footnote{See R.~Bhatia, \emph{Positive Definite Matrices}, Princeton (2007).}  
Moreover, $\log\det$ is concave:
\begin{equation}
\label{eq:logdet-concave}
\log\det(\theta A + (1-\theta)B) \;\ge\; \theta \log\det A + (1-\theta)\log\det B,
\qquad 0\le \theta \le 1,
\end{equation}
which implies the \emph{geometric mean} inequality
\begin{equation}
\label{eq:fan}
\det(\theta A+(1-\theta)B) \;\ge\; \det(A)^{\theta}\,\det(B)^{1-\theta}.
\end{equation}
\index{key}{Fan determinantal inequality}\footnote{A convenient compendium is Petersen--Pedersen, \emph{The Matrix Cookbook}, Sec.~9.}

\medskip

\paragraph{Block determinants.}
If 
\(
A=\begin{pmatrix} A_1 & B \\ B^\ast & A_2 \end{pmatrix}\succeq 0
\)
with $A_1,A_2\succ 0$, then
\begin{equation}
\label{eq:block-det}
\det(A) \;\le\; \det(A_1)\det(A_2),
\end{equation}
with equality iff $B=0$.\footnote{By Schur complements; see Horn--Johnson (1985).}  
More generally, if $B$ is square (no definiteness on $A$),
\begin{equation}
\label{eq:BBast}
\det(BB^\ast) \;\le\; \det(A_1)\det(A_2).
\end{equation}

\subsubsection{Determinant ratios (Ky Fan type)}
\index{key}{Ky Fan inequalities}

Let $H,K\succ 0$ be $n\times n$ and partitioned as
\(
qH=\begin{pmatrix}H_{11}&H_{12}\\ H_{21}&H_{22}\end{pmatrix},
\;
K=\begin{pmatrix}K_{11}&K_{12}\\ K_{21}&K_{22}\end{pmatrix}
\)
with $H_{11},K_{11}\in\mathbb{C}^{k\times k}$. Then
\begin{equation}
\label{eq:fan-ratio}
\left(\frac{\det(H+K)}{\det(H_{11}+K_{11})}\right)^{\!1/(n-k)}
\;\ge\;
\left(\frac{\det H}{\det H_{11}}\right)^{\!1/(n-k)}
+
\left(\frac{\det K}{\det K_{11}}\right)^{\!1/(n-k)}.
\end{equation}
This is a determinantal analogue of Minkowski for Schur complements.\footnote{See F.~Zhang, \emph{Matrix Theory}, Springer (2011), and references therein to Ky Fan.}

\medskip

\paragraph{Products/means of principal minors.}
Let $P_k$ be the product of all principal minors of order $k$ of $A$. Then (Szasz/Maclaurin–Newton)
\begin{equation}
\label{eq:szasz}
P_1 \;\ge\; P_2^{\,a_2} \;\ge\; \cdots \;\ge\; P_n,
\qquad a_k=\binom{n-1}{k-1}^{-1}.
\end{equation}
Equivalently, the sequence
\begin{equation}
\label{eq:Sk}
S_k^{(n)} \;=\; \frac{1}{\binom{n}{k}}
\sum_{|I|=k} \det(A[I])^{1/k}
\end{equation}
is decreasing:
\begin{equation}
\frac{\text{Tr}(A)}{n} \;=\; S_1^{(n)} \;\ge\; S_2^{(n)} \;\ge\; \cdots \;\ge\; S_n^{(n)} \;=\; \det(A)^{1/n}.
\end{equation}
\footnote{See Bhatia (1997), Ch.~III; also Petersen--Pedersen, \emph{Matrix Cookbook}.}

\medskip

\paragraph{Toeplitz determinants (brief).}
For Hermitian Toeplitz matrices $T_n(\varphi)$ generated by a nonnegative integrable symbol $\varphi$ on the unit circle, Szegő’s limit theorem gives
\begin{equation}
\lim_{n\to\infty} \det(T_n(\varphi))^{1/n} \;=\; 
\exp\!\left(\frac{1}{2\pi}\int_0^{2\pi}\log \varphi(e^{i\theta})\,d\theta\right).
\end{equation}
This underpins monotonicity patterns of principal Toeplitz minors in many classical cases.\footnote{See B.~Simon, \emph{Szegő’s Theorem and Its Descendants}, Princeton (2011).}

\subsection{Integral representations of the determinant}
\index{key}{Gaussian integral for determinant}

For $A\succ 0$,
\begin{equation}
\int_{\mathbb{R}^n} e^{-\xi^\top A\,\xi}\,d\xi \;=\; \pi^{n/2}\,\det(A)^{-1/2},
\end{equation}
and more generally (with source $J$),
\begin{equation}
\int_{\mathbb{R}^n} e^{-\xi^\top A\,\xi + J^\top \xi}\,d\xi 
\;=\; \pi^{n/2}\,\det(A)^{-1/2}\, e^{\tfrac{1}{4}J^\top A^{-1}J}.
\end{equation}
Thus
\begin{equation}
\det(A) \;=\; \pi^n \left(\int_{\mathbb{R}^n} e^{-\xi^\top A\,\xi}\,d\xi\right)^{-2}.
\end{equation}
(Analogous fermionic/Grassmann integrals represent $\det(A)$ directly.)\footnote{See standard Gaussian integral formulas; e.g. Horn--Johnson (1985) or many QFT texts.}

\paragraph{Differential identity and a path formula.}
If $A(\xi)$ is differentiable and nonsingular,
\begin{equation}
\label{eq:jacobi-logdet}
\frac{d}{d\xi}\log\det A(\xi) \;=\; \text{Tr}\!\big(A(\xi)^{-1}\, A'(\xi)\big).
\end{equation}
Let $B$ be any matrix and decompose $B=D+O$ with $D=\mathrm{Diag}(B)$ and $O=B-\mathrm{Diag}(B)$.  
Define $A(\xi)=D+\xi O$ so that $A(0)=D$ and $A(1)=B$. Integrating \eqref{eq:jacobi-logdet} from $0$ to $1$ yields
\begin{equation}
\det(B) \;=\; \exp\!\left(\int_0^1 \text{Tr}\!\big(A(\xi)^{-1} O\big)\,d\xi\right)\;\prod_{i=1}^n B_{ii}.
\end{equation}
This expresses $\det(B)$ as a diagonal product times an exponential ``interaction" correction.

\subsubsection{Operator monotonicity and convexity}
\index{key}{Operator monotone function}
\index{key}{Operator convex function}

Beyond Jensen-type inequalities, it is often useful to characterize classes of scalar
functions that preserve the L\"{o}wner order when applied to matrices.
Let $I \subset \mathbb{R}$ be an interval.

\paragraph{Operator monotone functions.}
A function $f:I\to\mathbb{R}$ is said to be \emph{operator monotone} if
\[
A \leq B \ \ \Rightarrow \ \ f(A) \leq f(B),
\]
for all Hermitian matrices $A,B$ with spectra contained in $I$.
For instance, $f(x)=\sqrt{x}$ is operator monotone on $[0,\infty)$,
as is $f(x)=\log x$ on $(0,\infty)$.  
By contrast, the function $f(x)=x^2$ is not operator monotone on all of $\mathbb{R}$,
although it is monotone on $[0,\infty)$ in the scalar sense.

\paragraph{Operator convex functions.}
A function $f:I\to\mathbb{R}$ is \emph{operator convex} if
\[
f(\theta A + (1-\theta)B) \ \leq\ \theta f(A) + (1-\theta) f(B),
\qquad 0\leq \theta \leq 1,
\]
for all Hermitian $A,B$ with spectra in $I$.
The prototypical examples are $f(x)=x^2$, which is operator convex on all of $\mathbb{R}$,
and $f(x)=-\log x$, which is operator convex on $(0,\infty)$.

\paragraph{Connections.}
These notions generalize classical scalar inequalities.
For example, Jensen’s operator inequality holds precisely because convex functions
in the operator sense preserve the ordering of averages.
The study of operator monotone and convex functions goes back to Löwner (1934),
and is now a central tool in matrix analysis and quantum information theory.\footnote{%
For background see R.~Bhatia, \emph{Matrix Analysis}, Springer (1997).}

\paragraph{Remarks.}
Note that:
\begin{itemize}
\item The set of operator monotone functions on $(0,\infty)$ is closed under
pointwise limits and under certain integral transforms (Pick–Nevanlinna functions).
\item Operator convexity implies scalar convexity, but not conversely.
\item Many important inequalities in matrix analysis, such as the Golden - Thompson
inequality and Lieb’s concavity theorem, rely on these classes of functions.
\end{itemize}

\subsection{Carleman linearization identities}
\index{key}{Carleman linearization}

Carleman linearization is a classical technique, introduced by Carleman in 1932, which embeds a nonlinear polynomial ODE into an infinite-dimensional linear system. By truncating this embedding, one obtains systematic finite-dimensional approximations of nonlinear dynamics. A modern treatment with explicit truncation error bounds is given in Forets and Pouly.\footnote{M.~Forets and A.~Pouly, ``Explicit Error Bounds for Carleman Linearization,'' \emph{arXiv preprint} arXiv:1711.02552 (2017).}

\paragraph{Setup.}
Consider a polynomial vector field
\begin{equation}
\dot{x}(t) = F_1 x(t) + F_2 x(t)^{[2]} + \cdots + F_k x(t)^{[k]},
\qquad x(0)=x_0 \in \mathbb{R}^n,
\end{equation}
where $F_j \in \mathbb{R}^{n\times n^j}$ and $x^{[j]}$ denotes the $j$-th Kronecker power,
\[
x^{[j]} = \underbrace{x \otimes x \otimes \cdots \otimes x}_{j \ \text{times}} \in \mathbb{R}^{n^j}.
\]

\paragraph{Kronecker identities.}
The Kronecker product $\otimes$ satisfies the crossnorm property
\[
\|A\otimes B\| = \|A\|\cdot \|B\|,
\]
and for vectors $\|x^{[j]}\| = \|x\|^j$.\footnote{Forets--Pouly (2017), Sec.~2.2.}  
Differentiating $x^{[j]}$ yields, by Leibniz’s rule,
\[
\frac{d}{dt}x^{[j]} = \sum_{\nu=1}^j I^{\otimes (\nu-1)}\otimes \dot{x}\otimes I^{\otimes (j-\nu)},
\]
so Kronecker powers evolve linearly once augmented with higher-order terms.

\paragraph{Infinite-dimensional embedding.}
Defining blocks $y_j = x^{[j]}$, one obtains the coupled linear system
\[
\dot{y}_i = \sum_{j=1}^k A_i^{\,i+j-1}\, y_{i+j-1}, \qquad i\in \mathbb{N},
\]
with structured transfer matrices $A_i^{\,i+j-1}$ built from the $F_j$. Concatenating the $y_j$ yields the infinite-dimensional linear system
\[
\dot{y} = A y,
\]
with $A$ block upper-triangular.\footnote{Forets--Pouly (2017), Sec.~3.1.}

\paragraph{Truncation.}
Truncating at order $N$ gives a finite linear system
\[
\dot{\hat{y}} = A_N \hat{y},
\]
where $\hat{y}=(y_1,\dots,y_N)$. This approximation converges to the true solution as $N\to\infty$ on bounded time intervals.\footnote{Forets--Pouly (2017), Sec.~3.2.}

\paragraph{Reduction to quadratic case.}
Any polynomial ODE can be rewritten as a quadratic system by introducing auxiliary Kronecker variables. Specifically,
\[
\dot{x} = F_1 x + F_2 x^{[2]} + \cdots + F_k x^{[k]}
\]
can be reformulated as
\[
\dot{\tilde{x}} = \tilde{F}_1 \tilde{x} + \tilde{F}_2 \tilde{x}^{[2]},
\]
with $\tilde{x}=(x,x^{[2]},\dots,x^{[k-1]})$.\footnote{Forets--Pouly (2017), Sec.~3.3.}

\paragraph{Error bounds.}
Explicit error bounds for the truncated Carleman embedding are available. One bound, requiring an a priori estimate of the solution norm $\alpha \ge \sup_{\tau\le t}\|x(\tau)\|$, is
\[
\|x(t)-\hat{x}(t)\| \leq \alpha^{N+1}\|F_2\|^N
\frac{(e^{\mu(F_1)t}-1)^N}{\mu(F_1)^N}.
\]
Another, depending only on the initial condition, is
\[
\|x(t)-\hat{x}(t)\| \leq \|x_0\| e^{\|F_1\|t}\,
\frac{(1+\beta_0) - \beta_0 e^{\|F_1\|t}}{\beta_0 (e^{\|F_1\|t}-1)^N},
\]
with $\beta_0=\frac{\|x_0\|\|F_2\|}{\|F_1\|}$.\footnote{Forets--Pouly (2017), Sec.~4.}

\paragraph{Combinatorial path sums.}
Higher-order derivatives of the embedded system can be expressed using \emph{path sums}, i.e.
\[
\frac{d^\nu}{dt^\nu} y_i = \sum_{j=0}^\nu C^{(\nu)}_{i,i+j}\, y_{i+j},
\]
where $C^{(\nu)}_{i,i+j}$ are combinatorial sums of products of the transfer matrices. This encodes all possible “paths” in the Kronecker tensor tower and provides a compact way to compute series coefficients.\footnote{Forets--Pouly (2017), Sec.~5.2.}

\clearpage
\addcontentsline{toc}{chapter}{General bibliography}

More citations are provided in footnotes throughout the book.
\addcontentsline{toc}{chapter}{Indices}
\printindex{key}{Index by keywords}

\end{document}